\documentclass[12pt,twoside]{mitthesis}
\usepackage{lgrind}
\usepackage{cmap}
\usepackage[T1]{fontenc}

\usepackage{amsmath}
\usepackage{amssymb}
\usepackage{blindtext}
\usepackage[sort&compress,numbers]{natbib}
\usepackage{doi}

\usepackage{siunitx}
\DeclareSIUnit\parsec{pc}
\DeclareSIUnit\electronvolt{e\kern-.05em V}

\usepackage{nicefrac}
\usepackage{graphicx}

\usepackage{multirow}
\usepackage{tabularx}
\usepackage{array}
\usepackage{pdflscape}

\usepackage{capt-of}
\usepackage{subfigure}

\usepackage{tikz}
\usepackage{tikz-feynman}
\usepackage{slashed}
 \usepackage{braket}

\usepackage{listings}
\usepackage{xcolor}

\usepackage{afterpage}
\usepackage{longtable}
\def\toprule{\hline\hline}%
\def\botrule{\hline\hline}%
\newcommand{\cL}{\mathcal{L}}

\begin{document}

%
%
%
%
%
%
%
%
%
%
%

\title{Dark Matter Energy Deposition and Production from the Table-Top to the Cosmos}

\author{Hongwan Liu}
\prevdegrees{B.A., Cornell University (2011)}
\department{Department of Physics}

\degree{Doctor of Philosophy}

\degreemonth{June}
\degreeyear{2019}
\thesisdate{May 16, 2019}


\supervisor{Tracy R. Slatyer}{Associate Professor of Physics}

\chairman{Nergis Mavalvala}{Associate Department Head}

\maketitle



\cleardoublepage
\setcounter{savepage}{\thepage}
\begin{abstractpage}
%
%
%

The discovery of nongravitational interactions between dark matter and the Standard Model would be an important step in unraveling the nature of dark matter. If such an interaction exists, it would have profound implications on how dark matter is produced in both the early universe and in collider experiments. In addition, it would also allow dark matter to deposit energy into Standard Model particles in unexpected ways. This thesis details some recent progress made in understanding these implications, including (i) a new freezeout mechanism for thermal dark matter dominated by a 3-to-2 process within a vector portal dark sector model; (ii) a study of how the existence of dark sector bound states can influence collider, direct and indirect searches for dark matter; (iii) a new axion dark matter interferometric search using a cavity that is sensitive to the axion-induced rotation of linearly polarized light; (iv) a definitive assessment of the potential contribution of dark matter annihilation and decay to cosmic reionization; (v) new constraints on dark matter annihilation rates and decay lifetimes from 21-cm cosmology, and (vi) a new numerical code, \texttt{DarkHistory}, which significantly improves the computation of the ionization and thermal histories of the universe in the presence of exotic sources of energy injection. These novel ideas span length scales ranging from table-top experiments to the entire cosmos, and represent just a few of the myriad of ways in which dark matter may yet surprise us.
\end{abstractpage}


\cleardoublepage

\section*{Acknowledgments}

Five years ago, when I was just starting to irreversibly commit to graduate school, it would have been impossible to imagine myself at this point. Certainly, it would have been impossible to picture myself contributing in any constructive way to the work presented in the next 380 or so pages of this thesis (to be honest, as I'm writing this, I'm still somewhat surprised). There are many people who have made all of this possible in one way or another. For many of you, words expressing my gratitude are entirely inadequate, but I'm going to try anyway. 

Having Tracy Slatyer as my Ph.D.\ advisor has been the greatest fortune of my academic life. In one of my first few meetings, she very patiently taught me the very basics of how dark matter redshifts. From that point of near zero knowledge, she gradually gave me more freedom to discover things for myself and to work independently. She has always been incredibly generous with her time, once spending a full hour listening in silence as I gave a rehearsal journal club talk; she knew that it was my first proper talk as a graduate student, and that it meant a lot to me at that point. Talking to her about research never fails to at least set you on the right direction, so much so that working with her gives you confidence that any problem can be solved. This is true even after a fatal flaw in our bound states paper was pointed out just a few hours before posting on the arXiv (solved after a month of learning a surprising amount about $Z$-toponium mixing), or after finding what seemed to be a hopelessly obscure bug creating a mysterious discrepancy between two supposedly identical calculations (resolved after five straight hours of debugging together). Tracy represents to me all that is right in academia, and I am honored every day by her support. I cannot express enough how thankful I am to her for teaching me not just about science, but about how science should be done.

Working with and learning from Jesse Thaler has also been a huge privilege. Jesse does physics with pure joy. His sheer creativity and also the burning clarity he has when he understands something --- and that's a lot of things! --- are an inspiration. I learnt, both directly from him and indirectly through his influence on everyone in the CTP, to hold my work to the highest possible standard, especially in clarity of thought, writing, making plots and giving talks. His advice on all things research has simply been invaluable.  

I also owe a big debt of gratitude to Jim Alexander, Julia Thom-Levy and Csaba Csaki. All of them helped me cobble together an unlikely plan to get me to graduate school in high-energy theory, and without their advice, I would probably not have realized that my path to this Ph.D. was possible. Jim in particular gave me my first experience in high-energy research, instiled in me a healthy respect for experimental work, and was incredibly supportive even long after I graduated from Cornell. I will never forget the kindness that all of you showed me.

To my other collaborators, you have all taught me so much about physics and more generally about how to be a good researcher, and the lessons I learnt from you form an important part of everything that I know about this field. I want to thank in particular Jim Cline, Yotam Soreq and Wei Xue for teaching me, among other things, the importance of parametric estimates, doing amplitude calculations with FeynCalc, how to write a referee response and much more, all while being kind, patient and incredibly supportive. I also want to thank Gregory Ridgway, without whom DarkHistory would never have been part of this thesis; the process would have been twice as long with much less than half the fun.

It's hard to imagine a better place than the CTP to be a graduate student in high-energy theory. The willingness of faculty like Tracy, Jesse and Iain Stewart to support graduate students is tremendous. The openness and friendliness shown by everyone here are absolutely essential to doing good science, and I hope to emulate it wherever I go. I want to thank Lina Necib and Ian Moult for being such great officemates in the first few years; Rebecca Leane for being an awesome conference companion; Nick Rodd, for being so great to talk to, and for being a senior graduate student to look up to; Patrick Fitzpatrick, who never fails to bring colour to everything; Gherardo Vita, Jasmine Brewer and Chih Liang Wu, for enduring with me the insanity that is Part 3, and finally Andrew Turner, for the fun we had puzzling through all of our classes together. I must also thank all of the physics and CTP administrators, especially Scott Morley, Charles Suggs, Sydney Miller and Cathy Modica, without whom things would very quickly fall apart.

Outside of the CTP, I've also had the pleasure of meeting some wonderful people who have made my time in graduate school a blast, including Pouya Asadi, Bi Ran, Laura Chang, Mariangela Lisanti, Matt Moschella, Aditya Parikh, Nadav Outmezguine, Diego Redigolo, Josh Ruderman, Oren Slone, Tan Tzer Han, Alex Tinguely, Yu-Dai Tsai, Harry Hengyun Zhou and many more. I'm thankful too for my long-time friends from Singapore, especially those who are/were here in Boston and made it feel so much more like home, including Jayce Cheng, Liang Kaicheng, Edwin Khoo, Apoorva Murarka and Kegon Teng Kok Tan. 

To my parents-in-law, Boon Chey and Thim Hee, thank you for welcoming me into your family, and for so generously supporting Liang Si and me over all these years. To my sister-in-law Liang Ying and her husband Rick, I am very thankful to have family in New York City, and cannot wait to meet the newest member of the family!

To my sister Lena and her husband Gary, as well as my wonderful niece Sophie and nephew Julian, you have all brought such joy into our lives, and seeing you happy makes me very happy too. My sister was a big part of my early years and beyond, and is a significant part of what made this possible. 

To my parents, Nam Shong and Swee Guat, what I owe to you is beyond measure and beyond words. This thesis could only have been written because you made me into who I am, and because of your unwavering support from the very beginning. I love you both very much. 

And finally, to my wife and best friend Liang Si, I dedicate this thesis to you. Thank you for your superhuman support of me, and for keeping me anchored with your love. Getting married to you is by far the best thing to have happened to me over the last five years, and whatever else is in our future, I am so happy knowing that we will live it together.


\pagestyle{plain}
\tableofcontents
\newpage



\chapter{Introduction}
\label{chap:intro}

\section{The Dark Matter Mystery}
\label{sec:the_dm_mystery}

The vast majority of all matter in our universe is unknown. Discovering this known unknown took several decades of work, work that has transformed cosmology from a once speculative discipline into the highly precise science that we are familiar with today. The existence of what we call ``dark matter'' forms a cornerstone of our modern understanding of the universe, and is one of the great triumphs of physics.

The question, however, still remains: what \textit{is} dark matter? To arrive at an answer, we must begin with what we do know. Thus far, we know that dark matter is: 

\begin{figure}[t]
    \centering
    \includegraphics[scale=0.7]{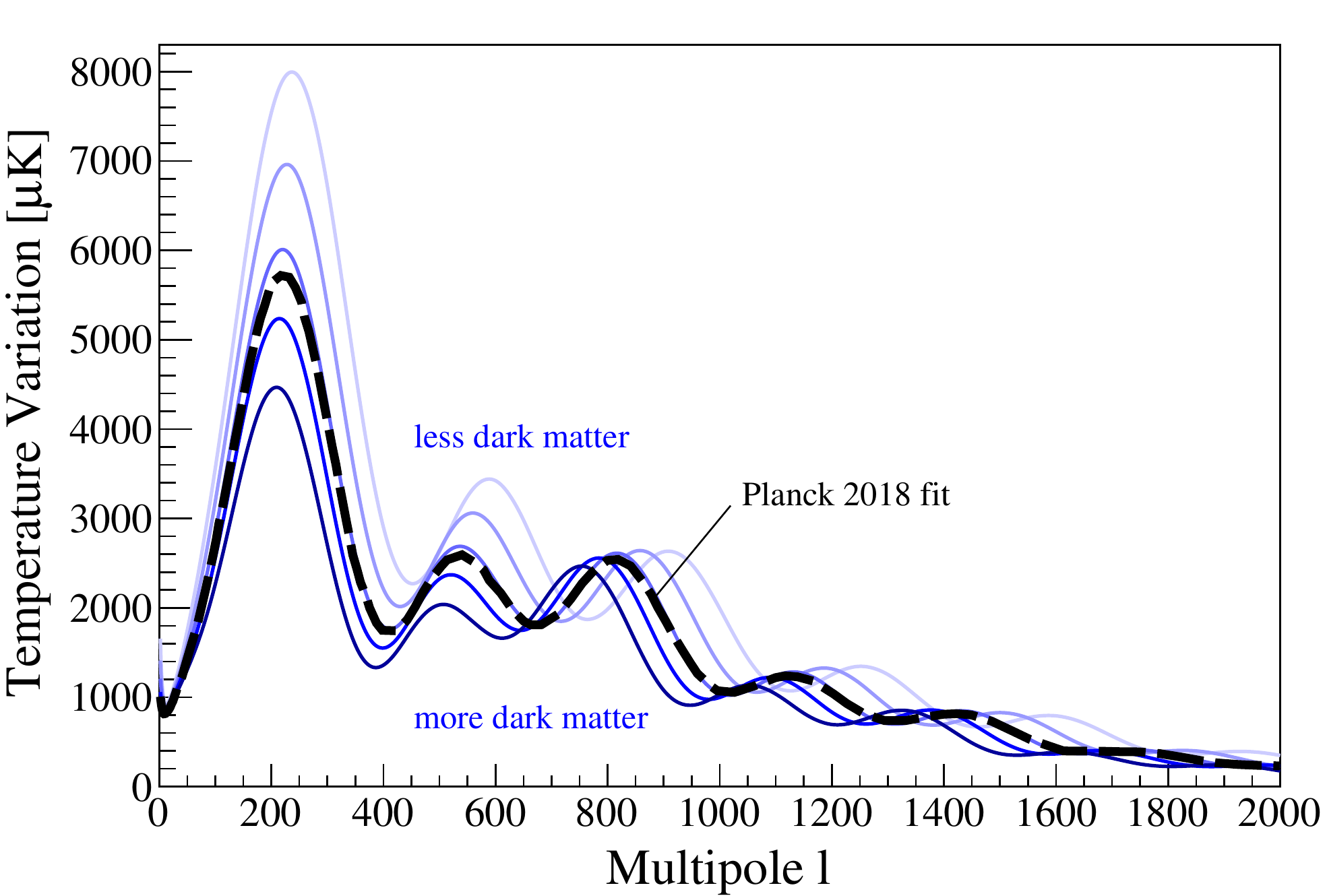}
    \caption{The CMB TT anisotropy power spectrum for dark matter densities $0.11 \leq \Omega_c \leq 0.43$, with all other cosmological parameters held constant. The best fit to the 2018 Planck data~\cite{Aghanim:2018eyx} is shown (black, dashed). This image was originally published in Ref.~\cite{Schumann:2019eaa}.}
    \label{fig:CMB_DM}
\end{figure}

\begin{enumerate}
    \item \textit{Abundant} --- The cosmic microwave background (CMB) anisotropy power spectrum gives a snapshot of the acoustic oscillations of the photon-baryon fluid at the point of recombination. Different modes, each with their characteristic length scales, oscillate at frequencies that depend on the gravitational potential set by the dark matter abundance, $\Omega_c$; the amplitude of each mode at recombination can therefore be turned into a measurement of $\Omega_c$. Fig.~\ref{fig:CMB_DM} illustrates how the power spectrum changes as a function of $\Omega_c$. The latest results~\cite{Aghanim:2018eyx} shows that $\Omega_c$ is given by
    \begin{alignat}{1}
        \Omega_c h^2 = 0.1200 \pm 0.0012 \,,
        \label{eqn:DM_abundance}
    \end{alignat}
    where $h$ is the Hubble parameter expressed in units of \SI{100}{\kilo\metre \per \second \per \mega \parsec}. This represents approximately 85\% of all matter in the universe, measured at an astounding relative uncertainty of about 2\%;

    \item \textit{Important for structure formation} --- the formation of large-scale structure is driven in large part by dark matter. If dark matter is a new, fundamental particle, then power at length scales smaller than the free-streaming length of dark matter becomes more suppressed as the temperature of dark matter increases. All of our observations of large-scale structure so far have been consistent with a cold dark matter (CDM)~\cite{Tegmark:2003uf}, and observations of the Lyman-$\alpha$ forest have set increasingly strong limits on how warm dark matter can be~\cite{Irsic:2017ixq}. There are, however, some features of structure formation that may pose a challenge to the CDM paradigm. These potential hints are collectively known as the ``small-scale structure problems''~\cite{DelPopolo:2016emo} and may be alleviated by warm dark matter, although it is still far from clear if these are really problems for CDM;

    \item \textit{Weakly interacting} -- There is thus far no evidence for any nongravitational interaction between dark matter and the Standard Model (SM). However, the fact that the abundance of dark matter is within an order of magnitude of the abundance of baryonic matter, together with other theoretical (e.g.\ the expectation of new physics at the weak scale, such as supersymmetry) and experimental hints of new physics, makes it reasonable to expect such interactions. This has motivated a broad range of experimental searches, including the following classes of searches:

    \begin{enumerate}

        \item Direct detection experiments, aimed at detecting momentum transfer between dark matter and the SM through the process $\chi + \text{SM} \to \chi + \text{SM}$ (see e.g.\ Refs.~\cite{Aprile:2018dbl,Agnese:2018col,Abramoff:2019dfb,Cui:2017nnn,Akerib:2016vxi,Angloher:2015ewa,Agnese:2015nto}, and Ref.~\cite{Schumann:2019eaa} for a recent review);

        \item Dark sector particle production at the Large Hadron Collider (LHC), aimed at detecting collider signatures associated with such processes, including final states with missing energy (e.g.\ from $p + p \to \chi + \chi$) and the production of dark sector-SM mediators (see Ref.~\cite{Aaboud:2019yqu} for a review of relevant ATLAS results, Ref.~\cite{Sirunyan:2016iap,Sirunyan:2017hci,Sirunyan:2017jix,Sirunyan:2019gfm} for some examples of relevant CMS results, and Ref.~\cite{Aaij:2017rft} for an LHCb search for dark photons);

        \item Indirect detection of dark matter annihilation ($\chi + \chi \to \text{SM} + \text{SM}$) or decay ($\chi \to \text{SM} + \text{SM}$) with gamma-ray telescopes (see e.g.\ Refs.~\cite{Gruber:1999yr,Bouchet:2008rp,COMPTEL,2005A&A...444..495S,Abeysekara:2017jxs,Johnson:2019hsm,TheFermi-LAT:2015kwa,Fermi-LAT:2016uux,TheFermi-LAT:2017vmf,Rinchiuso:2019rrh,Archambault:2017wyh}), neutrino telescopes (see e.g.\ Refs.~\cite{Aartsen:2016zhm,Albert:2016dsy,Aartsen:2018mxl}) and other cosmic ray experiments (e.g.\ Refs.~\cite{barger,Cui:2016ppb,Cuoco:2016eej,Yuan:2018rys,vonDoetinchem:2015zva}), or with cosmological probes such as the CMB anisotropy power spectrum~\cite{Aghanim:2018eyx}. 

    \end{enumerate}

\end{enumerate}

What we know about dark matter lays the foundation for how to uncover the nature of dark matter itself. In this thesis, we will focus on two aspects of the dark matter mystery:

\begin{enumerate}

    \item \textit{Dark matter production} --- We will examine a new way of producing the correct dark matter relic abundance in the early universe, and explore the possibility of producing dark sector bound states at colliders, and

    \item \textit{Dark matter energy deposition} --- We will look at a new experimental method to detect axion effects on photon polarization, and study the effects of dark matter annihilation or decay on the cosmic ionization and thermal history. 

\end{enumerate}

On both fronts, we will discuss novel ideas over length scales ranging from the table-top to the cosmos, and touching on all of the main search strategies discussed above.

\section{Dark Matter Production in the Early Universe}
\label{sec:dm_production_early_universe}

Since the abundance of dark matter is so well-established, understanding how it could have achieved this particular abundance is important.

\subsection{Lee-Weinberg Scenario}
\label{subsec:Lee_Weinberg_scenario}

Consider a dark matter particle $\chi$ with mass $m_\chi$ that is in thermal equilibrium with the Standard Model particle SM at some early time through the $2 \to 2$ process $\chi \chi \rightleftharpoons \text{SM} + \text{SM}$, assuming that the SM particle has a much lighter mass than $m_\chi$. This scenario is commonly known as the Lee-Weinberg scenario~\cite{Lee:1977ua}; the evolution of the abundance of dark matter is governed by the following Boltzmann equation,
\begin{alignat}{1}
    \dot{n}_\chi + 3 H n_\chi = \langle \sigma_{\chi\chi} v \rangle \left(n_{\chi,0}^2 - n_\chi^2 \right) \,,
    \label{eqn:2to2_boltzmann}
\end{alignat}
where $\langle \sigma_{\chi\chi} v \rangle$ denotes the thermally-averaged annihilation cross section, $n_\chi$ is the number density of $\chi$ and $\dot{n}_\chi$ its derivative with respect to time, with $n_{\chi,0}$ denoting the equilibrium number density. In principle, we should solve Eq.~(\ref{eqn:2to2_boltzmann}) numerically, but some analytic estimates will prove useful here. 

There are three energy scales that are relevant to this freezeout process. The dark matter mass $m_\chi$ sets the rate of interaction between SM and dark matter, and also determines the temperature at which dark matter first becomes nonrelativistic. The other two parameters are known or measured quantities that determine how the dark matter must dilute:

\begin{enumerate} 
    \item \textit{Reduced Planck mass}, $M_\text{pl}$ --- the reduced Planck mass, $M_\text{pl} = (8 \pi G)^{-\nicefrac{1}{2}}$, sets the Hubble expansion rate, which is given by $H \sim T^2/M_\text{pl}$ in the early universe, with $T$ being the temperature of the radiation bath, and
    \item \textit{Temperature at matter-radiation equality}, $T_\text{MRE}$ --- this is the temperature at which matter and radiation had equal energy density, and is set by the abundance of dark matter that we observe today.
\end{enumerate}

If such a $2 \to 2$ process governs thermal freezeout, we can derive a relationship between the unknown $m_\chi$ and the known values of $M_\text{pl}$ and $T_\text{MRE}$. The dark matter freezes out when the rate of annihilation falls below the Hubble rate $H$, i.e.\ when
\begin{alignat}{1}
    n_{\chi,f} \langle \sigma_{\chi\chi} v \rangle \sim H \sim \frac{m_\chi^2}{M_\text{pl}} \,,
    \label{eqn:freezeout_condition}
\end{alignat}
where $n_{\chi,f}$ is the number density of dark matter at freezeout. Here, we have used the fact that freezeout must occur when $T \sim m_\chi$, since the equilibrium number density of dark matter falls exponentially at lower temperatures once it becomes nonrelativistic. We have assumed that freezeout occurs deep in the radiation dominated era, and neglect all dependence on the number of relativistic degrees of freedom throughout.

After freezeout, the dark matter number density dilutes as $n_\chi \propto a^{-3} \propto T^3$, where $a$ is the scale factor. By definition, the mass density of dark matter is equal to the radiation energy density at matter-radiation equality. Since the radiation energy density is proportional to $T^4$, we see that $T_\text{MRE}$ sets the following relation at the point of matter-radiation equality:
\begin{alignat}{1}
    m_\chi n_\chi \sim m_\chi n_{\chi,f} \frac{T_\text{MRE}^3}{m_\chi^3} \sim T_\text{MRE}^4 \,.
\end{alignat}
Putting this together with Eq.~(\ref{eqn:freezeout_condition}), we obtain
\begin{alignat}{1}
    \langle \sigma_{\chi \chi} v \rangle \sim \frac{1}{M_\text{pl} T_\text{MRE}} \sim \SI{e-26}{\centi\meter\cubed\per\second} \,.
    \label{eqn:2to2_annxsec}
\end{alignat}
This relation shows that if a $2\to2$ process governs freezeout, then the required value of $\langle \sigma v \rangle$ is to first order independent of the dark matter model, and is set entirely by the expansion history and the dark matter abundance today. A careful calculation shows that $\langle \sigma v \rangle \approx \SI{3e-26}{\centi\meter\cubed\per\second}$ if $\chi$ is a Majorana fermion, with a weak dependence on $m_\chi$~\cite{Steigman:2012nb}. 

Suppose $\chi$ annihilates into the SM through an electroweak process, such that $\langle \sigma v \rangle \sim \alpha_W^2/m_\chi^2$. Plugging this into Eq.~(\ref{eqn:2to2_annxsec}), we obtain
\begin{alignat}{1}
    m_\chi \sim \alpha_W \sqrt{M_\text{pl} T_\text{MRE}} \sim \SI{}{\tera\eV} \,.
    \label{eqn:WIMP_miracle}
\end{alignat}
This coincidence between the geometric mean of $M_\text{pl}$ and $T_\text{MRE}$ and the electroweak scale, where new physics has long been expected, is known as the ``WIMP miracle''.

\subsection{Two Exceptions to the Lee-Weinberg Scenario}
\label{subsec:exceptions_to_Lee_Weinberg}

The standard 2-to-2 freezeout with a thermal dark matter annihilating into two particles is merely the simplest of many possible thermal production mechanism of dark matter; many important exceptions have been proposed and studied extensively in the literature (see e.g.\ Refs.~\cite{Griest:1990kh,DEramo:2010keq}). In this thesis, we will restrict ourselves to two exceptions that have implications for the preferred mass of the dark matter: the freezeout of dark matter through an annihilation channel that is kinematically forbidden at zero temperature, and an annihilation channel involving $3 \to 2$ interactions.

\subsubsection{Forbidden Dark Matter}
\label{subsubsec:forbidden_dark_matter}

Consider once again a thermal freezeout through a $2 \to 2$ process, $\chi \chi \rightleftharpoons \psi \psi$, but this time with a particle $\psi$ that has a mass $m_\psi > m_\chi$ and remains in equilibrium with the Standard Model throughout freezeout~\cite{Griest:1990kh,DAgnolo:2015ujb}. The Boltzmann equation governing this process is still similarly given by Eq.~(\ref{eqn:2to2_boltzmann}), and so Eq.~(\ref{eqn:2to2_annxsec}) still holds. However, the process $\chi \chi \to \psi\psi$ is now kinematically forbidden at zero temperature, and can only occur for particles at the high-velocity tail of the Boltzmann distribution of $\chi$. 

To work out what this means for the preferred value of $m_\chi$, we first note that detailed balance enforces the following relationship between the two cross sections: 
\begin{alignat}{1}
    \langle \sigma_{\psi \psi} v \rangle = \langle \sigma_{\chi \chi} v \rangle \frac{n_{\chi,0}^2}{ n_{\psi,0}^2} \sim \langle \sigma_{\chi \chi} v \rangle e^{2 \Delta/T} \,,
    \label{eqn:detailed_balance_forbidden_DM}
\end{alignat}
where $\Delta = m_\psi - m_\chi$, and we take $\Delta \lesssim m_\chi$. We have also used the fact that freezeout occurs in the nonrelativistic limit, and so the equilibrium number density of $\chi$ follows a Maxwell-Boltzmann distribution, and is given by
\begin{alignat}{1}
    n_{\chi,0} = g_\chi \left(\frac{m_\chi T}{2 \pi}\right)^{3/2} e^{-m_\chi/T} \,,
\end{alignat}
where $g_\chi$ is the degeneracy of states for $\chi$ (and similarly for $\psi$). If $\langle \sigma_{\psi \psi} v \rangle$ has the same parametric dependence as for the WIMP miracle, i.e.\ $\langle \sigma_{\psi \psi} v \rangle \sim \alpha^2 / m_\psi^2$ for some coupling $\alpha$, then Eq.~(\ref{eqn:WIMP_miracle}) gives
\begin{alignat}{1}
    m_\chi \sim \alpha e^{-\Delta/T_f} \sqrt{M_\text{pl} T_\text{MRE}} \,,
\end{alignat}
where $T_f$ is the temperature at freezeout. Comparing this with Eq.~(\ref{eqn:WIMP_miracle}), the exponential factor means that we now prefer lighter dark matter masses: for $\alpha = \alpha_W$, $\Delta = 0.5 m_\chi$ and $T_f = 0.05 m_\chi$, for example, we obtain $m_\chi \sim \SI{100}{\mega \eV}$, significantly lighter than the standard expectation from the WIMP miracle. 

\subsubsection{3-to-2 Freezeouts}
\label{subsubsec:3_to_2_freezeouts}

Freezeout may also be determined by a $3 \to 2$ process, instead of a $2 \to 2$ process~\cite{Hochberg:2014dra}. Conventionally, we expect the rate of $3 \to 2$ processes to be significantly suppressed compared to any available $2 \to 2$ process; however, dark matter models can be easily constructed where such $2 \to 2$ processes are suppressed by small couplings~\cite{Hochberg:2014dra} or kinematics, or where freezeout occurs entirely within the dark sector, and the $3 \to 2$ process is the only number changing process available~\cite{Hochberg:2014dra,Kuflik:2015isi,Kuflik:2017iqs}. 

Consider a $3 \to 2$ process $\chi \chi \chi \rightleftharpoons \chi \chi$ that determines the freezeout of $\chi$. The rate of the forward process would be proportional to $n_\chi^2$, multiplied by a velocity-averaged annihilation cross section denoted $\langle \sigma v^2 \rangle$. At freezeout, we must therefore have
\begin{alignat}{1}
    n_{\chi,f}^2 \langle \sigma v^2 \rangle \sim H \sim \frac{m_\chi^2}{M_\text{pl}} \,.
\end{alignat}
A similar exercise as the one we did to obtain Eq.~(\ref{eqn:2to2_annxsec}) now gives
\begin{alignat}{1}
    \langle \sigma v^2 \rangle \sim \frac{m_\chi^{-2}}{M_\text{pl} T_\text{MRE}^2} \,.
    \label{eqn:3to2_annxsec}
\end{alignat}
Unlike the Lee-Weinberg scenario, $\langle \sigma v^2 \rangle$ now depends strongly on the dark matter mass. On dimensional grounds, we expect $\langle \sigma v^2 \rangle \sim \alpha_\text{eff} / m_\chi^5$ parametrically, where $\alpha_\text{eff}$ is some dimensionless effective coupling constant. Putting these results together with Eq.~(\ref{eqn:3to2_annxsec}), we obtain
\begin{alignat}{1}
    m_\chi \sim \alpha_\text{eff} \left(M_\text{pl} T_\text{MRE}^2\right)^{1/3} \sim \left( \frac{\alpha_\text{eff}}{1.0} \right)  \SI{}{\giga\eV} \,.
\end{alignat}
Yet again, we find that the preferred range of $m_\chi$ lies below a \SI{}{\giga\eV}. 

\subsection{Not-Forbidden Dark Matter}
\label{subsec:not-forbidden_dm}

Dark sector models have been constructed that exhibit the freezeout mechanisms described above. The preferred mass range of $m_\chi \lesssim \SI{}{\giga\eV}$ is particularly appealing for two reasons. First, it avoids powerful constraints coming from the xenon-based direct detection experiments~\cite{Akerib:2016vxi,Aprile:2017iyp,Cui:2017nnn} while motivating a range of new experiments targeting the \SI{}{\mega \eV}--\SI{}{\giga \eV} mass range. Second, the smaller dark matter masses tend to also come with increased self-interaction rates, offering a potential solution to the so-called small-scale structure problems~\cite{Rocha:2012jg,Spergel:1999mh,Zavala:2012us}. The annihilation rate of these sub-\SI{}{\giga\eV} dark matter particles into Standard Model particles in these models can be small enough to evade the stringent CMB limits for dark matter in this mass range~\cite{Aghanim:2018eyx}. 

In the forbidden dark matter scenario, a simple $U(1)$ vector portal dark sector with a dark photon $A'$ that is slightly more massive than the Dirac fermion dark matter candidate $\chi$, with $m_{A'}/m_\chi \lesssim 1.5$, is sufficient to realize the forbidden dark matter scenario, allowing for $m_\chi \lesssim \SI{10}{\giga\eV}$~\cite{DAgnolo:2015ujb}. 

In the case of $3 \to 2$ freezeouts, the picture is significantly more complicated. In many dark matter models, the stability of dark matter is guaranteed by a $\mathbb{Z}_2$ symmetry: if the $3 \to 2$ freezeout involves only one particle species, this symmetry simply cannot be part of the theory. On the other hand, some interaction between the dark and Standard Model sectors is desirable in order to maintain kinetic equilibrium with the Standard model and ensure that $3 \to 2$ annihilations do not heat the dark sector. Some care must therefore be taken to ensure the stability of the dark matter candidate, while also ensuring that the $3 \to 2$ interaction is dominant during freezeout. Much of the focus prior to the work discussed in this thesis had been on five-point interactions in the dark sector from a Wess-Zumino-Witten term arising from chiral symmetry breaking of a flavor symmetry with nontrivial topological structure~\cite{Hochberg:2014kqa}, or on dark sectors with a $\mathbb{Z}_3$ or $\mathbb{Z}_5$ symmetry that stabilizes the dark matter and leads to $3 \to 2$ annihilation processes~\cite{Hochberg:2014dra,Choi:2016hid,Choi:2015bya}. 

In Chapter~\ref{chap:nfdm} (based on Ref.~\cite{Cline:2017tka}), I will discuss a novel way to achieve a thermal freezeout dominated by $3 \to 2$ processes, which can be realized even within the simple $U(1)$ vector portal dark sector. Such a scenario can be realized in the deeply forbidden region of parameter space in the forbidden dark matter model, i.e.\ with $1.5 \lesssim m_{A'}/m_\chi \lesssim 2$, where the annihilation cross section for $\chi \bar{\chi} \to A' A'$ becomes extremely suppressed. In this regime, the $3 \to 2$ process $\chi \chi \bar{\chi} \to A' \chi$ has a rate that becomes comparable to $\chi \bar{\chi} \to A' A'$ in the early universe, and can dominate freezeout if the kinetic mixing of the dark photon to the Standard Model is sufficiently small. $3 \to 2$ freezeouts are therefore possible even within simple models like the $U(1)$ vector portal model, simply due to kinematics; these models inherit all of the interesting properties of both exceptions to the Lee-Weinberg scenario, including a sub-\SI{}{\giga \eV} dark matter candidate and the possibility of alleviating the small-scale structure problems.

\section{Dark Bound States Production at Colliders}
\label{sec:dark_bound_states_at_colliders}

\subsection{Bound States}
\label{sec:bound_states}

Bound states are ubiquitous in the Standard Model. The lightest state with nonzero baryon number, the proton, is itself a bound state of three quarks, and for close to a billion years during the cosmic dark ages, almost all electrons were locked in a bound state with protons in the form of hydrogen and helium atoms. It is therefore natural to consider the possibility that dark matter itself can form bound states, and that this could lead to interesting phenomenological signatures in searches for dark matter. 

We begin our discussion by observing some simple parametric estimates for properties of hydrogenic, nonrelativistic bound states comprising two identical particles $\chi$ of mass $m_\chi$ with reduced mass $\mu_\chi = m_\chi/2$ in a potential given by the Coulomb potential, $V(r) = -\alpha_\mathcal{B}/r$, where $\alpha_\mathcal{B}$ is a coupling constant for the force between the constituents of $\mathcal{B}$.\footnote{Exact results can of course be obtained by solving the Schr\"{o}dinger equation.} First, let us determine the characteristic length scale $a_0$ defining the ground state known as the Bohr radius in the case of the hydrogen atom. Classically, there is no lower bound to the characteristic size of bound states, but we know that quantum effects act as a length scale cut-off for the ground state. In particular, the uncertainty principle applied to the ground state must enforce
\begin{alignat}{1}
    \langle p^2 \rangle a_0^2 \sim 1 \,,
\end{alignat}
where $\langle \cdot \rangle$ denotes the expectation value of an observable, with $p$ being the momentum of one of the particles relative to the other. The virial theorem, on the other hand, tells us that the ground state must have 
\begin{alignat}{1}
    2 \left< \frac{p^2}{2\mu_\chi} \right> = \left< \frac{\alpha_\mathcal{B}}{r}  \right> \,,
\end{alignat}
or in other words, that
\begin{alignat}{1}
    \frac{1}{a_0} \sim \alpha_\mathcal{B} \mu_\chi \,.
\end{alignat}
The quantity $1/a_0$, known as the Bohr momentum in the hydrogen atom, sets a momentum scale, above which individual particles within the bound state can be resolved. The ground state binding energy is then $E_B \sim \alpha_\mathcal{B}^2 \mu_\chi$. 

For a massive mediator $Y$ with mass $m_Y$, the potential between particles is given by the Yukawa potential, $V(r) = -\alpha_\mathcal{B} e^{- m_Y r}/r$. We can once again ask what kind of bound states such a potential would lead to, or even if bound states are allowed. The mass of the mediator introduces another length scale into the problem: for length scales well below $1/m_Y$, the mediator can effectively be treated as massless, and the Yukawa potential reduces to the $1/r$ Coulomb potential. For scales $r \gtrsim 1/m_Y$, however, the potential is suppressed by the exponential term. 

For bound states to exist in such a potential, we expect the ground state to have a size smaller than $1/m_Y$, where the force between the two particles is not subject to the exponential cut-off. In this regime, the particle experiences a potential that is close to $1/r$, and so the preceding estimate of the characteristic size of the ground state holds. In other words, for a bound state to exist in a Yukawa potential, we expect\footnote{Solving the Schr\"{o}dinger equation numerically with the Yukawa potential gives $\alpha_\mathcal{B} \mu_\chi / m_Y > 0.84$ instead~\cite{PhysRevA.1.1577}.}
\begin{alignat}{1}
    \frac{\alpha_\mathcal{B} \mu_\chi}{m_Y} \gtrsim 1 \,.
\end{alignat}
From this result, we learn that any dark sector model where bound states play an important role in its phenomenology must have both a dark matter candidate and a light mediator between dark sector states. 

\subsection{Bound State Production}
\label{subsec:bound_state_production}

Having understood the conditions under which bound states may exist, we now want to understand if the production of dark sector bound states can be phenomenologically interesting, particularly in the context of collider searches. 

To tackle this question, we first begin with a rudimentary example of how to calculate the cross section for the production of a bound state of two fermions $\chi$ with mass $m_\chi$. The nonrelativistic spin-1 bound state $\ket{\mathcal{B}}$ with mass $m_\mathcal{B}$ with its spin oriented up is given by~\cite{Peskin:1995ev}
\begin{alignat}{1}
    \ket{\mathcal{B}} = \sqrt{2 m_\mathcal{B}} \int \frac{d^3 \vec{k}}{(2\pi)^3} \tilde{\psi}( \vec{k} ) \frac{1}{\sqrt{2m_\chi}} \frac{1}{\sqrt{2m_\chi}} \ket{\vec{k} \uparrow, -\vec{k} \uparrow} \,,
    \label{eq:bound_state_expr}
\end{alignat}
where $\tilde{\psi}(\vec{k})$ is the momentum space wavefunction of the bound state, and the factors of $\sqrt{2m}$ are needed to obtain the correct normalization. Eq.~(\ref{eq:bound_state_expr}) simply states that $\ket{\mathcal{B}}$ is a linear combination of free particle states with the appropriate momentum, weighted by $\braket{\mathcal{B} | \vec{k} \uparrow, -\vec{k} \uparrow}$, which we can work out with nonrelativistic quantum mechanics. The spin state of the free fermions must have the appropriate combination of spins to produce the spin-1 state under consideration. 

It follows immediately that the matrix element for the production of $\mathcal{B}$ from some initial state $i$ is simply
\begin{alignat}{1}
    \mathcal{M}(i \to \mathcal{B}) = \frac{\sqrt{2 m_\mathcal{B}}}{2 m_\chi} \int \frac{d^3 \vec{k}}{(2\pi)^3} \tilde{\psi}(\vec{k}) \mathcal{M}(i \to \vec{k} \uparrow, - \vec{k} \uparrow)
\end{alignat}
More generally, we want to consider the production of bound states with some overall spin $S$, and project out the relevant final state spins for the fermions. For simplicity, we will now consider only the case of a spin-1 bound state $\mathcal{B}$, produced via $i \to V \to \chi \overline{\chi}$, where $V$ is a vector mediator. The matrix element can then be written
\begin{alignat}{1}
    \mathcal{M}(i \to \mathcal{B}) = \frac{\sqrt{2 m_\mathcal{B}}}{2 m_\chi} \int \frac{d^3 \vec{k}}{(2\pi)^3} \tilde{\psi}(\vec{k}) \mathcal{A}_i^\mu \sum_{s,s'} \overline{v}(k'; s') \gamma_\mu u(k; s) \braket{s,s' | J, J_z} \,,
    \label{eqn:matrix_element_bound_state}
\end{alignat}
where $\mathcal{A}_i^\mu$ is the part of the matrix element that excludes the fermion-antifermion vertex $\overline{v}\gamma_\mu u$, $k = (m_\chi, \vec{k})$ and $k' = (m_\chi, -\vec{k})$. We can now rewrite $\overline{v} \gamma^\mu u$ as $\text{Tr}[u \overline{v} \gamma^\mu]$, and expand the entire expression in the nonrelativistic limit~\cite{Bodwin:2016whr,Pal:2007dc,Guberina:1980dc}. Making the further approximation that $\tilde{\psi}(\vec{k})$ is suppressed away from the origin, and that $m_\mathcal{B} \approx 2 m_\chi$, we finally arrive at the following compact expression for the matrix element~\cite{Petrelli:1997ge,Guberina:1980dc}:
\begin{alignat}{1}
    \mathcal{M}(i \to \mathcal{B}) = \text{Tr} \left[\frac{\psi^*(0)}{\sqrt{8 m_\chi^3}} (\slashed{k} - m_\chi) \slashed{\epsilon}^* (\slashed{k}' + m_\chi) \gamma_\mu \mathcal{A}^\mu_i\right] \,,
\end{alignat}
where $\epsilon$ is the polarization vector for the massive spin-1 bound state $\mathcal{B}$.\footnote{Some helpful references for the algebra required are Ref.~\cite{Pal:2007dc} for spinor identities, and the identification $\xi_s \xi_{s'}^\dagger \to \vec{\epsilon}^{\,*} \cdot \vec{\sigma}$ where $\xi$ is the two-component spinor that forms the Dirac spinors $u$ and $v$, as explained in Ref.~\cite{Peskin:1995ev}.} The wavefunction at the origin for the ground state, which can be obtained by solving the Schr\"{o}dinger equation, is given by
\begin{alignat}{1}
    |\psi(0)|^2 = \frac{\alpha_\mathcal{B}^3 m_\chi^3}{8 \pi} \,.
\end{alignat}
%

\subsection{Experimental Signatures of Dark Sector Bound States}

The production of dark sector bound states, if they exist, can be of potential interest at colliders. First, the resonant production of bound states has historically been a powerful mode of finding new physics: it was deployed to find the charm quark~\cite{Aubert:1974js,Augustin:1974xw} and the bottom quark~\cite{Aubert:2009cp}, to name two examples. Compared to missing energy searches for dark matter~\cite{Abdallah:2014hon,Abdallah:2015ter}, resonance signatures are striking and usually make for a cleaner analysis. Since the production cross section scales as $|\psi(0)|^2$, it is possible that for tightly-bound bound states, the search for a bound state resonance may be the best way to discover certain dark matter models at colliders. 

In Chapter~\ref{chap:bound_states} (based on Ref.~\cite{Elor:2018xku}), I will explore precisely this possibility, and also examine the generic implications of such scenarios for indirect and direct detection. We show that resonance searches are complementary to monojet searches and can probe dark matter masses above 1 TeV with current LHC data. We argue that this parameter regime, where the bound-state resonance channel is the most sensitive probe of the dark sector, arises most naturally in the context of non-trivial dark sectors with large couplings, nearly-degenerate dark-matter-like states, and multiple force carriers. The presence of bound states detectable by the LHC implies a minimal Sommerfeld enhancement that is appreciable, and potentially also radiative bound state formation in the Galactic halo, leading to large signals in indirect searches. We calculate these complementary constraints, which favor either models where the bound-state-forming dark matter constitutes a small fraction of the total density, or models where the late-time annihilation is suppressed at low velocities or late times. We also discuss two concrete examples of models that satisfy all these constraints and where the LHC resonance search is the most sensitive probe of the dark sector.

\section{Axions and Light Polarization}
\label{sec:axions_and_light_polarization}

The QCD axion is a well-motivated solution to the strong CP problem which also provides a natural dark matter candidate~\cite{Peccei:1977hh,Peccei:1977ur,Weinberg:1977ma,Wilczek:1977pj,Preskill:1982cy,Abbott:1982af,Dine:1982ah}, particularly in the \SI{}{\micro\eV} to \SI{10}{\milli\eV} range~\cite{Graham:2015ouw}. More generally, however, dark matter could be made up of light, pseudoscalar axion-like particles (ALPs). These are generically predicted in string theory, and can have masses much less than a \SI{}{\micro\eV}~\cite{Jaeckel:2010ni,Svrcek:2006yi,Arvanitaki:2009fg,Acharya:2010zx,Cicoli:2012sz}. For brevity, we will often use the word ``axion'' to refer to any such ALP dark matter candidate.  

ALP dark matter would behave as a coherent, classical field $a$, and may couple weakly to photons through the following interaction term: 
\begin{alignat}{1}
    \mathcal{L} \supset -\frac{1}{4} g_{a\gamma\gamma} a F_{\mu\nu} \tilde{F}^{\mu\nu}.
    \label{eqn:axion_EM_interaction}
\end{alignat}
Such an interaction has motivated a large range of experimental searches for ALPs converting into photons (and vice versa) in the presence of a static magnetic field~\cite{Sikivie:1983ip,Wilczek:1987mv}. This includes proposed and ongoing searches for unexpected modifications to the amplitude, phase, or polarization of propagating light~\cite{Cameron:1993mr,Tam:2011kw,DellaValle:2015xxa}, light-shining-through-walls experiments~\cite{Chou:2007zzc,Robilliard:2007bq,Ehret:2010mh,Betz:2013dza,Ballou:2014myz,Ballou:2015cka}, the conversion of axion oscillations into electromagnetic waves~\cite{Horns:2012jf,Chaudhuri:2014dla,Kahn:2016aff,TheMADMAXWorkingGroup:2016hpc,Foster:2017hbq,Chaudhuri:2018rqn,Du:2018uak,Baryakhtar:2018doz,Ouellet:2018beu,Ouellet:2019tlz}, and axion helioscopes~\cite{Anastassopoulos:2017ftl,Armengaud:2014gea}. Among these, past and ongoing experiments~\cite{Cameron:1993mr,DellaValle:2015xxa,Chou:2007zzc,Robilliard:2007bq,Ehret:2010mh,Betz:2013dza,Ballou:2014myz,Ballou:2015cka,Horns:2012jf,Du:2018uak,Ouellet:2018beu,Anastassopoulos:2017ftl} have steadily improved constraints on $g_{a\gamma\gamma}$, and have even begun probing couplings that are relevant to the QCD axion~\cite{Du:2018uak}. 

One important effect that the interaction in Eq.~(\ref{eqn:axion_EM_interaction}) produces is the rotation of linear polarization as light propagates through an axion classical field. Eq.~(\ref{eqn:axion_EM_interaction}) modifies the inhomogeneous Maxwell's equations in vacuum, giving in Lorenz gauge
\begin{alignat}{1}
    \square A^\nu = - g_{a\gamma\gamma} \partial_\mu a \cdot \varepsilon^{\mu\nu\rho\sigma} \partial_\rho A_\sigma \,.
\end{alignat}
For a plane electromagnetic wave propagating along the $z$-direction, each polarization is given by
\begin{alignat}{1}
    \mathbf{A}_\pm = A_0 \boldsymbol{\epsilon}_\pm e^{i(k_0 z - \omega_0 t)} \,,
\end{alignat}
where $\boldsymbol{\epsilon}_\pm$ are the usual left- and right-polarization vectors, $\boldsymbol{\epsilon}_\pm = (1, \pm i, 0)/\sqrt{2}$. For this plane wave, the equations of motion become
\begin{alignat}{1}
    (k_0^2 - \omega_0^2) \frac{A_0}{\sqrt{2}} e^{i(k_0 z - \omega_0 t)} &= \mp g_{a\gamma\gamma} k_0 \frac{\partial a}{\partial t} \frac{A_0}{\sqrt{2}} e^{i(k_0 z - \omega_0 t)}
    \label{eqn:axion_dispersion}
\end{alignat}
which leads immediately to the dispersion relation obeyed by electromagnetic waves in the presence of the axion field:
\begin{alignat}{1}
    k_0^2 - \omega_0^2 \pm g_{a\gamma\gamma} k_0 \frac{\partial a}{\partial t} = 0 \,.
\end{alignat}
In the limit of small coupling, we see that the phase velocity for circularly-polarized light varies as
\begin{alignat}{1}
    \frac{\omega_0}{k_0} = 1 \pm \frac{g_{a\gamma\gamma}}{2k_0} \frac{\partial a}{\partial t} \,.
\end{alignat}
For a classical axion field that oscillates sinusoidally, this results in the oscillation in phase velocity of the two circular polarizations. In terms of linear polarization, this leads to an oscillating rotation in linear polarizations; ALP dark matter in essence transfers energy from one linear polarization to another. 

In Chapter~\ref{chap:adbc} (based on Ref.~\cite{Liu:2018icu}), I will discuss the use of interferometric techniques to detect these small oscillations in polarization. Inspired by earlier work in detecting axions through laser interferometry~\cite{Melissinos:2008vn,DeRocco:2018jwe,Obata:2018vvr}, we propose the Axion Detection with Birefringent Cavities (ADBC) experiment, a new axion interferometry concept using a cavity that exhibits birefringence between its two, linearly-polarized laser eigenmodes. This experimental concept overcomes several limitations of the designs currently in the literature, and can be practically realized in the form of a simple bowtie cavity with tunable mirror angles. Our design thereby increases the sensitivity of the axion-photon coupling over a wide range of axion masses.

\section{Energy Deposition in the Early Universe}
\label{sec:energy_deposition_early_universe}

The production of high-energy, electromagnetically interacting particles in the early universe is a generic outcome of many new physics processes, including the annihilation or decay of dark sector particles. These high-energy particles cool by interacting with baryons in the intergalactic medium, transferring their energy to the gas in the form of extra ionization and heating, affecting any cosmological observable that is sensitive to the ionization and temperature histories of the universe. Understanding the cooling process and its effects, in conjunction with cosmological probes of ionization and temperature, allows us to set powerful constraints on potential sources of new physics. 

\subsection{Three-Level Atom}
\label{subsec:TLA}

In the absence of any source of energy injection, the three-level atom (TLA) model, first derived in~\cite{Peebles1968,Zeldovich:1969en}, provides a pair of coupled differential equations for the matter temperature in the IGM and the hydrogen ionization fraction:
\begin{alignat}{1}
    \dot{T}_m^{(0)} &= -2 H T_m + \Gamma_C (T_\text{CMB} - T_m) \,, \nonumber \\
    \dot{x}_\text{HII}^{(0)} &= -\mathcal{C} \left[n_\text{H} x_e x_\text{HII} \alpha_\text{H} - 4(1 - x_\text{HII}) \beta_\text{H} e^{-E_{21}/T_\text{CMB}}\right] \,,
    \label{eqn:TLA}
\end{alignat}
where $H$ is the Hubble parameter, $n_\text{H}$ is the total number density of hydrogen (both neutral and ionized), $x_\text{HII} \equiv n_\text{HII}/n_\text{H}$ where $n_\text{HII}$ is the number density of free protons, $x_e \equiv n_e/n_\text{H}$ is the free electron fraction with $n_e$ being the free electron density, and $E_{21} = \SI{10.2}{\eV}$ is the Lyman-$\alpha$ transition energy. $T_m$ and $T_\text{CMB}$ are the temperatures of the IGM and the CMB respectively.\footnote{We follow the standard astrophysical convention in which H and H$^+$ are denoted HI and HII, while He, He$^+$ and He$^{2+}$ are denoted HeI, HeII and HeIII respectively.} $\alpha_\text{H}$ and $\beta_\text{H}$ are case-B recombination and photoionization coefficients for hydrogen respectively, and $\mathcal{C}$ is the Peebles-C factor that represents the probability of a hydrogen atom in the $n = 2$ state decaying to the ground state before photoionization can occur~\cite{Peebles:1968ja,AliHaimoud:2010dx}. The photoionization coefficient is evaluated at the radiation temperature, $T_\text{CMB}$, in agreement with Ref.~\cite{Chluba:2015lpa}. $\Gamma_C$ is the Compton scattering rate, given by
\begin{alignat}{1}
    \Gamma_C = \frac{x_e}{1 + \mathcal{F}_\text{He} + x_e} \frac{8 \sigma_T a_r T_\text{CMB}^4}{3 m_e} \,,
    \label{eqn:compton_rate}
\end{alignat}
where $\sigma_T$ is the Thomson cross section, $a_r$ is the radiation constant, $m_e$ is the electron mass, and $\mathcal{F}_\text{He} \equiv n_\text{He}/n_\text{H}$ is the relative abundance of helium nuclei by number. In the absence of helium, note that $x_e = x_\text{HII}$. The solutions to Eq.~(\ref{eqn:TLA}) --- i.e.\ without any sources of energy injection --- define what we will call the baseline temperature and ionization histories, $T_m^{(0)}(z)$ and $x_\text{HII}^{(0)}(z)$. More accurate calculations of $T_m$ and $x_\text{HII}$ such as \textsc{cosmorec}~\cite{Chluba:2010ca} and \textsc{hyrec}~\cite{AliHaimoud:2010dx} are routinely used for CMB analyses, but such a high degree of accuracy is not currently needed when also including potential energy injection sources, since the uncertainties associated with these processes and the cooling of the injected particles are relatively large.

Exotic sources may inject additional energy into the universe, altering the thermal and ionization evolution shown in Eq.~(\ref{eqn:TLA}). For example, the rate of energy injection from DM annihilating with some velocity averaged cross section $\langle \sigma v \rangle$, or decaying with some lifetime $\tau$ much longer than the age of the universe, is given by
\begin{alignat}{1}
    \left(\frac{dE}{dV \, dt}\right)^\text{inj} = \begin{cases} 
        \rho_{\chi,0}^2 (1 + z)^6 \langle \sigma v \rangle/m_\chi\,, & \text{annihilation}, \\
        \rho_{\chi,0} (1 + z)^3 /\tau \,, & \text{decay}, 
    \end{cases}
    \label{eqn:energy_injection}
\end{alignat}
where $\rho_{\chi,0}$ is the mass density of DM today, and $m_\chi$ is the DM mass. This injected energy, however, does not in general manifest itself instantaneously as ionization, excitation, or heating of the gas. Instead, the primary particles injected into the universe may cool over timescales significantly larger than the Hubble time, producing secondary photons that may redshift significantly before depositing their energy into the gas.

Although the primary particles injected into the universe may be any type of Standard Model particle, we will only need to consider the cooling of photons and electron/positron pairs \cite{Slatyer:2009yq}. This simplification occurs because either the primaries are stable particles like photons, electrons and positrons, neutrinos, protons and anti-protons, and heavier nuclei, or are unstable particles that resolve into these particles on time scales much shorter than the cosmological time scales under consideration.  For typical sources of energy injection we can neglect heavier nuclei because they are produced in negligible amounts, and neutrinos because they are very ineffective at depositing their energy. Protons and antiprotons generally form a subdominant component of stable electromagnetic particles across all possible Standard Model primaries~\cite{Cirelli:2010xx}, and deposit energy less effectively than electrons, positrons, and photons (although their effects are not completely negligible \cite{Weniger:2013hja}). We therefore only decompose the injection of any primary into an effective injection of photons, electrons, and positrons, in accordance with Ref.~\cite{Slatyer:2009yq} and subsequent works. Adding the contribution from protons and antiprotons may strengthen these constraints by a small amount.

A significant amount of work has been done on computing the cooling of high energy photons, electrons, and positrons~\cite{Slatyer:2009yq,Furlanetto:2009uf,Valdes:2009cq,Slatyer:2012yq,Evoli:2012zz,Galli:2013dna,Evoli:2014pva,Slatyer:2015kla,Kanzaki:2008qb,Kawasaki:2015peu}.
Once the cooling of injected primary particles is determined, the energy deposited into channel $c$ (hydrogen ionization, excitation, or heating) can be simply parametrized as

\begin{alignat}{1}
    \left(\frac{dE}{dV \, dt}\right)_c^\text{dep} = f_c(z) \left(\frac{dE}{dV \, dt}\right)^\text{inj} \,,
    \label{eqn:fz}
\end{alignat}
with all of the complicated physics condensed into a single numerical factor that is dependent on the redshift. These $f_c$ functions also depend on the energies and species of the injected particles, but for simplicity of notation we will not write these arguments explicitly. 

The effect of energy injection on the thermal and ionization history can now be captured by additional source terms,
\begin{alignat}{1}
    \dot{T}_m^\text{inj} &= \frac{2 f_\text{heat}(z)}{3(1 + \mathcal{F}_\text{He} + x_e) n_\text{H}} \left(\frac{dE}{dV\, dt}\right)^\text{inj} \,, \nonumber \\
    \dot{x}_\text{HII}^\text{inj} &= \left[\frac{f_\text{H ion}(z)}{\mathcal{R} n_\text{H}} + \frac{(1 - \mathcal{C}) f_\text{exc} (z)}{0.75 \mathcal{R} n_\text{H}}\right] \left(\frac{dE}{dV \, dt}\right)^\text{inj} \,,
\end{alignat}
where $\mathcal{R} = \SI{13.6}{\eV}$ is the ionization potential of hydrogen. The temperature and ionization evolution is then governed by
\begin{alignat}{1}
    \dot{T}_m &= \dot{T}_m^{(0)} + \dot{T}_m^\text{inj} \,, \quad \quad \dot{x}_\text{HII} = \dot{x}_\text{HII}^{(0)} + \dot{x}_\text{HII}^\text{inj} \,.
    \label{eqn:TLA_with_injection}
\end{alignat}
%

\subsection{Simple Parametric Estimates}
\label{subsec:simple_parametric_estimates}

In subsequent chapters, we will solve Eq.~(\ref{eqn:TLA_with_injection}) numerically, and make significant improvements to this model. For now, we will perform some simple parametric estimates to understand the baseline ionization and temperature histories. 

\subsubsection{Recombination}
\label{subsubsec:recombination}

In the early universe, the population of photons with energy above the ionization potential of hydrogen $\mathcal{R}$ is large, and any proton and electron that happened to recombine to form hydrogen were quickly ionized by a sufficiently energetic CMB photon. However, as the universe cooled and the photons gradually redshifted and cooled, neutral hydrogen eventually became viable, and universe transitioned rapidly from mostly ionized to mostly neutral. One might expect that the temperature $T_\text{re}$ at which this occurs would be given by $\mathcal{R}$; however, baryons are a factor of $\eta \approx 6 \times 10^{-10}$ less abundant than baryons~\cite{Tanabashi:2018oca}, and so a slightly better estimate of when recombination occurs is $e^{-\mathcal{R}/T_\text{re}} \sim \eta$, i.e.
\begin{alignat}{1}
     T_\text{re} \sim -\frac{\mathcal{R}}{\log \eta} \sim \SI{0.6}{\eV}\,.
\end{alignat}
A more accurate estimate, assuming that all species are in Saha equilibrium, yields an estimate of $T_\text{re} \sim \SI{0.34}{\eV}$ when half of all protons have recombined into neutral hydrogen. 

The process of recombination also freezes out once the recombination rate drops below the Hubble expansion rate, leaving a residual free electron and proton population, each with density $n_e$ by neutrality of the universe. The recombination rate can be experimentally measured for hydrogen, and turns out to be characterized by the case-B recombination coefficient
\begin{alignat}{1}
    \alpha_\text{B}(z_\text{re}) \sim \SI{e-12}{\centi\meter\cubed\per\second} 
\end{alignat}
around recombination.\footnote{The case-B coefficient describes the rate of recombination to the $n=2$ state of hydrogen. This is the relevant process here, since direct recombination to $n=1$ produces a photon that quickly photoionizes another hydrogen atom.} The rate of change of the electron number density, $\dot{n}_e$, is then given by $\dot{n}_e = n_e^2 \alpha_B$. Freezeout occurs when $n_e$ drops to the point where
\begin{alignat}{1}
    \frac{\dot{n}_e}{n_e} = n_e \alpha_B \sim H \,,
\end{alignat}
which happens soon after recombination. This translates to an electron number density of
\begin{alignat}{1}
    x_e \equiv \frac{n_e}{n_\text{H}} \sim 10^{-4} \,,
\end{alignat}
where we have introduced the free electron fraction $x_e$, and $n_\text{H}$ is the number density of hydrogen in all ionization states. This estimate of $x_e$ agrees very well with precision calculations of $x_e$ from codes like \textsc{recfast}~\cite{Seager:1999bc,Seager:1999km}, \textsc{hyrec}~\cite{AliHaimoud:2010dx} and \textsc{cosmorec}~\cite{Chluba:2010ca}. 

\subsubsection{Temperature Evolution}
\label{subsubsec:temperature_evolution}

The temperature history of baryons in the universe as shown in Eq.~(\ref{eqn:TLA}) is dominated by two processes: Compton scattering of free electrons off CMB photons, and adiabatic expansion. At early times, when the CMB energy density is large, Compton scattering keeps baryons in thermal contact with the CMB. This process keeps $T_m \propto (1+z)$ as the universe expands. Adiabatic expansion, on the other hand, causes the de~Broglie wavelength of particles to scale as $(1+z)^{-1}$, and so the matter temperature scales as $T_m \propto p^2 \propto (1+z)^2$. The two processes become equally important at
\begin{alignat}{1}
    \frac{4\pi^2 x_e \sigma_T T_\text{CMB}^5}{15 m_e} \sim 2HT_\text{CMB} \,.
\end{alignat}
At this point, the temperature dependence transitions from $(1+z)$ to $(1+z)^2$, with the baryons decoupling from the CMB. This redshift of thermal decoupling $z_\text{td}$ is given by
\begin{alignat}{1}
    1+z_\text{td} \sim \left( \frac{15 m_e H_0}{2\pi^2 \sigma_T x_e T_{\text{CMB},0}^4} \right)^{2/5} \sim 155 \,.
\end{alignat}
%

\subsubsection{Sensitivity to New Physics: Dark Matter Annihilation}
\label{subsubsec:sensitivity_to_new_physics}

Now suppose that the dark matter is made up of a particle that can undergo $s$-wave, velocity-independent annihilation to produce high-energy electrons and photons that can ultimately deposit their energy into the IGM. Let us make a parametric estimate of the impact of dark matter $s$-wave annihilation on the ionization history of the universe. The universe is fully ionized before recombination, which occurs at a redshift of $z_\text{re} \approx 1100$. Once recombination occurs, the universe transitions quickly into neutrality, and energy injected from annihilating dark matter has a possibility of eventually ionizing atoms in the gas. The rate at which annihilation occurs is given by $n_\chi \langle \sigma v \rangle$. Between recombination ($z_\text{re} \approx 1100$) and today, the probability of annihilation per dark matter particle is given by
\begin{alignat}{1}
    \int_{t_\text{re}}^{t_0} dt\, n_\chi \langle \sigma v \rangle = \frac{\rho_{\chi,0}}{m_\chi} \langle \sigma v \rangle \int_{z_\text{re}}^{0} dz  \, (1+z)^3 \frac{dt}{dz} \sim \frac{2 \rho_{\chi,0} \langle \sigma v \rangle}{3 m_\chi H_0} (1+z_\text{re})^{3/2} \,,
\end{alignat}
where $t_\text{re}$ and $t_0$ are the time at recombination and today respectively, and $\rho_{\chi,0}$ is the dark matter density today. For simplicity, we have made the approximation that the universe has been matter-dominated since recombination. 

From this, we immediately see that the expected energy released per baryon $\langle E \rangle$ due to annihilations from recombination until today is
\begin{alignat}{1}
    \langle E \rangle &\sim 2m_\chi \frac{n_\chi}{n_b} \cdot \frac{2 \rho_{\chi,0} \langle \sigma v \rangle}{3 m_\chi H_0} (1+z_\text{re})^{3/2} \sim \SI{10}{\eV} \left(\frac{\SI{1}{\giga\eV}}{m_\chi}\right) \left(\frac{\langle \sigma v \rangle}{\SI{3e-26}{\centi\meter\cubed\per\second}}\right) \,.
\end{alignat}
Keeping in mind that the ionization potential of hydrogen $\mathcal{R}$ is only \SI{13.6}{eV}, we have shown that the total expected number of annihilations since recombination of \SI{1}{\giga\eV} dark matter with a thermal cross section required to achieve relic abundance is sufficient to ionize an $\mathcal{O}(1)$ fraction of all hydrogen in the universe. This is driven by the fact that the release of the entire mass-energy of dark matter during annihilation typically dwarfs the atomic energy scales, allowing one annihilation event to provide sufficient energy to ionize roughly $ m_\chi/\mathcal{R}$ hydrogen atoms. 

A similar estimate holds for the thermal history as well. Baryons are held in thermal contact with the CMB until $z_\text{td} \sim 150$, after which Compton scattering between free electrons and CMB photons becomes too inefficient to maintain both the baryons and the CMB at the same temperature. Therefore, energy from dark matter annihilation after $z \lesssim z_\text{td}$ can potentially go toward heating the gas. 
\begin{alignat}{1}
    \langle E \rangle &\sim 2m_\chi \frac{n_\chi}{n_b} \cdot \frac{2 \rho_{\chi,0} \langle \sigma v \rangle}{3 m_\chi H_0} (1+z_\text{td})^{3/2} \sim \SI{0.5}{eV} \left(\frac{\SI{1}{\giga\eV}}{m_\chi}\right) \left(\frac{\langle \sigma v \rangle}{\SI{3e-26}{\centi\meter\cubed\per\second}}\right) \,.
\end{alignat}
The temperature of baryons just before the end of the dark ages is estimated to be \SI{e-3}{\eV}, which is much smaller than the energy that is potentially available to heat the gas from annihilations.

For dark matter annihilation or decay, these constraints are typically most useful for sub-\SI{}{\giga\eV} dark matter. For annihilation, the constraints are set mainly by $\langle \sigma v \rangle / m_\chi$, which is proportional to the energy injection rate from these annihilations; thermal dark matter candidates with $\langle \sigma v \rangle$ satisfying Eq.~(\ref{eqn:2to2_annxsec}) therefore inject more energy as they get lighter. Moreover, above the \SI{}{\giga\eV} scale, gamma-ray telescopes generally set more stringent constraints on these processes. 

Similar estimates also show that energy deposition from decaying or $p$-wave annihilating dark matter can have a significant impact on ionization and temperature levels in the early universe. For the case of $p$-wave annihilation, the annihilation cross section of dark matter scales as $\langle \sigma v \rangle \propto v^2$; this velocity dependence can arise from particle physics considerations, e.g.\ in the annihilation of millicharged scalar dark matter, and leads to a strong suppression of the annihilation rate during the cosmic dark ages. However, the formation of structures leads to an increase the the dark matter dispersion, which can result in a significant amount of energy deposition despite the velocity suppression. In general, $p$-wave annihilation and decay tend to result in large deviations in the cosmic ionization and temperature history at $z \lesssim 100$, and can be constrained effectively by probes of this epoch.

\subsection{New Directions in Cosmic Energy Deposition}

Eq.~(\ref{eqn:TLA_with_injection}) serves as the basis for understanding how dark matter processes affect the early universe, and also gives us a powerful way of constraining dark matter energy injection processes, setting the stage for Chapters~\ref{chap:DM_reionization},~\ref{chap:21cm_annihilation_decay} and~\ref{chap:DarkHistory}. 

The fact that dark matter processes can contribute to increased levels of ionization in the IGM naturally leads to the question of whether dark matter can contribute significantly to reionization~\cite{Chuzhoy2008,Natarajan:2010dc,Belikov:2009qx,Poulin2015,Lopez-Honorez:2013lcm,Mapelli:2006ej,Hansen:2003yj,Kasuya2004}. Conversely, the fact that reionization is known to be complete by $z \sim 6$ combined with constraints on the total optical depth to recombination can be used to constrain dark matter processes~\cite{Cirelli2009,Kaurov2015}. In Chapter~\ref{chap:DM_reionization} (based on Ref.~\cite{Liu:2016cnk}), I explain how to use the $f_c(z)$ results computed in Ref.~\cite{Slatyer:2015kla} together with a new calculation of boosts to the annihilation rate of dark matter due to structure formation to definitively study the potential contribution of dark matter annihilation ($s$-wave- or $p$-wave-dominated) or decay to cosmic reionization, via the production of electrons, positrons and photons. We map out the possible perturbations to the ionization and thermal histories of the universe due to dark matter processes, over a broad range of velocity-averaged annihilation cross-sections/decay lifetimes and dark matter masses. We extend the previous $f_c(z)$ calculations down to a redshift of $1+z = 4$ for two different reionization scenarios. We find that for dark matter models that are consistent with experimental constraints, a contribution of more than 10\% to the ionization fraction at reionization is disallowed for all annihilation scenarios. Such a contribution is possible only for decays into electron/positron pairs, for light dark matter with mass $m_\chi \lesssim \SI{100}{MeV}$, and a decay lifetime $\tau_\chi \sim 10^{24} - 10^{25}\SI{}{s}$. 

In Chapter~\ref{chap:21cm_annihilation_decay} (based on Ref.~\cite{Liu:2018uzy}), I will discuss the constraints on dark matter annihilation and decay that can potentially be set by the 21-cm global signal, a powerful new probe of the IGM temperature at the cosmic dawn. Motivated in part by recent claims of a detection of 21-cm absorption from $z \sim 17$ by the EDGES experiment~\cite{Bowman:2018yin}, we derive the constraints on dark matter annihilation and decay that can be placed in the presence of extra radiation backgrounds or effects that modify the gas temperature, such as dark matter-baryon scattering and early baryon-photon decoupling. We find that if the EDGES observation is confirmed, then constraints on light dark matter decaying or annihilating to electrons will in most scenarios be stronger than existing state-of-the-art limits from the cosmic microwave background, potentially by several orders of magnitude. More generally, our results allow mapping any future measurement of the global 21-cm signal into constraints on dark matter annihilation and decay, within the broad range of scenarios we consider. 

Finally, in Chapter~\ref{chap:DarkHistory} (based on Ref.~\cite{Liu:2019bbm}), I will present a new public Python package, \texttt{DarkHistory}, for computing the effects of dark matter annihilation and decay on the temperature and ionization history of the early universe. Prior to this work, $f_c(z)$ has largely been computed assuming the baseline ionization history, and are therefore applicable only so long as perturbations to the assumed ionization history from exotic sources of energy injection are sufficiently small. For $z \lesssim 100$, however, ionization levels that exceed the baseline value of $x_e \sim 2 \times 10^{-4}$ by several orders of magnitude are experimentally allowed~\cite{Liu:2016cnk}. \texttt{DarkHistory} simultaneously solves for the evolution of the free electron fraction and gas temperature, and for the cooling of annihilation/decay products and the secondary particles produced in the process. Consequently, we can self-consistently include the effects of both astrophysical and exotic sources of heating and ionization, and automatically take into account backreaction, where modifications to the ionization/temperature history in turn modify the energy-loss processes for injected particles. This represents a significant improvement over all previous calculations of this sort, which have used values of $f_c(z)$ calculated assuming the baseline recombination history with no exotic energy injection. We present a number of worked examples, demonstrating how to use the code in a range of different configurations, in particular for arbitrary dark matter masses and annihilation/decay final states. Possible applications of \texttt{DarkHistory} include mapping out the effects of dark matter annihilation/decay on the global 21cm signal and the epoch of reionization, as well as the effects of exotic energy injections other than dark matter annihilation/decay.



\usetikzlibrary{trees}
\usetikzlibrary{decorations.pathmorphing}
\usetikzlibrary{decorations.markings}
\tikzset{
    photon/.style={decorate, line width=0.15mm, decoration={snake,amplitude=3pt,segment length=8pt}, draw=black},
    wino/.style={draw=redwine},    
    fermion/.style={draw=black, line width=0.2mm, postaction={decorate},
        decoration={markings,mark=at position .55 with {\arrow[draw=black,scale=2,#1]{>}}}},
    scalar/.style={draw=black, dashed,postaction={decorate},
        decoration={markings,mark=at position .55 with {\arrow[draw=black,scale=2,#1]{>}}}},
    scalarline/.style={draw=black, postaction={decorate},
        decoration={markings,mark=at position .55 with {\arrow[draw=black,scale=2,#1]{>}}}},
    gluon/.style={decorate, draw=black,
        decoration={coil,amplitude=3pt, segment length=4pt}},
    graviton/.style={decorate, draw=black,
        decoration={zigzag,amplitude=3pt, segment length=4pt}}
}
\tikzstyle{blob}=[circle,
                              minimum size=20,
                              draw=black!80,
                              fill=black!80]   
\tikzstyle{redblob}=[circle,
                              thick,
                              minimum size=0.4cm,
                              draw=red!80,
                              fill=red!60]

\newcommand{\Ref}[1]{Ref.\:\cite{#1}}
\newcommand{\Refs}[1]{Refs.\:\cite{#1}}
\newcommand{\Fig}[1]{Fig.\:\ref{#1}}
\newcommand{\Figs}[2]{Figs.\:\ref{#1} and \ref{#2}}
\newcommand{\Tab}[1]{Table~\ref{#1}}
\newcommand{\Tabs}[2]{Tables~\ref{#1} and \ref{#2}}
\newcommand{\Sec}[1]{Sec.\:\ref{#1}}
\newcommand{\Secs}[2]{Secs.\:\ref{#1} and \ref{#2}}
\newcommand{\App}[1]{App.\:\ref{#1}}
\newcommand{\Apps}[2]{Apps.\:\ref{#1} and \ref{#2}}
\newcommand{\Eq}[1]{Eq.\:(\ref{#1})}
\newcommand{\Eqs}[2]{Eqs.\:(\ref{#1}) and (\ref{#2})}

\newcommand{\be}{\begin{eqnarray}}
\newcommand{\ee}{\end{eqnarray}}
\def\lsim{\mathrel{\rlap{\lower4pt\hbox{\hskip 0.5 pt$\sim$}}
\raise1pt\hbox{$<$}}}  
\newcommand{\dae}{DAE$\delta$ALUS}
\newcommand{\apr}{A^\prime}
\newcommand{\nova}{NO$\nu$A}
\newcommand{\pizero}{\pi^0}
\newcommand{\mpi}{m_{\pi^0}}
\newcommand{\DM}{\textrm{DM}}
\newcommand{\Br}{\textrm{Br}}
\newcommand{\alphaEM}{\alpha_{\rm{EM}}}
\newcommand{\ord}[1]{\mathcal{O}(#1)}

\newcommand{\cO}{\mathcal{O}}
\newcommand{\cE}{\mathcal{E}}
\newcommand{\cM}{\mathcal{M}}
\newcommand{\cR}{\mathcal{R}}
\newcommand{\BR}{{\rm BR}}
\newcommand{\sh}{\hat{s}}
\def\sfrac#1#2{{\textstyle{#1\over #2}}}
\newcommand{\sss}{\scriptscriptstyle}
\newcommand{\lra}{\leftrightarrow}

\newcommand{\dummyfig}[1]{
  \centering
  \fbox{
    \begin{minipage}[c][0.33\textheight][c]{0.5\textwidth}
      \centering{#1}
    \end{minipage}
  }
}

\newcommand{\dummyfiglong}[1]{
  \centering
  \fbox{
    \begin{minipage}[c][0.33\textheight][c]{0.9\textwidth}
      \centering{#1}
    \end{minipage}
  }
}

\newcommand{\mathbfit}[1]{\textbf{\textit{\textsf{#1}}}}
\def\pt         {\mbox{$p_{\rm T}$}\xspace}
\newcommand{\tev}{\ensuremath{\mathrm{\: Te\kern -0.1em V}}\xspace}
\newcommand{\gev}{\ensuremath{\mathrm{\: Ge\kern -0.1em V}}\xspace}
\newcommand{\mev}{\ensuremath{\mathrm{\: Me\kern -0.1em V}}\xspace}
\def\fb   {\ensuremath{\mbox{\,fb}}\xspace}
\def\invfb   {\ensuremath{\mbox{\,fb}^{-1}}\xspace}
\def\pythia     {\mbox{\textsc{Pythia}}\xspace}
\def\evtgen     {\mbox{\textsc{EvtGen}}\xspace}
\def\geant     {\mbox{\textsc{Geant}}\xspace}
\def\Dbar    {{\kern 0.2em\overline{\kern -0.2em \mathrm{D}}{}}\xspace}
\def\D       {{\ensuremath{D}}\xspace}
\def\Dz      {{\ensuremath{\D^0}}\xspace}
\def\Dzb     {{\ensuremath{\Dbar{}^0}}\xspace}
\def\A {{\ensuremath{A^{\prime}}}\xspace}
\def\mee {{\ensuremath{m_{e^+e^-}}}\xspace}
\def\epem {{\ensuremath{e^+e^-}}\xspace}
\def\smdecay {{\ensuremath{D^{*0}\!\to D^0\gamma}}\xspace}
\def\sigdecay {{\ensuremath{D^{*0}\!\to D^0\A}}\xspace}
\def\Adecay {{\ensuremath{\A \!\to e^+e^-}}\xspace}

\def\sigdecaypi {{\ensuremath{D^{*0}\!\to D^0\pi^0(\gamma \A)}}\xspace}
\def\jpsi     {{\ensuremath{{J\mskip -3mu/\mskip -2mu\psi\mskip 2mu}}}\xspace}
\def\kstarbar    {{\kern 0.2em\overline{\kern -0.2em K}{}^{*0}}\xspace}


\chapter{Enabling Forbidden Dark Matter}
\label{chap:nfdm}

\section{Introduction} \label{sec:intro}

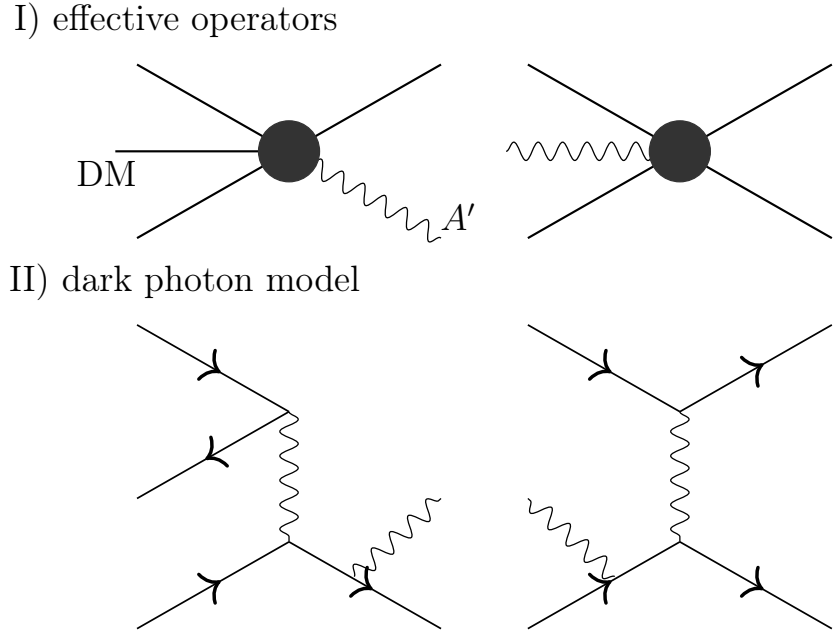
\begin{figure}
\begin{center}
\resizebox{0.75\textwidth}{!}{
\begin{tikzpicture}
\node at (-0.3,1.5) {I) effective operators};
\node at (-1.1,-0.25) {DM};
\node at (2.95,-0.75) {$A'$};
\draw[line width=0.25mm ] (-1,0)--(1,0);
\draw[line width=0.25mm ] (-0.75,1)--(1,0);
\draw[line width=0.25mm ] (-0.75,-1)--(2.75,1);
\draw[photon] (1,0)--(2.75,-1);
\node[blob] at (1,0) {};
\begin{scope}[shift={(4.5,0)}]
\draw[photon] (-1,0)--(1,0);
\draw[line width=0.25mm ] (-0.75,1)--(2.75,-1);
\draw[line width=0.25mm ] (-0.75,-1)--(2.75,1);
\node[blob] at (1,0) {};
\end{scope}
\begin{scope}[shift={(0,-4.5)}]
\node at (-0.2,3.) {II) dark photon model};
\draw[photon] (1,1.5)--(1,0);
\draw[fermion] (-0.75,2.5)--(1,1.5);
\draw[fermion] (1,1.5)--(-0.75,0.5);
\draw[fermion] (-0.75,-1)--(1,0);
\draw[fermion] (1,0)--(2.75,-1);
\draw[photon] (1.75,-0.40)--(2.75,0.5);
\end{scope}
\begin{scope}[shift={(4.5,-4.5)}]
\draw[photon] (1,1.5)--(1,0);
\draw[fermion] (-0.75,2.5)--(1,1.5);
\draw[fermion] (1,1.5)--(2.75,2.5);
\draw[fermion] (-0.75,-1)--(1,0);
\draw[fermion] (1,0)--(2.75,-1);
\draw[photon] (0.25,-0.40)--(-0.75,0.5);
\end{scope}
\end{tikzpicture} 
}
\end{center}
\caption{
Schematic description of {\it Not-Forbidden Dark Matter}~(NFDM) 
paradigm. 
I) effective operators for the $3\to 2$ scattering processes; 
II) explicit model described in the text:
vector-portal dark matter model.  
}
\label{fig:3to2plot}
\end{figure}

As we discussed in Chapter~\ref{chap:intro}, the observed relic abundance of dark matter (DM) may provide a clue to its non-gravitational interactions. In the conventional Lee-Weinberg scenario,   Many variations on the standard thermal freezeout scenario have recently been considered (e.g. \cite{Pospelov:2007mp,ArkaniHamed:2008qn, Hochberg:2014dra,Hochberg:2014kqa,Lee:2015gsa,
Hochberg:2015vrg, Bernal:2017mqb, D'Agnolo:2015pha, D'Agnolo:2015koa,Delgado:2016umt, Carlson:1992fn,Pappadopulo:2016pkp,Bernal:2015ova,Kuflik:2015isi,Bernal:2015xba,
Farina:2016llk, Dror:2016rxc, Okawa:2016wrr,
Bandyopadhyay:2011qm, D'Eramo:2010ep, Agashe:2014yua,Berger:2014sqa, Kopp:2015bfa, Berlin:2016vnh, Kopp:2016yji}); in this article, we point out that even for simple and weakly-coupled dark sectors, $3 \rightarrow 2$ annihilations -- as
illustrated in Fig.~\ref{fig:3to2plot} -- can play a critical role. 

For weakly-coupled DM, $3\rightarrow 2$ processes are usually considered to be
subdominant to their $2\rightarrow
2$ counterparts at the time of freezeout,  but if
the latter are kinematically suppressed while
$3\rightarrow 2$ is unsuppressed, the situation is more
complex. This can occur when the DM couples
to a ``mediator'' particle with a mass somewhat larger than
that of the DM itself, as might arise in a
hidden sector characterized by a single scale. 

Kinematic suppression
of $2\rightarrow 2$ annihilation, leading to a novel cosmological
history during freezeout, was previously invoked
in the 
``Forbidden DM'' \cite{D'Agnolo:2015koa} and ``Impeded DM''
\cite{Kopp:2016yji} scenarios; the new feature in our study is the
presence of a kinematically allowed dark-sector $3\rightarrow 2$ annihilation channel. We
refer to this scenario as {\it Not-Forbidden Dark Matter} (NFDM).
The $3\rightarrow 2$ channel is also important in 
the Strongly Interacting Massive Particle (SIMP) scenario
\cite{Hochberg:2014dra},  but work on SIMPs has focused on strongly
coupled theories with scalar DM \cite{Hochberg:2014kqa,Choi:2016tkj}, whereas 
NFDM is a more generic mechanism: it is potentially important in any situation where $2\rightarrow 2$ annihilations within the dark sector are kinematically suppressed, and has no obvious dependence on whether the DM is fermionic or bosonic or whether the dark sector coupling is strong or weak. Hidden sector or multicomponent DM models may have regions of parameter space where NFDM is an important mechanism to consider. 

We illustrate our paradigm with a Dirac fermion DM
charged under a hidden
$U(1)$ symmetry, with dark gauge boson $A'$.  This mediator can
provide a portal to the SM by having a small coupling to
the electromagnetic current $J_{\text{EM}}^\mu$ through a kinetic mixing term $(\epsilon/2)F'_{\mu\nu} F^{\mu\nu}$. In the mass basis, the Lagrangian becomes
\begin{alignat}{1}
  \mathcal{L} \supset -\frac{1}{4} F_{\mu\nu} F^{\mu\nu} - \frac{1}{4}F_{\mu\nu}' {F'}^{\mu\nu} + \frac{1}{2}
   { m_{A'}^2}\, {A'}_\mu {A'}^{\mu}
   + \bar{\chi} ( i \slashed{D} - m_\chi ) \chi   
         + e J_{\text{EM}}^\mu (A_\mu + \epsilon {A'}_{\mu})   \ . 
\label{model_def}
\end{alignat}
The gauge coupling is $\alpha' = g'^2/4\pi$, and $\slashed{D} \equiv \slashed{\partial} - ig' \slashed{A'}$. It is clear in this basis that there is no tree-level coupling between $\chi$ and the SM photon. We can also consistently assume that the dark Higgs boson giving mass to
$A'$ is very heavy and can be neglected in the effective
description \cite{Kahlhoefer:2015bea}. 
Depending upon the size of the kinetic mixing parameter $\epsilon$,
there are two possible regimes of interest:

(1) $\epsilon$ is
relatively large, such that the hidden sector and the SM sector 
have the same temperature before DM freezeout, 
while $\epsilon$ is still 
small enough so that $3 \to 2$ and $2 \to 2 $ reactions involving 
only hidden sector particles dominate over annihilation of $\chi$
to SM particles;

(2) For sufficiently small $\epsilon \lesssim 10^{-8}$,
the hidden sector will have its own temperature and in the limit
 $\epsilon \to 0$, 
it becomes secluded:  
both $\chi$ and $A'$ contribute to the ultimate DM density.

In
Sec.~\ref{Cosmology}, we  discuss the freezeout  history of
the NFDM model, 
and by solving the Boltzmann equations, we determine the dark sector
parameter values $\{m_\chi,\,m_{A'},\,\epsilon\}$ that
are consistent with the
observed relic density.   In Sec.~\ref{constraints} we incorporate
constraints from a variety of astrophysical and
laboratory searches, showing that a significant parameter region is 
allowed while realizing the NFDM mechanism.  Conclusions are given in Sec.~\ref{sec:Summary}. In the appendix,
we present a more detailed account of how the order of freezeout of
the various reactions determines the relic abundance;
the dependence of our results on the temperature of the dark sector; how 
the constraints change with $m_{A'}/ m_\chi$,
and cross sections for the relevant scattering processes.

\section{Cosmology}
\label{Cosmology}

Previous studies of the vector-portal DM model, shown in Eq.~(\ref{model_def}), 
have divided the parameter space into two broad regions: $m_\chi <
m_{A'}$ or  $m_\chi > m_{A'}$. In the latter case, the dominant
process at the epoch of thermal freezeout is $\chi \bar{\chi} \to A'
 A'$ followed by  $A' $ decays to SM particles, whereas when
$m_{\chi} < m_{A'}$, the $s$-channel annihilation  $\chi\bar\chi\to
f\bar f$ to SM particles  $f$ via off-shell $A'$ is dominant. 
This regime is ruled out for $m_\chi \sim$ MeV-GeV by
CMB constraints \cite{Adams:1998nr, Chen:2003gz, Padmanabhan:2005es, Ade:2015xua, Slatyer:2015jla}.

In the present work, however, we are interested in the intermediate region 
$m_{\chi } \lesssim m_{A'}$, where it is possible for  the $3\to 2$
scatterings $\chi \chi \bar{\chi} \to \chi A' $ or $\chi \bar\chi
 A' \to \chi  \bar{\chi}$ to have an important effect on the dark
matter abundance.  The system is governed by the coupled
Boltzmann equations for the $\chi$ and $A'$ densities. For $m_\chi \lesssim m_{A'}$, the relevant terms in these equations are
\begin{alignat}{2}
  \frac{ d n_\chi} { d t } + 3 H n_\chi &=&&  - \frac{1}{4} \langle \sigma v^2 \rangle_{\chi\chi\bar{\chi} \to \chi A'} \left(n_\chi^3 - n_{\chi,0}^2 n_\chi  \frac{n_{A'}}{n_{A', 0}}\right) \nonumber \\
  & &&+ \langle \sigma v \rangle_{A'A' \to \bar{\chi} \chi} \left(  n_{A'}^2  -  n_{A', 0}^2 \frac{n_\chi^2}{n_{ \chi,0}^2} \right) \,, \label{eq:boltz1} \\
  \frac{ d n_{A'}} { d t } + 3 H n_{A'} &=&& \,\, \frac{1}{8} \langle \sigma v^2 \rangle_{\chi\chi\bar{\chi} \to \chi A'} \left(  n_\chi^3 - n_{\chi,0}^2 n_\chi \frac{n_{A'}} {n_{A',0}}\right)  \nonumber \\
  & &&- \langle \sigma v \rangle_{A'A'\to \bar{\chi} \chi} 
  \left(  n_{A'}^2  -  n_{A',0}^2 \frac{n_\chi^2}{n_{\chi,0}^2}     \right) - \Gamma_{A'\to f\bar f}\left( n_{A'} -n_{A',0}\right) \,,
  \label{eq:boltz2}
\end{alignat}
where $n_\chi\,(n_{\chi,0})$ denotes the (equilibrium) density of 
$\chi+\bar\chi$,
and similarly $n_{A'}\,(n_{A',0})$ for the dark photon. Throughout this paper, we have assumed zero chemical potential for $\chi$ and $\overline{\chi}$, and take the densities of $\chi$ and $\overline{\chi}$ to be equal.
The $1/4$ in the first term of Eq.~(\ref{eq:boltz1}) is the symmetry factor for Dirac DM, taking into account the two identical particles in the initial state and the fact that each annihilation process removes a $\chi \overline{\chi}$ pair. The conjugate process $\chi \overline{\chi} \overline{\chi} \to \overline{\chi} A'$ is also accounted for in this factor. The relative numerical factors between the two equations are consistent with the way each process changes the number density of $\chi$ and $A'$; for example, the factor of $1/4$ and $1/8$ in the first terms of Eq.~(\ref{eq:boltz1}) and (\ref{eq:boltz2}) respectively are consistent with the fact that the $3 \to 2$ process has a net effect of removing a $\chi \overline{\chi}$ pair and producing a single $A'$. A detailed discussion of the derivation of the Boltzmann equation for $3 \to 2$ processes can be found in \cite{Bernal:2015bla}. 
 
Other $3 \to 2$ processes such as $\chi \overline{\chi} A' \to \chi \overline{\chi}$, $3A' \to \chi \overline{\chi}$ etc. are important only in the case of 
$m_{A'}/m_\chi < 1$ and $\epsilon=0$ in Sec.~\ref{sec:shd}. The complete Boltzmann equations containing all of these processes are shown in Eq. (\ref{eq:fullboltz1}) and (\ref{eq:fullboltz2}) in Appendix \ref{app:boltz}. All numerical results in this paper across the full range of $m_{A'}/m_\chi$ considered were obtained using the complete equations.  Expressions for 
the cross sections are given in Appendix \ref{app:xsects}. 

We will focus on the two regimes where (1) 
the hidden
sector and the SM remain in thermal equilibrium, requiring
values of the kinetic mixing $\epsilon \gtrsim 10^{-7}$ (but still
small enough to avoid dominance of the  $\chi \bar{\chi}  \to
e^+e^-$ process); (2) the hidden  sector is
secluded from the SM, $\epsilon \to 0$.

\subsection{Kinetic Equilibrium with the Standard Model}

\begin{figure*}[t!]
\centering
\includegraphics[scale=0.51]{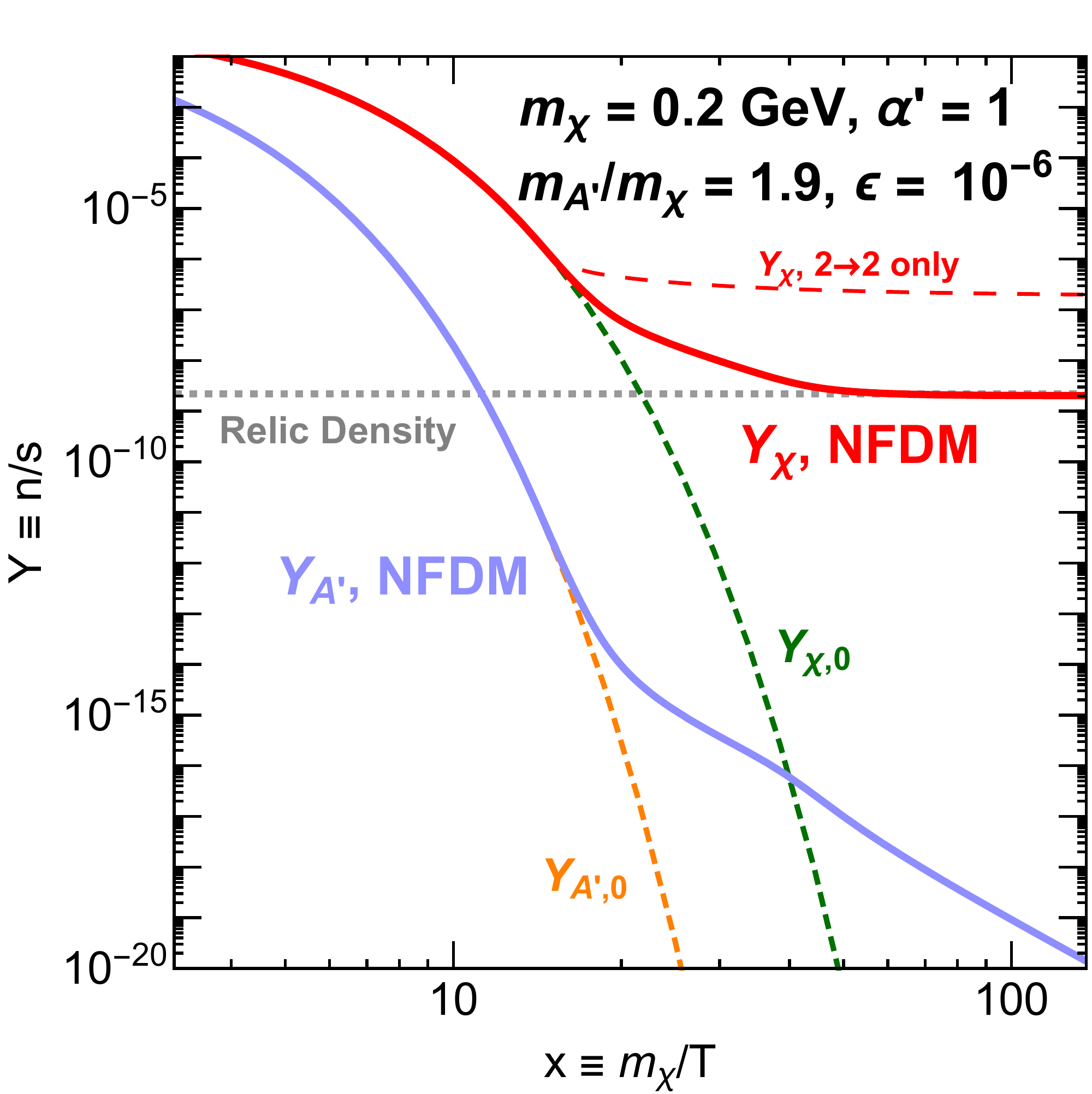}
\caption{Relic density in the NFDM scenario, assuming 
kinetic equilibrium of the dark sector with the SM. The evolution of the energy density of $\chi$ (red) and $A'$ (blue) for all processes (bold) and the corresponding energy density of $\chi$excluding the $3 \to 2$ process (red, dotted). The equilibrium distribution of $\chi$ (green) and $A'$ (orange) are also shown for
reference.}
\label{fig:23decay}
\end{figure*}

\begin{figure*}[t!]
\centering
\includegraphics[scale=0.66]{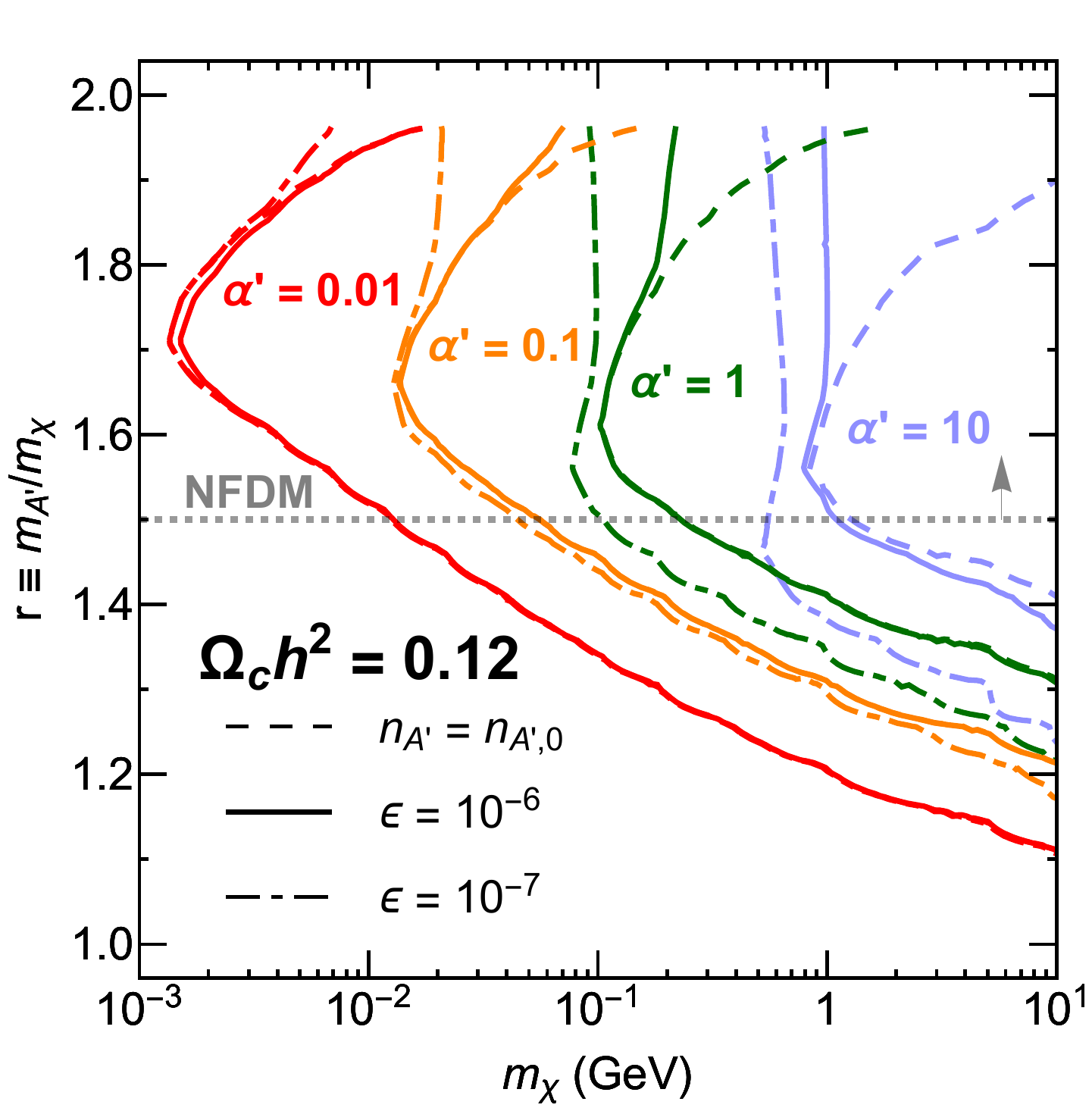}
\caption{Relic density in the NFDM scenario, assuming 
kinetic equilibrium of the dark sector with the SM. Contours of the observed present-day relic density in the $m_\chi$-$r$ plane for different values of the coupling constant $\alpha'$.}
\label{fig:23decaycontour}
\end{figure*}

For sufficiently large $\epsilon$, the scattering process
$\chi e^\pm\to \chi e^\pm$  
is fast enough to keep the dark and visible sectors in kinetic 
equilibrium, $T_{d} = T_{\text{SM}}$.  By comparing the  rate inferred from the $\chi e^\pm \to \chi e^\pm$ cross section to the Hubble rate $H$ at DM freezeout, we estimate the condition to be 
\begin{equation}   
   \epsilon \gtrsim 10^{-8} \left( \frac{0.1}{ \alpha'}    \right)^{1/2}
         \left( \frac{m_\chi} { 1\, \mathrm{GeV} } \right)^{1/2} \ , 
\label{eq:thermaleq}
\end{equation}   
taking $x_f \sim 20$, the $e^\pm$ to be
relativistic,  and 
$m_{A'} \simeq m_{\chi}$.
This leaves a significant range of 
$\epsilon \lesssim 10^{-8}-10^{-4}$, depending upon $m_\chi$, such 
that
$A'$-mediated annihilations $\chi\bar\chi\to e^+e^-$ are out of equilibrium
(a requirement of our scenario), as will be shown below.

We take the dark sector masses to be in the ranges $m_\chi \lsim 10 ~
\mathrm{GeV}$ and $ m_\chi \lsim m_{A'}  < 2 m_{\chi}$.  The lower
bound on $m_{A'}$ makes $\chi \bar{\chi} \to A' A'$ kinematically
inaccessible, while the upper bound forbids the $A'\to\chi\bar\chi$ decay
channel. If $m_{A'} > 2 m_\chi$, the number-changing process
$\chi\chi\bar\chi\to A' \chi$ effectively becomes number-conserving,
$\chi\chi\bar\chi\to \chi\chi\bar\chi$.
In terms of the parameter $r\equiv m_{A^\prime}/m_\chi$, 
the relevant range is thus $1 \lesssim r \lesssim 2$.

It is enlightening to compare the equilibrium  rates (per $\chi$
particle) of the $3\to 2$ and $2\to 2$ reactions in Eqs.
(\ref{eq:boltz1}) and~(\ref{eq:boltz2}),  $\Gamma_{\chi\chi\bar{\chi} 
\to \chi A'} \sim \langle \sigma v^2 \rangle_{\chi\chi\bar{\chi}
\to \chi A'} n_{\chi,0}^2$ and  $\Gamma_{A'A' \to \bar{\chi}
\chi}  \sim \langle \sigma v \rangle_{A'A' \to \bar{\chi} \chi} 
n_{A',0}^2 /n_{\chi,0}  $. From the exponential dependences in the
equilibrium number densities, $n_{i,0} \sim \exp (-m_i/T)$, we find that
if  $m_{A'} \gtrsim \frac{3}{2}  m_\chi$, the $3 \to 2 $ reaction will
be exponentially enhanced with respect to the $2\rightarrow 2$
reaction at low temperatures.

The Boltzmann equations are solved numerically, and the results shown
in Fig~\ref{fig:23decay}. As an example, Fig.~\ref{fig:3to2plot}
illustrates the evolution of the $\chi$ and $A'$ abundances as a function of
$x\equiv m_\chi/T$  with $m_\chi = 0.2\,$ GeV, gauge coupling $\alpha'=1$, 
kinetic mixing $\epsilon = 10^{-6}$ and the ratio $r = 1.9$. This example has been chosen to emphasize the importance of $3 \to 2$ scatterings, but similar results are obtained for $r \gtrsim 1.5$. Here, in the case with only $2\to2$ annihilation, the DM abundance would reach its relic value at $x_f \sim 20$; in our NFDM case, in contrast, the $3\to 2$ processes and decay of the $A'$ control the freezeout, and their interplay leads to an extended freezeout continuing to $x_f\sim 60$.
If we neglect the $3 \to 2$ process the resulting abundance is overestimated by
several orders of magnitude. It is noteworthy that $Y_{A'}$
departs from the equilibrium abundance at late times, even though the
rate for $A'\to e^+e^-$ exceeds the Hubble rate, because the $3 \to 2$ or $ 2 \to 2$ processes can also strongly affect the evolution of $n_{A'}$. 

In  Fig.~\ref{fig:23decaycontour} we plot the contours in the
$m_\chi$-$r$ plane matching the observed relic density 
\cite{Ade:2015xua}, for several values of $\alpha'$ and $\epsilon$. We consider values of $\alpha' \leq 4 \pi$, since every loop integral introduced in a Feynman diagram typically introduces an additional factor of $\alpha/4\pi$, and so perturbativity is naively maintained for this range of $\alpha'$.
$n_{A'} = n_{A',0}$ correponds to large $\epsilon$, where the rate for $A'\to e^+e^-$ 
dominates the rates for either of the two annihilation processes that generate $A'$s.
The region $r\lesssim 1.5$ corresponds to the Forbidden DM regime, and 
Ref.~\cite{D'Agnolo:2015koa} studied this regime with the assumption of $n_{A'} = n_{A',0}$: smaller values of $\epsilon$ show increasing deviation from the relic density contours obtained from this assumption, even for $r < 1.5$. For the rest of the paper, we will focus on the NFDM region $1.5\lesssim r<2$, where the $3\to 2$ process leads to a strong transition in the behavior of the relic density contour, with the exact value of $r$ for the transition depending on the coupling constant $\alpha'$. 

Normally the DM relic density is set by the strongest annihilation 
channel, which is also the last to freeze out, 
since only a single Boltzmann equation for DM is considered.  
This applies when $\epsilon$ is large, forcing 
$n_{A'} \simeq n_{A',0}$ (dashed contours). 
These contours turn to the right as
$r \to 2$ because the $3 \to 2 $ cross section
diverges, $\langle \sigma v^2 \rangle_{\chi\chi\bar{\chi} \to \chi A'  }
\propto {\alpha'}^3 m_\chi^{-5} ( r - 2 )^{-7/2} $, and
$Y_\chi \sim x_f^2 / [  m_{pl} m_\chi^2  \langle \sigma v^2
\rangle_{\chi\chi\bar{\chi} \to \chi A'}]^{1/2}$. Thus obtaining the correct relic density as $r \rightarrow 2$ requires a larger value of $m_\chi$.

In contrast, for moderate values of $\epsilon$, the NFDM mechanism
applies,  where the two coupled Boltzmann equations must be 
solved together. In general, we find that typically the {\it two strongest processes} (either annihilations or decays) keep the coupled system in equilibrium until the rate for one process (per $\chi$ particle) becomes comparable to the Hubble rate, and thus any weaker processes are not relevant for determining the relic abundance. In this regime, typically the decay of $A'\to e^+ e^-$ and either the $3\to2$ or $2\to2$ annihilation are the relevant processes. 
In particular, for $r\gtrsim 1.5-1.8$, the $3\to 2$ scatterings are faster than $2\to 2$, and so they dominate the freezeout, as shown in Fig.~\ref{fig:23decay}. The combination of $3\to 2$ scatterings and $A'$ decays can lead to a non-equilibrium density for the $A'$ particles during the freezeout of the $3\to 2$ process if $\epsilon$ is sufficiently small (e.g. $\epsilon \sim 10^{-6}-10^{-7}$), resulting in a lengthy freezeout and an $\epsilon$-dependent relic density. This behavior corresponds
to the divergence
of the dashed and solid contours in Fig.~\ref{fig:23decaycontour}
at large $r$.

\begin{figure*}[t!]
    \centering
    \includegraphics[scale=0.65]{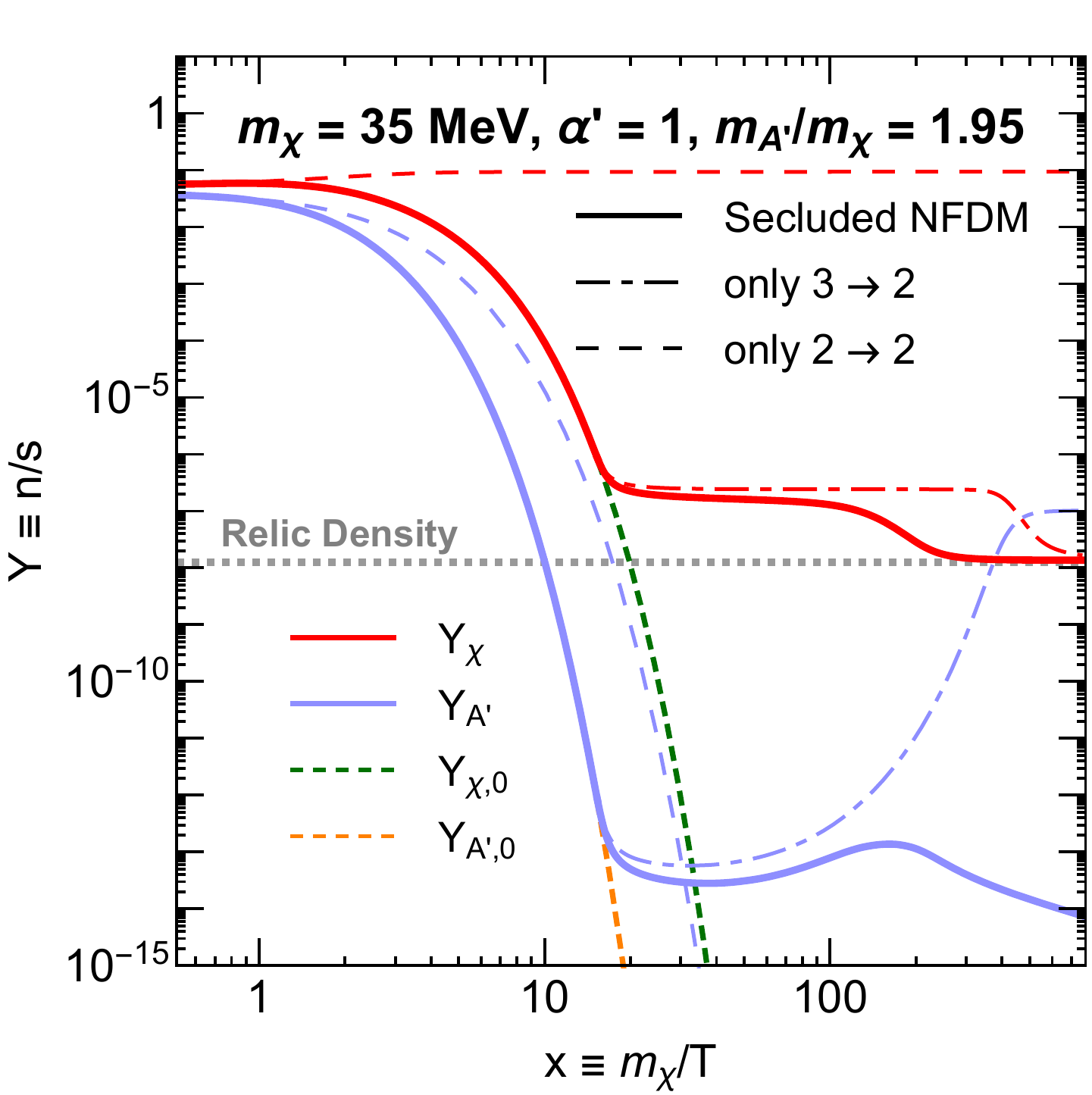}
    \caption{NFDM, secluded hidden sector. The evolution of energy
    density of $\chi$ (red) and $A'$ (blue) with (solid) all relevant
    processes; (dot-dashed) only $3 \to 2$ processes; (dashed) only $2 \to 2$ processes. The equilibrium distribution of $\chi$ (green) and $A'$ (orange) are also shown for reference.}
    \label{fig:eps0}
\end{figure*}

\begin{figure*}[t!]
    \centering
    \includegraphics[scale=0.48]{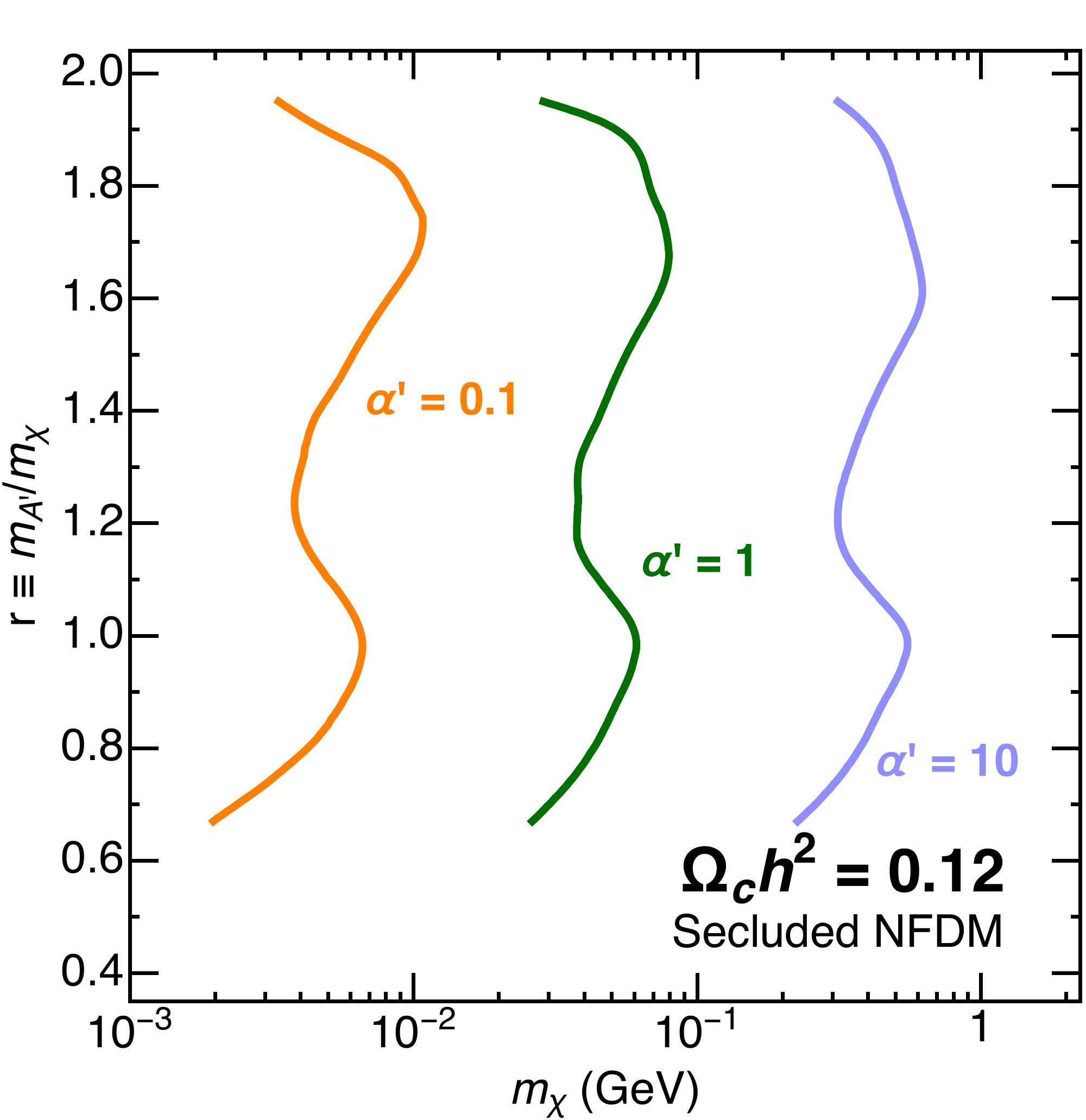}
    \caption{NFDM, secluded hidden sector. Contours of the observed
    present-day relic density in the $m_\chi$-$r$ parameter space for
    different coupling constants  $\alpha'$.}
    \label{fig:eps0contour}
\end{figure*}

\subsection{Secluded Hidden Sector}
\label{sec:shd}

Next we consider the limit of $\epsilon\to 0$, so that the dark 
photon is effectively stable, and the hidden sector is secluded.  
This analysis can be easily applied to multi-component DM models. Even though secluded hidden sectors are in general difficult to probe due to the lack of any interaction with the SM, they are not entirely impossible to study. Secluded hidden sectors can, for example, be constrained by the number of relativistic degrees of freedom during Big Bang Nucleosynthesis (BBN). Furthermore, in the U(1) theory considered here, the relic abundance is set by the coupling strength $\alpha'$, which in turn determines the self-interaction cross section in the dark sector. This cross section is a prediction of the model, and has observable consequences for structure formation, which can in principle be highly constraining.

Moreover, the secluded case is a useful limit that gives insight into the region of parameter space where $\epsilon$ is small but non-zero, so that kinetic equilibrium cannot be maintained with the SM. Despite the small couplings to the SM, this regime can still be effectively probed by observations of the cooling of SN1987a~\cite{Dent:2012mx}. The secluded limit is also highly instructive as an illustration of the rich behavior that can occur in the $U(1)$ vector portal DM model when the $2 \to 2$ and $3 \to 2$ annihilations are the dominant processes at freezeout. 

To avoid warm or hot dark matter \cite{Pappadopulo:2016pkp}, we assume that $\chi$ couples additionally to some relativistic degree of freedom $\phi$ until freezeout, strongly enough to maintain
thermal equilibrium in the dark sector so that the DM temperature
redshifts with the Hubble expansion in the conventional manner,
$T\sim 1/a$.  However, the coupling of $\phi$ to $\chi$ should be
sufficiently weak that annihilation of $\chi\bar\chi\to \phi\phi$ is
negligible during freezeout, to make the NFDM freezeout mechanism dominate over
conventional $2\to 2$ annihilation. 

For a concrete model of how this can be achieved, we take $\phi$ to be a light scalar charged under some additional $U(1)$ symmetry, interacting with the dark sector through the dimension-5 operator $(1/\Lambda) \overline{\chi} \chi \phi^* \phi$. The $T \sim 1/a$ dependence is maintained by $\chi \phi \to \chi \phi$ scatterings, which has a rate that scales as $n_\phi \langle \sigma v \rangle_{\chi\phi \to \chi \phi}$, while the $\chi \overline{\chi} \to \phi^* \phi$ rate scales as $n_\chi \langle \sigma v \rangle_{\chi \overline{\chi} \to \phi^* \phi}$. To obtain a parametric estimate for a value of $\Lambda$ that would maintain both $T \sim 1/a$ and subdominance to the $2 \to 2$ and $3 \to 2$ processes considered in Eq.~(\ref{eq:boltz1}) and (\ref{eq:boltz2}), we take $\langle \sigma v \rangle_{\chi\phi \to \chi \phi} \sim \langle \sigma v \rangle_{\chi \overline{\chi} \to \phi^* \phi} \sim 1/\Lambda^2$, and look for values of $\Lambda$ which ensure that the $\chi \overline{\chi} \to \phi^* \phi$ annihilation rate is subdominant up to the point of freeze-out of the two main processes. This condition is most difficult to satisfy in the case where $r = 2$, and the $2 \to 2$ rate becomes highly suppressed. Nevertheless, we find that in this limit, a suitable range of $\Lambda$ is $m_\chi^{4/3} m_{\text{pl}}^{2/3} \lesssim \Lambda^2 \lesssim m_\chi m_{\text{pl}}$, which for \SI{}{\giga\eV} dark matter corresponds to $10^6 \lesssim \Lambda/\SI{}{\giga\eV} \lesssim 10^9$. 

More generally, the dark sector has its own temperature $T_d$ which need not be the
same as that of the visible sector, $T_{\text{SM}}$; it is determined by
details of the thermal cosmological history such as the efficiency of
reheating into the dark sector after inflation.  The relic abundance
in this case depends upon the unknown parameter $\gamma \equiv T_d/T_{\text{SM}}$, 
but in a simple way: $Y_\chi \propto \gamma^{p(r)}$
where $p(r)\sim 1.6-1.8$ depends upon the mass ratio 
$r = m_{A'} / m_{\chi}$.  Here we illustrate the case of $\gamma=1$.

The evolution of $n_\chi$ and $n_{A'}$ for the secluded dark sector
is shown in 
Fig.\ \ref{fig:eps0}, taking $m_\chi =
\SI{35}{\mega\eV}$, $\alpha'=1$, $r=1.95$ as an example to illustrate the important interplay between the $2 \to 2$ and $3 \to 2$ interactions. 
Keeping only the $3\to 2$ reaction would predict that 
$A'$ becomes the dominant DM component, whereas in 
reality it remains highly subdominant.  
Again the freezeout process is prolonged, 
starting with the decoupling of $2\to 2$
scatterings at $x\sim 20$, while the $3\to 2$ reactions decouple
at $x\sim 150$.  Interestingly, $n_{A'}$ temporarily grows
between these two times, allowing the $2\to 2$ rate to
come back above $H$ just before freezeout completes.

In Fig.~\ref{fig:eps0contour}, we plot contours corresponding to
the observed thermal relic density in the $m_\chi$-$r$
plane for different values of $\alpha'$.  In the following we 
give a brief explanation of the contour shapes in the regions
$r \lesssim 1$, $1 \lesssim r \lesssim 1.5$ and  $1.5\lesssim r\lesssim 2$, which each show a distinct qualitative behavior:

(1) $ r \lesssim 1$. Being lighter than $\chi$, $A'$ is the dominant DM
constituent. The fastest process in this mass range is the $2 \to 2$ process $\chi \overline{\chi} \to A' A'$. Significantly below $r = 1$, the second fastest process is $3A' \to \chi \overline{\chi}$, since $n_{A',0} > n_{\chi,0}$. Near the threshold, with $n_{A',0} \sim n_{\chi,0}$, all of the other possible $3 \to 2$ processes ($\chi \chi A' \to \chi \chi$, $\chi \overline{\chi} A' \to \chi \overline{\chi}$, $\chi \overline{\chi} A' \to A' A'$, $\chi A' A' \to \chi A'$, as well as $\chi \chi \overline{\chi} \to \chi A'$ plus any conjugate processes) become important. The relic abundance curves in Fig.~\ref{fig:eps0contour} are computed with all of these processes taken into account in the complete Boltzmann equations shown in Eqs.~(\ref{eq:fullboltz1}) and~(\ref{eq:fullboltz2}).

(2) $1 \lsim r \lsim 1.5$. $\chi$ is the dominant DM 
component.  The fastest reaction is $A'A'\leftrightarrow\chi\bar\chi$,
and it enforces $n_{A'} = n_{A',0}n_\chi/n_{\chi,0}$ during the
freezeout, and the second fastest reaction is now
$\chi\chi\bar{\chi}  \to \chi A'$, which
determines the DM abundance.   The $3\to 2$ rate goes as $n_\chi^2
\langle\sigma v^2\rangle$, which depends only weakly on $r$ through
the phase space. Therefore there is no strong correlation between the
abundance and $r$ in this region. 

(3) $1.5\lsim r \lsim 2$. $\chi$ is the dominant DM  component, but
now its abundance is determined by the two freezeout events
$A'A'\to\chi\bar\chi$ (whose rate becomes comparable to Hubble at later times) followed by  $\chi\chi\bar{\chi}  \to \chi A'$.  At large
$r\lesssim 2$,  just before freezeout completes, both reactions are
faster than $H$, allowing one to estimate the freezeout times. Taking
the $2 \to 2 $ and $ 3 \to 2 $ rates $\sim H$, and  $n_{A'} \simeq
n_{A',0}\,n_\chi/n_{\chi,0}$ enforced by fast $3 \to 2$ scatterings, we
can analytically derive contours consistent with the numerical results.

\section{Constraints}
\label{constraints}
\begin{figure}[t!]
\centering
\includegraphics[scale=0.75]{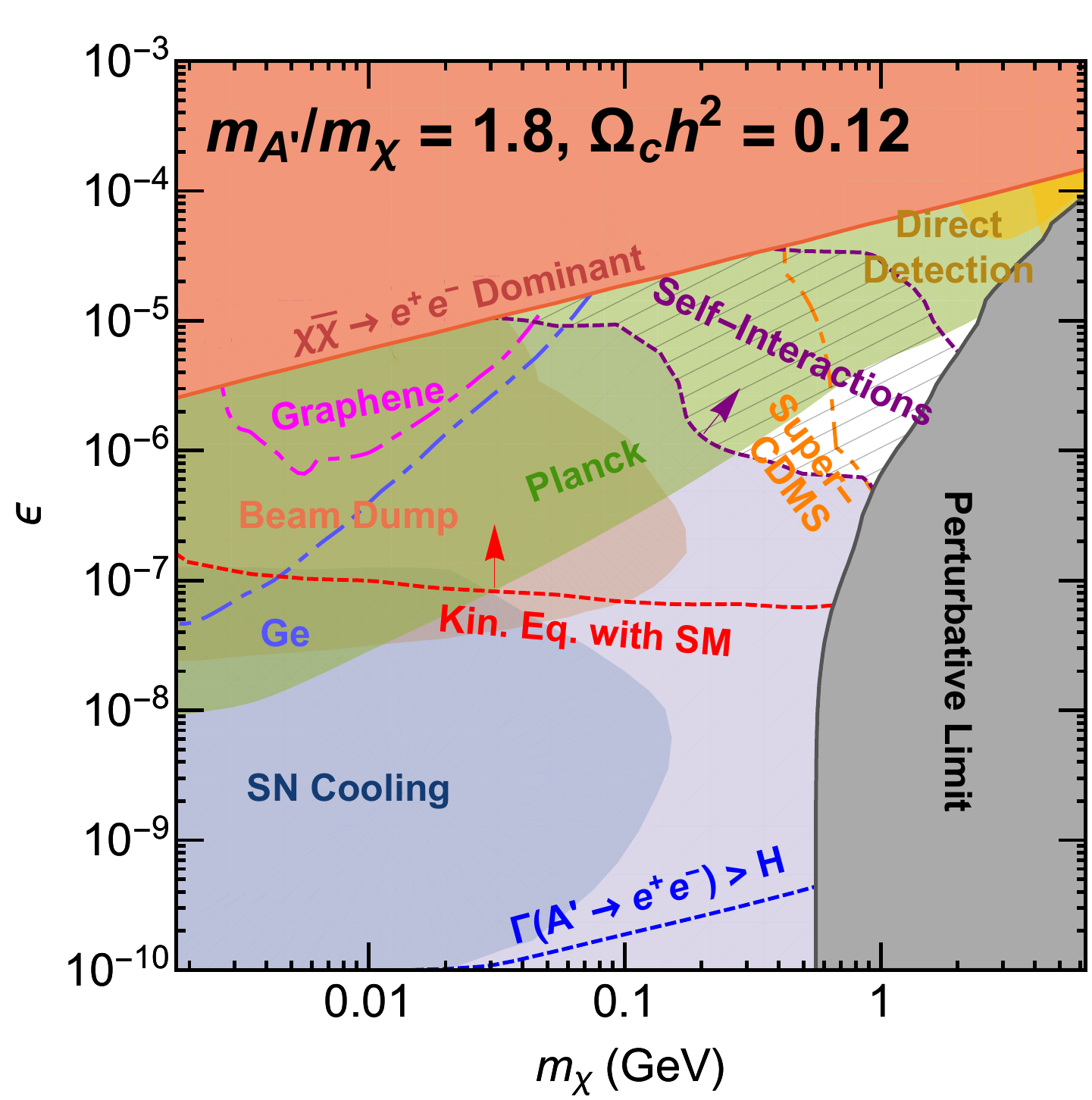}
\caption{Constraints in the $m_\chi$-$\epsilon$ plane for the case
of $m_{A'}/m_\chi = 1.8$, with  $\alpha'$ chosen to produce the
observed relic density. The allowed region is shown in white. The upper-left shaded region (red) indicates where freezeout
is dominated by the conventional $\chi\bar\chi\to e^+e^-$
annihilations.
Limits are derived from the
CMB power spectrum~\cite{Ade:2015xua} (green), beam dump 
experiments~\cite{Bjorken:2009mm,Andreas:2012mt} (pale orange),
SN1987a  cooling ~\cite{Dent:2012mx} (blue), direct 
detection~\cite{Tan:2016zwf,Akerib:2015rjg,Agnese:2015nto} (yellow) and
perturbativity, $\alpha'\geq 4\pi$ (gray). The projected
reach of SuperCDMS~\cite{Agnese:2016cpb} (orange
dot-dashed line), electron ionization of graphene~\cite{Hochberg:2016ntt} (magenta dot-dashed line) and germanium in a low-threshold experiment~\cite{Lee:2015qva} (blue dot-dashed line) are also shown. The curve near $\epsilon\sim 10^{-7}$ indicates where kinetic equilibrium  with
SM is established (red dashed line). The region of parameter space where the self-interaction cross section exceeds current limits ($\sigma/m_\chi > 1 \; {\rm cm^2/g}$) (purple), and the region where the self-interaction cross section can potentially solve the small-scale structure problems 
($0.1 \; {\rm cm^2/g} < \sigma/m_\chi < 1 \; {\rm cm^2/g}$) (purple dashed lines) are displayed. The purple arrow points into the region allowed by self-interaction bounds, above and to the right of the line. The $A'$ decay rate is
faster than $H$ at freezeout above the lowest (blue) curve.
}
\label{fig:constraints}
\end{figure}

The parameter space of NFDM is constrained by a variety of
experimental observations: i) dark photon limits coming from the cooling of
SN1987a~\cite{Dent:2012mx}; ii) similar bounds from beam dump
experiments~\cite{Bjorken:2009mm,Andreas:2012mt}; iii) limits on the
thermally-averaged cross section of $\chi \bar{\chi} \to e^+e^-$
deduced from the CMB power spectrum measured by
Planck~\cite{Ade:2015xua,Finkbeiner:2011dx,Slatyer:2009yq,Slatyer:2015jla,Liu:2016cnk},
and iv) direct detection constraints on the dark matter-nucleon
scattering cross section from PandaX-II~\cite{Tan:2016zwf}, LUX~\cite{Akerib:2015rjg} and
CDMSLite~\cite{Agnese:2015nto}. Although we have only assumed a coupling to electrons in much of this analysis for simplicity, these direct detection limits are relevant to the vector-portal DM model considered here.

Future direct detection experiments including SuperCDMS SNOLAB~\cite{Agnese:2016cpb}, as well as electron scattering off germanium~\cite{Derenzo:2016fse,Essig:2011nj,Lee:2015qva,Essig:2015cda} and graphene~\cite{Hochberg:2016ntt} are also shown in the same plot.  Other current limits from XENON10~\cite{Essig:2017kqs}, indirect detection~\cite{Essig:2013goa}
and lower bounds on $m_\chi$ from $N_\text{eff}$~\cite{Boehm:2013jpa}
are sub-dominant to the current constraints presented here and are not shown.

Fig.~\ref{fig:constraints} summarizes these constraints  in the
$m_\chi$-$\epsilon$ plane for the illustrative value of $r
= 1.8$, with $\alpha'$ fixed to give the correct present-day relic
density, subject to the perturbativity constraint $\alpha' \leq 4\pi$.
At a large $\epsilon$ and small $m_\chi$, conventional freezeout from 
$\chi \bar{\chi} \to e^+ e^-$ annihilations dominates over the NFDM
mechanism, but this is ruled out by the CMB constraint. The
approximately horizontal red dashed contour shows the minimum value of
$\epsilon$ for which the visible and dark sectors are in kinetic
equilibrium, estimated in Eq.~(\ref{eq:thermaleq}).

Self-interactions between dark matter particles with a cross section
$\sigma_\text{SI}\sim 0.1 \lesssim \sigma_\text{SI}/m_\chi \lesssim 1
\text{ cm}^2 \text{ g}^{-1}$ can potentially resolve the core-cusp and
the too-big-to-fail problems of small-structure formation with 
cold DM~\cite{Rocha:2012jg, Spergel:1999mh,
Zavala:2012us} while remaining consistent with experimental
constraints, which set an upper bound of between 1--\SI{2}{\centi\meter\squared\per\gram}~\cite{Harvey:2015hha,Robertson:2016xjh,Tulin:2017ara}. A DM mass of $m_\chi \sim (0.1 -
1)$\SI{}{\giga\eV} with $\epsilon \sim 10^{-7} - 10^{-6}$ in our model leads to a
velocity-independent self-interaction cross section that lies within this range, and can
provide a possible solution to both puzzles (though recent analysis of
clusters indicates some preference for a velocity-dependent cross
section~\cite{Kaplinghat:2015aga}).  The preferred region is
between the purple dashed lines in Fig.~\ref{fig:constraints}, while
the cosmologically constrained region is shown in purple.

\section{Summary and Outlook} 
\label{sec:Summary}

We have demonstrated a novel scenario called {\it Not-Forbidden Dark Matter}, where an allowed  $3 \to 2 $ 
annihilation process compensates for its conventional $2\to 2$
counterpart being kinematically forbidden during thermal freezeout.
This mechanism can be potentially important in a variety of hidden
sector models, including vector-portal, scalar-portal and composite
DM. The DM mass and the mediator (or second DM)
mass are of the same order, which would naturally arise in a hidden
sector characterized by a single scale.  

Taking the vector-portal DM model as an example,  we found that
in some parts of the NFDM  parameter space, the combined effect of $ 3
\to 2 $, $ 2 \to 2$ and $A'$ decay channels is to significantly prolong the period
of freezeout.  The commonly-neglected $3 \to 2$ annihilation channel
can change the predicted relic density by orders of magnitude. 
Although this model is restricted by an abundance of experimental
constraints, viable examples remain in the mass range $\sim
(0.1-1)\SI{}{\giga\eV}$, with a self-interaction cross section that is
coincidentally of the right order for solving the small scale
structure problems of  $\Lambda$CDM cosmological simulations. This is
a well-motivated target for future direct detection~\cite{Agnese:2016cpb,Essig:2011nj,Derenzo:2016fse} 
and dark photon searches~\cite{Ilten:2015hya,Ilten:2016tkc,Alekhin:2015byh,Gardner:2015wea,Moreno:2013mja}. 

While we were completing this work, Ref.~\cite{Dey:2016qgf} appeared, presenting a related idea.
Their work focuses on \SI{}{\kilo\eV}-\SI{}{\mega\eV} scalar DM and requires additional scalar
``assister'' particles.






\newcommand\cB{\mathcal{B}}

\chapter{Complementarity for Dark Sector Bound States}
\label{chap:bound_states}

\section{Introduction}
\label{sec:AltIntroduction}

The existence of dark matter~(DM) is well-established by observations of its gravitational effects. However, the particle nature of DM is still very much a mystery, despite the ongoing efforts of many complementary experimental searches. Constraints set by XENON~\cite{Aprile:2017iyp}, LUX~\cite{Akerib:2016vxi} and PandaX~\cite{Cui:2017nnn} have strongly ruled out generic DM candidates that interact in a spin-independent manner through a $Z$-exchange, and are now starting to probe Higgs-mediated interactions (e.g.~\cite{Escudero:2016gzx}).
These direct detection experiments are complemented by dark sector searches at colliders. 
The main DM search strategy at the Large Hadron Collider~(LHC) is based on missing transverse momentum (MET) balanced by a jet, electroweak~(EW) gauge boson or Higgs, known generically as mono-$X$ searches. 
Searches for dijet or dilepton resonances, while not directly probing the existence of DM, can also effectively constrain models where a mediator particle is responsible for interactions between the Standard Model~(SM) and a ``dark sector'' containing the DM, limiting the parameter space for the mediator. 
Finally, indirect searches for DM annihilation or decay to SM particles, as well as the well-measured relic abundance of the DM, set powerful limits on the strength and nature of the interaction of DM with the SM. Any model of DM must successfully contend with all of these constraints.

With no hint yet of what the dark sector may look like, we might look to the SM for clues as to its possible composition and structure. 
In this light, we should not be surprised to find bound states in the dark sector; after all, bound states are ubiquitous in the SM, and even the simplest dark sector models with a DM candidate and a force carrier can potentially support the existence of bound states. Asymmetric dark matter may even form atomic bound states, mediated by a hidden U(1) gauge symmetry, in complete analogy with the SM~\cite{Kaplan:2009de,Kaplan:2011yj,Baldes:2017gzu}.
Dark sector bound states, much like QCD bound states, may be produced when a pair of heavy dark sector particles are produced close to their kinematic threshold and have a sufficiently strong attractive interaction between them. The subsequent decay of these bound states into lighter SM particles can lead to distinctive signatures at the LHC. This strategy has been studied in the context of bound states formed by supersymmetric~(SUSY) particles, and has been shown to be a potential search channel at the LHC~\cite{Drees:1993uw,Martin:2008sv,Kats:2009bv,Kats:2012ym}, capable of probing regions of parameter space where traditional searches are challenging.

Dark sector bound states and their potential collider signatures have been studied extensively in the literature. 
Bound states formed from weakly-interacting massive particles~(WIMPs) that are charged under the SM SU(2)$_L$ $\times$ U(1)$_Y$ gauge group or non-SM forces, known as WIMPonium~\cite{Shepherd:2009sa}, can be detected at the LHC through resonant decays into a pair of leptons, provided the coupling to the mediator which supports the bound state is large enough. 
Other model-specific dark sector bound state collider searches that have been proposed include searches for higgsino bound states in $\lambda$-SUSY and bound states within the self-interacting DM framework~\cite{Tsai:2015ugz}; DM bound states in a U(1) vector portal model decaying into multilepton final states, which can be searched for at $B$-factories~\cite{An:2015pva}; dark sector baryons of a new confining gauge interaction~\cite{Mitridate:2017oky}, and asymmetric DM bound states in a Higgs portal model with decays to electrons \cite{Bi:2016gca}. Mono-photon searches at lepton colliders can also potentially be used to probe the full resonance structure of the dark sector~\cite{Hochberg:2017khi}. 
However, the large couplings typically required for detectable bound states often predict large signals in direct detection experiments, especially if the light force carrier responsible for the bound state formation also couples to the SM; likewise, in this light-mediator regime, searches for the mediator are often a more promising dark-sector discovery channel than searches for the bound states~\cite{Han:2007ae}.

In this chapter, we broadly explore the challenges of building a dark sector model which can be discovered through the production of a bound state at the LHC, in light of the current stringent and complementary experimental constraints. 
Direct detection limits can be evaded in models with TeV-mass DM if the DM candidate only has an off-diagonal coupling to the SM that couples the DM, the mediator and a heavier dark sector state, so that at tree-level, the DM only scatters into this heavier state when interacting with the SM~\cite{ TuckerSmith:2004jv,TuckerSmith:2001hy}. 
At the LHC, dark sector particles can be produced on their kinematic threshold and form a bound state $\mathcal{B}$, which can subsequently undergo annihilation decay into a pair of SM leptons, showing up as a dilepton resonance at the LHC.\footnote{Di-jets are also a plausible search strategy, but the backgrounds and triggers make this much more challenging to explore.}

We will show that in models where the mediator between the SM and the dark sector couples to two different states in the dark sector, it is possible to arrange for such a resonance to occur and have a substantial branching ratio into SM leptons. 
In these scenarios, searches for a dilepton resonance from $\mathcal{B}$ are complementary to the existing mono-$X$ and vector resonance searches that are already deployed for dark sector searches at the LHC, with the ability to probe higher mass scales for the mediator and DM. Since $\mathcal{B}$ can have the same quantum numbers as the SM mediator, we explore the importance of mixing between bound states and mediator particles with equal quantum numbers and similar masses.

Models with bound states that are detectable at the LHC can also possess large indirect signals, as the long-range potential implied by the existence of bound states generically enhances the annihilation cross section for slow-moving DM particles, and the bound state formation and decay can also serve as an annihilation  channel. We will study the constraints from indirect detection and cosmology that result from considering these effects.

The rest of the chapter is structured as follows. 
In Sec.~\ref{sec:PhenoBound}, we will make some remarks on the general features of dark sector models where the bound-state resonance search is the most sensitive channel.
We will discuss why the bound-state resonance search is complementary to the current dark sector search strategies used by the LHC experiments for such models, and discuss their general phenomenology in direct, indirect and collider DM searches. 
In Sec.~\ref{sec:darksector} we will lay out some specific models containing bound states in the dark sector and study their phenomenology. We will first discuss the MSSM in the pure wino/higgsino limit, which already meets some of the criteria needed for a successful model with bound states, although the production rate at the 13\,TeV LHC is too small for detection. We will then discuss two vector portal models which realize the requirements needed for a viable dark sector with bound states to be probed by the LHC. In Sec.~\ref{sec:experimentalConstraints} we compute and discuss the potential experimental signatures of these models.
Our conclusion will then follow in Sec.~\ref{sec:dis}. 

\section{Phenomenology of Bound States}
\label{sec:PhenoBound}

The existence of DM bound states has implications for the phenomenology of the dark sector, and for its signatures in direct, indirect and collider searches. In this section, we consider the circumstances under which collider searches for bound states can probe otherwise unexplored regions of DM parameter space. Aside from these searches, DM bound states with long lifetimes have also recently been shown to have potentially interesting implications for neutrino experiments~\cite{Grossman:2017qzw}.

As we will show, models where bound-state resonance searches at the LHC probe new regions of parameter space are most easily realized in the presence of several common features:
\begin{enumerate}
\item DM couples to at least two distinct force carriers; one of these, $Y$, is light and mediates the bound state formation, while the other, $V$, is heavier and couples appreciably to the SM. The constraints from LHC resonance searches of the bound state are most competitive when the SM mediator $V$ is heavier than twice the DM mass;
\item the coupling of the DM to the light mediator, which we denote $\alpha_\cB \equiv g_\cB^2/4\pi$, should be fairly large, as the bound state production rate is proportional to the third power of this parameter;
\item decay of $s$-wave bound states with the same spin as the heavy mediator into a pair of light mediators is suppressed, so that decays through the heavier mediator into two SM fermions dominate; and
\item the relevant spin-independent direct detection cross section is suppressed, e.g.\ by loops, by momentum-dependent factors, or by small couplings. This is particularly easy to achieve in models where the DM is part of a multiplet with small mass splittings, and the heavy mediator has an off-diagonal coupling to the mass eigenstates, so that elastic scattering off nuclei occurs only at one-loop level. 
An alternate approach to this criterion would be to consider flavor-dependent couplings between $V$ and the quarks. 
\end{enumerate}
The mono-$X$ process, resonant production of the mediator $V$ and the resonant production of the bound state $\mathcal{B}$ are the main collider signatures of this general setup, and are depicted in Fig.~\ref{fig:feynmanMonoXAndResonance}. When discussing generic models, we will denote the heavy mediator as $V$ and its mass by $m_V$, and the light mediator by $Y$ and its mass as $m_Y$ (for ``Yukawa''). 
In the example models we present, $V$ will be a vector in all cases, but $Y$ can be either a scalar or vector. 
In principle, $V$ could also be a scalar (or a scalar bound state can mix directly with the Higgs sector \cite{Bi:2016gca}), but we will leave the analysis of such scenarios to future work; as we will see, a vector mediator facilitates a sizable production cross section and a large branching ratio to leptons, while evading direct detection bounds.

\begin{table}[t!]
    \begin{center}
    \begin{tabular}{c}
    	\begin{tikzpicture}
    		\begin{feynman}
    			\vertex (a1) {\(q\)};
    			\vertex[below right=0.75cm and 0.4375cm of a1] (jet);
    			\vertex[above right=0.75cm and 0.75cm of jet] (jetEnd) {\(X\)};
    			\vertex[below right=1.2cm and 0.7cm of a1] (c1);
                \vertex[left=0.2cm of c1] (gqlabel) {\(g_q\)};
    			\vertex[below left =1cm and 0.5cm of c1] (a2) {\(\overline{q}\)};
    			\vertex[right=1.5cm of c1] (c2);
                \vertex[right=0.2cm of c2] (gchilabel) {\(g_\chi\)};
    			\vertex[above right=1cm and 0.5cm of c2] (b1) {\(\chi\)};
    			\vertex[below right=1cm and 0.5cm of c2] (b2) {\(\overline{\chi}\)};
    			\diagram*{
    				(a1) -- [fermion] (jet);
    				(jet) -- [fermion] (c1);
    				(jetEnd) -- [scalar] (jet);
    				(c1) -- [fermion] (a2); 
    				(c1) -- [boson, edge label'=\(V^{(*)}\)] (c2);
    				(c2) -- [fermion] (b1);
    				(b2) -- [fermion] (c2);
    			};
    		\end{feynman}
    	\end{tikzpicture} \\
    	\begin{tikzpicture}
    		\begin{feynman}
    			\vertex (a1) {\(q\)};
    			\vertex[below right=1.2cm and 0.7cm of a1] (c1);
    			\vertex[below left =1cm and 0.5cm of c1] (a2) {\(\overline{q}\)};
    			\vertex[right=1.5cm of c1] (c2);
    			\vertex[above right=1cm and 0.5cm of c2] (b1) {\(q, \ell^-\)};
    			\vertex[below right=1cm and 0.5cm of c2] (b2) {\(\overline{q}, \ell^+\)};
    			\diagram*{
    				(a1) -- [fermion] (c1);
    				(c1) -- [fermion] (a2); 
    				(c1) -- [boson, edge label'=\(V\)] (c2);
    				(c2) -- [fermion] (b1);
    				(b2) -- [fermion] (c2);
    			};
    		\end{feynman}
    	\end{tikzpicture} \\
        \begin{tikzpicture}
          \begin{feynman}
            \vertex (a1) {\(q\)};
            \vertex[below right=1.2cm and 0.7cm of a1] (c1);
            \vertex[below left =1cm and 0.5cm of c1] (b1) {\(\overline{q}\)};
            \vertex[right=0.9cm of c1] (c3);
            \vertex[above right=0.7cm and 0.5cm of c3] (a2);
            \vertex[above =0.1cm of a2] (alphaBlabel) {\(\alpha_{\mathcal{B}}\)};
            \vertex[right=0.7cm of a2] (a3);
            \vertex[right=0.3cm of a3] (a4) {\( \)};
            \vertex[below right=0.7cm and 0.5cm of c3] (b2);
            \vertex[right=0.7cm of b2] (b3);
            \vertex[right=0.3cm of b3] (b4) {\( \)};
            \vertex[below right=0.7cm and 0.8cm of a4] (c4) {\(\cdots\)};
            \vertex[right=0.5cm of a4] (d1);
            \vertex[right=0.8cm of d1] (d2);
            \vertex[right=0.5cm of b4] (e1);
            \vertex[right=0.8cm of e1] (e2);
            \vertex[below right=0.7cm and 0.5cm of d2] (f1);
            \vertex[right=0.9cm of f1] (g1);
            \vertex[above right=1cm and 0.5cm of g1] (h1) {\(q, \ell^-\)};
    		\vertex[below right=1cm and 0.5cm of g1] (h2) {\(\overline{q}, \ell^+\)};

            \diagram*{
                (a1) -- [fermion] (c1);
                (c1) -- [fermion] (b1);
                (c1) -- [boson, edge label'=\(V^{(*)}\)] (c3);
                (c3) -- [fermion] (a2);
                (b2) -- [fermion] (c3);
                (a2) -- [fermion] (a3);
                (a3) -- (a4);
                (b3) -- [fermion] (b2);
                (b4) -- (b3);
                (a2) -- [scalar, edge label=\(Y\)] (b2);
                (a3) -- [scalar] (b3);
                (d2) -- [scalar] (e2);
                (d1) -- [fermion, edge label=\(\chi\)] (d2);
                (e2) -- [fermion, edge label=\(\overline{\chi}\)] (e1);
                (d2) -- [fermion] (f1);
                (f1) -- [fermion] (e2);
                (f1) -- [boson, edge label'=\(V^{(*)}\)] (g1);
                (g1) -- [fermion] (h1);
                (h2) -- [fermion] (g1);
            };
            
            \draw[decoration={brace}, decorate] (a4.north east) -- (b4.south east) node[pos=0.5, right] {\(\cB\)};
          \end{feynman}
        \end{tikzpicture}
    \end{tabular}
    \end{center}
    
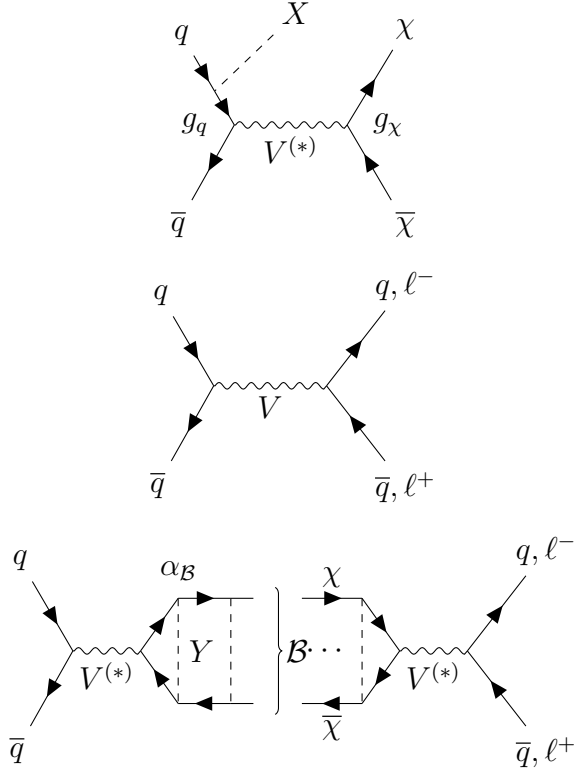
\captionof{figure}{Feynman diagrams for relevant dark sector processes at colliders. These processes are (top) the mono-$X$ process, (middle) the resonant production of $V$ decaying into a pair of jets or leptons, and (bottom) the resonant production of $\cB$, subsequently undergoing a similar decay. The coupling of the mediator between the dark sector and the SM to quarks ($g_q$) and to the DM ($g_\chi$), as well as the coupling responsible for the Yukawa potential that forms the bound state $\mathcal{B}$ ($\alpha_{\mathcal{B}}\equiv g_\cB^2/4\pi$) are shown. In our models, $V$ is always a vector, while $Y$ can be either a scalar or a vector.}
    \label{fig:feynmanMonoXAndResonance}
\end{table}

Many of the earlier works in the literature on bound states exhibit some of these features. 
Both~\cite{Shepherd:2009sa} and~\cite{Tsai:2015ugz} introduce an additional mediator to support the bound state formed from DM charged under the EW gauge group, so that the couplings between the DM to the light mediator can be made large. An additional mediator was also introduced in~\cite{Grossman:2017qzw} to alleviate the tension between a suitably light mediator that can support a bound state and the need for a massive enough SM mediator that can decay into electron pairs. In~\cite{An:2015pva}, direct detection limits are avoided by having sub-\SI{}{\giga\eV} DM. 
Furthermore, there is only one vector boson to mediate both the bound state formation and the interaction with the SM, at the cost of allowing the bound state to decay into 4- or 6-lepton final states. This is an important signature in $B$-factories for DM with a mass on the order of a GeV~\cite{An:2015pva}. In principle, this scenario can be probed at the LHC by multi-lepton searches, or by di-photon searches where two $e^+e^-$ pairs are detected as fake photons~\cite{Bi:2016gca}. However, multi-lepton signatures turn out to be relatively unimportant for the kinetic mixing models that we will study later.

We will demonstrate that the characteristics listed above can be achieved in Higgsed dark-sector models, with a vector portal between the dark sector and the SM. But before we give examples of such models, we will first discuss each of these criteria in more detail.

\subsection{General Model Building Considerations}
\label{sec:modelBuilding}

The existence of DM bound states in a Yukawa potential with range $1/m_Y$ is only possible if~\cite{PhysRevA.1.1577, An:2015pva}:
\begin{align}
	\label{eqn:boundStateRequirement}
	\frac{\alpha_\cB \, m_\chi}{m_Y} > 1.68,
\end{align}
where $m_\chi$ is the DM mass. Thus, the presence of a bound state supported by scalar or vector exchange requires a relatively light force carrier -- certainly lighter than the dark matter itself, for weak couplings. For more complex dark sectors with potentials that couple multiple two-particle states (e.g.\ the neutralino sector of supersymmetric models), the details of this criterion may be modified, but it is still generically true that there must be a force with range longer than the Bohr radius of the bound state, i.e. there should be at least one mediator with $m_Y \lesssim \alpha_\mathcal{B} m_\chi$.

If this force carrier is also the mediator between the DM and the SM, then searches for the force carrier will generally offer a more accessible probe of dark-sector physics than searches for the heavier DM, both because the force carrier is lighter and because it couples directly to SM particles (see e.g.~\cite{Han:2007ae}). 
This leads us to consider models where there are at least two distinct particles that couple to the DM, one which has appreciable interactions with the SM (and can be heavier than the DM itself), and the other of which mediates the bound state formation and so must be light.

One alternative to this structure is the case where the DM is charged under the SM SU(2)$_L$ EW gauge group, and the photon, $W$ and/or $Z$ support the bound state; this is possible, for example, for bound states consisting of neutralinos and/or charginos~\cite{Asadi:2016ybp}. 
However, as we will show later in this work, at present-day colliders the production rate for such EW bound states is undetectably low. 

Returning to dark-sector models with at least two mediators, the presence of the light mediator has some immediate implications. 
First, the DM will generically annihilate into the light mediators. If these mediators are absolutely stable, they will constitute some fraction of the DM relic density, which must be sufficiently small; if they are below the $\sim$ MeV scale in mass, they may be constrained by limits on the number of effective relativistic degrees of freedom in the early universe (e.g.~\cite{Nollett:2013pwa,Nollett:2014lwa}). 
We will generally assume that these mediators decay through some small mixing with the SM, on timescales less than one second, so that they do not affect Big Bang nucleosynthesis; in this case, while the coupling can be made small enough that these mediators do not contribute substantially to collider signals and direct detection, indirect detection constraints from this annihilation channel must be considered. For a weak enough mixing, a displaced vertex search can be a good search strategy if the bound state decays primarily into these light mediators \cite{Tsai:2015ugz,Li:2017xyf}.

Resonance searches for bound states will typically become difficult when there is a significant branching ratio of the bound state into dark sector states (while the dark sector states could decay promptly to SM particles as in Ref.~\cite{An:2015pva}, for the heavy DM models considered in this chapter, which are most relevant for LHC searches, we expect this signature to be relatively unimportant, as we will explain later). Thus, a model where collider searches for bound state resonances are effective must ensure that the branching ratio of the bound state (formed at a collider) into two light mediators is small relative to the decay of the bound state via the off-shell heavy mediator into a pair of SM particles. 
One way that this can be achieved is if both the heavy and light mediators are vectors, then suppression of the spin-1 bound state decay to two light mediators is automatic: charge parity symmetry forbids the decay of a spin-1 $s$-wave bound state into two vectors, so any decays into dark sector vectors must be a 3-body process.
In fact, decays into any number of the light mediators can be completely forbidden if the bound state is formed not from a particle-antiparticle pair, but from two different fermions in the bound state with nontrivial quantum numbers, which cannot be conserved if the bound state could decay into states containing only light mediators. This behavior is natural in cases where the mediator $V$ couples off-diagonally to the multiplet containing the DM. Models of this type have additional advantages in evading constraints from direct detection. 

Note that if the mediator to the SM is a vector, then the bound states formed at the LHC by resonant production will dominantly be spin-1 $s$-wave states; if the mediator is a scalar, they will instead dominantly be spin-0 $s$-wave states. The spin and angular momentum of the bound states determine their possible decays.

\subsection{Vector-Bound State Mixing}
\label{sec:VBMixing}

When $V$ and $\mathcal{B}$ have similar masses or the coupling between $V$ and the constituents of $\mathcal{B}$ is large, significant mixing can occur between the two states if they have the same quantum numbers. Both the $V$-resonance and $\mathcal{B}$-resonance diagrams in Fig.~\ref{fig:feynmanMonoXAndResonance}, together with higher order diagrams with more inter-conversions between $V$ and $\mathcal{B}$, need to be re-summed. The new mass eigenstates that result from the mixing have masses and widths that are shifted with respect to their unmixed values by an amount determined by the strength of the mixing.

The formalism that accounts for the mixing was used to study $Z$-toponium mixing \cite{Kuhn:1985eu,Kuhn:1987ty,Gusken:1985tf,Hall:1985jf,Franzini:1987jw}, and more recently to study Higgs-stoponium mixing \cite{Bodwin:2016whr}. The mixing shifts the masses and widths of the unmixed states, denoted $V_0$ and $\mathcal{B}_0$, to new values given by the eigenvalues of the following mass matrix:
\begin{alignat}{1}
    \mathcal{M} = \begin{pmatrix}
        m_{V,0}^2 - i m_{V,0} \Gamma_{V,0}(s) & -f \\
        -f & m_{\mathcal{B},0}^2 - i m_{\mathcal{B},0} \Gamma_{\mathcal{B},0}(s)
    \end{pmatrix},
\end{alignat}
where all masses and widths are for the unmixed states, and $f$ is a model-dependent parameter determined by the coupling between $V_0$ and $\mathcal{B}_0$.

If $f$ is small compared to the difference in the diagonal entries (see Eq.~(\ref{eqn:tan2theta}) below), the final mixed states $V$ and $\mathcal{B}$ are approximately their respective initial unmixed states, up to higher order corrections. The width of $V_0$ should be evaluated at the appropriate energy scale $\sqrt{s}$ at which the final mixed resonances $V$ or $\mathcal{B}$ are produced; this scale dependence is important especially when $m_\mathcal{B}$ lies below the $\chi \overline{\chi}$ open production threshold while $m_{V,0}$ lies above it. 
The width of $\mathcal{B}_0$ should not include decays through mixing with the $V$: such effects are exactly what the mixing accounts for. For the kinetic mixing models that we will consider later, we take $\Gamma_{\mathcal{B},0} = 0$, since the dark sector particles do not have any tree-level coupling to the SM, and the unmixed width of the bound state excluding mixing into the SM is always much smaller than $\Gamma_{V,0}$.

After mixing, the mixed mass eigenstates are rotated by a complex mixing angle $\theta$ with respect to the unmixed states, and the masses and widths are shifted by \cite{Hall:1985jf,Kuhn:1987ty}
\begin{alignat}{1}
    Q_V &= Q_{V,0} \cos^2 \theta + Q_{\mathcal{B},0} \sin^2 \theta + f \sin 2 \theta, \nonumber \\
    Q_\mathcal{B} &= Q_{V,0} \sin^2 \theta + Q_{\mathcal{B},0} \cos^2 \theta - f \sin 2 \theta,
    \label{eqn:massMixing}
\end{alignat}
where $Q_j \equiv m_j^2 - i m_j \Gamma_j$, with
\begin{alignat}{1}
    \tan 2 \theta = \frac{2f}{Q_{V,0} - Q_{\mathcal{B},0}}.
    \label{eqn:tan2theta}
\end{alignat}
The rotated mass eigenstate $\mathcal{B}$ therefore develops a coupling to the SM through its $V_0$ component.

When the mixed masses $m_V$ and $m_\mathcal{B}$ are nearly equal, a resonance search for each individual mass eigenstate becomes impossible, since the $s$-channel diagrams with intermediate $V$- and $\mathcal{B}$-states interfere with each other, and the end result is a cross section that may not have a Breit-Wigner form. However, if $\theta$ is small, Eq.~(\ref{eqn:massMixing}) shows that the mixed mass eigenstates are separated by $\Delta m^2 \sim 4f \text{Re}(\theta)$, where $\text{Re}(\theta)$ is the real part of $\theta$. Furthermore, the shift in the masses defined by Eq.~(\ref{eqn:massMixing}) and~(\ref{eqn:tan2theta}) always results in a mass eigenstate that is lighter than both $m_{V,0}$ and $m_{\mathcal{B},0}$, and is therefore always strictly below the threshold for open production of $\chi \overline{\chi}$. These two facts can ensure that the lighter resonance is always narrow, as it cannot decay into $\chi \overline{\chi}$, and is always well-separated from the heavier resonance. We have checked that this is always the case for the models that we consider later.

Finally, in the limit of small $\theta$, this mixing procedure gives a final decay width $\Gamma_\mathcal{B}$ that agrees with the perturbative calculation to $\mathcal{O}(\theta^2)$, i.e. with the result obtained by summing the partial widths of $\mathcal{B}_0$ decaying through mixing with $V_0$ (with $\Gamma_{V,0}$ evaluated at $s = m_{\mathcal{B},0}^2$), which then decays into SM final states \cite{Kuhn:1985eu,Kuhn:1987ty}. Throughout this chapter, we will therefore qualitatively discuss the nature of the $\mathcal{B}$ resonance using the perturbative picture, while taking the mixing fully into account quantitatively. We will also not make a distinction between $\mathcal{B}_0$ and $\mathcal{B}$ or $V_0$ and $V$, unless we are explicitly discussing the mixing. 

\subsection{Collider Signatures} 
\label{sec:Collider}

There are three important classes of collider signatures for models of the type we have discussed:
(i)~mono-$X$, where the DM state $\chi$ is produced and observed as MET recoiling against a SM final state such as $X=j,h,W,Z\,$;
(ii)~$V$ resonant production with decaying channels such as dilepton, dijet or any other SM final states, and
(iii)~$\cB$ resonant production with $m_\cB\approx 2m_\chi$, decaying into a pair of leptons or jets.

The three channels probe different physics, as well as different regions of the dark sector parameter space. The mono-$X$ channel is an unavoidable signature of DM. The properties of the $\mathcal{B}$ resonance are completely determined by the DM mass and its self-interaction through the light mediator $Y$; therefore, by analyzing its properties, we study the DM directly. The $V$ resonance on the other hand probes the structure of the dark sector, but is not directly related to the puzzle of DM.


The mono-$X$ signature has been discussed previously~\cite{Abdallah:2015ter,Abdallah:2014hon,Bai:2010hh,Goodman:2010yf,Fox:2011pm}, and there are on-going searches at the LHC. We will demonstrate that for the models we consider, mono-jet searches probe a different region of parameter space than bound state resonance searches.

The production rate for the bound state $\cB$ at a $pp$ collider is given by (as discussed in Chapter~\ref{chap:intro} and Refs.~\cite{Tsai:2015ugz,Kats:2009bv,Petrelli:1997ge})
\begin{alignat}{1}
	\label{eqn:qqbartoB}
	\sigma_{\cB} \approx 	\sum_{q} \zeta(3) \frac{8\pi^2 (2J+1)}{9m_{\mathcal{B}}^3}  \Gamma_{\mathcal{B} \to q \overline{q}} \mathcal{L}_{q\bar{q}}\left( \tau_\cB \right),
\end{alignat}
where $J$ is the spin of the bound state; for a bound state produced from a vector mediator, $J=1$. $\zeta(s)$ is the Riemann $\zeta$-function, which takes into account the cross section for the production of all of the excited states of the bound state. Here $\tau_\cB \equiv m^2_\cB/s$, $m_\cB$ is the mass of the bound state, and $\sqrt{s}$ is the collider center-of-mass energy. $\cL_{q\bar{q}}$ is the parton luminosity function defined as 
\begin{alignat}{1}
    \label{eqn:PDF}
    \mathcal{L}_{q \overline{q}}(\tau) = \tau  \int_\tau^1 \frac{dx}{x}  f_q(x) f_{\overline{q}} (\tau/x) \, ,
\end{alignat}
with $f_q(x)$ being the parton distribution functions~(PDF), taken from~\cite{Martin:2009iq} for calculations in this chapter.

In the perturbative limit, we can write
\begin{align}
    \label{eqn:qqbartoBPerturb}
    \sigma_{\cB} 
&\approx  \sum_{q} \frac{8 \pi \zeta(3)}{3 m_\cB}  \frac{g_q^2 g_\chi^2 |\psi(0)|^2 \mathcal{L}_{q\bar{q}}\left(\tau_\cB \right)}{(m^2_\cB - m^2_V)^2 +\Gamma_V^2(s = m_\mathcal{B}^2) m_V^2}  \, ,
\end{align}
where $g_q\,(g_\chi)$ sets the coupling of the mediator $V$ to quarks\,(DM) and $\psi(0)$ is the wave function of the bound state at the origin. 
For a Coulomb-like potential with coupling $\alpha_\cB$ (i.e. where the mass of the bound state mediator $m_Y$ can be neglected), $|\psi(0)|^2 = \alpha_\cB^3 m_\chi^3/8\pi$. Throughout this chapter, $\alpha_\cB$ is always evaluated at the dark matter mass scale, and we will only consider $\alpha_\cB < 1$. Although the relevant energy scale for bound state formation should strictly be the Bohr momentum $\alpha_\cB m_\chi$, we will only consider models where the hierarchy of energy scales is $m_Y < \alpha_\cB m_\chi < m_\chi$, with all other scales lying above $m_\chi$. As noted in Ref.~\cite{Tsai:2015ugz}, since there are no relevant energy scales between $m_Y$ and $m_\chi$, the running of the coupling is expected to be insignificant between $\alpha_\cB m_\chi$ and $m_\chi$, the primary scales of interest throughout.

This perturbative $\cB$ production cross section can be understood in three limits:
(i)~the heavy mediator limit, $m_V\gg m_\cB$;
(ii)~the light mediator limit, $m_V\ll m_\cB$, and
(iii)~$m_V\approx m_\cB$. 
The cross section in each limit is
\begin{alignat}{1}
	\label{eqn:qqbartoBlimit}
	\sigma_{\cB} = \frac{4 \pi \zeta(3)}{3 m_\chi} g_\chi^2 |\psi(0)|^2 \sum_{q} g_q^2 \mathcal{L}_{q \overline{q}} \left(\tau_{\mathcal{B}} \right) \begin{cases}
		\frac{1}{m_V^4}, & m_V \gg m_{\cB}, \\
		\frac{1}{m_\cB^4}, & m_V \ll m_{\cB}, \\
		\frac{1}{\Gamma_V^2 m_\cB^2}, & m_V \approx m_{\cB}.
	\end{cases}
\end{alignat}
These equations show that the $\cB$ production cross section is enhanced when its mass is close to the mediator mass, and suppressed in the other two limits. Thus, we expect stronger sensitivity in this channel when $m_V \approx m_\cB \,$. Moreover, if $m_\cB \gg m_V$, which is in the limit where $V$ can also support dark matter bound states, the $\cB$ production cross section is suppressed by $\Gamma_V^2/m^2_\cB$  relative to the $m_V\approx m_\cB$ region. We also can see that for models where $\mathcal{B}$ is heavy enough to decay primarily into two or three $V$'s which then decay into 4 or 6 leptons at the LHC, the production rate of the bound state is suppressed relative to the regime where $m_V \approx m_\cB$.  

The mediator production cross section, $V$, is
\begin{alignat}{1}
	\sigma_V \approx 	\sum_q  \frac{8\pi^2}{3}\frac{\Gamma_{V\to q\bar q}}{m^3_V} \cL_{q\bar q}\left( \tau_V \right),
\end{alignat}
where $\tau_V \equiv m^2_V/s$. 
As we pointed out above, the $V$ resonance search does not directly probe the dark matter content. Further searches must be used to uncover the dark sector after discovering the mediator between the SM and the dark sector. Most importantly, when $m_V > 2m_\chi$ and $g_\chi\gg g_q, g_\ell$~(the coupling to leptons), the branching ratio of $V$ to SM particles becomes small, and resonance searches for $V$ grow ineffective. The full mixing calculation also bears out this conclusion: once $m_{V,0} > 2 m_\chi$, the $V$ resonance is heavier and lies above the $\chi \overline{\chi}$ threshold and is a wide resonance, while the lighter $\mathcal{B}$ resonance remains narrow and below the threshold. 

The comparison between mono-$X$ and bound state production is more complicated as the backgrounds for the two searches are different, and a more detailed comparison is required; we will show results for some specific models below.  
On generic grounds, the mono-$X$ cross section is reduced because of the PDF price of the additional jet. However, the two production cross sections scale as $\alpha_s g^2_q g^2_\chi$ and $\alpha_\cB^3 g^2_q g^2_\chi$, for the mono-$X$ and bound state cases respectively. Thus for $\alpha^3_\cB \ll \alpha_s$ we expect a reduced sensitivity  in the bound-state searches; this suggests $\alpha_\cB$ rather close to 1 will be required to make bound-state searches competitive. Moreover, the mono-jet search becomes ineffective once $m_V < 2m_\chi$, since the mono-jet process must then proceed through an off-shell $V$. 

In summary, the mono-jet search probes the region of parameter space where $m_V > 2 m_\chi$, while the $V$ resonance search is more sensitive to the region where $m_V < 2 m_\chi$. The bound-state production cross section, on the other hand, is enhanced precisely in the intermediate region, and outperforms the other two searches when $m_V \gtrsim 2 m_\chi$. These three searches are thus complementary, and probe different parts of parameter space, as we will show explicitly in our models below.

\subsection{Direct Detection Limits}
\label{sec:DDL}

Direct detection searches are very sensitive probes of DM, especially for DM with substantial couplings to hadrons, and mass at the EW scale or higher. 
Thus, viable models of dark resonance signals at the LHC must evade direct detection bounds. 

A naive estimate of the DM-nucleon scattering cross section at tree level, in terms of the parameters discussed in the previous subsection, gives $\sigma \sim g_q^2 g_\chi^2 m_N^2 /m_V^4 \sim 10^{-40} \text{cm}^2 g_q^2 g_\chi^2 \left( \text{TeV}/m_V\right)^4 $,  assuming $m_V$ is much larger than the typical momentum transfer in the scattering, and $m_\chi$ is much larger than the nucleon mass $m_N$. 
For comparison, under standard assumptions, the limit from XENON\,1T on this scattering cross section is of order $10^{-45} \text{cm}^2 (m_\chi/\text{TeV})$ \cite{Aprile:2017iyp}. 
Thus, if the elastic scattering spin-independent cross section is unsuppressed, we infer that the product of couplings $g_q^2 g_\chi^2 \lesssim 10^{-5} m_V^4 m_\chi/\text{TeV}^5$. 
This simple estimate is broadly consistent with more carefully obtained limits on a dark sector interacting with nucleons through a vector mediator for current and future direct detection experiments~\cite{DEramo:2016gos,Buchmueller:2014yoa}. 
Reasonably large couplings and sufficiently low dark sector masses are necessary for the significant production of the bound state resonance, but this parameter region of interest ($g_q g_\chi \sim 1$ and $m_V \sim 2 m_\chi \sim 1-4$ TeV) is generically in tension with direct detection bounds.

\begin{table}
    \begin{center}
    \begin{tabular}{cc}
    \resizebox{0.375\textwidth}{!}{
        \begin{tikzpicture}
          \begin{feynman}
            \vertex (a1) {\(\chi_1\)};
            \vertex[right=1.5cm of a1] (a2); 
            \vertex[above=0.1cm of a2] (labelalphaD) {\(g_\chi\)};
            \vertex[right=1.3cm of a2] (a3) {\(\chi_{2}\)};
            \vertex[crossed dot, scale=1, below=1.3cm of a2] (b2) {\( \)};
            \vertex[below=0.3cm of b2] (labeleps) {\(g_q\)};
            \vertex[left=1.5cm of b2] (b1) {\(q\)};
            \vertex[right=1.5cm of b2] (b3) {\(q\)};
            \diagram*{
                (a1) -- [fermion] (a2);
                (a2) -- [fermion] (a3);
                (a2) -- [boson, edge label'=\(V\)] (b2);
                (b1) -- [fermion] (b2);
                (b2) -- [fermion] (b3);
            };
          \end{feynman}
        \end{tikzpicture}} &
    \resizebox{0.47\textwidth}{!}{    
        \begin{tikzpicture}
          \begin{feynman}
            \vertex (a1) {\(\chi_1\)};
            \vertex[right=1.1cm of a1] (a2); 
            \vertex[right=1.5cm of a2] (a3);
            \vertex[above right=0.1cm and 0.5cm of a2] (labelchi2) {\(\chi_2\)};
            \vertex[right=0.9 of a3] (a4) {\(\chi_{1}\)};
            \vertex[crossed dot, scale=1, below=1.3cm of a2] (b2) {\( \)};
            \vertex[crossed dot, scale=1, below=1.3cm of a3] (b3) {\( \)};
            \vertex[left=1.05cm of b2] (b1) {\(q\)};
            \vertex[right=1.05cm of b3] (b4) {\(q\)};
            \vertex[below=0.4cm of b2] (buffer) {\( \)};
            \diagram*{
                (a1) -- [fermion] (a2);
                (a2) -- [fermion] (a3);
                (a3) -- [fermion] (a4);
                (a2) -- [boson] (b2);
                (a3) -- [boson] (b3);
                (b1) -- [fermion] (b2);
                (b2) -- [fermion] (b3);
                (b3) -- [fermion] (b4);
            };
          \end{feynman}
        \end{tikzpicture}
    }
    \end{tabular}
    \end{center}
    
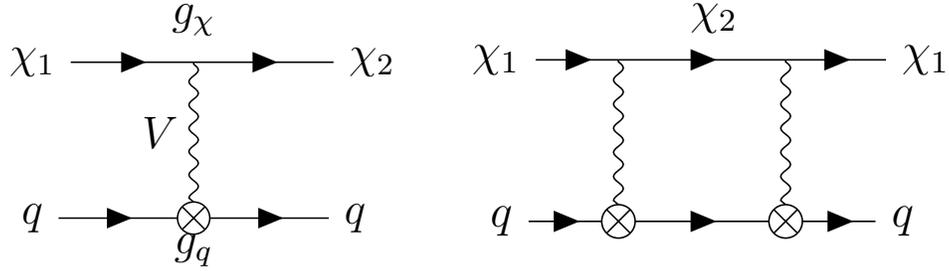
\captionof{figure}{Direct detection Feynman diagrams for inelastic DM models, with (left) tree-level inelastic scattering, and (right) one-loop, elastic scattering off nuclei in these experiments.}
    \label{fig:feynmanDirectDet}
\end{table}

However, any suppression to the naive tree-level cross section can alleviate this tension. As mentioned above, a simple scenario (``inelastic dark matter'') that leads to suppressed direct detection signals posits that the coupling between the DM $\chi_1$ and the mediator $V$ involves an unstable partner particle $\chi_2$, and the mass splitting between the DM and its partner is greater than the maximum kinetic energy of DM particles in the halo~\cite{TuckerSmith:2001hy, TuckerSmith:2004jv}. 
Fig.~\ref{fig:feynmanDirectDet} shows the relevant Feynman diagrams for direct detection of the DM particles. 
Such models also have interesting consequences for bound state formation at the LHC: if the bound state is produced in the $s$-channel from the mediator $V$, it will automatically be composed of the DM and its partner particle, or may only involve dark sector particles in the same multiplet as the DM, and not the DM at all. 

In such models, elastic scattering can still occur, but only at loop level. The direct-detection spin-independent cross section for scattering off a nucleon with target mass $m_T$ is given by~\cite{Kurylov:2003ra}
\begin{align}
	\sigma_{\rm SI} 
=	 \frac{4}{\pi} \left( \frac{m_\chi m_T}{m_\chi + m_T} \right)^2 (n_p f_p + n_n f_n)^2 \, ,
\end{align}
where $n_{p,n}$ are the number of protons and neutrons respectively, and $f_{p,n}$ are the corresponding matrix elements. 
We can generalize the effective operator analysis of~\cite{Hisano:2010fy}, to the dark sector models that will be of interest to us. Then 
\begin{align}
	\frac{f_N}{m_N} 
=	\sum_{q} \left(  f_q f_{T_q} + \frac{3}{4} (q(2) + \bar{q}(2)) (g_q^{(1)} + g_q^{(2)})  \right) \, ,
\end{align} 
where $N = n,p$, and we sum over $u,d,s$ quarks. Here the first term comes from a 1-loop diagram involving the Higgs, while the second term is a box diagram with two $V$ propagators. 
$q(2)$ and $\bar{q}(2)$ are the second moments of the quark and anti-quarks PDFs and $f_{T_q} = \langle N | m_q \bar{q}q | N\rangle / m_N$ is the nuclear form factor. For these we use the numeric values from~\cite{Hisano:2010fy}.
In the wino scenario discussed in \cite{Hisano:2010fy} these two contributions $f_q$ and $g_q^{1,2}$ are non-negligible but of opposite sign, thus leading to a cancellation. 
In the case of the dark sector models that we will consider later, the second contribution is suppressed by the small coupling between the SM mediator and the SM, while the first contribution will be negligible as the dark sector coupling to the SM Higgs will always be very small. 
Explicitly, we can write $g_q^{(1,2)} = (g_\chi^2 e \epsilon c_W Q / 4 \pi m_V^3) g_{T(1,2)} (m_V^2/m_\chi^2)$, where $g_{T(1,2)}$ are functions computed in~\cite{Hisano:2010fy}, $\epsilon$ is the small mixing parameter, and $Q$ is the charge of the quark. We find that in regions of parameter space of interest to the present work the contribution from loops to direct detection is thus no larger than $\sigma_{\rm SI} \sim \SI{e-48}{\centi\meter\squared}$ and is thus unconstrained.

\subsection{Overclosure and Indirect Searches}
\label{sec:OverInD}

In general, if annihilation to the light mediators that support the bound state is allowed, this process will tend to dominate freezeout. The same attractive potential that permits bound state formation will also generically enhance annihilation through the Sommerfeld enhancement~\cite{Hisano:2003ec,Hisano:2004ds}, potentially giving rise to large indirect signals in the present day. 
Formation of bound states followed by their decay can also significantly enhance indirect signals (e.g.~\cite{Pospelov:2008jd}), if the mediator supporting the bound state is light enough that radiative capture of two DM particles into the bound state is kinematically allowed.

Let us first note that there are several possible annihilation channels which are $p$-wave suppressed at late times~\cite{Kumar:2013iva}; if these processes dominate freezeout, the late-time indirect detection signals will generally be very suppressed. We will see an example of this when we consider a model where the dominant annihilation is of Majorana fermions to light scalars. Furthermore, if DM-DM scatterings experience a repulsive potential rather than an attractive one, the DM-DM annihilation will be exponentially suppressed at low velocities~\cite{Cirelli:2007xd}. Note that in order for bound-state searches to be interesting in such a scenario, there must be at least one other particle with which the DM can form a bound state, and the DM must have an attractive interaction with that particle. 
In this initial discussion, therefore, we assume the dominant annihilation is $s$-wave and experiences an attractive Sommerfeld enhancement, to explore when indirect searches can set interesting constraints.

Let us consider the simple case where the effective potential experienced by the DM is a Yukawa potential, as discussed above. The $s$-wave Sommerfeld enhancement for a Yukawa potential with coupling $\alpha_\cB$ and mediator mass $m_Y$ can be well approximated by~\cite{Cassel:2009wt, Slatyer:2009vg}:
\begin{alignat}{1}
	S =	\frac{2 \pi \alpha_\cB}{v_\text{rel}} 
	\frac{\sinh\left(\frac{6}{\pi} \frac{m_\chi v_\text{rel}}{m_Y}\right)}{\cosh\left(\frac{6}{\pi} \frac{m_\chi v_\text{rel}}{m_Y}\right) - \cos\theta} \ge \frac{2 \pi \alpha_B}{v_\text{rel}} 
	\frac{\sinh\left(\frac{6}{\pi} \frac{m_\chi v_\text{rel}}{m_Y}\right)}{\cosh\left(\frac{6}{\pi} \frac{m_\chi v_\text{rel}}{m_Y}\right) +1} \, , 
	\label{eq:sommerfeld} 
\end{alignat}
where $v_\text{rel}$ is the relative velocity between the DM particles, and in the second line we use $\cos\theta> -1$ with 
\begin{alignat}{1}
\theta = 2\pi\sqrt{\frac{6}{\pi^2} \frac{m_Y}{\alpha_\cB m_\chi} - \left(\frac{3 m_\chi v_\text{rel}}{\pi^2 m_Y}\right)^2}
\end{alignat}
being an angle controlling the resonance positions. The inequality is saturated for real values of $\theta$, at the minima between resonances where $\cos\theta = -1$. (It is also approximately saturated where $m_Y \rightarrow 0$ and $v_\text{rel} \ll \alpha_\cB$, where $S \approx   2 \pi \alpha_\cB/v_\text{rel}$.) 

Note that for fixed $m_\chi$, requiring the correct relic density fixes $\alpha_\cB$, if the assumptions are made that 
(a)~this channel dominates during freezeout, and 
(b)~the mass of the mediator is irrelevant during freezeout. 
The latter assumption is approximately true away from resonances and if $m_V/m_\chi$ is smaller than the typical velocity of particles around freezeout ($v\sim 1/3$). 
For large values of $m_Y$ (requiring large $\alpha_\cB$, since $m_Y/m_\chi < \alpha_\cB/1.68$), or values of $m_Y$ corresponding to resonant Sommerfeld enhancement ($\cos \theta \rightarrow 1$ as $v_\text{rel} \rightarrow 0$), freezeout may be more complicated and needs to be studied more carefully; we will include the full $m_Y$ dependence when we examine specific models.

However, if we consider $\alpha_\cB$ to be fixed given $m_\chi$, and hold $v_\text{rel}$ fixed, then our expression for the lower bound on the Sommerfeld enhancement is a monotonically decreasing function of $m_Y$; thus indirect detection will set a \textit{lower} bound on $m_Y$ (all values of $m_Y$ below this threshold will be ruled out). 
Since the requirement for bound state formation sets an \textit{upper} bound on $m_Y$, one can ask whether these two criteria are in conflict. 
Equivalently, requiring $m_Y < \alpha_\cB m_\chi/1.68$ implies that 
\begin{alignat}{1}
S > \frac{2 \pi \alpha_\cB}{v_\text{rel}}\sinh\left(\frac{3.21 v_\text{rel}}{\alpha_\cB}\right)\Big/\left( \cosh\left(\frac{3.21 v_\text{rel}}{\alpha_\cB}\right) +1\right) \,.
\end{alignat}
In order for the model to avoid exclusion by indirect detection (except possibly where $m_Y$ is important to freezeout), this minimal Sommerfeld enhancement must be permitted by the data. Note that for $\alpha_\cB \gg v_\text{rel}$ ( $v_\text{rel} \sim 10^{-3}$ in the present-day Milky Way halo), this minimal enhancement will reduce to an $\alpha_\cB$- and $v_\text{rel}$-independent prefactor of:
\begin{align} 
	S_\text{min} = 2\pi \times 3.21/2 \approx 10.
\end{align}
This minimal Sommerfeld-enhanced cross section is rather close to indirect detection bounds for a DM species that comprises 100\% of the DM and whose abundance is set by thermal freezeout, for DM masses below $\sim 1$ TeV (e.g. \cite{Fermi-LAT:2016uux, Abdallah:2016ygi}); permitting $m_Y \lesssim m_\chi v_\text{rel}$ would generally significantly overproduce limits from indirect detection, unless $Y$ decays primarily into invisible channels. If we assume $m_Y \gg m_\chi v_\text{rel}$ in the present day, then we can approximate $S \gtrsim 6 \alpha_\cB m_\chi/m_Y$, and thus if the maximum allowed Sommerfeld factor is $S_\text{max}$, then $m_Y \gtrsim 6 \alpha_\cB m_\chi / S_\text{max}$. Of course, smaller values for $m_Y$ are permissible if the species that forms bound states comprises only a small fraction of the overall dark matter density.

If the dominant annihilation channel consists of $s$-wave annihilation to mediators coupled to the DM with strength $\alpha_\mathcal{B}$, then the annihilation cross section at low velocities is of order $\langle \sigma v_\text{rel} \rangle \approx \pi \alpha_\mathcal{B}^2/m_\chi^2$ (this expression is exact for Dirac or pseudo-Dirac DM annihilating to U(1) dark gauge bosons). Requiring that this cross section fall below the thermal value of $\langle \sigma v_\text{rel} \rangle \approx 2\times10^{-26}$ cm$^3$/s $\approx 1.7 \times 10^{-9}$ GeV$^{-2}$ suggests an overclosure bound of $m_\chi \lesssim \alpha_\mathcal{B} \times 43 \, \text{TeV}$. As we will see, we will generally be interested in masses around a few TeV and $\alpha_\mathcal{B} \gtrsim 0.1$, so the overclosure bound will not typically be particularly constraining. 
This estimate ignores Sommerfeld enhancement and bound state formation during freezeout, which can be important \cite{Feng:2010zp,vonHarling:2014kha}. For $\alpha_B \gtrsim 0.1$, the Sommerfeld enhancement is non-negligible during the freezeout epoch; however, for attractive Sommerfeld enhancement, including this effect only reduces the late-time relic abundance. This further relaxes the overclosure bound, and since it reduces the abundance of the species in question, also weakens constraints from indirect detection. (However, it makes it more challenging to generate 100\% of the DM abundance by the same species that forms bound states.)

Likewise, radiative formation of bound states can also contribute to the depletion of DM at early times and indirect signals at late times \cite{Pospelov:2008jd, MarchRussell:2008tu,vonHarling:2014kha,An:2016kie}. These radiative processes are only kinematically unsuppressed if enough energy is available to produce an on-shell light mediator, i.e. the binding energy + kinetic energy of the particles is greater than $m_Y$. Bound state formation can also occur through radiation of an off-shell heavy mediator that decays to SM particles, but such processes will be suppressed by a small mixing with the SM and also by the mass of the heavy mediator. 
Thus, there are two distinct regimes for $m_Y$ from an indirect-detection perspective: $\alpha_\mathcal{B} m_\chi/1.68 \gtrsim m_Y \gtrsim \alpha_\mathcal{B}^2 m_\chi/4$, where bound states exist but radiative capture into them is suppressed, and $m_Y \lesssim \alpha_\mathcal{B}^2 m_\chi/4$, where radiative capture processes are unsuppressed. We will ignore bound-state effects in the former case, but account for their impact on indirect-detection signatures in the latter case.

However, we will ignore the effects of bound-state formation during freezeout. A careful treatment of bound-state effects during freezeout requires accounting for dissociation of the bound states through interactions with the light-mediator bath. If $m_Y \lesssim \alpha_\mathcal{B}^2 m_\chi/4$, then for $\alpha_\mathcal{B} \lesssim 0.5$ we expect the temperature at freezeout to be comparable to or larger than the binding energy (taking the standard estimate $T_\text{freezeout} \sim m_\chi/20$), and so dissociation effects could be substantial. Thus while the presence of radiative capture into bound states during freezeout may further deplete the DM abundance, relaxing both the overclosure and indirect limits further, a full calculation would require a careful analysis (as performed in e.g.\ Ref.~\cite{vonHarling:2014kha}).

We will show that the indirect detection constraints and overclosure limit cannot fully exclude the regions of parameter space relevant to collider searches for the bound states, even without taking the impact of bound-state effects on freezeout into account, for both models we consider. Since including the bound-state effects during freezeout would only relax these constraints further, we are justified in neglecting them for purposes of this work.

\subsection{Dark Matter Self-Interactions}
\label{sec:DMSI}

Constraints on DM self-interactions require $\sigma/m_\chi \lesssim \SI{1}{\centi\meter\squared\per\gram} \approx 1/(\SI{60}{\mega\eV})^3$~\cite{Robertson:2016xjh, Tulin:2017ara, Harvey:2015hha}. 
As we are interested in the regime where a long-range potential exists and can support bound states, we cannot use the Born approximation to estimate scattering rates. However, if $m_Y/m_\chi \gtrsim v_\text{rel}$ while still satisfying Eq.~\eqref{eqn:boundStateRequirement}, the typical relative velocity of DM particles in galaxies and galaxy clusters, then we can make the approximation that $s$-wave scattering dominates and use the analytic estimate for the scattering cross section derived in Refs.~\cite{Tulin:2013teo,Schutz:2014nka}.

The scattering cross section is approximated in the low-velocity limit by~\cite{Tulin:2013teo,Schutz:2014nka}
\begin{align} 
	\sigma_T 
&= 	\frac{4\pi}{(m_\chi v_\text{rel})^2} \left|1 - e^{2 i \delta} \right|^2 \, , 
\end{align}
where $\delta = -\left[2 \gamma + \ln(c) + \pi \cot(\pi\sqrt{c}) \right] a c$,  $a = v_\text{rel} / 2 \alpha_{\cB}$, $c  = \alpha_{\cB} m_\chi/1.6 m_Y$ and $\gamma \approx 0.577$ is the Euler-Mascheroni constant.
We see that away from resonances, which occur when $\cot(\pi \sqrt{c})$ diverges, the size of the phase shift is controlled by $a c = m_\chi (v_\text{rel}/2)/(1.6 m_Y)$. The regime where the $s$-wave contribution dominates is thus a regime where (away from resonances) this phase shift is small, and we can write:
\begin{align} \sigma_T \sim \frac{4 \pi}{(m_\chi v_\text{rel}/2)^2} a^2 c^2 \sim \frac{4 \pi}{ m_Y^2}, \end{align}
which is just the geometric cross section.

Assuming this geometric cross section, we see that the self-interaction bound will be satisfied provided $(m_\chi m_Y^2)^{1/3} \gtrsim 100$ \SI{}{\mega\eV}, which for \SI{1}{\tera\eV} DM requires only that $m_Y \gtrsim 1$ \SI{}{\mega\eV}. 
Thus, away from points in the parameter space where there is a near-zero-energy bound state, we expect the self-interaction rate to be undetectable, despite the rather large couplings we invoke.

\section{Dark Sector Models With Detectable Bound States}
\label{sec:darksector}

We now consider two examples of phenomenologically viable DM models containing bound states, which can lead to interesting signatures at the LHC. These models serve as examples of how to build non-supersymmetric models that realize the requirements laid out in Sec.~\ref{sec:PhenoBound}. 
We will show that in these models, the search for bound-state resonances can probe parameter space which is not accessible to mono-jet searches and resonance searches for the mediator.

In the first model, which we label as the ``pseudo-Dirac'' model, the dark sector consists of a pair of almost-degenerate Weyl fermions that are charged under a dark-sector U(1)$_D$ gauge group, which is broken by a dark Higgs-like scalar. These fermions can form bound states with the dark Higgs as the mediator. 
The second model, which we refer to as the ``triple Higgs'' model, is based on a completely broken SU(3)$_D$ gauge theory, with the dark matter candidate being a Dirac fermion in the fundamental of the gauge group. Much of the phenomenology of this model, including bound state formation and couplings to the SM, is derived from the symmetry breaking pattern of the theory, with both the mediator that supports the bound state and the mediator to the SM being massive gauge bosons of the broken SU(3)$_D$ group. 
In both cases, the dark sector interacts with the SM via a vector portal with kinetic mixing, and the DM direct detection cross section is suppressed by the fact that at tree-level the DM scatters into a heavier state.

Before introducing these models, however, we will consider a simpler scenario that is familiar from SUSY, that of pure wino/higgsino DM (Sec.~\ref{sec:winoHiggsino}). We will show that the production rate of wino/higgsino-onium bound states at the LHC is too small to be constraining, but this scenario shares many of the properties of our more complicated dark-sector models and thus has pedagogical value. We will then review the details of kinetic mixing between new dark gauge bosons and the SM neutral gauge bosons (Sec.~\ref{sec:KineticMixing}), since this mechanism describes the leading interaction of the SM with the dark sector in both dark-sector scenarios we consider, before describing in detail the two models (Sec.~\ref{sec:PseudoDirac} and~\ref{sec:TripleHiggs}).


\subsection{A Weakly Interacting Example: SU(2)\texorpdfstring{$_L$}{L} Minimal Dark Matter}
\label{sec:winoHiggsino}

Sub-TeV superpartners of the EW bosons and of the two Higgs doublets in SUSY theories can potentially be produced and detected at the LHC, with the lightest neutralino being a particularly well-motivated, weakly-interacting DM candidate. Outside of SUSY theories, models of ``minimal dark matter'' where the DM transforms under a low-dimensional representation of SU(2)$_L$ have similar phenomenology to neutralino DM~\cite{Cirelli:2005uq, Cirelli:2007xd, Fukuda:2017jmk}. Pure wino or higgsino DM corresponds to the lowest-lying mass eigenstates from, respectively, an SU(2)$_L$ triplet of Majorana fermions or an SU(2)$_L$ Dirac fermion doublet with hypercharge $1/2$. The hypercharge-zero SU(2)$_L$ quintuplet is also a viable ``minimal dark matter'' candidate.

If the DM transforms as part of a SU(2)$_L$ multiplet, then it will be accompanied by heavier charged partner particles in the same multiplet. After EW symmetry breaking, the wino triplet separates into a lighter neutral Majorana fermion $\chi^0$ and a heavier charged Dirac fermion $\chi^\pm$; the higgsino multiplet gives rise to two neutral Majorana states $\chi^1$, $\chi^2$ and a charged Dirac fermion $\chi^\pm$. These charged partners can always form Coulombic bound states; when the DM is sufficiently heavy, $W$ and $Z$ exchange may also support bound states including the DM itself (e.g.~\cite{Asadi:2016ybp}). Numerical calculations indicate that for wino DM there is a crossover point at a DM mass of around 5\,TeV, where the ground state transitions from being primarily composed of $\chi^+ \chi^-$ bound by photon exchange, to being composed of an admixture of $\chi^0 \chi^0$ and $\chi^+ \chi^-$ bound by the gauge bosons of an approximately unbroken SU(2)$_L$ symmetry.

SU(2)$_L$-DM models have many attractive features of the type discussed in Sec.~\ref{sec:PhenoBound}, and behave as prototypes for the models of interest to us. They naturally possess multiple mediators, one of which is massless and supports bound states, while the other massive mediators are all known particles in the SM. The SU(2)$_L$ multiplets contain several states nearly-degenerate with the DM; the couplings of the gauge bosons with the DM and its partners are naturally off-diagonal, and so the elastic scattering relevant to direct detection proceeds only at one-loop level (and also suffers from additional cancellations which suppress the rate further \cite{Hill:2013hoa}). Direct and indirect constraints on wino and higgsino DM have been studied extensively; thermal wino DM constituting 100\% of the DM is in tension with H.E.S.S. observations of the Galactic Center (e.g.~\cite{Cohen:2013ama, Fan:2013faa,Hryczuk:2014hpa,Ovanesyan:2016vkk,Cuoco:2017iax,Abramowski:2013ax,Baumgart:2017nsr}), but a subdominant wino DM contribution at lower wino masses is difficult to exclude. 
Pure higgsino DM is not currently experimentally testable by either direct or indirect detection~\cite{Krall:2017xij,Baumgart:2015bpa}. Finite temperature effects on the freezeout of weakly-interacting DM have also been studied~\cite{Kim:2016kxt,Biondini:2017ufr}.

A complicating factor in SU(2)$_L$ DM models is the presence of multiple mediators that can potentially support a bound state, which become most important if the DM is heavy enough that $\alpha_W m_\chi \gtrsim m_W, m_Z$ with $\alpha_W=g_W^2/4\pi\approx1/30$. 
In this case, there is a long-range potential that mixes the two-body DM--DM state with other particle anti-particle states, i.e. $\chi^0 \chi^0$ mixes with $\chi^+ \chi^-$ in the wino case, and $\chi^1 \chi^1$ can mix with $\chi^2 \chi^2$ and $\chi^+ \chi^-$ in the higgsino case. 
This can lead to $\chi^+ \chi^-$ states that are only pseudo-bound, despite the presence of the photon-mediated Coulomb potential: if the combined $W/Z/\gamma$-exchange potential is not deep enough to also bind the $\chi^0 \chi^0$ component, then the $\chi^+ \chi^-$ state (or e.g. $\chi^{++} \chi^{--}$ in representations, such as the quintuplet, where higher-charge states exist) may decay rapidly to unbound $\chi^0 \chi^0$ through $t$-channel exchange of $W$ bosons. Parametrically, the cross section for $\chi^+ \chi^- \rightarrow \chi^0 \chi^0$ through this channel, for heavy DM with $m_\chi \gg m_W$, is $\sigma v_\text{rel} \sim \sqrt{\Delta/m_\chi} \alpha_W^2 m_\chi^2/m_W^4$, where $\Delta$ is the available energy (i.e. the splitting between the mass of the $\chi^+ \chi^-$ two-body state, including any binding energy, and the mass of the final $\chi^0 \chi^0$ state). By comparison, the cross section for annihilation to SM quarks, leptons and gauge bosons is of order $\sigma v_\text{rel} \sim \alpha_W^2/m_\chi^2$. Thus we expect the former to dominate over the latter when $\sqrt{\Delta/m_\chi} \gtrsim (m_W/m_\chi)^4$.

However, there is an important caveat to this argument: in fermionic models of this type, this mixing between $\chi^0 \chi^0$ and $\chi^+ \chi^-$ occurs only in the states with even $L+S$ (where $L$ and $S$ are the quantum numbers describing the total orbital angular momentum and total spin of the bound state); states with odd $L+S$ have symmetric wavefunctions and cannot support two identical fermions. Since the mediator to the DM is an EW gauge boson, the bound state dominantly produced at colliders has $L=0$ and $S=1$; these are true bound states, not pseudo-bound, and cannot decay rapidly to pairs of identical DM particles. In particular, in the pure wino case the $\chi^+ \chi^-$ bound state with $L=0$, $S=1$, denoted $\mathcal{B}_w$, decays dominantly via an $s$-channel $\gamma/Z$ to SM fermion pairs or through a $t$-channel exchange of a $\chi^0$ into a $W^+ W^-$ final state~\cite{Asadi:2016ybp}; final states involving the DM are suppressed. We will see this behavior arise again in our example dark-sector models.

The pure higgsino limit serves as an example of a model where there are two neutral mass eigenstates that can be close in mass, denoted as $\chi^0_1$ and $\chi^0_2$, the lighter of which ($\chi^0_1$) is the DM. 
In this case, the decay of $\chi^+ \chi^-$ to $\chi^0_1 \chi^0_2$ may be allowed. If~$\Delta_{+0} \equiv 2m_{\chi^\pm} - m_{\chi_1^0} - m_{\chi_2^0} < 0$, the $\chi^+ \chi^-$ bound state never mixes into the $\chi_1^0 \chi_2^0$ from kinematic considerations. When $\Delta_{+0} > 0$ however, the $\chi^+ \chi^-$ can simply decay into free $\chi_1^0 \chi_2^0$, and if the width for this decay is significantly larger than the width of the $\chi^+ \chi^-$ bound state, the bound state is effectively never formed.\footnote{When the widths are comparable, bound state decays into $\chi_1^0 \chi_2^0$ becomes an additional decay channel, together with decays to $W^+W^-$ or SM fermions.}
Thus, for the pure higgsino case the sign of the parameter $\Delta_{+0}$ is critical to the bound state phenomenology, at least for DM masses below the TeV scale. 
This parameter is positive when the lightest neutralino is a pure higgsino and both the wino and the bino are taken to be infinitely massive, but there exists a range of SUSY-breaking parameters which can produce a lightest neutralino that is almost purely higgsino with a significantly more massive bino and wino, while having $\Delta_{+0} < 0$~\cite{Han:2014kaa}. 
With this choice, a $\chi^+ \chi^-$ higgsino bound state $\mathcal{B}_h$ can be formed and can decay in the same way as $\mathcal{B}_w$, albeit with different coupling constants to the EW bosons. 

\begin{figure}
    \centering
    \includegraphics[scale=0.78]{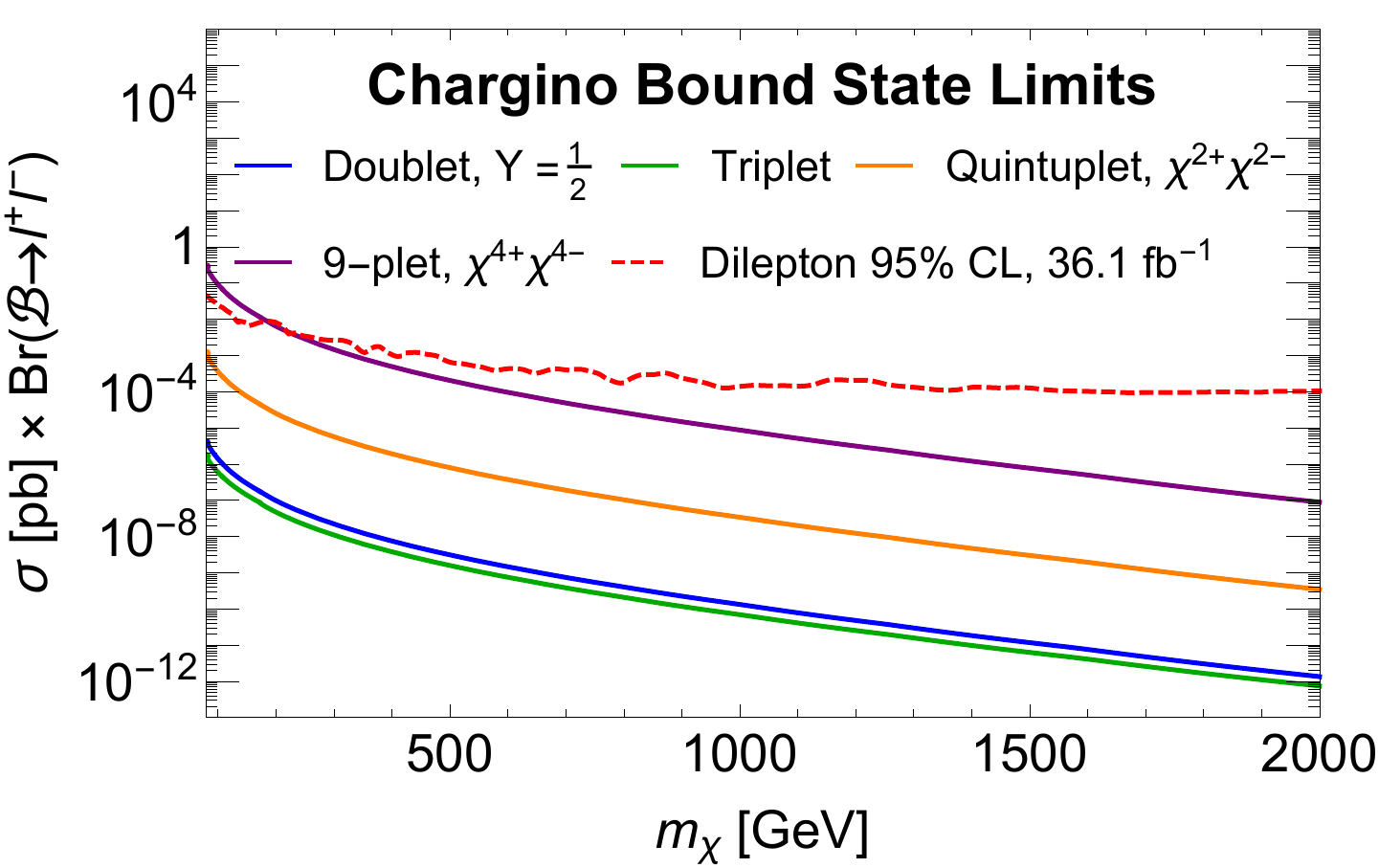}
    \caption{The production cross section times branching ratio into leptons for chargino-onium states made up of: (green) pure higgsino $\chi^+ \chi^-$; (blue) pure wino $\chi^+ \chi^-$; (orange) SU(2)$_L$ quintuplet, zero hypercharge $\chi^{2+} \chi^{2-}$, and (purple) SU(2)$_L$ 9-plet, zero hypercharge $\chi^{4+} \chi^{4-}$. The 95\% confidence limits from a dilepton resonance search for $Z'$ with \SI{36}{\per\femto\barn} of data are also shown (red, dashed).}
    \label{fig:charginoonium}
\end{figure}

Unfortunately, if the DM is part of an SU(2)$_L$ doublet or triplet, the bound state production rate at the LHC is too small to be observed. This is due to the smallness of the EW couplings, which controls the production rate. Figure~\ref{fig:charginoonium} shows the production cross section times branching ratio into leptons of chargino-onium states for fermions charged under the EW gauge group in different representations. Chargino-onia from both pure winos and pure higgsinos have production cross sections that are far too small for dilepton searches at the LHC to be effective. However, for DM in a larger representation of SU(2)$_L$, fermions having large electromagnetic charges $Q$ can be produced. The production cross section of these states scales rapidly with $Q$, while the partial widths into SM particles remain unchanged. The enhancement factor relative to the pure wino is $Q^8$, with $Q^6$ coming from the wavefunction of the bound state at the origin $|\psi(0)|^2$, and an additional $Q^2$ from the coupling of these fermions to the $\gamma$ and $Z$. For charginos with $Q = 4$ in an SU(2)$_L$ 9-plet, the production cross section for the $\chi^{4+} \chi^{4-}$ chargino-onium becomes large enough to be probed by the current dilepton resonance search results. Such large representations are generally disfavored since they lead to non-perturbative values of $\alpha_W$ below the Planck scale~\cite{Cirelli:2005uq}; however, these results more broadly demonstrate that models with large coupling constants or large charges are particularly suited for bound state searches at the LHC. Searches for multi-charged lepton bound states decaying into two photons, for example, have been shown to be effective in searches for leptons with a sufficiently large hypercharge~\cite{Barrie:2017eyd}. 
%

\subsection{Kinetic Mixing}
\label{sec:KineticMixing}

We now turn our attention the dark sector models that we briefly described above. Both of the models we will consider interact with the SM through a vector portal, with kinetic mixing with the SM U(1)$_Y$:
\begin{alignat}{1}
	\mathcal{L}_{\text{kin-mix}} 
= 	-\frac{1}{4} V_{\mu\nu} V^{\mu\nu} - \frac{\epsilon}{2} B_{\mu\nu} V^{\mu\nu} - \frac{1}{4}B_{\mu\nu} B^{\mu\nu} + \frac{1}{2} m_{V}^2 V_\mu V^\mu + \frac{1}{2} m_Z^2 Z_\mu Z^\mu \, ,
\end{alignat}
where $V_{\mu\nu}$ ($B_{\mu\nu}$) is the field strength of the dark gauge boson (SM hypercharge), and we have included the mass term for both $V$ and the SM $Z$. Here, $V_{\mu\nu}$ can be non-abelian: such a mixing term appears in the triple Higgs model in the form of a dimension-5 operator $H^a_D V_{\mu\nu}^a B_{\mu\nu}$ where $H^a_D$ is an adjoint scalar that acquires a VEV, and $a = 1, \cdots, 8$ is an SU(3)$_D$ color index. 

This interaction can be diagonalized in the mass basis; a detailed description of this diagonalization procedure is discussed in~\cite{Cassel:2009pu,Hook:2010tw}, and reviewed in Appendix~\ref{app:kineticMixing}. In the non-abelian case, only the abelian portion of the field strength is diagonalized, with the non-abelian portion remaining as an interaction term in the model. The diagonalization introduces an $\epsilon$-suppressed coupling between the physical dark gauge boson and the SM electromagnetic, $J^\mu_{\text{EM}}$, and weak-neutral, $J^\mu_{Z}$, currents, as well as an $\epsilon$-suppressed coupling between the SM $Z$-boson and the dark sector current, $J^\mu_D$: 
\begin{alignat}{1}
	J^\mu_{\text{EM}} A_\mu &\to J^\mu_{\text{EM}} \left(A_\mu - \epsilon c_W V_\mu\right), \nonumber \\ 
	J^\mu_Z Z_\mu &\to J_Z^\mu \left(Z_\mu + \frac{\epsilon s_W }{1- r^2}V_\mu \right), \nonumber \\
	J^\mu_D V_\mu &\to J_D^\mu \left(V_\mu - r^2\frac{\epsilon s_W}{1 - r^2} Z_\mu \right),
    \label{eqn:currents}
\end{alignat}
where $A$ is the SM photon, $s_W$ ($c_W$) is the sine (cosine) of the weak mixing angle, and $r \equiv m_Z/m_{V}$. 
All of the fields are given in the mass basis: note that the DM fermionic current couples directly to the $Z$, so both $V$ and $Z$ mediate the production of dark sector particles with $q \overline{q}$ interactions, and both must be included in amplitude calculations.

The mixing between $V$ and $Z$ also shifts their masses by a fraction of $\mathcal{O}(\epsilon^2)$: the shift in the $Z$-mass has important consequences for EW precision constraints on these models which we will discuss below, but otherwise these shifts will be neglected for the rest of the chapter. We will always assume that $r \ll1$ throughout in both models.

\subsection{U(1)\texorpdfstring{$_D$}{D} Pseudo-Dirac Dark Matter}
\label{sec:PseudoDirac}

We now consider a simple, viable dark matter model, where the bound state signature gives complementary information about the dark sector and probes different region of the parameter space than the mono-$X$ searches. 
Our model is based on the ``minimal model'' of ~\cite{Finkbeiner:2010sm} (loosely based on the ``excited dark matter'' scenario of ~\cite{Finkbeiner:2007kk}), but we use an ordinary Yukawa interaction between the dark Higgs and the fermions in the dark sector instead of a dimension-5 operator. 

This model contains a gauged U(1)$_D$ field, $V$, kinetically mixed with the SM U(1)$_Y$, a Dirac fermion $\Psi$ and a dark Higgs, which in unitary gauge can be written as $\Phi_D = (v_D + h_D)/\sqrt{2}$, with $v_D$ as its VEV. The U(1)$_D$ charges for the fermion $\Psi$ and $\Phi_D$ are 1 and 2 respectively. The Lagrangian is
\begin{alignat}{2}
	\cL_{\rm dark-Maj} 
&=&& \, i \overline{\Psi} \slashed{D} \Psi + (D_\mu \Phi_D)^\dagger(D^\mu \Phi_D)  - m_D \overline{\Psi}\Psi - y_D \left( \overline{\Psi}^C \Psi \Phi_D^* + \text{h.c.} \right) 
	+\mathcal{L}_{\text{kin-mix}} \, ,
\label{eqn:pseudoDiracL}
\end{alignat}
where $D_\mu \equiv \partial_\mu - i g_D V_\mu$ is the covariant derivative for $\Psi$ and $D_\mu \equiv \partial_\mu - 2i g_D V_\mu$ is the covariant derivative for $\Phi_D$, with $C$ denoting charge conjugation. 
Following~\cite{Dreiner:2008tw}, we write $\Psi$ as a  Weyl fermion pair $(\chi, \eta^\dagger)$. Thus, the Yukawa interaction becomes
\begin{align}
	\cL_{Y_D}
=	-y_D \left(\chi\chi \Phi_D^* + \eta\eta \Phi_D +  \text{h.c.} \right) \, .
\end{align}
After the dark Higgs gets a VEV, the Yukawa interaction generates a fermion mass splitting. The fermion mass matrix is
\begin{align}
	\frac{1}{2}(\chi \ \eta)
	\begin{pmatrix}
		m_M & m_D \\
		m_D & m_M
	\end{pmatrix}
	\begin{pmatrix} \chi \\ \eta \end{pmatrix} + \text{h.c.}
\end{align}
with $m_M=\sqrt{2}y_D v_D$. The mass eigenstates are then given by $\chi_1 = (\eta + \chi)/\sqrt{2}$ and $\chi_2 = i(\eta - \chi)/\sqrt{2}$, with masses $m_{1,2} = m_M \pm m_D$. 
In the mass basis, the dark Yukawa interaction terms can be written as
\begin{align}
	\cL_{Y_D} = 
	- \frac{y_D}{\sqrt{2}}(v_D + h_D)\left(\chi_1 \chi_1 -\chi_2 \chi_2 + \text{h.c.} \right) \, ,
\end{align}
and the interaction with the dark photon is then given by
\begin{alignat}{1}
	-i g_D \left(V_\mu - r^2\frac{\epsilon s_W}{1 - r^2 } Z_\mu \right) \left(\chi_1^\dagger \overline{\sigma}^\mu \chi_2 - \chi_2^\dagger \overline{\sigma}^\mu \chi_1 \right) \, .
\end{alignat}
The interaction with the SM is thus off-diagonal, and the direct detection constraint is significantly relaxed because the $\chi_1$ - $\chi_2$ mass splitting means the elastic scattering cross section is suppressed at one-loop (and the one-loop contribution is expected to be small as previously discussed).

In this model, a DM bound state can be produced at the LHC through the process shown in Fig.~\ref{fig:feynmanMonoXAndResonance}, supported by the exchange of either a dark Higgs or a dark photon. We will focus on the case where the dark Higgs is light and supports the bound state, while the dark photon is heavier and is the principal mediator to the SM, in order to ensure a one-loop suppression in the direct detection cross section while maintaining a large coupling between the quarks and the mediator to the SM and a sizable branching ratio of the bound state to leptons. The dark Higgs is assumed to have some small mixing with the SM Higgs that allows it to decay.

\begin{figure}
    \centering
    \includegraphics[scale=0.33]{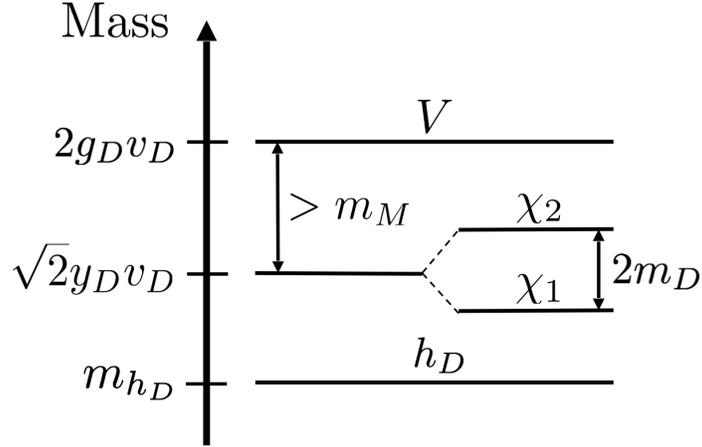}
    \caption{Spectrum of particles in the pseudo-Dirac model.}
    \label{fig:spectrum_PD}
\end{figure}

Because of the symmetry breaking pattern, there are only three independent parameters among~$\{m_\chi, m_V, \alpha_D, y_D\}$. The mass hierarchy required above can be achieved by choosing $m_D \ll m_M$, so that $m_D$ is the small mass splitting, and $m_{1,2} \simeq m_M$. The spectrum of particles in this model is shown in Fig.~\ref{fig:spectrum_PD}.

A large value of $\alpha_D$ leads to a Landau pole in a broken U(1) theory at a scale above $m_V$~\cite{Davoudiasl:2015hxa}. However, since we are mainly interested in the phenomenology of bound states below the scale $m_V$, we assume that a UV completion of the model will avoid the Landau pole. We will later discuss another model with a non-abelian gauge group in the dark sector which will avoid the need for a UV completion. 

If the dark bound state, $\mathcal{B}$, is produced from SM initial states, it must be produced from a $Z$ or $V$ exchange. 
Since the couplings of these gauge bosons to the dark Majorana fermions are off-diagonal, the resulting bound state must be composed of a $\chi_1$ and a $\chi_2$ particle, and for an $s$-wave state, it must have spin-1.
Moreover, since $h_D$ only couples $\chi_1$ to $\chi_1$ and $\chi_2$ to $\chi_2$, decays into final states containing only $h_D$ are forbidden, and if $m_V > m_\mathcal{B}$, the only available decay modes for $\mathcal{B}$ are through $V$ back into the SM particles. 

\subsection{SU(3)\texorpdfstring{$_D$}{D} Triple Higgs Model}
\label{sec:TripleHiggs}

We now consider a dark sector model based on a completely broken SU(3)$_D$ gauge theory, where all of the phenomenologically desirable properties of the dark sector emerge from the breaking pattern of the gauge symmetry. This model has some similarities with the non-abelian DM models of~\cite{Chen:2009ab}, featuring small mass splittings among the components of the DM multiplet that suppress the direct detection cross section. Because the mediator supporting the bound state is a vector in this model as opposed to a scalar in the pseudo-Dirac model above, the indirect detection constraints of the two models turn out to be quite different.

A completely broken SU(3) gauge group was chosen to allow for a sufficiently large gauge coupling, which is favorable for the production of bound states that are supported by gauge bosons.\footnote{DM models with an unbroken gauge group are constrained by the fact that dark matter is effectively collisionless in galactic dynamics~\cite{Ackerman:mha}.} A broken U(1) theory, such as the one found in the pseudo-Dirac model, with a coupling strength $\alpha_D \gtrsim 0.5$ at momentum scales above the gauge boson mass quickly runs into a Landau pole. Thus a broken U(1) theory with a large coupling constant is likely to have emerged from a larger, nonabelian gauge group in the first place~\cite{Davoudiasl:2015hxa}. 
We choose an SU(3) gauge group rather than SU(2), because for a fermion in the fundamental of a completely broken SU(2) theory with an off-diagonal coupling to the SM, the gauge boson corresponding to the diagonal generator produces a repulsive potential between the two components of the fermion, making it difficult for a phenomenologically viable bound state to exist without introducing additional light mediators. 

As in the previous model, the coupling between the dark sector and the SM is mediated by the mixing of the dark and SM gauge bosons; in this non-abelian case, the mixing operator is non-renormalizable. 
Bound states in this model are supported by the exchange of one of the SU(3)$_D$ gluons, which acquires a relatively small mass during the symmetry breaking. 

The dark sector contains a triplet of Dirac fermions $\chi = (\chi_1, \chi_2, \chi_3)$ charged under SU(3)$_D$, with a Dirac mass,~$m_\chi$. After symmetry breaking, the components acquire a small mass splitting, so that $m_{\chi_1} < m_{\chi_2} = m_{\chi_3}$, with $\chi_1$ and $\chi_2$ ultimately forming an $s$-wave, spin-1 bound state, $\mathcal{B}$, which can be produced at colliders. $\chi_1$, being the lightest fermion in this theory, serves as our DM candidate.

The SU(3)$_D$ breaking occurs via three Higgs-like fields: two scalars in the adjoint representation of SU(3)$_D$, $H_1$ and $H_2$, and another scalar in the fundamental, $H_8$. The dark sector Lagrangian is given by
\begin{alignat}{2}
    \mathcal{L}_{\text{dark}} &=&& \sum_{1,2} \frac{1}{2} D_\mu H_i^a D^\mu H_i^a + \frac{1}{2}|D_\mu H_8|^2 - V(H_1, H_2, H_8) \nonumber \\
    & &&+ \overline{\chi} \left(i \slashed{D} - m_\chi \right)\chi - \frac{1}{4} V^{\mu\nu} V_{\mu\nu}\,,
\end{alignat}
where $V_{\mu\nu}$ is the SU(3)$_D$ field strength of the dark gluons, $a = 1, \cdots , 8$ is an SU(3)$_D$ index and $\tau^a \equiv \lambda^a/2$ with $\lambda^a$ being the Gell-Mann matrices. $D_\mu \equiv \partial_\mu - i g_D V_\mu^a \tau^a$ for fields in the fundamental and $D_\mu H_i^a \equiv \partial_\mu H_i^a + g_D f^{abc} V_\mu^b H_i^c$ for the two adjoint Higgs fields. 
The structure of the Gell-Mann matrices is such that $V^1$, $V^2$ couple $\chi_1$ to $\chi_2$, $V^4$, $V^5$ couple $\chi_1$ to $\chi_3$, and $V^6$, $V^7$ couple $\chi_2$ to $\chi_3$; $V^3$ couples diagonally to $\chi_1$ and $\chi_2$, while $V^8$ couples diagonally to all three fermions; the interaction vertices are shown in Appendix~\ref{app:triplehiggs}. The scalar potential  $V(H_1, H_2, H_8)$ can be chosen to satisfy the symmetry breaking pattern that we will describe below.  

We impose a $\mathbb{Z}_2 \times \mathbb{Z}_2$ symmetry at the renormalizable level; each adjoint Higgs transforms under the corresponding $\mathbb{Z}_2$ as $H_i^a \to - H_i^a$ for $i=1, 2$. This forbids any marginal interaction terms between the Higgs sector and the fermion sector, including a Yukawa interaction term. Therefore, we can treat both sectors as decoupled to first order. However, the following dimension-5 operator is allowed:
\begin{alignat}{1}
    \mathcal{L}_{\text{mass}} = \frac{1}{\Lambda_m} \left(H_8^\dagger \tau^a H_8 \right) \left(\overline{\chi} \tau^a \chi \right),
\end{alignat}
so that after $H_8$ acquires a suitable VEV, a mass splitting occurs among the components of $\chi$. Finally, we introduce the following operators that encapsulate the mixing of the dark sector with the SM:
\begin{alignat}{1}
    \mathcal{L}_{\text{mix}} = - \frac{1}{\Lambda_1} H_1^a V_{\mu\nu}^a B^{\mu\nu} - \frac{1}{\Lambda_8^2} \left(H_8^\dagger \tau^a H_8 \right) V^a_{\mu\nu} B^{\mu\nu}
    \label{eqn:tripleHiggsKineticMixing}
\end{alignat}
Notice that the first term introduces a small breaking of the $\mathbb{Z}_2$ symmetry. This term can originate from a dimension-6 operator that respects this discrete symmetry, such as $\phi H_1^a V_{\mu\nu}^a B^{\mu\nu}$, with $\phi$ being a scalar field that is odd under $\mathbb{Z}_2$, which acquires a VEV as well. 
The details of the origin of this operator are unimportant, as we will focus instead on the phenomenology resulting from the kinetic mixing.\footnote{One can in principle include the interaction term $H_2^a V_{\mu\nu}^a B^{\mu\nu}$, but this term does not affect the main features of this model. With the symmetry breaking pattern discussed later, the gauge bosons $V^1$ and $V^2$ couple to the same dark fermions, $\chi_1$ and $\chi_2$. We will leave this term out from the Lagrangian for simplicity.} 

At the point of symmetry breaking, $H_1$ and $H_2$ acquire a VEV $v_1$ and $v_2$ in the 1- and 2-component respectively, and $H_8$ acquires a VEV given by $\langle H_8 \rangle = v_8 (\cos \theta, 0, \sin \theta)$, with $v_8 \lesssim m_\chi \ll v_1, v_2$ and some arbitrary angle $\theta$. This symmetry breaking pattern can be accomplished by choosing an appropriate Higgs potential. Note that similar phenomenology can be obtained even if the VEVs of $H_1$ and $H_2$ are not orthogonal and the second component of $\langle H_8 \rangle$ of the order of $v_8$. This choice of the breaking pattern is therefore not fine-tuned, but is made to avoid unnecessary complications. Further details on the Higgs potential and the symmetry breaking pattern can be found in Appendix~\ref{app:triplehiggs}. The VEV of $H_1$ in the first term of $\mathcal{L}_{\text{mix}}$ leads to the conventional kinetic mixing term discussed above, with $\epsilon \equiv 2 v_1/\Lambda_1$, and $V^1$ as the mediator to the SM. The second term in $\mathcal{L}_{\text{mix}}$ guarantees the prompt decay of the other dark gluons through small mixings into the SM: details are discussed further in Appendix~\ref{app:triplehiggs}. The choice of $\langle H_8 \rangle$ gives a small mass splitting to the Dirac fields in $\chi$, leading to the following fermion masses: 
\begin{align}
    m_{\chi_1} = m_\chi - \frac{v_8^2}{3 \Lambda_m}\, , \qquad m_{\chi_2} = m_{\chi_3} = m_\chi + \frac{v_8^2}{6 \Lambda_m} \,.
\end{align}
We will always neglect the mass splitting when not considering its role in suppressing the direct detection of DM, so that $m_{\chi_1} \simeq m_{\chi_2} = m_{\chi_3} \simeq m_\chi$. The lightest fermion $\chi_1$ is the DM candidate and it is stable; the other particles in the theory decay promptly. More details are provided in Appendix~\ref{app:triplehiggs}.  

\begin{figure}
	\centering
	\includegraphics[scale=0.33]{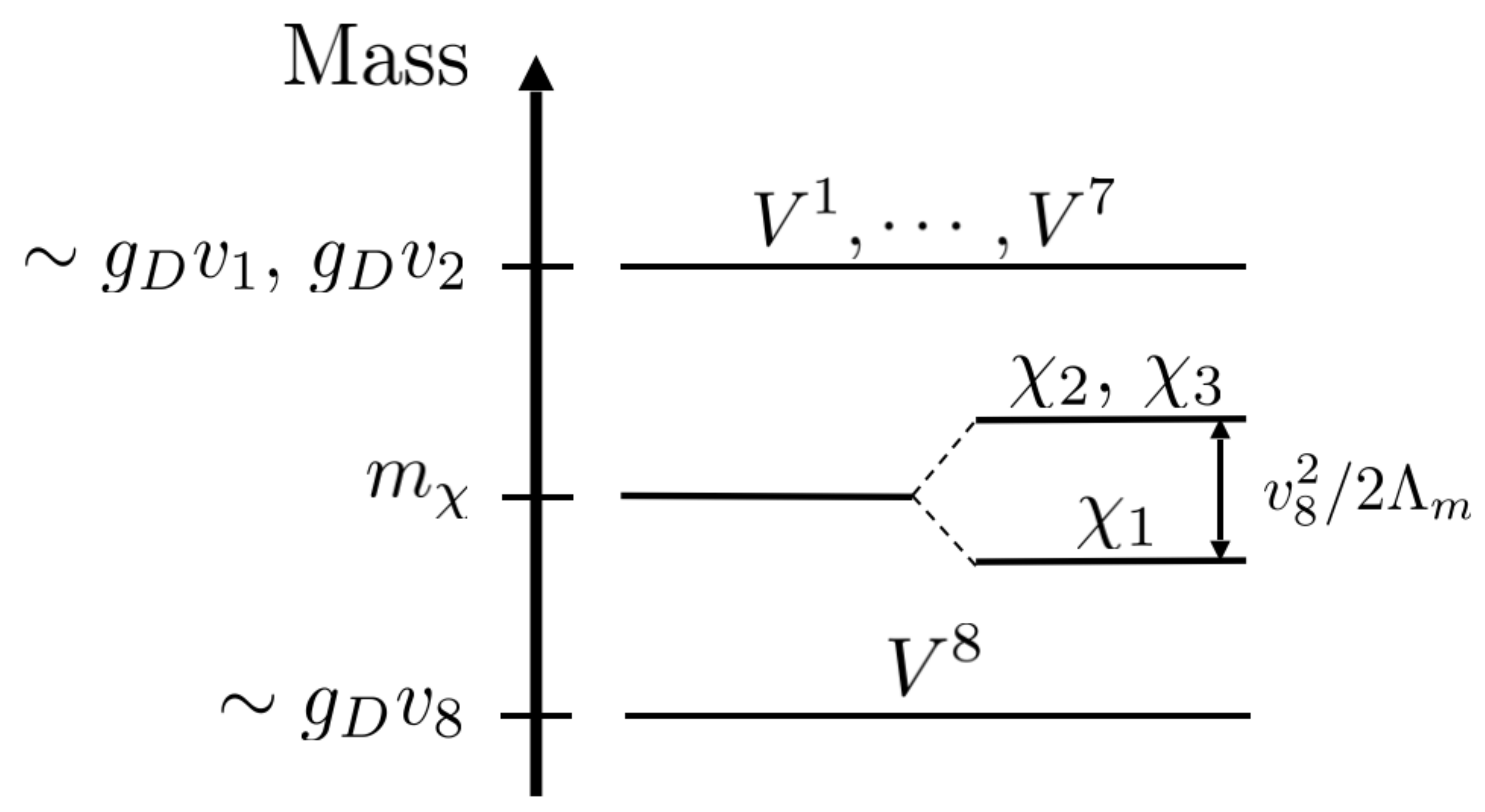}
	\caption{Spectrum of particles in the triple Higgs model.}
	\label{fig:spectrum}
\end{figure}

Finally, the dark gluons remain approximately diagonal after the symmetry breaking, with squared masses (up to order $g_D^2 v_8^2 \ll g_D^2 v_{1,2}^2$) given by:
\begin{gather}
    m_1^2 = g_D^2 v_2^2 \, , \qquad m_2^2 = g_D^2 v_1^2 \, ,  \qquad m_3^2 = g_D^2(v_1^2 + v_2^2) \, , \nonumber \\
    m_4^2 = m_5^2 = m_6^2 = m_7^2 = \frac{1}{4} g_D^2 (v_1^2 + v_2^2) \,, \qquad m_8^2 = \frac{1}{24} g_D^2 v_8^2 (5 - 3 \cos 2 \theta) \, .
\end{gather}
$m_1$ also receives $\mathcal{O}(\epsilon^2)$ corrections from the kinetic mixing with $Z$, which we will neglect as was explained above. Thus, the dark gluon masses satisfy the hierarchy
\begin{alignat}{1}
    m_8 < m_\chi < m_{1, \cdots, 7},
\end{alignat}
and $V^8$ serves as a good candidate for a bound state mediator. Fig. \ref{fig:spectrum} illustrates the spectrum of particles in this model. 

As in the Majorana case, if the dark bound state $\mathcal{B}$ arises from SM processes, then it must be produced from the mediator $V^1$; the resulting bound state must be $\overline{\chi}_1 \chi_2$ or its antiparticle equivalent $\overline{\mathcal{B}}$. Again, since the mediator is spin-1, $s$-wave bound states must be in the spin-triplet configuration. 

In the mass basis, the interaction term responsible for the production is (all fields now denote their mass eigenstate)
\begin{alignat}{1}
    \mathcal{L} \supset \frac{g_D}{2} \overline{\chi}_1 \gamma^\mu \left[V_\mu^1 - r^2 \frac{\epsilon s_W}{1 - r^2 } Z_\mu \right] \chi_2 + \text{h.c.}
\end{alignat}
with $r \equiv m_Z/m_1$. With $m_8 < m_\chi$ and the other gluons being significantly more massive than $m_\chi$, $\mathcal{B}$'s are mediated by $V^8$ through the interaction terms
\begin{alignat}{1}
    \mathcal{L}_{\mathcal{B}} = \frac{g_D}{2 \sqrt{3}} \gamma^\mu \left(\overline{\chi}_1 V^8_\mu \chi_1 + \overline{\chi}_2 V^8_\mu \chi_2 \right),
    \label{eqn:V8chichi}
\end{alignat}
which leads to an attractive potential between the constituents of $\mathcal{B}$. The coupling between $V^8$ and the fermions in $\mathcal{B}$ is therefore $\alpha_{\mathcal{B}} = \alpha_D/12$. The mass hierarchy of this model forbids decays into any of the dark gluons $V^a$ for $a = 1,\cdots,7$. Furthermore, the decay of $\mathcal{B}$ into any number of $V^8$ is forbidden by the conservation of the SU(3)$_D$ color charge in the unbroken SU(3): $V^8$ only couples $\chi_1$ to $\chi_1$ and likewise $\chi_2$ to $\chi_2$, and cannot carry away the net color charge of $\mathcal{B}$. 

\section{Experimental Constraints}
\label{sec:experimentalConstraints}

In this section, we will first discuss in Sec.~\ref{sec:modelParameters} the range of viable model parameters in each of the dark sector models detailed above. We will then study the phenomenology of each of these models at the LHC in Sec.~\ref{sec:LHC}, and their cosmology and indirect detection signatures in Sec.~\ref{sec:id}.

\subsection{Viable Model Parameters}
\label{sec:modelParameters}

In both models, the bound state $\overline{\chi}_1 \chi_2$ (or its antiparticle equivalent, if applicable) is formed from a stable dark matter candidate $\chi_1$ and an unstable fermion $\chi_2$. In order for the decay of $\chi_2$ to not dilute the production of the bound state, we must ensure that the decay width of $\chi_2$ is much smaller than the decay width of $\mathcal{B}$. In both models, $\chi_2$ decays through an off-shell SM mediator to $\chi_1$ and two SM particles. In the pseudo-Dirac model, this three-body decay width is parametrically $\Gamma_{\chi_2} \sim \epsilon^2 g_D^2 g_{\text{SM}}^2 (\Delta m)^5/m_V^4$, where $g_{\text{SM}}$ is a coupling constant to the SM which depends on the actual SM particle considered, and $\Delta m=m_{\chi_2}-m_{\chi_1}$, which we always take to be small. On the other hand, the bound state decay width is $\Gamma_{\mathcal{B}} \sim \epsilon^2 g_D^2 g_{\text{SM}}^2 m_\chi^2 |\psi(0)|^2 / m_V^4$. The relative ratio of these widths is therefore
\begin{alignat}{1}
    \frac{\Gamma_{\chi_2}}{\Gamma_{\mathcal{B}}} \sim \frac{(\Delta m)^5}{\alpha_{\mathcal{B}}^3 m_\chi^5} \ll 1,
    \label{eqn:Gammachi2OverGammaB}
\end{alignat}
where $\alpha_\mathcal{B} = y_D^2/4 \pi \sim \mathcal{O}(0.1 - 1)$ for situations where LHC production of bound states is important. An identical relationship holds for the triple Higgs model, with $\alpha_{\mathcal{B}} = \alpha_D/12$. 

In both of the models we have presented, the interaction between the dark sector and the SM is controlled by a single vector boson: $V$ in the pseudo-Dirac model of Sec.~\ref{sec:PseudoDirac} and $V^1$ in the triple Higgs model of Sec.~\ref{sec:TripleHiggs}. 
The mixing of the SM and dark sectors shifts the $Z$ mass, and is thus constrained by EW precision tests~(EWPT). In particular, the $\rho$ parameter is shifted by an amount~\cite{Cassel:2009pu}
\begin{alignat}{1}
    \Delta \rho = -\frac{m_W^2}{m_V^2} t^2_W \cdot \epsilon^2 + \mathcal{O}\left( \frac{m_W^4}{m_V^4} \right),
\end{alignat}
where $m_V$ is the mass of the SM mediator in either model, and $t_W$ is the tangent of the weak mixing angle. The global fit for the central value of $\rho$ to EWPT data is $\rho_0 = 1.00037 \pm 0.00023$~\cite{Patrignani:2016xqp}. Constraints are set by requiring that any choice of $\epsilon$ leads to a minimum value of $m_V$ such that $\Delta \rho$ is consistent with the 2$\sigma$ limit for the value of $\rho_0$. 

Next, in order for a bound state to be possible, the constraint given in Eq.~\eqref{eqn:boundStateRequirement} must be satisfied. This condition can be satisfied by ensuring that the mass of the particle supporting the bound state is sufficiently small. For the pseudo-Dirac model, this means choosing a sufficiently small dark Higgs mass such that $y_D^2 m_\chi > 21.1 m_{h_D}$, and for the triple Higgs model, ensuring that $\alpha_D m_\chi > 20.16 m_8$.

Finally, to avoid direct detection constraints, the mass splitting must exceed the typical kinetic energy of DM in the solar circle. Taking the velocity dispersion of DM to be $v \sim 10^{-3}$, this means that the mass splitting has to exceed approximately $10^{-6} m_\chi$. A small mass splitting, albeit large enough to be consistent with this lower bound, can be achieved by picking suitable values for the Dirac bare mass $m_D$ in the pseudo-Dirac model and $\Lambda_m$ in the triple Higgs model. 

In both theories, there are two parameters ($m_D$ and $m_{h_D}$ for the pseudo-Dirac model, $m_8$ and $\Lambda_m$ for the triple Higgs model) that can be set to naturally satisfy both the criterion for bound states and avoid direct detection constraints, while having little impact on the LHC phenomenology. However, these parameters can have some influence on the relic abundance of DM in these theories, as well as on indirect detection bounds. This will be discussed after the next section.

\subsection{LHC Phenomenology}
\label{sec:LHC}

We now turn our attention to the production and detection of bound states at the LHC for both theories. In the perturbative picture, bound states $\mathcal{B}$ are produced by quark anti-quark parton interactions through an $s$-channel $V$ and $Z$ (mass-eigenstate) boson, with the only available decay mode of $\cB$ being an off-shell $V$ or $Z$ back into SM particles, leading to resonance signatures. The more accurate procedure of taking into account the mixing of $V$ and $\mathcal{B}$ yields a qualitatively similar result; we use the full mixing calculation in all of the plots shown, but focus our qualitative discussion primarily on the perturbative picture.\footnote{We neglect any mixing between $V$, $\mathcal{B}$ with $Z$, since we will usually take $V$ and $B$ to be much heavier than $Z$, and the coupling between $Z$ and the dark sector particles is suppressed by $\epsilon$.}

The mono-$X$ + MET search can be effective in setting constraints on these dark models, particularly in the range of parameter space where $2m_\chi < m_V$, the region of interest for both dark sector models. To study the constraints that mono-jet + MET searches can place on our models, we use \texttt{FeynRules}~\cite{Alloul:2013bka} and \texttt{MadGraph}~\cite{Alwall:2014hca} to obtain the MET distribution for a wide range of $m_\chi$ and $m_V$. The distribution is then compared to the observed 95\,\% confidence upper limit on the number of mono-jet + MET events in 10 inclusive MET bins obtained by ATLAS with \SI{36.1}{\per\femto\barn} of data~\cite{Aaboud:2017phn}. Any value of $m_\chi$ and $m_V$ with a MET distribution that has more events in any inclusive bin than the 95\% upper limit is deemed to be ruled out by the experiment. 

Next, we recast bounds from a search for resonance in dilepton events in \SI{36.1}{\per\femto\barn} of \SI{13}{\tera\eV} ATLAS data~\cite{ATLAS:2017wce} to set constraints on the production of $\mathcal{B}$. 
In the models considered here, $\mathcal{B}$ decays entirely into SM particles with a significant branching ratio to pairs of leptons, making the dilepton resonance search a particularly powerful probe. This search constrains the production cross section times branching ratio of a $Z'$ boson assuming some minimal vector couplings to the SM fermions, which allows us to directly interpret these constraints as a limit on the production of cross section times branching ratio of the bound state $\mathcal{B}$. 

These searches are also sensitive to the resonant production of the vector mediator $V$ itself, which tends to be significantly more constraining than mono-jet + MET searches when the coupling of the mediator to SM quarks are comparable to the coupling to DM. However, in portal models like the ones we are considering, the mixing into the SM $\epsilon$ is small while the coupling to DM $\alpha_D$ can be large. In the range of parameter space where the mediator mass $m_V \gtrsim 2m_\chi$, $V$ overwhelmingly decays into $\overline{\chi}_1 \chi_2$ or $\overline{\chi}_2 \chi_1$, which correspond to final states with MET and are vetoed in dilepton resonance searches to suppress $W$ and $Z$ backgrounds~\cite{ATLAS:2017wce}. The search for $\mathcal{B}$, however, faces no such limitation in this region of parameter space.

The production cross section of $\mathcal{B}$ (and equivalently of $V$) can be computed from Eq.~\eqref{eqn:qqbartoB}, assuming the narrow width approximation. 
In the perturbative picture, $\mathcal{B}$ decays through an $\epsilon$-suppressed coupling to the $Z$, or through $V$, which has an $\epsilon$-suppressed coupling to both $J^\mu_{\text{EM}}$ and $J^\mu_Z$. The resulting expression for the bound state width to quarks is
\begin{alignat}{2}
    \label{eqn:Bwidth}
    \Gamma_{\mathcal{B} \to q \overline{q}} &=&& \frac{16 \pi N_m r^4 m_\chi^2}{c_W^2(m_Z^2 - 4 m_\chi^2)^2}  \frac{\alpha \alpha_D \epsilon^2 |\psi(0)|^2}{(m_Z^2 - 4 r^2 m_\chi^2)^2 + r^2 m_Z^2 \Gamma_V^2(s = m_\mathcal{B}^2)} \nonumber \\
    & &&\times \bigg[ \left(c_W^2 Q(m_Z^2 - 4m_\chi^2) + 4 g_V m_\chi^2\right)^2 + 16 g_A^2 m_\chi^4  \nonumber \\
    & && \qquad + \frac{r^2}{(1 - r^2)^2}  \Gamma_V^2(s = m_\mathcal{B}^2)(g_V^2 + g_A^2) \bigg] \,,
\end{alignat}
%
%
%
where $\alpha$ is the EM fine structure constant, $Q$ is the electric charge of the quark, $g_V$ and $g_A$ are the vector and axial couplings of $q$ to the $Z$ respectively, given by  $g_V = \{0.25, -0.0189, 0.0959, -0.1730 \}$ and $g_A = \{0.25, -0.25, 0.25, -0.25\}$ for $\{\nu_e, e, u, d\}$ and for the other 2 generations respectively. $N_m = 4$ for the pseudo-Dirac model and $N_m = 1$ for the triple Higgs model, which accounts for the difference in coupling and fermion types. As previously, $|\psi(0)|^2$ is the squared amplitude of the wave function of the bound state at the origin, given explicitly by
\begin{alignat}{1}
    |\psi(0)|^2 = \begin{cases}
        \left( \frac{y_D^2}{4\pi} \right)^3 \frac{m_\chi^3}{8 \pi}, &\text{Pseudo-Dirac}, \\
        \left(\frac{\alpha_D}{12}\right)^3 \frac{m_\chi^3}{8\pi},  &\text{Triple Higgs}.
    \end{cases}
    \label{eqn:psi02}
\end{alignat}
Note that we have assumed throughout that the bound state is well-approximated by non-relativistic quantum mechanical results, which is a valid assumption so long as the binding energy of $\mathcal{B}$ is far less than $m_\chi$. For this bound state, we thus require
\begin{alignat}{1}
    \frac{1}{4}\alpha_\mathcal{B}^2 m_\chi \ll 2 m_\chi,
\end{alignat}
where $\alpha_\mathcal{B} = y_D^2/4\pi$ for the pseudo-Dirac case, and $\alpha_\mathcal{B} = \alpha_D/12$ for the triple Higgs model. 

For sufficiently large values of $\alpha_{\mathcal{B}}$, next-to-leading order (NLO) corrections may be significant. For our benchmark values $y_D = 2.5$ and $\alpha_D = 3.0$, the NLO corrections can be roughly estimated to be of order $\alpha_{\mathcal{B}}/4\pi \sim 5\%$. Even if the NLO corrections turn out to be larger, we do not expect our results to change qualitatively, since the parameter space that is both probed by the bound-state dilepton search and unconstrained by indirect detection for our benchmark couplings is significant. A proper NLO calculation is thus beyond the scope of our work.

As we argued earlier, the production cross section of $\mathcal{B}$ crucially depends on the total width of $V$; this means that the total width of $V$ should be included in the computation of the width shown in Eq.~\eqref{eqn:Bwidth}. Importantly, the width of $V$ should be evaluated at $s = m_\mathcal{B}^2$, since $\mathcal{B}$ lies below the $\chi \overline{\chi}$ open production threshold \cite{Patrignani:2016xqp}. The perturbative partial widths of $\mathcal{B}$ as well as $V$ into all possible SM final states are shown in Appendix~\ref{app:widths}. 

In the mixing picture, the partial widths of $V$ calculated here correspond to $\Gamma_{V,0}$. We take $\Gamma_{\mathcal{B},0} = 0$, since $\Gamma_{\mathcal{B},0} = \Gamma_{\chi_2} \ll \Gamma_{\mathcal{B}}$, as shown in Eq.~(\ref{eqn:Gammachi2OverGammaB}). In the pseudo-Dirac model, there is only one bound state, and the mixing calculation proceeds in the same fashion as described in Sec.~\ref{sec:VBMixing}. The sum of the perturbative partial widths of $\mathcal{B}$, calculated in Appendix~\ref{app:widths}, is numerically a good approximation to the width after mixing, $\Gamma_{\mathcal{B}}$. For the triple Higgs model, there are two bound states, $\mathcal{B}$ and $\overline{\mathcal{B}}$, and so all three states need to be simultaneously diagonalized. However, $\mathcal{B}$ and $\overline{\mathcal{B}}$ maximally mix to form two CP eigenstates,
\begin{alignat}{1}
    \mathcal{B}_\pm = \frac{\ket{\mathcal{B}} \pm \ket{\overline{\mathcal{B}}}}{\sqrt{2}}.
\end{alignat}
Since $V^1$ is a CP-even state, it does not mix with the CP-odd combination $\mathcal{B}_-$, and the diagonalization is performed over $V^1$ and the CP-even $\mathcal{B}_+$; the CP-odd state $\mathcal{B}_-$ does not interact with the SM.
In both models, the unmixed mass matrix given in Eq.~(\ref{eqn:massMixing}), with the mixing parameter $f$ given by \cite{Franzini:1987jw,Kuhn:1985eu,Kuhn:1987ty}
\begin{alignat}{1}
    f = 4 N_f \psi(0) \sqrt{\pi \alpha_D m_{\mathcal{B},0}}\, ,
\end{alignat}
where $N_f = 1$ for the pseudo-Dirac model, and $N_f = 1/\sqrt{2}$ for the triple Higgs model, which accounts for the differences in coupling and bound-state mixing.

In both models, $\mathcal{B}$ cannot decay into final states that only contain the mediator which supports the bound state: this is because both the dark Higgs in the pseudo-Dirac model and $V^8$ in the triple Higgs model have couplings with the DM fermion number that conserves the number of each of $\chi_1$ and $\chi_2$. 

Decays of $\mathcal{B}$ into dark sector final states become possible once $m_{\mathcal{B}} \gtrsim m_V$ in the pseudo-Dirac model, or $m_{\mathcal{B}} \gtrsim m_1$ in the triple Higgs model: the final states are $Vh_D$ and $V^1 V^8 V^8$ respectively. Because of the large coupling between the DM fermions and the mediators, these dark sector decays are the main decay modes of $\mathcal{B}$, rendering the dilepton resonance search for $\mathcal{B}$ ineffective. These dark sector final states all mix with the SM, and can in principle lead to multilepton signatures at the LHC. Earlier studies have exploited this signature to look for bound states \cite{An:2015pva,Bi:2016gca}, but we do not explore this possibility here for two reasons. First, once $m_{\mathcal{B}}$ becomes significantly greater than the SM mediator mass, the resonant enhancement derived in Eq.~(\ref{eqn:qqbartoBlimit}) becomes ineffective, and the cross section for producing $\mathcal{B}$ drops quickly away from $m_{\mathcal{B}} \sim m_V$ or $m_1$. Furthermore, the branching ratio of these mediators to leptons is small, since they kinetically mix through the U(1)$_Y$ and decay predominantly into quarks. Second, a direct search for the $V$ or $V^1$ resonance is significantly more constraining, since the mediator is lighter than the bound state, and there is one fewer factor of the branching ratio to leptons to contend with. In both models, the mediator dilepton resonance search rules out all of the parameter space for $m_\mathcal{B} > m_1$ or $m_V$ once the coupling to the SM is sufficiently large.

At tree level, we are therefore only interested in the decay modes of $\mathcal{B}$ and $V$ into the SM: both particles can decay into a pair of SM fermions, as well as $W^+W^-$ and $Zh$ where $h$ is the SM Higgs, through the mixing of $V$ with $Z/\gamma$. Neither particle can decay into $ZZ$ or $\gamma \gamma$ final states, since these processes are forbidden by charge conjugation symmetry. 

The sensitivity of the dilepton resonance search depends strongly on the width of the resonance, and the \SI{13}{\tera\eV} ATLAS limits with \SI{36.1}{\per\femto\barn} of data as a function of the ratio of the width of the resonance to its mass $\Gamma/m$ are presented in \cite{ATLAS:2017wce}. The total widths of both states are fully taken into account when computing the limits of the search, and the search is assumed to be completely ineffective once $\Gamma/m > 0.32$.

\begin{figure*}[t!]
    \centering
    \includegraphics[scale=0.68]{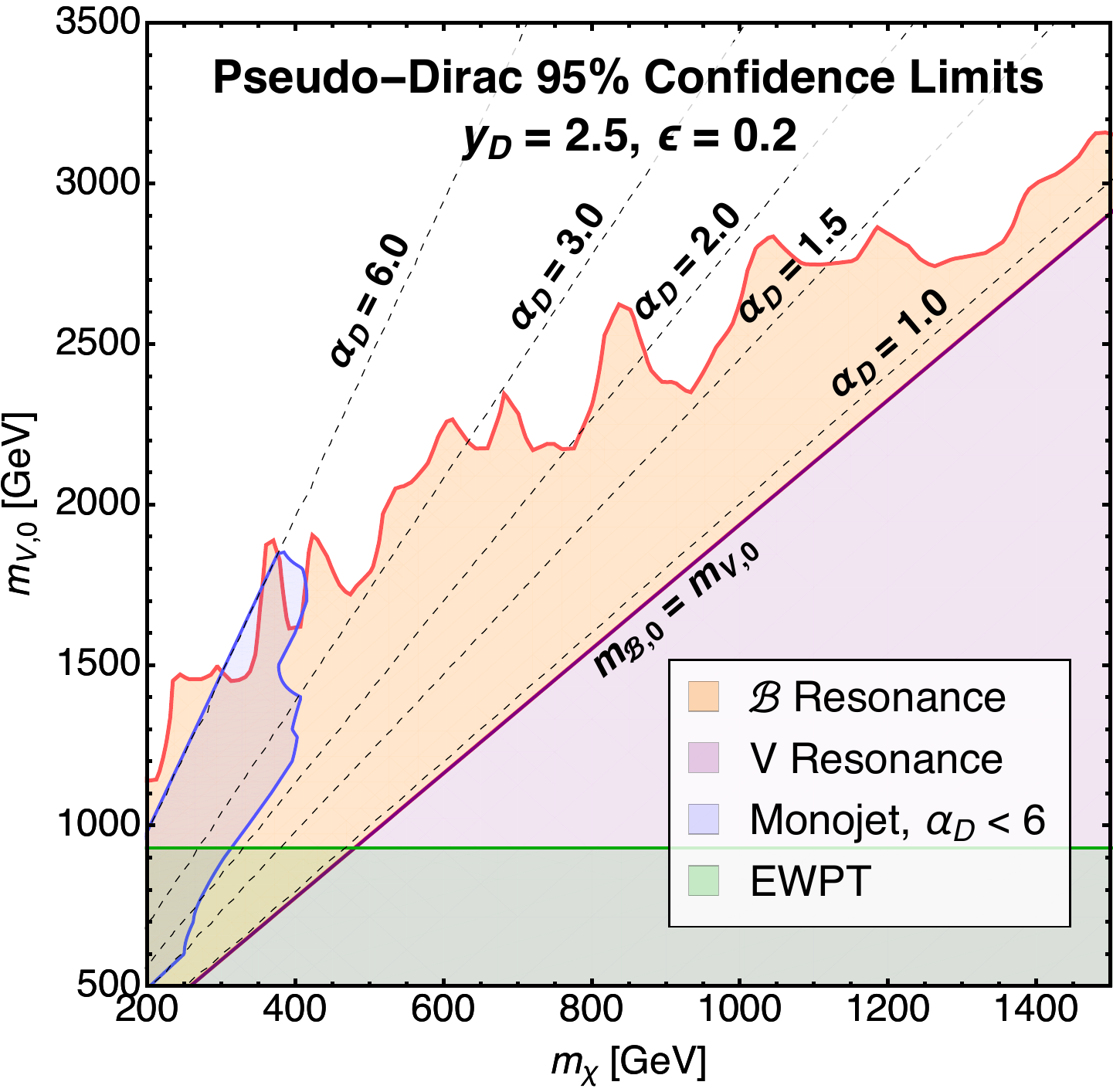}
    \caption{95\% confidence limits in the $m_\chi - m_{V,0}$ plane of the pseudo-Dirac model. $m_{V,0}$ and $m_{1,0}$ are the unmixed masses of the mediator in each respective model. All resonance calculations are made using the full mixing calculation.
    Experimental constraints from mono-jet + MET~(blue), dilepton $\mathcal{B}$ resonance~(orange), dilepton $V$ resonance~(purple) and EWPT constraints~(green) are shown for $y_D = 2.5$, $\epsilon = 0.2$ for the pseudo-Dirac model, and $\alpha_D = 3$, $\epsilon = 0.3$ for the triple Higgs model. All dilepton resonance searches are for the lighter mass eigenstate after mixing. The dark sector coupling $\alpha_D$ is completely fixed by a choice of $\{m_\chi, m_V, y_D\}$; contours (black, dashed) indicate the value of $\alpha_D$ on the $m_\chi - m_V$ plane when $y_D = 2.5$.
    }
    \label{fig:LHCPseudoDirac}
\end{figure*}

\begin{figure*}[t!]
    \centering
    \includegraphics[scale=0.68]{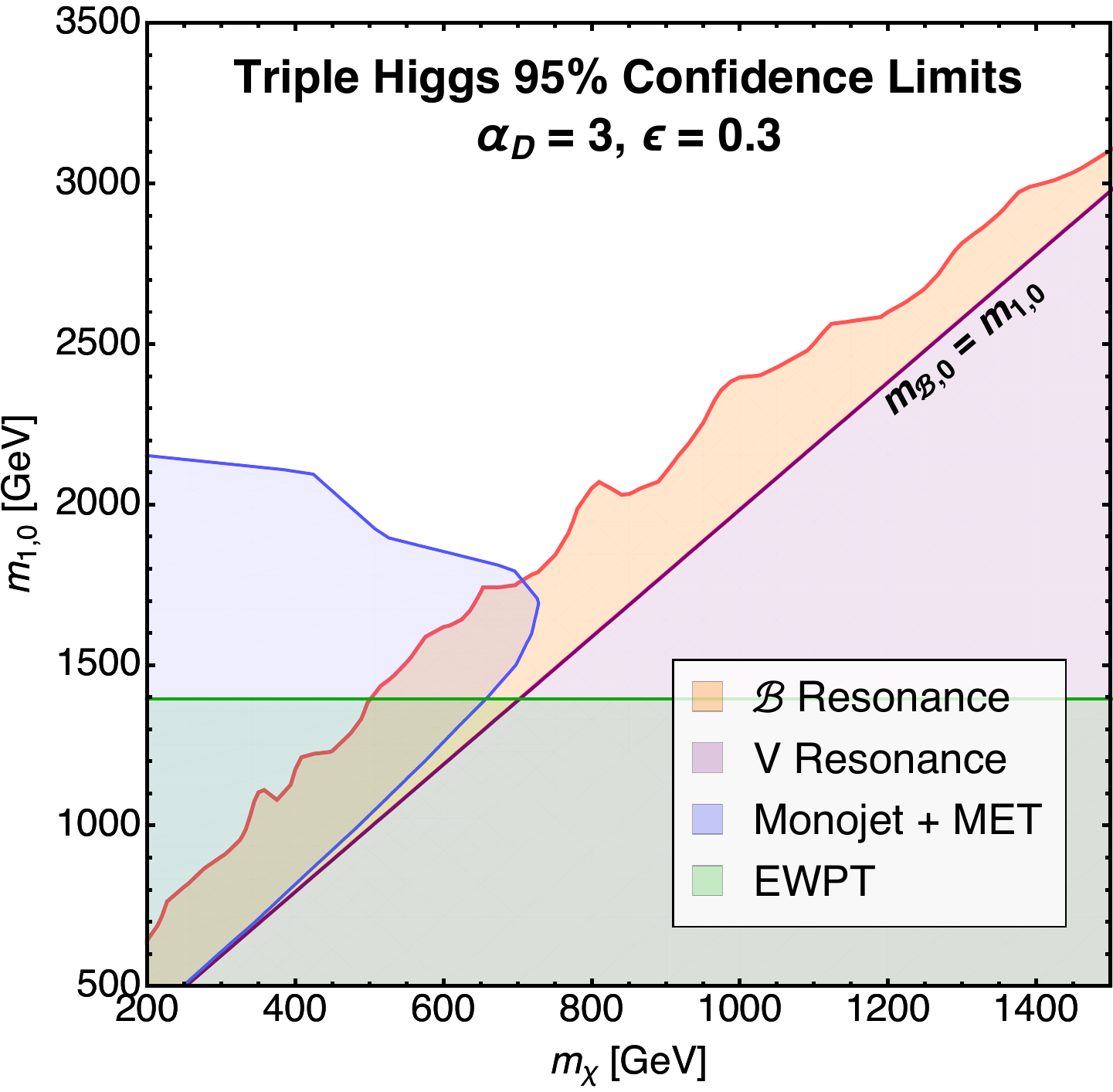}
    \caption{95\% confidence limits in the $m_\chi - m_{1,0}$ plane for the triple Higgs model. $m_{V,0}$ and $m_{1,0}$ are the unmixed masses of the mediator in each respective model. All resonance calculations are made using the full mixing calculation.
    Experimental constraints from mono-jet + MET~(blue), dilepton $\mathcal{B}$ resonance~(orange), dilepton $V$ resonance~(purple) and EWPT constraints~(green) are shown for $y_D = 2.5$, $\epsilon = 0.2$ for the pseudo-Dirac model, and $\alpha_D = 3$, $\epsilon = 0.3$ for the triple Higgs model. All dilepton resonance searches are for the lighter mass eigenstate after mixing.
    }
    \label{fig:LHCTripleHiggs}
\end{figure*}

\begin{figure*}[t!]
    \centering
    \includegraphics[scale=0.68]{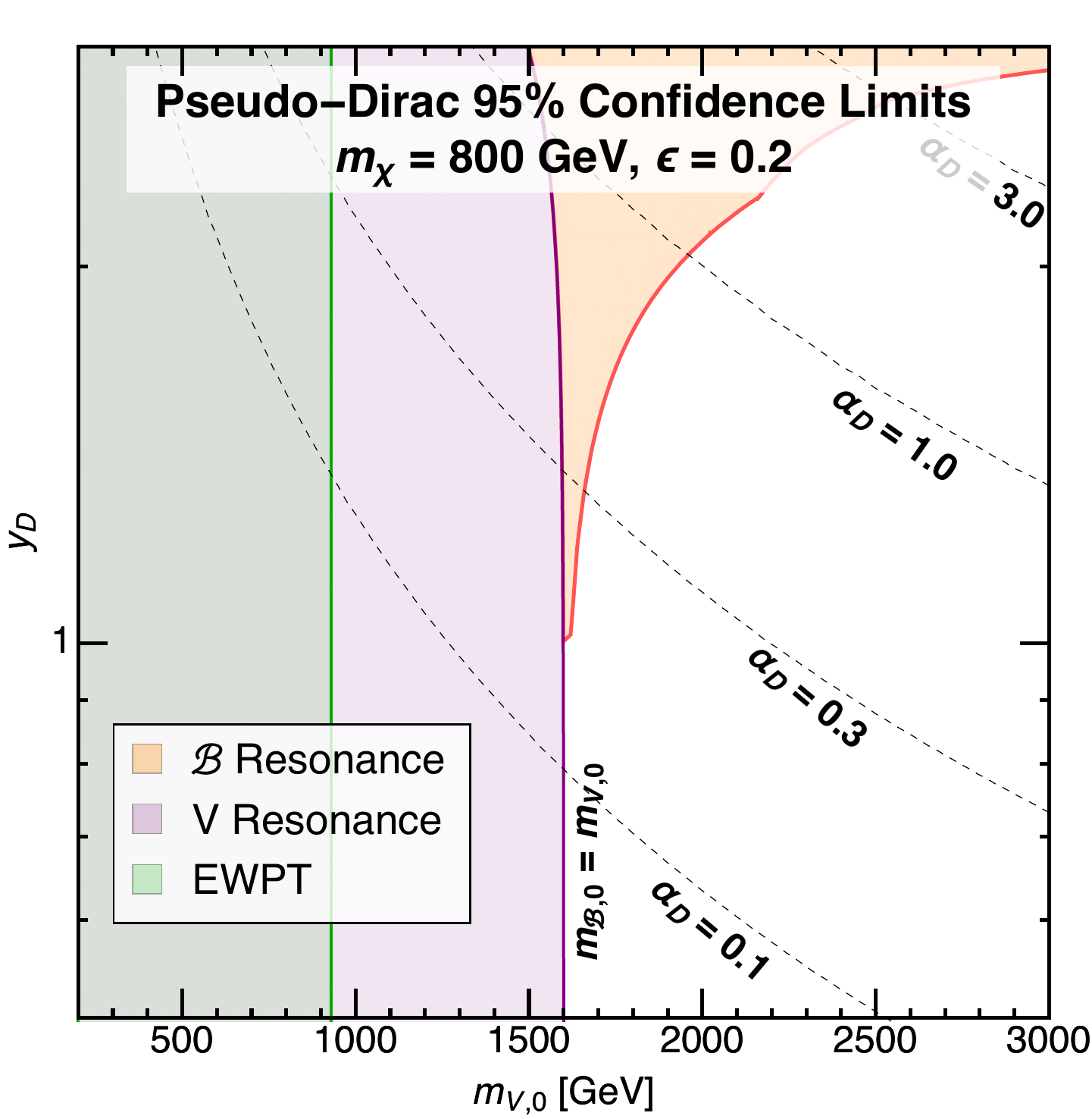}
    \caption{95\% confidence limits in the $m_{V,0} - y_D$ plane of the pseudo-Dirac model, similar to Fig.~\ref{fig:LHCPseudoDirac}. Experimental constraints from dilepton $\mathcal{B}$ resonance~(orange), dilepton $V$ resonance~(purple) and EWPT constraints~(green) are shown for $m_\chi = 800$ GeV, $\epsilon = 0.2$ for the pseudo-Dirac model, and $m_\chi = 800$ GeV, $\epsilon = 0.3$ for the triple Higgs model. Contours (black, dashed) of $\alpha_D$, which is fixed for a given choice of $\{m_\chi, m_V, y_D\}$ for the pseudo-Dirac model, are also shown. 
    }
    \label{fig:LHCPseudoDirac2}
\end{figure*}

\begin{figure*}[t!]
    \centering
    \includegraphics[scale=0.69]{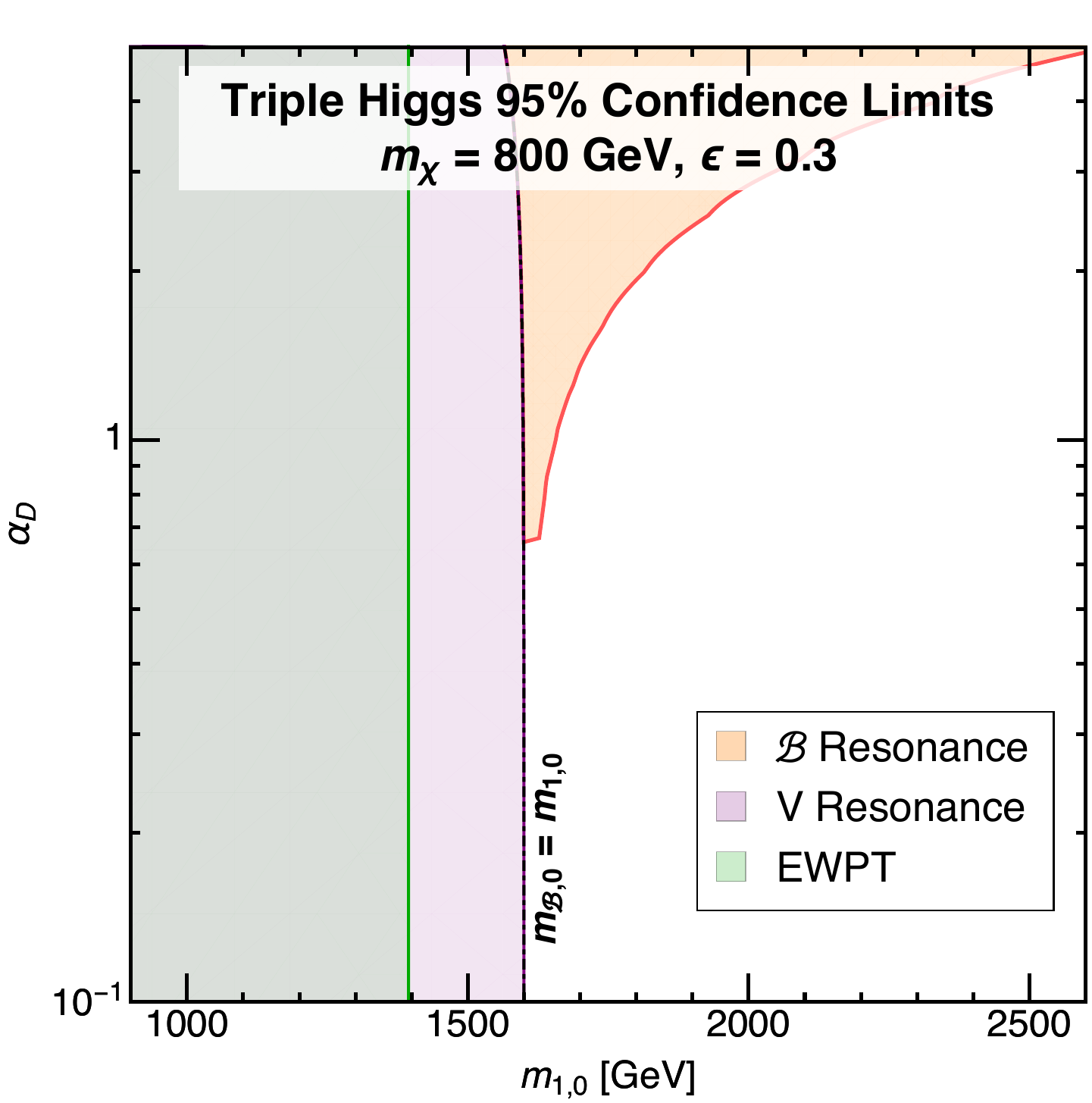}
    \caption{95\% confidence limits in the $m_{1,0} - \alpha_D$ plane of the triple Higgs model, similar to Fig.~\ref{fig:LHCTripleHiggs}. Experimental constraints from dilepton $\mathcal{B}$ resonance~(orange), dilepton $V$ resonance~(purple) and EWPT constraints~(green) are shown for $m_\chi = 800$ GeV, $\epsilon = 0.2$ for the pseudo-Dirac model, and $m_\chi = 800$ GeV, $\epsilon = 0.3$ for the triple Higgs model.
    }
    \label{fig:LHCTripleHiggs2}
\end{figure*}

The resulting 95\% confidence limits from mono-jet + MET, dilepton resonance and EWPT are shown in Figs.~\ref{fig:LHCPseudoDirac} and~\ref{fig:LHCTripleHiggs} in the $m_\chi - m_V$ plane and in Figs.~\ref{fig:LHCPseudoDirac2} and~\ref{fig:LHCTripleHiggs2} in the $m_\chi$ - bound-state coupling plane for both models.

The dilepton resonance search results presented in both figures are searches for the lighter resonance state in the mixing picture; the search switches from $V$ to $\mathcal{B}$ along the line $m_{V,0} = m_{\mathcal{B},0}$, where the lighter resonance changes rapidly from being mostly $V_0$ to mostly $\mathcal{B}_0$ as one moves from below to above this line.\footnote{In spite of this, the partonic cross section including both $V$ and $\mathcal{B}$ is continuous across this line; it is only the particle that should be identified with the narrow Breit-Wigner signal at the low mass eigenvalue that changes.} As we argued earlier, since the mass eigenstates are always well-separated and the lighter resonance is always narrow, we can simply assume that the total cross section is given by a Breit-Wigner profile with a width given by either the $V$ ($m_{V,0} < m_{\mathcal{B},0}$) or the $\mathcal{B}$ ($m_{\mathcal{B},0} > m_{V,0}$) and neglect interference effects. In both cases, the search for the $\mathcal{B}$ resonance when $m_{V,0} > m_{\mathcal{B},0}$ extends the reach of experimental constraints significantly into this region of parameter space, as compared to what we might expect from just the vector resonance search and the mono-jet + MET search combined.

For the values of $\alpha_D$ and $y_D$ selected in these benchmark models, the separation between the ground state and the first excited state of the bound state is about 2 - 10\% of the DM mass. These states may be resolvable into different lepton resonances, since the mass resolutions for the dielectron and dimuon channels are $\sim 1\%$ and $\sim 5\%$ respectively. The cross section of production of the ground state in this case is still given byEq.~(\ref{eqn:qqbartoB}), but without the factor of $\zeta(3) \approx 1.202$. Resolvable resonances would be a strong signature of bound states, but will come after an initial discovery of a new resonance, which is the main focus of this chapter.


\subsection{Freezeout and Indirect Detection}
\label{sec:id}

We now turn our attention to the freezeout process for the DM in each model, as well as constraints derived from indirect detection experiments. Let us focus on the annihilation channels that do not suffer a suppression by $\epsilon$, in order to be as model-independent as possible. In the pseudo-Dirac model, the potential kinematically available final states (at late times) are $h_D h_D$ and $\mathcal{B}^\prime h_D$, with the latter channel corresponding to radiative formation of a bound state, $\mathcal{B}^\prime$ (which may be spin-1 or spin-0). The $Vh_D$ final state is forbidden, since $V$ couples $\chi_1$ to $\chi_2$, and $h_D$ couples $\chi_1$ to $\chi_1$. In the triple Higgs model, if all the gauge bosons and Higgses except $V^8$ are heavier than the DM, the only open final states are $V^8 V^8$ and the radiative bound state formation. Note that in the limit where the DM is slow-moving, radiative bound state formation requires not merely that the mediator be light compared to $\alpha_\cB m_\chi$, as required for a bound state, but that it satisfy the stronger condition that the mediator mass is smaller than the binding energy, $m_Y \lesssim \alpha_\cB^2 m_\chi/4$. 
Thus, this process can be forbidden by increasing the mediator mass, and indeed we will see that indirect detection limits are much easier to satisfy in regions of parameter space where $\alpha_\mathcal{B} m_\chi \gtrsim m_Y \gtrsim \alpha_\mathcal{B}^2 m_\chi/4$. In this regime, the DM annihilation products will thus be determined by the decays of the bound state mediator.

During freezeout, the partner particles $\chi_2$ (in the pseudo-Dirac model) and $\chi_2$, $\chi_3$ (in the triple Higgs model) are also present, and their annihilation and co-annihilation channels may also relevant.

\begin{figure*}[t!]
    \centering
	\includegraphics[scale=0.68]{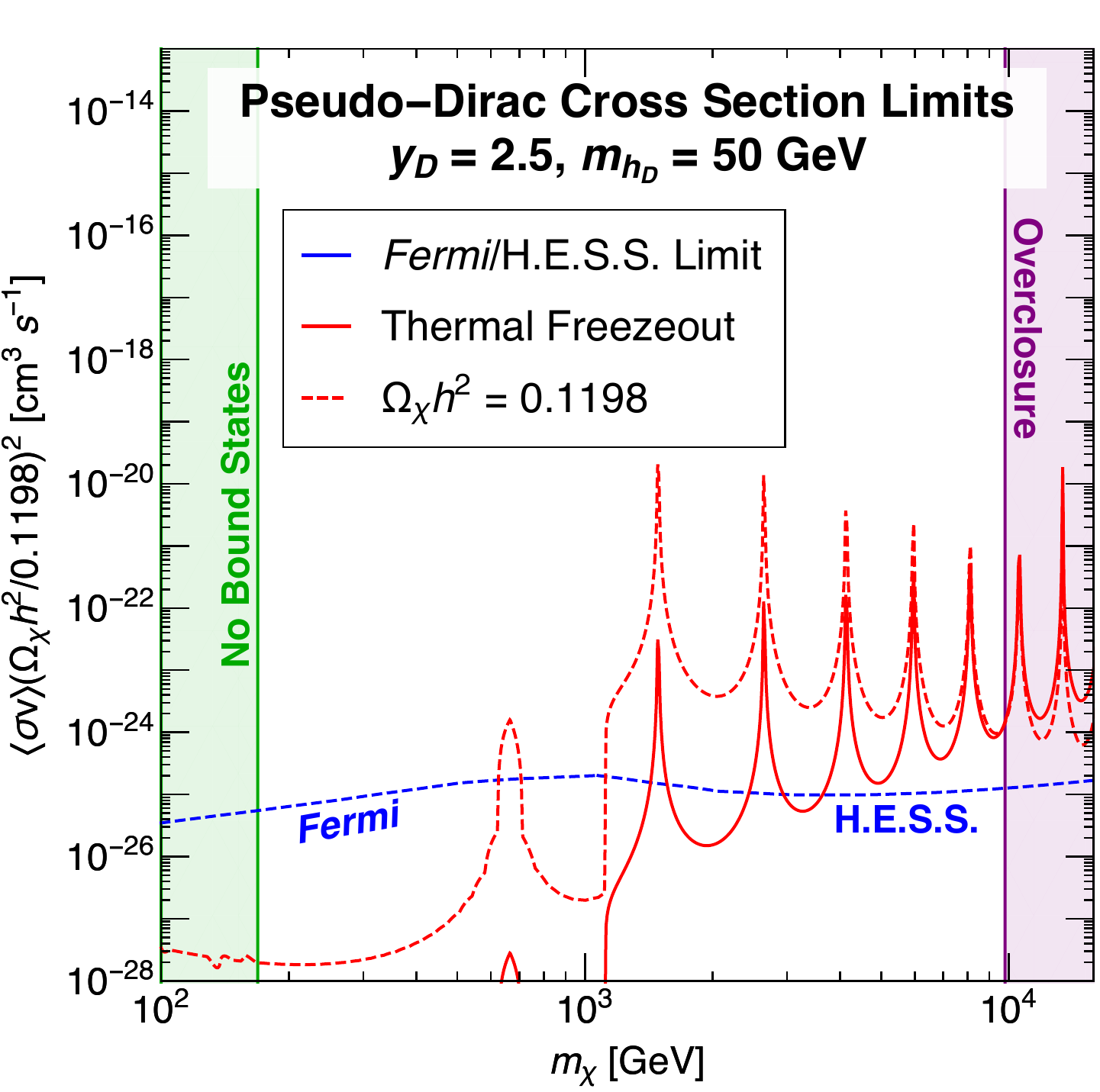}
    \caption{Comparison of predicted DM annihilation rates (including Sommerfeld enhancement and radiative bound state formation) to constraints on the $\bar{b} b$ channel from \textit{Fermi} observations of dwarf galaxies~\cite{Fermi-LAT:2016uux} and H.E.S.S. observations of the Galactic center region~\cite{Abdallah:2016ygi}. The red solid line indicates the predicted cross section, rescaled by the fraction of DM squared, for thermally produced DM. For the total DM abundance we take $\Omega_\chi h^2 = 0.1198$ \cite{Ade:2015xua}. The red dashed line shows the predicted cross section only, corresponding to the assumption that the annihilating species constitutes 100\% of the DM. The region to the right of the vertical purple line is ruled out by overproducing the DM abundance. We show the result for the pseudo-Dirac model with $y_D=2.5$.}
    \label{fig:indirectPseudoDirac}
\end{figure*}

\begin{figure*}[t!]
    \centering
    \includegraphics[scale=0.68]{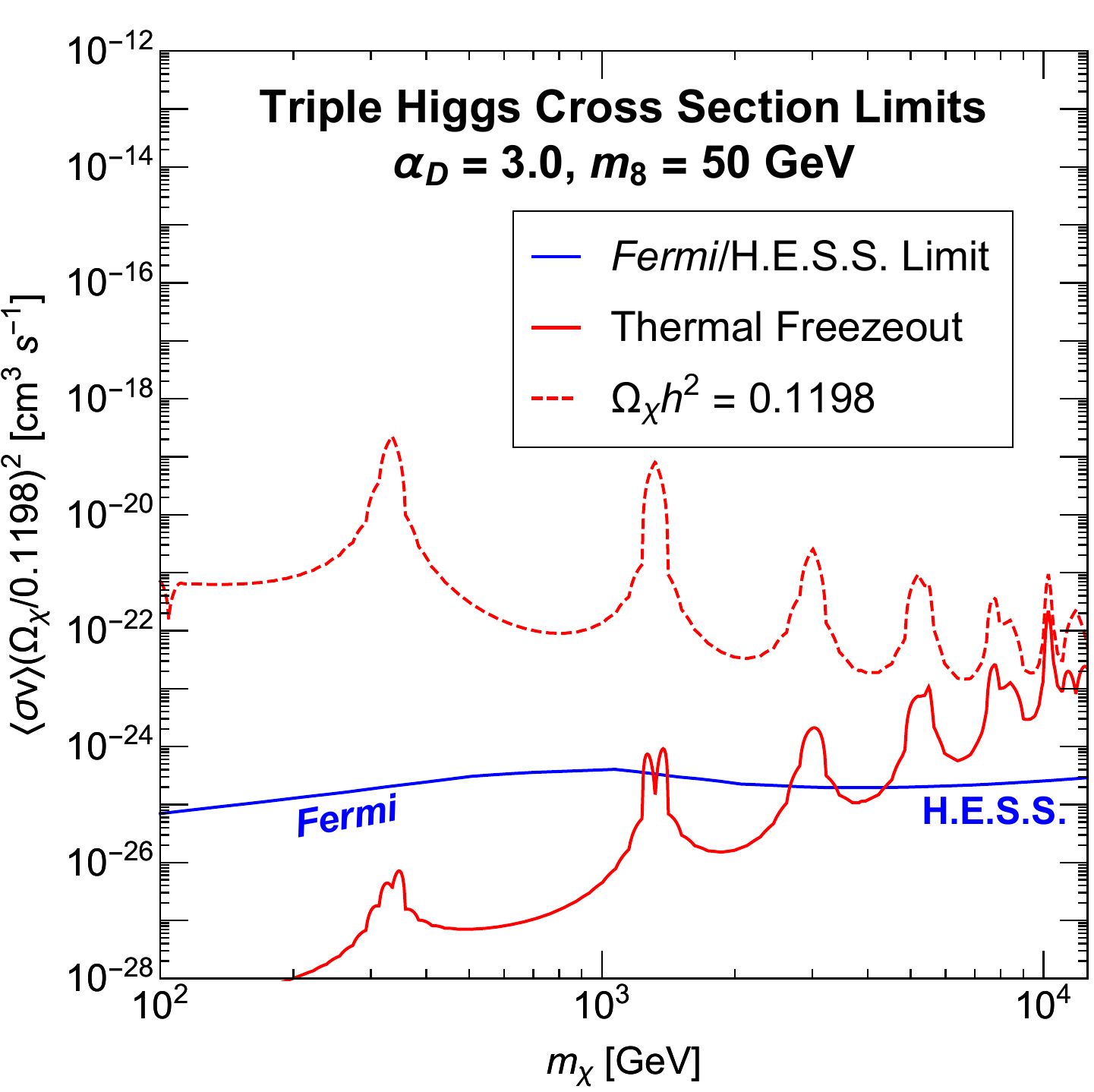}
    \caption{Comparison of predicted DM annihilation rates (including Sommerfeld enhancement and radiative bound state formation) to constraints on the $\bar{b} b$ channel from \textit{Fermi} observations of dwarf galaxies~\cite{Fermi-LAT:2016uux} and H.E.S.S. observations of the Galactic center region~\cite{Abdallah:2016ygi}. The red solid line indicates the predicted cross section, rescaled by the fraction of DM squared, for thermally produced DM. For the total DM abundance we take $\Omega_\chi h^2 = 0.1198$ \cite{Ade:2015xua}. The red dashed line shows the predicted cross section only, corresponding to the assumption that the annihilating species constitutes 100\% of the DM. The region to the right of the vertical purple line is ruled out by overproducing the DM abundance. We show the result for the triple-Higgs model with $\alpha_D = 3.0$.}
    \label{fig:indirectTripleHiggs}
\end{figure*}

\begin{figure*}[t!]
	\centering
	\includegraphics[scale=0.68]{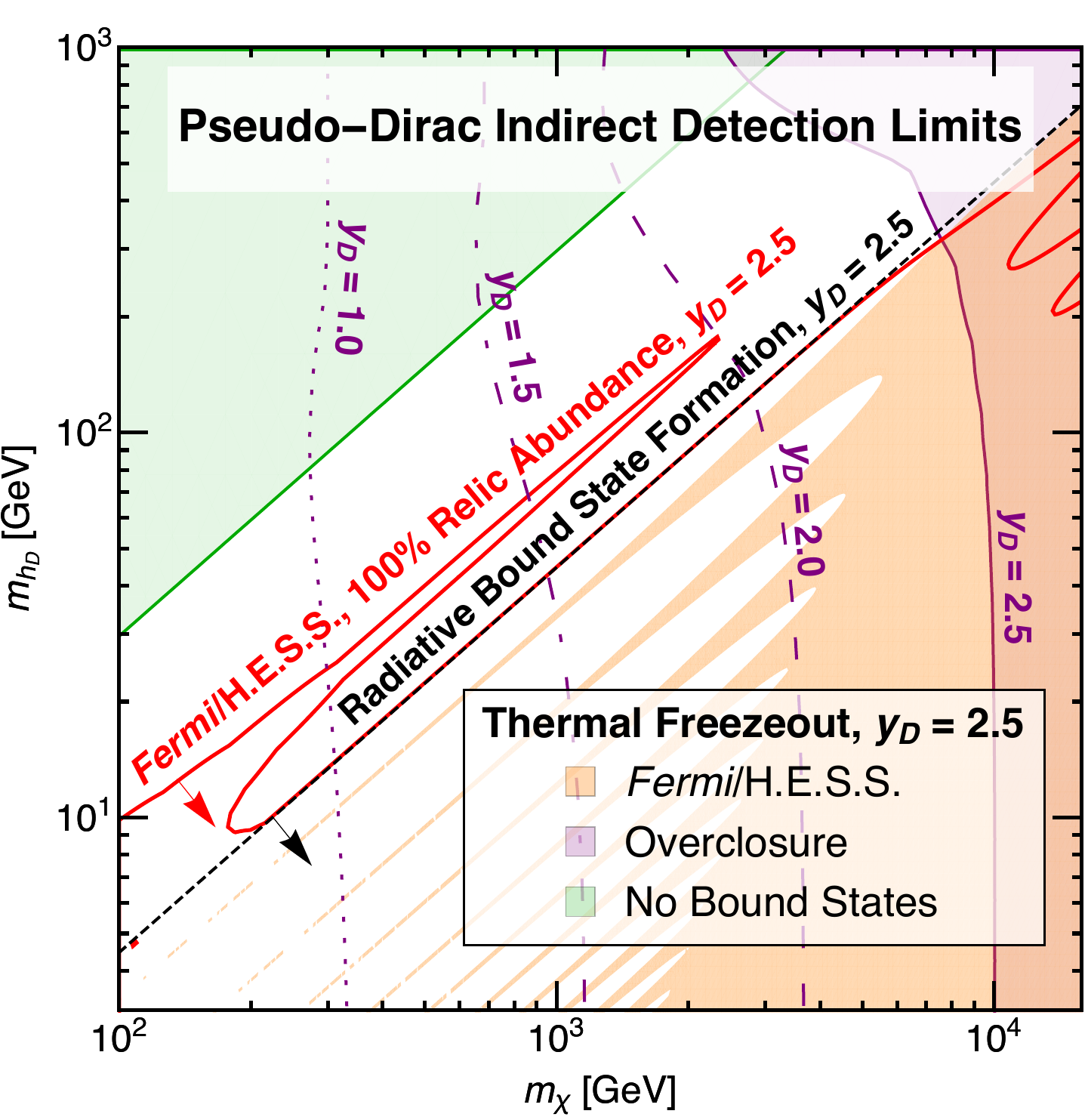}
    \caption{Indirect detection and overclosure limits on the $m_\chi - m_{h_D}$ plane of the pseudo-Dirac model. Shaded regions indicate excluded regions where no bound states exist~(green), the cosmological DM abundance is overproduced~(purple), and the estimated gamma-ray signal exceeds bounds from the \textit{Fermi} and H.E.S.S. telescopes~(orange), for $y_D = 2.5$. In the region below the dashed black line, bound state formation can proceed in the Milky Way halo through emission of an on-shell $h_D$ or $V^8$, and contributes to the indirect detection signal. Dashed, dot-dashed and dotted purple lines indicate the more stringent overclosure limits for smaller values of the coupling. The region below the solid red line is excluded by gamma-ray bounds if the DM candidate of the model is assumed to be symmetric and to comprise 100\% of the DM (from a non-thermal origin).
    }
    \label{fig:indirectPseudoDirac2}
\end{figure*}

\begin{figure*}[t!]
   \centering
   \includegraphics[scale=0.68]{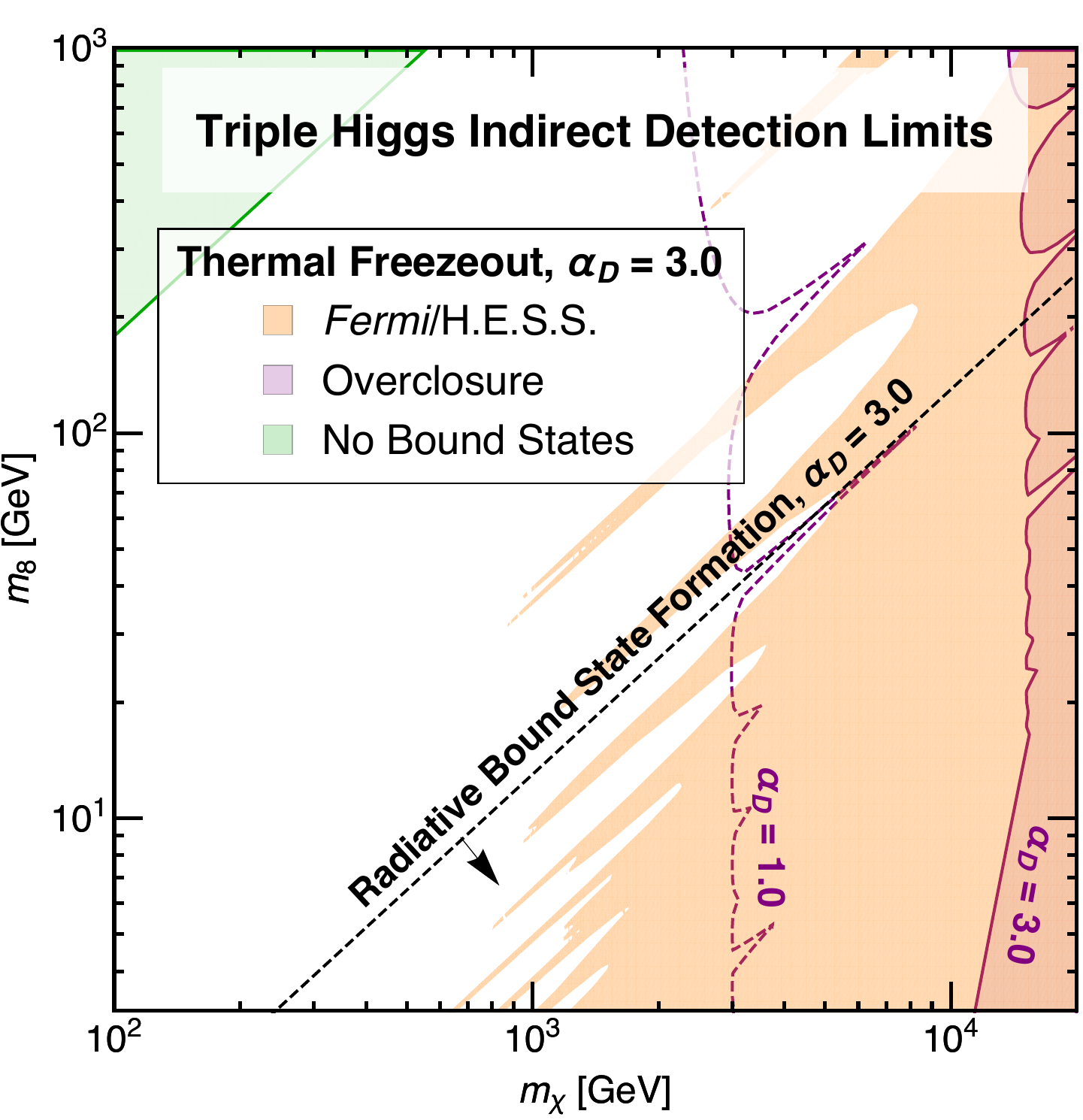}
    \caption{Indirect detection and overclosure limits on the $m_\chi - m_{h_D}$ plane of the $m_\chi - m_8$ plane of the triple Higgs model. Shaded regions indicate excluded regions where no bound states exist~(green), the cosmological DM abundance is overproduced~(purple), and the estimated gamma-ray signal exceeds bounds from the \textit{Fermi} and H.E.S.S. telescopes~(orange), for $\alpha_D = 3.0$. In the region below the dashed black line, bound state formation can proceed in the Milky Way halo through emission of an on-shell $h_D$ or $V^8$, and contributes to the indirect detection signal. Dashed, dot-dashed and dotted purple lines indicate the more stringent overclosure limits for smaller values of the coupling. For the case where we assume that the DM candidate comprises 100\% of the DM, the entire parameter space for $\alpha_D = 3.0$ is excluded.
    }
    \label{fig:indirectTripleHiggs2}
\end{figure*}

\subsubsection{Pseudo-Dirac Model}

If $m_V > 2 m_\chi$ and the bound-state mediator is too heavy for radiative bound state formation, then the only annihilation channel not suppressed by $\epsilon$ or kinematically forbidden is annihilation to $h_Dh_D$\,. Both $\chi_1 \chi_1$ and $\chi_2 \chi_2$ pairs can annihilate in this fashion, but there is no tree-level coannihilation; $\chi_1 \chi_2 \rightarrow h_D h_D$ does not occur for the same reason that the $\chi_1 \chi_2$ bound state does not decay into the dark sector. The cross section for $\chi_i \chi_i \rightarrow h_D h_D$ in the limit of low DM velocity, before accounting for the Sommerfeld enhancement, is given by:
\begin{align} \sigma v_\text{rel}
& = \frac{\pi}{6} v_\text{rel}^2 \left( \frac{ y_D^2}{ 4 \pi} \right)^2 \frac{(9 - 8x_h^2 + 2x_h^4) \sqrt{1 - x_h^2}}{(2 - x_h^2)^4 m_\chi^2},
 \end{align}

where $x_h \equiv m_{h_D}/m_\chi$. We will assume that during freezeout the mass splitting between $\chi_2$ and $\chi_1$, set by $m_D$, is small compared to the freezeout temperature; for $\mathcal{O}$(TeV) DM this corresponds to requiring a mass splitting at the GeV scale or below, which is not in tension with the requirement that the mass splitting be large enough to prevent elastic scattering in the present-day halo (where typical kinetic energies for a TeV DM particle are of order 1\,MeV or less). 
In this case, the abundances of $\chi_1$ and $\chi_2$ remain equal during freezeout, as their equilibrium abundances are equal and their annihilation channels are identical. 
Consequently, each of $\chi_1$ and $\chi_2$ must constitute half the DM abundance, with the $\chi_2$ subsequently decaying to $\chi_1$ ({this occurs through emission of an off-shell $V$).

Since $p$-wave processes can dominate during freezeout, to compute the full rate we will need the Sommerfeld enhancement factor for higher-$l$ processes. The Sommerfeld enhancement for multipole $l$ due to a Yukawa potential can be numerically approximated by~\cite{Cassel:2009wt}:
 \begin{multline}
     S_l 
     \approx 
     \frac{\pi}{\epsilon_v} \frac{\sinh\left(2\pi \delta \right)}{\cosh\left(2\pi \delta \right) - \cosh\left(2\pi \delta \sqrt{1 - \epsilon_\phi^*/\epsilon_v^2}\right)} \prod_{k=1}^l  \frac{k^4 \epsilon_\phi^{*2} + 2 k^2 (2 \epsilon_v^2  - \epsilon_\phi^*) + 1}{k^2 \epsilon_\phi^{*2} + 4 \epsilon_v^2},  
 \end{multline}
 where $\delta \equiv \epsilon_v / \epsilon_\phi^*$, $\epsilon_v = v_\text{rel}/(2\alpha_\mathcal{B})$ and $\epsilon_\phi^*= (\pi^2/6) m_Y/(\alpha_\mathrm{B} m_\chi)$.
 
 We determine the relic density by numerically solving the Boltzmann equation (following the method of~\cite{Steigman:2012nb}) for the $\chi_1$ state and then doubling the result to account for the contribution from $\chi_2$. We integrate the Sommerfeld-enhanced velocity-dependent cross section over the thermal velocity distribution (assuming a Maxwell-Boltzmann distribution) for the DM at each timestep. As discussed earlier, we neglect radiative bound state formation during freezeout. We define overclosure to occur when $\Omega_\chi h^2 > 0.1228$, corresponding to the $2 \sigma$ upper limit ($0.1198 + 2 \times 0.0015$) from Ref.~\cite{Ade:2015xua}.
 
To estimate the signal in indirect detection, we first compute the Sommerfeld-enhanced cross section for $\chi_1 \chi_1 \rightarrow h_D h_D$ in the Milky Way halo, assuming the local DM velocity distribution in the Galactic frame follows a Maxwellian distribution, $f(v) = \sqrt{2/\pi} v^2 e^{-v^2/2\sigma^2}/\sigma^3$, with $\sigma = 150$\,km/s. This choice corresponds to $v_c = \sqrt{2} \sigma \sim 220$ km/s, consistent with the standard halo model~\cite{1986MNRAS.221.1023K,Reid:2009nj,2010MNRAS.402..934M,Herzog-Arbeitman:2017zbm}.

For $m_{h_D}$ smaller than the binding energy, we also account for radiative formation of $\chi_1 \chi_1$ bound states (followed by decay into SM particles). To estimate the bound state formation rate via light scalar emission at low velocities, we add to this rate the analytic low-velocity estimate of~\cite{An:2016kie} for the cross section for capture into the ground state (which dominates the overall capture rate),
\begin{equation} 
	\sigma v_\text{rel} 
	\approx 
	\frac{1}{2} \frac{ \pi \alpha_\mathcal{B}^2}{m_\chi^2} \frac{2^6\pi^2 \alpha_\mathcal{B}^2 e^{-4}}{9 \epsilon_\phi^* \sin^2(\pi/\sqrt{\epsilon_\phi^*})} \, .
	\label{eq:scalarbound}
\end{equation}
Note that Ref.~\cite{An:2016kie} derives this expression from the Hulth\'{e}n potential, so in Eq.~\eqref{eq:scalarbound} we have replaced $m_\phi/(\alpha_D m_D)$ in their result with the parameter $\epsilon_\phi^*$; the Hulth\'{e}n potential with this rescaled mass parameter gives a better approximation to the Yukawa potential~\cite{Cassel:2009wt}. Furthermore, we have included an extra factor of $1/2$ to account for the fact that our annihilating particles are identical fermions, and thus only spin-singlet configurations contribute to this $s$-wave process (yielding a factor of $1/4$), but the overall cross section is increased by a factor of 2, as discussed in Ref.~\cite{Asadi:2016ybp}.

The experimental sensitivity to this cross section will depend on the final state to which the $h_D$ particles eventually decay, which in turn depends on $m_{h_D}$ and whether $h_D$ mixes with the SM-Higgs. However, in general hadronic decays will dominate the signal (due to the larger number of hadronic degrees of freedom), and the photon spectra from decays to different quark species are rather similar, as they arise largely from the decays of neutral pions produced in hadronic showers~\cite{Cirelli:2010xx}. 
Thus, we can estimate the sensitivity of indirect detection by examining the constraints set by assuming a $b\bar{b}$ final state.

Figs.~\ref{fig:indirectPseudoDirac} and~\ref{fig:indirectTripleHiggs} we show limits on the annihilation cross section to $b\bar{b}$ for Majorana DM from the \textit{Fermi}~\cite{Fermi-LAT:2016uux} and H.E.S.S.~\cite{Abdallah:2016ygi} gamma-ray telescopes, and sample results for the predicted annihilation rate from our two models. 
The H.E.S.S. limit, which dominates for DM masses above 1\,TeV, is based on a study of the region within 300\,pc of the Galactic Center, and assumes an Einasto density profile for the dark matter; if the Milky Way possesses a large core, these limits might be substantially weakened.
The \textit{Fermi} limits are based on a study of Milky Way dwarf spheroidal galaxies. The intermediate step of light mediator production will further broaden the photon spectrum, but Ref.~\cite{Elor:2015bho} demonstrated that the effect on the constraints is modest for hadronic final states where the spectrum is already quite broad. Thus to obtain an estimate of the constraints, we simply adopt the cross section limits for annihilation to $b\bar{b}$. We compare the maximum allowed cross section $\langle \sigma v_\text{rel} \rangle_\text{max}$ to the predicted cross section scaled by the fraction of DM in the $\chi_1$ state, $\langle \sigma v_\text{rel} \rangle (\Omega_{\chi_1} h^2/0.1198)^2$; examples for the pseudo-Dirac model with $y_D=2.5$ and the triple-Higgs model with $\alpha_D=3.0$ are shown in Figs.~\ref{fig:indirectPseudoDirac} and~\ref{fig:indirectTripleHiggs}, both for $m_V(m_8)=50\,$GeV.

In Fig.~\ref{fig:indirectPseudoDirac2}, we plot the regions in $m_\chi-m_{h_D}$ plane where bound states exist, the universe is not overclosed, and indirect limits are not violated. 
We see that there are almost no indirect constraints for DM masses below a few TeV and $m_{h_D}$ larger than the binding energy (when $m_{h_D}$ is below the binding energy, there remain allowed regions, but they must be chosen to avoid resonant Sommerfeld enhancement). We also plot the regions allowed by indirect detection bounds if a non-thermal history is assumed to ensure that $\chi_1$ constitutes 100\% of the DM, with $\Omega_\chi h^2 = 0.1198$. In this case, the indirect constraints are much more stringent, but the bulk of the region where $m_{h_D}$ exceeds the binding energy is still unconstrained.

\subsubsection{Triple-Higgs Model}

If the vector bosons other than $V^8$ are all at a heavy mass scale, then the dominant DM annihilation process (not involving bound states) both during freezeout and in the present day is tree-level annihilation to two $V^8$ bosons. This channel is available for $\bar{\chi}_i \chi_i$, where $i=1,2,3$; if $\sigma_i$ denotes the cross section for $\bar{\chi}_i \chi_i \rightarrow V^8 V^8$, then we have:
\begin{align} 
	\sigma_1 v_\text{rel} & = \sigma_2 v_\text{rel} = \frac{\pi (\alpha_D/12)^2}{m_\chi^2} = \frac{1}{16} \sigma_3 v_\text{rel}.
\end{align}
This channel furthermore experiences an attractive $s$-wave Sommerfeld enhancement, which for purposes of this estimate we approximate using Eq.~(\ref{eq:sommerfeld}).

Potential exchanges of $V^8$ bosons, which have large rates compared to processes involving the heavier gauge bosons, do not couple the $\bar{\chi}_i \chi_i$ and  $\bar{\chi}_j \chi_j$ states for $i \ne j$. Likewise, there is no (tree level) coannihilation to the $V^8 V^8$ final state. Thus, we can treat the $\chi_i$ species as evolving independently from each other, annihilating only with their own antiparticles, each experiencing its own long-range attractive Yukawa potential due to $V^8$ exchange. The effective couplings are $\alpha_D/12$ for $\chi_1$ and $\chi_2$ and $\alpha_D/3$ for $\chi_3$.

However, one important question is whether the different $\chi_i$ fields truly evolve independently, and in particular, whether decays and scatterings that interconvert between the $\chi_i$ states are rapid enough to keep the various state populations coupled during freezeout. An example process is $\bar{\chi}_1 \chi_1 \leftrightarrow \bar{\chi}_3 \chi_3$ scattering via $t$-channel $V^4$or $V^5$ exchange (see Appendix~\ref{app:triplehiggs}). As all such processes involve the heavier gauge bosons, they are slow compared to annihilation into a $V^8 V^8$ final state near the time of freezeout. For the $\overline{\chi}_1 \chi_1 \to \overline{\chi}_3 \chi_3$ process, the cross section for this scattering process is approximately $\sigma_{\chi_1 \overline{\chi}_1 \to \chi_3 \overline{\chi}_3} \sim \alpha_D^2 m_\chi^2/M^4$ where $M\sim m_{4,5}$. Compared with $\overline{\chi}_i \chi_i \to V^8 V^8\sim\alpha^2_D/m^2_\chi$, the rate of processes that scatter one type of fermion into another is suppressed by a factor of $\sim m_\chi^4/M^4$. Thus, the process $\chi_1\bar{\chi}_1 \to \chi_3\bar{\chi}_3$ freezes out before the $\chi_1\bar{\chi}_1 \to V^8V^8$ and is therefore not relevant to determining the relic abundance. 
This estimate assumes that all of the gluons other than $V^8$ are more massive than the DM; if this assumption breaks down, the three dark-matter-like populations will no longer evolve independently, and freezeout will be modified.

Under this assumption, we solve separate Boltzmann equations for each of the $\chi_i$ species (accompanied by their antiparticles), and require that the resulting mass density $2 (m_{\chi_1} n_{\chi_1} + m_{\chi_2}  n_{\chi_2} + m_{\chi_3} n_{\chi_3})$ matches the cosmological density of dark matter. The masses of the three states are assumed to be equal, with mass splittings small compared to the temperature at freezeout. The greater annihilation rate of $\bar{\chi}_3 \chi_3$ causes its abundance to be depleted faster than $\bar{\chi}_1 \chi_1$ and $\bar{\chi}_2 \chi_2$.

To estimate the late-time indirect detection limits, we proceed as for the pseudo-Dirac case above, and present our results in Figs.~\ref{fig:indirectTripleHiggs} and~\ref{fig:indirectTripleHiggs2}. The allowed cross section for DM annihilation is doubled as the DM $\chi_1$ is a Dirac fermion in this case. Since there is an unsuppressed $s$-wave annihilation channel, there are useful constraints from indirect detection even when radiative bound-state formation is kinematically forbidden. To estimate the contribution from bound-state formation, we numerically calculate the cross section for capture into the ground state by dipole photon emission, and also add the contribution from an $s$-wave initial state transitioning into the first excited state by emission of a dipole photon. The former process dominates when the mediator mass can be neglected \cite{Asadi:2016ybp}, but is suppressed in the very-low-velocity regime as it corresponds to a $p$-wave initial state \cite{An:2016gad}; thus we add the latter process to properly include the leading contribution at very small velocities. We follow the numerical method described in Ref.~\cite{Asadi:2016ybp}.

Note that in this case, the scenario where $\chi_1$ constitutes 100\% of the DM at late times is essentially completely excluded by indirect detection, for $\alpha_D=3.0$ and $m_\chi$ below 10 TeV; such a scenario requires a non-thermal origin for the DM, as the annihilation cross section is well above the thermal value and would deplete the DM density efficiently during freezeout. If non-thermal processes produce more DM at late times, then the large bare annihilation cross section and accompanying Sommerfeld enhancement (and possibly radiative bound-state formation) gives rise to very strong indirect detection signals, as shown in Fig.~\ref{fig:indirectTripleHiggs}.

Both the overdepletion of the DM density and the large direct detection signals may be avoided if the Dirac-fermion DM possesses some tiny asymmetry, similar to the baryon asymmetry of the SM. The large annihilation cross sections found in these models can readily deplete the DM abundance to the point where the asymmetry sets the residual relic density, and then the indirect-detection signals are suppressed by the absence of the symmetric component. Note that if no such asymmetry is present, indirect detection limits may also pose challenges for sub-\SI{}{\giga\eV} DM and mediators as studied by Ref.~\cite{An:2015pva}; thermal relic DM can be quite generically excluded for sub-\SI{}{\giga\eV} mediators and sub-\SI{}{\tera\eV} DM \cite{Cirelli:2016rnw}.

This behavior does not occur for the pseudo-Dirac model because the principal annihilation channel is $p$-wave suppressed; this both makes it possible for \SI{}{\tera\eV}-scale $\chi_1$ particles to constitute 100\% of the DM with a thermal history, and ensures that large regions of parameter space remain that are not excluded by current indirect detection bounds (although bound state formation can provide indirect detection signals, as in Ref.~\cite{An:2016kie}).

\section{Discussion}
\label{sec:dis}

The resonant production of dark sector bound states at the LHC can be an important complementary search channel to the missing energy and mediator resonance searches. 
Unlike a mediator resonance search, a bound state resonance search directly probes the properties of the DM, and can be more effective when the mediator decays primarily to invisible DM particles. In addition, a bound state resonance search can be more sensitive than a missing energy search strategy at high DM masses.

We have studied the general features of models that can be probed by bound state resonance searches at the LHC while remaining consistent with other powerful experimental constraints. These models generally require a sufficiently strong coupling to a light mediator that can support the bound state, and a heavy mediator that couples the dark sector to the SM. We also carefully take into account the mixing between the heavy mediator and the dark matter bound state. Bound state decays into the light mediator should also be suppressed to allow for a significant partial width into SM particles. 

These requirements must be reconciled with constraints from both direct and indirect detection experiments. Spin-independent direct detection cross sections can be suppressed by having only loop-level interactions between the dark sector and nucleons, which can result from an off-diagonal coupling between the SM and two DM states with a mass splitting between them. Constraints from gamma-ray experiments and overclosure must be carefully considered, taking into account Sommerfeld enhancement due to the presence of a light mediator and radiative bound state formation both during freeze-out and in the present day. 

The SU(2)$_L$ minimal DM models possess many of the properties that we have discussed above, but pure wino or higgsino DM chargino bound states have a production cross section that lies well below the sensitivity of dilepton resonance searches, although DM particles in larger representations of SU(2)$_L$ with a large electric charge forming a deeply-bound electromagnetic bound state can potentially be discovered. 

We propose two dark sector models with kinetic mixing into the SM that contain bound states that can be probed effectively through bound state resonance searches at the LHC, while remaining consistent with direct and indirect detection constraints. The pseudo-Dirac model contains two Weyl fermions with a small mass splitting between them, capable of forming bound states through a light Higgs mediator, while the triple Higgs model is an SU(3) gauge theory with a single Dirac fermion in the fundamental representation, with all of the properties required for a viable model being generated by symmetry breaking of the gauge group. We study the LHC phenomenology of these models and find that bound states searches are complementary to both missing energy and vector mediator resonance searches, and are particularly powerful at high DM masses. A simple rescaling of our constraints indicates that future  \SI{27}{\tera\eV} or \SI{100}{\tera\eV} $pp$ colliders could potentially probe DM bound states with masses in the $\mathcal{O}(10)$\SI{}{\tera\eV} range.

We find that these models naturally avoid overclosure of the universe, and broad swaths of parameter space exist where they also evade limits from indirect detection searches under the assumption of a thermal history, despite the presence of the bound state implying model-independent large enhancements to the low-velocity annihilation rate. The indirect limits are most easily satisfied when radiative capture to the bound state in the local DM halo is kinematically forbidden, because the mass of the mediator supporting the bound state exceeds the binding energy. 

If the bound-state-forming species is required to constitute 100\% of the DM, through a non-thermal history, but is symmetric in the present day, then the indirect searches are sufficient to rule out almost all of the parameter space of interest in the LHC bound-state resonance search for the triple Higgs model; the pseudo-Dirac model evades this fate through a late-time velocity suppression of its annihilation rate. Where a DM species has a greater-than-thermal annihilation cross section but still constitutes 100\% of the observed DM density, a viable model that can be first detected by resonance searches at the LHC should possess some suppression to the annihilation cross section at late times, due e.g. to a dominant $p$-wave annihilation channel or a small primordial asymmetry.

To summarize, dark sectors with bound states can be probed at the LHC through resonance decays to SM particles. Models with multiple force carriers and DM-like states, where the DM scatters inelastically off SM quarks at tree-level, can naturally give rise to a sufficiently large production cross section while evading direct detection constraints. The presence of a light mediator, needed to support the bound state, modifies freezeout and leads to stringent indirect detection limits; however, these constraints leave a wide region of parameter space open, while suggesting a preferred mass spectrum where the mediator mass exceeds the binding energy. DM models with bound states possess a rich phenomenology, allowing complementary probes from many different experimental directions.





\def\polarc{horizontal }
\def\polars{vertical }
\def\Polarc{Horizontal }
\def\Polars{Vertical }
\def\polarcly{horizontally }
\def\polarsly{vertically }
\def\Polarcly{Horizontally }
\def\Polarsly{Vertically }
\def\Ec{E_0^\rightarrow}
\def\Ep{E_+^\uparrow}
\def\Em{E_-^\uparrow}
\def\Epm{E_\pm^\uparrow}
\def\Ein{E_{0,\text{in}}^\rightarrow}
\def\rc{r_\rightarrow}
\def\rs{r_\uparrow}
\def\tc{t_\rightarrow}
\def\ts{t_\uparrow}

\def\Ref{Ref.~}



\chapter{
    Searching for Axion Dark Matter with Birefringent Cavities
}
\label{chap:adbc}
\section{Introduction}

Laser interferometry without a strong, static magnetic field has been shown to be an effective way of searching for axions~\cite{Melissinos:2008vn,DeRocco:2018jwe,Obata:2018vvr}. The interaction term in Eq.~(\ref{eqn:axion_EM_interaction}) causes a difference in phase velocity between left- and right-handed circularly polarized light, and an appropriately designed high-finesse Fabry-Perot cavity can be used to accumulate the resulting phase difference. These studies have shown how to exploit the exquisite sensitivity of interferometry to small phase differences to obtain new limits on low mass axions.

Despite their ingenuity, these designs face two key limitations. First, they are limited by the non-ideal behavior of optical elements. The introduction of a beam splitter~\cite{Melissinos:2008vn} or quarter-wave plates~\cite{DeRocco:2018jwe} inside a cavity leads to losses and imperfect phase shifts between polarization modes that accumulate with each pass of laser light in the cavity. \Ref\cite{Obata:2018vvr} attempts to overcome this difficulty by using a bowtie cavity; however, circularly polarized light is not in general a bowtie eigenmode, as reflection off any surface at a nonzero angle of incidence does not preserve circular polarization. These difficulties would have to be addressed for an actual realization of these proposals.

Second, and more importantly, these proposed experiments rely on the coherent build-up of the phase difference over the entire light storage time in the cavity. The sensitivity of these experiments starts to deteriorate once the axion oscillation period becomes comparable to the storage time, i.e.\ when $m_a \ell \sim 1/\mathcal{F}$, where $\ell$ is the length of the cavity, $\mathcal{F}$ is the finesse, and $m_a$ is the mass of the axion. Increasing $\mathcal{F}$ therefore restricts the experimental sensitivity to lower axion masses, even though a large value of $\mathcal{F}$ is desirable to maximize a possible axion signal. 

In this chapter, we propose a new axion interferometry experimental design that simultaneously overcomes both of these limitations. The presence of ALP dark matter results in a rotation of \polarcly polarized laser light propagating with frequency $\omega_0$ in a cavity, causing a small, \polars polarization to develop in the frequency sidebands $\omega_0 \pm m_a$. We exploit the fact that oblique reflection generally results in a phase difference between different linear polarizations to design a cavity that is resonant at $\omega_0$ in the \polarc (carrier) polarization, and $\omega_0 \pm m_a$ in the \polars (signal) polarization. The signal sidebands can then be detected using conventional interferometry techniques. Our design is sensitive to axion masses $m_a \lesssim 1/\ell$ independent of the finesse of the cavity, significantly improving the reach in $m_a$ without compromising on the strength of the axion signal. All of this can be achieved by a simple, practical cavity design requiring only that light reflects off multiple mirrors at oblique angles. 

\section{Axions and Light Polarization}

Consider two orthogonal, circular polarizations of a laser beam (denoted by $\circlearrowright$ and $\circlearrowleft$) propagating with frequency $\omega_0$ and wavenumber $k_0$ in the presence of an axion field $a(t) = a_0 \cos(m_a t - k_a z)$, starting at some time $t_0$. The axion momentum is $k_a = m_a v$, where $v \sim 10^{-3}$ is the typical dark matter velocity at the Earth. We will only be interested in $m_a \ell \lesssim 1$, so that $k_a \ell \ll 1$, allowing us to neglect spatial gradients in the axion field.

The interaction term in Eq.~(\ref{eqn:axion_EM_interaction}) leads to the following dispersion relation for the two polarizations, which we showed in Eq.~(\ref{eqn:axion_dispersion}):
\begin{alignat}{1}
    - \omega_0^2 + k_0^2 = \pm k_0 g_{a\gamma\gamma} \frac{\partial a}{\partial t} \,.
\end{alignat}
After some time $t_{\circlearrowright, \circlearrowleft}$, each polarization travels a distance $\ell$, given by
\begin{alignat}{1}
    \ell = \int_{t_0}^{t_0 + t_{\circlearrowright, \circlearrowleft}} \left[ 1 \mp \frac{G}{\omega_0} \cos(m_a t) \right] \, dt \,,
    \label{eqn:polarization_travel_time}
\end{alignat}
where $G \equiv g_{a\gamma\gamma} \sqrt{2 \rho_\text{DM}}/2$, and $\rho_\text{DM} = m_a^2 a_0^2/2$ is the local density of dark matter. Equating the result from each polarization on the right-hand side of Eq.~(\ref{eqn:polarization_travel_time}), and working out the phase difference between the two polarizations $\Delta \alpha \equiv \omega_0 (t_{\circlearrowright} - t_{\circlearrowleft})$ to first order in $G/m_a$, we obtain  
\begin{alignat}{1}
    \Delta \alpha \simeq \frac{iG}{m_a} \left[ e^{i m_a t_0} \left(e^{i m_a \ell} - 1\right) + e^{-i m_a t_0} \left(1 - e^{-i m_a \ell} \right)\right]\, .
    \label{eqn:phase_difference}
\end{alignat}
Eq.~(\ref{eqn:phase_difference}) makes it clear that the axion field takes a carrier wave with frequency $\omega_0$ and generates signal sidebands with frequencies $\omega_0 \pm m_a$. 

This phase difference between circular polarizations is equivalent to a rotation of linearly polarized light. Writing the complex electric field in each circular polarization as a vector $(E^\circlearrowright,\, E^\circlearrowleft)$ and keeping track of the relative phase difference only, the translation matrix over a distance $\ell$ can be expressed as $\text{diag}(e^{i \Delta \alpha/2},\, e^{-i \Delta \alpha/2})$. The circular polarizations are related to the linear polarizations via $E^{\circlearrowright,\circlearrowleft} = E^\rightarrow \mp i E^\uparrow$, so that in the linear polarization basis $(E_\uparrow, E_\rightarrow)$, the matrix for translation is 
\begin{alignat}{1}
    P = \begin{pmatrix}
        \cos \frac{\Delta \alpha}{2} & - \sin \frac{\Delta \alpha}{2} \\
        \sin \frac{\Delta \alpha}{2} & \cos \frac{\Delta \alpha}{2} 
    \end{pmatrix} \simeq \begin{pmatrix}
        1 & - \frac{\Delta \alpha}{2} \\
        \frac{\Delta \alpha}{2} & 1 
    \end{pmatrix} \,.
    \label{eqn:2x2_translation_matrix}
\end{alignat}

\section{Axion Interferometry} 

The basic principle of axion interferometry is summarized in Fig.~\ref{fig:axion_interferometry_cartoon}. A carrier wave with electric field $\Ec$ in the \polarc polarization is injected into a cavity that is tuned to be resonant in the \polarc polarization at the laser carrier frequency $\omega_0$. As the field propagates in the presence of axions, signal sidebands in the \polars polarization are generated, with frequencies $\omega_0 \pm m_a$. The amplitude of the sidebands can be enhanced using an appropriately tuned high-finesse Fabry-Perot cavity. At each end of the cavity, a reflection occurs at a mirror with some real reflectivity coefficient, and a phase difference $\Delta \varphi_{1,2}$ between horizontally and vertically polarized light.
\begin{figure}
    \centering
    \includegraphics[scale=0.54]{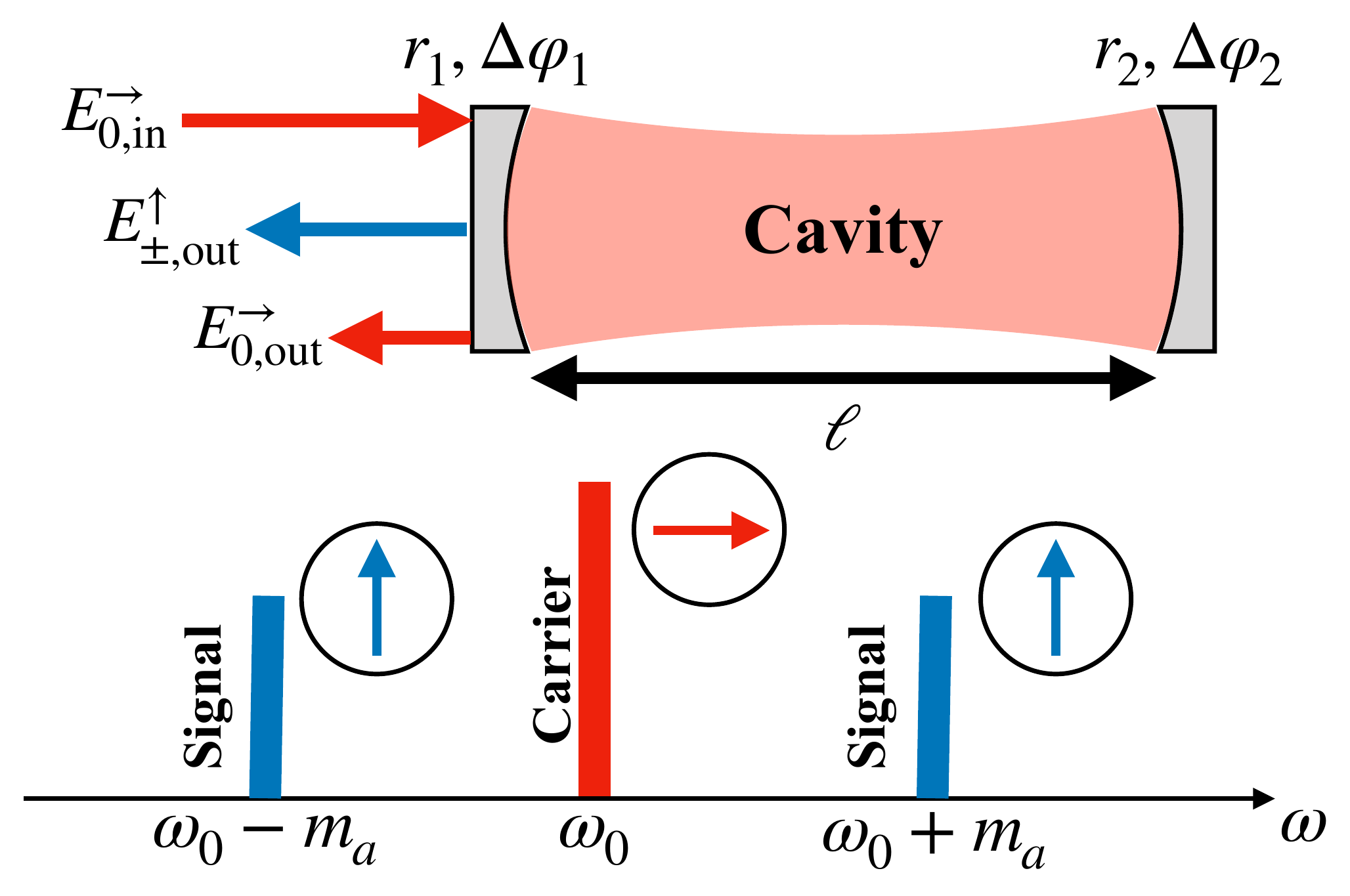}
    \caption{Summary of axion interferometry. A \polarcly polarized laser fed into a cavity with reflectivities $r_{1,2}$ and a relative phase shift between horizontally and vertically polarized light $\Delta \varphi_{1,2}$ at each end leads to the generation of frequency sidebands in the \polars polarization.}
    \label{fig:axion_interferometry_cartoon}
\end{figure}

In order to distinguish between the two sidebands, we split the \polars signal polarization into its two frequency components by writing the electric field in the cavity as the complex column vector $\mathbf{E}_\text{cav} = (\Em,\, \Ec,\, \Ep)$. The subscripts indicate that the components have different frequencies $(\omega_0 - m_a,\, \omega_0,\, \omega_0 + m_a)$, respectively. The transfer matrix for translation in our 3-component notation follows from Eq.~(\ref{eqn:2x2_translation_matrix}):
\begin{alignat}{1}
    T \simeq \begin{pmatrix}
        e^{-i m_a \ell} & \frac{iG}{2m_a}(e^{-i m_a \ell} - 1) & 0 \\
        0 & 1 & 0 \\
        0 & \frac{iG}{2m_a} (1 - e^{i m_a \ell}) & e^{i m_a \ell}
    \end{pmatrix}.
    \label{eqn:3x3_translation_matrix}
\end{alignat}

For reflection at each end, the transfer matrix is given by the expression $R_{1,2} = r_{1,2} \, \text{diag} \left(e^{i \Delta \varphi_{1,2}},\, 1,\, e^{i \Delta \varphi_{1,2}} \right)$. The signal field in the cavity is then given by the solution to the following equation~\cite{Maggiore:1900zz}:
\begin{alignat}{1}
    \mathbf{E}_\text{cav} = t_1\, \mathbf{E}_0 + R_1 T R_2 T \, \mathbf{E}_\text{cav} \,,
    \label{eqn:E_cav}
\end{alignat}
where $\mathbf{E}_0 = (0, \Ein, 0)$ is the electric field of the laser fed into the cavity, and $t_X = \sqrt{1 - r_X^2}$ is the field amplitude transmission coefficient. 

Axion interferometry shares many parallels with conventional microwave cavity experiments like ADMX~\cite{Du:2018uak}. In both, the axion converts a frequency mode pumped to a large energy density (a DC magnetic field in microwave cavities, $\omega_0$ in our set-up) into another mode related to the original by $m_a$ (a standing electromagnetic mode of frequency $m_a$ in microwave cavities, and the signal sidebands $\omega_0 \pm m_a$ in our set-up). This conversion between electromagnetic modes is a generic property of ALPs coupled to photons through Eq.~(\ref{eqn:axion_EM_interaction}), as studied more generally in \Ref\cite{Goryachev:2018vjt}. 

The parallel extends to the power stored in both cavities. In the laser cavity, the power stored in the signal sidebands within the cavity is $P_\pm \propto |\Epm|^2 w^2$, where $w$ is the laser beam width. Solving Eq.~(\ref{eqn:E_cav}) gives $P_\pm \sim g_{a\gamma\gamma}^2 (\rho_\text{DM}/m_a) E_0^{\rightarrow 2} V Q_\pm$, where $V \sim w^2 \ell$ is the volume encompassed by the beam, and $Q_\pm$ is a quantity dependent on the geometry of the cavity, and is analogous to the quality factor for microwave cavities. This reproduces the scaling of the signal power produced in ADMX, with $E_0^\rightarrow = B_0^\rightarrow$ for the laser. 

\section{Birefringent Cavities}

We now turn our attention to the importance of the phase difference between horizontally and vertically polarized light in the cavity, $\Delta \varphi_{1,2}$. Birefringence in a cavity has been used by the PVLAS experiment~\cite{DellaValle:2015xxa} to look for axion-induced changes in the polarization of a propagating laser beam in the presence of a large, static magnetic field due to the Primakoff effect. In contrast, our set-up relies on light transitioning between polarizations due to the absorption or emission of axions. 

In Ref.~\cite{Melissinos:2008vn}, a single beam passes through a polarizing beam splitter so that each beam propagates over a different path length along two different, perpendicular arms, effectively introducing birefringence between the two polarizations. However, a cavity with two perpendicular arms and a beam splitter at its center is unlikely to have a high finesse. More recent work has always ensured that $\Delta \varphi_1 = \Delta \varphi_2 \simeq 0$ either by using quarter-wave plates in front of mirrors with near-zero transmission~\cite{DeRocco:2018jwe}, or by performing two reflections at each end of the cavity, separated by an optical path length that is much shorter than the cavity length~\cite{Obata:2018vvr}. The signal generated by the axion builds constructively as long as the axion field value does not change significantly during the storage time, i.e.\ $m_a \mathcal{F} \ell \ll 1$. Once $m_a \sim 1/(\mathcal{F} \ell)$, the cavity loses sensitivity to the axion signal. 

An equivalent way of understanding this criterion is to observe that setting the phase difference $\Delta \varphi_{1,2} = 0$ means that light in both polarizations are resonant at the laser frequency $\omega_0$. The full-width half-maximum of the cavity transmission band is $\delta \lambda \sim 1/(\mathcal{F} \ell)$, and so we must have $m_a \ll \delta \lambda \sim 1/(\mathcal{F} \ell)$ in order for the signal sidebands (produced by axion-driven polarization modulation) to lie within the transmission band. 

Now consider the case where $\Delta \varphi_{1,2} = \Delta \varphi \neq 0$ (we take $r_2 = 1$ in the following discussion for simplicity). When the resonance condition in the signal polarization $m_a \ell = |\Delta \varphi|$ is met, the signal polarization builds constructively in the cavity. With a phase shift of $\Delta \varphi = \pm \pi/2$, the cavity is resonant at $m_a = \pi/(2\ell)$, the maximum mass reach, for the sidebands $\omega_0 \pm m_a$. Axion masses up to this maximum value can be scanned by increasing $\Delta \varphi$ in steps from 0 to $\pi/2$. Since a larger finesse $\mathcal{F}$ is desirable for producing a large signal field, this represents a significant improvement in axion mass reach without affecting the sensitivity in coupling. Although higher frequency resonances exist for each choice of $\Delta \varphi$, the axion field value at these higher frequencies oscillates more than once over the cavity length $\ell$, suppressing the sensitivity by $\text{sinc}(m_a \ell)$~\cite{Maggiore:1900zz}. 
\begin{figure}
    \centering
    \includegraphics[scale=0.45]{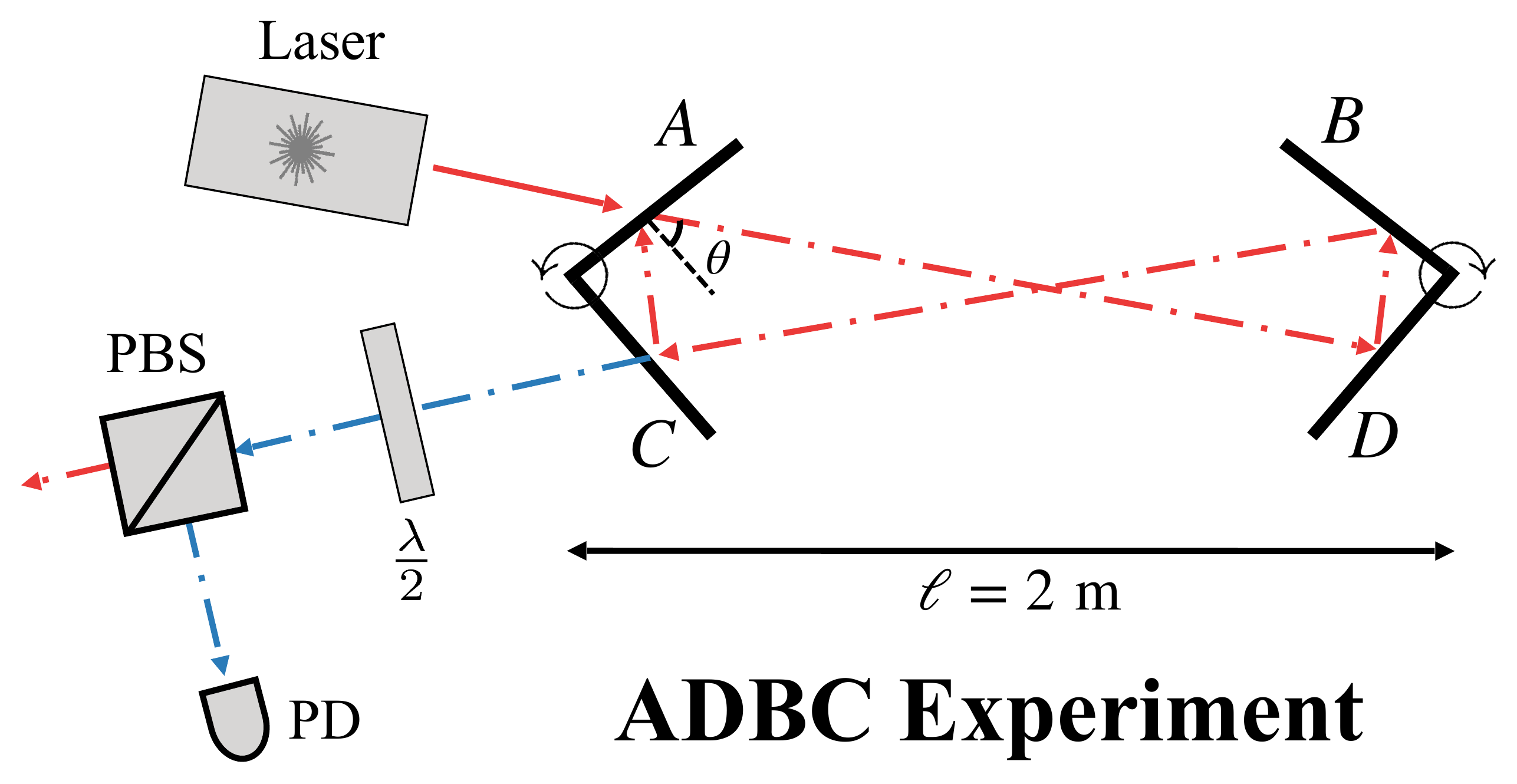}
    \caption{Schematic of the ADBC experiment. The red optical path is that of the input and cavity, while the blue optical path is read-out. The beam enters at $A$ and is read out after $C$. Two sets of mirrors $A$, $C$ and $B$, $D$ can be rotated to change the angle of incidence $\theta$ while roughly maintaining cavity alignment and length. To produce an electrical signal, the leakage fields from mirror $C$ pass through a half-waveplate ($\lambda/2$) before reflecting off a polarizing beam splitter (PBS) and arriving at a photodetector (PD).}
\label{fig:design}
\end{figure}

\section{Experimental Set-Up}
\label{experimentalsetup}

Fig.~\ref{fig:design} gives a schematic of the proposed Axion Detection with Birefringent Cavities (ADBC) experiment, featuring a practical cavity design with the necessary birefringence. The Fresnel equations~\cite{hecht2016optics} show that orthogonal, linear polarizations reflecting off a dielectric surface at an oblique angle of incidence $\theta$ in general develop a relative phase shift $\Delta \varphi$. By rotating the mirrors to adjust the angles of incidence, we can thus tune the cavity birefringence to make the axion-induced, vertically polarized sidebands resonant in our cavity at $m_a \ell = |\Delta \varphi|$.

The proposed design consists of two sets of two mirrors spaced \SI{2}{\meter} apart, with each set acting as a retroreflector that can pivot independently. The angle between the mirrors in a set should be fixed at slightly less than $90^\circ $ so that the angles of incidence are roughly $\theta$ and $90^\circ - \theta$. This allows us to vary the angle of incidence while roughly maintaining optical path-length and cavity alignment. The short dimension of the cavity (e.g.\ $\ell_{DB}$) is of order \SI{10}{\centi\meter}. One set, $A$ and $C$, will be taken as our input and output ports respectively, so that the optical path goes in the order $ADBC$. 

The Fresnel equations show that the reflectivity of the horizontal polarization will be lower than the vertical. Placing the carrier in the horizontal polarization (lower finesse) therefore reduces the accumulation of experimental noise in the cavity, while simultaneously placing the signal in the vertical polarization (higher finesse) leads to a larger signal-to-noise ratio (SNR). 

To prevent appreciable leakage of the carrier from the cavity, the cavity should be optimally coupled, meaning the transmissivity of $A$ in the carrier polarization, $\tc^A$, must be matched to the total losses in the cavity. This would almost entirely eliminate any reflection off $A$.  To allow a significant signal field to be read out, we also need $\ts^C$ to be larger than the total losses from the other mirrors. However, the Fresnel equations force $\tc > \ts$, and as a result, cavity loss for the carrier will be dominated by $\tc^C$, leaving us with $\tc^A \simeq \tc^C$.
To maintain high finesse in the signal and carrier, all other transmissivities should be smaller than the cavity optical loss.

To maximize the axion mass reach, the mirrors should cover as much of the range $0 \leq \Delta \varphi \leq \pi/2$ as possible. $\Delta \varphi$ increases with more oblique angles of incidence, but large optical surfaces are required near grazing incidence.

\section{Experimental Sensitivity}
\label{experimentalsensitivity} 

The sensitivity of ADBC to $g_{a\gamma\gamma}$ is ultimately dependent on the finesse of the cavity $\mathcal{F}_\uparrow$ and $\mathcal{F}_\rightarrow$ in each polarization, and on $t_{\rightarrow,\uparrow}^C$, for which we will use benchmark values of $\mathcal{F}_\uparrow = 2.25 \times 10^5$, $\mathcal{F}_\rightarrow = 2700$, $t_\uparrow^C = 0.0037$, and $t_\rightarrow^C = 0.030$ (recall that $t_X$ is the \textit{amplitude} transmission coefficient). These finesse values are typical of mirrors used in the LIGO cavity and other experimental studies. The reach in axion mass is determined by $\Delta \varphi$, which in turn depends on the mirror properties. We find that a range of  $0 < \Delta \varphi \lesssim \pi/5$ can typically be probed over a $6^\circ$ range in angle of incidence $\theta$, with $\theta \lesssim 65^\circ$. Over this small range of angles, the finesse of the cavity does not vary significantly in either polarization, and we have adopted the smallest values in this range for simplicity. 

The signal field inside the cavity can be found by solving this cavity's equivalent of Eq.~(\ref{eqn:E_cav}) for $\mathbf{E}_\text{cav}$. For simplicity, we neglect the translation matrix for the short legs (i.e.\ $\ell_{DB}$ and $\ell_{CA}$), and take the same matrix $R$ for both sets of mirrors.
The reflection matrix has the form $R = \text{diag} (\rs e^{i \Delta \varphi}, \, \, \rc,  \,\, \rs e^{i \Delta \varphi})$, with $\rc^2$ and $\rs^2$ being the product of the reflectivities of all 4 cavity mirrors. These quantities are related to the finesse by $\mathcal{F}_{\uparrow, \rightarrow} \simeq \pi/(1 - r_{\uparrow,\rightarrow}^2)$.

\begin{figure}[t!]
    \centering
    \includegraphics[scale=0.78]{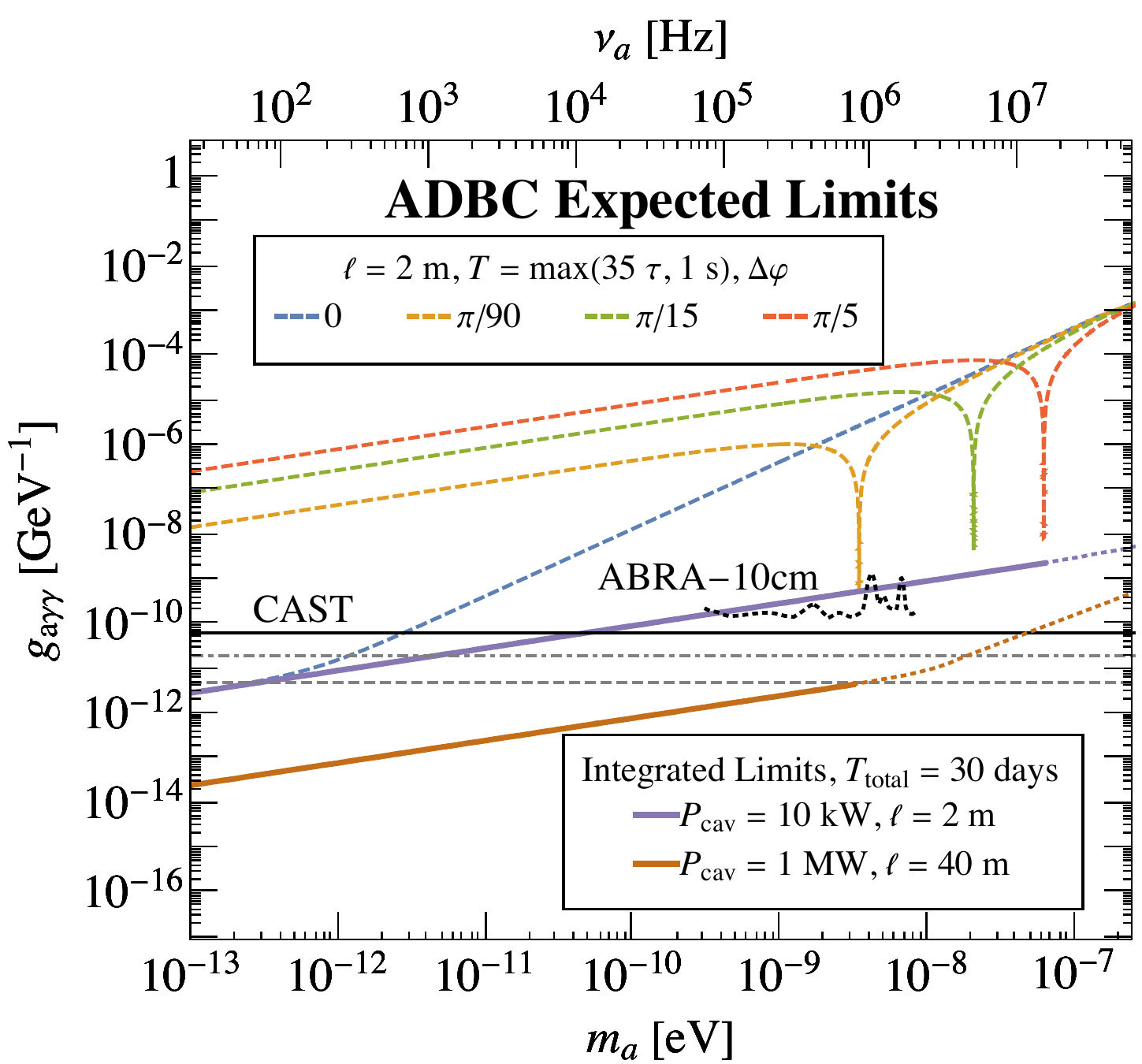}\caption{
    Expected ADBC limits on the axion coupling $g_{a\gamma\gamma}$. The limit for a phase shift of 0 (blue, dashed), $\pi/90$ (orange, dashed), $\pi/15$ (green, dashed) and $\pi/5$ (red, dashed) are shown. The integrated limits obtained by scanning through a phase shift of 0 to $\pi/5$ are shown for a 2 m cavity with 10 kW laser power in the cavity (purple) and for a 40 m cavity with 1 MW laser power in the cavity (brown), with a total integration time of 30 days. The envelope of the reach can be extended in both the 2 m (purple, dotted) and 40 m cavity (brown, dotted) to higher axion masses if the optics were improved to scan up to $\Delta \varphi = \pi/2$. Limits from CAST~\cite{Anastassopoulos:2017ftl} (black) and ABRACADABRA-10cm~\cite{Ouellet:2018beu} (black, dotted), together with projected limits from ALPS-II~\cite{Bahre:2013ywa} (gray, dot-dashed) and IAXO~\cite{Irastorza:2013dav} (gray, dashed) are shown for comparison. }
    \label{fig:ADBC_reach}
\end{figure}
The signal sidebands emerging from the cavity are read out using a heterodyne detection scheme. The carrier and signal are passed through a half-wave plate with its fast axis rotated by a small angle $\varepsilon$ away from vertical, after which a polarizing beam splitter (PBS) is used to isolate the vertical polarization for readout by a photodetector. This mixes a small amount of the DC (carrier) component into the AC (sideband) component modulated at the frequency corresponding to the axion mass. If the phase difference is tuned so that $\Delta \varphi = m_a \ell$, the cavity is resonant in the vertical (signal) polarization at a frequency sideband $\omega_0 - m_a$, giving an output AC power at the heterodyne readout of
\begin{alignat}{1}
	P_\text{AC} &= 4 \sqrt{2} G \varepsilon P_\text{cav} \frac{\sin \left(m_a \ell/2 \right)}{m_a}  \frac{t^C_\rightarrow  t^C_\uparrow}{1 - r_\uparrow^2} \,,
    \label{eqn:ACPower}
\end{alignat}
where we have assumed that all reflectivity coefficients are approximately 1. The sensitivity is estimated by finding the value of $g_{a\gamma\gamma}$ that sets the SNR to 1, with~\cite{Budker:2013hfa}
\begin{alignat}{1}
	\text{SNR} = \frac{P_\text{AC}}{S_\text{shot}^{1/2}} (T \tau)^{1/4} \,, 
    \label{eqn:SNR}
\end{alignat}
 where $S_\text{shot} = 2 P_\text{DC} \omega_0$ is the laser shot noise power spectral density with the DC power given by $P_\text{DC} = (2 \varepsilon t^C_\rightarrow)^2 P_\text{cav}$, $T$ is the integration time for this step in $\Delta \varphi$, $\tau \equiv 2\pi/(m_a v^2)$ is the coherence time of the axion field, and we assume $T \gg \tau$. The expression for the SNR in Eq.~(\ref{eqn:SNR}) is independent of $\varepsilon$; this is consistent with the fact that shot noise power always scales as the square root of the carrier power regardless of the amount of filtering performed, and is contrary to the result obtained in Ref.~\cite{Melissinos:2008vn}. 

Another source of noise in our set-up is laser technical noise, which leads to finite laser frequency width and decreases the sensitivity of ADBC as $m_a \to 0$. In order to probe axion masses down to $m_a \sim 10^{-13}$ eV, technical noise must be subdominant to shot noise down to $\nu_a \sim $ 10--100 Hz. ADBC will adopt many of the same techniques used by the LIGO collaboration to achieve isolation at these frequencies~\cite{Cook:2012tu}. A LIGO-like suspension system mounted on a rotating platform will be used for both pairs of mirrors. Since only a single beam is used in a cavity which is held on resonance via feedback to $\omega_0$, radiation pressure and other displacement noises are less relevant. Thermal noise in the mounted optics, for instance, will dominate over other non-technical noises (e.g.\ quantum radiation pressure noise),
with an estimated magnitude of~\cite{Gonzalez1994}
\begin{alignat}{1}
    S_{\Delta \varphi}^{1/2} \sim \frac{d \Delta \varphi}{d \theta} \frac{S_x^{1/2}}{\ell_{DB}} \sim 10^{-14} \frac{\SI{100}{Hz}}{\nu_a}\,.
    \label{eqn:thermal_noise}
\end{alignat}
This places requirements on the experimental design for small values of $m_a$ where $G/m_a \sim S_{\Delta \varphi}^{1/2}$ ($g_{a\gamma\gamma} \sim 10^{-14}\,\si{GeV^{-1}}$ at $m_a \sim 10^{-13}\,\si{eV}$).

Several steps can be taken to ascertain that a signal is indeed due to the presence of axion dark matter. First, the axion always produces a signal for both $\Delta \varphi = \pm m_a \ell$, even though the configuration of the mirrors may be very different in each case. Second, we can tune the cavity so that $\Delta \varphi_1 + \Delta \varphi_2 = 2 m_a \ell$, but with $\Delta \varphi_1 \neq \Delta \varphi_2$, where $\Delta \varphi_{1,2}$ are the phase differences generated at mirrors $A,C$ and $B,D$ respectively. With such a tuning, $P_\text{AC} \propto 1 + \cos(m_a \ell - \Delta \varphi_2)$. This dependence arises only due to the off-diagonal entries in the translation matrix $T$ shown in Eq.~(\ref{eqn:3x3_translation_matrix}), and serves as an explicit check that the signal is produced from the conversion of laser power from the carrier to the signal polarization.

The expected sensitivity for a \SI{2}{\meter}, $P_\text{cav} = \SI{10}{\kilo\watt}$ and a \SI{40}{m}, $P_\text{cav} = \SI{1}{\mega\watt}$ version of ADBC is given in Fig.~\ref{fig:ADBC_reach}. The \SI{2}{\meter} benchmark can currently be achieved in the laboratory, while the \SI{40}{\meter} version is similar in size to the \SI{40}{\meter} LIGO prototype at Caltech~\cite{Abramovici:1996dz} or the Fermilab Holometer~\cite{Chou:2015sle}, using an optical configuration similar to advanced LIGO~\cite{TheLIGOScientific:2014jea}.  In order to cover the range $0 < m_a \lesssim \pi/(5\ell)$, the experiment must be run a number of times given by $\mathcal{F}_\uparrow/5 \sim 5 \times 10^{4}$, each with a different value of $\theta$.
$\mathcal{F}_\uparrow/5$ is chosen so that the peak of each resonance in $m_a$ falls on the half-maximum for the previous resonance, starting from $m_a = 10^{-13}$~eV. Given a total integration time of $T_\text{tot} = $ 30 days, we integrate each step for $T = \max(N_\ell \, \tau, \text{ 1 sec})$, where $N_\text{2\,m} = 35$ and $N_\text{40\,m} = 4$. This choice is equivalent to allocating the integration time logarithmically among bins of $m_a$, as recommended by Ref.~\cite{Chaudhuri:2018rqn}, and in agreement with Refs.~\cite{Chaudhuri:2014dla,Kahn:2016aff}. The envelope of the sensitivity to $g_{a\gamma\gamma}$ can be obtained analytically from Eq.~(\ref{eqn:SNR}), giving
\begin{alignat}{2}
	g_{a\gamma\gamma} &> && \,\, \SI{6.13e-11}{\per\giga\eV} \frac{N_\ell^{-1/4}}{\text{sinc}(m_a \ell/2)} \nonumber \\
    & && \times \left(\frac{\SI{0.3}{\giga\eV \per \centi\meter\cubed}}{\rho_\text{DM}} \frac{\SI{10}{\kilo\watt}}{P_\text{cav}} \frac{\SI{1.064}{\micro\meter}}{\lambda_0} \right)^{1/2} \left( \frac{\SI{2}{\meter}}{\ell} \frac{10^{-3}}{t_\uparrow^C} \frac{10^5}{\mathcal{F}_\uparrow} \right) \sqrt{\frac{m_a}{\SI{e-13}{\eV}}} \,\,,
\end{alignat}
with $\lambda_0$ being the laser wavelength. For a given $m_a$, adding up the SNR in quadrature from every step may improve the reach by up to a factor of 2. A  \SI{40}{m} cavity with a circulating laser power of \SI{1}{MW} in the cavity improves upon CAST limits~\cite{Anastassopoulos:2017ftl} by almost four orders of magnitude for  $m_a \sim 10^{-13}$ eV. Ultimately, full-sized versions of ABRACADABRA~\cite{Kahn:2016aff,Ouellet:2018beu,Ouellet:2019tlz} and DM-Radio~\cite{Chaudhuri:2014dla,Battesti:2018bgc} may eventually cover much of the parameter space shown in Fig.~\ref{fig:ADBC_reach}. ADBC can, however, serve as a powerful complementary search to these experiments, relying on a completely different strategy in looking for axions. In particular, ADBC's ability to obtain two separate resonances at $\Delta \varphi = \pm m_a \ell$ is a striking experimental signature that would bolster any potential evidence for axions in other experiments.

\section{Conclusion}

We proposed a new axion interferometry experimental design that exploits the birefringence of a bowtie cavity in order to generate axion-modulated, vertically polarized sidebands from a horizontally polarized laser beam carrier. This design is practical to implement and can improve on the reach of previous interferometry designs from $m_a \sim 1/(\mathcal{F} \ell)$ up to $m_a \sim 1/\ell$, with the sensitivity improving with finesse. The sensitivity and mass range of our experiment can both be improved by a careful design of the mirrors used in the cavity, so that the cavity is optimally coupled with minimal loss, and the phase shift at each end extends to $\Delta \varphi = \pi/2$. We look forward to implementing this design and beginning the search for axions with the ADBC experiment.

\chapter{Contributions to Cosmic Reionization from Dark Matter Annihilation and Decay}
\label{chap:DM_reionization}

\section{Introduction}
\label{sec:Introduction}

The epoch of reionization and the emergence of the universe from the cosmic dark ages is a subject of intense study in modern cosmology. As baryonic matter began to collapse around initial fluctuations in the dark matter (DM) density seeded by inflation, the earliest galaxies in our universe began to form. These structures, perhaps accompanied by other sources, eventually began to emit ionizing radiation, creating local patches of fully ionized hydrogen gas around them. These patches ultimately grew to encompass the entire universe, leading to the fully ionized intergalactic medium (IGM) that we observe today. 

While the process of reionization is broadly understood, the exact details of how and when reionization occurred are still somewhat unclear. Quasars and the earliest stars certainly played a part in reionization, but their relative energy contributions to the process are still a matter of ongoing research. Some studies have found~\cite{Fan:2001ff} that a significant population of dim and unobserved quasars must be present in order for them to completely reionize the universe. Similar conclusions have  been drawn for star-forming galaxies~\cite{Robertson2013}. This uncertainty has resulted in some interest in other sources of energy that might contribute to reionization. 

DM provides a particularly compelling candidate, and has been considered several times in the literature. Many models allow DM to annihilate or decay into Standard Model particles, which in turn can deposit energy into the IGM through ionization, heating or other processes. The annihilation rate, which scales as the square of the density, rises substantially with the onset of structure formation and the collapse of DM into dense haloes, potentially yielding a large energy injection in the reionization epoch.

Our current knowledge of reionization can already place interesting constraints on DM properties. Constraints from optical depth and the temperature of the IGM placed strong constraints on DM models~\cite{Cirelli:2009bb} that could generate the cosmic ray excesses observed by PAMELA~\cite{Adriani:2008zr} and Fermi+HESS~\cite{Abdo:2009zk,Collaboration:2008aaa,Aharonian:2009ah}. IGM temperature data as well as CMB power spectrum measurements can also be used to constrain the properties of $p$-wave annihilating and decaying DM~\cite{Diamanti2014}. More recently, it has been shown that with improved measurements of the optical depth to the surface of last scattering and near-future probes of the cosmic ionization history, it should be possible to set new and significant constraints on the properties of annihilating or decaying DM~\cite{Kaurov2015}. 

Turning the question around, the potential role that DM may have played in reionization has also been broadly explored. Earlier papers in the literature were able to find possible scenarios in which annihilating DM could contribute significantly to reionization, once structure formation was taken into account~\cite{Chuzhoy2008,Natarajan:2008pk}. Subsequently,~\cite{Belikov:2009qx} included the important effect of inverse Compton scattering off the cosmic microwave background (CMB) photons, and showed that weakly interacting massive particle (WIMP) DM candidates could play a dominant role in reionization. More recently, studies of $s$-wave annihilation of dark matter using an analytic description for the boost to the DM density during structure formation found that an unrealistic structure formation boost to the annihilation rates or an overly large cross section was required for a DM-dominated reionization scenario consistent with existing experimental results from the CMB~\cite{Poulin2015,Lopez-Honorez:2013lcm}. Multiple authors~\cite{Mapelli:2006ej,Hansen:2003yj,Kasuya2004} have also shown that a significant contribution from decaying DM to reionization in a manner consistent with WMAP results is possible using specific DM decay rates and products.

In this chapter, we examine the potential contribution of dark matter toward reionizing the universe, but improve on previous results in four crucial ways:

\begin{enumerate}
\item We consider an extremely wide range of DM masses, from 10 keV to TeV scales, and rather than selecting specific annihilation/decay channels, we consider the impact of electrons, positrons and photons injected at arbitrary energies. This allows us to place general, model-independent constraints on DM annihilation or decay, beyond the WIMP paradigm;

\item In addition to $s$-wave annihilation, we consider energy injection into the IGM through $p$-wave annihilation and decay. Energy injections in these scenarios have a different dependence on redshift and on the details of structure formation compared to the case of $s$-wave annihilation: consequently, different constraints dominate. We improve on these earlier results by performing a more accurate calculation of the energy injection/deposition rates and by taking into account the relevant constraints in each energy injection channel; 

\item The details of structure formation and its uncertainties are critical in determining the $s$-wave and $p$-wave annihilation rates~\cite{Mack2014}. We use a detailed and up-to-date prescription of structure formation for our calculations, including the contribution of substructure in haloes (previous studies on substructure include~\cite{Bartels:2015uba,Moline:2016pbm}). By calculating the boost factor to DM annihilation assuming two different halo profiles (consistently applied to both haloes and subhaloes) as well as the difference to the boost factor that results from including substructure effects, these results also allow us to estimate the uncertainties associated with structure formation, including uncertainties related to the subhalo boost factor;

\item We use the latest results presented in~\cite{Slatyer2015} to determine how energy injection from annihilations or decays is eventually deposited into the IGM via ionization and heating. We have extended the code to be applicable even when the universe is completely ionized, allowing us to determine how energy is deposited into the IGM at redshifts below $1+z=10$ (the previous lower limit for the code) assuming different reionization scenarios. This improvement allows us to use astrophysical constraints from $z \lesssim 6$ with confidence, and to estimate the sensitivity of our constraints to the details of the (re)ionization history.
\end{enumerate} 

This chapter is structured as follows: in Section \ref{sec:ExptConstraints}, we will review the main existing results that will be used to set constraints on the DM contribution to reionization. Section \ref{sec:EnergyInjection} gives a brief overview of energy injection from $s$-wave annihilation, $p$-wave annihilation and decays, for an unclustered/homogeneous distribution of DM. Our structure formation prescription is detailed in Section \ref{sec:StructureFormation}, while Section \ref{sec:fz} explains how we determine the heating and ionization deposited to the IGM, given an energy injection history and a structure formation model. Section \ref{sec:FreeEleFrac} outlines the three-level atom model for hydrogen used to determine the ionization and IGM temperature history from the energy deposition history. Finally, Section \ref{sec:Constraints} shows our derived constraints for each of the DM processes considered here, with our conclusions following in Section \ref{sec:Conclusion}. 

Throughout this chapter, we make use of the central values for the cosmological parameters derived from the TT,TE,EE+lowP likelihood of the Planck 2015 results~\cite{Ade:2015xua}. This is obtained from a combination of the measured TT, TE and EE CMB spectra for $l \geq 30$ and a temperature and polarization pixel-based likelihood for $l<30$. Specifically, our choice of parameters are $H_0 = \SI{67.27}{\kilo\meter\per\second\per\mega\parsec}$, $\Omega_m = 0.3156$, $\Omega_b h^2 = 0.02225$ and $\Omega_c h^2 = 0.1198$. These values give a present day atomic number density of $n_A = 0.82 \rho_c \Omega_b/m_p = \SI{2.05E-7}{\per\centi\meter\cubed}$.

\section{Constraints from Experimental Results}
\label{sec:ExptConstraints}

To understand how significant a role DM can play in the process of reionization, we must first examine the current experimental constraints on both reionization and DM. 

Extensive astrophysical observations of early quasars and the IGM around them have enhanced our understanding of the process of reionization. By studying quasars at redshift $z \sim 6$ and hydrogen Ly$\alpha$ absorption in their spectra due to the Gunn-Peterson effect, multiple groups have shown that reionization of hydrogen was mostly complete by $z \sim$ 6~\cite{,Becker:2001ee,Fan2006,Ota2008}. Observations from even larger redshifts $z\sim 7-8$ indicate that hydrogen reionization occurred relatively quickly, with the neutral hydrogen fraction rising to 0.34 at $z\sim 7$ and exceeding $0.65$ at $z \sim 8$~\cite{Schenker2014}. Neutral helium became reionized at a similar time compared to hydrogen due to their relatively similar ionization energies, but a harder spectrum of ionizing radiation is required to doubly-ionize neutral helium atoms~\cite{Loeb2013,Choudhury2006}. Work done on the helium Ly$\alpha$ spectra for quasars at lower redshifts has shown that helium was completely reionized by $z\sim 3$~\cite{Zheng2004}, when quasars could produce the required ultraviolet spectrum. 

Another quantity important to understanding reionization is the IGM temperature, $T_m$. Energy deposited into the IGM can both ionize and heat the gas, and the rate of ionization and heating are both highly dependent on $T_m$. Measurements of $T_m$ place interesting constraints on processes that inject energy into the IGM at redshifts $z \lesssim 6$, since a large injection of energy at these redshifts would result in excessive heating of the IGM. For example, in the case of potential DM contributions,~\cite{Diamanti2014} made use of $T_m$ measurements to constrain the velocity-averaged cross section of MeV-TeV DM undergoing $p$-wave annihilation into lepton pairs, as well as the decay lifetimes for MeV-TeV DM decaying into lepton pairs. They found that bounds from $T_m$ considerably improved the constraints set by measurements from the CMB and from baryon acoustic oscillations, strengthening the constraints set for the $p$-wave annihilation cross section by more than an order of magnitude over the full range of DM masses considered. 

Several measurements of $T_m$ as a function of redshift have been performed in the last two decades. Earlier studies~\cite{Schaye2000} measured the distribution of widths in Ly$\alpha$ absorption spectra from quasars in the redshift range $z = 2.0 - 4.5$ to determine the history of $T_m$ in this range, and determined that $\SI{5100}{\kelvin} \leq T_m(z=4.3) \leq \SI{20000}{\kelvin}$. More recent studies~\cite{Becker2011,Bolton2011} of the IGM temperature from the Lyman-$\alpha$ forest~\cite{Becker2011} and from quasars~\cite{Bolton2010,Bolton2011} have pushed these measurements back to $z \sim 6$, with the two measurements of $T_m$ at the largest redshifts given by (errors reflect 95\% confidence):

\begin{alignat}{1}
	\log_{10} \left( \frac{T_m(z=6.08)}{\text{K}} \right) &= 4.21^{+0.06}_{-0.07}, \nonumber \\
	\log_{10} \left( \frac{T_m(z=4.8)}{\text{K}} \right) &= 3.9 \pm 0.1.
	\label{eqn:TIGMConstraints}
\end{alignat}

The first measurement, discussed in~\cite{Bolton2011}, is almost certainly an overestimate of the true IGM temperature at that redshift: this result does not account for photo-heating of HeII around the quasar being measured, which would result in the measured temperature being significantly higher than the temperature of the IGM away from these quasars. Nonetheless, it serves as a conservative upper bound on $T_m$. 

Aside from direct astrophysical measurements, the CMB can also reveal much about reionization. One important aspect of this epoch that can be measured from the CMB is the total optical depth $\tau$ since recombination, given by
\begin{align}
	\tau = -\int_0^{z_\text{CMB}} dz \, n_e(z) \sigma_T \frac{dt}{dz},
\label{eqn:OpticalDepth}
\end{align}
where $n_e$ is the number density of free electrons, $\sigma_T$ is the Thomson scattering cross section and $z_{\text{CMB}}$ is the redshift of recombination. Scattering of CMB photons off free electrons present after reionization suppresses the small-scale acoustic peaks in the power spectrum by a factor of $e^{-2\tau}$. The Planck collaboration reports the measured optical depth to be~\cite{Adam:2016hgk}
\begin{align}
	\tau = 0.058 \pm 0.012.
\label{eqn:measuredOpticalDepth}
\end{align}
Planck has also been able to determine a reionization redshift $z_{\text{reion}}$, assuming a step-like reionization transition modeled by a $\tanh$ function and characterized by some width parameter $\delta z = 0.5$ (referred to as the ``redshift-symmetric'' parameterization in~\cite{Adam:2016hgk}). $z_{\text{reion}}$ is the redshift at which the free electron fraction $x_e \equiv n_e/n_{\text{H}} = 0.54$. Here $n_{\text{H}}$ is the number density of hydrogen (both neutral and ionized) and $n_e$ is the number density of free electrons. $x_e=1.08$ upon complete reionization after taking into account the complete (single) ionization of helium as well. Based on the measured optical depth, the derived $z_{\text{reion}}$ assuming a redshift-symmetric parameterization of the reionization is
\begin{alignat}{1}
  z_{\text{reion}} = 8.8 \pm 0.9.
\end{alignat}
We can factor out the uncertainty associated with reionization after $z = 6$ and its contribution to the optical depth by writing:
\begin{alignat}{2}
	\tau &=&& -\int_0^3 dz \left[n_{\text{H}}(z) + 2n_{\text{He}}(z) \right] \sigma_T \frac{dt}{dz} - \int_3^6 dz\, [n_{\text{H}} (z) + n_{\text{He}}(z)]  \sigma_T \frac{dt}{dz} \nonumber \\
	& &&- \int_6^{z_{\text{CMB}}} dz\, n_e(z) \sigma_T \frac{dt}{dz},
\end{alignat}
where $n_{\text{He}}$ is the redshift-dependent number density of helium (both neutral and ionized). The first two terms are the contribution to the optical depth from reionized hydrogen and helium, while the last term is the contribution from the unknown ionization history of the universe above $z = 6$. The first two terms can be directly evaluated given the baryon number density today, and give a total contribution of $\delta \tau_0 = 0.038$. The remaining measured optical depth must therefore have come from contributions prior to $z=6$, i.e. 
\begin{align}
	\delta \tau = -\int_6^{z_{\text{CMB}}} dz\, n_e(z) \sigma_T \frac{dt}{dz} \leq 0.044,
\label{eqn:ExcessOpticalDepth}
\end{align}
in order for $\tau$ to be within the experimental uncertainty of Eq.~(\ref{eqn:measuredOpticalDepth}) at the 95\% confidence level.

For the case of $s$-wave annihilation, the CMB power spectrum also provides a robust constraint on the velocity-averaged annihilation cross section $\langle \sigma v \rangle$, since additional ionization of the IGM at high redshifts induces a multipole-dependent modification to the temperature and polarization anisotropies~\cite{Padmanabhan:2005es}. The Planck collaboration~\cite{Ade:2015xua} has placed an upper bound on $p_{\text{ann}}$, defined as
\begin{alignat}{1}
	p_{\text{ann}} (z) = f_{\text{eff}} \frac{\langle \sigma v \rangle}{m_\chi},
\end{alignat}
where $f_{\text{eff}}$ is a constant proxy for $f(z)$, the efficiency parameter that describes the ratio of total energy deposited to  total energy injected at a particular redshift $z$, and $m_\chi$ is the mass of the DM particle. The CMB power spectra are most sensitive to redshifts $z \sim 600$ (for $s$-wave annihilation), and so the constraint on $\langle \sigma v \rangle$ can be estimated from that redshift~\cite{Finkbeiner2012}. Using the TT,TE,EE+lowP Planck likelihood, the 95\% upper limit on this parameter at $z=600$ was found to be:
\begin{alignat}{1}
	p_{\text{ann}}(z = 600) < \SI{4.1E-28}{\centi\meter\cubed\per\second\per\giga\eV}.
	\label{eqn:pann}
\end{alignat}

Given $f_{\text{eff}}$ for $s$-wave annihilation, which in turn is obtained from $f(z)$, this leads immediately to a constraint on $\langle \sigma v \rangle$ as a function of $m_\chi$. $f(z)$ has been calculated for arbitrary injections of electrons, positrons  and photons in the 10 keV-TeV range; in this chapter we will thus refer to injections of electron/positron pairs ($e^+e^-$) and photon pairs ($\gamma \gamma$), while keeping in mind that more general DM annihilation/decay channels can be represented as linear combinations of photons/electrons/positrons at different energies.\footnote{See~\cite{Slatyer2012,Slatyer2015} and the publicly available results and examples found at \texttt{http://nebel.rc.fas.harvard.edu/epsilon} for further information on how this is done.} This approach neglects the contribution of protons and antiprotons, which is generally quite small~\cite{Weniger2013}.

In Section \ref{sec:fz}, we will give a brief summary of our calculation of $f(z)$, which is based on the work detailed in~\cite{Slatyer2012,Slatyer2015}. The full details of obtaining an actual value for $f_{\text{eff}}$ from our calculation of $f(z)$ across a large range of DM masses can be found in~\cite{Slatyer2015a}. Figure~\ref{fig:excludedXSec} shows the constraints on $s$-wave annihilation into $e^+e^-$ (upper panel) and $\gamma \gamma$ (lower panel), based on the CMB power spectrum data from Planck. 

\begin{figure*}
    \centering
	\subfigure{
		\includegraphics[scale=0.6]{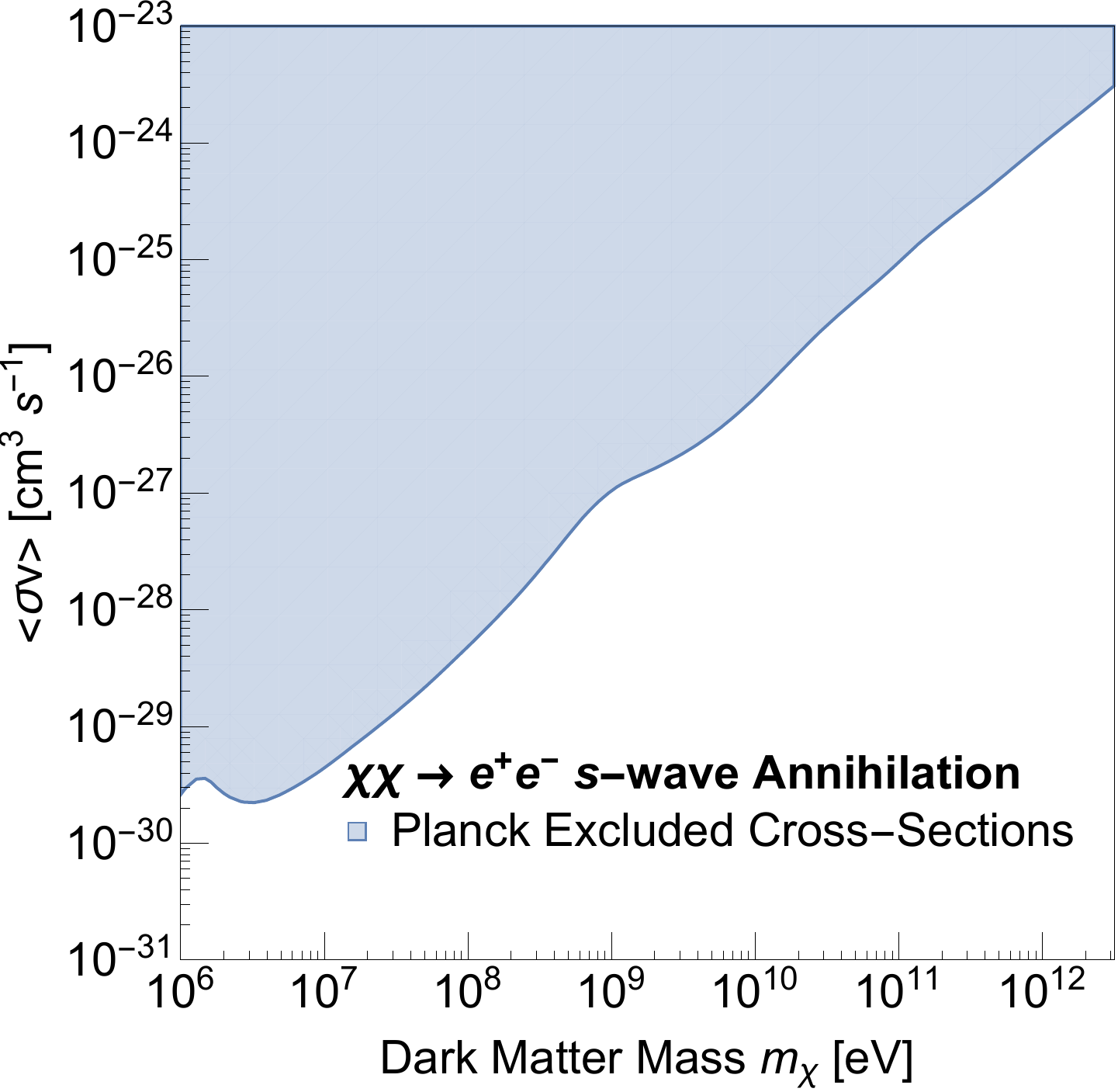}
	} \\
	\subfigure{
		\includegraphics[scale=0.6]{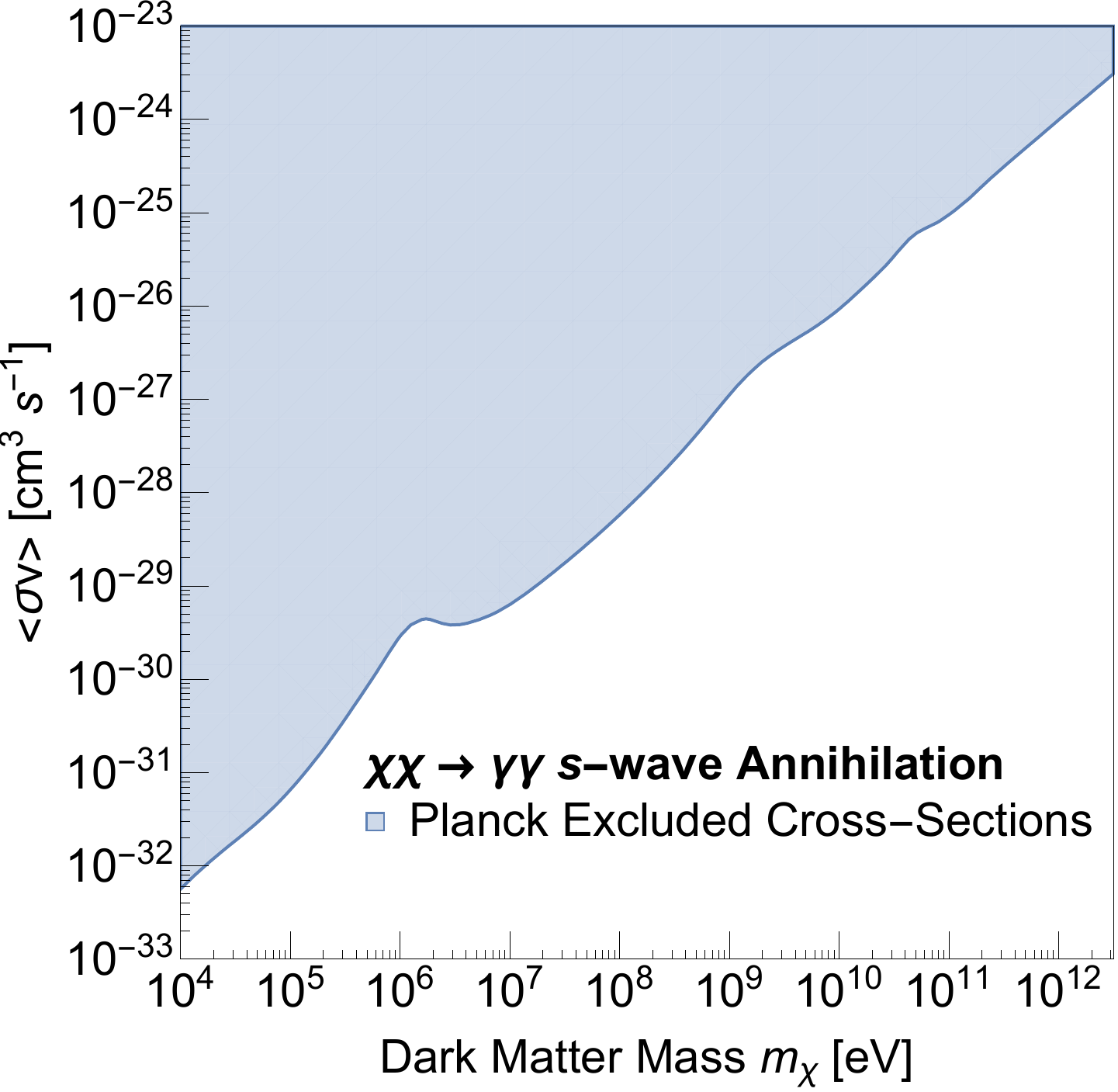}
	}
	\caption{The $95$\% excluded cross section based on Planck's upper limit given by Eq.~(\ref{eqn:pann}) for (left) $ \chi \chi \to e^+e^-$ and (right) $\chi \chi \to \gamma \gamma$ $s$-wave annihilation.}
	\label{fig:excludedXSec}
\end{figure*}

\section{Unclustered Dark Matter Energy Injection}
\label{sec:EnergyInjection}

In this chapter, three scenarios by which DM can inject energy into the IGM are considered: $s$-wave annihilation, $p$-wave annihilation and decay. The total energy injected by both $s$- and $p$-wave annihilation of uniformly distributed DM is given by
\begin{alignat}{1}
	\left( \frac{dE}{dV dt} \right)_{\text{ inj}} = \rho^2_{\chi,0} (1+z)^6 \frac{\langle \sigma v \rangle}{m_{\chi}},
	\label{eqn:injRateSmooth}
\end{alignat}
 where $m_{\chi}$ is the DM particle mass and $\rho_{\chi,0} = \rho_{c} \Omega_c$ is the overall smooth density of DM today, with $\rho_c$ being the critical density of the universe today. In $s$-wave annihilation, $\langle \sigma v \rangle$ is constant, while in $p$-wave annihilation, $\sigma v \propto v^2$. This velocity dependence can be factored out by assuming a Maxwellian velocity distribution, which
simplifies the calculation since we can take the 1D velocity dispersion ($\sigma_{1\text{D}}$) as a proxy for the velocity enhancement/suppression in the thermal average:
\begin{alignat}{1}\label{proxy_p}
	\langle \sigma v \rangle_p \propto\int_0^1v^2f_{\rm MB}(v)dv=\sigma_{1D}^2.
\end{alignat}
We can then write, by picking a reference dispersion velocity $\sigma_{1\text{D,ref}}$:
\begin{alignat}{1}
	\langle \sigma v \rangle_{p,B} = \left(\frac{\sigma_{1\text{D,B}}}{\sigma_{1\text{D,ref}}}\right)^2  (\sigma v)_{\text{ref}},
\end{alignat}
where $\sigma_{1\text{D,B}}$ is the one-dimensional characteristic dispersion velocity of unclustered DM. This quantity is redshift dependent, but assuming thermal equilibrium of the DM distribution, $ \sigma_{1\text{D,B}}^2 \propto T$, which for non-relativistic DM scales as $T \propto (1+z)^2$. Thus the energy injection rate for $p$-wave annihilation for uniformly distributed DM can be written as
\begin{alignat}{1}
	\left(\frac{dE}{dV dt} \right)_{p\text{ inj}} = \rho^2_{\chi,0} (1+z)^8 \frac{(\sigma v)_{\text{ref}}}{m_\chi} \left(\frac{\sigma_{1\text{D,B}} (z=0)}{\sigma_{1\text{D,ref}}}\right)^2,
	\label{eqn:smoothpwave}
\end{alignat}
where $\sigma_{1\text{D,B}}(z=0)$ is the present-day value of $\sigma_{1\text{D,B}}$. Throughout this chapter, we choose $\sigma_{1\text{D,ref}} = 100 \mathrm{km/s}$ (a value consistent with~\cite{Diamanti2014}), which is roughly the present-day DM dispersion velocity in haloes with a mass comparable to the Milky Way ($\lesssim10^{12}$M$_\odot$) today.

Finally, the energy injected from the decay of DM is given by
\begin{alignat}{1}
	\left(\frac{dE}{dV dt} \right)_{d \text{ inj}} = \rho_{\chi,0}(1+z)^3 \frac{1}{\tau_{\chi}},
\end{alignat}
where $\tau_\chi$ is the decay lifetime, which is taken to be much longer than the age of the universe so that the change in DM density due to decay is negligible. This assumption is valid given known limits on the decay lifetime deduced from Planck and WMAP~\cite{Diamanti2014} as well as gamma-ray experiments~\cite{Dugger:2010ys,Essig2013} for a large range of decay channels.

We have thus far only considered unclustered DM distributions, where the comoving DM density is constant, but structure formation causes the local density and velocity dispersion of DM to deviate strongly from the expected value for a homogeneous distribution. The onset of structure formation thus significantly changes the energy injection history due to $s$- and $p$-wave annihilations. However, the previous notation is still useful: once we have obtained a structure formation history, we can characterize the energy injection from a realistic DM distribution by replacing Eqs.~(\ref{eqn:injRateSmooth}) and~(\ref{eqn:smoothpwave}) with effective multipliers to the unclustered DM density. A realistic structure formation history is thus crucial in calculating the energy injection rate from DM.

\section{Structure Formation}
\label{sec:StructureFormation}

In the Cold Dark Matter (CDM) scenario, DM clusters into gravitationally self-bound haloes across a very large range of scales, from the (model-dependent) minimum limit set by DM kinetic decoupling ($10^{-11}-10^{-3}$M$_\odot$ for WIMPs \citep[e.g.][]{Bringmann2009}) to $10^{15}$M$_\odot$ cluster-size haloes. $N$-body simulations can accurately follow DM structure formation but only in a limited mass range: it is not yet possible to cover the full dynamical range corresponding to CDM particles. In order to explore the unresolved regime, hybrid approaches which have a core analytical model calibrated against numerical simulations must be used, e.g., the well-known halo model \citep[e.g.][]{Seljak_00}, or the recently introduced $P^2SAD$ (clustering in phase space) \citep{Zavala2015}. We will follow these two
approaches in this chapter, describing their most relevant elements.
 
We assume that after recombination, structure formation is described by linear perturbation theory followed by the immediate formation (collapse) of haloes. In this scenario, haloes collapse (form) at a redshift $z_{\rm col}$ with an average overdensity $\bar{\rho}_h=\Delta\rho_{c}(z_{\rm col})$, where $\rho_{c}$ is the critical density of the universe. The choice of the overdensity $\Delta$ varies in the literature, but for simplicity we will use the redshift independent, widely used value of $\Delta=200$. 
The formation redshift is given by the spherical collapse model, which connects the linear power spectrum with the epoch of collapse, resulting in a hierarchical picture of structure formation. In particular, the halo collapses when the 
rms linear overdensity $\sigma(M,z)$ (mass variance) crosses the linear overdensity threshold $\delta_c\sim1.686$:
\begin{equation}\label{sigma_rms}
	\sigma^2(M,z)=\int d^3{\bf k}\,P(k,z)W^2(k,M),
\end{equation}
where $W(k,M)$ is a filter function in Fourier space, and $P(k,z)$ is the linear CDM power spectrum. For the spherical collapse model, the window function is a top-hat filter in real space. 
We compute the primordial matter power spectrum with the code CAMB \citep{2000ApJ...538..473L} with a cosmology consistent with Planck data.

\subsection{Halo Model}

{\bf (i) Flux multiplier.} For the purposes of this work, we are interested in computing the excess DM annihilation over the contribution from the smooth background due to the collapse of DM into haloes. Following the notation of~\cite{Taylor2003},\footnote{To avoid conflicting with notation used in later sections, we use the letter $\mathcal{B}$ to refer to the flux multiplier instead of the letter $f$ as in~\cite{Taylor2003}.} we write this excess (flux multiplier) for a particular redshift as:
\begin{eqnarray}\label{flux_cosmic}
	\mathcal{B}(z)&=&\frac{1}{\rho_B^2V_B}\int_{m_{\rm min}}^{\infty}\left(V_B\frac{dn}{dM}dM\right)\bar{\rho}^2_hV_h(M)B_{h}(M)\nonumber\\
	&=&\frac{\Delta}{\Omega_m^2\rho_{\rm crit}}\int_{m_{\rm min}}^{\infty}MB_h(M)\frac{dn}{dM}dM,
\end{eqnarray}
where $\left(V_B\frac{dn}{dM}dM\right)$ is the number of haloes in the cosmic volume $V_B$, with a background matter density $\rho_B=\Omega_m\rho_c$. Each halo is assumed to be spherical with a radial density profile $\rho(r)$ truncated at a virial radius $r_{200}$. The annihilation rate in the halo is enhanced over the rate based on the average DM density by an amount
\begin{equation}\label{flux_halo}
	B_h(M)=\frac{4\pi}{\bar{\rho}^2_hV_h(M)}\int_0^{r_{200}}\rho^2(r)r^2\,dr.
\end{equation}

{\bf (ii) Density profile.} In most of the resolved mass regime of current simulations, haloes are well-fitted by a {\it universal} two-parameter NFW density profile \citep{Navarro:1996gj}. An even better fit is that of a three-parameter Einasto profile \citep{Einasto}. The simplicity of the NFW profile and, more importantly, its reduction to an almost one-parameter profile makes it an appealing choice in analytic studies. We will consider these two profiles for this study except at very low halo masses near the filtering mass scale, where recent simulations of the formation of the first haloes (microhaloes) indicate that their inner density profiles might be cuspier than the NFW profile \citep[e.g.][]{Anderhalden2013,2014ApJ...788...27I}. Although these simulations can follow the evolution
of microhaloes only until $z\sim30$ (due to limited resolution, since long wavelength perturbations comparable to the box size cannot be neglected at lower redshifts), we assume that the density profile of these microhaloes can be described by these results all the way down to $z=0$. 

{\it NFW profile and microhaloes.} We use the density profile given by
\begin{equation}\label{rho_smooth}
\rho(x)=\frac{\rho_s}{x^\alpha(1+x)^{3-\alpha}},
\end{equation}
 where $x\equiv r/r_s$, and $r_s$ and $\rho_s$ are the scale radius and density, respectively. Setting $\alpha=1$ gives the NFW profile, which adopt for haloes and subhaloes. For haloes near the filtering mass scale, we follow~\cite{2014ApJ...788...27I}, which states that $\alpha$ scales as a power law of the halo mass:
 \begin{equation}\label{alpha_micro}
 	\alpha=-0.123~{\rm log}\left(\frac{M}{10^{-6}M_\odot}\right)+1.461
 \end{equation}
for $M<10^{-3}$$M_\odot$. Above this scale, we set $\alpha=1$. Substituting Eq.~(\ref{rho_smooth}) into Eq.~(\ref{flux_halo}), we have:
\begin{equation}\label{flux_halo_power}
B_h(M)=\frac{c^3}{3m^2(c)}\int_0^c\frac{x^2 dx}{x^{2\alpha}(1+x)^{6-2\alpha}},
\end{equation}
where $c\equiv r_{200}/r_s$ is the concentration parameter, which is a function of halo mass (see below), and:
\begin{equation}\label{flux_halo_power_2}
	m(c)=\int_0^c\frac{x^2 dx}{x^\alpha(1+x)^{3-\alpha}}.
\end{equation}
Equations (\ref{flux_halo_power}) and (\ref{flux_halo_power_2}) both have analytic solutions.

{\it Einasto profile.} The density profile is given by:
\begin{equation}\label{einasto_eq}
\rho(r)=\rho_{-2}\,{\rm exp}\left(\frac{-2}{\alpha_e}\left[\left(\frac{r}{r_{-2}}\right)^{\alpha_e}-1\right]\right),
\end{equation}
where $\rho_{-2}$ and $r_{-2}$ are the density and radius at the point where the logarithmic density slope is -2, and $\alpha_e$ is the Einasto
shape parameter. This three-parameter profile is reduced to only two parameters once the total mass $M\equiv M_{200}$ of a halo is fixed. In particular
we can write:
\begin{alignat}{1}
	M_{200} = \frac{4\pi r_{-2}^3\rho_{-2}}{\alpha_e}{\rm exp}\left(\frac{3{\rm ln\alpha_e}+2-{\rm ln} 8}{\alpha_e}\right) \gamma\left[\frac{3}{\alpha_e},\frac{2}{\alpha_e}\left(\frac{r_{200}}{r_{-2}}\right)^{\alpha_e}\right].
\end{alignat} 
The parameter $\alpha_e$ and the ``concentration'' $c_e=r_{200}/r_{-2}$ are connected to $M_{200}$ through $\sigma(M,z)$ as we describe below. 
Once these parameters are known, we can compute the boost to the annihilation rate over the average in a halo by solving Eq.~(\ref{flux_halo}) numerically.

The cosmic annihilation flux multiplier given by Eq.~(\ref{flux_cosmic}) due to the population of haloes above a minimum mass $M_{\rm min}$ is fully determined once we specify the halo mass function $dn/dM$ and the properties of the density profiles. In the Extended Press-Schechter (EPS) formalism, both of these are fully determined for a given halo mass. More specifically, they can be written as formulae that depend on $\sigma(M,z)$.

{\bf (iii) Mass function.} The mass function in the case of ellipsoidal collapse is given by \citep{ST1999}:
\begin{align}\label{eq_mf}
	\frac{dn}{d{\rm ln}M}&=\frac{1}{2}f(\nu)\frac{\rho_B}{M}\frac{d{\rm ln} (\nu)}{d{\rm ln}M},\\
	f(\nu)&=A\sqrt{\frac{2q\nu}{\pi}}\left[1+\left(q\nu\right)^{-p}\right]{\rm exp}^{-q\nu^2},
\end{align}
 with $A=0.3222$, $p=0.3$, and $q=1$, and:
 \begin{equation}
 	\nu\equiv\frac{\delta_c(z)^2}{\sigma(M,z)^2},
 \end{equation}
 where $\delta_c(z)=1.686/D(z)$ is the linearly extrapolated threshold for spherical collapse, with $D(z)$ being the growth factor normalized to unity at $z=0$.
 
Free-streaming of DM particles prevents the formation of haloes below a (filtering) scale, which depends on the mass of the DM particle. This results in a cutoff to the primordial power spectrum at the filtering scale. The difference between a CDM power spectrum with a filtering scale and without (i.e.\ setting the mass of the DM particles effectively to zero) is typically given in terms of the transfer 
function $T^2_\chi=P_{\rm m_\chi}/P_{\rm m_\chi\rightarrow0}$, which for neutralino DM has the form \citep{Green2005}:
\begin{equation}\label{transfer_func}
	T_\chi(k)=\left[1-\frac{2}{3}\left(\frac{k}{k_A}\right)^2\right]{\rm exp}\left[-\left(\frac{k}{k_A}\right)^2-\left(\frac{k}{k_B}\right)^2\right],
\end{equation}
where
\begin{alignat}{1}\label{transfer_func_2}
	k_A= 2.4\times10^6\left(\frac{m_\chi}{100~{\rm GeV}}\right)^{1/2} \frac{(T_{\rm kd}/30~{\rm MeV})^{1/2}}{1+{\rm ln}(T_{\rm kd}/30~{\rm MeV})/19.2}~{\rm Mpc}/h,
\end{alignat}
\vspace{-0.6cm}
\begin{alignat}{1}
	k_B&=5.4\times10^7\left(\frac{m_\chi}{100~{\rm GeV}}\right)^{1/2}\left(\frac{T_{\rm kd}}{30~{\rm MeV}}\right)^{1/2}~{\rm Mpc}/h, 
\end{alignat}
and $T_{\rm kd}$ is the (model-dependent) kinetic decoupling temperature.

To include the effect of free-streaming into the mass function, we use the code provided by \citep{2013MNRAS.433.1573S}, which computes the mass function following Eq.~(\ref{eq_mf}) using a {\it sharp-k} window function for the mass variance calibrated to match the results of simulations that include a cutoff in the power spectrum as given by the transfer function in Eq.~(\ref{transfer_func}). We note that $T_{\rm kd}$ and $m_\chi$ together determine the minimum self-bound halo mass $M_{\rm min}$. Choosing a different $M_{\min}$ changes the global contribution of (sub)haloes by some overall factor in a redshift-independent manner.
We take $m_\chi=100$~GeV and $T_{\rm kd}=28$~MeV to compute the cutoff to the primordial power spectrum given by Eqs.~(\ref{transfer_func}-\ref{transfer_func_2}).\footnote{ For neutralino dark matter, the kinetic decoupling temperature generally increases with particle mass, although a broad range of values for a fixed mass is allowed. Based on Fig. 2 of~\cite{Bringmann2009} we have chosen a typical value within that range for $m_\chi=100$~GeV.} This results in a damping scale due to free streaming with a characteristic mass of $M_{\rm min}=10^{-6}$M$_\odot$ \citep[see Eq.~(13) and Fig. 3 in Ref.][]{Bringmann2009}, which is the canonical value for WIMPs. The impact of choosing different values of $M_{\rm min}$ will be studied later in this section.

{\bf (iv) Parameters of the density profiles.} The median density profile of haloes with a given mass is fully specified by one parameter, typically the halo mass. Since CDM haloes form hierarchically, low mass haloes are more concentrated than more massive ones. This specifies the second parameter (concentration) of the profile. Ultimately, this parameter is connected to the density of the Universe at the (mass-dependent) time of collapse for a given halo. 

{\it NFW profile and microhaloes}. The concentration of an NFW halo is a  strong function of halo mass that has been explored in great detail in the literature using analytical and numerical methods. We use the model by~\cite{2012MNRAS.423.3018P} to compute the concentration-mass 
relation. The model is calibrated to recent simulations down to their resolution limit ($M\sim10^{10}$~M$_\odot$), but more importantly, it is physically motivated since it uses $\sigma(M,z)$ as the main quantity connected to the concentration. In this way, it takes into account the flattening of the linear power spectrum towards smaller halo masses. We refer the reader to Section 5 of 
\cite{2012MNRAS.423.3018P} for the formulae that lead
to the computation of $c(M,z)$. We only consider haloes with a ``peak-height'' $\nu\equiv\delta_c/\sigma$ up to $3\sigma$. The larger $\nu$ is, the rarer and the more massive the halo is relative to the characteristic clustering mass defined by $\nu=1$. 
 
 For microhaloes, we make a correction to the NFW concentrations given by the Ref.~\cite{2012MNRAS.423.3018P} model to take into account the steeper profiles of microhaloes. To do so, we follow
 the results from~\cite{2014ApJ...788...27I} (see their Figure 9). In particular, for $\alpha=1.5,1.4,1.3,1.0$ in Eq.~(\ref{alpha_micro}), they find $c_{\rm NFW}=2.0c_{\rm micro},1.67c_{\rm micro}, 1.43c_{\rm micro}, 1.0c_{\rm micro}$; we use these values to interpolate for a given microhalo mass.
 
 {\it Einasto profile.} In this case we follow the work by~\cite{Klypin2014} to connect the parameters $\alpha_e$ and $c_e$ (concentration) with $\sigma(M)$. These authors use a similar
 analysis as that of~\cite{2012MNRAS.423.3018P}, and find the following empirical relations:
 \begin{eqnarray}
 	\alpha_e&=&0.015+0.0165\nu^2,\nonumber \\
	r_{200}/r_{-2}&=&6.5\nu^{-1.6}(1+0.21\nu^2).
 \end{eqnarray}
 Note that $\alpha_e$ approaches a constant value asymptotically for low $\nu$ (i.e. low halo masses), which implies that low mass haloes of a given mass only differ in one parameter, their concentration (as in the NFW case).
 
 {\bf (v) Substructure.} Each DM halo is composed of a smooth DM distribution and a hierarchy of subclumps that merged into the main halo at some point in the past and have been subjected to 
 tidal disruption. The modeling of the abundance of main haloes and their inner smooth structure have been described previously, and we now consider the impact of substructure on the annihilation rate.
 
 To account for the self-annihilation of DM in substructures, we define a {\it subhalo boost} over the flux multiplier of a main halo (i.e. over $B_h(M)$ in Eq.~(\ref{flux_halo})):
 \begin{alignat}{1}\label{sub_boost}
 \mathcal{B}(m_{\rm sub})=\frac{1}{B_h(M)} \int_{m_{\rm min}}^{m_{\rm max}}\frac{\bar{\rho}_{\rm sub}(m_{\rm sub})}{\bar{\rho}_h}  B_{\rm sub}(m_{\rm sub})m_{\rm sub} \frac{dN}{dm_{\rm sub}}dm_{\rm sub},
 \end{alignat}
 where $dN/dm_{\rm sub}$ is the subhalo mass function and $\bar{\rho}_{\rm sub}$ and $B_{\rm sub}$ are the average density within a subhalo and its flux multiplier of mass $m_{\rm sub}$, respectively. Because of tidal disruption, these quantities depend in principle on the distance 
 of the subhalo relative to the halo center, but since we are interested in the total subhalo boost to the annihilation rate, we can assume that most of the boost comes from subhaloes near the virial radius of the host. This is a good approximation since tidal disruption considerably reduces the abundance of subhaloes near the halo center. For instance, looking at Figure 3 of 
Ref.~\cite{Springel:2008cc}, we
 see that only $\sim30\%$ of the annihilation rate in subhaloes comes from within 100 kpc ($\sim0.4r_{200}$) of a Milky Way-sized halo. On the other hand, near the virial radius of a host with an assumed NFW profile, the tidal radius for a
 subhalo of mass $m_{\rm sub}$ is approximately given by \citep[e.g. Eq.~(12) of][]{Springel:2008cc}
 \begin{alignat}{2}
 	r_t &=&& \left(\frac{m_{\rm sub}}{\left[2-\frac{d{\rm ln}M}{d{\rm ln} r}\right]M(<r)}\right)^{1/3}r \nonumber \\
 	&\sim&& \left(\frac{m_{\rm sub}}{M}\right)^{1/3} r_{200} \left(2-\frac{c^2}{(1+c)^2}\frac{1}{{\rm ln}(1+c)-c/(1+c)}\right)^{-1/3},
 \end{alignat}
where $c\equiv c(M,z)$ is the concentration of the host. We can then substitute $\frac{\bar{\rho}_{\rm sub}}{\bar{\rho}_h}$ for the following in Eq.~(\ref{sub_boost}):
\begin{eqnarray}
	\left.\frac{\bar{\rho}_{\rm sub}(<r_t)}{\bar{\rho}_h}\right\vert_{r_{200}}=
	2-\frac{c^2}{(1+c)^2}\frac{1}{{\rm ln}(1+c)-c/(1+c)}.\nonumber\\
\end{eqnarray}
This density ratio has only small variations around 2 with low mass haloes being more overdense on average than more massive subhaloes. 

The {\it subhalo mass function} is in principle also a function of halocentric distance, but it becomes the global subhalo mass function under the approximation that subhaloes near the virial radius dominate the annihilation rate. The subhalo mass function
has a similar functional form as the halo mass function. In particular, it is approximately a power law (except at very large masses) with a similar slope to the halo mass function, 
$dN/dm_{\rm sub}\propto m_{\rm sub}^{-1.9}$ \citep{Springel:2008cc}; the normalization however is different. This functional form is nearly universal if $m_{\rm sub}$ is scaled to the host mass.\footnote{This universality is even clearer if the ratio of maximum circular velocities is used instead of the masses to define the subhalo mass function \citep[e.g.][]{Cautun2014}.} We use the fitting formulae for the subhalo mass function given by~\cite{Gao2011}, which is based on a suite of high resolution simulations covering a large dynamical range of masses and is valid for $z\leq2$; for higher redshift we assume that the formulae at $z=2$ holds (our results are actually not very sensitive to this assumption). We assume also that these formulae are preserved in the unresolved regime, down to the filtering mass scale, and apply the same cutoff at low masses due to free streaming (or kinetic decoupling) as that for the halo mass function.
 
 To calculate the subhalo flux multiplier $B_{\rm sub}$, we assume the same density profiles as in the case of main haloes, i.e. we use Eqs.~(\ref{flux_halo_power}) and (\ref{flux_halo_power_2}) in the case of the NFW profile and the microhaloes, and find the result numerically in the case of the Einasto profile. 
 This is a good approximation since, as we mentioned before, the subhaloes that contribute most to the signal are those near the virial radius of the host. Thus, tidal disruption would not have transformed their
 inner structure significantly, particularly their inner regions, which strongly dominate the annihilation rate. However, in the case of the NFW profile, we do account for a slight modification to the concentration-mass relation in the form of an upscaling of a factor of 2.6 to the characteristic density $\rho_s$ (which is roughly a $30\%$ increase in concentration, see Figure 28 of Ref.~\cite{Springel:2008cc}). This modification is because for a given mass, subhaloes (even near the virial radius) are slightly more concentrated than isolated haloes. For the case of the Einasto profile, we do not
 make this correction since there is no systematic study about this. We note however that this correction to the overall flux multiplier $\mathcal{B}(z)$ is relatively small. 
 
 \subsection{The Particle Average Phase Space Density Approach}

Instead of modeling the clustering of DM indirectly as a collection of haloes (and subhaloes) with a certain internal DM distribution, one can model it directly by looking at the DM two point correlation function $\xi(\Delta x)$ (or its Fourier transform, the power spectrum). It has been shown that the flux multiplier, defined in Eq.~(\ref{flux_cosmic}), is equal to the limit of $\xi$ when the separation between particles $\Delta x$ goes to zero~\cite{Serpico2012}:
\begin{equation} \label{eq_p2sad}
	\mathcal{B}={\rm lim}_{\Delta x\rightarrow 0} \xi(\Delta x).
\end{equation}
Thus, if one can directly obtain a prediction of the DM power spectrum in the deeply non-linear regime, then it is possible to directly compute the flux multiplier without the many steps and approximations involved in the halo model.

This approach has been developed recently by analyzing the coarse-grained phase space distribution directly from DM simulations. In particular, by measuring the two dimensional particle phase space average density ($P^2SAD\equiv\Xi(\Delta x, \Delta v)$, where $\Delta x$ and $\Delta v$ are the distance and relative speed between particles) in high resolution simulations, it has been possible to physically model this new statistic of DM clustering and predict the right hand side of Eq.~(\ref{eq_p2sad})~\cite{Zavala2014a,Zavala2014b,Zavala2015}. In particular one can write:
\begin{equation}\label{real_2pcf_std}
	\xi(\Delta x)_{{\cal V}_6} = \frac{\langle\rho\rangle_{{\cal V}_6}}{\rho_B^2}\int d^3{\bf \Delta v}~\Xi(\Delta x, \Delta v)_{{\cal V}_6} - 1,
\end{equation}
where $\langle\rho\rangle_{{\cal V}_6}$ is the average DM density within the phase space volume (${\cal V}_6$) over which $P^2SAD$ is averaged. In a cosmic volume $V_B$ we can write:
\begin{equation}\label{normalization}
	\frac{\langle\rho\rangle_{{\cal V}_6}}{\rho_B^2}=\frac{1}{\rho_B}\frac{M_{V_B}}{\rho_B V_B}=\frac{\mathcal{F}_{\rm subs}(V_B)}{\rho_B},
\end{equation}
where $\mathcal{F}_{\rm subs}(V_B)$ is the mass fraction contained in substructures within the cosmic volume $V_B$ that is calculated using the subhalo and halo mass functions, described above
in the halo model section:
\begin{equation}\label{norm_p2sad}
	\mathcal{F}_{\rm subs}(V_B)=\frac{1}{\rho_B}\int_{M_{\min}}^{\infty}M\frac{dn}{dM}\mathcal{F}_{\rm s,h}(M)dM,
\end{equation} 
where $\mathcal{F}_{\rm s,h}(M)$ is the mass fraction within subhaloes in a halo of mass $M$ (computed from the subhalo mass function). 

$P^2SAD$ can be described with a physically motivated model that combines the stable clustering hypothesis in phase space, the spherical collapse model and tidal 
disruption of subhaloes~\cite{Zavala2014b,Zavala2015}. This model has 7 free parameters, which have been calibrated in~\cite{Zavala2015}  for DM particles inside
subhaloes exclusively. Since the clustering of DM at very small scales is dominated precisely by these particles, we can use this model to predict the global flux
multiplier in a cosmic volume. We note that although $P^2SAD$ has remarkably universal structural properties (this is the reason why it is a powerful 
statistic to predict the nonlinear power spectrum at unresolved scales), the parameters of its modeling have only been calibrated at relatively low redshifts. We therefore
warn that above $z=1$, its predictions remain uncertain at this point. Since we are particularly interested in DM annihilation at higher redshift in this chapter, we assume
that the parameters of the physical model of $P^2SAD$ calibrated at $z=0$ remain unchanged. 

Overall, because of its direct connection with the annihilation signal, there is significantly less uncertainty associated with $P^2SAD$ compared to the more traditional halo models used to calculate the boost factor described earlier. With proper calibration at higher redshifts, $P^2SAD$ could have been used as the main method in this chapter, but owing to the current limitations, we use it only as a sanity check on the results obtained from the halo model approach, and as a brief introduction to a powerful new method of obtaining boost factors that may become useful in future work. 

\subsection{The Effective Density for Dark Matter Annihilation due to Structure Formation}

\begin{figure}[t!]
\center{
\includegraphics[height=10.0cm,width=10.0cm]{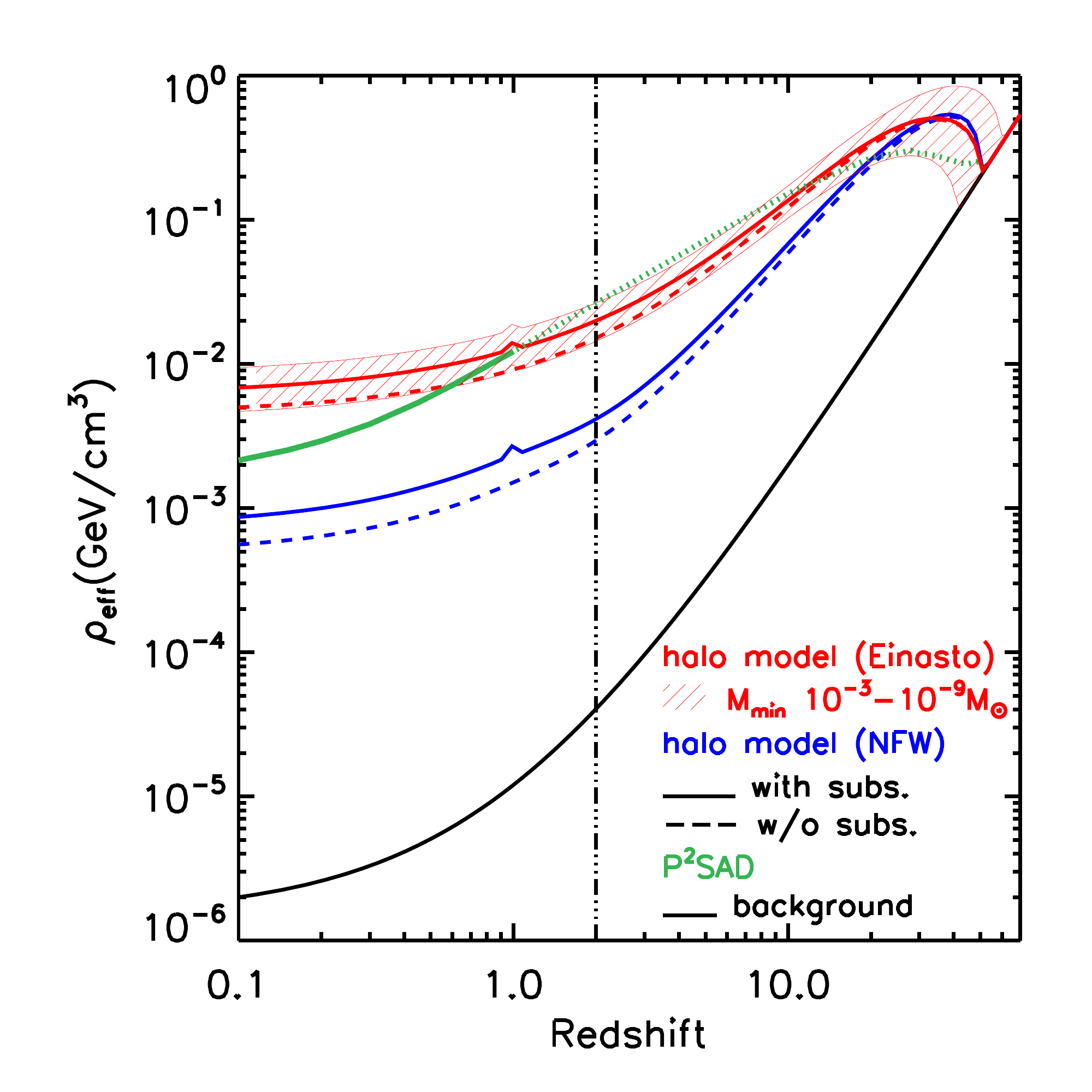} 
}
\caption{The effective DM density as a function of redshift (relevant for $s$-wave annihilation). The blue and red lines show the predictions from the {\it halo model} of structure formation with (solid) and
without (dashed) substructures. The blue (red) line uses an NFW (Einasto) profile for the haloes with parameters given by Ref.~\cite{2012MNRAS.423.3018P} (\cite{Klypin2014}). The green line shows the prediction by a new approach based on the clustering of phase space ($P^2SAD$, Ref.~\cite{Zavala2015}). This approach has only been calibrated at
low redshifts, and thus is uncertain for $z>1$ (green dotted line). The vertical dot-dashed line marks the maximum redshift where the subhalo mass function we have used has been calibrated.  In the case of the halo model with the Einasto profile, we also show with a hatched area the impact of varying $M_{\rm min}$ by 6 orders of magnitude, from $10^{-3}$M$_\odot$ (lower contour) to $10^{-9}$M$_\odot$ (upper contour). For all the other cases, we have used $M_{\rm min}=10^{-6}$M$_\odot$. The solid black line shows the average smooth background density.}
\label{fig_rho_eff} 
\end{figure}

Having described our modeling of the flux multiplier, we can finally write the effective DM density $\rho_{\text{eff}}$ as a boost over the background due to structure formation, which we will then use to compute
the DM annihilation rate as a function of redshift:
\begin{equation}
	\rho_{\rm eff}(z)=\rho_B(z)\left(1+\mathcal{B}_s(z)\right)^{1/2},
	\label{eqn:rhoeff}
\end{equation}
where $\rho_B(z)= \rho_{\chi,0}(1+z)^3$ and $\mathcal{B}_s=\mathcal{B}$ (defined in Eq.~(\ref{flux_cosmic})).

The predictions for $\rho_{\text{eff}}$ for the two structure formation models are shown in Figure~\ref{fig_rho_eff}. The predictions of the  {\it halo model} are in blue (``conservative'', or low-boost) and red (``stringent'', or high-boost), corresponding to the cases where (sub)haloes are modeled with an NFW profile with a concentration mass relation as given by the model in~\cite{2012MNRAS.423.3018P}  and with an Einasto profile with parameters given in~\cite{Klypin2014} respectively. In the plot we show these cases with (solid) and without (dashed) substructure. Beyond $z=2$ (vertical dot-dashed line), the 
parameters of the fitting formulae for the subhalo mass function have not been calibrated and the predictions are thus more uncertain, but at higher redshifts the impact of substructure on the global annihilation rate is minimal. The large difference between the red and blue curves is actually not caused directly by the use of different density profiles (Einasto vs NFW), but by the relatively different concentrations of low mass haloes predicted by the formulae in Refs.~\cite{2012MNRAS.423.3018P} and~\cite{Klypin2014}. We have also explored variations over the minimum self-bound halo mass, varying $M_{\rm min}$ by 6 orders of magnitude. The impact of this on $\rho_{\text{eff}}$ is shown by the hatched area for the Einasto halo model with substructures (the other cases show a similar variation). Although $M_{\rm min}$ plays a role in setting the value of $\rho_{\rm eff}$, varying $M_{\rm min}$ between $10^{-9}$ to $10^{-3} M_\odot$ changed $\rho_{\rm eff}$ by only a factor of approximately 2.15, with the effect being larger at larger redshifts, since a larger value of $M_{\text{min}}$ leads to a delay in the onset of structure formation. This effect is relatively minor compared to the uncertainties in the halo model, at least at $z<10$. We have also found that for both $s$-wave and $p$-wave annihilation, the level of variation in $M_{\rm min}$ explored here produced only percent-level variations in the ionization and thermal histories, and consequently none of our subsequent results are sensitive to our choice of $M_{\min}$. We therefore adopt the canonical value of $M_{\rm min}=10^{-6}$M$_\odot$ for the rest of this chapter.

The approach based on the DM clustering in phase space, $P^2SAD$, is shown with a solid green line, and with a dotted green line beyond the reach where it has been calibrated. It predicts a behavior for $\rho_{\rm eff}$
that lies in between the {\it halo model} predictions. It does seem to favor a larger annihilation rate (i.e. ultimately larger halo concentrations) than the model with the smallest structure formation boost (blue), given that it lies closer to the model with the largest structure formation boost (red). This approach is however only certain close to $z=0$, where the green line is lower than the red one by a significant amount. We will take the difference between the red and the blue line as our degree of uncertainty in the predictions of the structure formation prescriptions.

Equation~(\ref{eqn:rhoeff}) is the quantity of relevance for the case of $s$-wave annihilation, where the astrophysical part of the signal scales as $\rho_{\rm eff}^2$. In the case of $p$-wave annihilation, given the velocity dependence of the
astrophysical signal, we can write instead
\begin{equation}
	(\rho v/c)_{\rm eff}(z)=\rho_B(z)(\sigma_{\rm 1D, B}(z)/c)\left(1+\mathcal{B}_p(z)\right)^{1/2},
	\label{eqn:rhoeff_p}
\end{equation}
where we assume that the velocity distribution of the DM particles is Maxwellian, as in Eq.~(\ref{proxy_p}). In particular, $\sigma_{\rm 1D, B}(z)=\sigma_{\rm 1D, B}(z=0)(1+z)=10^{-11}c({\rm GeV}/m_\chi)^{1/2}(1+z)$ is the velocity dispersion of unclustered DM, and $\mathcal{B}_p$ is given by multiplying the halo and subhalo flux multipliers by $(\sigma_{1D, h}/c)^2$. We have approximated the average 1D velocity dispersion of the (sub)halo by $\sigma_{1D, h}\sim V_{\rm max,h}/\sqrt{3}$, with $V_{\rm max, h}$ being the maximum circular velocity of the (sub)halo computed from its density profile.

Notice that while we have characterized the structure formation contribution as a boost factor multiplying the smooth background contribution, in reality this is an additive contribution: $(\rho v/c)_{\text{eff}}$ within the haloes does not depend on $\sigma_{\text{1D,B}}(z)$, since once structure formation sets in, the characteristic velocity of dark matter particles is set by gravity and not by the primordial thermal motion of unclustered dark matter. Thus the exact value of $\sigma_{\text{1D,B}}(z)$ is important only before the onset of structure formation at $z \gtrsim 50$. Throughout this chapter, we have used the value of $\sigma_{\text{1D,B}}(z = 0)$ computed with $m_\chi = \SI{100}{GeV}$ and $T_\mathrm{kd} = 28$ MeV. This choice results in a highly suppressed annihilation rate prior to structure formation, and results in ionization histories that are indistinguishable from an ionization history with no dark matter at redshifts $z \gtrsim 50$. We have also investigated the effects of adopting larger values of $\sigma_{\text{1D,B}}(z=0)$ corresponding to smaller $m_\chi$ or $T_\mathrm{kd}$, but have found that our present choice is optimistic for producing significant ionization just prior to reionization in a manner that is consistent with the optical depth constraints. Further discussion of this matter can be found in Section \ref{sec:Constraints}.

We show the effective DM density $\times$ velocity in Figure~\ref{fig_rho_eff_pwave}, defined in Eq.~(\ref{eqn:rhoeff_p}). The uncertainties in the structure formation scenario in this case are minimal since annihilation in massive, resolved haloes dominates the overall flux. The uncertain contribution for haloes below the resolution limit of current simulations is minimal. This is why the predictions from the halo model for the two cases we have considered nearly overlap each other, and is the reason why there is a negligible impact of substructures (the lines showing the effect overlap completely with those without substructures in Figure~\ref{fig_rho_eff_pwave}). A different value of $M_{\rm min}$ is only important at the redshifts closest to the onset of structure formation. Still, within the 6 orders of magnitude of variation of $M_{\rm min}$, we have found no important changes in our main results.

\begin{figure}[t!]
\center{
	\includegraphics[height=10.0cm,width=10.0cm]{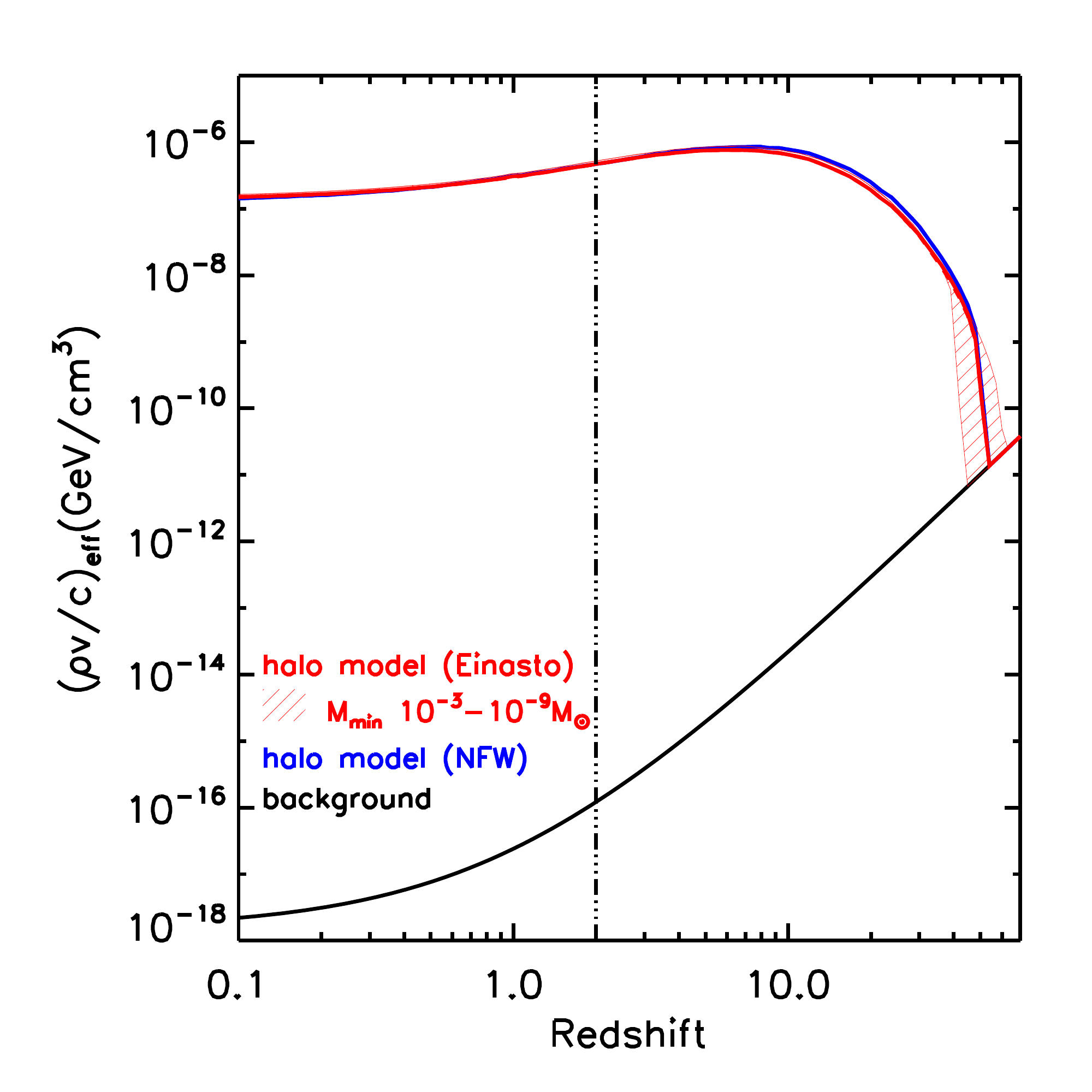} 
}
\caption{The effective DM density $\times$ velocity as a function of redshift (equivalent to Fig.~\ref{fig_rho_eff} but for the case of $p$-wave annihilation given by Eq.~(\ref{eqn:rhoeff_p})). All the line styles and colors are as in Fig.~\ref{fig_rho_eff}.  In the case of the halo model with the Einasto profile, we also show with a hatched area the impact of varying $M_{\rm min}$ by 6 orders of magnitude, from $10^{-3}$M$_\odot$ (lower contour) to $10^{-9}$M$_\odot$ (upper contour). For all the other cases, we have used $M_{\rm min}=10^{-6}$M$_\odot$. The background is normalized to the thermal velocity dispersion of DM particles with $m_\chi=100$~GeV.}
\label{fig_rho_eff_pwave} 
\end{figure}

\section{Effective Deposition Efficiency}
\label{sec:fz}

\subsection{\texorpdfstring{$f_c(z)$}{fc(z)} for Smooth Dark Matter Distributions}

As we discussed in Chapter~\ref{sec:Introduction}, energy injected by DM annihilation or decay at any given redshift is not immediately deposited into the IGM. At certain redshifts and input energies, the characteristic time for a photon to completely deposit its energy can be comparable to or greater than the Hubble time, making the `on-the-spot' approximation for the deposition of energy problematic~\cite{Slatyer2009}. Moreover, the efficiency at which injected energy is deposited into various channels (e.g. ionization of the IGM vs. heating of the IGM) is generically a complicated function of redshift, the energy of the injected particles, and the background level of ionization.

The details of the deposition process can be distilled into a single quantity $f_c(z)$, the ratio between energy deposited in channel $c$ and the injected energy at a given redshift $z$, i.e. 
\begin{alignat}{1}
	\left(\frac{dE}{dV \, dt} \right)_{c,\text{dep}} = f_c(z) \left(\frac{dE}{dV \, dt} \right)_{\text{inj}}
	\label{eqn:fcz}
\end{alignat}
where the channels considered are ionization of H (H ion), ionization of He (He ion), Lyman-$\alpha$ excitation of H atoms (Ly$\alpha$), heating of the IGM (heat), and energy converted into continuum photons that we observe as distortions to the CMB energy spectrum (cont). 

To calculate $f_c(z)$, we first need to calculate $T_c(z_{\text{inj}},z_{\text{dep}},E) \, d \log(1+z_{\text{dep}})$, the fraction of energy injected at redshift $z_{\text{inj}}$ that is deposited at redshift $z_{\text{dep}}$ into channel $c$ due to an injection of particles with individual energy $E$, discretized into redshift bins of size $d \log(1+z_{\text{dep}})$. This is done using the code developed in~\cite{Slatyer2012,Slatyer2015}, and only a brief summary of the code is given here. Starting with some injection of an $e^+ e^-$ or $\gamma \gamma$ pair at $z_{\text{inj}}$, the code tracks the cooling of particles and all of the secondary particles produced in these cooling processes in steps of $d \log(1 + z_{\text{dep}}) = 10^{-3}$. Photons that can efficiently photoionize HI, HeI and HeII in the IGM are removed from the main code and are considered to be ``deposited'', together with all electrons (including secondary electrons from photoionization) below \SI{3}{keV}. The proportion of energy deposited into each channel $c$ from the deposited photons and electrons is then determined by a separate low-energy code, which is described in full detail in~\cite{Slatyer2015}. The code assumes only small modifications to the ionization history of the universe from DM, since large modifications are ruled out by observational constraints. With this assumption, any arbitrary injection history with an arbitrary energy spectrum of particles can then be treated as a linear combination of individual injections of fixed energy at particular redshifts.

In the original code, $T_c (z_{\text{inj}},z_{\text{dep}},E) \, d \log(1+z_{\text{dep}})$ was computed from $1+z = 3000$ to $1+z=10$ for both injection and deposition redshift, over a large range of particle kinetic energies ($E \sim 10$ keV to $\SI{}{TeV}$).  Below $1+z_{\text{dep}} = 10$, the ionization history becomes much less certain due to the process of reionization. The exact details of the ionization history can have a significant impact on our calculation of $f_c(z)$: $f_{\text{H ion}}$, for example, should decrease significantly when $x_e \equiv n_e/n_{\text{H}}$ is close to 1. However, in order to make use of constraints on $T_m$ and $\delta \tau$, the code has to be extended down to lower redshifts. Given this uncertainty, we defer a discussion of how these results are extended down to $1+z_{\text{dep}} = 4$ to the following sub-section.

At the end of the calculation, we would have determined the fraction of energy injected at $z_{\text{inj}}$ that is deposited at some deposition redshift $z_{\text{dep}}$, broken down by deposition channel. Determining the total deposited energy at some redshift $z_{\text{dep}}$ therefore requires knowledge of the full injection history. To relate the deposited energy to the current injected energy and obtain $f_c(z)$ as defined in Eq.~(\ref{eqn:fcz}), we have to integrate $T_c(z_{\text{inj}},z_{\text{dep}},E) d\log(1+z_{\text{dep}})$ over all injection redshifts prior to $z_{\text{dep}}$. For any arbitrary DM energy injection process, the spectrum of particles injected has a typical redshift dependence $dN/(dE\, dV\, dt) \propto (1+z)^\alpha$, where $\alpha = 6$ for $s$-wave annihilation, $\alpha = 8$ for $p$-wave annihilation and $\alpha = 3$ for decay. In each case, we can factor the spectrum into a redshift-dependent factor multiplied by an energy spectrum $d\bar{N}/dE$ that is independent of redshift. Doing this, one can show~\cite{Slatyer2012} that
\begin{alignat}{1}
  f_c(z) = \frac{H(z)}{(1+z)^{\alpha-3} \sum\limits_{\text{species}}\int E \frac{d\bar{N}}{dE} dE} \sum_{\text{species}} \int \frac{(1+z')^{\alpha-4}}{H(z')} dz' \int T_c(z',z,E) E \frac{d\bar{N}}{dE} dE,
\end{alignat}
where the sum over species indicates that we are combining effects from all species produced in the annihilation process. For this chapter, we only consider the case where DM annihilates or decays into $e^+e^-$ or $\gamma \gamma$, with each particle having fixed, identical total energy $E=m_\chi$ for annihilations or $E = m_\chi/2$ for decays. In this case, $f_c(z)$ further simplifies to
\begin{alignat}{1}
	f_{c}(z,E) = \frac{H(z)}{(1+z)^{\alpha-3}} \int \frac{(1+z')^{\alpha-4}}{H(z')} T_{c}(z',z,E)\, dz'
\end{alignat}
for each of the injection species being considered. The quantity $f_c(z,E)$ for the injection species $e^+e^-$ and $\gamma\gamma$ will be denoted by a subscript $e$ and $\gamma$, respectively. While the spectrum of particles associated with any DM injection process may be significantly more complicated, ultimately any such process deposits energy into the IGM via $e^+e^-$ pairs or photon pairs. Understanding the energy deposition efficiency through $e^+e^-$ or $\gamma\gamma$ is thus sufficient to understand the effect of DM annihilation/decay on the IGM, since the energy deposition efficiency of any annihilation/decay process is simply an appropriate sum over $f_{c,e/\gamma}(z,E)$ over injection species and all relevant energies.

\subsection{\texorpdfstring{$f_c(z)$}{fc(z)} at Low Redshifts}

We defer a full treatment of calculating $f_c(z)$ to low redshifts to Chapter~\ref{chap:DarkHistory}, and instead give a brief summary of the method here. We have computed $f(z)$ down to a redshift of $1+z = 4$ in three different scenarios: (i) instantaneous and complete reionization at $z = 6$, which is close to the expected redshift of reionization from astrophysical measurements of $T_m$; (ii) instantaneous and complete reionization at $z = 10$, which is close to the expected redshift of reionization from measurements of the CMB power spectrum; and (iii) no reionization. These different reionization conditions were used not just for the deposition of energy by low-energy photons and electrons, but also for the high-energy code which tracks high-energy electrons and photons as they cool over time, since the photoionization rate of high-energy photons depend strongly on the ionization history. Previous studies typically assume that $f_c(z)$ can be written as a redshift- and model-dependent efficiency function $f(z)$, which describes the efficiency with which high-energy particles are degraded to low energies and is independent of the deposition channel. This function multiplies a channel-dependent factor $\chi_c(x_e(z))$ that depends only on the free electron fraction and describes the absorption of low-energy particles into each of the deposition channel.\footnote{One popular choice is the scheme called the ``SSCK approximation'' in~\cite{Slatyer2015a}, where a fraction $(1-x_e)/3$ is deposited into ionization and excitation each, with the remaining $(1+2x_e)/3$ going into heating.} However, our calculation of $\chi_c(z)$ depends on the low-energy photon spectrum at each redshift, and so depends on both $x_e$ and the injection history in a non-trivial way. The $f_c(z)$ results found in~\cite{Slatyer2015} took these effects into account assuming the standard \texttt{RECFAST} ionization history, and can be used for small perturbations about that scenario. However, when considering reionization and markedly different reionization scenarios, $f_c(z)$ must be re-computed in each case by re-calculating the cooling in both the high-energy and low-energy regimes.

In order to perform these calculations, we also assume simultaneous reionization of neutral helium (HeI) at the same redshift as HI reionization. After HI and HeI reionization, low-energy photons can deposit their energy through (i) the ionization of singly-ionized helium (HeII); (ii) excitations to HeII; or (iii) distortions of the CMB energy spectrum. 

After reionization, the high energy code tags photons as deposited only when they can efficiently photoionize HeII. Thus any ``deposited'' photon with energy $E > \SI{54.4}{eV}$ corresponds to a HeII ionization and consequently gives rise to a secondary low-energy electron spectrum. Photons below this threshold cannot ionize anything else, and are assigned to the excitation or distortion channels. Low-energy electrons, including the secondary spectrum produced by photoionizing photons, deposit energy according to the same model used in~\cite{Slatyer2015}, which is in turn based on~\cite{Valdes:2007cu,Valdes:2009cq,MNR:MNR20624}. In accordance with these results, once full reionization occurs, the electrons deposit their energy into the IGM solely through heating, since there are no longer any neutral hydrogen atoms to ionize or excite.

We note here that prior to the instantaneous reionization, the code assumes a standard ionization history computed by the recombination code \texttt{RECFAST}. Furthermore, we have assumed the instantaneous reionization of HeII at $1+z = 4$, which is not a fully realistic model. Once the contribution to $x_e$ from DM annihilations become significant enough, our calculation for $f_c(z)$ based on the \texttt{RECFAST} result will not reflect the true $f_c(z)$ for the new ionization history that includes the DM contribution, and likewise for a HeII reionization scenario that differs significantly from instantaneous reionization at $1+z = 4$. 

In principle, this means that $f_c(z)$ should be calculated iteratively: after calculating $x_e(z)$ for a certain DM model using the $f_c(z)$ obtained from the \texttt{RECFAST} ionization history, $f_c(z)$ should be recalculated with the new $x_e(z)$, with this process repeated until convergence of $x_e(z)$ is achieved. However, we stress that such a computationally intensive process is unnecessary, since calculating $f_c(z)$ assuming a \texttt{RECFAST} ionization history results in an $x_e$ ($T_m$) prior to reionization that is always larger (smaller) than what we would get with an iterative calculation. This ensures that we have not unintentionally ruled out any DM model with a significant contribution to reionization consistent with the $T_m$ constraints, even without performing an iterative calculation of $f_c(z)$. This behavior can be seen in Figure~\ref{fig:freeEleFracDecayAllowedRegion}, which shows a comparison of the ionization and thermal history computed with $f_c(z)$ after one iteration with the default $f_c(z)$ used in the rest of the chapter. This point will be discussed further in Section \ref{sec:Constraints}.

\subsection{\texorpdfstring{$f_c(z)$}{fc(z)} Including Structure Formation}

The formation of structures at late times gives rise to local densities that greatly exceed the cosmological DM density $\rho_{\chi,0}$, accompanied by an increase in the velocity dispersion of DM particles within haloes. This has no effect on the rate of energy injection from DM decay, since the average rate of decays per unit volume across the universe remains the same. 
In the case of DM $s$-wave annihilation, however, the increased density increases the rate of interaction, while for $p$-wave annihilation both the increased density and increased velocity dispersion dramatically enhance the annihilation rate. These effects cause a significant deviation from the expected energy injection due to a smooth/homogeneous DM distribution. 

The increase in the density can be parameterized by an effective density $\rho_{\text{eff}}(z)$ for $s$-wave annihilation (Eq.~(\ref{eqn:rhoeff}) and Figure~\ref{fig_rho_eff}), and an effective density times velocity dispersion $(\rho v/c)_{\rm eff}(z)$ for $p$-wave annihilation (Eq.~(\ref{eqn:rhoeff_p}) and Figure~\ref{fig_rho_eff_pwave}). 

With these effective quantities, the energy injection rate can be written as a boost factor multiplied by the unclustered distribution injection rate:
\begin{alignat}{1}
	\left(\frac{dE}{dV dt}\right)_{\text{inj}} &= \left(\frac{dE_s}{dV dt}\right)_{\text{inj}}[1 + \mathcal{B}_{s,p}(z)],
\end{alignat}
where the subscript $s$ in $E_s$ indicates the energy injection due to a smooth distribution of DM given by Eqs.~(\ref{eqn:injRateSmooth}) and
(\ref{eqn:smoothpwave}) for the $s$- and $p$-wave cases, respectively. The effective deposition efficiency can now be re-defined as
\small
\begin{alignat}{1}
	f_c(z) &= \frac{H(z)}{(1+z)^{\alpha-3}}  \int \frac{(1+z')^{\alpha-4}}{H(z')} T_c(z',z,E) [1 + \mathcal{B}_{s,p}(z')] \, dz',
	\label{eqn:fz}
\end{alignat}
\normalsize
so that
\begin{alignat}{1}
	\left(\frac{dE}{dV dt} \right)_{c,\text{dep}} = f_c(z) \left(\frac{dE_s}{dV dt} \right)_{\text{inj}}.
    \label{eqn:f_z_structure_formation}
\end{alignat}
$f_c(z)$ is now the ratio of the energy deposited in channel $c$ including structure formation effects to the injected energy due only to the smooth DM distribution, which has a simple analytic form. For $s$-wave annihilation, the boost factor is 
\begin{alignat}{1}
	1 + \mathcal{B}_s(z) = \frac{\rho_{\text{eff}}^2(z)}{(1+z)^6 \rho_{\chi,0}^2},
\end{alignat}
where $\rho_{\text{eff}}$ is shown in Figure~\ref{fig_rho_eff}. For $p$-wave annihilation, the effect of structure formation is parametrized not only by an effective density $\rho_{\text{eff}}$, but also by the characteristic one-dimensional velocity of the DM particles. The boost factor is:
\begin{alignat}{1}
	1 + \mathcal{B}_p(z) = \frac{(\rho v/c)_{\text{eff}}^2 (z)} {(1+z)^8 \rho_{\chi,0}^2(\sigma_{1D,B}(z=0)/c)^2}.
	\label{eqn:pwaveInj}
\end{alignat}
where $(\rho v/c)_{\text{eff}}$ is shown in Figure~\ref{fig_rho_eff_pwave}.

\section{Free Electron Fraction and IGM Temperature History}
\label{sec:FreeEleFrac}

As we discussed in Chapter~\ref{chap:intro}, we solve Eq.~(\ref{eqn:TLA_with_injection}) numerically to obtain the free electron fraction and IGM temperature history with dark matter energy injection. Aside from DM and the instantaneous reionization scenarios considered, no further sources of heating or reionization (e.g. star-forming galaxies and other stellar phenomena) are included in these equations.\footnote{See~\cite{Poulin2015} and Chapter~\ref{chap:DarkHistory} for an example of how heating from astrophysical sources can be included in a similar analysis.} This simplification is consistent with our computation of $f_c(z)$ using the standard ionization history, which overestimates the true contribution of $x_e(z)$ from DM, while underestimating the corresponding $T_m(z)$ contribution. A full treatment including astrophysical sources of heating and ionization would require a better understanding of $f_c(z)$ in situations where reionization is gradual, and we defer such a study to Chapter~\ref{chap:DarkHistory}.

The initial conditions used for the integration are $x_e(z=1700)=1$ and $T_m = T_{\text{CMB}}(z=1700)$, corresponding to the state of baryonic matter prior to recombination. The contribution to the optical depth by DM annihilation/decay $\delta \tau$, at a given $\langle \sigma v \rangle$ or $\tau_\chi$ and mass $m_\chi$ is then determined by integrating Eq.~(\ref{eqn:OpticalDepth}) up to $z=1700$ and subtracting the residual integrated optical depth that is already present when there is no DM. Note that when we consider reionization at $z = 10$, we do not include the contribution to $\delta \tau$ from $x_e$ between $z = 6$ and 10.\footnote{Note that the optical depth contribution from instantaneous reionization at $z = 10$ exceeds the Planck optical depth measurement, and thus would leave no room for any contribution from DM at all. However, we do not use the optical depth constraint in this manner.} We will discuss the calculation of $\delta \tau$ and the use of the optical depth constraints given by Eq.(~\ref{eqn:ExcessOpticalDepth}) further in Section~\ref{sec:Constraints}.


\section{Results}
\label{sec:Constraints}

We now calculate the integrated free electron fraction $x_e$ and IGM temperature $T_m$ as a function of redshift in each of the three DM energy injection scenarios considered ($s$-wave annihilation, $p$-wave annihilation and decay), for a wide range of $\langle \sigma v \rangle$ and decay lifetimes $\tau_\chi$, and $m_\chi$ between $\sim 10$ keV and $\sim 1$ TeV. As we discussed in Section \ref{sec:fz}, we have neglected any additional $x_e$ contribution from DM processes in our computation of $f_c(z)$, even though DM energy injection can produce significant deviations from the standard ionization history prior to reionization. Moreover, even after reionization occurs, the prescription for HeII reionization could affect the energy deposition. Thus the $f_c(z)$ curves we compute may not be completely accurate for an ionization history that is significantly different from the \texttt{RECFAST} result, or where HeII reionization cannot be approximated as occurring instantaneously at $1+z=4$.

Fortunately, our $f_c(z)$ calculations underestimate the contribution of DM to reionization, as more realistic ionization histories would generally have \emph{higher} ionization fractions, which in turn would suppress the additional ionization from DM. With a higher ionization fraction for HI (HeII), the energy deposited into ionization of HI (HeII) decreases, since there are fewer HI (HeII) atoms to ionize or excite prior to reionization (after reionization), while energy going into heating increases in both cases. This intuitive explanation of the behavior of $f_c(z)$ is consistent with the results used in our low-energy code to assign deposited energy from low-energy electrons into the various channels, where the MC results show that all of the energy from low-energy electrons go into collisional heating processes as $x_e$ tends to 1. Thus the $f_c(z)$ curves calculated under our assumptions consistently overestimate the rate of energy deposition into ionization, while underestimating the rate of energy deposited as heat. 

This means that if the contribution to reionization is small with the $f_c(z)$ values used here for a given cross section/lifetime and mass, then a more accurately computed $f_c(z)$ assuming an elevated $x_e$ will have an even smaller contribution to $x_e$ and a larger contribution to $T_m$, making the result more constrained by the $T_m$ limits. Similarly, including other conventional sources of ionization would only decrease the contribution that DM can make to reionization: the presence of other sources would produce a larger $x_e$ than we have assumed, which again suppresses the energy deposition fraction into ionization while enhancing the fraction into heating. 

To check the robustness of our constraints, we have also repeated our calculations considering: 

\begin{enumerate}
	\item Different reionization conditions, namely (i) instantaneous and complete reionization at $z=6$; (ii) instantaneous and complete reionization at $z=10$; and (iii) no reionization, to see how sensitive our results are to the uncertainty in the specifics of reionization and in particular in the redshift at which reionization occurs. For each reionization condition, $\delta \tau$ is integrated appropriately over $x_e(z)$, after which the optical depth from $x_e(z)$ without DM is subtracted. This includes the optical depth contribution from redshifts after reionization, where $x_e = 1.08$. Each reionization scenario results in a different $T_m(z)$ evolution after reionization occurs, and also has a different redshift at which we assess the contribution of DM to reionization (more details below); 

	\item A range of structure formation scenarios that bracket the uncertainties on the properties of low-mass (sub)haloes, below the resolution of current cosmological simulations; and

	\item Two different IGM temperature constraints as shown in Eq.~(\ref{eqn:TIGMConstraints}), namely (i) $T_m(z=6.08) = \SI{18621}{K}$; (ii) $T_m(z = 4.8) = \SI{10000}{K}$, where we have taken the upper bound at 95\% confidence. We do not make use of the lower bound, since $f_{\text{Heat}} (z)$ is likely to be an underestimate for reasons outlined above. The second temperature measurement is more constraining and will be used as the main temperature constraint, but constraints obtained from both temperature limits will be shown for the main $p$-wave result.   
\end{enumerate}

The three main quantities of interest are: (i) $x_e$ at a redshift just prior to the assumed instantaneous reionization at $z=6$ or $z=10$, or at $z=6$ for the case of no reionization, since hydrogen reionization is known to be complete by then; (ii) $T_m$ at $z=6.08$ and $z=4.8$ for comparison with the results shown in Eq.~(\ref{eqn:TIGMConstraints}); and (iii) the total integrated optical depth $\delta \tau$. If DM with a given $\langle \sigma v \rangle$ or $\tau_\chi$ and $m_\chi$ can produce $x_e > 0.1$ just before reionization (or at $z=6$ for the case of no reionization) we consider this a possible scenario in which DM can contribute significantly to reionization. The 10\% level used in this chapter is arbitrary, and we will also present results for contributions ranging from 0.025\% to 90\% in the form of color density plots for all injection species and all DM processes.

A few remarks should be made about the calculation of optical depth and the use of the optical depth constraints in this chapter. To compute $\delta \tau$, we integrate the optical depth due to DM annihilation/decay from $z_{\text{reion}}$ to recombination.\footnote{When there is no reionization, we start integrating from $z = 6$, making $\delta \tau$ identical to the case with $z_{\text{reion}} = 6$.} We then compare $\delta \tau$ to the bound on excess optical depth from redshifts $z > 6$, assuming full ionization for $z \leq 6$; that is, for the purposes of computing the maximum allowed exotic contribution to optical depth, we essentially treat $z_{\text{reion}} = 6$ for all scenarios, even when $\delta \tau$ includes only DM contributions from $z > 10$. This allows us to understand how our limits could weaken if the reionization history were different: including gradual reionization from astrophysical sources between $z = 6$ and $z = 10$, for example, would likely suppress the contribution to reionization and hence optical depth from DM annihilation during this period, resulting in a smaller contribution from DM to reionization than would have been determined with instantaneous reionization at $z_{\text{reion}} = 6$. By taking $z_{\text{reion}} = 10$ and not considering the contribution to optical depth for $z < 10$, we obtain the weakest constraints from the $\delta \tau$ bound given in Eq.~(\ref{eqn:ExcessOpticalDepth}). In this way, these two reionization scenarios bracket the possible contribution of DM to reionization. Thus, although including the optical depth due to complete, instantaneous reionization at $z = 10$ would exceed the Planck optical depth measurement, we still consider this scenario in order to study the DM contribution to reionization in a model-independent way. Assuming two different instantaneous reionization scenarios also allows us to probe the possible effects of earlier reionization on the DM contribution to the temperature evolution. 

We will choose as our benchmark the scenarios where the largest $x_e$ just prior to reionization can be obtained from the {\it smallest} $\langle \sigma v \rangle$ or {\it longest} decay lifetimes, since various experimental constraints set upper bounds on the cross sections and lower bounds on the decay lifetimes. In all cases, reionization at $z = 6$ is more realistic than no reionization and is also more easily achieved than at $z = 10$, making it the main reionization scenario to consider. The structure formation scenario with the largest boost factor allows for reionization with a smaller cross section, and thus we choose this as our benchmark (for $s$-wave annihilation this is the ``stringent'' case shown with a solid red line in Figure~\ref{fig_rho_eff}, while for $p$-wave annihilation all scenarios give the same boost).

\subsection{\texorpdfstring{$s$}{s}-wave Annihilation}

Figure~\ref{fig:freeEleFracsWave} shows the integrated free-electron fraction $x_e$ for the particular case of DM with $m_\chi = \SI{100}{MeV}$ undergoing $s$-wave annihilation into a pair of $\SI{100}{MeV}$ photons with a cross section ranging from $\SI{3E-27}{}$ to $\SI{3E-25}{\centi\meter\cubed\per\second}$, as well as the case with no DM for comparison. These curves show the result with no reionization: different reionization conditions are identical up to the redshift of reionization $z_{\text{reion}}$, whereupon $x_e$ instantaneously becomes 1 until the present day. These curves are representative of the $x_e$ histories across all DM masses and cross sections for $s$-wave annihilation. At $z \sim 20$, structure formation becomes important, which greatly increases $f_c(z)$ in all channels, leading to an increase in $x_e$. $s$-wave annihilation of the smooth distribution of DM results in a larger baseline $x_e$ after recombination, which is higher for larger $\langle \sigma v \rangle$ at the same $m_\chi$. 

\begin{figure*}[t!]
    \centering
	\subfigure{
		\includegraphics[scale=0.64]{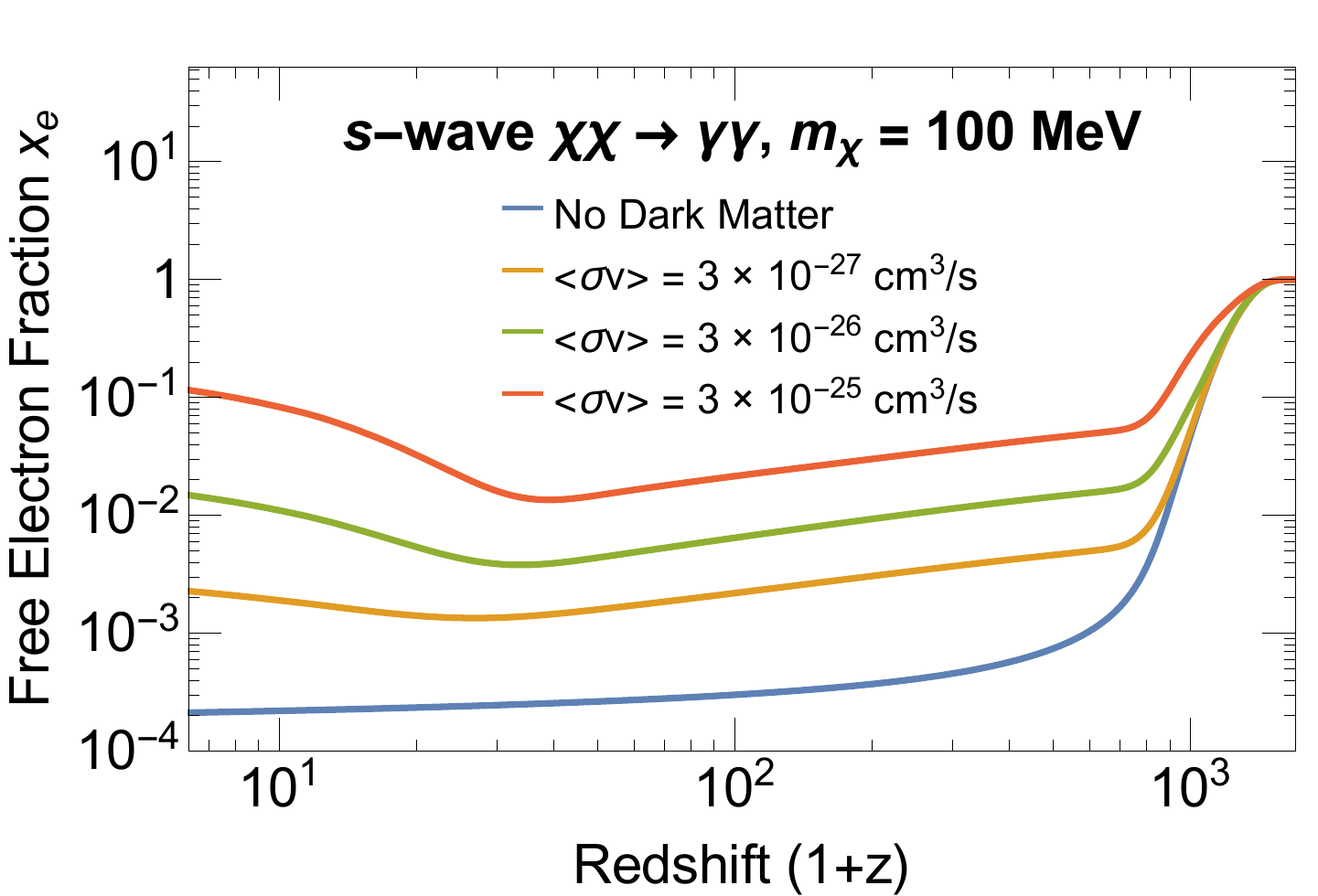}
	}
	\subfigure{
		\includegraphics[scale=0.64]{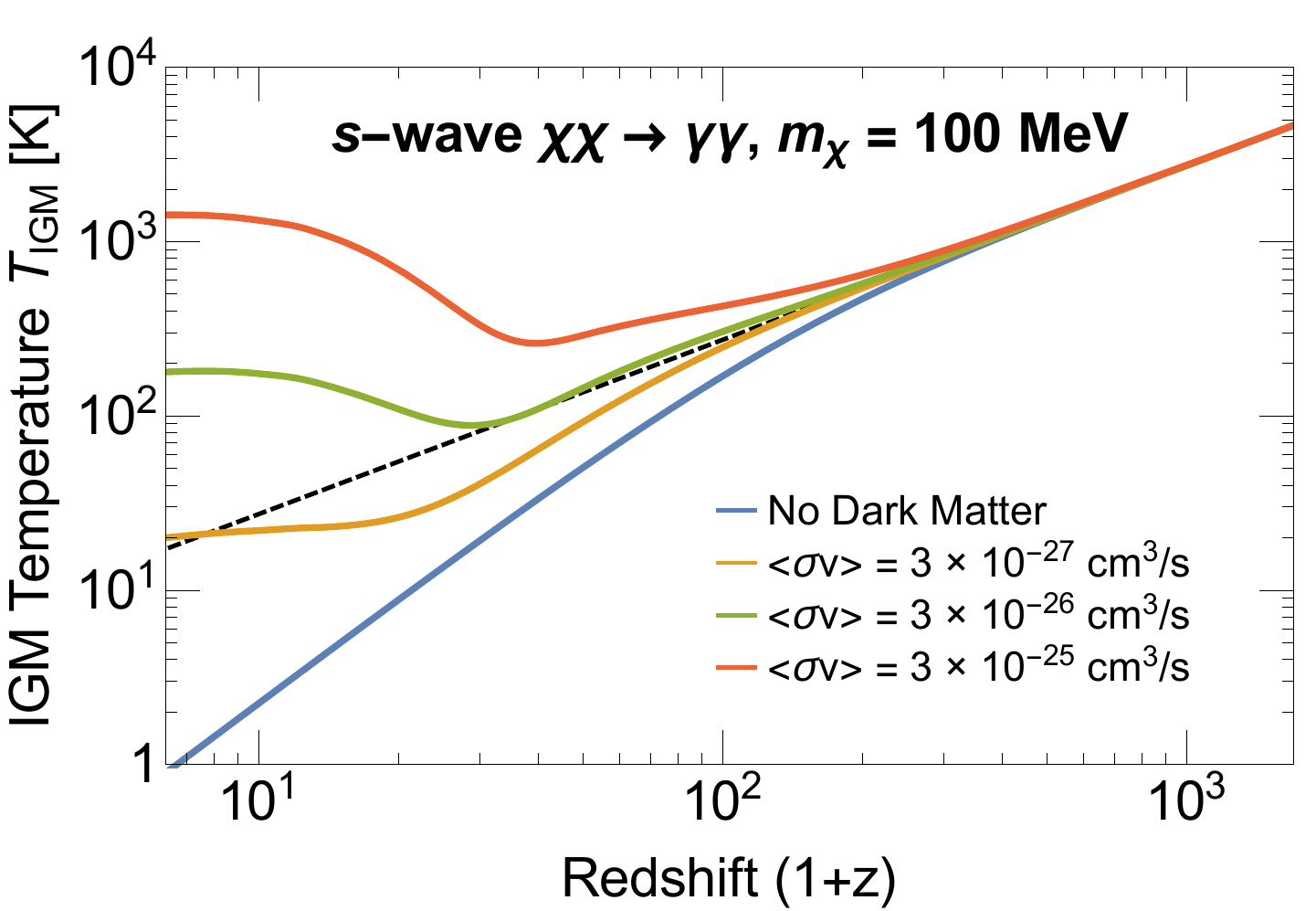}
	}
	\caption{Integrated free electron fraction $x_e$ and IGM temperature $T_m$ for $\chi \chi \to \gamma \gamma$ $s$-wave annihilation for $m_\chi = \SI{100}{\mega \eV}$ with (from bottom to top): no DM; $\left<\sigma v\right> = $ \SI{3E-27}{\centi\meter\cubed\per\second}; \SI{3E-26}{\centi\meter\cubed\per\second} and \SI{3E-25}{\cm^3 \s^{-1}} respectively. The CMB temperature is shown as a dashed line for reference. No reionization is assumed.}
	\label{fig:freeEleFracsWave}
\end{figure*}

Along with $x_e$, the IGM temperature history $T_m(z)$ is also simultaneously integrated. The IGM temperature curves for DM undergoing $s$-wave annihilation into \SI{100}{MeV} photons for cross sections ranging from $\SI{3E-27}{}$ to $\SI{3E-25}{\centi\meter\cubed\per\second}$ are shown in the same figure and are also representative of IGM temperature histories across a broad range of $\langle \sigma v \rangle$ and $m_\chi$. The CMB temperature is included for reference. The IGM is initially coupled to the CMB, but once recombination occurs, the temperature starts to fall more rapidly than the CMB temperature. DM $s$-wave annihilations decrease the fall-off in temperature at relatively large redshifts. At $z\sim 20$, the impact of structure formation once again increases the IGM temperature significantly relative to the case with no DM.

The contribution of DM to reionization through $s$-wave annihilation is significantly constrained by the CMB power spectrum measurements derived by Planck 2015~\cite{Ade:2015xua}, as well as by the measured total integrated optical depth. The cross section for annihilation must be large enough for significant ionization to occur at redshifts near reionization; however, increasing the cross section also increases the residual free electron fraction during the cosmic dark ages. This residual $x_e$ is constrained severely by the CMB anisotropy spectrum, which is sensitive to any additional ionization near redshifts $z \sim 600$. A large $x_e$ during the cosmic dark ages also contributes significantly to the optical depth. Since $n_e(z) \propto x_e(z)(1+z)^3$ and $dt/dz \propto (1+z)^{-5/2}$, the integrand in Eq.~(\ref{eqn:OpticalDepth}) is proportional to $x_e(z)(1+z)^{1/2}$. The significantly elevated $x_e$ baseline means that the dominant contribution to $\delta \tau$ comes from early times when $z$ is large: since structure formation is relevant at later times, it does not add significantly to $\delta \tau$.  

We performed the integration of $x_e(z)$ and $T_m(z)$ over a broad range of masses and cross sections, and computed the optical depth from $x_e(z)$ using Eq.~(\ref{eqn:OpticalDepth}). Figure~\ref{fig:xeConstraintsPlot_sWave} shows the free electron fraction just prior to reionization $x_e(z=6)$ for the benchmark scenario of both $\chi \chi \to e^+e^-$ and $\chi \chi \to \gamma \gamma$, as well as the excluded cross sections due to constraints from the CMB power spectrum as measured by Planck and from the integrated optical depth. Constraints from $T_m$ are presented in Appendix \ref{app:additionalConstraints}. These bounds are less constraining, but unlike the CMB and optical depth constraints, they are sensitive to the low redshift behavior of $s$-wave annihilations: increasing the boost from structure formation beyond the value used here may relax the CMB and optical depth bounds, but this would strengthen the $T_m$ constraints. 

Although we have shown the results for these two processes ($\chi \chi \to e^+e^-$ and $\chi \chi \to \gamma \gamma$) as a function of $\langle \sigma v \rangle$ and $m_\chi$, we stress that these constraints go beyond these two annihilation channels. We discuss this point and present bounds on $\langle \sigma v \rangle/m_\chi$ as a function of the injection energy of the final products (which may in general be very different from $m_\chi$) in Appendix \ref{app:additionalConstraints}.
\begin{figure*}[t!]
    \centering
	\subfigure{
		\includegraphics[scale=0.63]{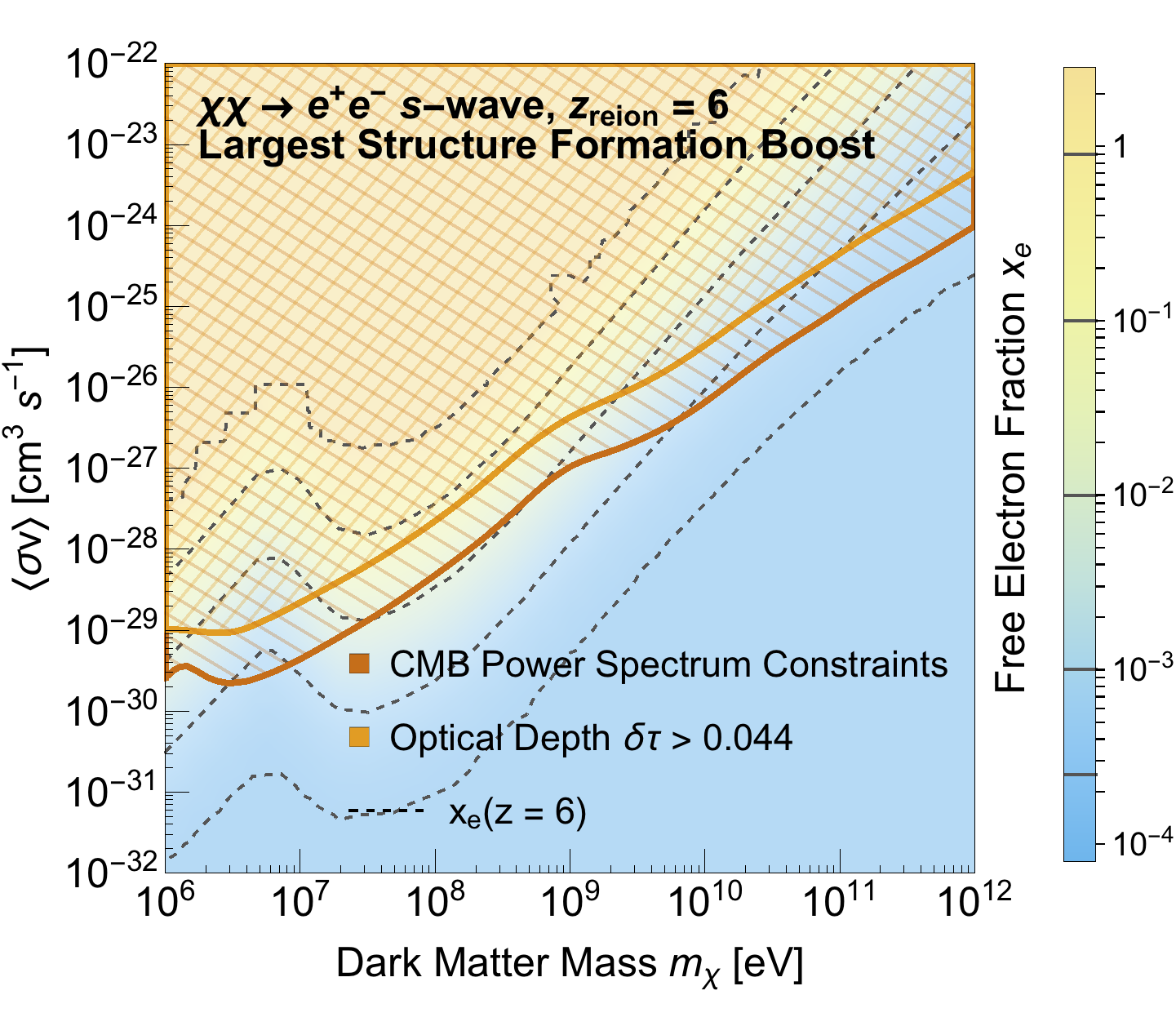}
	}
	\subfigure{
		\includegraphics[scale=0.63]{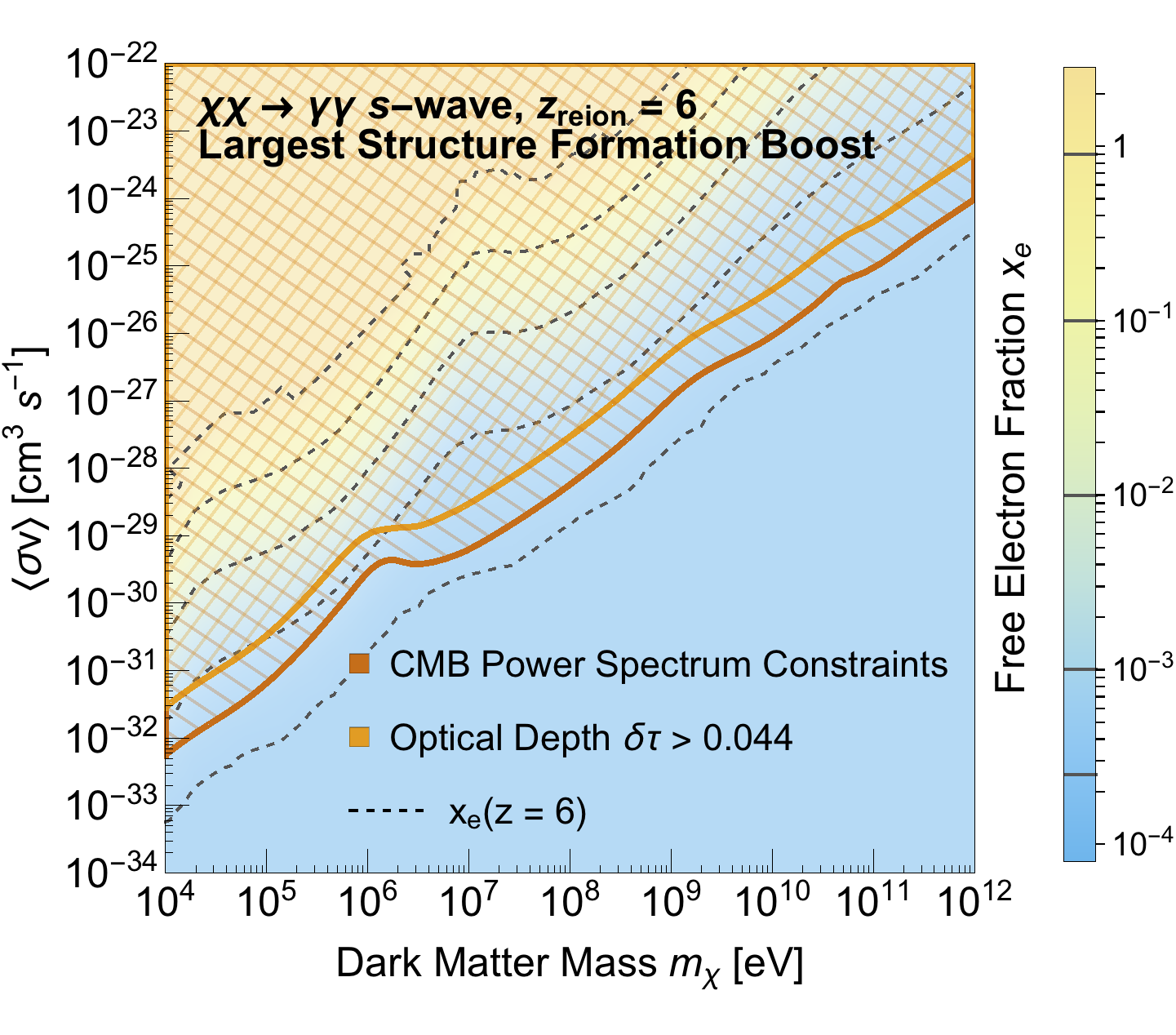}
	}
	\caption{DM contribution to reionization for $\chi \chi \to e^+e^-$ (left) and $\chi \chi \to \gamma \gamma$ (right) $s$-wave annihilation, benchmark scenario. The hatched regions correspond to parameter space ruled out by the CMB power spectrum constraints as measured by Planck (red) and optical depth constraints (orange) respectively. The color density plot shows the DM contribution to $x_e$ just prior to reionization at $z = 6$, with contours (black, dashed) shown for a contribution to $x_e(z = 6) = $ 0.025\%, 0.1\%, 1\%, 10\% and 90\% respectively.}
	\label{fig:xeConstraintsPlot_sWave}
\end{figure*}

In both annihilation channels, there is no parameter space where a significant contribution to reionization occurs while being consistent with either the CMB power spectrum or optical depth bounds, with the CMB power spectrum bounds being approximately one order of magnitude stronger than the optical depth bounds. We stress that the optical depth constraints are similar regardless of reionization conditions, since $\delta \tau$ is the additional contribution from DM only, and is therefore not affected by the period where $x_e = 1$ after reionization. As a result, the true optical depth limits for reionization at $z = 10$ are likely stronger than what is shown here, since we do not include the additional contribution to optical depth from the fully ionized universe between $z = 6$ and $z = 10$. Furthermore, $\delta \tau$ is dominated by contributions from larger redshifts ($z \gtrsim 100$) and is relatively insensitive to the exact details of reionization and structure formation at $z \lesssim 20$. At the maximum $\langle \sigma v \rangle$ allowed by the CMB power spectrum bound, the DM contribution to $x_e$ just prior to reionization is below 2\% for $\chi \chi \to e^+e^-$ and below 0.1\% for $\chi \chi \to \gamma \gamma$ across all $m_\chi$ considered. These results are shown in Figure~\ref{fig:xeMaxConstraints} in the conclusion.

Figure~\ref{fig:xeConstraintsStructSysPlot_sWave} shows the reionization constraints on $s$-wave annihilation for the structure formation prescriptions with the smallest and largest boost factor (used as the benchmark). As expected, significant ionization prior to reionization can be achieved at lower cross sections in the benchmark model, making it the most likely structure formation prescription for evading the constraints. Differences in structure formation can increase the value of $\langle \sigma v \rangle$ at which ionization becomes significant by less than an order of magnitude, and all of the regions with a significant contribution to reionization in either structure formation scenario are firmly ruled out by the Planck constraints.

\begin{figure*}[t!]
    \centering
	\subfigure{
		\includegraphics[scale=0.63]{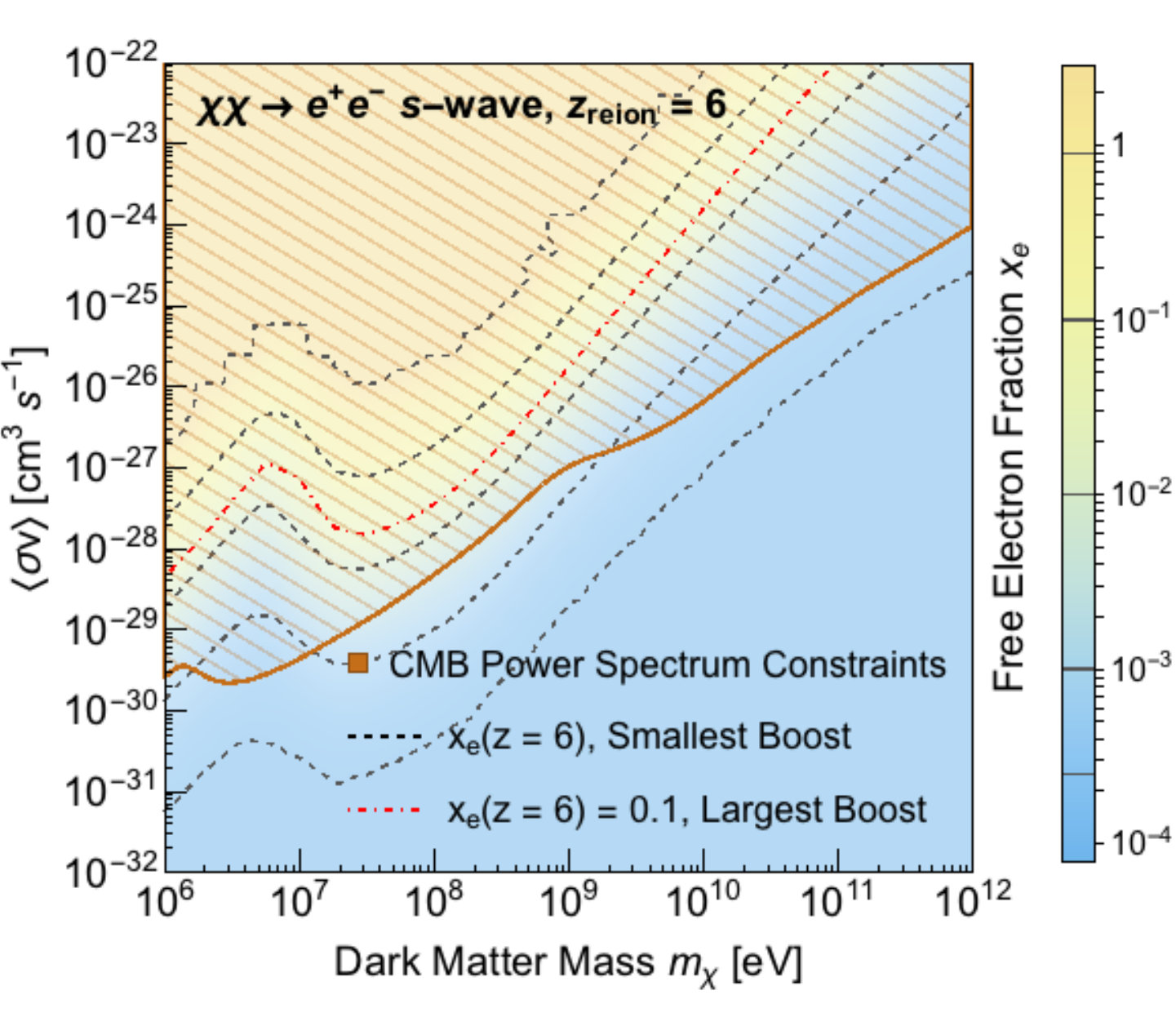}
	}
	\subfigure{
		\includegraphics[scale=0.63]{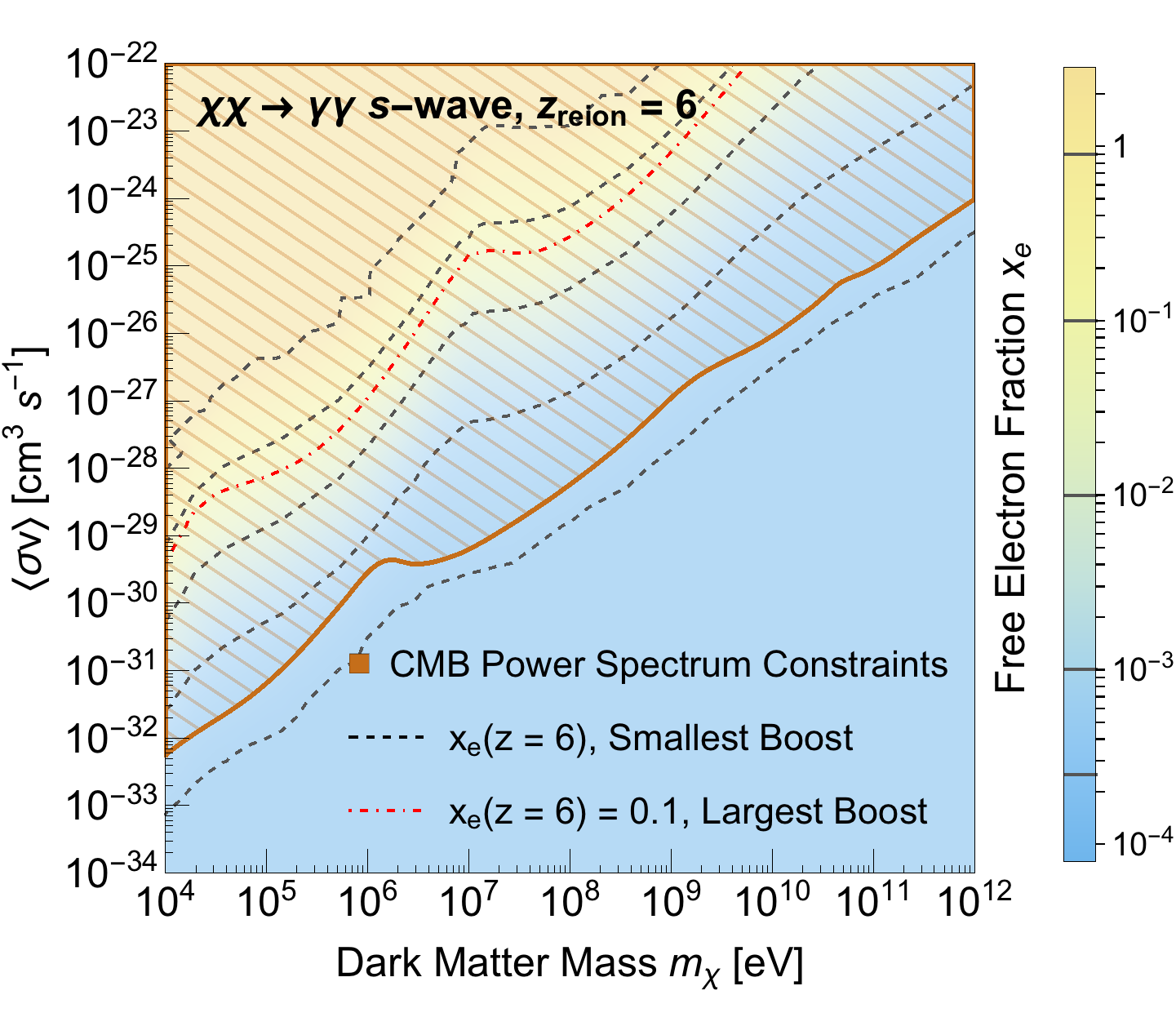}
	}
	\caption{DM contribution to reionization for $\chi \chi \to e^+e^-$ (left) and $\chi \chi \to \gamma \gamma$ (right) $s$-wave annihilation assuming a different structure formation prescription. The color density plot shows the DM contribution to $x_e$ just prior to reionization at $z = 6$ assuming an NFW profile without subhaloes, with contours (black, dashed) shown for a contribution to $x_e(z = 6) = $ 0.025\%, 0.1\%, 1\%, 10\% and 90\% respectively. The red, dot-dashed contour for $x_e(z = 6) = 0.1$ assuming the benchmark Einasto profile with subhaloes, which has the largest boost factor at all redshifts, is also shown for comparison. The CMB power spectrum constraints obtained by Planck are shown by the hatched red region. }
	\label{fig:xeConstraintsStructSysPlot_sWave}
\end{figure*}

Similarly, differences in reionization redshifts do little to change the result. Since $x_e(z)$ is identical in all three reionization scenarios until the point of reionization, there is no difference between $x_e(z=6)$ with reionization at $z=6$ and no reionization. With reionization at $z=10$, $x_e(z=10)$ is always less than $x_e(z=6)$ as $x_e$ increases rapidly between $z = 6$ and $10$, and so the region in parameter space where significant contribution to reionization occurs decreases when choosing an earlier redshift of reionization. Figure~\ref{fig:xeConstraintsReionSysPlot_sWave} summarizes these results. 

\begin{figure*}[t!]
    \centering
	\subfigure{
		\includegraphics[scale=0.63]{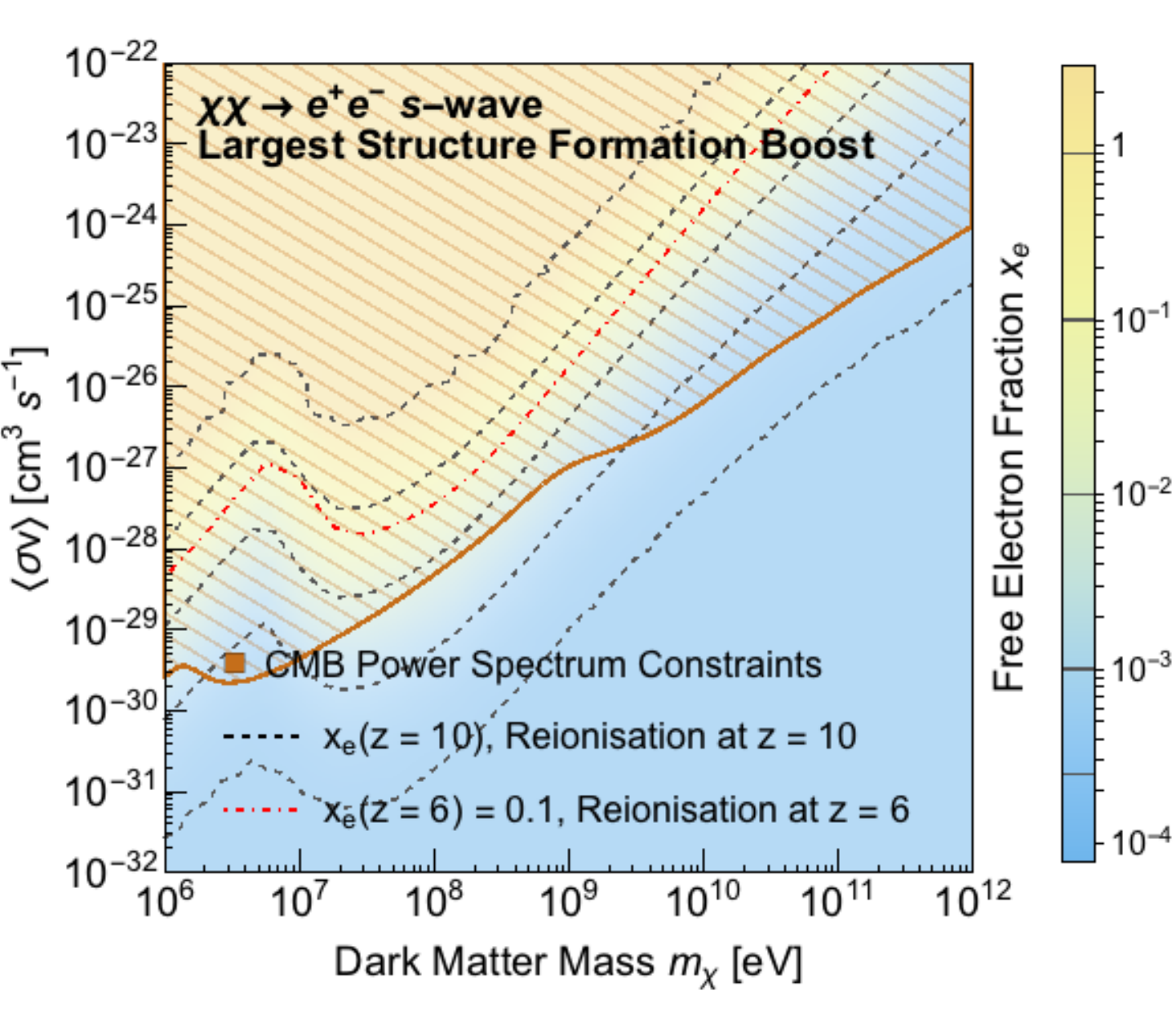}
	}
	\subfigure{
		\includegraphics[scale=0.63]{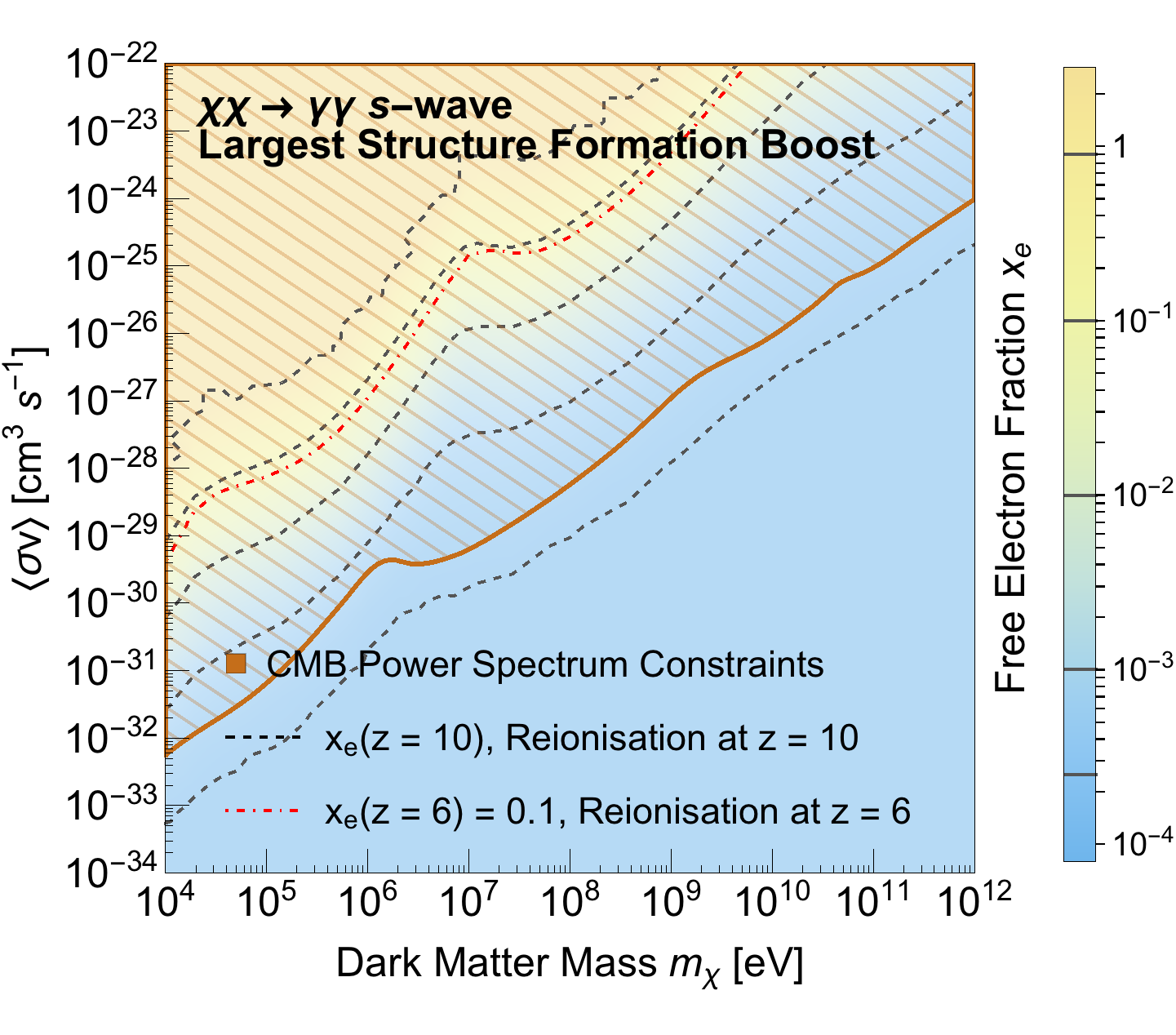}
	}
	\caption{DM contribution to reionization for $\chi \chi \to e^+e^-$ (left) and $\chi \chi \to \gamma \gamma$ (right) $s$-wave annihilation, assuming a different reionization scenario.The color density plot shows the DM contribution to $x_e$ just prior to reionization at $z = 10$, with contours (black, dashed) shown for a contribution to $x_e(z = 10) = $ 0.025\%, 0.1\%, 1\%, 10\% and 90\% respectively. The red, dot-dashed contour shows $x_e(z = 6) = 10\%$ with reionization at $z = 6$ for comparison. The CMB power spectrum constraints obtained by Planck are shown by the hatched red region. }
	\label{fig:xeConstraintsReionSysPlot_sWave}
\end{figure*}

To conclude, any significant contribution to reionization through $s$-wave DM annihilation is severely constrained by the cross section bounds from the Planck CMB power spectrum measurement as well as the expected integrated optical depth to the surface of last scattering. For values of $\langle \sigma v \rangle$ that are consistent with the Planck CMB power spectrum constraints, we can only expect a contribution of no more than 2\% of the total ionization just prior to reionization (see Figure~\ref{fig:xeMaxConstraints}). Our results are consistent with the conclusion reached in~\cite{Poulin2015}. We have also shown that these results are robust to our assumptions on the structure formation scenario and on the redshift of reionization.

\subsection{\texorpdfstring{$p$}{p}-wave Annihilation}\label{sec_pwave}

In $p$-wave annihilation, the $v^2$ dependence of the cross section results in a $v^2/v_{\text{ref}}^2$ suppression of the energy injection rate, given in Eq.~(\ref{eqn:pwaveInj}). Figure~\ref{fig:freeEleFracpWave} shows the integrated $x_e$ for the case of $\chi \chi \to \gamma \gamma$ $p$-wave annihilation with $(\sigma v)_{\text{ref}}$ between \SI{3E-24}{\centi\meter\cubed\per\second} and \SI{3E-22}{\centi\meter\cubed\per\second}. Prior to the relevance of structure formation, the velocity suppression is a large effect, resulting in no additional contribution to $x_e$ unless the cross section is exceptionally large. Once structure formation occurs, however, the velocity dispersion of DM particles within haloes increases significantly, increasing in turn the energy injection rate from $p$-wave annihilation. This results in a sudden and large increase in both $x_e$ and $T_m$ at $z \sim 20$. 

\begin{figure*}[t!]
    \centering
	\subfigure{
		\includegraphics[scale=0.64]{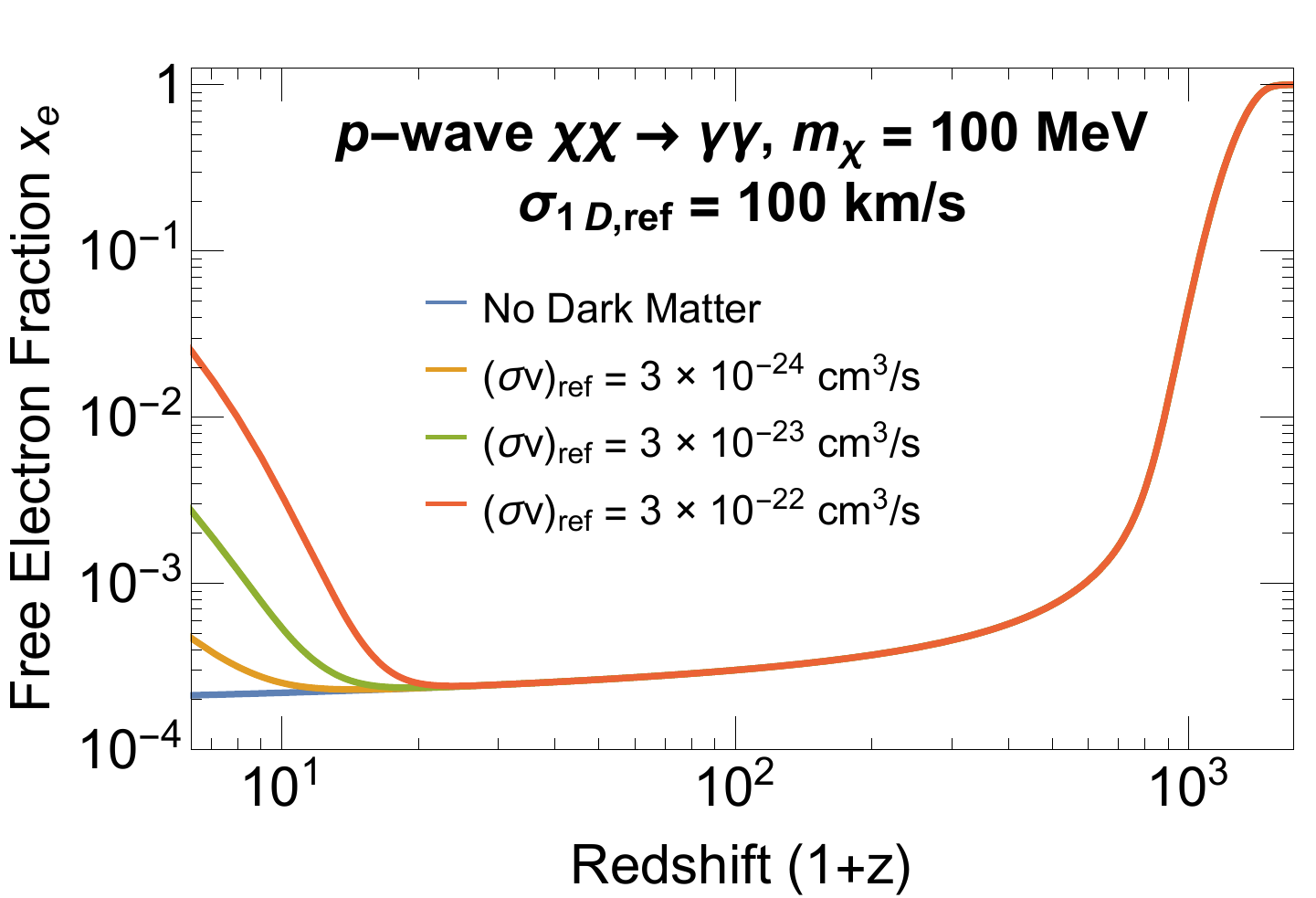}
	}
	\subfigure{
		\includegraphics[scale=0.64]{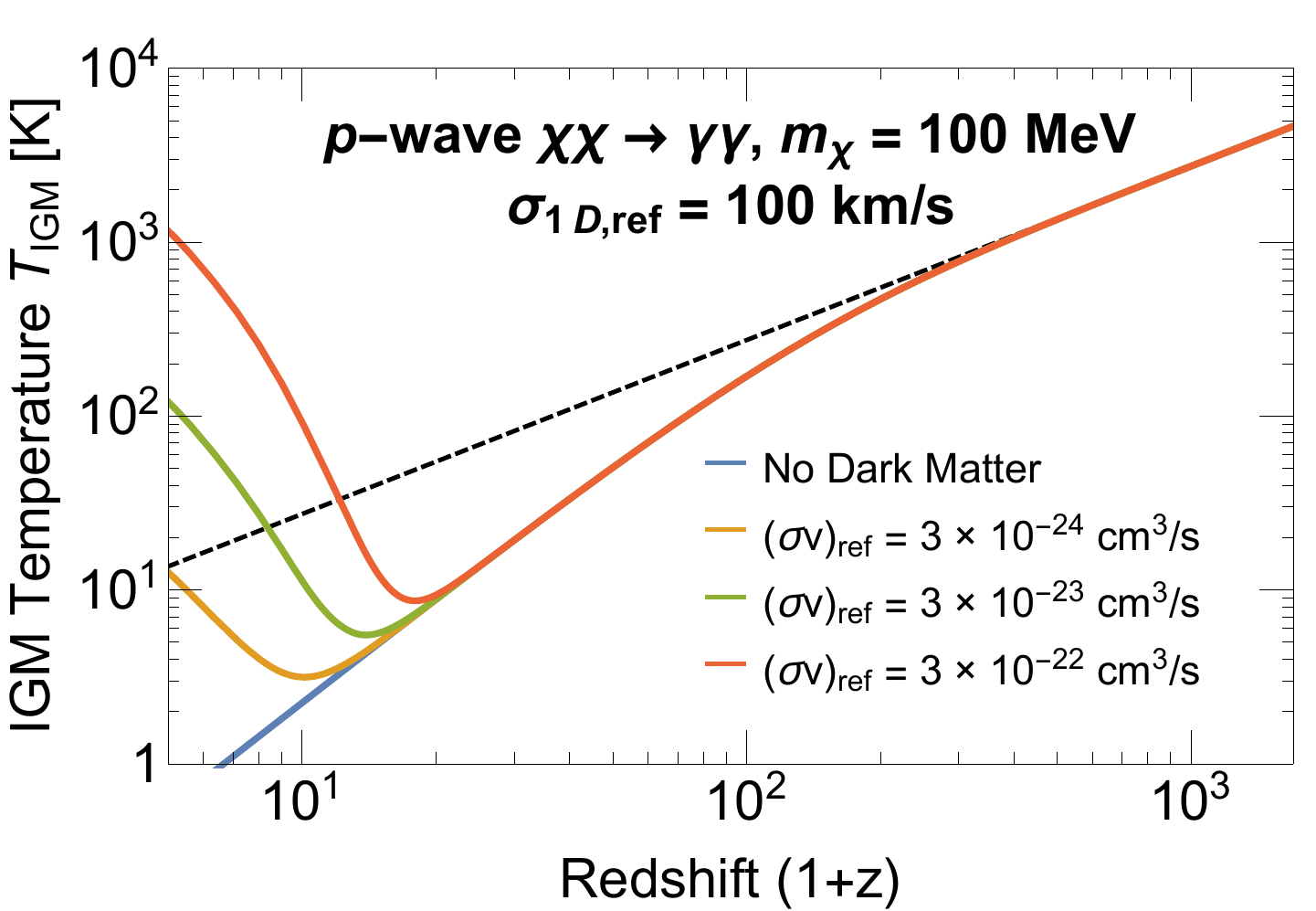}
	}
	\caption{Integrated free electron fraction $x_e$ and IGM temperature $T_m$ for $\chi \chi \to \gamma \gamma$ $p$-wave annihilation for $m_\chi = \SI{100}{\mega \eV}$ with (from bottom to top): (blue) no DM; $(\sigma v)_{\text{ref}} = \SI{3e-24}{\centi\meter\cubed\per\second}$, $(\sigma v)_{\text{ref}} = \SI{3e-23}{\centi\meter\cubed\per\second}$ and $(\sigma v)_{\text{ref}} = \SI{3e-22}{\centi\meter\cubed\per\second}$ respectively. The CMB temperature is shown as a dashed line. No reionization is assumed. }
	\label{fig:freeEleFracpWave}
\end{figure*}

As we discussed earlier in section \ref{sec:StructureFormation}, the annihilation rate prior to structure formation is dependent on our choice of $\sigma_{\text{1D,B}}$, which we have taken to be the velocity dispersion for unclustered DM with $m_\chi = \SI{100}{GeV}$ and $T_{\text{kd}} = \SI{28}{MeV}$. Choosing a significantly smaller value of $m_\chi$ or $T_{\text{kd}}$ increases $\sigma_{\text{1D,B}}$, which in turn increases the annihilation rate prior to structure formation. With a sufficiently small value of $m_\chi$ and/or $T_{\text{kd}}$, $x_e$ will stay at a value significantly above the expected $x_e$ with no dark matter, similar to the ionization histories typical of $s$-wave dark matter shown in Figure~\ref{fig:freeEleFracsWave}. While this leads to an increase in $x_e$ just prior to reionization, the optical depth bounds that we considered for $s$-wave annihilations become very constraining, particularly with the sharp increase in $x_e$ after structure formation that is not present in the $s$-wave case. Decreasing $m_\chi$ and/or $T_{\text{kd}}$ therefore makes it harder for a significant contribution to be made to reionization in a way that is consistent with the optical depth limits, making our unclustered velocity dispersion choice an optimistic one.

Unlike $s$-wave annihilation, constraints from the CMB power spectrum on the contribution of DM to reionization for $p$-wave annihilation are velocity-dependent, and depend strongly on the ``coldness'' of DM particles, i.e. on their unclustered velocity dispersion. Significant $x_e$ at low redshifts can be achieved without any significant increase in the free electron fraction at redshift $z \sim 600$ by choosing a  small enough $m_\chi$ so that the velocity dispersion prior to structure formation is small. Optical depth constraints are also weaker since there is no increase in the baseline ionization during the cosmic dark ages, unlike in $s$-wave annihilation. 
Instead, the IGM temperature after reionization has been shown to be a significantly more important constraint on the $p$-wave annihilation cross section than bounds obtained from the CMB power spectrum~\cite{Diamanti2014}. Once the effect of structure formation becomes relevant, the late-time energy injection results in significant heating of the IGM. Figure~\ref{fig:freeEleFracpWave} shows this behavior for the case of $\chi \chi \to \gamma \gamma$ $p$-wave annihilation with $\sigma v_{\text{ref}}$ between \SI{3E-24}{\centi\meter\cubed\per\second} and \SI{3E-22}{\centi\meter\cubed\per\second}. At large enough cross sections, $T_m$ after reionization exceeds the limits set by Eq.~(\ref{eqn:TIGMConstraints}). 

\begin{figure*}[t!]
    \centering
	\subfigure{
		\includegraphics[scale=0.63]{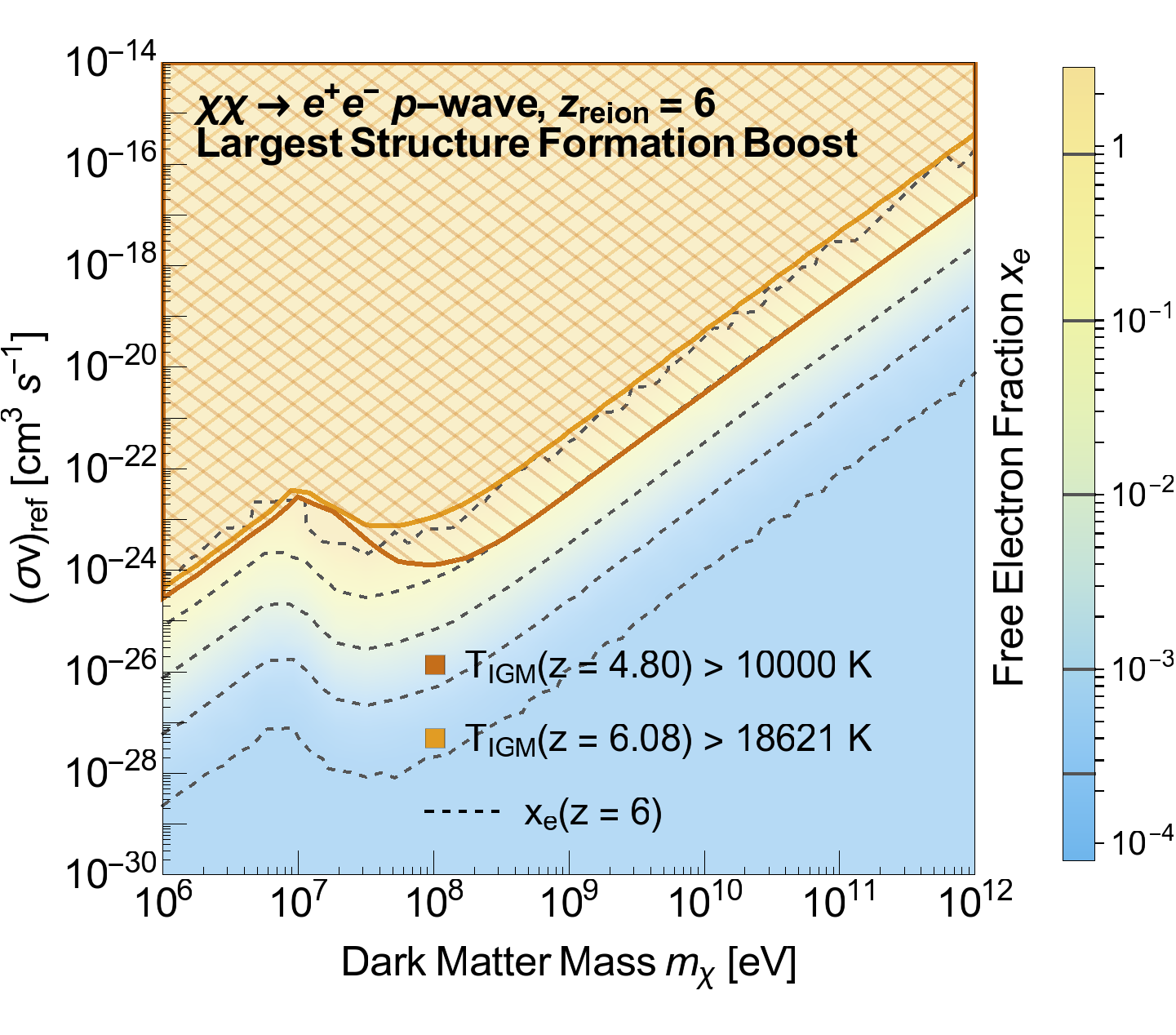}
	}
	\subfigure{
		\includegraphics[scale=0.63]{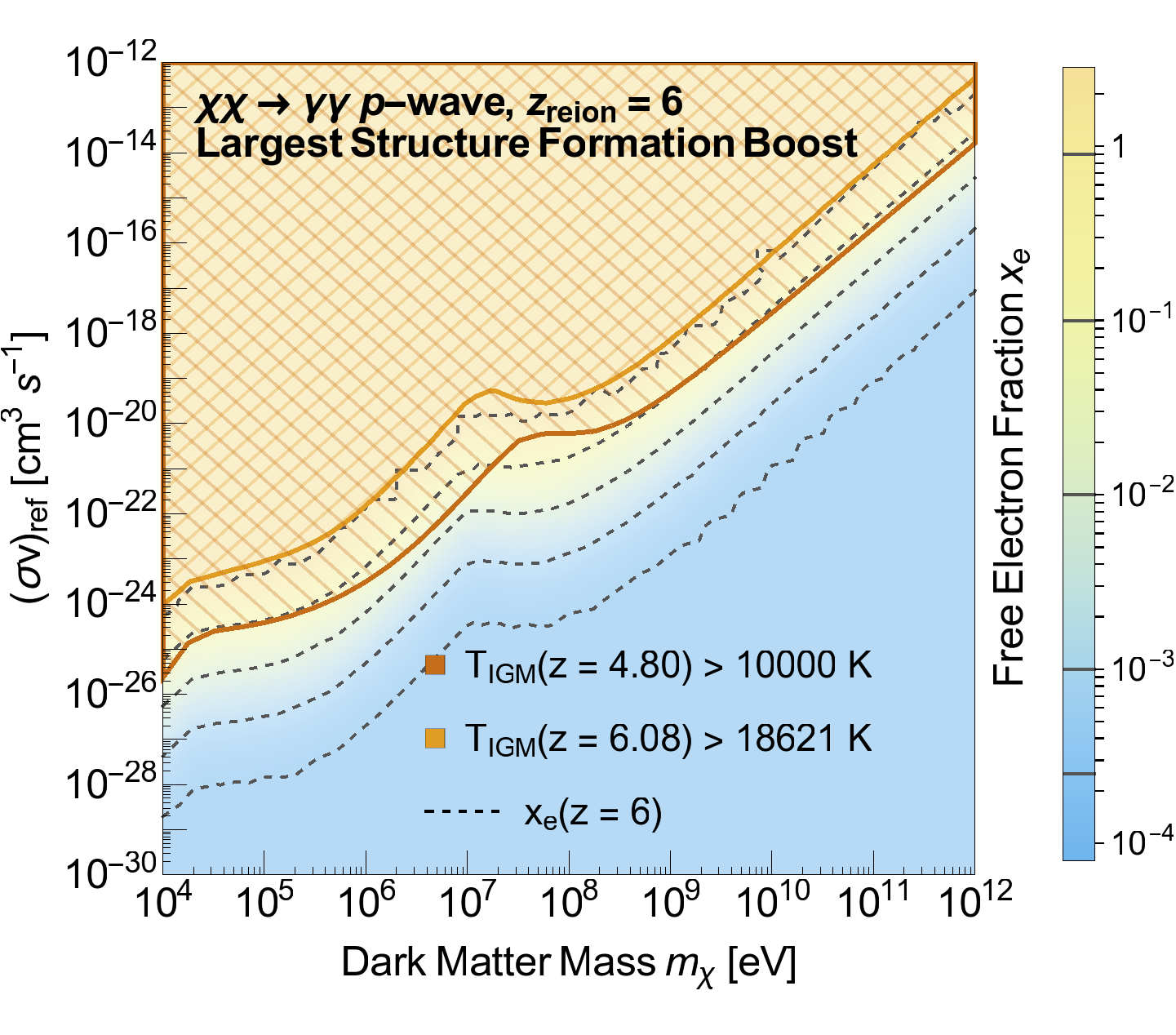}
	}
	\caption{DM contribution to reionization for $\chi \chi \to e^+e^-$ (left) and $\chi \chi \to \gamma \gamma$ (right) $p$-wave annihilation, benchmark scenario. The hatched regions correspond to parameter space ruled out by $T_m(z = 4.80) < \SI{10000}{K}$ (red) and $T_m(z = 6.08) < \SI{18621}{K}$ (orange) respectively. The color density plot shows the DM contribution to $x_e$ just prior to reionization at $z = 6$, with contours (black, dashed) shown for a contribution to $x_e(z = 6) = $ 0.025\%, 0.1\%, 1\%, 10\% and 90\% respectively.}
	\label{fig:xeConstraintsPlot_pWave}
\end{figure*}

Figure~\ref{fig:xeConstraintsPlot_pWave} shows $x_e(z=6)$ just prior to reionization for our benchmark scenario in the $(\sigma v)_{\text{ref}}$ - $m_\chi$ parameter space, as well as the excluded parameter space due to constraints from $T_m(z=6.08)$ and $T_m(z=4.8)$. The same results on the parameter space of $(\sigma v)_{\text{ref}}/m_\chi$ and injection energy of the annihilation products are shown in Appendix \ref{app:additionalConstraints}. Masses above \SI{100}{MeV} for $\chi \chi \to e^+ e^-$ and almost all $m_\chi$ for $\chi \chi \to \gamma \gamma$ are excluded by the benchmark IGM temperature constraint, $\log_{10} T_m(z=4.8) < 4.0$. The most likely region in parameter space that can still result in reionization is in the $\chi \chi \to e^+e^-$ channel with $m_\chi < \SI{100}{MeV}$ and $(\sigma v)_{\text{ref}}$ between $10^{-25}$ and $10^{-23} \SI{}{\centi\meter\cubed\per\second}$, and in the $\chi \chi \to \gamma \gamma$ channel with $m_\chi \sim \SI{100}{MeV}$ and $(\sigma v)_{\text{ref}} \sim 10^{-21} \SI{}{\centi\meter\cubed\per\second}$. These cross sections are much larger than a thermal relic cross section, but can be accommodated in a large variety of DM models, including any non-thermally produced DM or forbidden DM~\cite{DAgnolo2015}. 

The sudden relaxation of the $T_m$ constraints below $m_\chi \sim \SI{100}{MeV}$ and the corresponding decrease in $x_e(z=6)$ for $\chi \chi \to e^+e^-$ deserve a special mention here. DM particles with $m_\chi < \SI{100}{MeV}$ annihilating into electrons lose their energy principally through inverse Compton scattering off CMB photons, which by $z \sim 10$ mainly produces photons close to or below the ionizing threshold for hydrogen. After reionization, photoionization by these secondary photons is suppressed further, as the only remaining neutral species is HeII, which has a larger ionization energy. Thus, only a small fraction of the energy goes into collisional heating (due to secondary electrons) of the IGM, with most of the energy from the DM annihilation being deposited as continuum photons. This results in a decrease in IGM temperature after the reionization redshift. At higher DM masses, in contrast, the lower-redshift IGM temperature bound is significantly more constraining, as the IGM temperature invariably continues to increase even after reionization: the $e^+e^-$ pair produced by the annihilation can now upscatter photons to energies above the ionization threshold of HeII. These photoionization events produce low-energy secondary electrons even after reionization, which in turn can collisionally heat the IGM.

Next, we present our results assuming different reionization redshifts in Figure~\ref{fig:xeConstraintsReionSysPlot_pWave}. These results show that the allowed region for $\chi \chi \to e^+e^-$ is shifted upward in cross section, since a larger cross section is required to reionize the universe at an earlier redshift, while $T_m$ actually becomes less constraining as the IGM temperature now has more time to decrease after reionization. This suggests that the region that permits significant reionization is relatively independent of the reionization condition. The same is not true for the case of $\chi \chi \to \gamma \gamma$: the IGM temperature constraints remain fairly similar, but since we are now extracting $x_e$ at a higher redshift, the overall contribution to $x_e$ by DM decreases. With reionization at $z = 10$, for the $\gamma \gamma$ channel, there is no allowable $m_\chi$ where the contribution to $x_e$ prior to reionization exceeds 10\%.

\begin{figure*}[t!]
    \centering
	\subfigure{
		\includegraphics[scale=0.63]{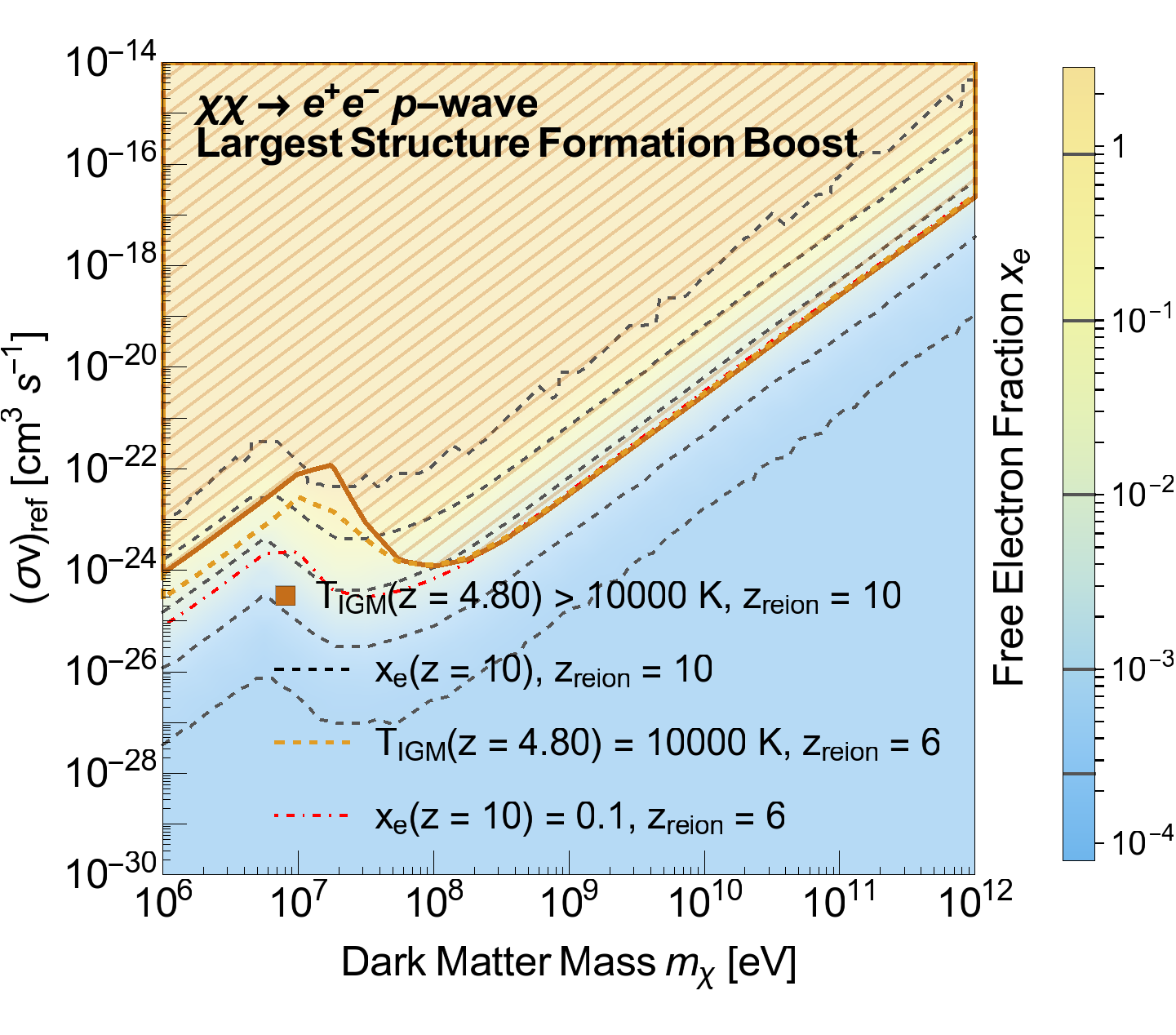}
	}
	\subfigure{
		\includegraphics[scale=0.63]{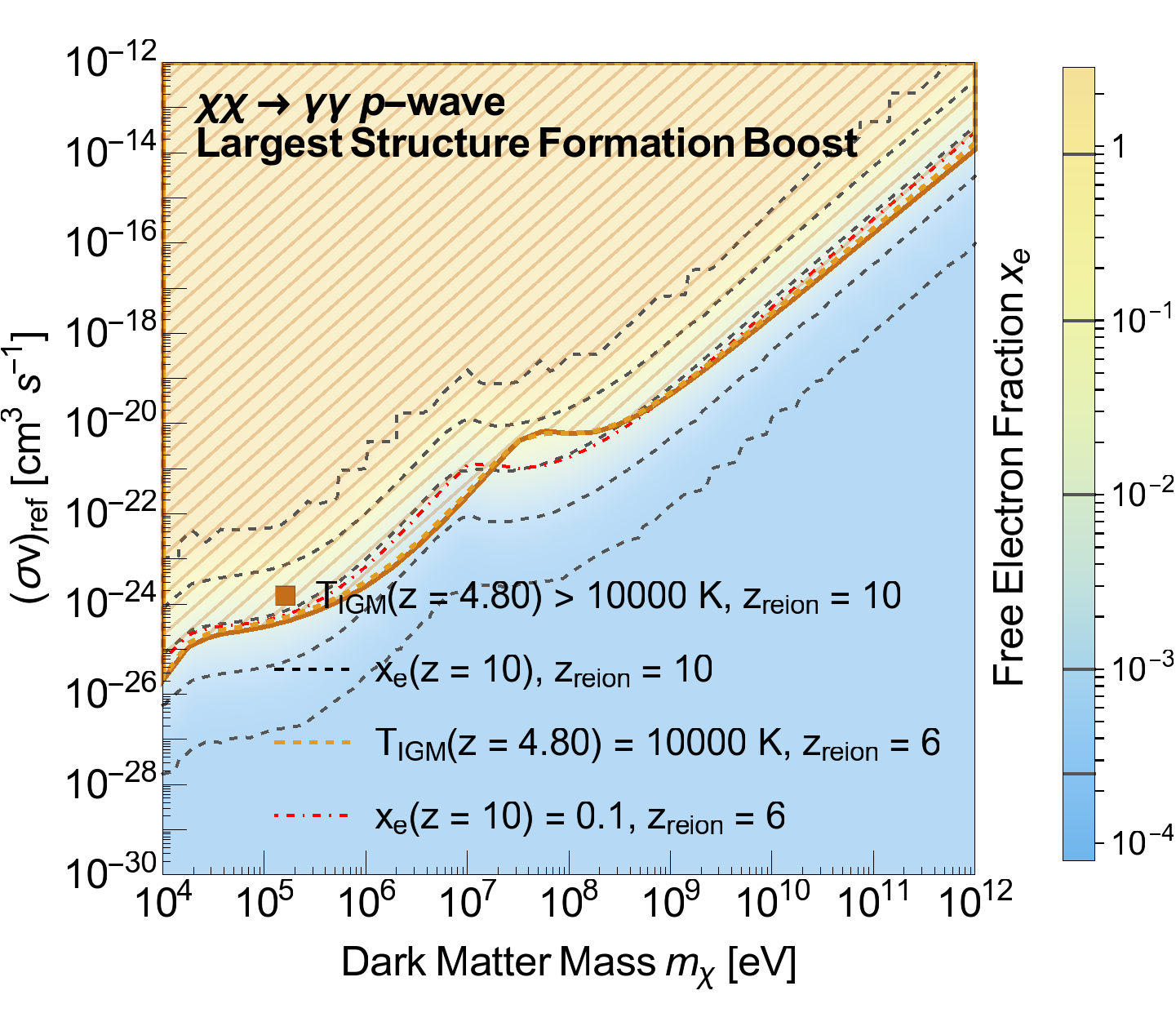}
	}
	\caption{DM contribution to reionization for $\chi \chi \to e^+e^-$ (left) and $\chi \chi \to \gamma \gamma$ (right) $p$-wave annihilation assuming a different reionization scenario. The color density plot shows the DM contribution to $x_e$ just prior to reionization at $z = 10$, with contours (black, dashed) shown for a contribution to $x_e(z = 10) = $ 0.025\%, 0.1\%, 1\%, 10\% and 90\% respectively. The regions ruled out by the benchmark $T_m$ constraint $T_m(z=4.80) < \SI{10000}{K}$ assuming reionization at $z = 10$. The red, dot-dashed contour shows $x_e(z = 6) = 10\%$ and the dashed, bold orange contour shows $T_m(z=4.80) = \SI{10000}{K}$, both assuming reionization at $z = 6$, for comparison: the region above the IGM temperature contour is ruled out in this case. Note that the 10\% line for reionization at $z = 6$ lies close to the 1\% line for reionization at $z = 10$ in both cases.}
	\label{fig:xeConstraintsReionSysPlot_pWave}
\end{figure*}

So far, there is still a range of DM masses with appropriate cross sections that can reionize the universe at at least the 10\% level through $p$-wave annihilations into $e^+e^-$ ($ m_\chi \lesssim \SI{100}{\mega \eV}$, $(\sigma v)_{\text{ref}} \sim 10^{-24}$ - $10^{-23} \SI{}{ \centi\meter\cubed\per\second}$), and into $\gamma \gamma$ ($m_\chi \sim \SI{100}{\mega \eV}$, $(\sigma v)_{\text{ref}} \sim 10^{-21}$ - $10^{-20} \SI{}{ \centi\meter\cubed\per\second}$) with reionization at $z = 6$. We turn our attention now to two further bounds on $(\sigma v)_{\text{ref}}$ that are relevant to these regions in parameter space. 

First, we consider the cross section constraints from the CMB power spectrum measurements. Although the results shown in Figure~\ref{fig:excludedXSec} are bounds on $\langle \sigma v \rangle$ for $s$-wave annihilation, they also serve as an estimate for the bound on the cross section $\langle \sigma v \rangle = (\sigma v)_\text{ref} v^2/v_{\text{ref}}^2$ in the case of $p$-wave annihilations, since the results are only sensitive to the rate of energy deposition into ionization of the IGM during the cosmic dark ages. The main difference with $p$-wave annihilations is that the bound now depends on $v^2$ after recombination and during the cosmic dark ages. $v^2$ is strongly dependent on the primordial ``coldness'' of DM, which in turn depends on the nature of the DM particles, i.e. mass and kinetic decoupling temperature. While DM is coupled to photons, $v^2 \sim 3 T_\gamma/m_\chi$, whereas after decoupling, $v^2 \propto (1+z)^2$. Taking the limit $L(m_\chi)$ on $\langle \sigma v \rangle$ set by the CMB spectrum at a particular DM mass $m_\chi$ as shown in Figure~\ref{fig:excludedXSec},
\begin{alignat}{1}
	(\sigma v)_\text{ref} \lesssim 3.7 L(m_\chi) \left(\frac{m_\chi}{\SI{1}{MeV}} \right)^2 \left( \frac{x_{\text{kd}}}{10^{-4}} \right) \left(\frac{\SI{1}{eV}}{T_\gamma} \right)^2,
\end{alignat}
where $x_{\text{kd}} \equiv T_{\text{kd}}/m_\chi$. $T_\gamma$ is some representative CMB temperature after recombination such that the CMB power spectrum is most sensitive to energy injections at the redshift $z$ corresponding to $T_\gamma$ ($z \sim 600$ in the $s$-wave case). 

In the case of $\chi \chi \to e^+e^-$, in the region of parameter space where a significant contribution to reionization can be made, the CMB bounds can rule out these regions if $x_{\text{kd}} \lesssim 10^{-2} - 10^{-1}$ for $m_\chi \sim \SI{1}{MeV}$ and $x_{\text{kd}} \lesssim 10^{-6}$ for $m_\chi \sim \SI{100}{MeV}$ (we have set $T_\gamma = \SI{0.14}{eV}$ as a representative value), while for 100 MeV DM annihilating $\chi \chi \to \gamma \gamma$, we have $x_{\text{kd}} \sim 10^{-3} - 10^{-2}$. Thus for the CMB bounds to exclude these regions, we would need to have $T_\mathrm{kd} \lesssim 100$ keV, and in some cases it would need to be much lower (at the sub-keV scale).

Values of $T_{\text{kd}}$ higher than these bounds are consistent (and expected) in a large variety of DM models, e.g. $T_{\text{kd}} \sim \SI{}{MeV} (m_\chi/\SI{}{GeV})^{2/3}$ for neutralino DM~\cite{Chen2001}, and $T_{\text{kd}} \sim \SI{2.02}{MeV} (m_\chi/\SI{}{GeV})^{3/4}$ for DM-lepton interactions of the form $(1/\Lambda^2)(\bar{X} X)( \bar{l} l)$ for some interaction mass scale $\Lambda$, giving rise to $p$-wave suppressed cross sections~\cite{Shoemaker2013,Diamanti2014}. In general, $T_\mathrm{kd}$ below the scale of the electron mass is unusual, as the only relativistic species available to maintain kinetic equilibrium are photons and neutrinos.\footnote{Models such as neutrinophilic DM~\cite{Shoemaker2013,VandenAarssen2012} can, however, exhibit such a behavior.} The CMB bounds therefore place few constraints on our parameter space for $p$-wave annihilation, in stark contrast to the $s$-wave case. 

\begin{figure*}[t!]
    \centering
	\subfigure{
		\includegraphics[scale=0.63]{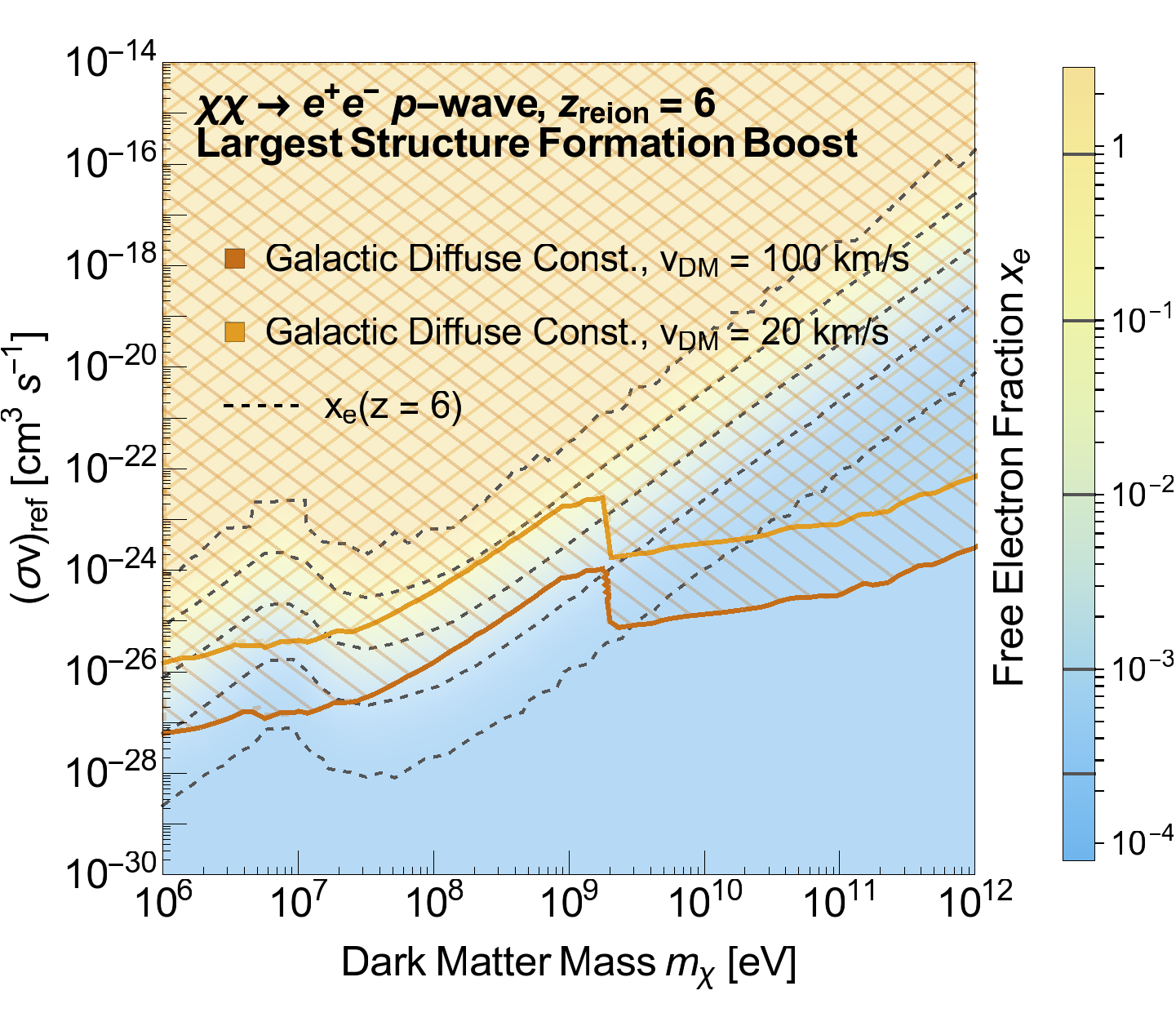}
	}
	\subfigure{
		\includegraphics[scale=0.63]{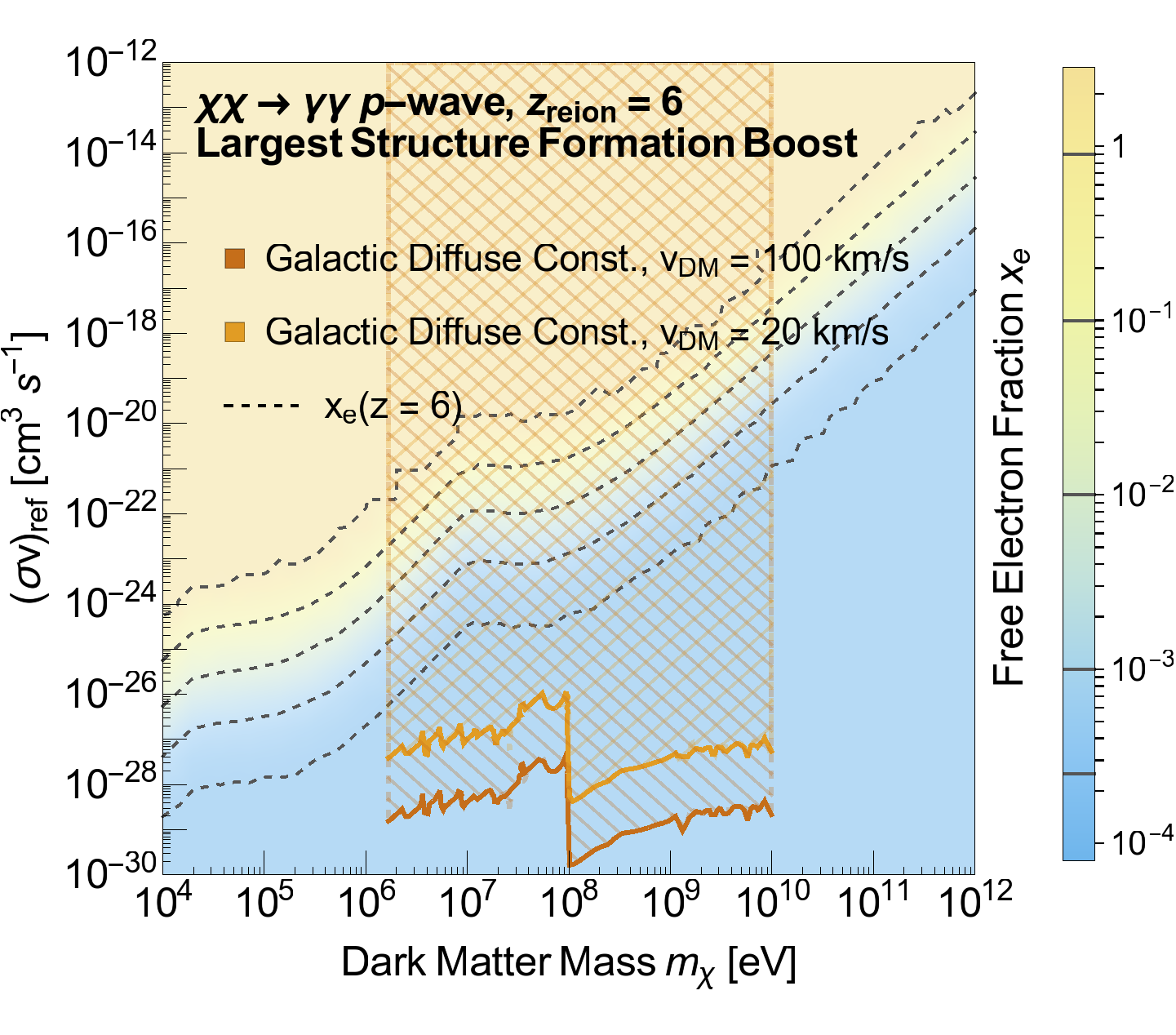}
	}
	\caption{DM contribution to reionization for $\chi \chi \to e^+e^-$ (left) and $\chi \chi \to \gamma \gamma$ (right) $p$-wave annihilation, together with limits from the galactic diffuse background. The color density plot shows the DM contribution to $x_e$ just prior to reionization at $z = 6$, with contours (black, dashed) shown for a contribution to $x_e(z = 6) = $ 0.025\%, 0.1\%, 1\%, 10\% and 90\% respectively. These constraints are dependent on the dispersion velocity $v_{\text{DM}}$: we show the constraints obtained assuming that $v_{\text{DM}} = \SI{100}{km s^{-1}}$ (red hatched region) and \SI{20}{km s^{-1}} (orange hatched region). The $\chi \chi \to e^+e^-$ constraints are obtained from~\cite{Essig2013,Massari2015}, while the $\chi \chi \to \gamma \gamma$ limits are from~\cite{Boddy2015,Albert2014}.}
	\label{fig:xeConstraintsGalacticPlot_pWave}
\end{figure*}

Next, we look at $p$-wave constraints from gamma ray flux measurements of the galactic diffuse background. The derived constraints from the galactic diffuse background are shown in Figure~\ref{fig:xeConstraintsGalacticPlot_pWave}. For $\chi \chi \to e^+ e^-$, final state radiation produced as part of the annihilation process in the Milky Way halo produces gamma ray photons that can be measured by these experiments, placing an upper bound on the rate of $p$-wave annihilation into $e^+e^-$ for DM masses of up to \SI{10}{GeV} in the Milky Way. Constraints derived in~\cite{Essig2013}  from a combination of data from INTEGRAL, COMPTEL and Fermi set a limit of $\langle \sigma v \rangle \lesssim 10^{-27}\SI{}{\centi\meter\cubed\per\second}$ for $m_\chi \lesssim \SI{100}{\mega \eV}$. This was derived assuming an NFW profile, which is a relatively conservative choice for these experiments: the constraints fluctuate by a factor of a few if different DM halo profiles are chosen. All of the measured photon flux is conservatively attributed to DM annihilation in the galaxy halo only, without accounting for extragalactic DM annihilation or other more conventional sources like inverse Compton scattering off starlight or synchrotron radiation. 

The translation of these velocity-averaged cross section bounds to constraints on $(\sigma v)_{\text{ref}}$ depends on the velocity dispersion $v_{\text{DM}}$ around the solar circle. Given a measured photon flux, a larger $v_{\text{DM}}$ would place a stronger constraint on $(\sigma v)_{\text{ref}}$, since the photon flux is proportional to the annihilation rate, which is in turn proportional to $(\sigma v)_{\text{ref}} v_{\text{DM}}^2$ in a $p$-wave process. Because of this, the constrained $(\sigma v)_{\text{ref}}$ is proportional to $1/v_{\text{DM}}^2$. However, in order for some region of parameter space with more than a 10\% contribution to reionization from DM to be allowed, the dispersion velocity in the solar circle needs to satisfy $v_{\text{DM}} < \SI{20}{km s^{-1}}$, which is significantly smaller than the local velocity of the solar circle and is hence unrealistic~\cite{Cerdeno:2010jj}. 

Similar results hold for $\chi \chi \to \gamma \gamma$, where searches for sharp spectral features such as lines or boxes in the galactic diffuse gamma ray background place strong bounds on the annihilation cross section of this process. By requiring the number of counts from $\chi \chi \to \gamma \gamma$ in each energy bin in the spectrum to not exceed the measured number of counts by $2 \sigma$, the gamma ray spectrum from COMPTEL and EGRET can be used to set an upper limit of $\langle \sigma v \rangle \lesssim 10^{-27} \SI{}{\centi\meter\cubed\per\second}$ for $m_\chi \sim \SI{100}{MeV}$~\cite{Boddy2015}, with a similar analysis using Fermi data~\cite{Albert2014} giving a limit of $\langle \sigma v \rangle \lesssim 10^{-29} \SI{}{\centi\meter\cubed\per\second}$ for $m_\chi \gtrsim \SI{100}{MeV}$. This means that the dispersion velocity required for a 10\% contribution to reionization is $v_{\text{DM}} \sim \SI{0.1}{km s^{-1}}$, which is once again unrealistic.

Although we have freely used the constraints for $\langle \sigma v \rangle$ to directly set constraints on $(\sigma v)_{\text{ref}}$, some caution must be taken when doing so. The contribution of DM annihilations to the observed photon flux measured by a detector is due to annihilations all along the line-of-sight. In order to set constraints on DM annihilation from gamma ray flux measurements, the appropriate function of the DM density and velocity must therefore be averaged along the line-of-sight. $\langle \sigma v \rangle$ bounds are frequently set by averaging over the DM density, but without taking into account the velocity dispersion of the Milky Way halo. Without performing this average, $\langle \sigma v \rangle$ bounds are implicitly assumed to be for $s$-wave processes only. 

However, as we demonstrate in Appendix \ref{app:JFactor}, averaging over the velocity dispersion as well as the density appears to change the $\langle \sigma v \rangle$ bounds for $p$-wave annihilation by less than a factor of 2 under many different assumptions. These bounds would need to relax by at least 2 orders of magnitude for $\chi \chi \to e^+e^-$ and 4 orders of magnitude for $\chi \chi \to \gamma \gamma$ to allow any significant contribution to reionization at all.

Overall, the possible contribution of $p$-wave DM annihilation to reionization appears to be constrained to the $<10\%$ level across all of the masses and injection species considered here. At $m_\chi \gtrsim \SI{10}{GeV}$, this contribution is limited by $T_\text{IGM}$ measurements, while for $m_\chi \lesssim \SI{10}{GeV}$, any allowed parameter space with more than 10\% contribution to reionization after accounting for $T_m$ appears to be ruled out by observations of the galactic diffuse emission gamma ray spectrum.

\subsection{Decay}

Figure~\ref{fig:freeEleFracDecay} shows $x_e(z)$ and $T_m(z)$ for $m_\chi = \SI{100}{MeV}$ DM undergoing $\chi \to \gamma \gamma$ decays (each photon now has an energy of \SI{50}{MeV}) with various representative decay lifetimes, which are typical for other masses and decay modes. Compared to $s$-wave annihilation, the energy injection rate in decays is not dependent on structure formation, and the $(1+z)^3$ redshift dependence for decays (compared to $(1+z)^6$ for $s$-wave annihilation) means that the energy injection is less weighted toward earlier redshifts. This leads to a steady rise in $x_e$ from immediately before recombination to the present day. 

\begin{figure*}[t!]
    \centering
	\subfigure{
		\includegraphics[scale=0.64]{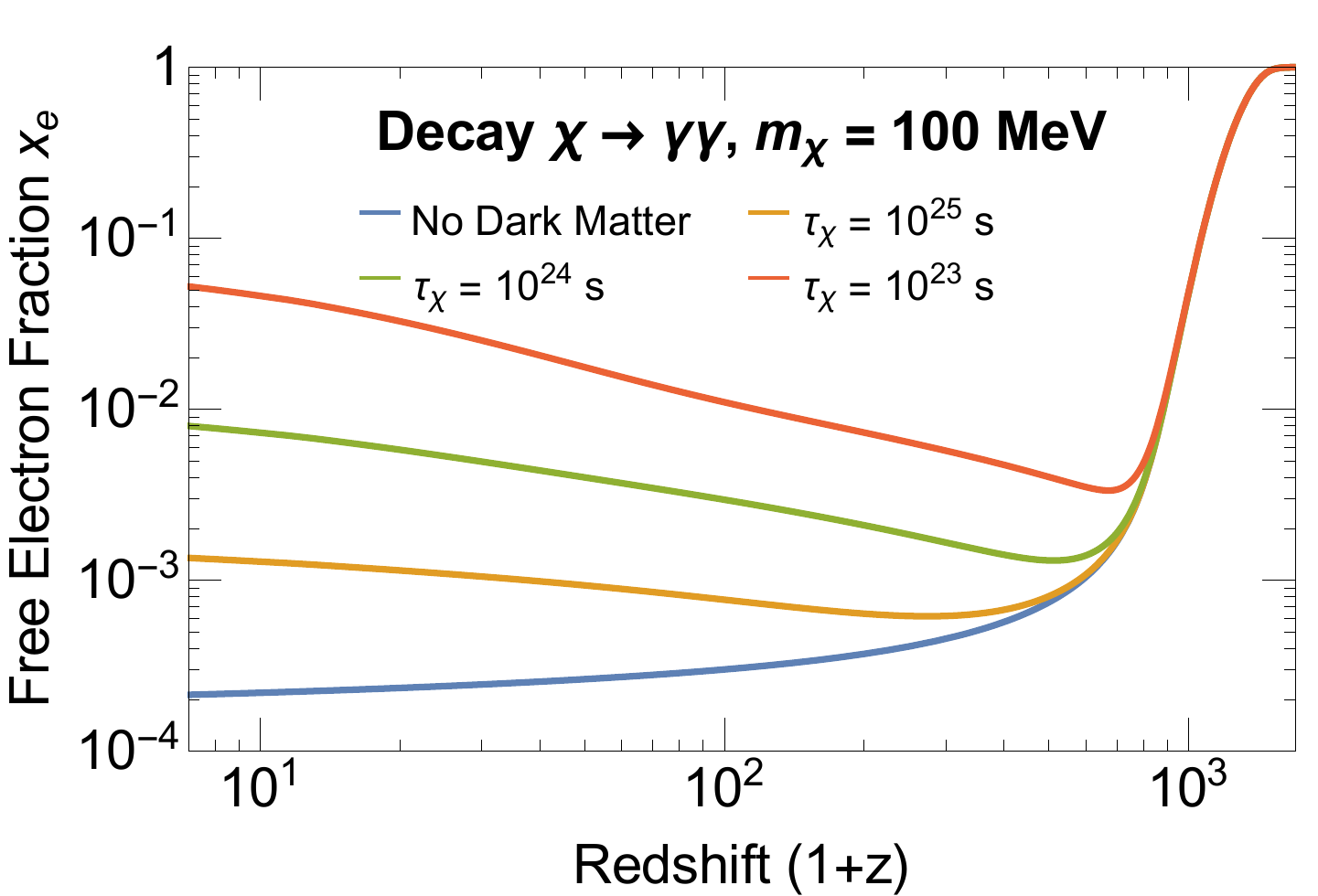}
	}
	\subfigure{
		\includegraphics[scale=0.64]{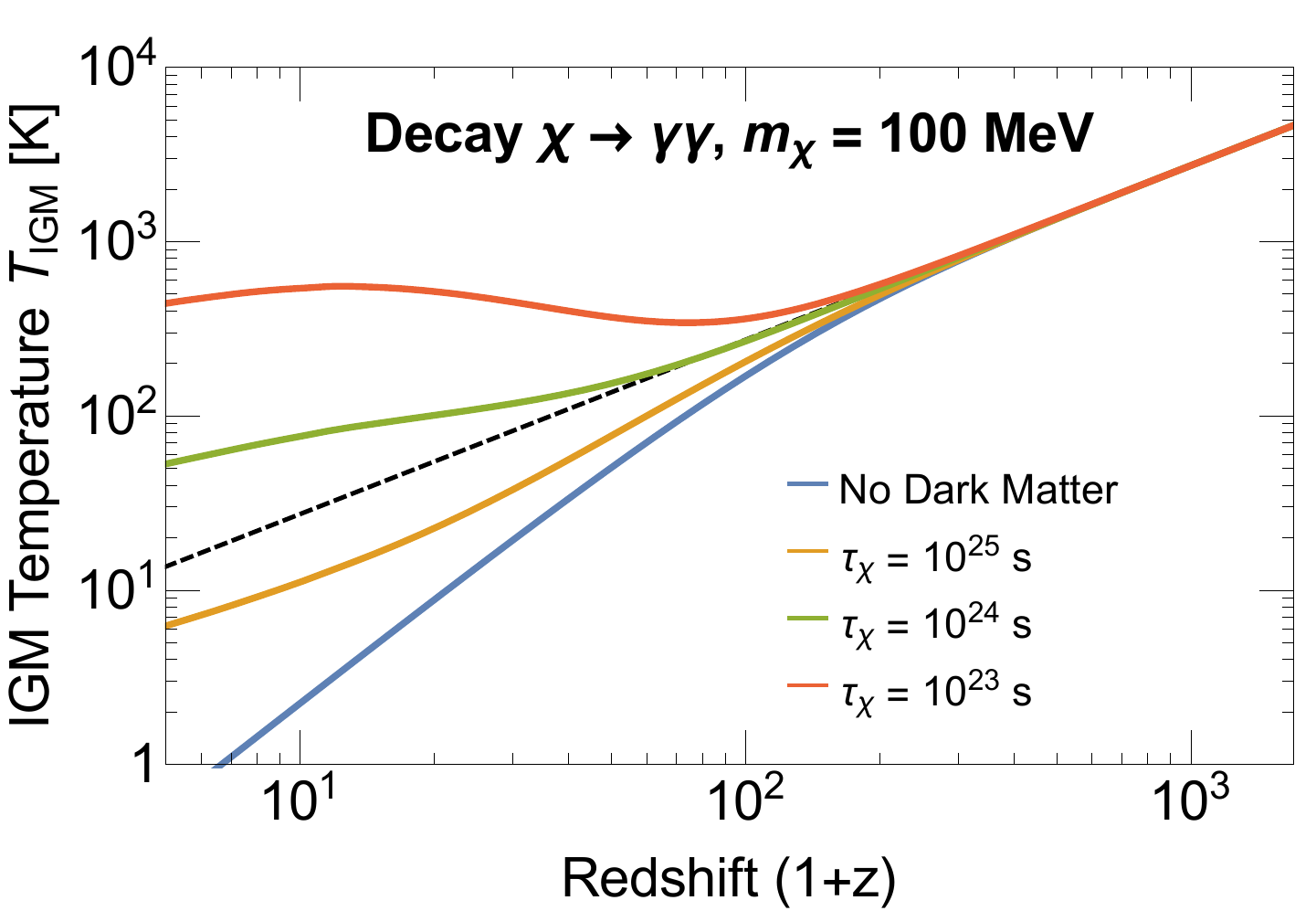}
	}
	\caption{Integrated free electron fraction $x_e$ and IGM temperature $T_m$ for $\chi \to \gamma \gamma$ decays ($m_\chi = \SI{100}{\mega \eV}$) with (from bottom to top): no DM, $\tau_\chi =  10^{25}$ s, $10^{24}$ s and $10^{23}$\SI{}{ s} respectively. The CMB temperature is shown as a dashed line for reference. No reionization is assumed. }
	\label{fig:freeEleFracDecay}
\end{figure*}

Optical depth constraints play an important role in placing bounds on the decay lifetime: with no structure formation boost, the only way for significant ionization at low redshifts to occur is for $x_e$ to be relatively high throughout the cosmic dark ages, contributing significantly to the optical depth. Figure~\ref{fig:xeConstraintsPlot_decay} shows the region of the ($\tau_\chi$,$m_\chi$) parameter space where DM can contribute significantly to reionization, as well as the constraints on the decay lifetime coming from IGM temperature and the optical depth. Significant reionization occurs for relatively longer decay lifetimes for masses where $f_{\text{H ion.}}(z)$ is large at low redshifts. However, both optical depth and IGM temperature constraints rule out large parts of the allowed parameter space for $\chi \to e^+e^-$ and all of the parameter space for $\chi \to \gamma \gamma$ at the 10\% level of contribution to reionization, with the $T_m$ bounds being more effective than optical depth for the $m_\chi \sim \SI{100}{MeV} - \SI{10}{GeV}$ range for $\chi \chi \to e^+e^-$. 

\begin{figure*}[t!]
    \centering
	\subfigure{
		\includegraphics[scale=0.63]{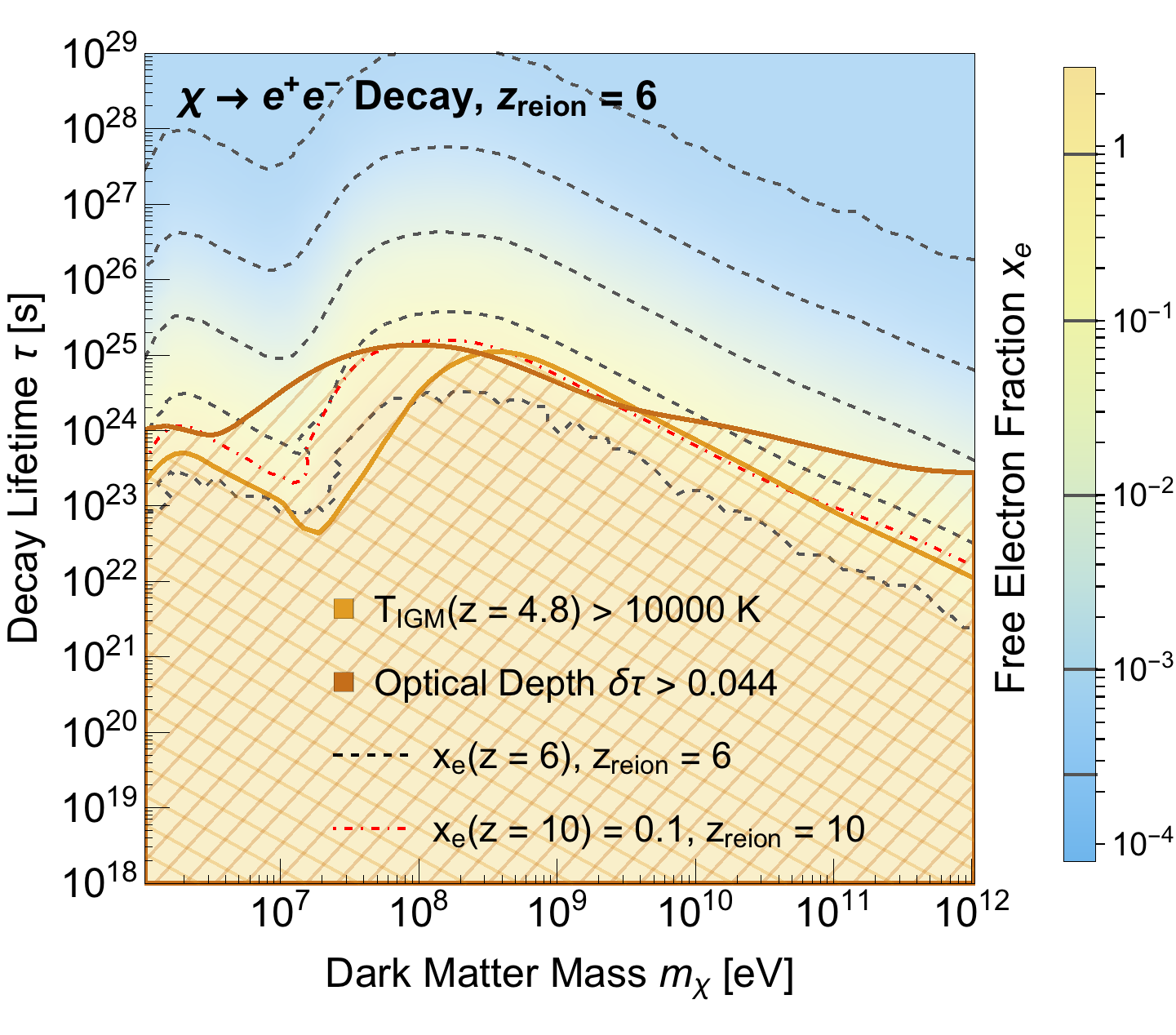}
	}
	\subfigure{
		\includegraphics[scale=0.63]{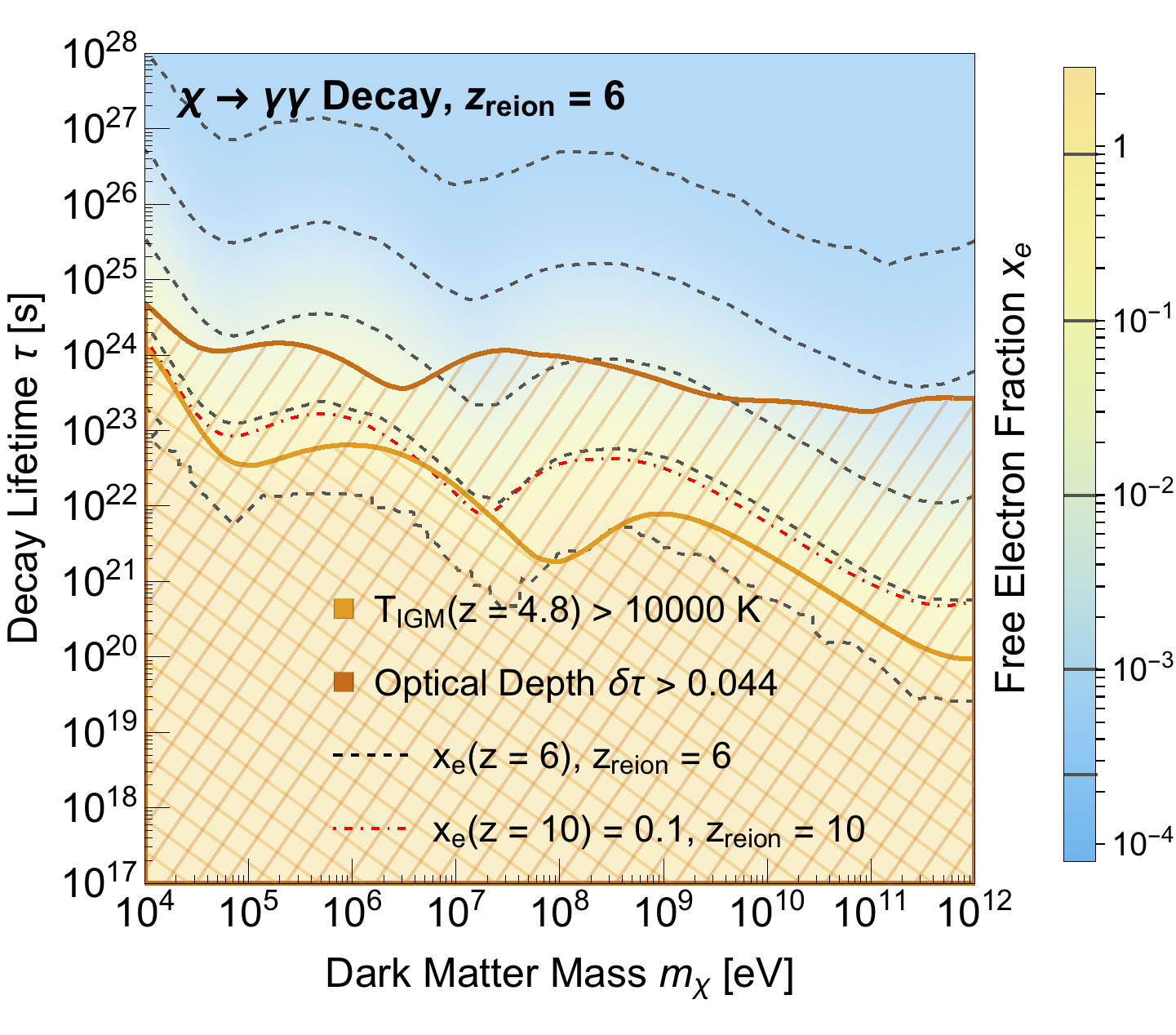}
	}
	\caption{DM contribution to reionization for $\chi \to e^+e^-$ (left) and $\chi \to \gamma \gamma$ (right) decays, benchmark scenario. The hatched regions correspond to parameter space ruled out by the optical depth (red) and the IGM temperature constraint $T_m(z=4.80) < \SI{10000}{K}$ (orange) respectively. The color density plot shows the DM contribution to $x_e$ just prior to reionization at $z = 6$, with contours (black, dashed) shown for a contribution to $x_e(z = 6) = $ 0.025\%, 0.1\%, 1\%, 10\% and 90\% respectively. We have also shown $x_e(z = 10) = 10\%$ when reionization occurs at $z = 10$ (red, dot-dashed contour). The optical depth limits are similar in both reionization scenarios, while the $T_m$ limits are similar between \SI{100}{MeV} and \SI{10}{GeV}, where they are more constraining than the optical depth limits.}
	\label{fig:xeConstraintsPlot_decay}
\end{figure*}

Figure~\ref{fig:xeConstraintsPlot_decay} also shows the same results after considering different reionization conditions. Once again, the optical depth constraints change very little with respect to reionization redshift, while the $T_m$ constraints are very similar in both reionization scenarios in the region where they are stronger than the optical depth, and we can hence simply compare the $x_e$ contributions with the $\delta \tau$ and $T_m$ constraints at $z_{\text{reion}} = 6$. As before, earlier reionization makes it more difficult for DM to contribute to $x_e$ just prior to reionization. For $\chi \to e^+e^-$, almost all decay lifetimes and masses which previously resulted in a 10\% contribution to reionization now result in a contribution below 10\% when the redshift of reionization is changed to $z = 10$, while the results for $z = 6$ and $z = 10$ for $\chi \to \gamma \gamma$ are similar. 

Nevertheless, a contribution to $x_e$ just prior to reionization at more than the 10\% level still remains possible for $\chi \to e^+e^-$ at a DM mass of $m_\chi \sim \SI{100}{MeV} - \SI{10}{GeV}$, $\tau_\chi \sim 10^{24} - 10^{25} \SI{}{s}$, as well as $m_\chi \sim \SI{1}{MeV}$, $\tau_\chi \sim 10^{24} \SI{}{s}$ in the benchmark reionization scenario. As with $p$-wave annihilation, the galactic diffuse background provides an additional constraint on the decay lifetime. These constraints are derived in a similar way to the $p$-wave case, i.e. by conservatively assuming that all of the diffuse gamma ray background comes from FSR from the DM decay. However, unlike with $p$-wave annihilation, the diffuse background constraints are of the same order as the optical depth bounds that we have set here. Figure~\ref{fig:xeConstraintsGalacticPlot_electron_decay} shows these constraints superimposed on Figure~\ref{fig:xeConstraintsPlot_decay}, showing that none of the experimental constraints are able to rule out the possibility of a more than 10\% contribution to $x_e$ prior to reionization in the $m_\chi \sim \SI{10}{} - \SI{100}{MeV}$, $\tau_\chi \sim 10^{25}\SI{}{s}$ and $m_\chi \sim \SI{1}{MeV}$, $\tau_\chi \sim 10^{24} \SI{}{s}$ regions of parameter space. This conclusion still holds true for a different redshift of reionization for $m_\chi \sim \SI{100}{MeV}$. 

\begin{figure}[t!]
	\centering
	\includegraphics[scale=0.63]{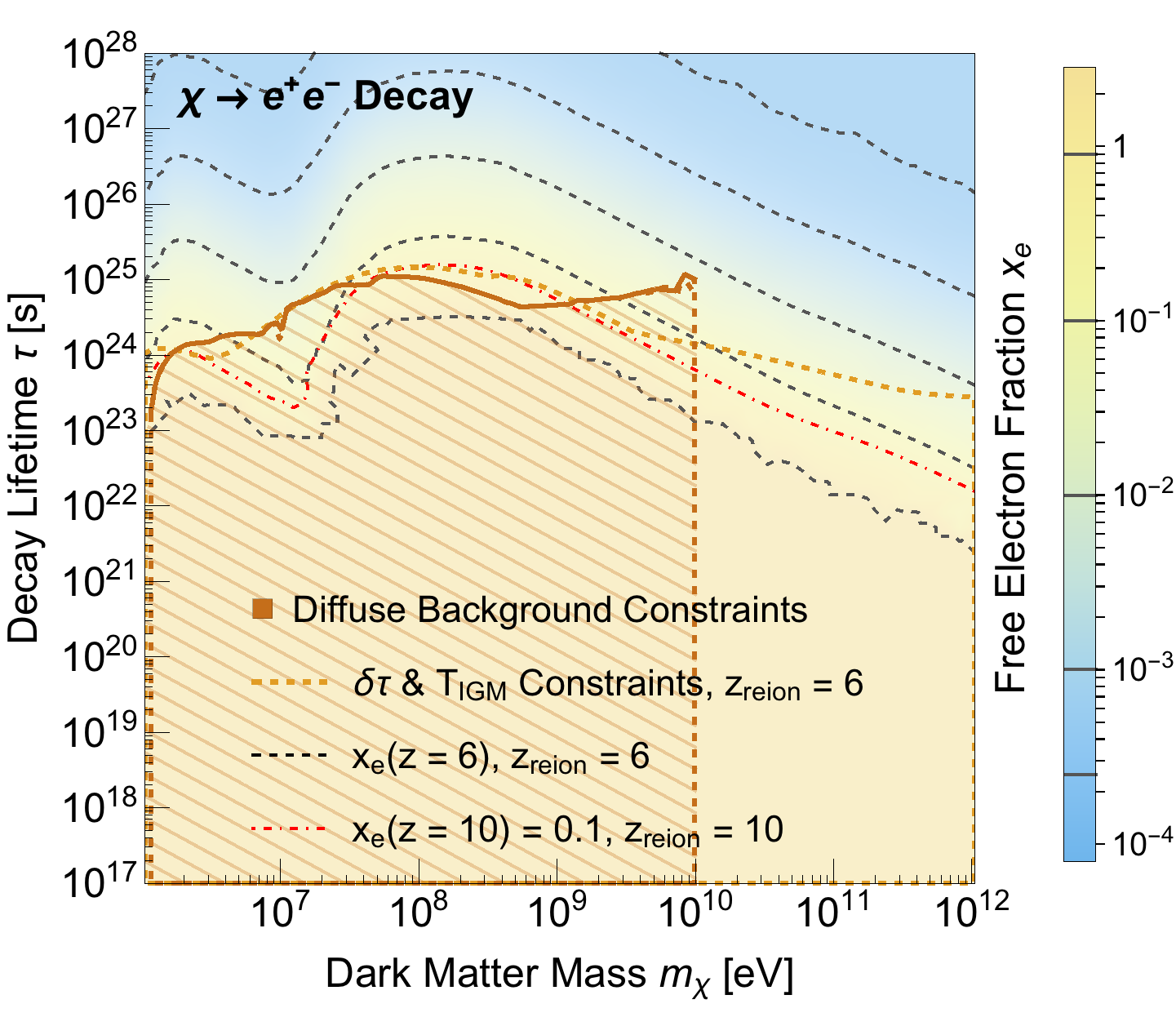}
	\caption{DM contribution to reionization for $\chi \to e^+e^-$ decays, benchmark scenario, including constraints from the galactic diffuse background (red contour, hatched) derived from~\cite{Essig2013}. The color density plot shows the DM contribution to $x_e$ just prior to reionization at $z = 6$, with contours (black, dashed) shown for a contribution to $x_e(z = 6) = $ 0.025\%, 0.1\%, 1\%, 10\% and 90\% respectively. We have also shown $x_e(z = 10) = 10\%$ when reionization occurs at $z = 10$ (red, dot-dashed contour) for comparison. The combined constraint from both optical depth $\delta \tau < 0.044$ and IGM temperature $T_m(z = 4.8) < \SI{10000}{K}$ (orange, dashed contour) is shown as well, with regions below this contour ruled out. These limits are almost identical in either reionization scenario.}
	\label{fig:xeConstraintsGalacticPlot_electron_decay}
\end{figure}

The blue curve in Figure~\ref{fig:freeEleFracDecayAllowedRegion} shows $x_e(z)$ and $T_m(z)$ assuming reionization at $z = 6$, with $m_\chi = \SI{100}{MeV}$ and $\tau_\chi = \SI{1.5e25}{s}$, parameters which lie in one of the allowed regions found above. Reionization at $z = 6$ causes the behavior of $T_m$ to change abruptly due to the instantaneous change of $x_e$. Just before reionization, $x_e(z = 6) \sim 0.2$, with the integrated optical depth being $\delta \tau = 0.040$, which lies within the allowed limit. $T_m(z = 4.8)$ lies below the lower limit of the $T_m$ constraint, but as we have previously explained, $T_m$ is always underestimated with the default ionization history. 

We have also performed the integration of $x_e(z)$ and $T_m(z)$ with $f_c(z)$ derived from the ionization history that we obtained above. Since $f_c(z)$ as calculated from the default ionization history overestimates $x_e(z)$, using this new $f_c(z)$ ensures that the allowed regions are not ruled out by a more accurate estimate of $x_e(z)$. The result is also shown in orange in Figure~\ref{fig:freeEleFracDecayAllowedRegion}. As we expect, this more accurate $f_c(z)$ increases $T_m(z)$ and decreases $x_e(z)$ slightly. The contribution to reionization remains the same, while still staying consistent with the $T_m(z = 4.8)$ and the optical depth bounds. 

Figure~\ref{fig:freeEleFracDecayAllowedRegion} also shows two measurements of $x_e$ from just before reionization obtained by~\cite{Schenker2014}, corresponding to
\begin{alignat}{1}
	x_e(z=7) &= 0.66^{+0.12}_{-0.09}, \nonumber \\
	x_e(z=8) &< 0.35.
\label{eqn:schenkerxe}
\end{alignat}
The ionization history for $m_\chi = \SI{100}{MeV}$ and $\tau_\chi = \SI{1.5e25}{s}$ is consistent with the bound from $z=8$, and can be made consistent with the $z=7$ bound with the addition of other sources of ionization between these two redshifts.  

\begin{figure*}[t!]
    \centering
	\subfigure{
		\includegraphics[scale=0.64]{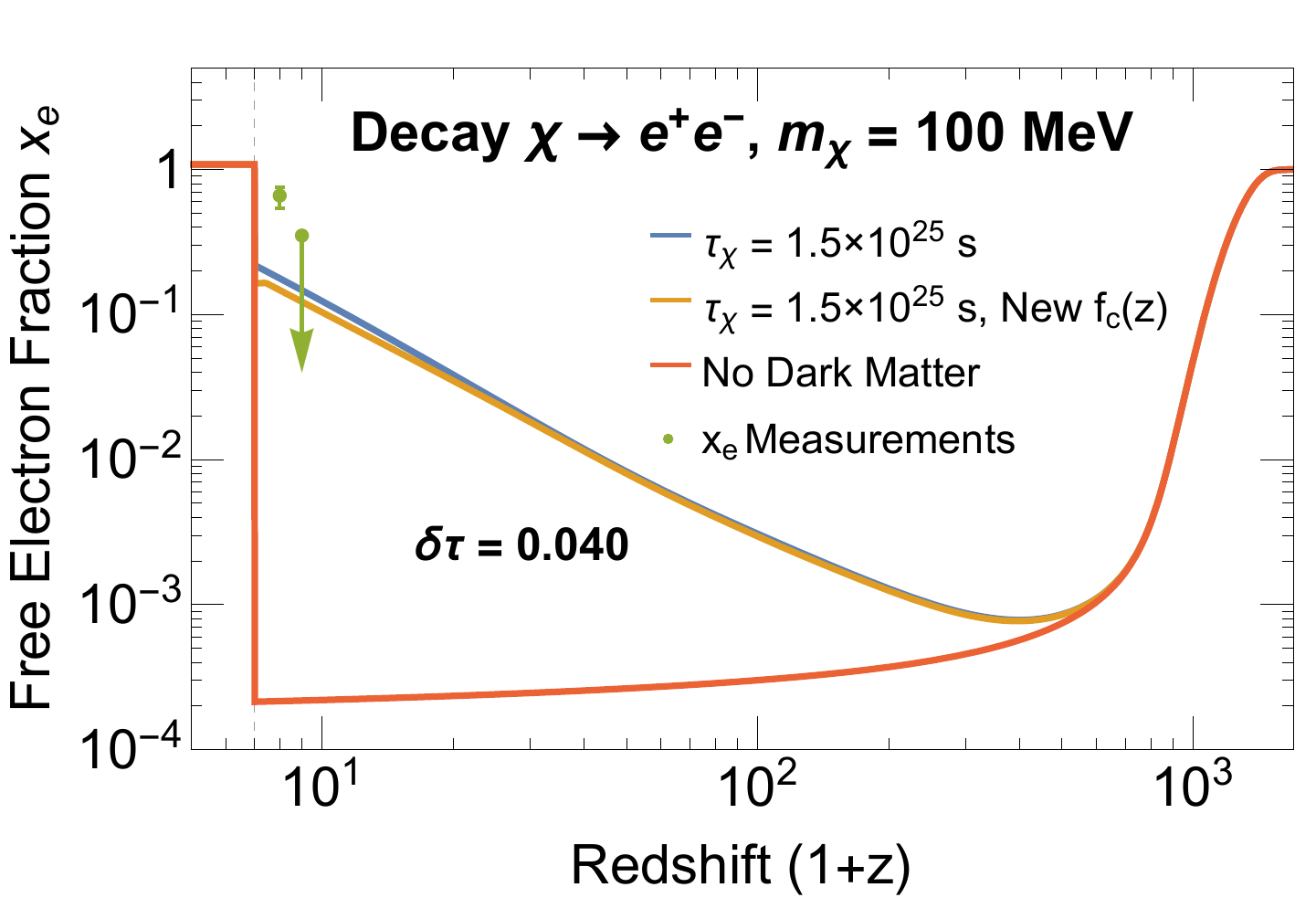}
	}
	\subfigure{
		\includegraphics[scale=0.64]{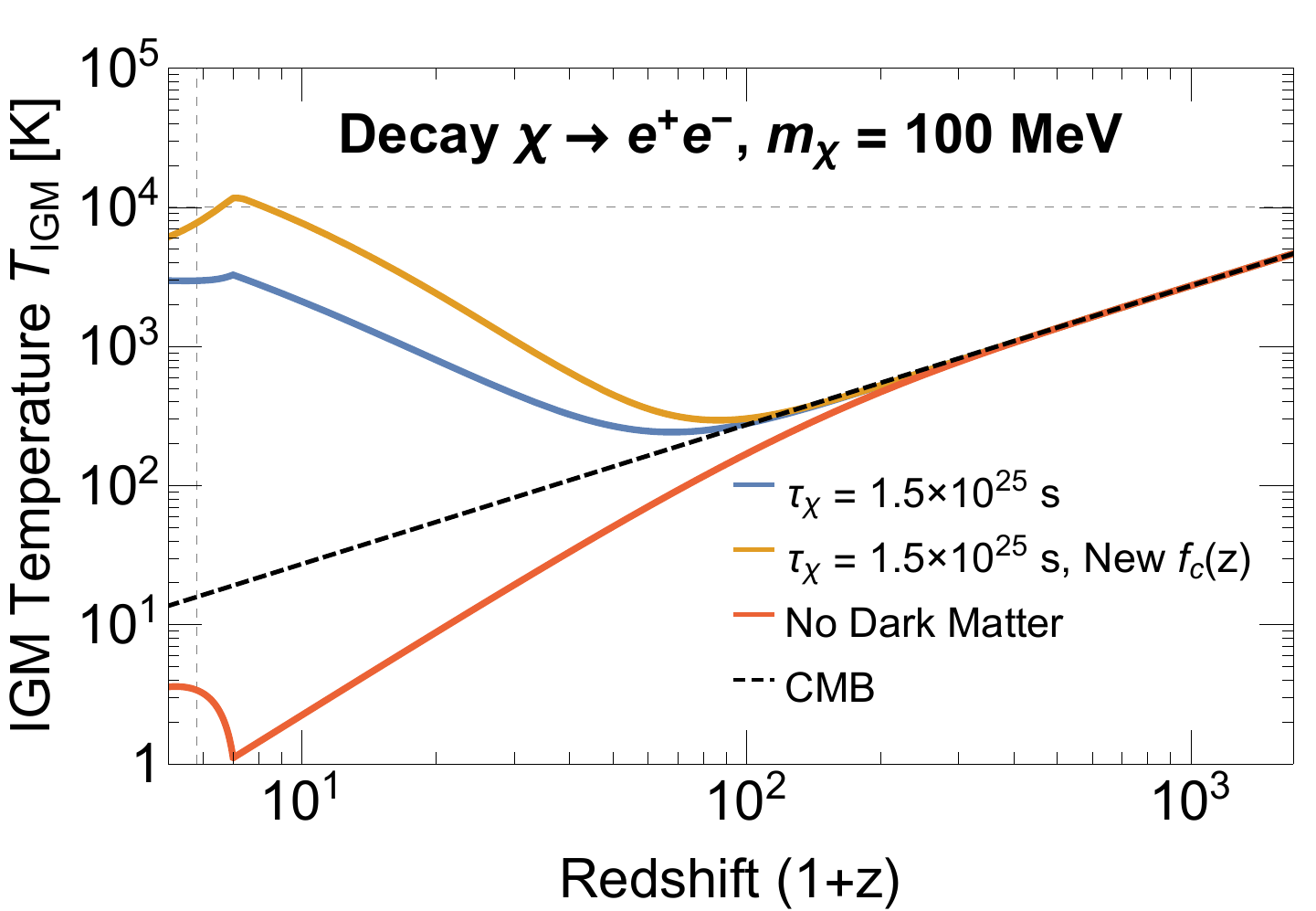}
	}
	\caption{Integrated free electron fraction $x_e$ and IGM temperature $T_m$ for $\chi \to e^+e^-$ decays ($m_\chi = \SI{100}{\mega \eV}$) with: (red) no DM; (blue) $\tau_\chi = \SI{1.5e25}{s}$ with the default $f_c(z)$; (orange) $  \SI{1.5e25}{s}$ with $f_c(z)$ computed using $x_e(z)$ obtained from the default $f_c(z)$ shown in blue. The green points and error bars show the observational limits for $x_e$ near reionization~\cite{Schenker2014}. The CMB temperature (bold, dashed line) and $T_m(z = 4.8) = \SI{10000}{K}$ (dashed line) are shown for reference. Reionization at $z = 6$ is assumed.}
	\label{fig:freeEleFracDecayAllowedRegion}
\end{figure*}

In summary, optical depth constraints as well as bounds from the galactic diffuse background constraints rule out reionization from $\chi \to \gamma \gamma$ and almost rules out reionization from $\chi \to e^+e^-$ at the 10\% level, except for $m_\chi \sim \SI{10}{} - \SI{100}{MeV}$, $\tau_\chi \sim 10^{25} \SI{}{s}$ and $m_\chi \sim \SI{1}{MeV}$, $\tau_\chi \sim 10^{24} \SI{}{s}$. The former region remains viable even under the different reionization scenarios considered here. 

\section{Conclusion}
\label{sec:Conclusion}

We have studied the potential impact of $s$-wave annihilation, $p$-wave annihilation and decay of DM to $e^+e^-$ and $\gamma \gamma$ on the process of reionization. Using the latest calculations for the fraction of the energy deposition rate in channel $c$ to the energy injection rate at redshift $z$, $f_c(z)$, we have determined the free electron fraction $x_e$ and IGM temperature $T_m$ as a function of redshift. We have extended the $f_c(z)$ calculation from $1+z = 10$ down to $1+z = 4$ by assuming three different reionization scenarios and determining the total amount of energy deposited as ionization of HeII, IGM heating and continuum photons once reionization occurs. 

We have also considered multiple detailed structure formation models in order to accurately calculate the $s$-wave and $p$-wave annihilation rates. This modeling accounts for the formation of DM haloes and their subhaloes, with abundance and internal properties that are consistent with current cosmological simulations. It also considers the uncertainties at the smallest scales (corresponding to low-mass haloes, $<10^8$~M$_\odot$, devoid of gas and stars) that cannot be resolved in current simulations in a full cosmological setting, but that are very relevant in predicting the annihilation rate in the case of $s$-wave self-annihilation. This is particularly important at low redshifts: at $z\sim10$, the uncertainty in $\rho_{\rm eff}^2$ is $\sim5$ for the case of $s$-wave self-annihilation (see Figure~\ref{fig_rho_eff}). On the other hand, for $p$-wave self-annihilation, the uncertainties in the unresolved regime are irrelevant since the signal is dominated by massive haloes (see Figure~\ref{fig_rho_eff_pwave}).

The integrated free electron fraction $x_e(z)$ and IGM mean temperature $T_m(z)$ were both computed using a pair of coupled differential equations derived from a three-level atom model, modified to include energy injection from DM. This simplified model agrees well with \texttt{RECFAST}, and enables us to compute these two quantities and set constraints across a large range of annihilation cross sections/decay lifetimes and DM masses $m_\chi$. For each process, we obtained constraints for different assumptions on the redshift of reionization, structure formation prescriptions as well as $T_m$ constraints to check the robustness of the constraints. 

For $s$-wave annihilation, constraints from measurements on the CMB power spectrum and on the integrated optical depth $\tau$ rule out any possibility of DM contributing significantly to reionization, with the CMB power spectrum constraints on $\langle \sigma v \rangle$ being approximately an order of magnitude stronger at a given $m_\chi$. The maximum allowed value of $\langle \sigma v \rangle$ can at most contribute to 2\% of $x_e$ at reionization for $\chi \chi \to e^+e^-$, and less than 0.1\% for $\chi \chi \to \gamma \gamma$. These results are largely independent of reionization redshift and structure formation prescription. 

In the case of $p$-wave annihilation, the velocity suppression at early times greatly relaxes the CMB constraints compared to $s$-wave annihilation, since the former are mainly dependent on the cross section immediately after recombination. However, the sudden increase in energy deposition once structure formation becomes important leads to a sharp rise in $T_\text{IGM}$, making astrophysical measurements of $T_m$ at redshifts $z \sim 4$ to 6 important. The most optimistic assumptions appear to allow for significant contributions to reionization, but much of the allowed parameter space is ruled out with the stricter $T_m$ constraint and earlier reionization. The sole exception to this is in the channel $\chi \chi \to e^+e^-$ with $m_\chi$ between \SI{1}{MeV} and \SI{100}{MeV}, but this region is in turn ruled out by constraints from the photon flux from the galactic diffuse background emission. Overall, we find that only a $\sim 0.1\%$ contribution to $x_e$ at reionization is permitted for $p$-wave annihilation dominantly to $e^+ e^-$ pairs; for annihilation dominantly to photons, a $\sim 5\%$ contribution is possible.

Finally, for DM decay, optical depth constraints rule out any large contribution from decays into $\gamma \gamma$, with the strongest bounds occurring for heavier DM (a contribution to $x_e$ at the $\sim 10\%$ level is viable for the lightest DM we consider, around 10 keV). Contributions at the 20-40\% level from decays into $e^+e^-$ are possible for $m_\chi \sim \SI{10}{} - \SI{100}{MeV}$, $\tau_\chi \sim 10^{25} \SI{}{s}$ and $m_\chi \sim \SI{1}{MeV}$, $\tau_\chi \sim 10^{24} \SI{}{s}$, with this result being independent of our assumptions on the redshift of reionization.

Overall, we find that DM is mostly unable to contribute more than 10\% of the free electron fraction after reionization across most of the DM processes and annihilation or decay products considered in this chapter, even after allowing for different structure formation prescriptions, reionization scenarios and choice of constraint. The one exception to this is found in $\chi \chi \to e^+e^-$, with a possible contribution of up to 40\% near $m_\chi = \SI{100}{MeV}$. Figure~\ref{fig:xeMaxConstraints} summarizes the maximum $x_e$ achievable prior to reionization that is consistent with all of the constraints considered in this chapter.

\begin{figure*}[t!]
    \centering
	\subfigure{
		\includegraphics[scale=0.64]{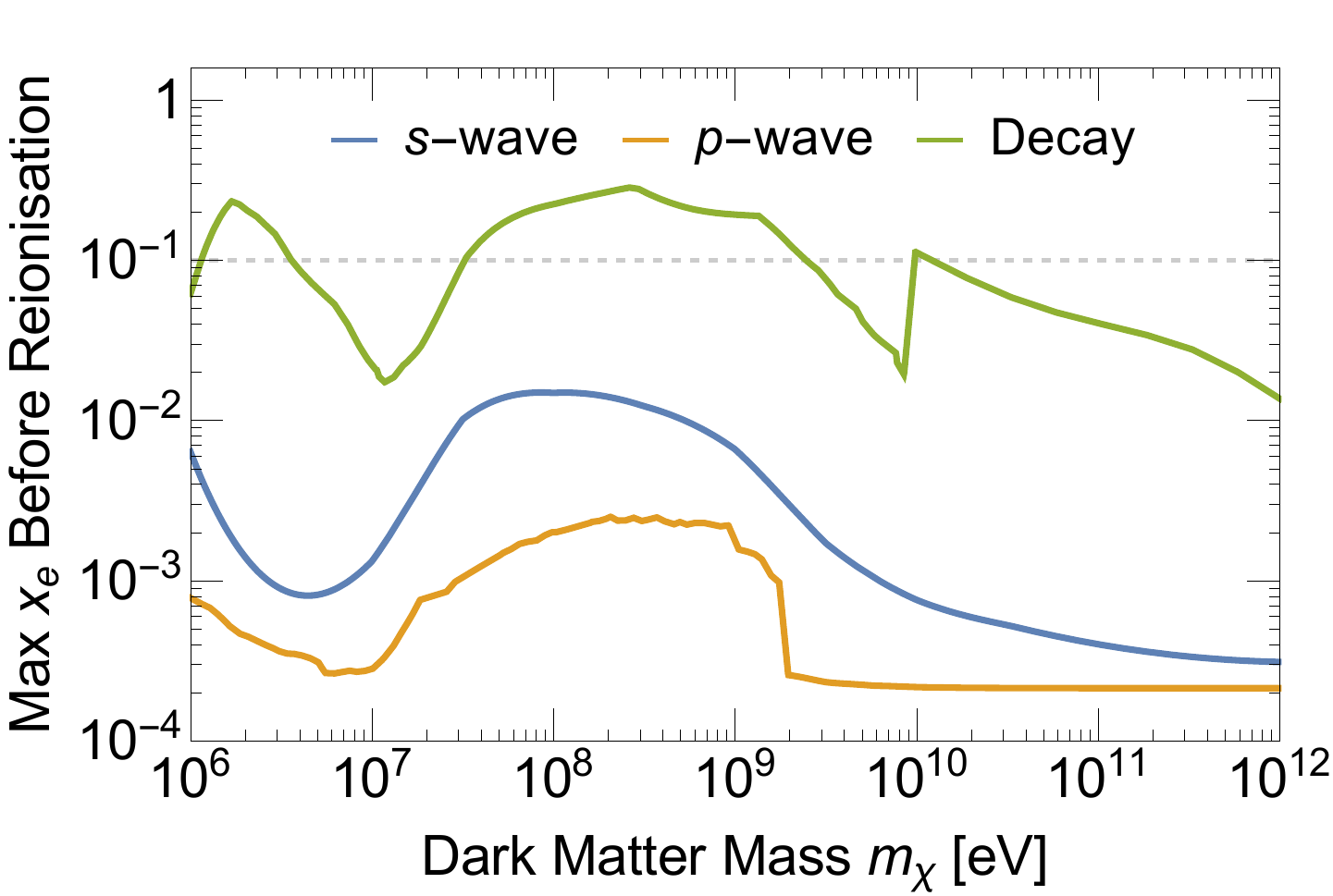}
	}
	\subfigure{
		\includegraphics[scale=0.64]{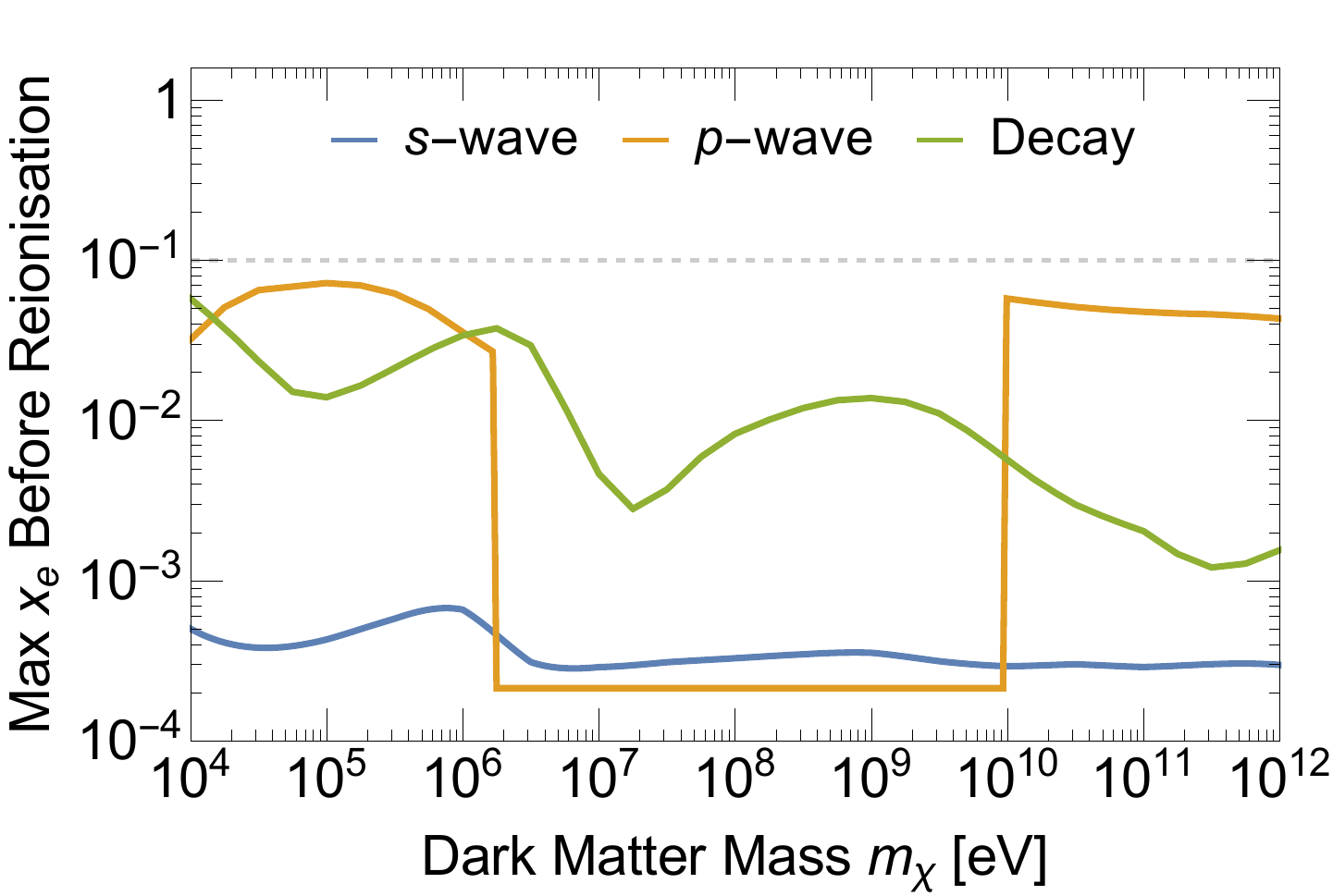}
	}
	\caption{The maximum free electron fraction $x_e$ just prior to reionization consistent with all constraints used in this chapter for $s$-wave annihilations (blue), $p$-wave annihilations (yellow) and decays (green) into $e^+e^-$ (left) and $\gamma \gamma$ (right). }
	\label{fig:xeMaxConstraints}
\end{figure*}

With potential input from 21 cm tomography and improved measurements of the IGM at large redshift and the CMB, we expect our understanding of the process of reionization and the end of the cosmic dark ages to improve dramatically in the near future. These future results may be sensitive to a contribution to reionization by DM at well below the 10\% level, and may serve as a good probe of the properties of DM.\footnote{See~\cite{Lopez-Honorez2016} for recent work in understanding the impact of DM annihilations on the 21 cm signal, using methods that are similar to those used here.} The continued relevance of DM to reionization and vice-versa serves as strong motivation to improve on the results developed here. Future work may include new ways to calculate $f_c(z)$ at $1+z \leq 10$ with greater accuracy by taking into account the ionization and thermal history of the universe near reionization, as well as understanding the potential impact of DM annihilation products on the haloes in which they are generated, building on results from~\cite{Schon2014}.

\chapter{
Implications of a 21-cm Signal for Dark Matter Annihilation and Decay
}
\label{chap:21cm_annihilation_decay}

\section{Introduction}
\label{sec:Introduction}

Between thermal decoupling of baryons from the CMB and star formation at $z \sim 20$, the universe can function as a sensitive calorimeter for exotic sources of energy injection. In the conventional $\Lambda$CDM model, the matter temperature $T_m$ and hydrogen ionization fraction as a function of redshift are simple and well-understood \cite{AliHaimoud:2010dx,Chluba:2010ca}. After recombination, Compton scattering between the residual free electrons and cosmic microwave background (CMB) photons keeps $T_m$ at the CMB temperature $T_\text{CMB}$ down to $z \sim 150$. Subsequently, the energy transfer rate becomes too small to prevent thermal decoupling, and soon after, $T_m$ is determined solely by adiabatic expansion, evolving as $T_m(z) \propto (1+z)^2$. Deviations in temperature from this well-understood standard history are thus a strong indication of new sources of heating or cooling in the universe. 

The recent measurement of an absorption profile at 78 MHz in the sky-averaged spectrum by the Experiment to Detect the Global Epoch of reionization Signature (EDGES) Collaboration \cite{Bowman:2018yin} opens a new window into the cosmic dark ages, shedding new light on the ionization and thermal history at precisely this period of interest. Radiation with a frequency near the hyperfine transition of hydrogen illuminates neutral hydrogen clouds during this epoch, and gets absorbed as they redshift into the transition frequency. The brightness temperature of the 21-cm hydrogen absorption line relative to the background radiation is given by \cite{Zaldarriaga:2003du}
\begin{alignat}{1}
    T_{21}(z) \approx x_\text{HI}(z) \left(\frac{0.15}{\Omega_m}\right)^{1/2} \left(\frac{\Omega_b h}{0.02}\right) \left(\frac{1+z}{10}\right)^{1/2} \left[1 - \frac{T_R(z)}{T_S(z)}\right] \SI{23}{\milli\kelvin},
    \label{eqn:T_21}
\end{alignat}
where $x_\text{HI}$ is the neutral hydrogen fraction, $\Omega_m$ and $\Omega_b$ are the matter and baryon energy density as a fraction of the critical density, $h$ is the Hubble parameter today in units of \SI{100}{\kilo\meter\per\second\per\mega\parsec}, and $T_R(z)$ is the effective temperature of the background 21-cm radiation at redshift $z$. $T_S(z)$, the spin temperature, determines the ratio of neutral hydrogen in the higher-energy spin-triplet state to the lower-energy spin-singlet state. 

The expected value of $T_S$ as a function of redshift has been studied extensively (see e.g. \cite{Furlanetto:2006jb} for a review). At $z \sim 30$, we expect $T_S = T_R$, with the radiation temperature commonly assumed to be $T_\text{CMB}$. Once the first stars start forming at $z \sim 20$ and begin to emit UV radiation, downward transitions from the spin-triplet to the spin-singlet state through the Wouthuysen-Field effect \cite{1952AJ.....57R..31W,1959ApJ...129..536F,1959ApJ...129..551F} start to occur, driving the spin temperature toward $T_m$. The combination of the background 21-cm radiation, UV radiation from stars and collisional hyperfine excitation/de-excitation ensures that well before reionization,
\begin{alignat}{1}
    T_m \lesssim T_S \lesssim T_R.
    \label{eqn:spin_temperature_bound}
\end{alignat}
A measurement of a negative $T_{21}(z)$ at this time indicates that $T_S$ lies below $T_R$, and also sets an upper bound on $T_m$ if $T_R$ is known.

The EDGES collaboration measured a strong 21-cm absorption trough in the redshift range $14 < z < 20$, reporting a value of $T_{21}$ at $z \sim 17.2$ of  $T_{21} = -500^{+200}_{-500}\text{ mK}$ \cite{Barkana:2018lgd}, with 99\% confidence limits specified. This result, together with Eq.~(\ref{eqn:spin_temperature_bound}), sets the following constraint on the matter and radiation temperature at $z = 17.2$ at the 99\% confidence level:
\begin{alignat}{1}
    \frac{T_m}{T_R}(z = 17.2) \lesssim 0.105.
    \label{eqn:T_m_T_R_ratio}
\end{alignat}
Precise calculations of the temperature evolution after recombination assuming the $\Lambda$CDM model \cite{AliHaimoud:2010dx,Chluba:2010ca} give $T_m(z = 17.2) \sim \SI{7}{\kelvin}$; however, assuming $T_R = T_\text{CMB}$ in Eq.~(\ref{eqn:T_m_T_R_ratio}), we obtain $T_m \lesssim \SI{5.2}{\kelvin}$, which lies well below the expected value.

Since the publication of the EDGES result, this discrepancy has been explained by either a colder-than-expected gas temperature or an additional source of 21-cm photons at $z \sim 20$. In both cases, the effect is to reduce the expected value of the ratio $T_m/T_R$. Models with interactions between baryons and cold dark matter (DM) with a Rutherford-like cross section have been explored~\cite{Barkana:2018lgd} as a mechanism to cool the gas, particularly in the context of millicharged DM models~\cite{Munoz:2018pzp,Berlin:2018sjs,Fraser:2018acy,Barkana:2018qrx}. These models have been shown to be highly constrained, with millicharged DM likely to only make up a subdominant component of DM. Modifications to the redshift of thermal decoupling of baryons from the CMB can also result in a cooler-than-expected gas temperature. Such a scenario can occur due to an imbalance between the proton and electron number densities~\cite{Falkowski:2018qdj} or early dark energy~\cite{Hill:2018lfx} (although the latter scenario appears difficult to reconcile with other observations). The possibility that interacting dark energy or other effects could modify the evolution of the Hubble parameter and change the 21-cm brightness temperature was proposed in~\cite{Costa:2018aoy}, but the change to the Hubble parameter required at $z \lesssim 20$ is large. Finally, models which inject additional 21-cm radiation through light DM decays~\cite{Fraser:2018acy,Pospelov:2018kdh} or radio emission from black holes~\cite{Gong:2018sos,Ewall-Wice:2018bzf} have been studied as a means of raising $T_R$.

In any model of DM, the annihilation and decay rates into Standard Model (SM) particles are important quantities to understand. Models with DM-baryon scattering are likely to imply the existence of DM annihilation to SM particles by crossing symmetry, and these annihilation processes could potentially set the relic abundance of DM via thermal freezeout at early times. Even if DM-baryon scattering does not occur or is not strong enough to markedly affect the matter temperature, new constraints on annihilation and decay can be set using the information on the thermal history provided by 21-cm measurements of this epoch.

Previous studies \cite{Poulin:2016anj,Furlanetto:2006wp,Lopez-Honorez:2016sur,Valdes:2007cu,Evoli:2014pva} have explored such constraints under the assumption that there are no other modifications to the conventional thermal history. However, any attempt to explain the EDGES result mandates the presence of additional effects, and such modifications could also be present even if the EDGES result is not confirmed.

In this chapter, we will study the implications that a confirmed 21-cm absorption measurement from $z \sim 20$ would have for DM annihilation and decay, in conjunction with three general mechanisms that could deepen an absorption signal: (i) non-standard recombination histories; (ii) baryon-DM scattering; and (iii) an additional source of 21-cm photons at $z \sim 20$. We will use the EDGES result as a benchmark; if it is confirmed, the forecast limits in this work can be applied as constraints on the DM parameter space.

Throughout this chapter, all algebraic expressions will be written in natural units with $\hbar = c = k_B = 1$, and we adopt cosmological parameters that are equal to the Planck 2015 TT,TE,EE+lowP central values \cite{Ade:2015xua}.

\section{Ionization and Thermal History}
\label{sec:IGMHistory}


\begin{figure}
    \centering
    \subfigure{
        \label{fig:example_temp_histories}
        \includegraphics[scale=0.35]{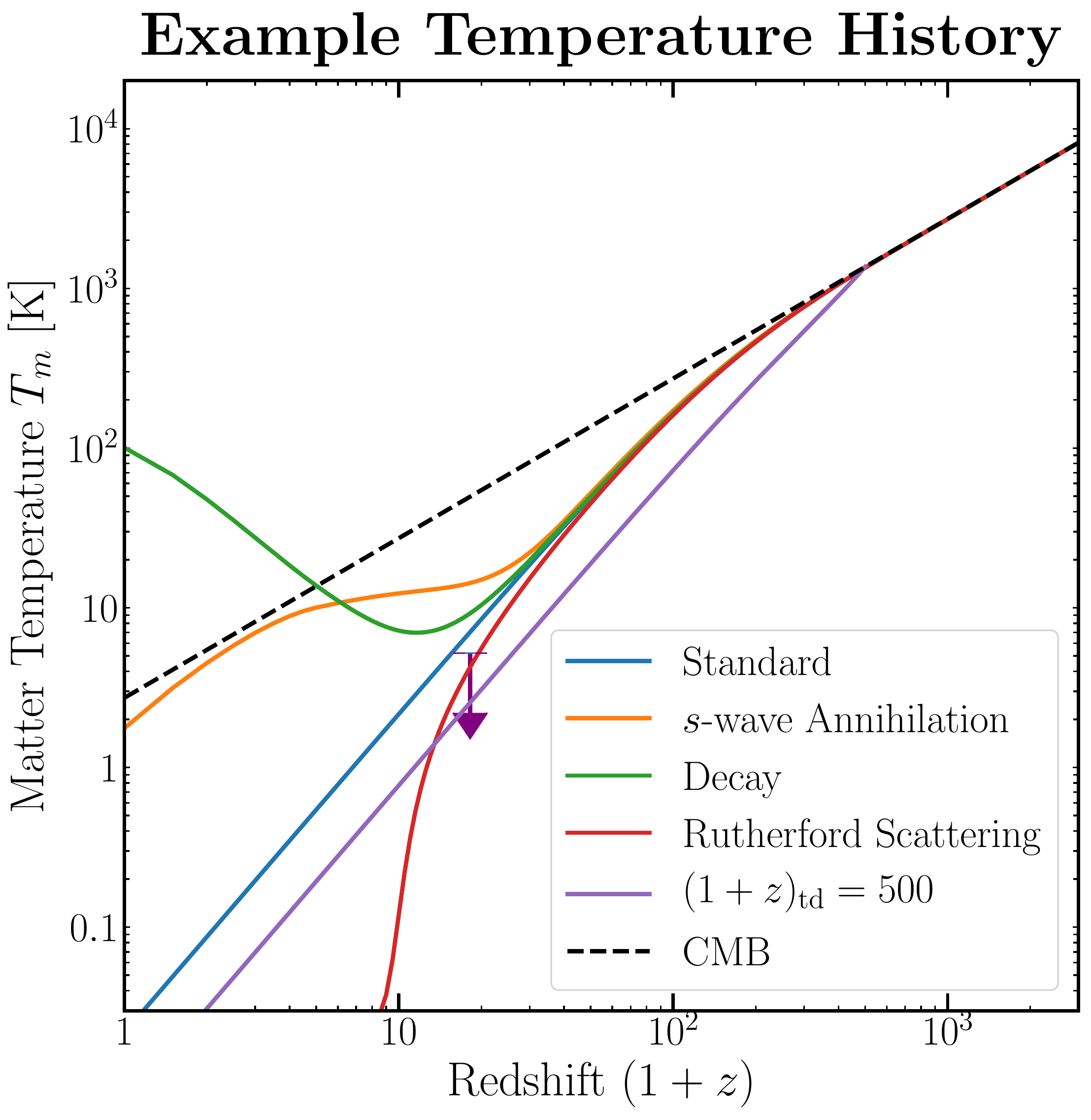}
    }
    \subfigure{
        \label{fig:example_ion_histories}
        \includegraphics[scale=0.35]{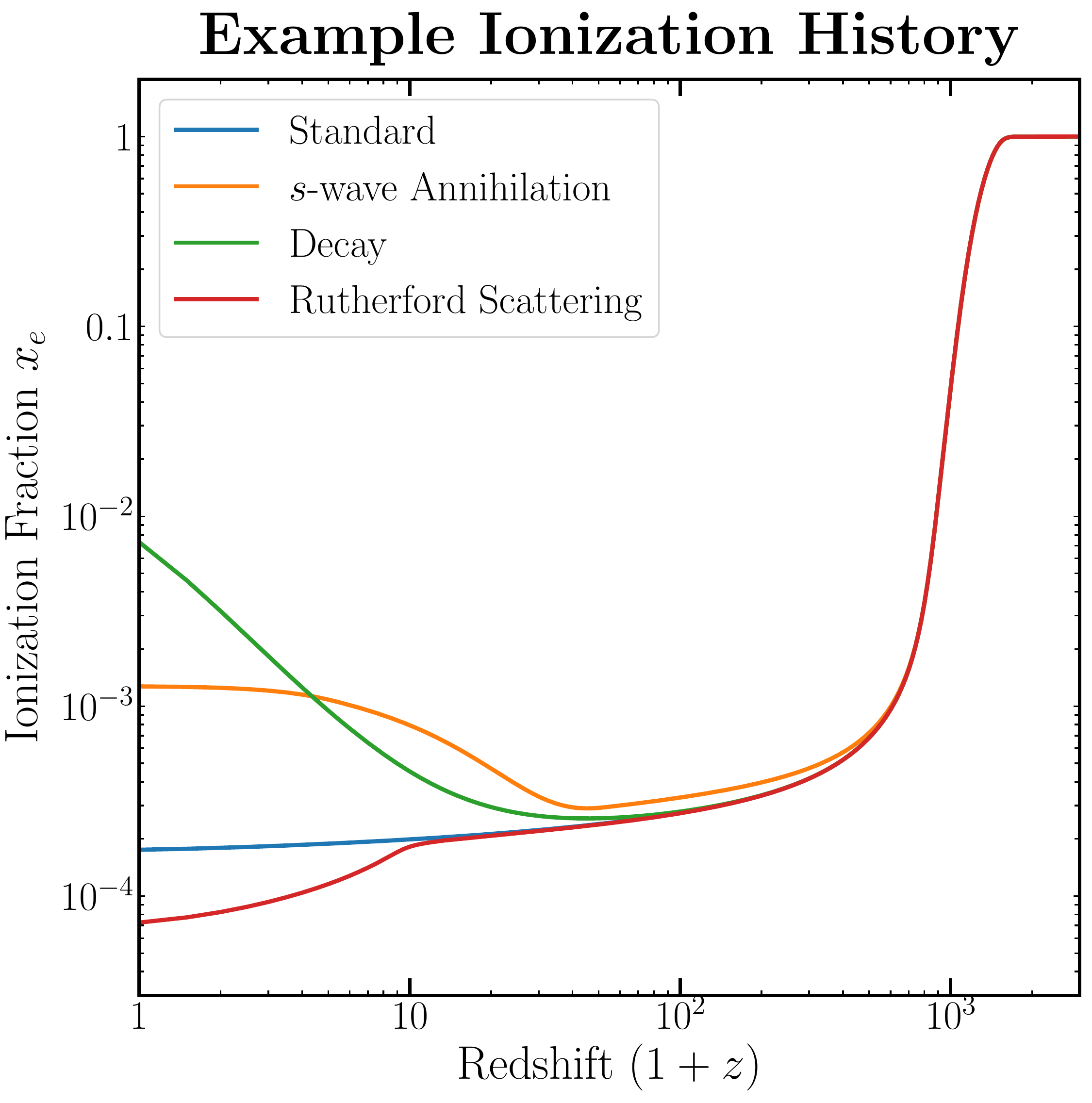}
    }
  \caption{Example thermal (left) and ionization (right) histories, for $m_\chi$ = 100 MeV. The standard history (blue), with DM $s$-wave annihilation ($\langle \sigma v \rangle = 10^{-29} \text{cm}^3 \text{ s}^{-1}$) (orange), with DM decays ($\tau = 10^{28}$ s) (green) and with DM-baryon Rutherford scattering (red) are shown. The thermal history with an earlier redshift of thermal decoupling at $(1+z)_\text{td} = 500$ (light purple) is also shown, with the CMB temperature (black, dashed) plotted for reference. The purple arrow indicates the EDGES temperature limit. 
  }
  \label{fig:example_histories}
\end{figure}

As we discussed in Chapter~\ref{chap:intro}, we integrate Eq.~(\ref{eqn:TLA_with_injection}) to obtain the thermal and ionization histories in the presence of dark matter energy injection processes. The injection rates are given by
\begin{alignat}{1}
    \left(\frac{dE}{dV \, dt}\right)^\text{inj} = \begin{cases}
        f_{\chi,\text{ann}}^2 \rho_{\chi,0}^2 (1+z)^6 \frac{\langle \sigma v \rangle}{m_\chi}, & \text{annihilation}, \\
        f_{\chi,\text{dec}} \rho_{\chi,0}(1+z)^3 \frac{1}{\tau}, & \text{decay},
    \end{cases}
    \label{eqn:energy_injection_21cm}
\end{alignat}
where we have introduced $f_{\chi,\text{ann}}$ and $f_{\chi,\text{dec}}$, the fraction of dark matter by mass density that annihilates and decays respectively.

Fig.~\ref{fig:example_histories} shows the baseline history, as well as two examples with DM $s$-wave annihilation and decay. The EDGES upper limit on the matter temperature if we take $T_R = T_\text{CMB}$ in Eq.~(\ref{eqn:T_m_T_R_ratio}) is also indicated. We have also included two of the new interactions that we will examine later: Rutherford-like interactions between the dark sector and hydrogen, as well as a temperature history with early decoupling of the photon and baryon temperatures. Throughout the chapter, no star-formation or reionization models are included in this analysis: excluding these effects, which would only raise the matter temperature near $z \sim 20$, leads to annihilation cross section or decay lifetime limits that are less constraining and thus conservative. The impact of $s$-wave annihilation becomes significantly enhanced beginning at $z \sim 40$ due to structure formation, which greatly increases the local DM density. We discuss the systematics associated with structure formation in Appendix~\ref{app:systematics}, and refer readers to Chapter~\ref{chap:DarkHistory} for a proper treatment of reionization in such calculations.

The authors of~\cite{Venumadhav:2018uwn} have recently pointed out that Lyman-$\alpha$ radiation at $z\sim 20$ is able to mediate a transfer of energy from the 21-cm CMB photons to the thermal motion of the gas, providing an additional and significant source of heating during this epoch. Although the inclusion of this effect would ultimately be important in setting precise DM annihilation and decay constraints, we neglect this effect in this chapter, and leave a proper treatment of this process to future work. This is consistent with our omission of the process of reionization, and leads to limits that are less constraining than they would be in a more complete treatment.

\section{Additional 21-cm Sources}
\label{sec:additional_sources}

A large absorption trough can be explained by the existence of an additional 21-cm source, which would raise $T_R$, the effective radiation temperature at a wavelength of 21 cm  at $z \sim 17.2$, above the CMB temperature. If $T_R$ is large enough so that Eq.~(\ref{eqn:T_m_T_R_ratio}) is satisfied with $T_m \gtrsim \SI{7}{\kelvin}$, no additional sources of cooling are required to explain the EDGES result. For $T_m = \SI{7}{\kelvin}$, we require $T_R = \SI{67}{\kelvin}$, compared to the CMB temperature at this redshift, $T_\text{CMB} = \SI{50}{\kelvin}$. 

\begin{figure}
    \centering
    \subfigure{
        \label{fig:source_elec_decay}
        \includegraphics[scale=0.34]{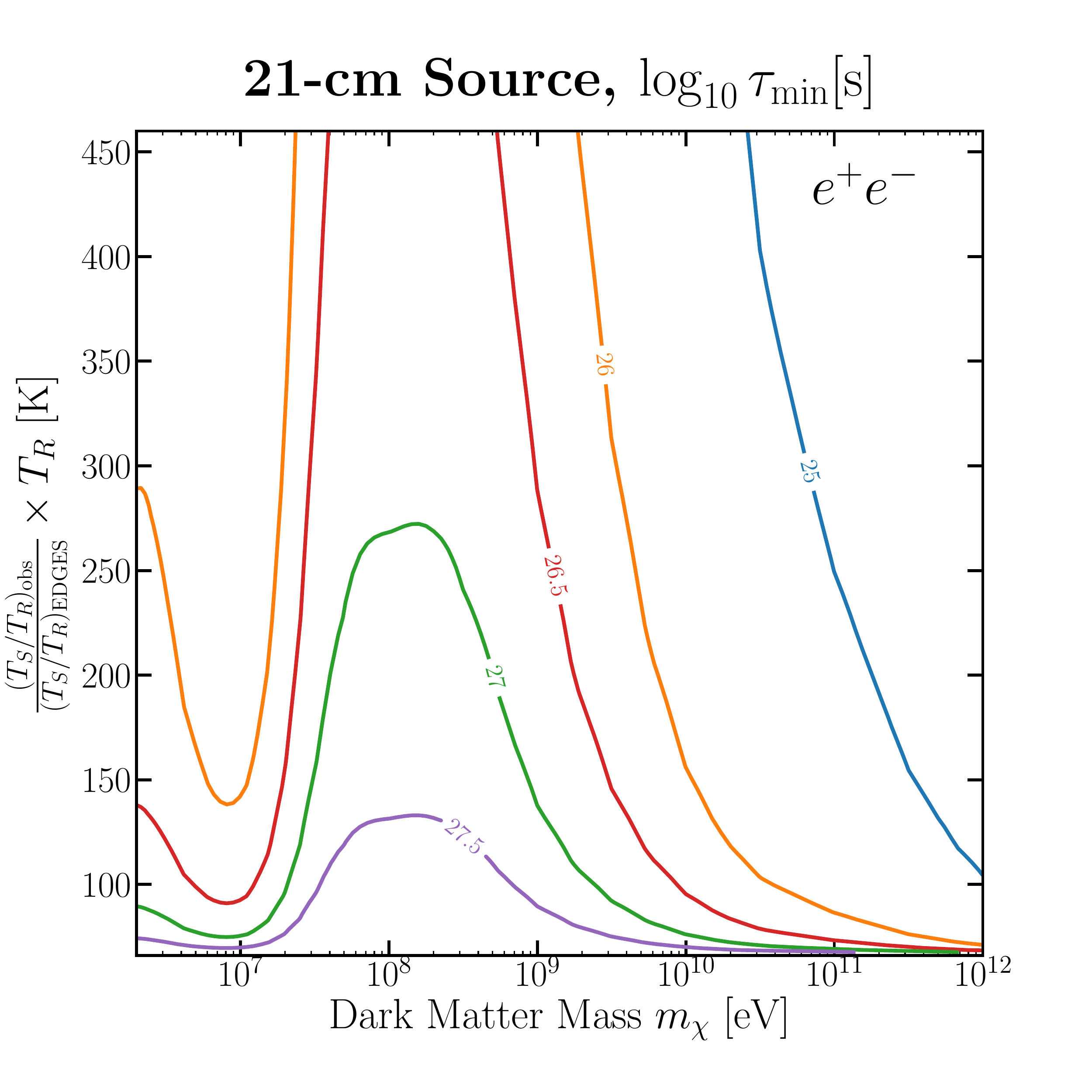}
    }
    \subfigure{
        \label{fig:source_phot_decay}
        \includegraphics[scale=0.34]{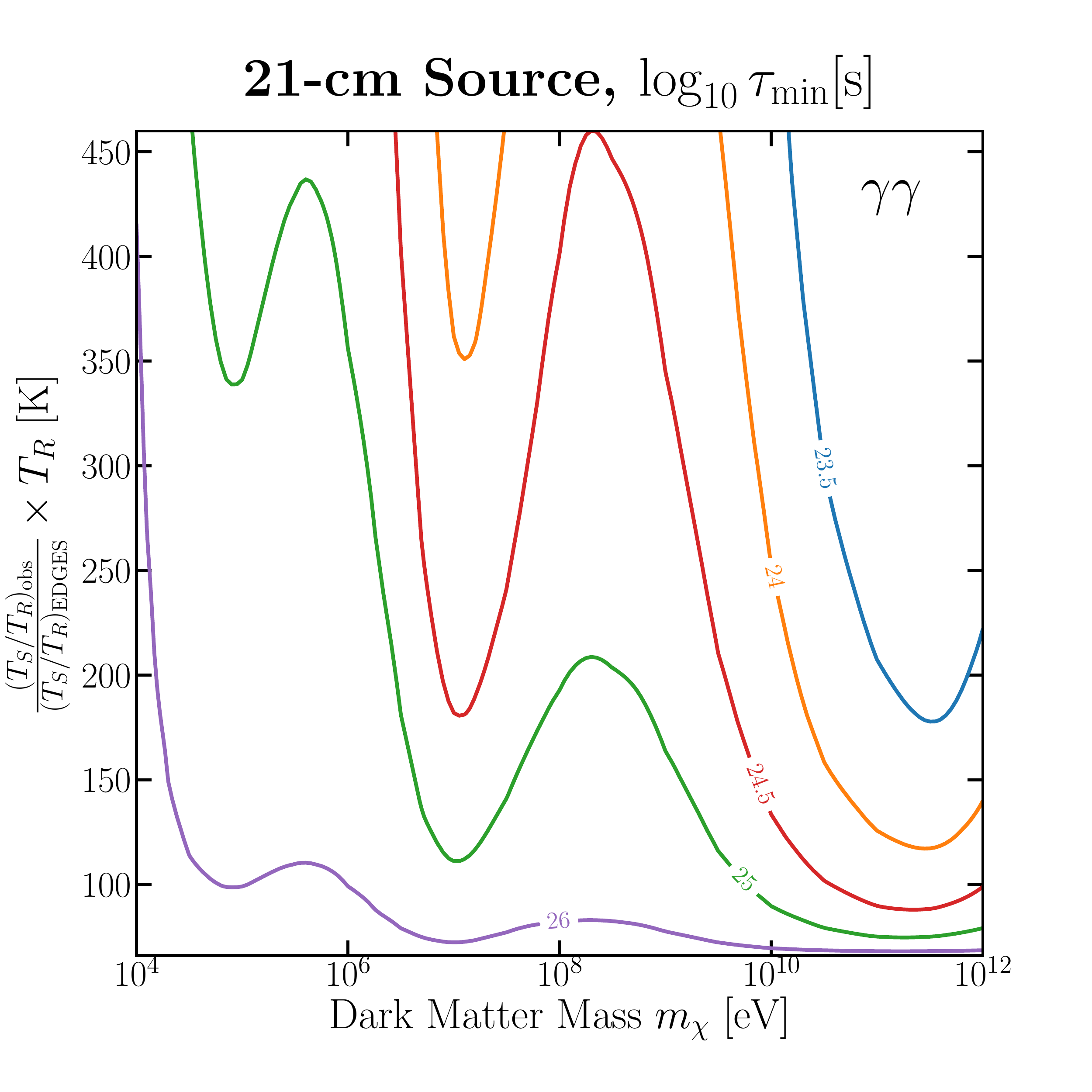}
    }
  \caption{Decay lifetime constraints with an additional 21-cm source with $\chi \to e^+e^-$ (left) and $\chi \to \gamma \gamma$ (right), as a function of $m_\chi$ and $(T_S/T_R)_\text{obs}/(T_S/T_R)_\text{EDGES} \times T_R$. Contour lines of constant minimum $\log_{10}\tau$ (in seconds) are shown. 
  }
  \label{fig:source_decay}
\end{figure}

\begin{figure}
    \centering
    \subfigure{
        \label{fig:source_elec_swave}
        \includegraphics[scale=0.34]{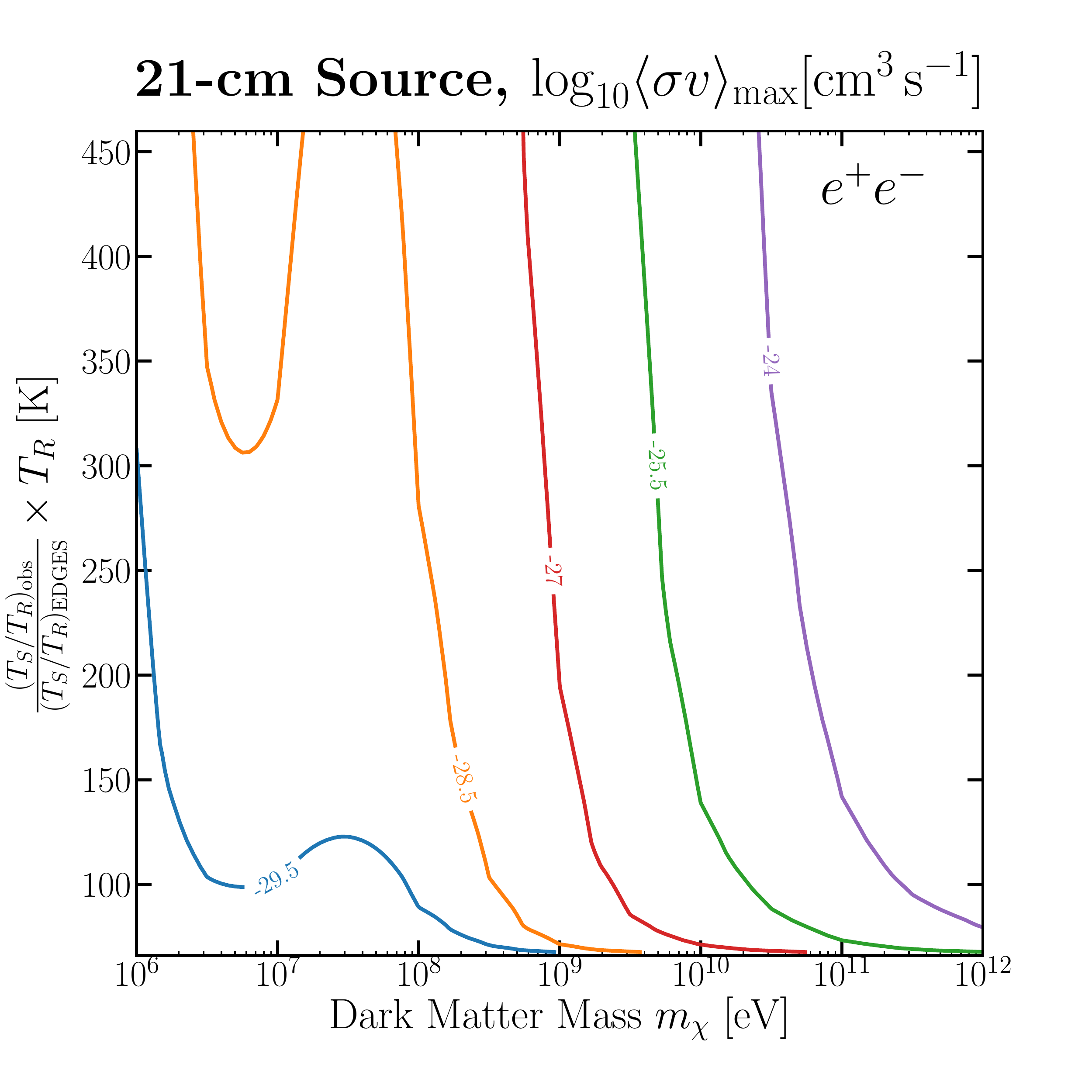}
    }
    \subfigure{
        \label{fig:source_phot_swave}
        \includegraphics[scale=0.34]{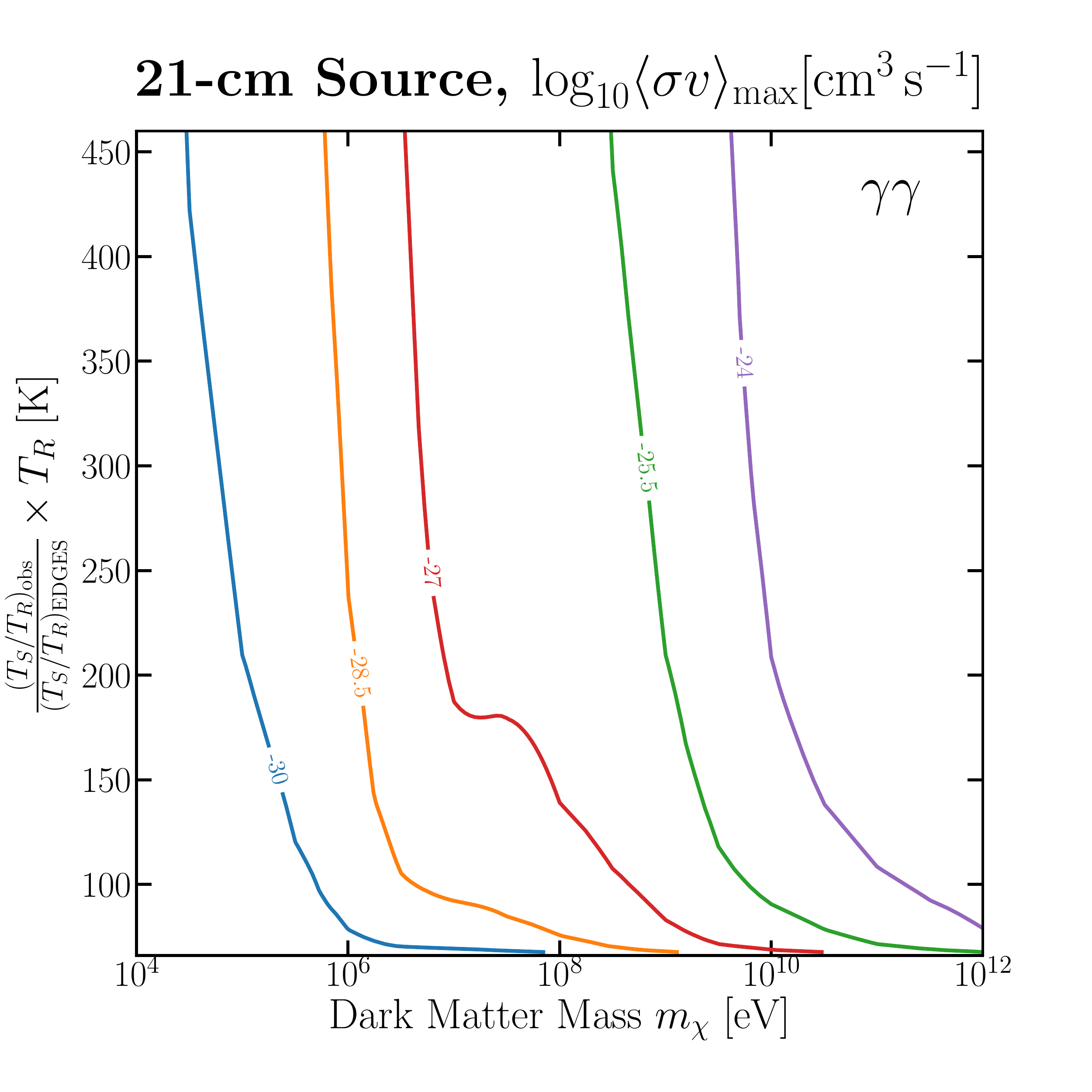}
    }
  \caption{Annihilation cross section constraints with an additional 21-cm source with $\chi \chi \to e^+e^-$ (left) and $\chi \chi \to \gamma \gamma$ (right), as a function of $m_\chi$ and $(T_S/T_R)_\text{obs}/(T_S/T_R)_\text{EDGES} \times T_R$. Contour lines of constant maximum $\log_{10} \langle \sigma v \rangle$ (in \SI{}{\centi\meter\cubed\per\second}) are shown. The green contour corresponds to the canonical relic abundance cross section of $\SI{3e-26}{\centi\meter\cubed\per\second}$.
  }
  \label{fig:source_swave}
\end{figure}

This required value of $T_R$ is large: since $T_R \gg \nu_{21}$, where $\nu_{21} = \SI{1.42}{\giga\hertz}$ is the hyperfine transition frequency, the effective temperature is directly proportional to the intensity, and so the additional 21-cm source must have approximately 35\% of the intensity of the CMB at this frequency, at $z=17.2$. The additional source must be confined to frequencies well below the peak of the CMB spectrum in order to be consistent with FIRAS observations of the CMB blackbody spectrum in the present day, which rule out distortions to the CMB spectrum greater than one part in $10^4$--$10^5$ for frequencies \SI{60}{\giga\hertz} and greater ~\cite{Chluba:2014sma,Fixsen:1996nj}. It is interesting to note that the ARCADE 2 experiment has reported a deviation from the CMB blackbody spectrum in the (present day) \SI{22}{\mega\hertz}--\SI{10}{\giga\hertz} range~\cite{Fixsen:2009xn}. 

Models where such a strong photon emission comes directly from DM decay or de-excitation run into stringent electroweak precision constraints \cite{Fraser:2018acy}, but models where DM decays into dark photons which oscillate into $21$-cm photons are viable, and can provide an order one or larger contribution to the intensity compared to the CMB \cite{Pospelov:2018kdh}. This large temperature also cannot be explained by uncertainties in the radio emission from astrophysical sources \cite{Barkana:2018lgd}, but may be possible to explain with optimistic black hole formation rates \cite{Ewall-Wice:2018bzf}.

We shall therefore set constraints on DM annihilation and decay as a function of $T_R$, assuming that $1.3 T_\text{CMB} \lesssim T_R \lesssim 10 T_\text{CMB}$ at $z = 17.2$. If we take the EDGES 99\% confidence limit on $T_m$ shown in Eq.~(\ref{eqn:T_m_T_R_ratio}), this corresponds to gas temperatures in the range $\SI{7}{\kelvin} \lesssim T_m \lesssim \SI{50}{\kelvin}$ at the same redshift. Lower values of $T_R$ lead to values of $T_m$ that are below the standard matter temperature at this redshift, in which case no additional heating would be allowed.

The evolution equations we solve are
\begin{alignat}{1}
    \dot{T}_m &= \dot{T}_m^{(0)} + \dot{T}_m^\chi, \nonumber \\
    \dot{x}_e &= \dot{x}_e^{(0)} + \dot{x}_e^\chi.
\end{alignat}
Note that the CMB temperature used in these equations remains unchanged, as we do not expect significant modifications to the overall energy density of the CMB. Since the evolution equations are essentially the same as the TLA with DM annihilation or decay, these constraints are related to those derived in~\cite{DAmico:2018sxd}, but are broadly applicable to measurements with $T_{21} \lesssim -\SI{200}{\milli\kelvin}$, including the EDGES measurement. We also use the structure formation prescription described in \cite{Slatyer:2012yq,Liu:2016cnk}, with the boost factor included in $f_c(z)$, to account for any delayed deposition of energy.

Figs.~\ref{fig:source_decay} and~\ref{fig:source_swave} show the constraints on the minimum decay lifetime and maximum annihilation cross section with an additional source of 21-cm radiation. The limits are presented as a function of $m_\chi$ and the ratio $(T_S/T_R)_\text{obs}/(T_S/T_R)_\text{EDGES} \times T_R$, with $(T_S/T_R)_\text{EDGES} = 0.105$ as given in Eq.~(\ref{eqn:T_m_T_R_ratio}); these limits can be rescaled if future 21-cm measurements alter or improve the measurement of $(T_S/T_R)_\text{obs}$. The constraints found in~\cite{DAmico:2018sxd} for $T_{21} = \SI{100}{\milli\kelvin}$ and \SI{50}{\milli\kelvin} are equivalent to the constraints obtained with $(T_S/T_R)_\text{obs}= 0.26$ and 0.41 respectively, and setting $T_R = T_\text{CMB}$ at $z = 17.2$. Zoomed-in versions of these plots for lower temperatures are shown in Figs.~\ref{fig:source_decay_zoom}-\ref{fig:source_swave_zoom}. 

For a given measurement of $T_{21}$, Eq.~(\ref{eqn:T_m_T_R_ratio}) permits a higher matter temperature for larger values of $T_R$, which weakens the constraints that can be set. Taking the observed EDGES measurement of $T_S/T_R$, a radiation temperature of $T_R \sim \SI{100}{\kelvin}$ constrains the decay lifetime for $\chi \to e^+e^-$ to more than $10^{25}$ s across all DM masses, which is significantly stronger than the existing Planck power spectrum limits \cite{Slatyer:2016qyl}. Cross section constraints similarly strengthen considerably with respect to the Planck limits for $T_R < \SI{100}{\kelvin}$.

\section{Non-Standard Recombination}
\label{sec:non_standard_recombination}

As we saw in Chapter~\ref{chap:intro}, thermal decoupling occurs when the Compton scattering rate becomes comparable to the adiabatic cooling rate, marking the point where the matter temperature transitions from $T_m \propto (1+z)$ to $T_m \propto (1+z)^2$. The standard redshift of thermal decoupling without additional sources of heating or cooling $(1+z)_{\text{td},0}$ is therefore obtained by setting $2HT_m = \Gamma_C T_m$, giving
\begin{alignat}{1}
    (1+z)_{\text{td},0} \approx \left[\frac{45 m_e H_0 \sqrt{\Omega_m}}{4 \pi^2 \sigma_T x_e T_{\gamma,0}^4}\right]^{2/5}.
    \label{eqn:td_redshift}
\end{alignat}
Substituting a value of $x_e = 3 \times 10^{-4}$, a typical value for $x_e$ during the dark ages, we get $(1+z)_{\text{td},0} \approx 155$.

In non-standard models of recombination, the ionization history can be altered in such a way that the $x_e$ evolution equation Eq.~(\ref{eqn:TLA}) is modified while leaving the temperature evolution unchanged; this can happen, for example, if the background radiation at energies on the order of the ionization potential for hydrogen deviates significantly from a blackbody distribution during recombination \cite{DeBernardis:2008tk}. 
Another example will be discussed in Sec.~\ref{sec:rutherford_cooling}: if a small fraction of DM couples strongly to baryons, it can act as an additional heat sink and likewise modify the ionization and thermal history. Other mechanisms for early thermal decoupling, which have been recently proposed to explain the EDGES measurement, include the influence of early dark energy ~\cite{Hill:2018lfx}, or charge sequestration~\cite{Falkowski:2018qdj}, where the number density of protons and electrons are unequal owing to the presence of an additional dark charged species. In this work, we remain agnostic as to the cause of early decoupling, parametrizing it by the modified redshift of decoupling.

While the ionization history in such a situation would be model-dependent, once thermal decoupling occurs, the evolution of the thermal history without DM energy injection is completely specified by $\dot{T}_m = -2 H T_m$. The full evolution equation that we will thus solve is
\begin{alignat}{1}
    \dot{T}_m = -2 H T_m + \dot{T}_m^\chi
\end{alignat}
starting from the redshift of thermal decoupling. In reality, the non-zero value of $x_e$ would still provide some additional Compton heating, but limits set by ignoring this effect are less constraining and thus conservative.

These modifications to the thermal history can therefore be parametrized by the redshift of decoupling $(1+z)_{\text{td}}$. An earlier redshift of decoupling, occurring when the condition specified in Eq.~(\ref{eqn:td_redshift}), results in a lower temperature at later times: The EDGES result can be explained, for example, by a modified ionization and thermal history of this sort \cite{Bowman:2018yin}. Without considering specific models for increasing $(1+z)_\text{td}$, we can set constraints on DM energy injection processes as a function of this quantity, as long as heating from these processes are unimportant relative to adiabatic expansion prior to thermal decoupling.

\begin{figure}
    \centering
    \subfigure{
        \label{fig:recomb_elec_decay}
        \includegraphics[scale=0.34]{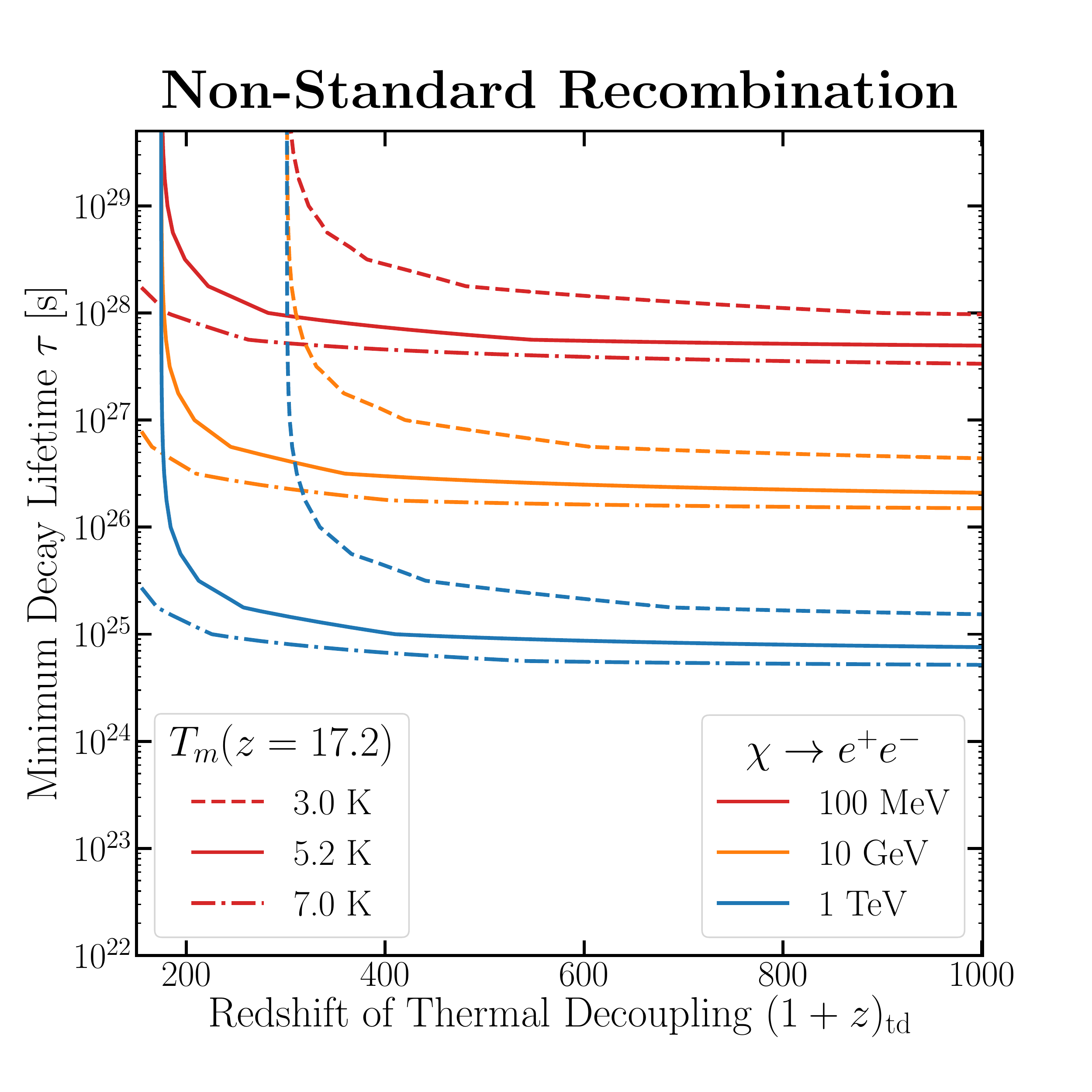}
    }
    \subfigure{
        \label{fig:recomb_phot_decay}
        \includegraphics[scale=0.34]{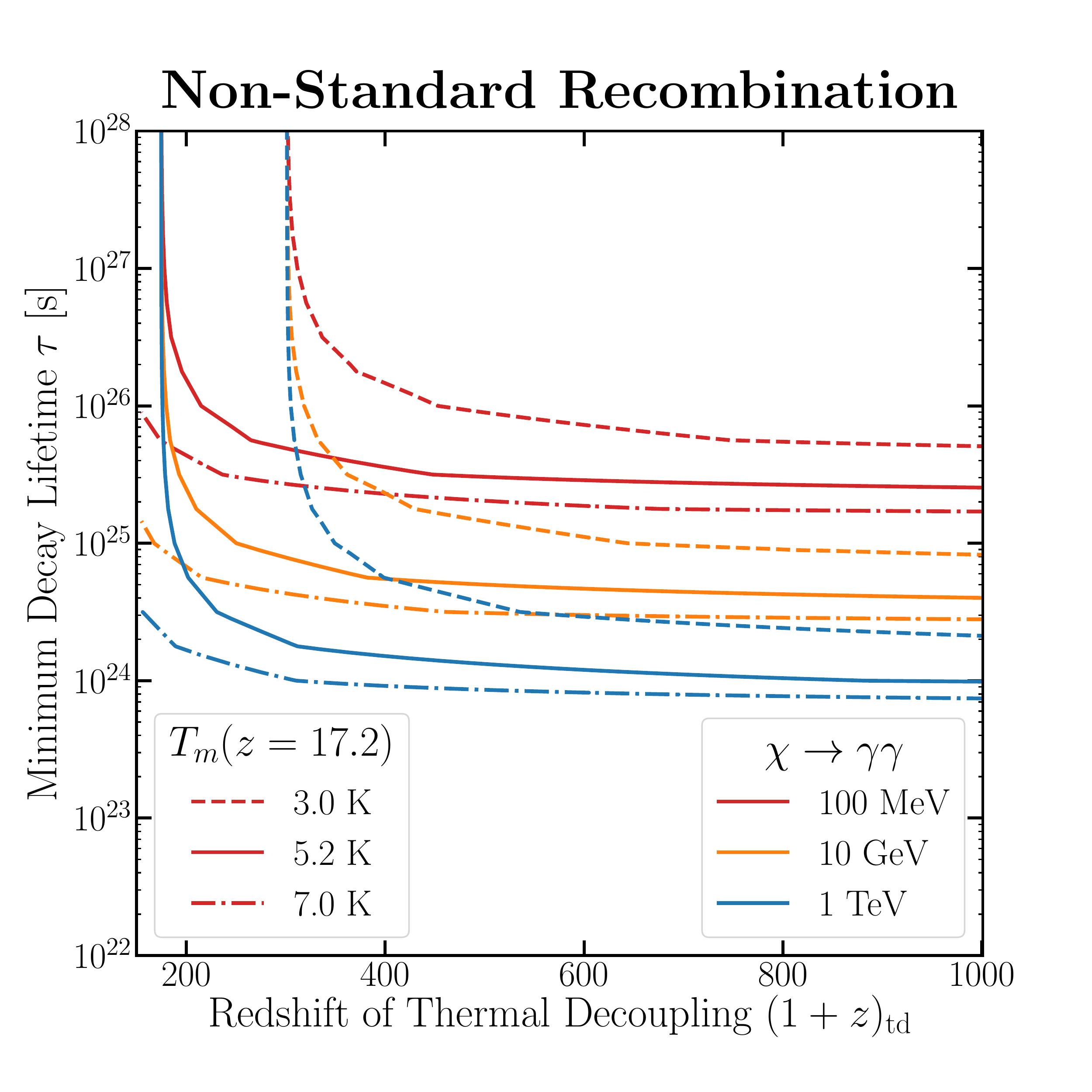}
    }
  \caption{Decay lifetime constraints for non-standard recombination as a function of the redshift of thermal decoupling, with $\chi \to e^+e^-$ (left) and $\chi \to \gamma \gamma$. In both plots, we show the limits for 100 MeV (red), 10 GeV (orange) and 1 TeV (blue) DM, assuming a measured upper limit on the matter temperature at $z = 17.2$ of 3.0 K (dashed), 5.2 K (solid) and 7.0 K (dot-dashed). The 5.2K value corresponds to the EDGES limit.
  }
  \label{fig:recomb_decay}
\end{figure}

Fig.~\ref{fig:recomb_decay} shows the constraints set on the decay lifetime of a DM particle $\chi$ decaying to $e^+e^-$ and $\gamma \gamma$ respectively as a function of $(1+z)_\text{td}$, for different possible values of $T_m$ at $z = 17.2$. For temperatures below 7 K, the temperature for standard recombination, thermal decoupling must occur at a sufficiently high redshift before adiabatic cooling can bring $T_m$ to that value. For the EDGES value of 5.2 K, this corresponds to $(1+z)_\text{td} \sim 175$; additional heating from DM energy injection is only allowed when the thermal decoupling occurs above this value. 

Once $(1+z)_\text{td}$ exceeds the critical value for sufficient cooling, the constraints on the minimum decay lifetime depends only weakly on $(1+z)_\text{td}$. To understand this, note that if the baryon temperature in the absence of heating is well below the observed temperature limit, then the temperature including heating is solely determined by the energy injection rate, and is relatively independent of the baseline baryon temperature without heating and hence $(1+z)_\text{td}$. 

With $T_m(z = 17.2) = $ 5.2 K, the constraints set by this temperature measurement for $m_\chi $ = 100 MeV is $\sim 5 \times 10^{27}$ s for decays to $e^+e^-$ and $\sim 10^{25}$ s for $\gamma \gamma$, which is both at least an order of magnitude stronger than limits set by the Planck CMB power spectrum measurement \cite{Slatyer:2016qyl}. These limits are valid assuming only no additional sources of cooling for the matter temperature after thermal decoupling. 

\begin{figure*}
    \centering
    \subfigure{
        \label{fig:recomb_elec_swave}
        \includegraphics[scale=0.34]{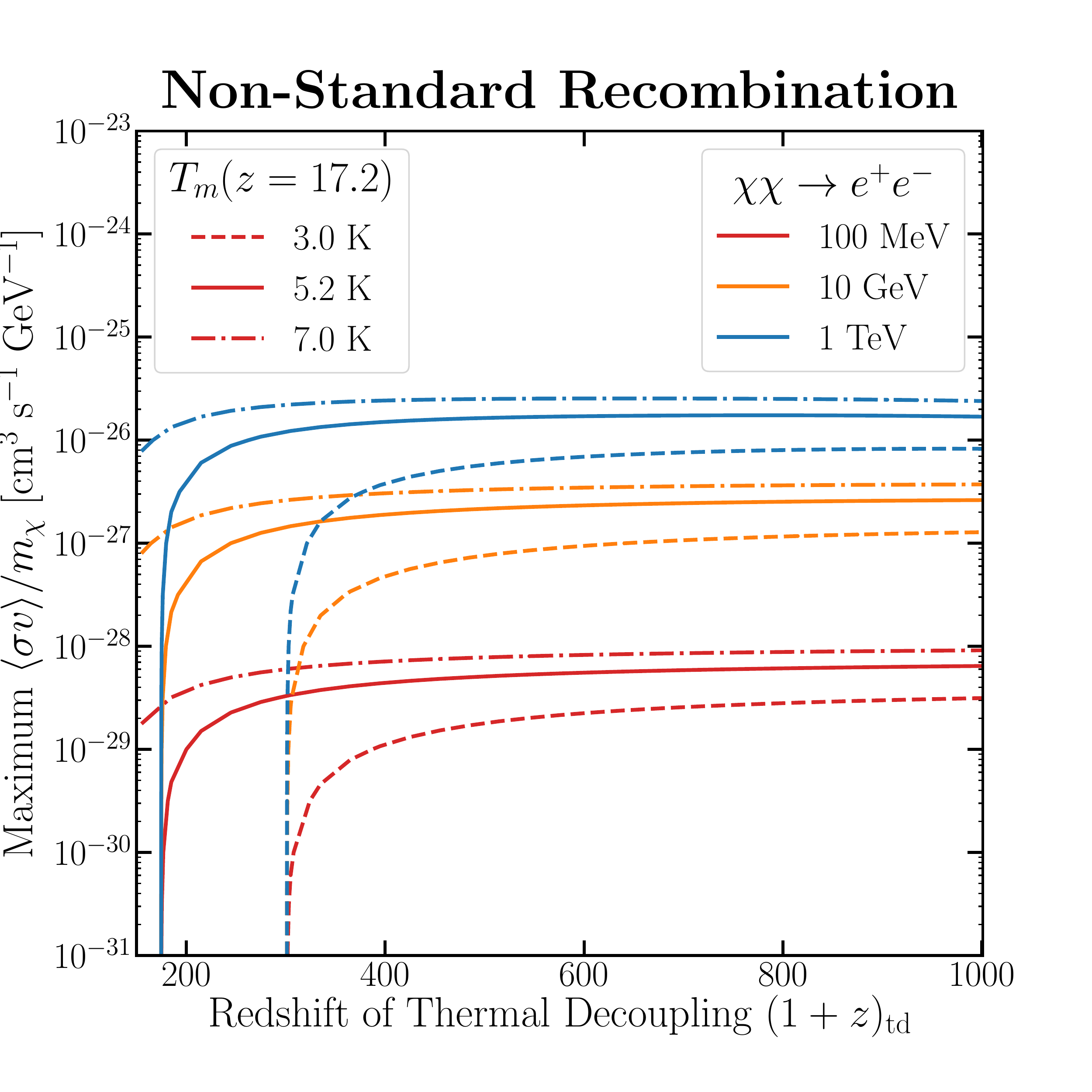}
    }
    \hfil
    \subfigure{
        \label{fig:recomb_phot_swave}
        \includegraphics[scale=0.34]{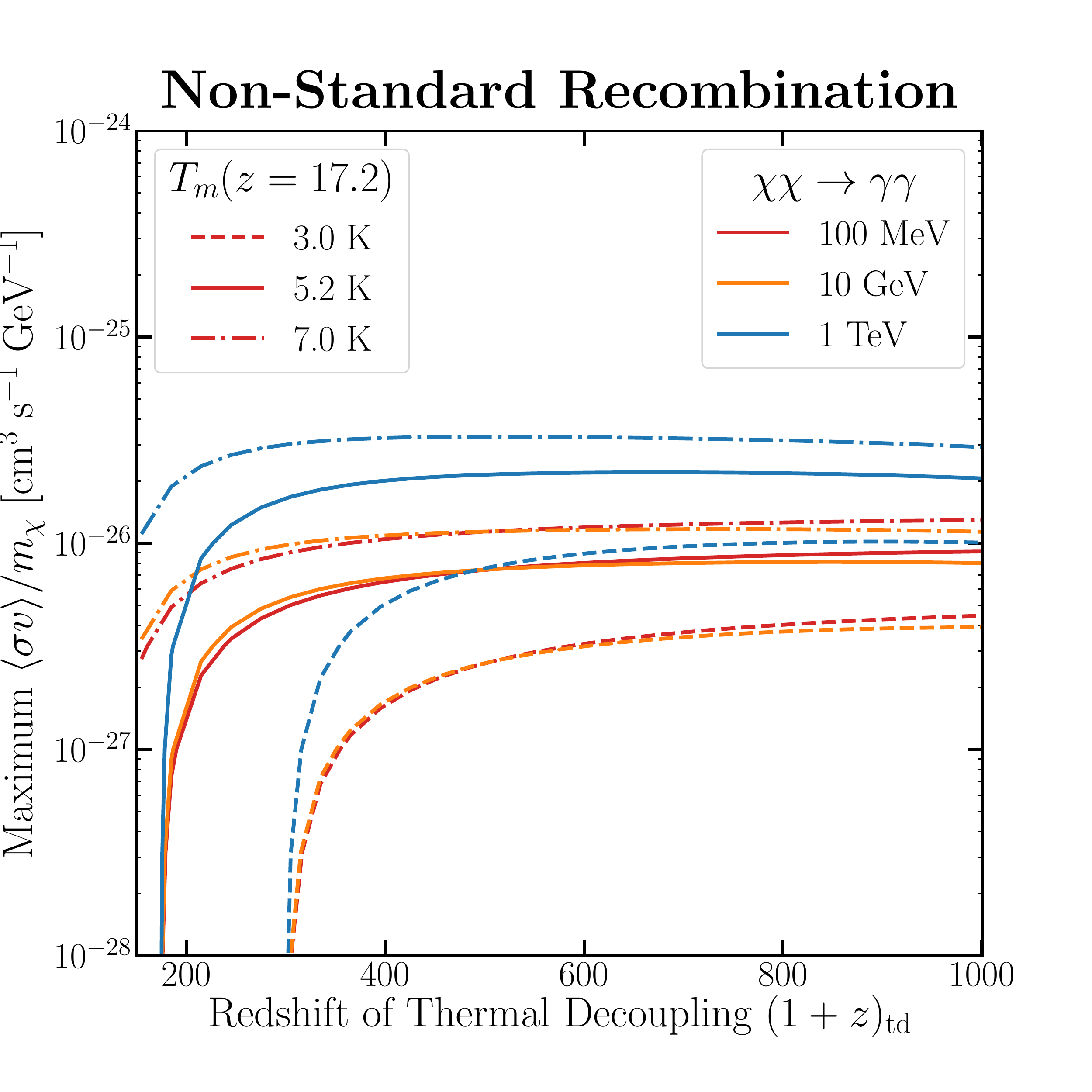}
    }
  \caption{Annihilation cross section constraints for non-standard recombination as a function of the redshift of thermal decoupling, with $\chi \chi \to e^+e^-$ (left) and $\chi \chi \to \gamma \gamma$. In both plots, we show the limits for 100 MeV (red), 10 GeV (orange) and 1 TeV (blue) DM, assuming a measured upper limit on the matter temperature at $z = 17.2$ of 3.0 K (dashed), 5.2 K (solid) and 7.0 K (dot-dashed). The 5.2K value corresponds to the EDGES limit.}
  \label{fig:recomb_swave}
\end{figure*}

Fig.~\ref{fig:recomb_swave} shows a similar plot for the constraints on the annihilation cross section, with the main features of these constraints being similar to the result for decays. The constraints set by $T_m(z = 17.2) = $ 5.2 K are once again stronger than the current Planck constraints \cite{Slatyer:2015jla,Ade:2015xua} by about an order of magnitude, with little dependence on $(1+z)_\text{td}$.

\begin{figure}
    \centering
    \subfigure{
        \label{fig:recomb_chan_decay}
        \includegraphics[scale=0.34]{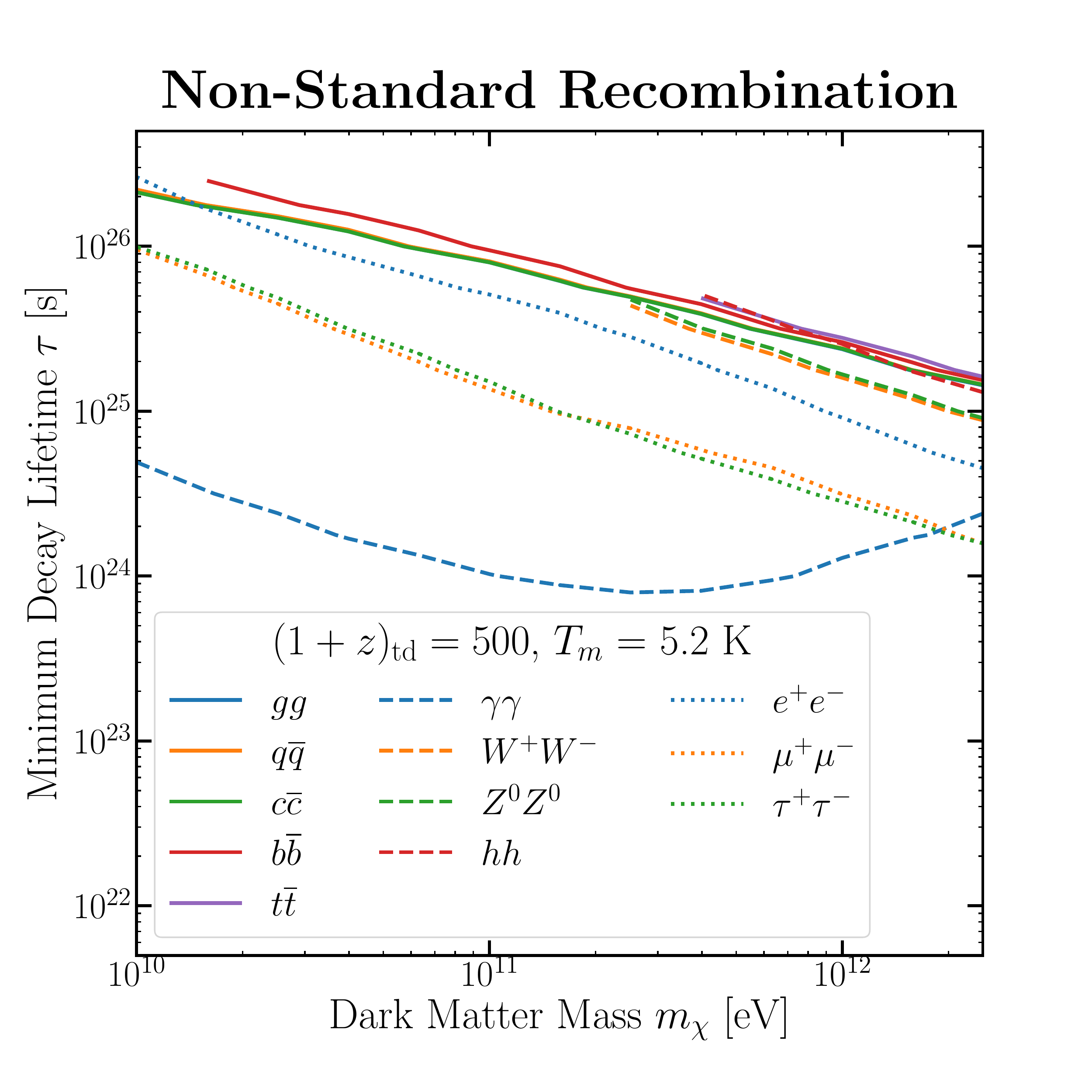}
    }
    \hfil
    \subfigure{
        \label{fig:recomb_chan_swave}
        \includegraphics[scale=0.34]{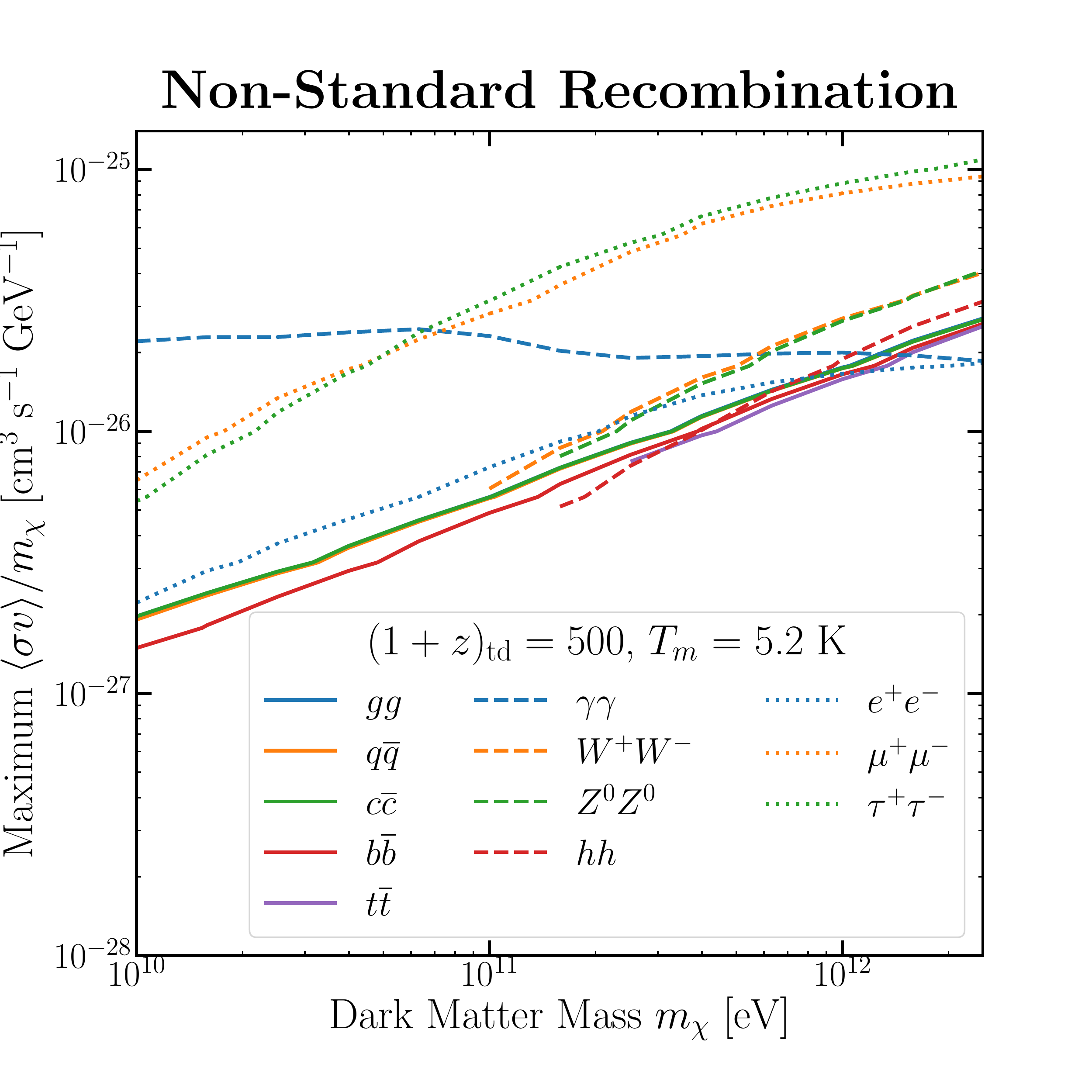}
    }
  \caption{Minimum decay lifetime (left) and maximum annihilation cross section (right) constraints for non-standard recombination as a function of the redshift of thermal decoupling for several SM channels. In both plots, $(1+z)_\text{td} = 500$, and $T_m(z = 17.2) = $ 5.2 K. 
  }
  \label{fig:recomb_chan}
\end{figure}

Fig.~\ref{fig:recomb_chan} shows the constraints for several decay and annihilation channels into SM particles for the case where $(1+z)_\text{td} = 500$, and $T_m(z = 17.2) = $ 5.2 K. These constraints apply across a large range of $(1+z)_\text{td} \gtrsim 250$ due to the weak dependence on $(1+z)_\text{td}$. To obtain these constraints, electron and photon spectra were computed using the \texttt{PPPC4DMID} \cite{Cirelli:2010xx}, and the corresponding $f_c(z)$ computed by taking an energy-weighted sum of the spectra \cite{Slatyer:2015jla}. The behavior of these limits are set mainly by the ability of the secondary photons and electrons of these decays to heat baryons at $z \sim 20$. The universe between $z \sim 20$ and recombination is mostly transparent to high energy photons, which explains the relatively weak constraints for $\chi \to \gamma \gamma$. The limits of the other channels relative to $\chi \to e^+e^-$ weaken when more neutrinos are produced during the cascade, and strengthen when soft electrons/positrons with energy $\sim \SI{100}{\mega\eV}$ are produced: electrons and positrons in this energy range are particularly effective at depositing their energy into baryons during this epoch \cite{Liu:2016cnk,Slatyer:2016qyl}.

\section{Rutherford Cooling}
\label{sec:rutherford_cooling}

\subsection{Evolution Equations}

To ensure that the matter temperature at $z \sim 17$ satisfies Eq.~(\ref{eqn:T_m_T_R_ratio}) while taking $T_R = T_\text{CMB}$, several groups \cite{Barkana:2018lgd,Munoz:2018pzp,Berlin:2018sjs,Fraser:2018acy,Barkana:2018qrx} have examined the possibility of a new DM-baryon or DM-electron scattering cross section that has a Rutherford-like behavior, i.e. $\sigma = \sigma_0 v^{-4}$. This interaction may occur between only a fraction of DM and the SM. Both the difference in temperature between matter and DM as well as their bulk relative velocity $V_{\chi b}$ from earlier DM clustering can affect $T_m$, which evolves according to~\cite{Munoz:2018pzp,Munoz:2015bca} 
\begin{multline}
    \dot{T}_m^c = \sum_j \frac{2}{3(1 + f_\text{He} + x_e)n_\text{H}} \frac{f_{\chi,\text{int}} \rho_\text{DM} \rho_j}{(m_\chi + m_j)^2} \frac{\sigma_{0,j}}{u_j} \left[\sqrt{\frac{2}{\pi}} \frac{e^{-r_j^2/2}}{u_j^2} (T_\chi - T_m) + m_\chi \frac{F(r_j)}{r_j}\right],
    \label{eqn:T_m_cooling}
\end{multline}
where the sum is over all species $j$ that can interact with the DM: this may be over all baryons \cite{Barkana:2018lgd}, or over free protons and electrons in millicharged DM models \cite{Munoz:2018pzp}. $\rho_\text{DM}$ and $\rho_j$ are the mass densities of all DM and species $j$ respectively, with $f_{\chi,\text{int}}$ being the fraction of DM interacting with the SM by mass. $m_\chi$ and $T_\chi$ is the mass and temperature of the interacting DM respectively, $u_j \equiv (T_m/m_j + T_\chi/m_\chi)^{1/2}$ is the thermal sound speed of the DM-$j$ fluid, and $r_j \equiv V_{\chi b}/u_j$. The function $F(r)$ is
\begin{alignat}{1}
    F(r) \equiv \text{erf} \left(\frac{r}{\sqrt{2}}\right) - \sqrt{\frac{2}{\pi}} e^{-r^2/2} r.
\end{alignat}

To solve for the full evolution, we must also evolve the temperature of the interacting DM \cite{Munoz:2015bca},
\begin{multline}
    \dot{T}_\chi = -2HT_\chi + \sum_j \frac{2}{3} \frac{m_\chi \rho_j}{(m_\chi + m_j)^2} \frac{\sigma_{0,j}}{u_j} \left[\sqrt{\frac{2}{\pi}} \frac{e^{-r_j^2/2}}{u_j^2} (T_m - T_\chi) + m_j \frac{F(r_j)}{r_j}\right],
    \label{eqn:T_chi_cooling}
\end{multline}
as well as the bulk relative velocity \cite{Munoz:2018pzp}
\begin{alignat}{1}
    \dot{V}_{\chi b} = -H V_{\chi b} - \left(1 + \frac{f_{\chi,\text{int}} \rho_\text{DM}}{\rho_b}\right) \sum_j \frac{m_j n_j \sigma_{0,j}}{m_\chi + m_j} \frac{F(r_j)}{V_{\chi b}^2}.
    \label{eqn:V_chi_b}
\end{alignat}
When $f_{\chi, \text{int}} < 1$, Eq.~(\ref{eqn:T_chi_cooling}) assumes that the interacting component of DM has a temperature that is separate from the rest of the dark sector; relaxing this assumption would mean that the energy flow into the dark sector is distributed among more particles, with the exact effect on the thermal history determined by the masses of both the interacting and non-interacting components. 

To set constraints on DM energy injection in the presence of this scattering process, the full set of rate equations which should be integrated are Eqs.~(\ref{eqn:T_chi_cooling}), ~(\ref{eqn:V_chi_b}), together with
\begin{alignat}{1}
    \dot{T}_m &= \dot{T}_m^{(0)} + \dot{T}_m^\chi + \dot{T}_m^c, \nonumber \\
    \dot{x}_e &= \dot{x}_e^{(0)} + \dot{x}_e^\chi.
    \label{eqn:cooling_full_eqns}
\end{alignat}
For simplicity, we will restrict our discussion to the case of DM-hydrogen scattering (both neutral and ionized, with no scattering on helium or free electrons) until we discuss the millicharged DM model, where scattering occurs between DM and free charged particles. From a model building perspective, the existence of a DM-baryon interaction with a Rutherford-like scattering cross section is hard to accomplish without invoking millicharged DM or introducing a new long-range force; the latter scenario is extremely constrained by fifth force experiments~\cite{Adelberger:2006dh,Kapner:2006si}, and the former leads to scattering on charged particles rather than neutral hydrogen. However, constraints on DM-hydrogen interactions from the CMB power spectrum~\cite{Boddy:2018kfv,Gluscevic:2017ywp,Dvorkin:2013cea,Xu:2018efh,Slatyer:2018aqg} or from forecasts of 21-cm measurements~\cite{Munoz:2015bca} have been derived for a range of velocity-dependent cross sections, including Rutherford scattering, and it is informative to compare our constraints with the existing literature.

\subsection{Weak and Strong Coupling Regimes}

The magnitude of $\sigma_0$ determines how tightly coupled the interacting DM and neutral hydrogen are. In the weakly coupled regime, the DM temperature $T_\chi$ remains well below the matter temperature $T_m$, and the interacting DM component is able to collapse into structures well before recombination, leading to a non-zero bulk relative velocity $V_{\chi b}$. However, for a sufficiently large $\sigma_0$, the temperature of the interacting DM becomes close to the matter temperature, and collapse into structures becomes impossible. For $f_{\chi,\text{int}} = 1$, i.e. all of the DM interacts with the SM, this scenario is highly constrained by the damping effect this would have on the CMB power spectrum \cite{McDermott:2010pa,Dvorkin:2013cea,Xu:2018efh,Slatyer:2018aqg}. However, a subdominant component ($f_{\chi,\text{int}} \lesssim 0.01$) of DM can have significant interactions with the SM at recombination without contradicting precision CMB measurements: the interacting DM component would essentially be an additional, small contribution to the baryon fluid, while leaving structure formation due to the bulk of DM unaffected \cite{Dolgov:2013una}. 

In the weak-coupling regime, the interacting component of DM remains cold and collapses efficiently, and $V_{\chi b}$ is expected to have an rms velocity of \SI{29}{\kilo\meter\per\second} at photon decoupling, $z = 1010$, the value expected for cold, non-interacting DM \cite{Ali-Haimoud:2013hpa}. From Eq.~(\ref{eqn:T_m_cooling}), while $T_m \gg T_\chi$, the effect of a non-zero value of $V_{\chi b}$ is generally to increase $T_m$. This additional source of heating forces the energy injection from DM annihilation or decay to be smaller than if we set $V_{\chi b} = 0$, leading to tighter cosmological constraints. For the rest of the results in this section, we will show only results with $V_{\chi b} = 0$, which leads to the most robust constraints: the effect of fully evolving $V_{\chi b}$ starting at a non-zero value at recombination will be shown in Appendix~\ref{app:systematics}. We integrate Eqs.~(\ref{eqn:T_chi_cooling}) to~(\ref{eqn:cooling_full_eqns}), with $T_\chi = 0$, $T_m = T_\text{CMB}$ and $x_e = 1$ starting from before recombination.

In the strong-coupling regime, the interacting component of DM is in thermal equilibrium with baryons and the CMB, and cannot collapse into structures. In this case, $V_{\chi b} = 0$ at the point of recombination, and the strong coupling between the two sectors ensures $T_m = T_\chi$ throughout. We can therefore integrate Eq.~(\ref{eqn:T_chi_cooling}) and~(\ref{eqn:cooling_full_eqns}), with $T_\chi = T_m = T_\text{CMB}$ and $x_e = 1$ starting from before recombination, with $V_{\chi b} = 0$. 

We delineate the two regimes by requiring the rate of DM heating due to DM-hydrogen scattering to be larger than $H T_\chi$ at recombination in the strong-coupling limit, so that DM and baryons remain at the same temperature at this point. This leads to the criterion
\begin{alignat}{1}
    \sigma_0^\text{strong} \gtrsim \frac{H}{n_H} \frac{(m_\chi + m_p)^2}{m_\chi m_p} \left(\frac{T_\text{CMB}}{m_p} + \frac{T_\text{CMB}}{m_\chi}\right)^{3/2}
\end{alignat}
at recombination for strong coupling to be valid, and we take the weak-coupling regime to be $\sigma_0 < 0.1 \sigma_0^\text{strong}$. 

\subsection{CMB Power Spectrum Limits}

\begin{figure}[t]
    \centering
    \includegraphics[scale=0.34]{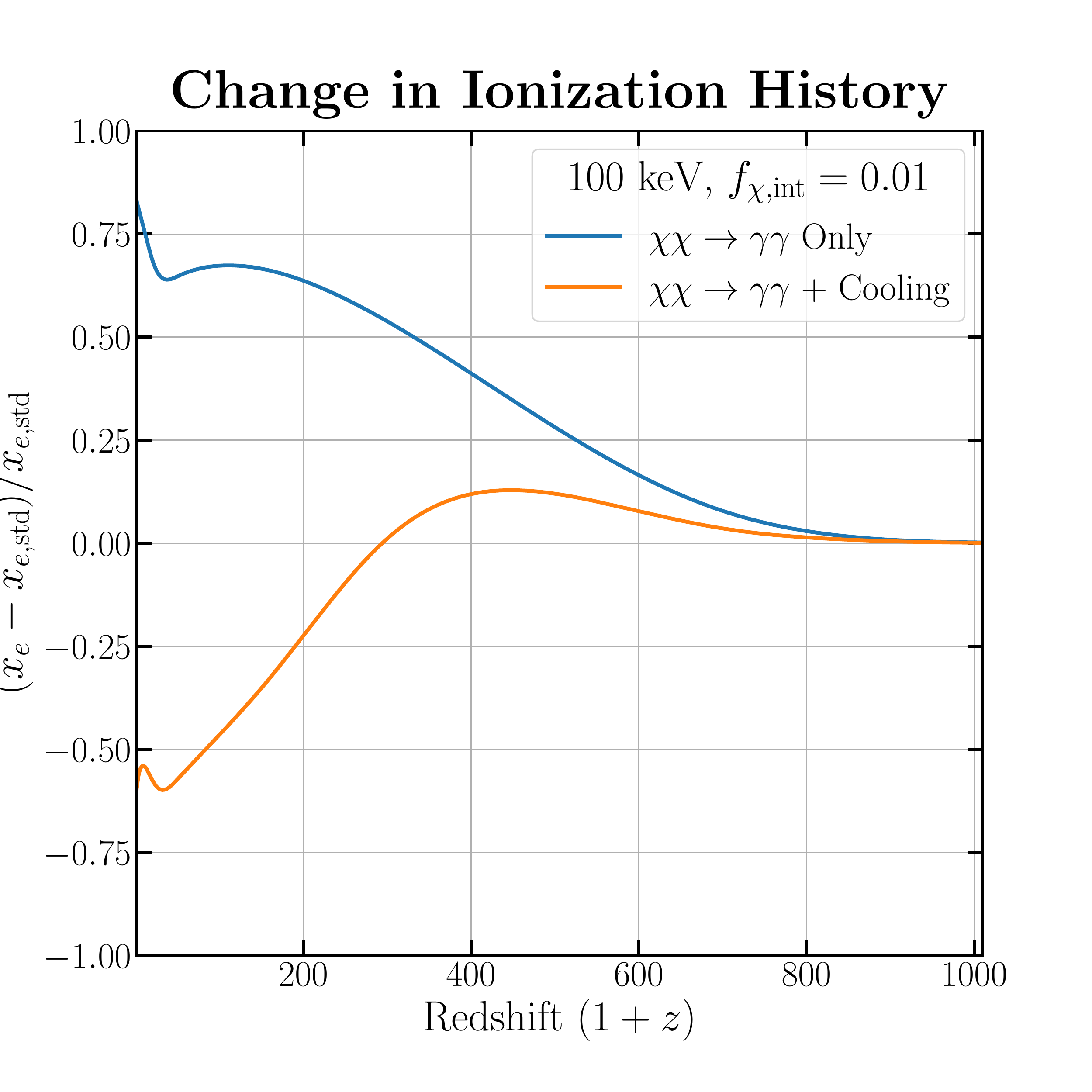}
    \caption{The change in ionization histories for $\chi \chi \to \gamma \gamma$ annihilation, with (yellow) and without (blue) Rutherford cooling, with respect to the standard ionization history (with no DM energy injection), $x_{e,\text{std}}$. Here, $m_\chi$ = \SI{100}{\kilo\eV} and $f_{\chi,\text{int}} = 0.01$. The chosen value of $\langle \sigma v \rangle = \SI{6.6e-32}{\centi\meter\cubed\per\second}$ is the maximum allowed from the Planck CMB limits in the absence of scattering; this scenario with scattering may evade these limits.}
    \label{fig:altered_CMB_const_history}
\end{figure}

DM annihilation and decay during the cosmic dark ages increase the residual ionization of the universe after recombination as compared to the standard history, and this change to the ionization history can be constrained by considering its impact on the CMB power spectrum. The presence of an additional source of cooling of the matter temperature, however, also modifies the ionization history during this time. If the rate of cooling is sufficiently large to decouple baryons from the CMB at a time earlier than $(1+z)_{\text{td},0}$ given in Eq.~(\ref{eqn:td_redshift}), then $T_m$ becomes smaller than expected, which in turn increases the recombination rate, decreasing the residual ionization. 

\begin{figure}
    \centering
    \subfigure{
        \label{fig:weak_coupling_Tm_history}
        \includegraphics[scale=0.34]{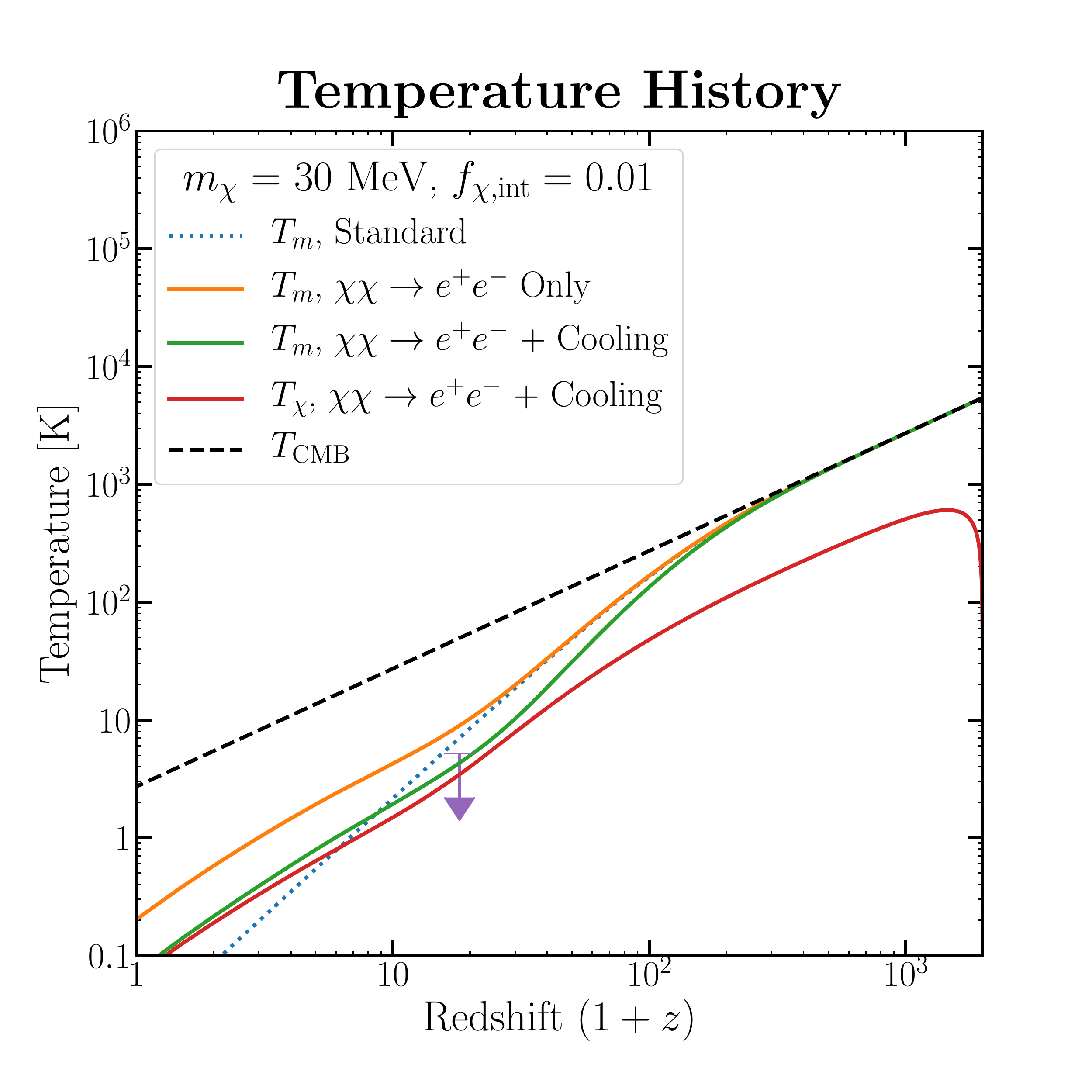}
    }
    \subfigure{
        \label{fig:weak_coupling_xe_history}
        \includegraphics[scale=0.34]{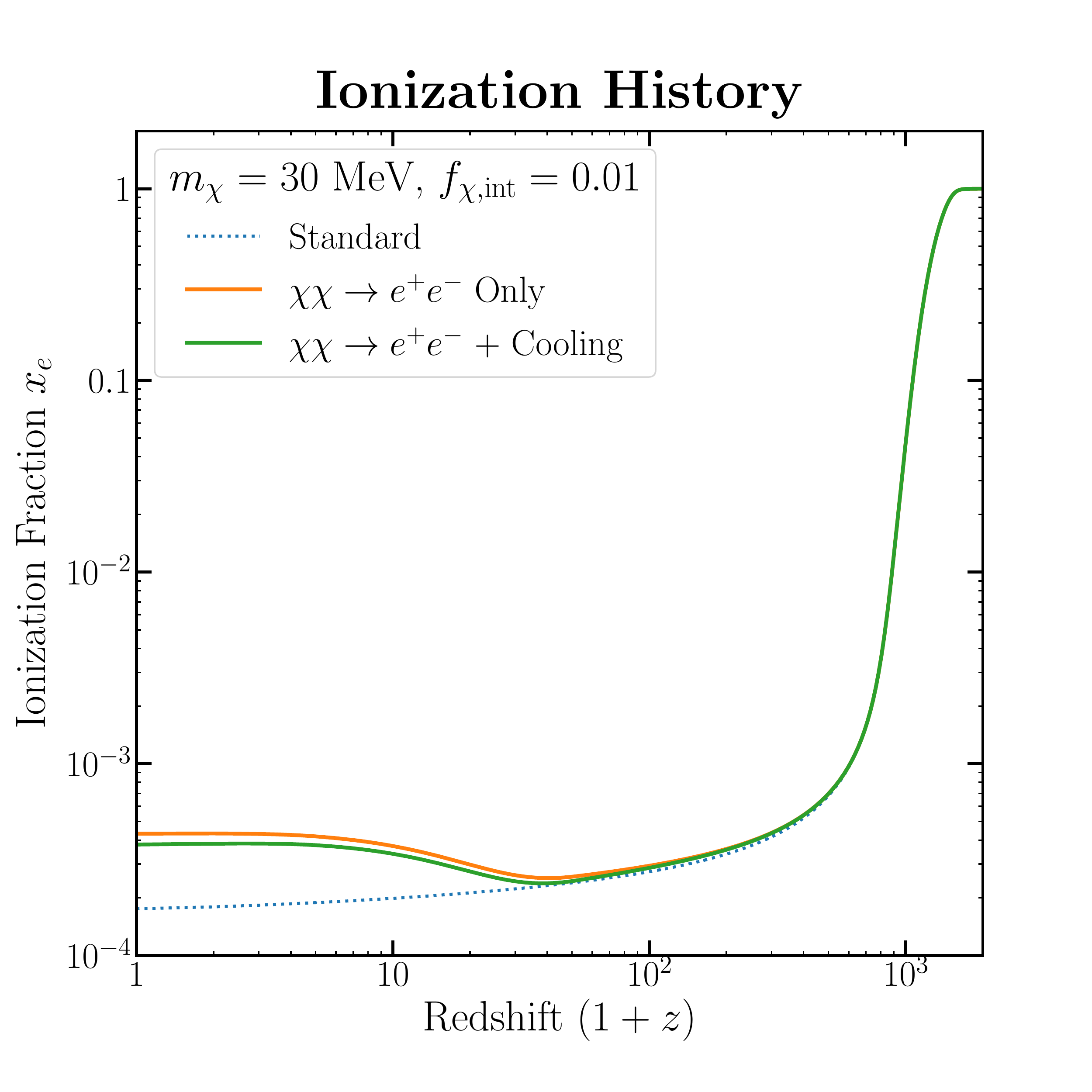}
    }
    \caption{Thermal (left) and ionization (right) histories with $\chi \chi \to e^+e^-$ annihilation and Rutherford cooling, $m_\chi$ = 30 MeV and $f_{\chi,\text{int}} = 0.01$. The standard history (blue, dotted), DM annihilation only with $\langle \sigma v \rangle = 10^{-30} \, \text{cm}^3 \text{ s}^{-1}$ (orange), and DM annihilation and DM-baryon Rutherford scattering with $\sigma_0 = 10^{-38} \, \text{cm}^2$ (green) are shown. The DM temperature evolution (red), CMB temperature (black, dashed) and the EDGES temperature limit (purple arrow) are also shown.}
    \label{fig:weak_coupling_history}
\end{figure}

Fig.~\ref{fig:altered_CMB_const_history} shows an example of the change in ionization history with respect to the standard ionization history. We have chosen an annihilation cross section for $\chi \chi \to \gamma \gamma$ that is naively ruled out by the Planck CMB limits. Due to the presence of Rutherford cooling, however, the ionization at $z \sim 600$ ($z \sim 300$) relative to the standard history is reduced by a factor of 2 (more than 10). Since the $s$-wave annihilation and decay constraints from the CMB power spectrum are most sensitive to energy injection at $z \sim 600$ and $z \sim 300$ respectively \cite{Slatyer:2016qyl}, we conclude that \emph{the CMB power spectrum constraints on energy injection during this epoch can be significantly relaxed if additional sources of cooling lead to thermal decoupling of baryons during or before these redshifts}. 

To estimate when thermal decoupling of baryons occurs in the presence of Rutherford cooling, we can compare the heat transfer rate due to cooling from DM to the Compton scattering term. For DM-hydrogen scattering, this gives the following condition for thermal decoupling to occur at $(1+z)_\text{td}$:

\begin{alignat}{1}
    \sigma_{0,\text{td}} \lesssim \sigma_T (1+z)_\text{td}^{5/2} \frac{x_e}{f_{\chi,\text{int}}} \left[\frac{ T_{\text{CMB},0}^{11/2} m_\chi^2}{\mu_{\chi p}^2 m_p^{1/2} m_e \rho_{\text{DM},0}}\right],
\end{alignat}
where $\mu_{\chi p}$ is the reduced mass of DM and protons, $T_{\text{CMB},0}$ is the CMB temperature today, and $\rho_{\text{DM},0}$ is the DM density today. Numerically, 
\begin{alignat}{1}
    \sigma_{0,\text{td}} \lesssim \left(\frac{m_\chi}{\mu_{\chi p}}\right)^2 \left(\frac{(1+z)_\text{td}}{600}\right)^{5/2} \frac{10^{-40}\,\text{cm}^2}{f_{\chi,\text{int}}},
    \label{eqn:thermal_decoupling_condition}
\end{alignat}
where we have taken $x_e \approx 3 \times 10^{-4}$. Thus, for $(1+z)_\text{td} = 300$ and $(1+z)_\text{td} = 600$, the CMB power spectrum constraints for decays and $s$-wave annihilation may become inapplicable for $\sigma_0 > \sigma_{0,\text{td}}$ due to the enhanced recombination from cooling at these redshifts. A sufficiently large $T_\chi$ can relax this condition, but we neglect this effect; CMB constraints on all plots are therefore only shown in regions where their validity is assured. A comprehensive study of how CMB constraints on DM annihilation relax under these circumstances is left to future work.

\subsection{Weak Coupling Results}

Fig.~\ref{fig:weak_coupling_history} shows a typical ionization and temperature history in the weak-coupling limit with both cooling and DM annihilation. Thermal decoupling of matter from the CMB occurs slightly earlier than $(1+z)_\text{td} \sim 155$ due to the additional cooling, but not significantly earlier. Since the matter temperature is locked to the radiation temperature until well after $z \sim 600$, the ionization history, even in the presence of DM annihilation, differs very little from the expected history without cooling. As a result, constraints on $s$-wave annihilation set by the CMB spectrum, which is most sensitive to energy injection at $z \sim 600$, are still applicable. 

Fig.~\ref{fig:cooling_decay} shows the constraints for DM decays to $e^+e^-$ and $\gamma \gamma$ respectively as a function of $\sigma_0$ for DM-hydrogen scattering in the weak coupling limit (set by the dashed lines), for the case where $f_{\chi,\text{int}} = 0.01$ and $f_{\chi,\text{dec}} = 1$. We also assume that the decaying DM component has the same mass as the interacting DM for simplicity. The CMB power spectrum constraints are shown up to $\sigma_0 = \sigma_{0,\text{td}}$, after which the constraints may not be applicable. Even without any energy injection from decay, a minimum scattering cross section of $\sigma_0 \sim 10^{-40} \text{ cm}^2$ is required for sufficient cooling to bring $T_m$ down to 5.2 K, absent any additional heat source. This minimum value is marked by the vertical contours of constant $\sigma_0$ at large decay lifetimes. Over a large range of $\sigma_0$, the temperature constraint set by the EDGES $21$-cm measurement is more constraining than the CMB limits for parts of parameter space. For 10 - 100 keV DM decaying to photons, thermal decoupling as given in Eq.~(\ref{eqn:thermal_decoupling_condition}) occurs earlier than $z \sim 300$ even in the weak coupling regime, and at large scattering cross sections, only the temperature measurement can effectively constrain the decay lifetime.

\begin{figure}
    \centering
    \subfigure{
        \label{fig:cooling_elec_decay}
        \includegraphics[scale=0.34]{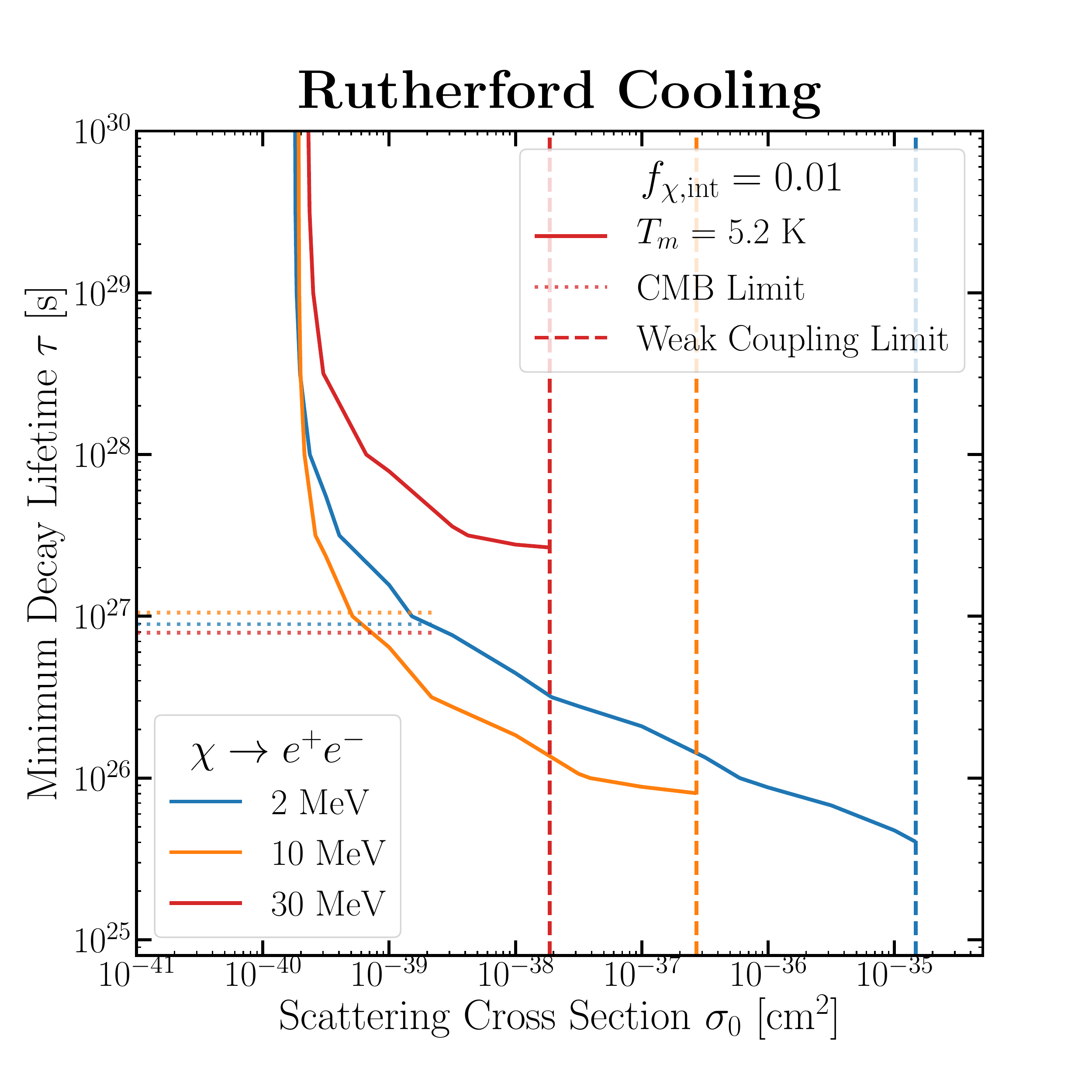}
    }
    \hfil
    \subfigure{
        \label{fig:cooling_phot_decay}
        \includegraphics[scale=0.34]{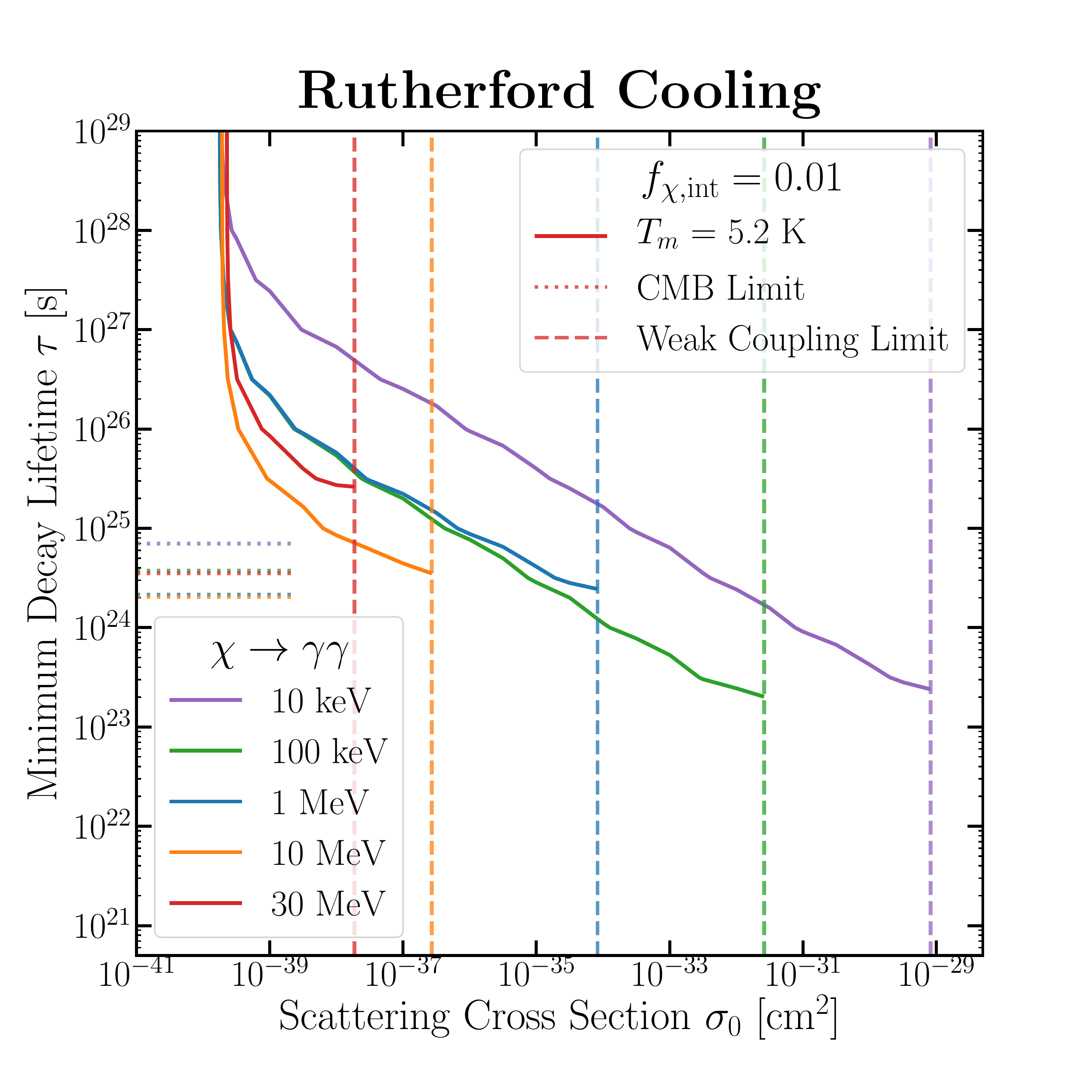}
    }
  \caption{Rutherford cooling constraints on the minimum decay lifetime for $\chi \to e^+e^-$ (left) and $\chi \to \gamma \gamma$ (right) from the matter temperature $T_m(z = 17.2) = $5.2 K (solid), $f_{\chi,\text{int}} = 0.01$. Limits from the Planck measurement of the CMB power spectrum are also shown up to $\sigma_0 = \sigma_{0,\text{td}}(z = 300)$ (dotted), together with the maximum scattering cross sections for the weak coupling limit to hold (dashed). The vertical part of the contours marks the minimum value of $\sigma_0$ required to cool the gas down to \SI{5.2}{\kelvin} in the absence of any source of energy injection.
  }
  \label{fig:cooling_decay}
\end{figure}

Fig.~\ref{fig:cooling_swave} shows similar constraints on the $s$-wave annihilation cross section. The temperature limits in both cases are relatively insensitive to the actual value of $T_m$ at $z \sim 20$.  The exact value of $T_m$ sets the minimum scattering cross section for cooling with no energy injection, but at higher cross sections/shorter decay lifetimes, the constraints are essentially set by having the large amount of heating almost entirely cancelled by Rutherford cooling.

Analogous plots for the case where $f_{\chi,\text{int}} = 1$ are shown in Figs.~\ref{fig:cooling_decay_f_1} and~\ref{fig:cooling_swave_f_1} in Appendix~\ref{app:supplemental_plots}.

\begin{figure}
    \centering
    \subfigure{
        \label{fig:cooling_elec_swave}
        \includegraphics[scale=0.34]{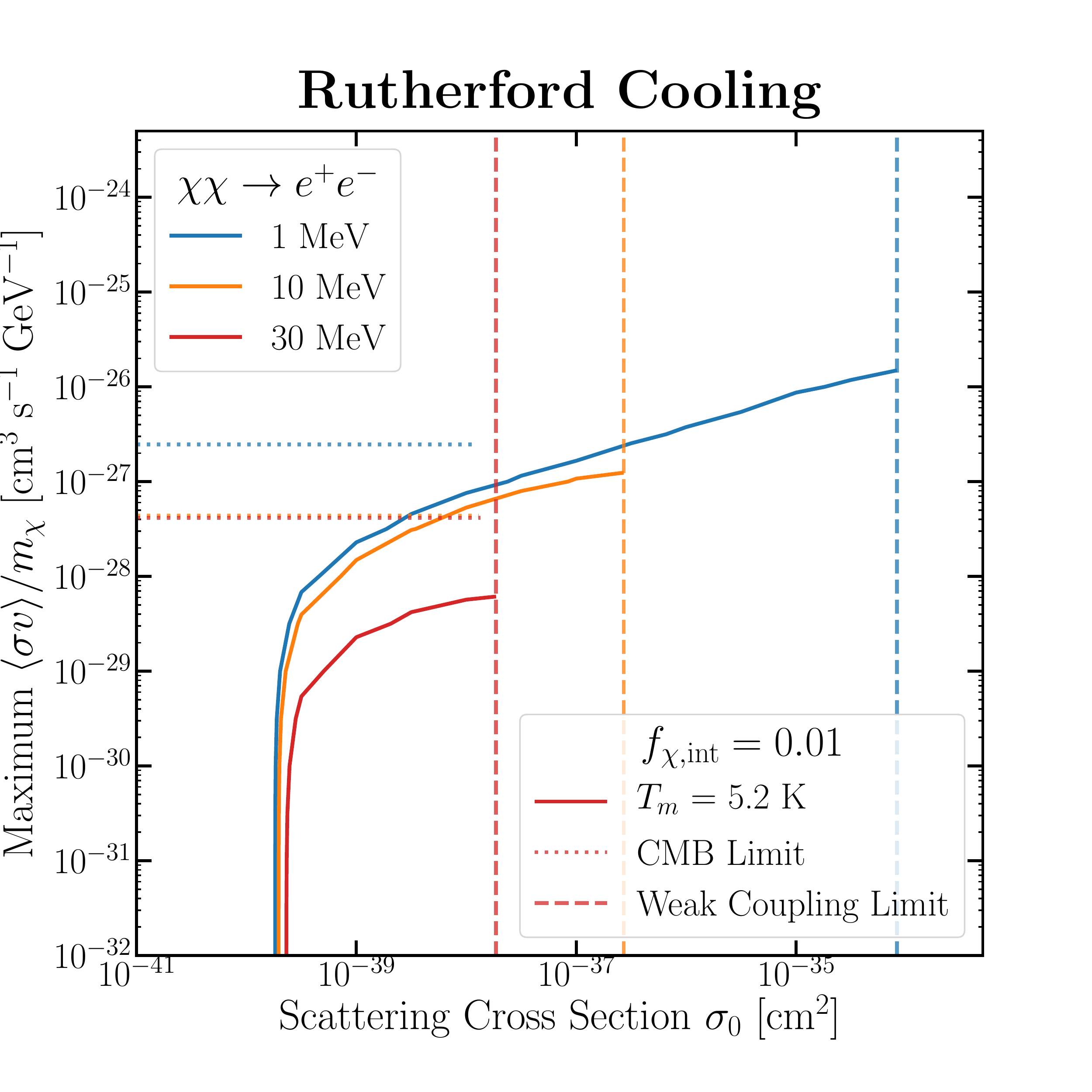}
    }
    \hfil
    \subfigure{
        \label{fig:cooling_phot_swave}
        \includegraphics[scale=0.34]{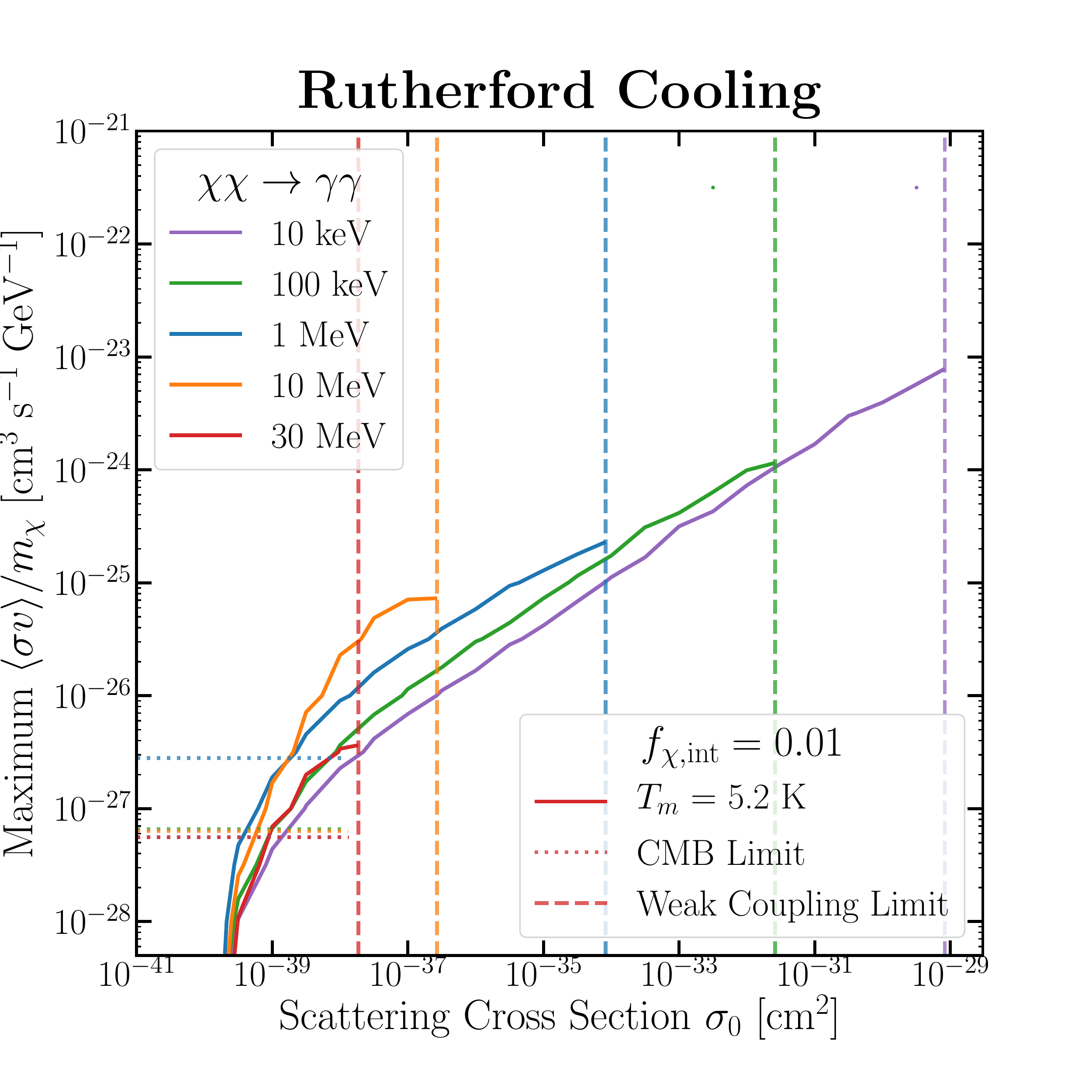}
    }
  \caption{Rutherford cooling $s$-wave annihilation constraints for $\chi \chi \to e^+e^-$ (left) and $\chi \chi \to \gamma \gamma$ (right) from the matter temperature $T_m(z = 17.2) = $ 5.2 K (solid), $f_{\chi,\text{int}} = 0.01$. Limits from the Planck measurement of the CMB power spectrum are also shown up to $\sigma_0 = \sigma_{0,\text{td}}(z = 600)$ (dotted), together with the maximum scattering cross section for the weak coupling limit to hold (dashed). The vertical part of the contours marks the minimum value of $\sigma_0$ required to cool the gas down to \SI{5.2}{\kelvin} in the absence of any source of energy injection.
  }
  \label{fig:cooling_swave}
\end{figure}

\subsection{Strong Coupling Results}

The ionization history with both cooling and DM annihilation in the strong-coupling limit, on the other hand, exhibits important differences from the history with no cooling, especially when the interacting component of DM is light. Since the transfer of energy between baryons and the interacting DM is efficient, Compton heating from the CMB must also be able to efficiently heat all of the interacting DM particles in order to keep the matter temperature at the CMB temperature. The additional heating needed means that thermal decoupling between the CMB and baryons can occur at a much higher redshift, if the DM mass is sufficiently light. After decoupling, since $T_m = T_\chi$, Eq.~(\ref{eqn:T_m_cooling}) shows that both of these temperatures simply evolve through adiabatic cooling in the absence of DM energy injections. The strong-coupling limit therefore reduces to a non-standard recombination history with early thermal decoupling, discussed in Sec.~\ref{sec:non_standard_recombination}. 

We can obtain the redshift of thermal decoupling between photons and the coupled baryon-DM fluid by replacing the Compton heating term in Eq.~(\ref{eqn:TLA}) and~(\ref{eqn:compton_rate}) by
 \begin{alignat}{1}
     \Gamma_C \to \Gamma_C \frac{n_\text{H}}{n_\text{H} + n_\chi},
 \end{alignat}
since energy from Compton heating must be redistributed into the dark sector as well. This gives
\begin{alignat}{1}
    (1 + z)_\text{td}^\text{strong} &\approx \left[\frac{45 m_e H_0 \sqrt{\Omega_m}}{4 \pi^2 \sigma_T x_e T_{\gamma,0}^4} \left(1 + \frac{f_{\chi,\text{int}} \rho_\chi}{n_\text{H} m_\chi}\right)\right]^{2/5} \nonumber \\
    &\approx 155 \left(1 + 5 f_{\chi,\text{int}} \frac{m_p}{m_\chi}\right)^{2/5},
\end{alignat}
with the redshift of thermal decoupling being independent of the scattering cross section. Note that we limit $(1+z)_\text{td}^\text{strong}$ to a maximum value of 1090, corresponding to the redshift of recombination, since thermal decoupling cannot occur before that, owing to the strong coupling between the fully ionized plasma and the CMB. In the limit of strong coupling, the exact details of how this coupling comes about is not important in determining the thermal and ionization history of the baryons.

For a given $\langle \sigma v \rangle$, the heating rate of baryons in the presence of a strongly-coupled DM is less than without DM, since some amount of the heating is transferred to the dark sector. If $x_e$ is small, we can account for this difference by replacing
\begin{alignat}{1}
    \langle \sigma v \rangle \to \frac{n_\text{H}}{n_\text{H} + n_\chi} \langle \sigma v \rangle,
\end{alignat}
and similarly for $1/\tau$ with decays. The constraints for the strong coupling limit can be easily determined from the non-standard recombination constraints: if $\langle \sigma v \rangle_{\max,(1+z)_\text{td}}$ is the maximum annihilation cross section from early thermal decoupling at redshift $(1+z)_\text{td}$, then the corresponding constraint from the strong coupling limit is
\begin{alignat}{1}
    \langle \sigma v \rangle_{\max, \text{strong}} &= \left(1 + \frac{f_{\chi,\text{int}} \rho_\chi}{n_\text{H} m_\chi} \right) \langle \sigma v \rangle_{\max,(1+z)_\text{td}^\text{strong}} \nonumber \\
    &\approx \left(1 + 5f_{\chi,\text{int}} \frac{m_p}{m_\chi}\right)\langle \sigma v \rangle_{\max,(1+z)_\text{td}^\text{strong}}.
\end{alignat}
This has been explicitly checked by directly solving Eqs.~(\ref{eqn:T_chi_cooling}) and~(\ref{eqn:cooling_full_eqns}) with $V_{\chi b} = 0$. 

\subsection{Millicharged DM}

We now turn our attention to the millicharged DM model discussed in \cite{Munoz:2018pzp,Berlin:2018sjs}, focusing on the case where $f_{\chi,\text{int}} = 0.01$, which evades the DM-baryon scattering CMB limits. We emphasize that we only allow the millicharged DM to scatter off free electrons and protons. The charge of the proton in neutral hydrogen is screened throughout the cosmic dark ages from any millicharged DM, since the typical momentum transfer between DM and neutral hydrogen lies well below the Bohr momentum of the electron in a hydrogen atom.

 \begin{figure}
    \centering
    \subfigure{
        \label{fig:millicharged_elec_decay}
        \includegraphics[scale=0.28]{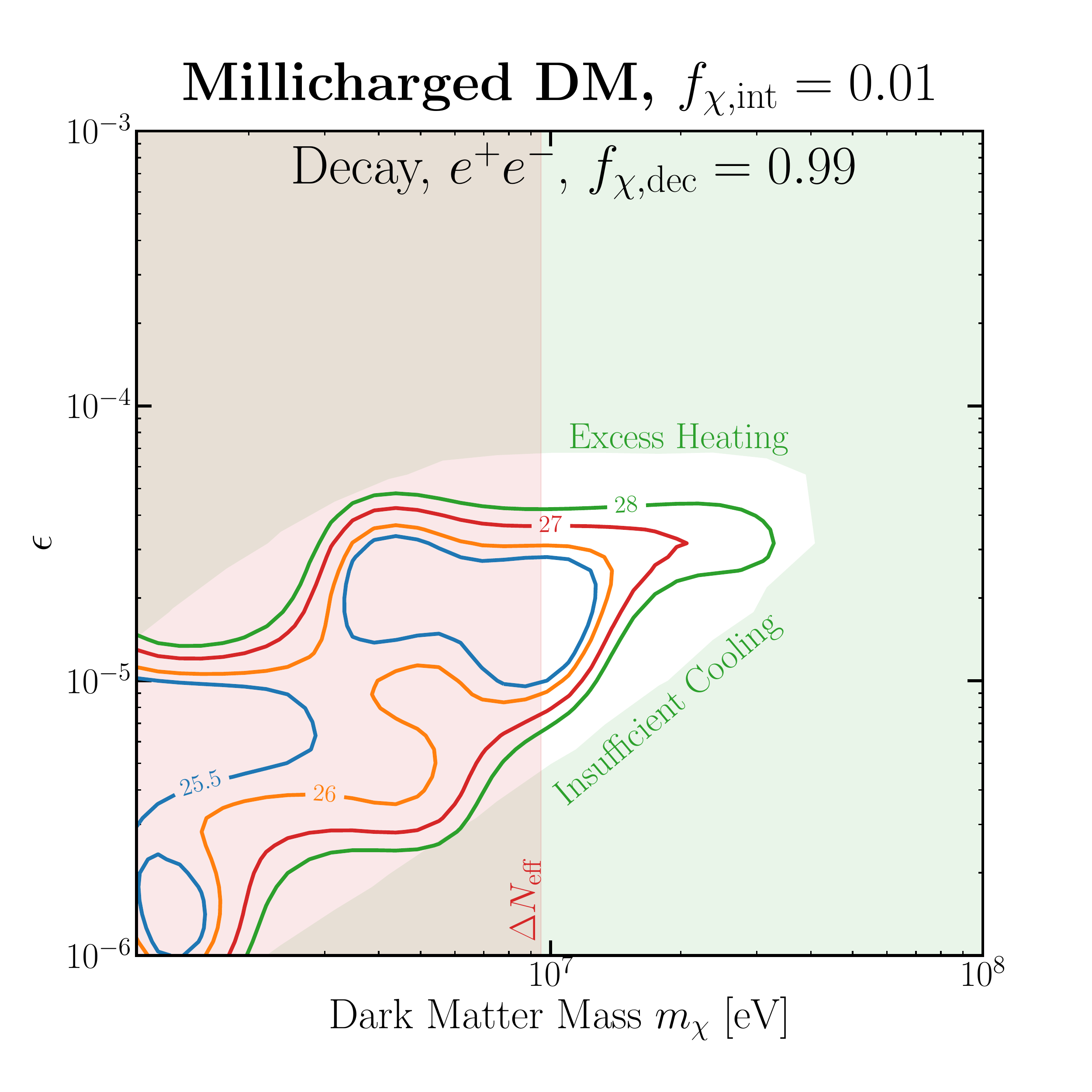}
    }
    \subfigure{
        \label{fig:millicharged_phot_decay}
        \includegraphics[scale=0.28]{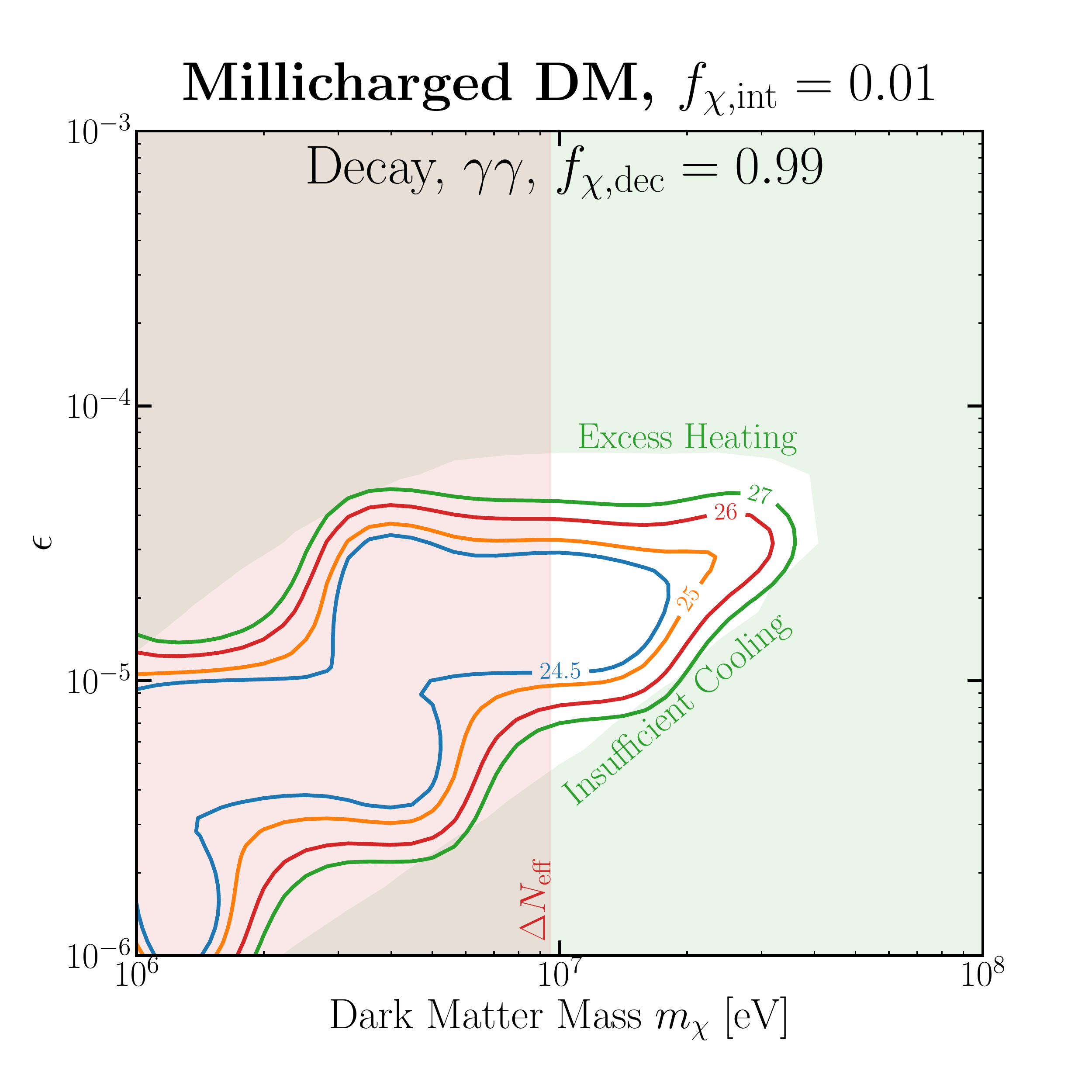}
    } \\
    \subfigure{
        \label{fig:millicharged_elec_swave}
        \includegraphics[scale=0.28]{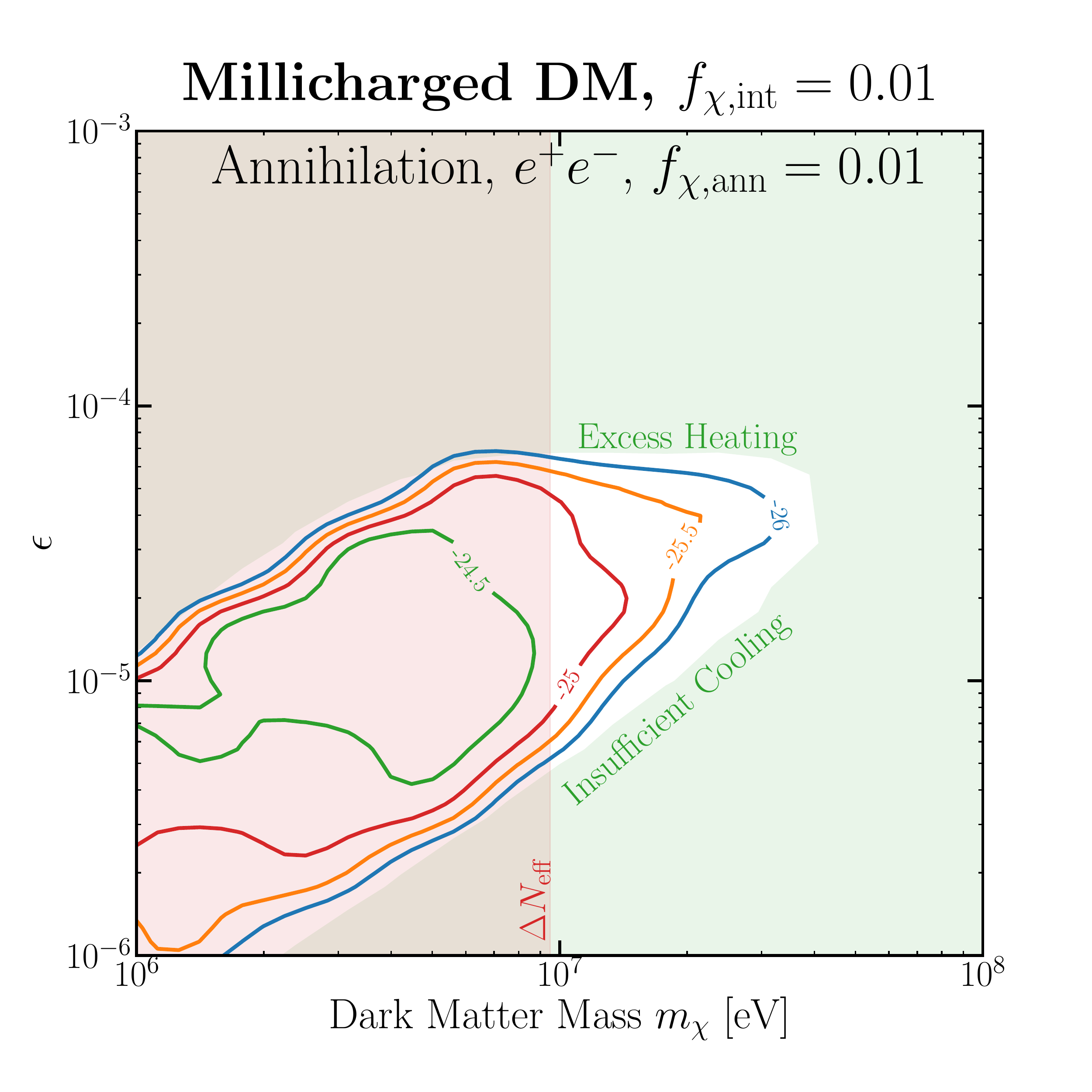}
    }
    \subfigure{
        \label{fig:millicharged_phot_swave}
        \includegraphics[scale=0.28]{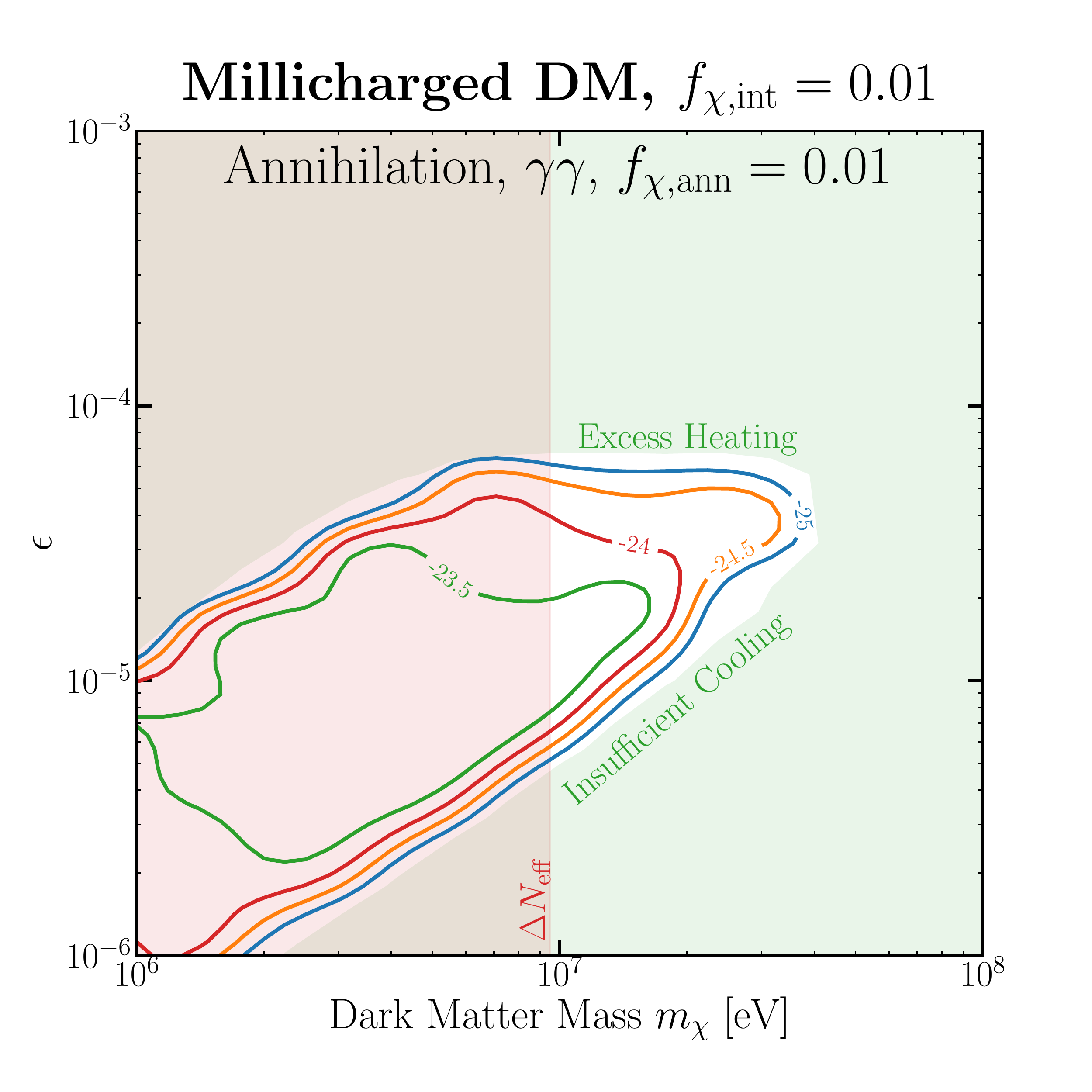}
    }
  \caption{Constraints on the millicharged DM, with an additional source of DM decay (upper panels) or annihilation (lower panels) to $e^+e^-$ (left panels) and $\gamma \gamma$ (right panels), with $f_{\chi,\text{dec}}$ = 0.99 and $f_{\chi,\text{ann}} = 0.01$ respectively. The region of parameter space ruled out by changes to $N_\text{eff}$ from the CMB power spectrum measurement (red) is shown. Charges $\epsilon$ that are not large enough for efficient cooling of baryons (green, below) or so large that excess heating occurs at $z \sim 20$ are excluded. Contours of constant minimum $\log_{10} \tau$ in seconds for decay and maximum $\log_{10} \langle \sigma v \rangle$ in $\text{cm}^3 \text{ s}^{-1}$ so that $T_m(z = 17.2) < 4$ K are drawn.
  }
  \label{fig:millicharged}
\end{figure}

For DM masses of interest ($\gtrsim$ 1 MeV), a symmetric, Dirac fermion\footnote{The cross section for annihilation of complex scalars to $e^+e^-$ pairs is $p$-wave suppressed, and while $p$-wave annihilation also leads to significant heating at $z \sim 20$, we defer a proper treatment of this process to future work.} millicharged DM has an unavoidable $s$-wave annihilation channel into $e^+e^-$, with a velocity-averaged cross section given by
 \begin{alignat}{1}
     \langle \sigma v \rangle = \frac{\pi \alpha_\text{EM}^2 \epsilon^2}{m_\chi^2} \sqrt{1 - \frac{m_e^2}{m_\chi^2}} \left(1 + \frac{m_e^2}{2m_\chi^2}\right),
 \end{alignat}
 where $\epsilon$ is the charge of the millicharged DM, and $\alpha_{\text{EM}}$ is the electromagnetic fine-structure constant. The tree-level annihilation to $\gamma \gamma$ also exists, but this is suppressed by a factor of $\epsilon^2$ relative to the annihilation to $e^+e^-$ and can be safely neglected. Raising $\epsilon$ therefore both increases the rate of DM-baryon scattering and the rate of DM annihilation to electrons, and the opposing effects on $T_m$ should be properly taken into account when considering the viability of this model. Since annihilation takes place between Dirac millicharged particles, the annihilation rate given in Eq.~(\ref{eqn:fz}) must have the additional factor of 1/2.

Fig.~\ref{fig:millicharged} show a plot of the $m_\chi - \epsilon$ parameter space of this model with several relevant constraints. Sufficiently light millicharged DM remains in thermal equilibrium with electrons and photons until after neutrinos decouple from the SM, altering the $N_\text{eff}$ measurement from the CMB power spectrum~\cite{Boehm:2013jpa}. Since the irreducible annihilation to $e^+e^-$ heats the baryons, by requiring $T_m(z = 17.2) \leq \SI{4}{\kelvin}$,\footnote{We choose 4 K in this section for consistency with existing results in the literature.} we can set an upper limit of $\epsilon \lesssim 5 \times 10^{-5}$, cutting the remaining parameter space down to a narrow window between $m_\chi \sim 10 - 100$ MeV and $\epsilon \sim 5 \times 10^{-6}$ to $5 \times 10^{-5}$. The region of parameter space that is ruled out is labeled ``Excess Heating'' in Fig.~\ref{fig:millicharged}, and is a new constraint set for the first time on this model. These limits are stronger than the conventional CMB power spectrum limits, since at large values of $\epsilon$, DM and baryons become strongly coupled early on, and the temperature evolution is mostly dominated by adiabatic cooling until structure formation; once structure formation starts, a small perturbation on the order of a few kelvins from millicharged DM annihilation to $e^+e^-$ is all that is required to raise the temperature above the EDGES measurement. Other experimental constraints set by the SLAC millicharge experiment \cite{Prinz:1998ua} and observations of the cooling of SN1987a \cite{Chang:2018rso} set limits that are already ruled out by a combination of the two limits shown, and have been left out.

We now consider an additional source of DM-related energy injection through $s$-wave annihilation or decay. This need not come from annihilation or decay of the millicharged DM itself; in principle, other particles in the dark sector could contribute such an energy injection. However, the existence of an additional annihilation channel for the millicharged DM could potentially allow it to obtain its correct relic abundance through thermal freezeout, since the cross section of the irreducible annihilation to $e^+e^-$ is too small in the allowed region for this to happen.

We set $f_{\chi,\text{ann}} = 0.01$ in Eq.~(\ref{eqn:energy_injection_21cm}) when including a new source of energy injection from annihilation. If the millicharged DM has an additional annihilation channel, the contours set an upper limit on the annihilation cross section of this channel. These results can also be rescaled for other values of $f_{\chi,\text{ann}}$, which may be useful for models where the dominant component of DM has an annihilation channel to the SM instead, assuming the dominant component has the same mass as the millicharged component. For decays, since the millicharged DM is expected to be stable, we choose $f_{\chi,\text{dec}} = 0.99$ corresponding to the remaining, dominant component of DM; the constraints would apply to decays of this component into the SM.

For $s$-wave annihilation, we find that $\langle \sigma v \rangle \lesssim 2 \times 10^{-25} \text{ cm}^3 \text{ s}^{ -1}$ for $\chi \overline{\chi} \to e^+e^-$ and $\langle \sigma v \rangle \lesssim 7 \times 10^{-24} \text{ cm}^3 \text{ s}^{ -1}$ for $\chi \overline{\chi} \to \gamma \gamma$. Since the cross section to produce the correct relic abundance of the millicharged DM with $f_{\chi,\text{int}} = 0.01$ is $\langle \sigma v \rangle \sim 6 \times 10^{-24} \text{ cm}^3 \, \text{s}^{ -1}$, it is unlikely that any additional source of $s$-wave annihilation to $e^+e^-$ (on top of the irreducible $s$-channel annihilation through the SM photon) can produce the correct relic abundance while remaining consistent with the EDGES $T_m$ measurement at $z \sim 20$. There is a small parameter space allowed for annihilation to photons to get the correct relic abundance without late-time suppression, but this requires a small branching ratio to electrons at the same time. 

\section{Conclusion}
\label{sec:conclusion}

We have computed the constraints that can be set on annihilating/decaying DM by a measurement of the 21-cm line of neutral hydrogen from the end of the cosmic dark ages. The recent claimed observation of an absorption trough by EDGES motivates the inclusion of some mechanism beyond the simplest scenario to explain the unexpectedly low inferred gas temperature; however, even if a future experiment found a weaker absorption signal, such additional mechanisms could still potentially be present and should be included to obtain conservative constraints.

We have considered three general scenarios that could weaken constraints from 21-cm observations on exotic energy injection from heating in the cosmic dark ages: (1) additional radiation backgrounds in the frequency range surrounding 21-cm, (2) non-standard recombination allowing the gas to decouple thermally from the CMB earlier, and (3) cooling of the gas through DM-baryon scattering. We have demonstrated that the strong-coupling limit of scenario (3) implements scenario (2) as a corollary, and that scenario (3) can generically weaken previously studied constraints on exotic energy injections from modifications to the ionization history during the cosmic dark ages.

We have mapped out the constraints on DM annihilation/decay in these three scenarios. We have found that in cases (2) and (3), there is an asymptotic behavior where the constraints become nearly independent of the redshift of decoupling (in case (2)) or the interaction cross section (in case (3)) for sufficiently early decoupling/large cross sections (see Fig.~\ref{fig:recomb_decay} and~\ref{fig:recomb_swave}). In these scenarios, we can thus present robust constraints that do not depend on the exact redshift of decoupling in case (2) or the size of the cross section in case (3). 

In the case where a small fraction of light DM (below 100 MeV) is millicharged and scatterings on this component are responsible for cooling of the gas, we have demonstrated that if this component has additional annihilation channels sufficient to obtain its relic density through thermal freezeout, then the energy injection from those channels will generically overheat the gas. Thus such a component would likely need to possess a non-thermal origin, or if a thermal relic, have annihilation channels in the early universe that are suppressed at late times (or have a large branching ratio for annihilation directly to neutrinos).

In Fig.~\ref{fig:summary} we summarize the constraints that can be obtained on keV-TeV DM annihilation or decay into $e^+ e^-$ pairs or photons, in these three scenarios, if the EDGES result is confirmed, and compare these limits with the Planck CMB constraints; other limits from indirect detection also exist for both channels, and may be more constraining at higher DM masses (e.g. \cite{Cohen:2016uyg,Fermi-LAT:2016uux}). These particles are the main stable, electromagnetically interacting byproducts of more general annihilation/decay channels (other than annihilation/decay directly to neutrinos), and consequently the constraints on more general channels can be estimated by combining these results. To set a limit in the case of additional radiation backgrounds, we assume that the effective radiation temperature $T_R$ at the 21-cm wavelength is not more than twice the temperature of the CMB at $z = 17.2$. 

In the case of DM-baryon scattering (scenario (3)), the cooling is only sufficient to reduce the gas temperature below 5.2K if the DM mass is below a certain critical scale (depending on the fraction of the DM that is interacting); consequently, the constraints cut off above a certain mass scale because even for zero energy injection from annihilation/decay, the proposed mechanism cannot explain the data. The other two scenarios are in principle viable at all DM mass scales. We find that in these scenarios, for decaying DM, these constraints would generically be stronger than previously derived early-universe bounds, and in the case of decay primarily to electrons (as is expected for sub-100-MeV DM), these limits are stronger by up to two orders of magnitude. For DM annihilating to electrons, the constraints in these scenarios are generally stronger than CMB-based limits for sub-GeV DM (without taking into account that the CMB constraints may weaken due to a modified ionization history).

\begin{figure}
    \centering
    \subfigure{
        \label{fig:summary_elec_decay}
        \includegraphics[scale=0.28]{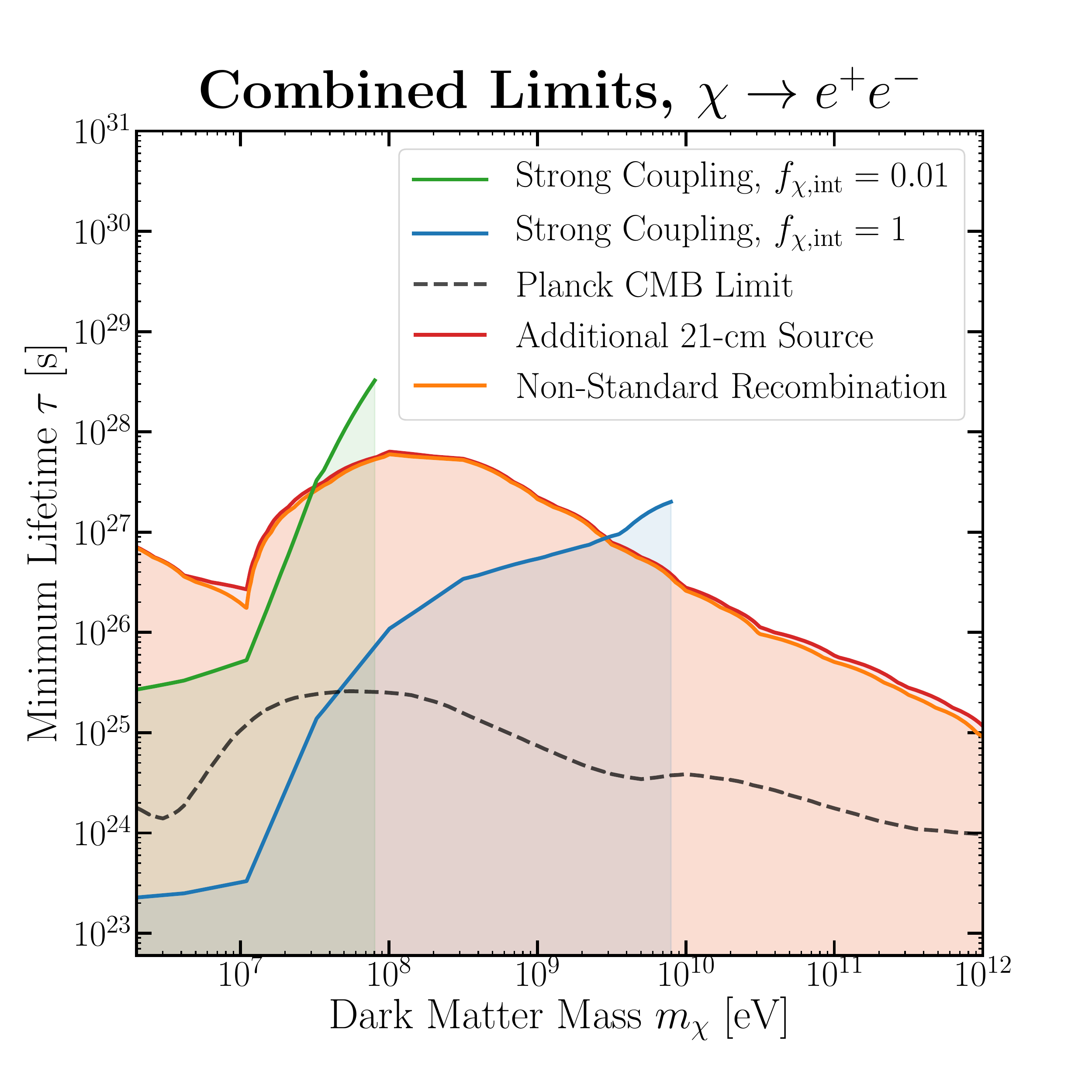}
    }
    \subfigure{
        \label{fig:summary_phot_decay}
        \includegraphics[scale=0.28]{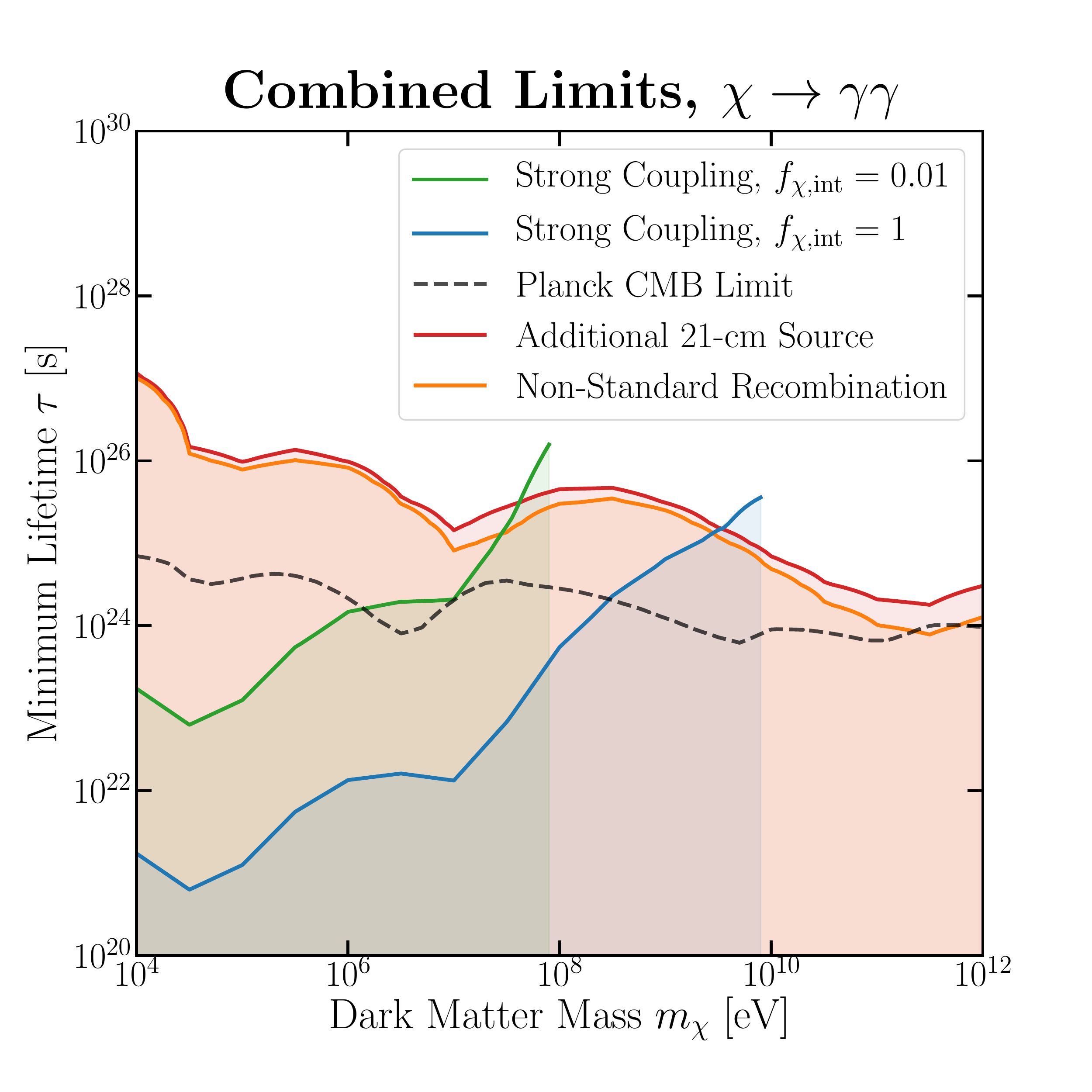}
    } \\
    \subfigure{
        \label{fig:summary_elec_swave}
        \includegraphics[scale=0.28]{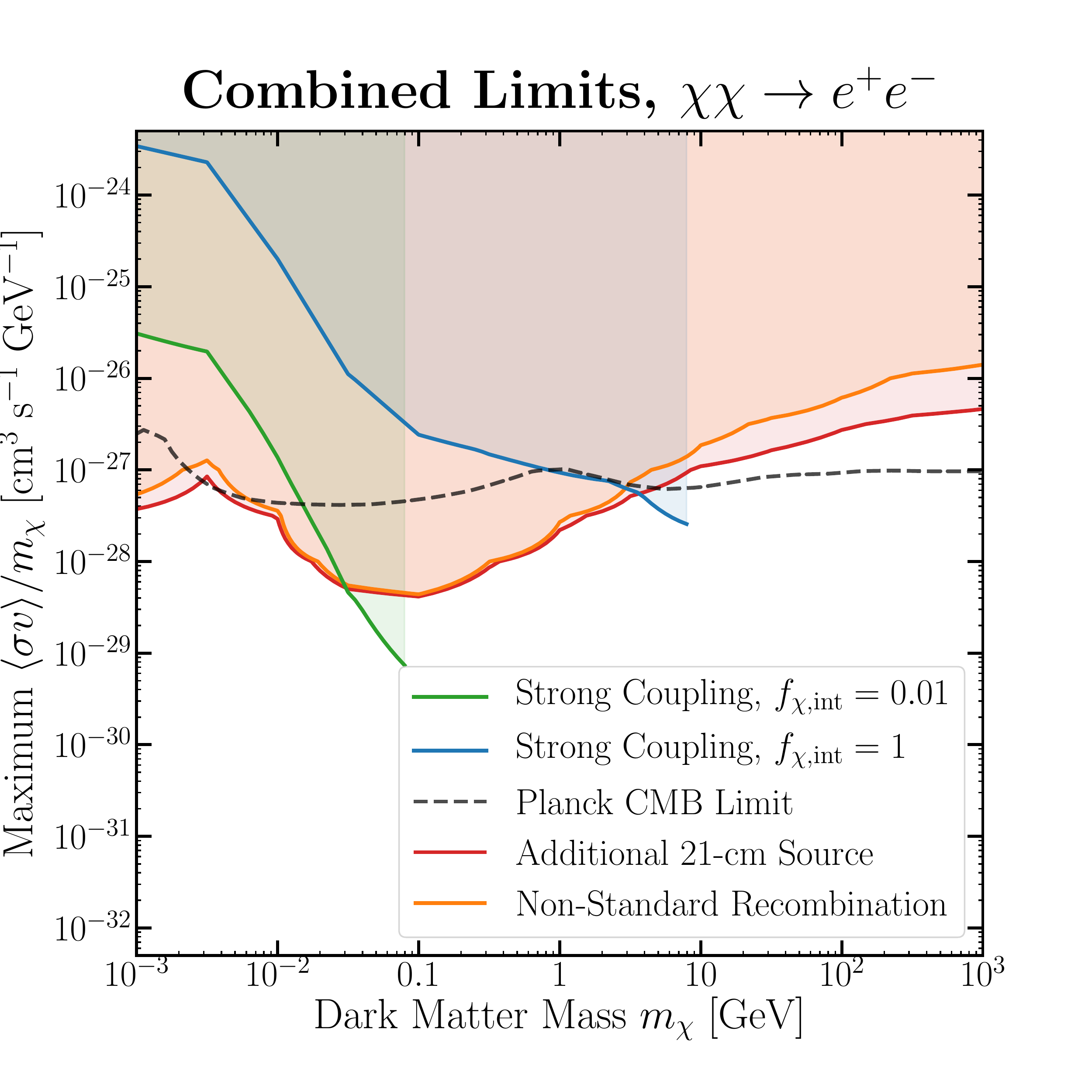}
    }
    \subfigure{
        \label{fig:summary_phot_swave}
        \includegraphics[scale=0.28]{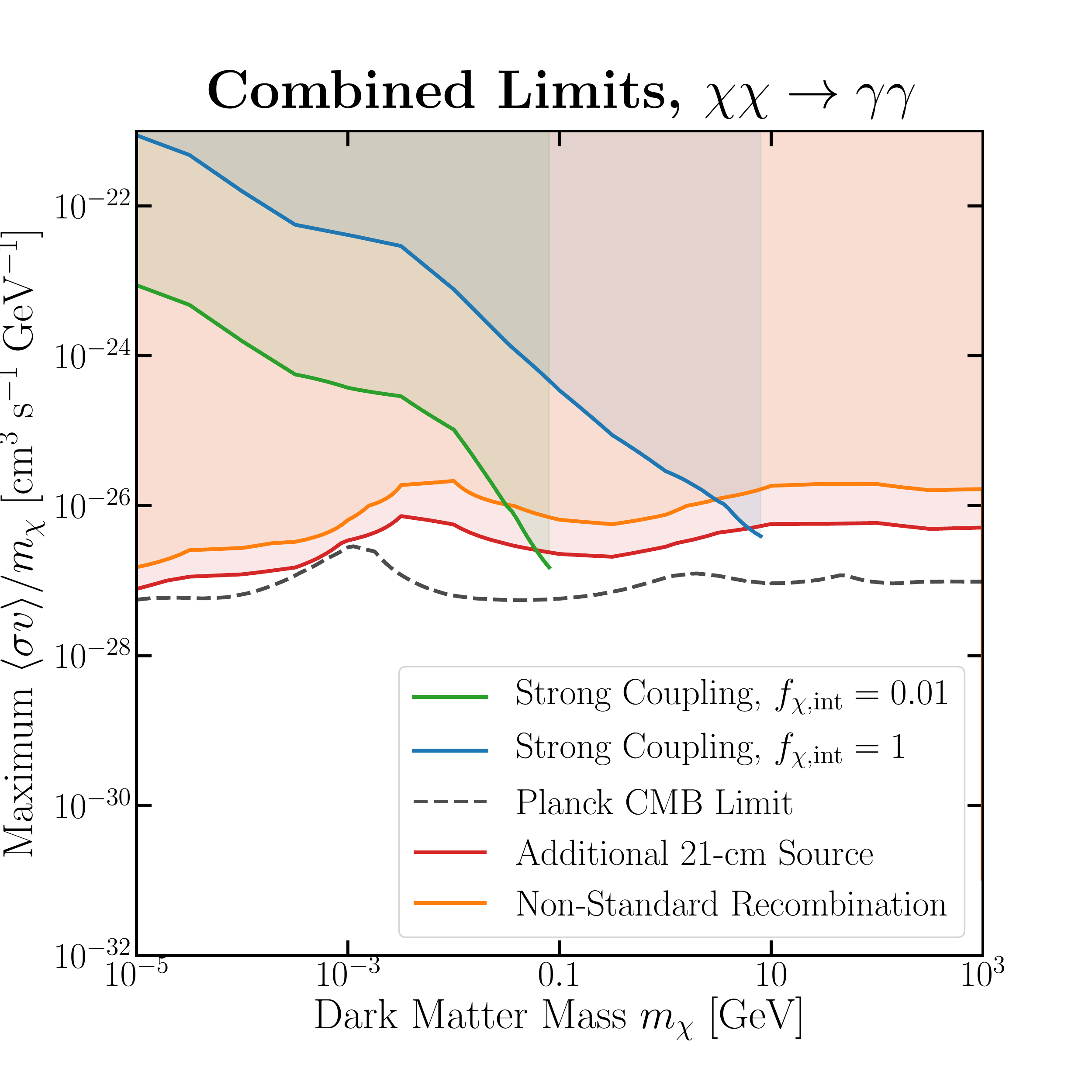}
    }
  \caption{Lower limits on the DM decay lifetime (upper panels) and upper limits of annihilation cross section (lower panels) from requiring $T_m/T_R(z=17.2) \le 0.105$ \cite{Bowman:2018yin}, for decay to $e^+e^-$ pairs (left panels) and photons (right panels). In the presence of an additional 21-cm radiation source with number density (at that frequency) smaller than or equal to that of the CMB number density, constraints are shown by the red solid line. In the limit of early baryon-photon decoupling, constraints are shown by the orange solid line. The solid green and solid blue lines  represent the constraints in the presence of DM-baryon scattering, in the limit of large cross section, for respectively 1\% and 100\% of the DM interacting with the baryons (these mechanisms cannot sufficiently cool baryons to match the data above critical mass scales, represented by the vertical cutoffs on the right-hand-side of the green/blue regions). The black dashed line represents previously derived constraints on the decay lifetime  \cite{Slatyer:2016qyl} (upper panels) or annihilation cross section (lower panels) \cite{Slatyer:2015jla} from measurements of the CMB.
  }
  \label{fig:summary}
\end{figure}

Simultaneous with and slightly after the release of this work, several other authors also studied the sensitivity of 21-cm measurements to DM annihilation or decays~\cite{Cheung:2018vww,Clark:2018ghm,Mitridate:2018iag}. In particular, the authors of~\cite{Clark:2018ghm,Mitridate:2018iag} set decay lifetime limits in a similar manner as~\cite{DAmico:2018sxd}, assuming an absorption signal that is smaller than the EDGES signal, with either $T_{21} = \SI{100}{\milli\kelvin}$ or \SI{50}{\milli\kelvin}. We reiterate that these limits are equivalent to our additional 21-cm source limits, with $(T_S/T_R)_\text{obs}= 0.26$ and 0.41 respectively, and setting $T_R = T_\text{CMB}$ at $z = 17.2$. Our work is more general than these other studies as we consider new effects that must be present to account for the large negative value of $T_{21}$ for the EDGES measurement. Consequently, our results are not merely a sensitivity study, and are immediately applicable to the various scenarios that have been suggested to explain the claimed EDGES detection. Even if future 21-cm measurements report a less negative value of $T_{21}$, the effects that we study here could potentially be present and are important to consider in setting future limits on DM annihilation and decay.

\newcolumntype{L}[1]{>{\raggedright\let\newline\\\arraybackslash\hspace{0pt}}m{#1}}
\newcolumntype{C}[1]{>{\centering\let\newline\\\arraybackslash\hspace{0pt}}m{#1}}
\newcolumntype{R}[1]{>{\raggedleft\let\newline\\\arraybackslash\hspace{0pt}}m{#1}}






\definecolor{codegreen}{rgb}{0,0.6,0}
\definecolor{codegray}{rgb}{0.5,0.5,0.5}
\definecolor{codepurple}{rgb}{0.58,0,0.82}
\definecolor{backcolour}{rgb}{0.95,0.95,0.92}

\lstdefinestyle{mystyle}{
    language = Python,  
    commentstyle=\color{codegreen},
    keywordstyle=\color{magenta},
    keywordstyle=[1]\color[rgb]{0,0,0.75},
    keywordstyle=[2]\color[rgb]{0.5,0.0,0.0},
    keywordstyle=[3]\color[rgb]{0.127,0.427,0.514},
    keywordstyle=[4]\color[rgb]{0.4,0.4,0.4},
    commentstyle=\color[rgb]{0.133,0.545,0.133},
    numberstyle=\tiny\color{codegray},
    stringstyle=\color{codepurple},
    basicstyle=\ttfamily,
    breakatwhitespace=false,         
    breaklines=true,                 
    captionpos=b,                    
    keepspaces=false,                 
    numbers=none,                    
    numbersep=5pt,                  
    showspaces=false,                
    showstringspaces=false,
    showtabs=false,                  
    tabsize=2,
    morekeywords={True, False, len},
    columns=flexible
}

\lstset{literate=%
   *{0}{{{\color{red!20!violet}0}}}1
    {1}{{{\color{red!20!violet}1}}}1
    {2}{{{\color{red!20!violet}2}}}1
    {3}{{{\color{red!20!violet}3}}}1
    {4}{{{\color{red!20!violet}4}}}1
    {5}{{{\color{red!20!violet}5}}}1
    {6}{{{\color{red!20!violet}6}}}1
    {7}{{{\color{red!20!violet}7}}}1
    {8}{{{\color{red!20!violet}8}}}1
    {9}{{{\color{red!20!violet}9}}}1
    {e-}{{{\color{red!20!violet}e-}}}2
    {e0}{{{\color{red!20!violet}e0}}}2
    {e1}{{{\color{red!20!violet}e1}}}2
    {e2}{{{\color{red!20!violet}e2}}}2
    {e3}{{{\color{red!20!violet}e3}}}2
    {e4}{{{\color{red!20!violet}e4}}}2
    {e5}{{{\color{red!20!violet}e5}}}2
    {e6}{{{\color{red!20!violet}e6}}}2
    {e7}{{{\color{red!20!violet}e7}}}2
    {e8}{{{\color{red!20!violet}e8}}}2
    {e9}{{{\color{red!20!violet}e9}}}2
    {e-}{{{\color{red!20!violet}e-}}}2
    {-0}{{{\color{red!20!violet}-0}}}2
    {-1}{{{\color{red!20!violet}-1}}}2
    {-2}{{{\color{red!20!violet}-2}}}2
    {-3}{{{\color{red!20!violet}-3}}}2
    {-4}{{{\color{red!20!violet}-4}}}2
    {-5}{{{\color{red!20!violet}-5}}}2
    {-6}{{{\color{red!20!violet}-6}}}2
    {-7}{{{\color{red!20!violet}-7}}}2
    {-8}{{{\color{red!20!violet}-8}}}2
    {-9}{{{\color{red!20!violet}-9}}}2
}
 
\lstset{style=mystyle}

\newcommand\beq{\begin{alignat}{1}}
\newcommand\eeq{\end{alignat}}
\newcommand{\dhis}{\texttt{DarkHistory} }

\newcommand*\bbar[1]{%
  \vbox{%
    \hrule height 0.5pt
    \kern-0.4ex
    \hbox{%
      \kern-0.2em
      \ifmmode#1\else\ensuremath{#1}\fi
      \kern-0.1em
    }
  }
}



\chapter{\texttt{DarkHistory}: A Code Package for Calculating Modified Cosmic Ionization and Thermal Histories with Dark Matter and Other Exotic Energy Injections}
\label{chap:DarkHistory}

\section{Introduction}
\label{sec:Introduction}

Dark matter annihilation or decay and other exotic sources of energy injection can significantly alter the ionization and temperature histories of the universe. In this chapter we describe a new public code package, \texttt{DarkHistory}, that allows fast and accurate computation of these possible effects of exotic energy injection on astrophysical and cosmological observables.
 
In particular, we will focus on interactions that allow dark matter (DM) to decay or annihilate into electromagnetically interacting Standard Model particles. This case has been studied extensively in the literature: stringent constraints on the dark matter annihilation cross section and decay lifetime have been derived from the way these Standard Model products would distort the anisotropies of the cosmic microwave background (CMB)~\cite{Slatyer:2009yq,Slatyer:2012yq,Slatyer:2016qyl,Kanzaki:2009hf}, or increase the temperature of the Inter-Galactic Medium (IGM), consequently affecting 21-cm and Lyman-$\alpha$ line emission \cite{Liu:2016cnk, Lopez-Honorez:2016sur, Liu:2018uzy, Diamanti:2013bia}. 

\dhis facilitates the calculation of these observables and the resulting constraints. In particular, \dhis makes the temperature constraint calculations significantly more streamlined, self-consistent, and accurate. It has a modular structure, allowing users to easily adjust individual inputs to the calculation -- e.g.\ by changing the reionization model, or the spectrum of particles produced by dark matter annihilation/decay. Compared to past codes developed for such analyses~\cite{Stocker:2018avm}, \dhis has a number of important new features:
\begin{itemize}
\item the first fully self-consistent treatment of exotic energy injection. Exotic energy injections can modify the evolution of the IGM temperature $T_\text{IGM}$ and free electron fraction $x_e$, and previously this modification has been treated perturbatively, assuming the backreaction effect due to these modifications on the cooling of injected particles is negligible. This assumption can break down toward the end of the cosmic dark ages for models that are not yet excluded \cite{Liu:2016cnk}. \texttt{DarkHistory} solves simultaneously for the temperature and ionization evolution and the cooling of the injected particles, avoiding this assumption;
\item a self-contained treatment of astrophysical sources of heating and reionization, allowing the study of the interplay between exotic and conventional sources of energy injection;
\item a large speed-up factor for computation of the full cooling cascade for high-energy injected particles (compared to the code employed in e.g.\ \cite{Liu:2016cnk}), via pre-computation of the relevant transfer functions as a function of particle energy, redshift and ionization level;
\item support for treating helium ionization and recombination, including the effects of exotic energy injections; and
\item a new and more correct treatment of inverse Compton scattering (ICS) for mildly relativistic and non-relativistic electrons; previous work in the literature has relied on approximate rates which are not always accurate.
\end{itemize}
Due to these improvements, \texttt{DarkHistory} allows for rapid scans over many different prescriptions for reionization, either in the form of photoheating and photoionization rates, or a hard-coded background evolution for $x_e$. The epoch of reionization is currently rather poorly constrained, making it important to understand the observational signatures of different scenarios, and the degree to which exotic energy injections might be separable from uncertainties in the reionization model. Previous attempts to model the effects of DM annihilation and decay into the reionization epoch have typically either assumed a fixed ionization history~\cite{Stocker:2018avm} -- requiring a slow re-computation of the cooling cascade if that history is changed \cite{Liu:2016cnk} -- or made an approximation for the effect of a modified ionization fraction on the cooling of high-energy particles~\cite{Lopez-Honorez:2013lcm,Diamanti:2013bia,Poulin:2015pna,Poulin:2016anj,Lopez-Honorez:2016sur}. 

Despite our emphasis on dark matter annihilation and decay, \texttt{DarkHistory} can be used to explore the effect of other forms of exotic particle injection. Other such possible sources include Hawking radiation from black holes~\cite{Poulin:2016anj,Clark:2018ghm}, radiation from accretion onto black holes~\cite{Hektor:2018qqw}, and processes from new physics such as de-excitation of dark matter or decay of meta-stable species~\cite{Hektor:2018lec}.

In Section~\ref{sec:histories} we review the physics of the ionization and temperature evolution, in the context of the three-level-atom (TLA) approximation, including the possibility of exotic energy injections. In Section~\ref{sec:code_structure} we discuss the overall structure of \texttt{DarkHistory}, which self-consistently combines the TLA evolution of the ionization and gas temperature with the cooling of particles injected by exotic processes. This section also describes the implementation of various physical processes in the code, in particular the treatment of cooling and production of secondaries by electrons and photons. In Section~\ref{sec:modules} we relate these processes to the various modules of \texttt{DarkHistory}, before providing a number of worked examples in Section~\ref{sec:examples}. We present our conclusions and discuss some future directions in Section~\ref{sec:conclusion}. We discuss our improved treatment of ICS in detail in Appendix~\ref{app:ICS}, provide the photon spectra from positronium annihilation in Appendix~\ref{app:positronium_annihilation_spec}, discuss a series of cross checks in Appendix~\ref{app:cross_checks}, and provide a table of definitions used throughout this chapter in Appendix~\ref{app:table}.

\section{Ionization and Thermal Histories}
\label{sec:histories}
\texttt{DarkHistory} computes the ionization and temperature evolution of the universe in the presence of an exotic source of energy injection, such as dark matter annihilation or decay, using a modified version of the three-level atom (TLA) model for both hydrogen and helium, based on \texttt{RECFAST}~\cite{Seager:1999km,Seager:1999bc}. The reader may refer to Ref.~\cite{AliHaimoud:2010dx} for a detailed derivation of the unmodified TLA equations with hydrogen only, and Refs.~\cite{Seager:1999km,Seager:1999bc,Wong:2007ym} for the treatment of helium recombination in \texttt{RECFAST}. A simplified treatment neglecting the evolution of helium has already been discussed in Sec.~\ref{sec:energy_deposition_early_universe}; we defer a detailed discussion of our treatment of helium to Sec.~\ref{sec:helium}. 

In Eq.~(\ref{eqn:fz}), we saw that once the cooling of injected primary particles is determined, the energy deposited into channel $c$ (hydrogen ionization, excitation, or heating) can be parametrized by $f_c(z)$. In fact, energy deposition is also dependent on the ionization fractions of all of the relevant species in the gas, which we denote $\mathbf{x} \equiv (x_\text{HII}, x_\text{HeII}, x_\text{HeIII})$. When helium is neglected, the ionization dependence of these $f_c(z,\mathbf{x})$ functions simplifies to a dependence on $x_\text{HII} = x_e$. These $f_c$ functions also depend on the energies and species of the injected particles, but for simplicity of notation we will not write these arguments explicitly. 


Prior to this work, $f_c(z, \mathbf{x})$ has largely been computed assuming the standard ionization history computed by recombination codes $\mathbf{x}_\text{std} (z)$, essentially making $z$ the only independent variable of $f_c$ as a function. These calculations are therefore applicable only so long as any perturbations to the assumed ionization history (e.g.\ by additional sources of energy injection) are sufficiently small. This is generally a good approximation near recombination: at these redshifts, the ionization history is well-constrained by CMB power spectrum measurements, and therefore large perturbations to $x_e$ are highly disfavored. For $z \lesssim 100$, however, ionization levels that exceed the standard value of $x_e \sim 2 \times 10^{-4}$ by several orders of magnitude are experimentally allowed~\cite{Liu:2016cnk}. Moreover, star formation during the process of reionization rapidly ionizes and heats the universe at $z \lesssim 20$, causing the ionization and thermal history to diverge from the baseline histories.

The primary effect of an increase in ionization levels is to decrease the number of neutral hydrogen and helium atoms available to ionize, decreasing the fraction of injected power that goes into ionization of these species; on the other hand, increasing $x_e$ increases the number of charged particles available for low-energy electrons to scatter off and heat the IGM, increasing the fraction of power going into heating. Since energy injection processes generally increase $x_e$ with time, the power into heating increases at an accelerated rate at late times, making a proper calculation of $f_c(z, \mathbf{x})$ crucial for an accurate computation of the temperature history.

Computing the full $\mathbf{x}$-dependence of $f_c(z, \mathbf{x})$ also allows us to perform, for the first time, a consistent calculation of the temperature and ionization histories with both exotic energy injection processes and reionization. At the onset of reionization, stars begin to form, and the ionizing radiation emitted by these objects injects a large amount of energy into the IGM. There remains a large degree of uncertainty regarding how reionization proceeds, but given some model for the photoionization and photoheating rates, and including other important energy transfer processes such as collisional ionization and excitation, additional terms $\dot{T}_m^\text{re}$ and $\dot{x}_\text{HII}^\text{re}$ (as well as the corresponding terms for helium) can be included in Eq.~(\ref{eqn:TLA}) to model reionization. These terms are discussed in much greater detail in Sec.~\ref{sec:TLA_and_reionization}. 

To summarize, \texttt{DarkHistory} computes the ionization and thermal history in the presence of exotic sources of energy injection, with the evolution equations in the absence of helium given by

\begin{alignat}{1}
    \dot{T}_m &= \dot{T}_m^{(0)} + \dot{T}_m^{\text{inj}} + \dot{T}_m^\text{re} \,, \nonumber \\
    \dot{x}_\text{HII} &= \dot{x}_\text{HII}^{(0)} + \dot{x}_\text{HII}^{\text{inj}} + \dot{x}_\text{HII}^\text{re} \,.
    \label{eqn:TLA_DarkHistory}
\end{alignat}
In the rest of the chapter, we will describe how we calculate the inputs required to integrate these equations, i.e.\ $f_c(z,\mathbf{x})$, $\dot{T}_m^\text{re}$, $\dot{x}_\text{HII}^\text{re}$ and the modifications necessary to include helium.

\section{Code Structure and Content}
\label{sec:code_structure}

In this section we discuss the structure and physics content of the \texttt{DarkHistory} package.

\subsection{Overview}
\label{sec:overview}

Fig.~\ref{fig:flowchart} shows a flowchart depicting the overall structure of \texttt{DarkHistory}. The overall goal of the code is to take in some injected spectrum of photons and electron/positron pairs at a given redshift, and partition the energy into several categories as they lose their energy over a small redshift step: 

\begin{enumerate}

    \item \textit{High-energy deposition}. This is the total amount of energy deposited into ionization, excitation and heating by any high-energy (above \SI{3}{\kilo\eV}) electron generated during any of the cooling processes;

    \item \textit{Low-energy electrons}. These are electrons that have kinetic energy below \SI{3}{\kilo\eV} where atomic cooling processes typically dominate over ICS after recombination. These electrons are separated out at each step in order to treat their energy deposition (which occurs in a timescale much shorter than the time step) more carefully;

    \item \textit{Low-energy photons}. These are photons with energies below \SI{3}{\kilo\eV} that either photoionize within the redshift step, or lie below \SI{13.6}{\eV}. Such photons either lose all their energy within the redshift step, or cool only through redshifting, and thus can be treated in a simplified manner; and

    \item \textit{Propagating photons}. These are photons that are present at the end of the redshift step and are not included in the low-energy photons category.
    
\end{enumerate}

Throughout the chapter, we use the word ``electrons'' to refer to both electrons and positrons. Although the interactions of electrons and positrons with the gas differ, the ICS cross-sections are identical, and ICS dominates the energy losses down to energy scales where the positron is nonrelativistic \cite{MEDEAII}.  For nonrelativistic positrons, their mass energy is converted into photons through annihilation with electrons. Since the positron mass is much larger than the kinetic energy in this regime, neglecting differences in kinetic energy loss between electrons and positrons is unlikely to be important. In a future version of \dhis we plan to include a more sophisticated treatment of low energy electrons and positrons.

The outputs in the first three categories are used to compute the evolution of the ionization and temperature history at this redshift step, before the code moves on to the next step and performs the same calculation again. A brief description of a step in this loop is as follows:

\begin{figure*}[t!]
    \centering
    \includegraphics[scale=0.3]{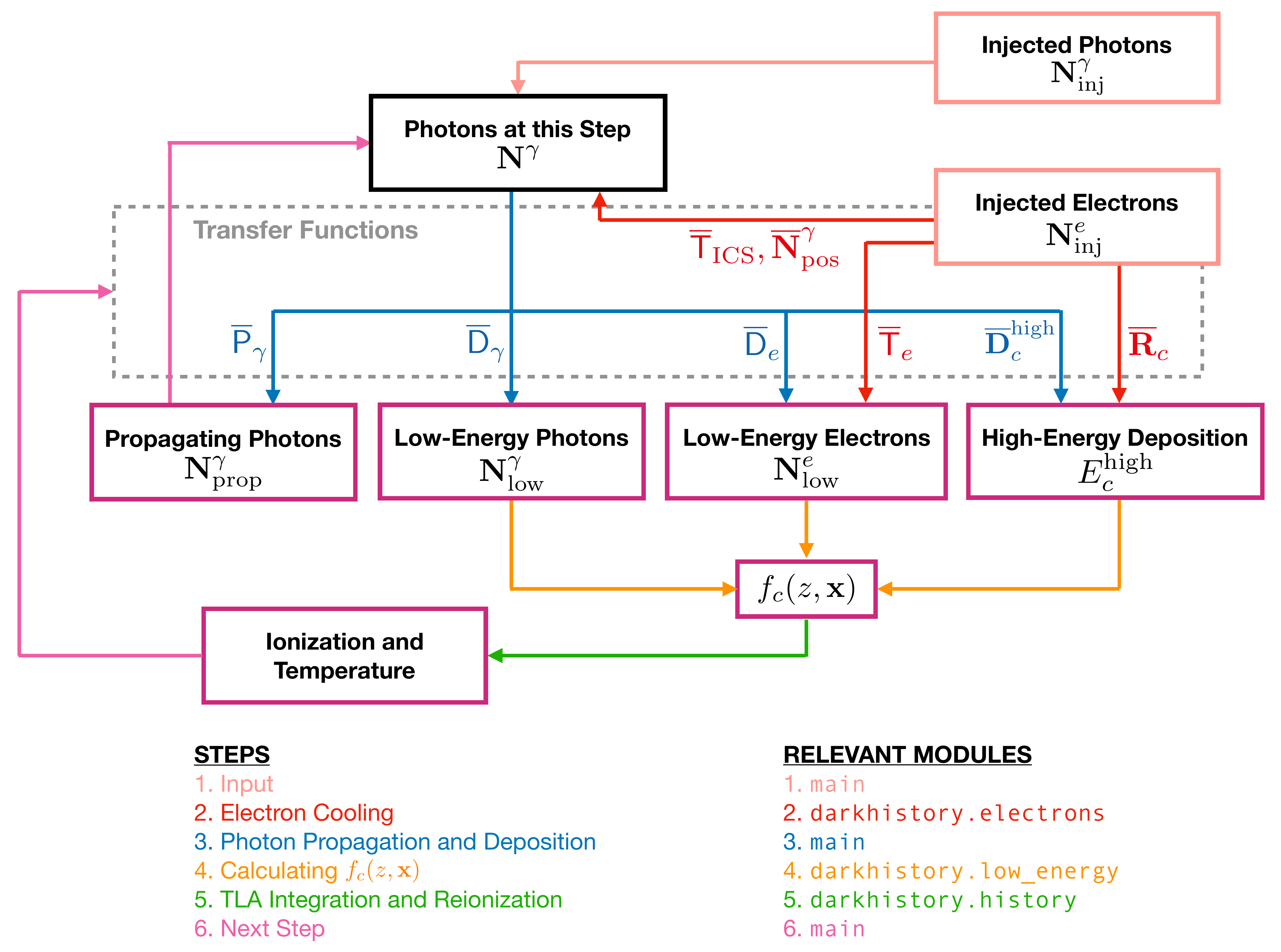}
    \caption{Flowchart showing schematically how the calculation of ionization and thermal histories in \texttt{DarkHistory} proceeds. Solid boxes represent input spectra (light pink), intermediate spectra used in calculations (black) and output spectra and quantities (purple), while arrows indicate numerical calculations that take place within the corresponding color-coded modules. The dashed grey box encloses all of the transfer functions for electron cooling (blue) and photon propagation and deposition (red), which are defined in Sec.~\ref{sec:electron_cooling} and~\ref{sec:photon_cooling} respectively. The calculation of $f_c(z)$ (orange) and the integration of the TLA (green) are explained in Sec.~\ref{sec:calculating_f} and~\ref{sec:TLA_and_reionization} respectively. Propagating photons and ionization/temperature values, which are used in calculating the transfer functions, are used as inputs for the next step (purple). All notation used here are defined in the text, and a summary table with their definitions can be found in Appendix~\ref{app:table}. Each step is outlined in Sec.~\ref{sec:overview}, and then explained in more detail in subsequent subsections within Sec.~\ref{sec:code_structure}. The modules shown here will also be outlined in Sec.~\ref{sec:modules}.}
    \label{fig:flowchart}
\end{figure*}

\begin{enumerate}
    \item \textit{Input}. Before the code begins, the user specifies a DM energy injection model or some other redshift-dependent energy injection rate, as well as the photon and $e^+e^-$ spectra produced per energy injection event. By default, \texttt{DarkHistory} starts from an initial redshift of $1+z = 3000$, ensuring that the spectra of particles present at and after recombination (at $z \sim 1000$) are accurate. Details are provided in Sec.~\ref{sec:input}. Inputs to the code are provided to the function \texttt{evolve()} found in the module \texttt{main}; some tools for obtaining spectra from an arbitrary injection of Standard Model particles can be found in the \texttt{pppc} module;

    \item \textit{Injected electron cooling}. Injected electrons (and positrons) cool through a combination of atomic processes and ICS. Transfer functions that map these injected electrons to high-energy deposition, secondary photons from ICS and positron annihilation, and low-energy electrons are computed and applied to the injected electrons. A discussion of these calculations can be found in Sec.~\ref{sec:electron_cooling} and in the \texttt{electrons} module of the code.
    
    The sum of the secondary photons produced by electron cooling, photons injected on this timestep, and propagating photons from the previous timestep are used as input to the photon cooling transfer functions, which we describe next;

    \item \textit{Photon propagation and energy deposition}. At this stage, we have a spectrum of photons that can undergo a range of  cooling processes to lose their energy over this redshift step. The effect of these cooling processes on the photon spectrum can be reduced to three transfer functions that we will describe in detail in Sec.~\ref{sec:photon_cooling}. These transfer functions have been pre-computed separately and can be downloaded at \href{https://doi.org/10.7910/DVN/DUOUWA}{https://doi.org/10.7910/DVN/DUOUWA}, together with all the other data required to run the code. These transfer functions determine how photons in this redshift step turn into propagating photons that continue on to the next redshift step, and low-energy photons and low-energy electrons that undergo further processing. All of these computations occur in the \texttt{main} module;

    \item \textit{Calculating }$f_c(z, \mathbf{x})$. The low-energy photons and low-energy electrons from this redshift step deposit their energy into ionization, heating and excitation of atoms, and the value of $f_c(z, \mathbf{x})$ at this step is computed by comparing the energy deposited in each channel to the energy injection rate for this timestep. Details of this computation are given in Sec.~\ref{sec:calculating_f}, and can be found in the \texttt{low\_energy} module;

    \item \textit{TLA integration and reionization}. With $f_c(z,\mathbf{x})$ at this step, we can now integrate the TLA across this redshift step. We can also include a reionization model, or track helium ionization, both of which add more terms to the TLA, as detailed in Sec.~\ref{sec:TLA_and_reionization}. We now know the $\mathbf{x}$ and $T_m$ that are reached at the end of this step. These calculations are done in the \texttt{history} module; and

    \item \textit{Next step}. The $\mathbf{x}$ and $T_m$ values computed above are passed to the next redshift step, so that all transfer functions at the next step can be computed at the appropriate ionization level. The propagating photons found above are also passed to the next step, and the loop repeats.
\end{enumerate}

Because $f_c(z, \mathbf{x})$ is computed by integrating the TLA at each step, and all transfer functions are evaluated at the value of $\mathbf{x}$ in the step, the backreaction of increased ionization levels is now fully accounted for. 

In the next several sections, we will describe both the physics and numerical methods that go into the loop.

\subsection{Discretization}
\label{sec:discretization}

Before describing in detail each part of \texttt{DarkHistory}, we will first describe how discretization occurs in our code, and the notation we will use throughout this chapter. Typically, we will deal with some smooth spectrum of particles $dN/dE(E, A, B, \cdots)$, which is a function of the energy abscissa $E$, and several other variables that we denote here as $A, B, \cdots$. Smooth functions that are derivatives will always use `$d$' to denote differentiation, and parentheses to denote functional dependence. We shall always discretize such spectra as
\begin{alignat}{1}
    \frac{dN}{dE}(E_i, A_j, B_k, \cdots) \approx \mathsf{S}[E_i, A_j, B_k, \cdots] \,.
    \label{eqn:discretize_dNdE}
\end{alignat}
The discretized spectrum $\mathsf{S}$ is a matrix of dimension equal to the number of variables it depends on, where $i,j,k,...$ index discrete values of these variables. Throughout this chapter, we will denote vectors (quantities which depend on a single variable) by a bold typeface and matrices (quantities that depend on multiple variables) by a sans-serif typeface. Discrete steps or changes are denoted by `$\Delta$', and discrete functional dependencies are written in square brackets. 

$\mathsf{S}$ times the bin width should always be regarded as a matrix of number of particles inside some bin, all with energy given by $E_i$. This matrix is mathematically defined as
\begin{alignat}{1}
    \mathsf{N}[E_i, A_j, B_k, \cdots] \equiv \mathsf{S}[E_i, A_j, B_k, \cdots] \times E_i \Delta \log E_i \,,
    \label{eqn:discretize_N}
\end{alignat}
where $\Delta \log E_i$ is the log-energy bin width. We will always take $E_i \Delta \log E_i$ to be the bin width by convention. In \texttt{DarkHistory}, spectra are binned into energy values that are evenly log-spaced. $E_i$ should be regarded as the bin center, with the bin boundaries occurring at the geometric mean of adjacent energy values, and the boundaries of the first and last bin are taken to be symmetric (in log-space) about the bin centers. 

\subsection{Input}
\label{sec:input}

To initialize the loop described above, the user must specify the discretized photon and electron spectra produced per injection event, which we denote $\overline{\mathbf{N}}^\gamma_\text{inj}[E_j']$ and $\overline{\mathbf{N}}^e_\text{inj}[E_j']$. Bars denote spectra or transfer functions that have been normalized by some process or quantity, while spectra without any markings denote a number of particles per baryon from here on, unless otherwise specified.

Given the redshift-dependent rate of injection events per volume $\left(dN/dV \, dt\right)^\text{inj}$ we can determine the spectrum of particles $\mathbf{N}_\text{inj}^\alpha$ injected within a log-redshift step of width $\Delta \log (1+z)$ per baryon by
\begin{alignat}{1}
	\mathbf{N}^\alpha_\text{inj}[E_i', z] = \overline{\mathbf{N}}^\alpha_\text{inj}[E_i'] \left(\frac{dN}{dVdt}\right)^\text{inj} G(z) \, ,
  \label{eqn:injected_discretized_spec}
\end{alignat}
where $\alpha$ take on values $\gamma$ or $e$, and 
\begin{alignat}{1}
  G(z) \equiv \frac{\Delta \log(1+z)}{n_B(z) H(z)} \,,
  \label{eqn:per_baryon_to_dVdt}
\end{alignat}
where $n_B$ is the number density of baryons. $G(z)$ converts between the rate of injection events per volume and the number of injection events per baryon in the log-redshift step.

In the following sections, we will be mostly concerned with log-redshift steps, and so it is convenient to define
\begin{alignat}{1}
  y \equiv \log(1+z) \,,
\end{alignat}
and likewise $\Delta y \equiv \Delta \log (1+z)$.

\subsection{Injected Electron Cooling}
\label{sec:electron_cooling}

After specifying the injected spectra, the next step of the code is to resolve the injected electron/positron pairs, $\mathbf{N}^e_\text{inj}$. High-energy electrons and positrons cool through atomic processes (collisional ionization, collisional excitation and Coulomb heating), as well as ICS off CMB photons. After losing their kinetic energy to these processes, positrons ultimately annihilate with free electrons in the IGM, producing high-energy photons. All of these processes occur within a timescale much shorter than the timesteps considered in \texttt{DarkHistory}. Because of this, the code converts all input high-energy electrons into energy deposited into ionization, excitation, heating, scattered photons from ICS, and low-energy electrons (below \SI{3}{\kilo \eV}), which we treat separately. The photons produced from ICS are added to those that are injected promptly from the DM energy injection process, as well as propagating photons from the previous step.

We will first briefly discuss our calculation of the scattered photon and electron spectra from ICS, and then move on to describe the numerical method used to compute electron cooling.  

\subsubsection{Inverse Compton Scattering}
\label{sec:ICS}

ICS off CMB photons is an important energy loss mechanism for electrons/positrons over a large range of energies and redshifts. The efficiency of ICS as a cooling mechanism relative to atomic cooling processes has been the subject of some confusion in the literature, with some earlier studies~\cite{Valdes:2009cq,Evoli:2012zz} underestimating the cooling rate of the electrons. ICS becomes more important relative to atomic processes as the electron energy increases, but a correct treatment shows that even nonrelativistic electrons can have ICS as the main cooling mechanism in the early universe; at $z \sim 600$, for example, it is the primary energy loss mechanism for electrons with kinetic energy $\gtrsim \SI{10}{\kilo\eV}$~\cite{Galli:2013dna,Slatyer:2015kla}. Existing work on electron cooling has focused on the highly nonrelativistic regime (electron kinetic energy below \SI{3}{\kilo\eV})~\cite{Furlanetto:2006jb}, where ICS is unimportant compared to atomic cooling processes, or on the relativistic regime~\cite{Valdes:2009cq,Evoli:2012zz,Hansen:2003yj}. 

Earlier work by one of the authors~\cite{Slatyer:2012yq,Slatyer:2015kla} already incorporates ICS cooling for electrons across both the Thomson and the relativistic regimes. \texttt{DarkHistory} improves the accuracy of the calculation in the Thomson regime by using the full expression for the spectrum of scattered photons, with no further approximation. As a result, the code is able to accurately calculate the scattered photon spectrum and the energy loss spectrum of electrons. This means that we fully cover all relevant regimes for ICS for electrons of arbitrary energy scattering off the CMB at all redshifts $z \sim 10^9$ and below.\footnote{Above this redshift, photons have energies comparable to the electron mass $m_e$, and Klein-Nishina scattering can occur between photons and non-relativistic electrons, which falls outside of the two regimes considered here.} These calculations are fast and numerically stable even for nonrelativistic electrons, where conventional numerical integration can be unreliable due to the presence of catastrophic cancellations between large terms. 

We leave a full discussion of how \texttt{DarkHistory} treats ICS to Appendix~\ref{app:ICS}. In summary, the code is able to compute the scattered photon and electron spectra that are produced per unit time due to ICS off the CMB across all relevant kinematic regimes. These spectra are then taken as inputs for the numerical computation of how an electron cools taking into account all processes, which is described below.

\subsubsection{Numerical Method}
\label{sec:elec_cooling_numerical_method}

Consider an injected electron (or positron) with kinetic energy $E'$ (all quantities associated with injected particles throughout this chapter will be denoted with $'$).
Let $R_c(E')$ be the energy eventually deposited into some channel $c$ by this electron, once it has lost all of its initial energy. Within a short time interval $\Delta t$ (taken to be \SI{1}{\second} in our calculation), the electron undergoes all possible cooling processes with some probability, producing the (averaged) secondary electron spectrum $dN/dE$. Within this same interval $\Delta t$, some portion of the energy $P_c(E')$ is also deposited promptly into the channel under consideration. The secondary electron spectrum then deposits its energy according to $R_c$ for energies lower than $E'$. We can thus write the following recursive equation:
\begin{alignat}{1}
    R_c(E') = \int dE \, R_c(E) \frac{dN}{dE} + P_c(E') \,.
    \label{eqn:elec_cooling_analytic}
\end{alignat}
Note that $R_c(E')$ does not include deposition to the channel $c$ via secondary photons from ICS or positron annihilation; because the cooling times of secondary photons can be much longer than a timestep, they must be treated separately. $R_c(E')$ as defined here is the ``high-energy deposition'' from electrons within the timestep, as described in Section~\ref{sec:overview}. The relevant channels are $c = \{\text{`ion'}, \text{`exc'}, \text{`heat'} \}$ for deposition into collisional ionization, collisional excitation and heating respectively. The `ion' and `exc' channels include ionization and excitation off all species.

As long as the time step $\Delta t$ is much shorter than the characteristic interaction timescale of all of the interactions, $dN/dE$ is simply the sum of all of the scattered electron spectra due to each process within $\Delta t$, normalized to a single injected electron. A detailed accounting of the relevant cross sections and secondary spectra is provided in Ref.~\cite{Slatyer:2009yq}, and these results can be used to calculate $dN/dE$ and $P_c$. We will denote the discretized version of the normalized scattered electron spectra by $\overline{\mathsf{N}}$, since it is normalized to one electron. 

Numerically, we would like to compute $\overline{\mathbf{R}}_c$, a vector containing the energy deposited into channel $c$, with each entry corresponding to a single electron with initial kinetic energy $E'$. The overline notation serves as a reminder that the quantity is normalized to one injected electron. The discretized version of Eq.~(\ref{eqn:elec_cooling_analytic}) reads
\begin{alignat}{1}
    \overline{\mathbf{R}}_c[E'_i] = \sum_j \overline{\mathsf{N}}[E'_i, E_j] \overline{\mathbf{R}}_c[E_j] + \overline{\mathbf{P}}_c[E'_i] \,,
    \label{eqn:elec_cooling_dep_tf}
\end{alignat}
where $\overline{\mathbf{P}}_c$ is the vector of the prompt energy deposition in channel $c$ per electron. This is a linear system of equations, and we can solve for each $\overline{\mathbf{R}}_c$ given $\overline{\mathsf{N}}$ and $\overline{\mathbf{P}}_c$.

A similar procedure also works for finding the ICS photon spectrum after an electron completely cools. Let the discretized spectrum be $\overline{\mathsf{T}}_{\text{ICS},0} [E_{e,i}', E_{\gamma,j}]$, where $E'_e$ is the initial electron kinetic energy, and $E_\gamma$ is the photon energy. Then the ICS photon spectrum produced after complete cooling of a single electron satisfies
\begin{alignat}{2}
    \overline{\mathsf{T}}_{\text{ICS},0} [E'_{e,i}, E_{\gamma,j}] &=&&  \sum_k\overline{\mathsf{N}}[E_{e,i}', E_{e,k}] \overline{\mathsf{T}}_{\text{ICS},0} [E_{e,k}, E_{\gamma, j}] + \overline{\mathsf{N}}_\text{ICS} [E_{e,i}', E_{\gamma,j}] \,,
    \label{eqn:ics_photons}
\end{alignat}
with $\overline{\mathsf{N}}_\text{ICS}$ being the discretized version of the scattered photon spectrum defined in Eq.~(\ref{eqn:ics_scat_phot_spec}) within $\Delta t$, and indices $e$ and $\gamma$ have been inserted to clarify the difference between electron and photon energies. This spectrum consists of CMB photons that are upscattered by the injected electron; in order to be able to track energy conservation, we also need to keep track of the initial energy of the upscattered photons. We therefore also need to solve
\begin{alignat}{1}
  \overline{\mathbf{R}}_\text{CMB}[E_i'] = \sum_j \overline{\mathsf{N}}[E_i', E_j] \overline{\mathbf{R}}_\text{CMB}[E_j] + \overline{\mathbf{P}}_\text{CMB} [E_i'] \,,
  \label{eqn:elec_cooling_cmbloss}
\end{alignat}
where $\overline{\mathbf{P}}_\text{CMB}$ is the total initial energy of photons upscattered in $\Delta t$.\footnote{We do not have to track the photon spectrum, since the initial CMB photon energy is only significant for nonrelativistic injected electrons, which are always in the Thomson regime and hence scatter in a frequency-independent manner. For relativistic electrons, the initial CMB photon energy is neglected, as the photon is overwhelmingly upscattered to a much higher final energy.} At this point, we now define $\overline{T}_\text{ICS}$ to be the ICS photon spectrum with the upscattered CMB spectrum subtracted out, so that $\overline{T}_\text{ICS}$ now represents a \textit{distortion} to the CMB spectrum: 
\begin{alignat}{2}
  \overline{\mathsf{T}}_\text{ICS} [E'_{e,i}, E_{\gamma,j}] &=&& \overline{\mathsf{T}}_{\text{ICS},0} [E'_{e,i}, E_{\gamma,j}] - \overline{\mathbf{R}}_\text{CMB}[E_{e,i}'] \overline{\mathbf{N}}_\text{CMB}[E_{\gamma,j}]\,,
\label{eqn:elec_cooling_ics}
\end{alignat}
where $\overline{\mathbf{N}}_\text{CMB}$ is the CMB spectrum normalized to unit total energy. The total of energy of $\overline{T}_\text{ICS}$ for each $E_{e,i}'$ therefore gives the energy lost by the incoming electron through ICS.

Finally, the low-energy electron spectrum produced is similarly given by
\begin{alignat}{2}
    \overline{\mathsf{T}}_e [E'_{e,i}, E_{e,j}] &=&& 
    \sum_k\overline{\mathsf{N}}_\text{high}[E_{e,i}', E_{e,k}] \overline{\mathsf{T}}_e [E_{e,k}, E_{e, j}] + \overline{\mathsf{N}}_\text{low} [E_{e,i}', E_{e,j}] \,,
    \label{eqn:elec_cooling_lowengelec}
\end{alignat}
where $\overline{\mathsf{N}}_\text{high}$ ($\overline{\mathsf{N}}_\text{low}$) is $\overline{\mathsf{N}}$ with only high-energy (low-energy) $E_{e,k}$ included. 

In \texttt{DarkHistory}, we choose a square matrix $\overline{\mathsf{N}}$ with the same abscissa for both injected and scattered electron energies. As a result, $\overline{\mathsf{N}}$ has diagonal values that are very close to 1, since most particles do not scatter within $\Delta t$. Because of this, we find that it is numerically more stable to solve the equivalent equation
\begin{alignat}{1}
  \frac{\widetilde{E}[E_i']}{E_i'} \overline{\mathbf{R}}_c[E_i'] = \sum_j \widetilde{\mathsf{N}}[E_i', E_j] \overline{\mathbf{R}}_c[E_j] + \overline{\mathbf{P}}_c [E_i'] \,,
  \label{eqn:elec_cooling_actual}
\end{alignat}
where
\begin{alignat}{1}
  \widetilde{\mathsf{N}}[E_i', E_j] &\equiv \begin{cases}
    \overline{\mathsf{N}}[E_i', E_j] \,, & E_i' < E_j \,, \\
    0 \,, & \text{otherwise},
  \end{cases}  \\ \nonumber \\ \widetilde{E}[E_i'] &\equiv \sum_j \widetilde{\mathsf{N}} [E_i', E_j] E_j + \sum_c \overline{\mathbf{R}}_c[E_i'] + \sum_j \overline{\mathsf{T}}_\text{ICS} [E_i', E_{\gamma,j}] E_{\gamma,j} \,.
\end{alignat}
The variables $\widetilde{\mathsf{N}}$ and $\widetilde{E}$ are simply the number of electrons and total energy excluding electrons that remained in the same energy bin after $\Delta t$. Eqs.~(\ref{eqn:elec_cooling_ics}) and~(\ref{eqn:elec_cooling_lowengelec}) can be similarly transformed in the same way as Eq.~(\ref{eqn:elec_cooling_actual}) and solved. Since $\widetilde{N}$ is a triangular matrix, the SciPy function \texttt{solve\_triangular()} is used for maximum speed.\footnote{The upscattering of electrons during ICS is negligible: see Appendix~\ref{app:ICS} for more details.}

Having calculated $\overline{\mathbf{R}}_c$, $\overline{\mathsf{T}}_\text{ICS}$ and $\overline{\mathsf{T}}_e$, all normalized to a single electron, the final result when an arbitrary electron spectrum $\mathbf{N}^e_\text{inj}[E'_{e,i}]$ completely cools is simply given by contracting these quantities with $\mathbf{N}^e_\text{inj}$. Note that all of these quantities are also dependent on redshift: we have simply suppressed this dependence for notational simplicity in this section.

Finally, after positrons have lost all of their kinetic energy, they are assumed to form positronium and annihilate promptly, producing a gamma ray spectrum that also gets added to the propagating photon spectrum. The positronium spectrum is given simply by
\begin{alignat}{1}
  \mathbf{N}^\gamma_\text{pos}[E_i] = \frac{1}{2} \overline{\mathbf{N}}^\gamma_\text{pos}[E_i] \sum_j \mathbf{N}^e_\text{inj}[E'_{j}] \,,
  \label{eqn:positronium_photons}
\end{alignat}
where $\overline{\mathbf{N}}^\gamma_\text{pos}$ is the positronium annihilation spectrum normalized to a single positron, shown in Appendix~\ref{app:positronium_annihilation_spec}. The factor of $1/2$ accounts for the fact that $\mathbf{N}^e_\text{inj}$ contains both electrons and positrons in equal number. 

Since all calculated quantities depend on $z$ and $\mathbf{x}$, all quantities discussed in this section have to be computed at each redshift step. This allows us to properly capture the effect of changing ionization levels on the energy deposition process. 

\subsection{Photon Propagation and Energy Deposition}
\label{sec:photon_cooling}
After resolving the injected electrons and obtaining the photons produced from their cooling, the spectrum of photons that have been newly injected per baryon per log-redshift can be discretized as
\begin{alignat}{1}
  \frac{dN^\gamma_\text{new}}{dE'_j \, dy}(E'_j) \times  E'_j \log \Delta E'_j \times \Delta y  \approx \mathbf{N}_\text{new}^\gamma[E_j'] \,,
\end{alignat}
where $\mathbf{N}^\gamma_\text{new}$ is the sum of photons injected directly by the injection event, and photons produced by the cooling of injected electrons, i.e.
\begin{alignat}{2}
  \mathbf{N}_\text{new}^\gamma[E_j'] &=&& \,\, \mathbf{N}^\gamma_\text{inj} [E'_j] + \mathbf{N}^\gamma_\text{pos}[E_j'] + \sum_i \overline{\mathsf{T}}_\text{ICS} [E_{e,i}', E_j'] \mathbf{N}^e_\text{inj}[E'_{e,i}] \,.
  \label{eqn:new_inj_photons}
\end{alignat}
These photons can cool through a number of processes, including redshifting, pair production, Compton scattering and photoionization. Within a particular log-redshift step, low-energy photons and low-energy electrons are produced, and some high-energy deposition from high-energy electrons produced by $\mathbf{N}^\gamma_\text{new}$ occur. On the other hand, some part of the photon spectrum lies above \SI{13.6}{\eV} and does not photoionize within the log-redshift step; instead, these photons propagate forward to the next step.

The resulting deposition into low-energy photons and electrons was used to compute $f_c$ in Ref.~~\cite{Slatyer:2015kla}, assuming the fixed baseline ionization history. In order to capture the dependence on ionization history, however, we need to be able to calculate the propagation and deposition processes at any ionization level, redshift and injected particle energy. 

One of the main ideas of \texttt{DarkHistory} is to capture the photon cooling processes as precomputed transfer functions with injection energy, redshift and ionization levels as the dependent variables. These transfer functions then act on some incoming spectrum and produce a spectrum of propagating particles, a spectrum of deposited particles or some amount of deposited energy within a log-redshift step. These transfer functions can be evaluated at various points in injection energy, redshift, and ionization levels, and interpolated at other points. With a given injection model, we can then string together these transfer functions to work out the propagation of photons and the deposition of energy, over an extended redshift range, given any exotic source of energy injection. 

\subsubsection{Propagating Photons}
\label{sec:propagating_photons}

Consider a spectrum of photons per baryon denoted $dN^\gamma/dE'$ that is present in the universe at some log-redshift $y$. As these photons propagate, various cooling processes result in these photons being scattered into energies below \SI{13.6}{\eV}, or they may photoionize on an atom in the gas. Those particles that do not undergo either process within a redshift step are called ``propagating photons'', and continue to propagate into the next redshift step.  

We define the transfer function for propagating photons $ \overline{P}^\gamma(E', E, y', y)$ through the following relation:
\begin{alignat}{2}
  \left. \frac{dN_\text{prop}^\gamma}{dE} \right|_y &=&& \int dE' \, \overline{P}^\gamma (E', E, y', y) \left. \frac{dN^\gamma}{dE'} \right|_{y'} \,.
\end{alignat}
$\overline{P}^\gamma$ takes a spectrum of photons that are present at $y'$ and propagates them forward to a spectrum of propagating photons at $y$. $\overline{P}^\gamma(E', E, y', y)$ is exactly the number of propagating photons per unit energy that results from a single photon injected at log-redshift $y'$ with energy $E'$ cooling until log-redshift $y$. The $\overline{P}^\gamma$ functions are calculated separately using the code described in Ref.~\cite{Slatyer:2009yq,Slatyer:2015kla}. 

We distinguish between two different sources of photons between two redshifts $y'$ and $y$ (with $y' > y$): propagating photons at $y'$, $dN_\text{prop}^\gamma/dE'$, and the newly injected photons between the redshifts $y'$ and $y$, defined in discretized form in Eq.~(\ref{eqn:new_inj_photons}). With these sources, we can write the spectrum of propagating photons at $y$ as
\begin{alignat}{2}
  \left. \frac{dN_\text{prop}^\gamma}{dE} \right|_y &=&& \int dE'\, \overline{P}^\gamma(E', E, y', y) \left. \frac{dN_\text{prop}^\gamma}{dE'} \right|_{y'} \nonumber \\
  & && + \int dE' \int_y^{y'} d\eta \, \overline{P}^\gamma (E', E, \eta, y) \left. \frac{dN^\gamma_\text{new}}{dE'\, d\eta} \right|_\eta \,.
  \label{eqn:prop_tf_def}
\end{alignat}

We discretize this expression by defining the following discrete quantities according to the conventions set down in Eqs.~(\ref{eqn:discretize_dNdE}) and~(\ref{eqn:discretize_N}):
\begin{alignat}{2}
  \overline{\mathsf{P}}^\gamma[E'_i, E_j, y', \Delta y] E_i' \Delta \log E_i' &\approx&& \,\, \overline{P}^\gamma (E'_i, E_j, y', y' - \Delta y) \,, \nonumber \\
  \mathbf{N}^\gamma_\text{prop} [E_i', y'] &\approx&&  \left. \frac{dN_\text{prop}^\gamma}{dE'} \right|_{y'} E_i' \, \Delta \log E_i' \,,
\end{alignat}
where we have chosen some fixed value of $\Delta y$, so that the final redshift is $y = y' - \Delta y$. In \texttt{DarkHistory}, the default value is $\Delta y = 10^{-3}$, although this can be adjusted by the process of coarsening, described in Sec.~\ref{sec:coarsening}. Dropping the dependence on $\Delta y$ for simplicity, the discretized version of Eq.~(\ref{eqn:prop_tf_def}) reads
\begin{alignat}{2}
  \mathbf{N}_\text{prop}^\gamma[E_j, y] &=&& \sum_i \overline{\mathsf{P}}^\gamma [E_i', E_j, y'] \mathbf{N}^\gamma[E_i',y'] \,,
  \label{eqn:discretized_prop_tf}
\end{alignat}
where we have defined
\begin{alignat}{1}
  \mathbf{N}^\gamma[E_i', y] \equiv \mathbf{N}_\text{prop}^\gamma [E_i', y] + \mathbf{N}_\text{new}^\gamma [E_i', y] \,.
  \label{eqn:N_prop_plus_new}
\end{alignat}
%

\subsubsection{Energy Deposition}
\label{sec:energy_deposition}

Aside from $\overline{\mathsf{P}}^\gamma$, we also have three deposition transfer functions describing the energy losses of $\mathbf{N}^\gamma$ into high-energy deposition, low-energy electrons and low-energy photons, as defined in Sec.~\ref{sec:overview}. These transfer functions are defined by their action on the discretized photon spectrum, $\mathbf{N}^\gamma$, and are discretized in a similar manner.

The low-energy electron deposition transfer matrix, $\overline{\mathsf{D}}^e$, yields the low-energy electrons produced via cooling of $\mathbf{N}^\gamma$. Adding the low-energy electrons produced directly from the injected electrons $\mathbf{N}^e_\text{inj}$, we obtain the full low-energy electron spectrum $\mathbf{N}^e_\text{low}[E_j, y]$ at a particular redshift step:
\begin{alignat}{2}
	 \mathbf{N}^e_\text{low}[E_{e,j},y] &=&&  \sum_i \overline{\mathsf{D}}^e[E_{\gamma,i}', E_{e,j}, y'] \mathbf{N}^\gamma[E'_{\gamma,i}, y'] + \mathbf{N}^e_\text{low,inj}[E_{e,j}, y] \,,
   \label{eqn:lowengelec_tf}
\end{alignat}
where
\begin{alignat}{1}
   \mathbf{N}^e_\text{low,inj}[E_{e,j}, y] = \sum_i \overline{\mathsf{T}}^e[E_{e,i}', E_{e,j}, y] \mathbf{N}^e_\text{inj}[E_{e,i}', y] \,,
\end{alignat}
while the deposition transfer matrix $\overline{\mathsf{D}}^\gamma$ yields the low-energy photons,
\begin{alignat}{1}
	 \mathbf{N}^\gamma_\text{low}[E_j, y] =  \sum_i \overline{\mathsf{D}}^\gamma[E_i', E_j, y'] \mathbf{N}^\gamma[E'_i, y'] \,.
   \label{eqn:lowengphot_tf}
\end{alignat}
$\mathbf{N}^\gamma_\text{low}$ is computed as a \textit{distortion} to the CMB spectrum, with $\overline{\mathsf{D}}^\gamma$ computed with the initial spectrum of upscattered CMB photons subtracted, in the same way as $\overline{T}_\text{ICS}$, as shown in Eq.~(\ref{eqn:elec_cooling_ics}). 

As the propagating photons cool over a single log-redshift step, they generate high-energy electrons along the way. These are handled in a similar manner to injected high-energy electrons as described in Sec.~\ref{sec:electron_cooling}, but instead of performing the calculation at each step, we simply provide transfer functions  $\overline{\mathbf{D}}_\text{c}^\text{high}$ that act on propagating photons and return the high-energy deposition into the channels $c = $\{`ion', `exc', `heat'\}.\footnote{For legacy reasons, \texttt{DarkHistory} actually computes the transfer function that returns the high-energy deposition per second; this is just a difference in convention.} We can then combine this with the result from electron cooling to obtain the high-energy deposition per baryon within a log-redshift step into each channel $c$:
\begin{alignat}{2}
  E_c^\text{high}[y] &=&& \sum_i \overline{\mathbf{D}}_\text{c}^\text{high}[E_{\gamma,i}', y'] \mathbf{N}^\gamma[E_{\gamma,i}', y'] + \sum_i \overline{\mathbf{R}}_c [E_{e,i}', y'] \mathbf{N}^e_\text{inj}[E_{e,i}', y']  \,.
  \label{eqn:highengdep_tf}
\end{alignat}

To summarize, we have defined the following transfer functions: $\overline{\mathsf{P}}^\gamma$ for propagating photons, and $\overline{\mathsf{D}}^\gamma$, $\overline{\mathsf{D}}^e$ and $\overline{\mathbf{D}}^\text{high}_c$ for deposition into low-energy photons, low-energy electrons and high-energy deposition channels respectively. These transfer functions act on the spectrum of photons $\mathbf{N}^\gamma$ (from both the injection source and the cooling of injected electrons). Together with the transfer functions for the cooling of injected electrons, we have all the information needed to propagate injected particles and compute their energy deposition as a function of redshift. 

\subsubsection{Coarsening}
\label{sec:coarsening}

The propagating photons transfer function $\overline{\mathsf{P}}^\gamma$ can always be evaluated with the same input and output energy abscissa, so that the 2D transfer matrix at each $y$ is square. If the transfer function $\overline{\mathsf{P}}^\gamma$ does not vary significantly over redshift, then in the interest of computational speed, we can make the following approximation of Eq.~(\ref{eqn:discretized_prop_tf}) for propagation transfer matrices:
\begin{alignat}{1}
  \mathbf{N}^\gamma_\text{prop}[E_j, y - n \Delta y] \approx \left( \overline{\mathsf{P}}^\gamma_{1/2} \right)^n_{ji} \mathbf{N}_i^\gamma[y] \,,
  \label{eqn:prop_tf_coarsening}
\end{alignat}
where repeated indices are summed. $i$ and $j$ index input and output energies, and $\overline{\mathsf{P}}^\gamma_{1/2}$ is $\overline{\mathsf{P}}^\gamma$ evaluated at log-redshift $y - n \Delta y/2$ to minimize interpolation error. When making this approximation, we also have to ensure that we redefine
\begin{alignat}{1}
  \mathbf{N}_\text{inj}^\alpha[E_i', y] \to n \mathbf{N}_\text{inj}^\alpha[E_i',y]
\end{alignat}
for both channels $\alpha = e$ and $\gamma$, so that we (approximately) include all of the particles injected between $y$ and $y - n \Delta y$. 

Likewise, if both the deposition and propagation matrices do not vary significantly over redshift, we can approximate Eq.~(\ref{eqn:lowengphot_tf}) as
\begin{alignat}{2}
  \mathbf{N}^\gamma_\text{low}[E_j, y - n \Delta y] &\approx&& \left( \overline{\mathsf{D}}^\gamma_{1/2} \right)_{jk}\sum_m \left(\overline{\mathsf{P}}^\gamma_{1/2} \right)^m_{ki} \mathbf{N}_i^\gamma[y] \,,
  \label{eqn:dep_tf_coarsening}
\end{alignat}
with repeated indices once again being summed over. $\overline{\mathsf{D}}^\gamma_{1/2}$ is defined in the same manner as $\overline{\mathsf{P}}^\gamma_{1/2}$. This equation essentially applies the deposition transfer matrix at $y - n \Delta y/2$ to all $n$ steps of the propagation of the spectrum $\mathbf{N}^\gamma$ from $y$ to $y - \Delta y$, which itself is approximated by $\overline{\mathsf{P}}^\gamma_{1/2}$. In our code, we call these approximations ``coarsening'', and the number $n$ in both Eqs.~(\ref{eqn:prop_tf_coarsening}) and~(\ref{eqn:dep_tf_coarsening}) the ``coarsening factor''.

\subsubsection{Different Redshift Regimes}
\label{sec:redshift_regimes}

In \dhis we separate our transfer matrices into three redshift regimes: redshifts encompassing reionization ($z  < 50$), redshifts encompassing the times between recombination and reionization ($50 \leq z \leq 1600$), and redshifts well before recombination ($z > 1600$).  During the redshifts encompassing reionization, we allow our transfer functions to be functions of $x_\text{HII}$ and $x_\text{HeII}$, enabling the use of reionization models that evolve hydrogen and helium ionization levels separately. We only consider singly-ionized helium in the current version of \dhis since we expect $x_\text{HeIII}$ not to play an important role until $z \sim 6$.  We compute the transfer functions on a grid of $z^k$, $x^m_\text{HII}$, and $x^n_\text{HeII}$, and linearly interpolate over the grid of pre-computed transfer functions.

Between recombination and reionization, the helium ionization level lies at or below the hydrogen ionization level, since helium has a larger ionization potential at \SI{24.6}{\eV}. After recombination, current experimental constraints typically forbid a large ionization fraction, i.e.\ we expect $x_\text{HII} \lesssim 0.1$~\cite{Liu:2016cnk}. As such, setting $x_\text{HeII} = 0$ is a good approximation for the photon propagation and deposition functions: since $\mathcal{F}_\text{He} \sim 8\%$, neglecting helium ionization only results in $\lesssim 8\%$ error to $x_e$, and $\lesssim 10\%$ error in the density of neutral helium. We therefore follow the same procedure as before, except we now calculate and interpolate the transfer functions over a grid of $z^k$ and $x_\text{HII}^m$ values while holding the helium ionization level fixed to zero.

Finally, well before recombination, we expect the universe to be close to 100\% ionized and tightly coupled thermally to the CMB. Any extra source of exotic energy injection that is consistent with current experimental constraints will likely have a negligible effect on the ionization and thermal histories. We thus calculate and interpolate our transfer functions over a grid of $z^k$ values while holding the hydrogen and helium ionization levels to the baseline values provided by \texttt{RECFAST}~\cite{Wong:2007ym}.

The actual grid values $z^k$, $x^m_\text{HII}$, and $x^n_\text{HeII}$ in each of these regimes can be found in the code, and have been chosen so that interpolation errors are at the sub-10\% level when $f_c(z)$ is calculated using the same method detailed in Ref.~\cite{Slatyer:2015kla}. Our results for $f_c(z)$ without taking into account backreaction, including some improvements over Ref.~\cite{Slatyer:2015kla}, can be found in Appendix~\ref{app:cross_checks}. 

\subsection{Calculating \texorpdfstring{$f_c(z)$}{fc(z)}}
\label{sec:calculating_f}
The low-energy photons $\mathbf{N}^\gamma_\text{low}[E_i, z]$ and low-energy electrons $\mathbf{N}^e_\text{low}[E_i, z]$, defined in Sec~\ref{sec:overview}, 
transfer their energy into ionization and excitation of atoms, heating of the IGM, and free-streaming photons to be added to the CMB continuum.
 In \dhis we keep track of how much energy low energy photons and electrons deposit into channels c $\in $ \{`H$_\text{ion}$', `He$_\text{ion}$', `exc', `heat', `cont'\}, which represent hydrogen ionization, helium ionization, hydrogen excitation, heating of the IGM, and sub-\SI{10.2}{\eV} continuum photons respectively.  
 The energy deposition fractions $f_c(z)$ are then found by normalizing the total energy deposited into channel c within a redshift step by the total energy injected within that step according to Eq~(\ref{eqn:fz}).  We closely follow the method for computing $f_c(z)$ described in Ref.~\cite{Slatyer:2015kla}.

Before calculating $f_c(z)$ for each channel, it is instructive to see how to calculate the total amount of energy deposited per unit time and volume, $\left(dE/dV\, dt\right)^\text{dep}$. The low-energy photon and electron spectra $\mathbf{N}^\gamma_\text{low}[E_i]$ and $\mathbf{N}^e_\text{low}[E_i]$ as defined above contain a number of particles per baryon deposited within each log-redshift bin (the $z$-dependence has been suppressed since all calculations in this section occur at the same redshift step). We can convert between these and spectra containing the number of particles produced per unit volume and unit time using the conversion factor $G(z)$ introduced in Eq.~(\ref{eqn:per_baryon_to_dVdt}). For example, to obtain the total amount of energy deposited at a given redshift per unit time and volume, one simply sums over low-energy particle type and applies the conversion factor,

\begin{alignat}{1}
    \left(\frac{dE}{dVdt}\right)^\text{dep}_\text{low} = \frac{1}{G(z)} \sum_\alpha \sum_i  \,E_i' \, \mathbf{N}_\text{low}^\alpha[E'_i] \,.
\end{alignat}
To calculate the total amount of energy deposited we must also add the amount deposited by high energy electrons and photons, which we computed in Eq.~(\ref{eqn:highengdep_tf}):
\begin{alignat}{1}
	\left(\frac{dE}{dVdt}\right)_\text{high}^{\text{dep}} = \frac{1}{G(z)} \sum_c E_c^\text{high} \,.
\end{alignat}
Then the total deposited energy summed over all channels is given by
\begin{alignat}{1}
    \left(\frac{dE}{dVdt}\right)^\text{dep} = \left(\frac{dE}{dVdt}\right)^\text{dep}_\text{low} + \left(\frac{dE}{dVdt}\right)^\text{dep}_\text{high} \,.
\end{alignat}
With this example in mind, we are now ready to understand how to split the energy deposition into the different channels.

\subsubsection{Photons}
\label{sec:low_energy_phot}

We first compute $f_c(z)$ for low-energy photons, starting with energy deposition into continuum photons.  These are photons with energy below $3 \mathcal{R}/4 = \SI{10.2}{\eV}$ that are unable to effectively transfer their energy to free electrons or atoms, so they just free stream.
The energy of these photons constitutes deposition into the continuum channel, i.e.
\begin{alignat}{1}
    \left(\frac{dE^\gamma}{dV \,dt}\right)^\text{dep}_\text{cont} = \frac{1}{G(z)} \sum_{E_i=0}^{3 \mathcal{R}/4}  \,E_i\, \mathbf{N}^\gamma_\text{low}[E_i] \,.
    \label{eqn:f_cont_phot}
\end{alignat}

To calculate the total amount of energy deposited into hydrogen excitation, we make the approximation that all photons with energies between $3\mathcal{R}/4 = \SI{10.2}{\eV}$ and $\mathcal{R} = \SI{13.6}{\eV}$ deposit their energy instantaneously into hydrogen Lyman-$\alpha$ excitation, following \cite{Slatyer:2015kla}:
\begin{alignat}{1}
    \left(\frac{dE^\gamma}{dVdt}\right)^\text{dep}_\text{exc} = \frac{1}{G(z)} \sum_{E_i=3 \mathcal{R}/4}^{\mathcal{R}} \!\!\!\! E_i \, \mathbf{N}_\text{low}^\gamma[E_i] \,.
    \label{eqn:f_exc_phot}
\end{alignat}
A more complete treatment of excitation would involve keeping track of sub-\SI{13.6}{\eV} energy photons as they redshift into the Lyman-$\alpha$ transition region at \SI{10.2}{\eV}, and should also include two-photon excitation into the $2s$ state.\footnote{Two-photon $1s \to 2s$ transitions are in fact as important as Lyman-$\alpha$ transitions near recombination in determining the ionization history, due to the fact that the Lyman-$\alpha$ line is optically thick at this time.} Finally, helium excitation has been neglected, since the de-excitation of helium atoms, which occurs quickly, produces photons that can eventually photoionize hydrogen. We therefore expect almost no net deposition of energy into helium excitation. Energy injection through helium excitation would mainly affect the process of helium recombination, when the probability of ionization after excitation to a higher state is significant due to the photon bath. However, we do not track this small effect, since the change to $x_e$ would be very small. We leave a more careful treatment of excitation that can correctly take into account all of these effects to future work.

We now move on to ionization. All photons above $\mathcal{R} = \SI{13.6}{\eV}$ that are included in $N^\gamma_\text{low}$ have photoionized one of the atomic species (HI, HeI and HeII). However, after photoionizing a helium atom, the resulting ion may quickly recombine with an ambient free electron, producing an $\mathcal{R}_\text{He} = \SI{24.6}{\eV}$ or $4 \mathcal{R} = \SI{54.4}{\eV}$ photon, which may then go on to photoionize hydrogen instead.\footnote{The photoionization rate on neutral hydrogen is much faster than the Hubble rate for $x_\text{HII} \lesssim 0.9999$ for $z > 3$.}

We can handle low-energy photons with energy $E_\gamma$ that photoionize neutral helium in one of the following three ways:

\begin{enumerate}
  \item if helium is completely ignored, the photon is assumed to photoionize hydrogen, producing a low-energy electron with energy $E_\gamma - \mathcal{R}$ from photoionization and depositing $
  \mathcal{R}$ into hydrogen ionization. This is the approach used in previous calculations of $f_c(z)$~\cite{Slatyer:2015kla}, but leaves us unable to self-consistently track $x_\text{HeII}$ if desired;

  \item the photon produces a low-energy electron with energy $E_\gamma - \mathcal{R}_\text{He}$ from photoionization, depositing $\mathcal{R}$ into hydrogen ionization from the recombination photon (with energy $\mathcal{R}_\text{He}$) and producing an electron with energy $\mathcal{R}_\text{He} - \mathcal{R}$, which ultimately deposits energy into hydrogen excitation, heating and sub-\SI{10.2}{\eV} photons. This approach was previously discussed in Ref.~\cite{Galli:2013dna}, and found to result in very little difference when compared to method (1); or
  \item the photon produces a low-energy electron with energy $E_\gamma - \mathcal{R}_\text{He}$ from photoionization and deposits $\mathcal{R}_\text{He}$ into helium ionization. 
\end{enumerate}

The most accurate accounting of helium ionization lies somewhere between methods (2) and (3); however, either method will likely lead to very similar results in terms of $x_e$ and $T_m$, since the bulk of the energy is deposited by the electron from the initial photoionization for photon energies $E_\gamma \gg \mathcal{R}_\text{He}$, and the remaining energy always leads to one ionization event overall. \texttt{DarkHistory} offers the choice of these three options for implementing helium ionization.

We have checked that all three methods lead to similar ionization and temperature histories for DM models over a large range of masses decaying to both $e^+e^-$ and $\gamma \gamma$; these checks are shown in Appendix~\ref{app:cross_checks}. We recommend simply using method~(1) with helium turned off if the user is interested in ionization and temperature histories well before reionization, and using both method~(2) and~(3) with helium turned on to bracket the uncertainties associated with energy deposition on helium if the user is interested in the epoch of reionization. 

To summarize, the amount of deposited energy into hydrogen per unit time and volume is given by
\begin{alignat}{1}
    \left(\frac{dE^\gamma}{dVdt}\right)^\text{dep}_{\text{H}_\text{ion}} =  \frac{\mathcal{R} }{G(z)} \sum_{E_i > \mathcal{R}} \!\! q_\text{H}^\gamma [E_i] \mathbf{N}^\gamma_\text{low}[E_i] \,,
    \label{eqn:H_ion_dep_phot}
\end{alignat}
and into helium ionization by:
\begin{alignat}{1}
  \left(\frac{dE^\gamma}{dV \, dt}\right)^\text{dep}_{\text{He}_\text{ion}} = \frac{\mathcal{R}_\text{He}}{G(z)} \sum_{E_i > \mathcal{R}_\text{He}} \!\! q_\text{He}^\gamma [E_i] \mathbf{N}^\gamma_\text{low}[E_i] \,,
  \label{eqn:He_ion_dep_phot}
\end{alignat}
producing a low-energy electron spectrum after photoionization of
\begin{alignat}{2}
  \mathbf{N}^e_\text{ion}[E_i] &=&& \,\, q_\text{H}^e(E_i + \mathcal{R}) \mathbf{N}^\gamma_\text{low}[E_i + \mathcal{R}]  \nonumber \\
  & &&+ q_\text{He,a}^e(E_i + \mathcal{R}_\text{He}) \mathbf{N}^\gamma_\text{low}[E_i + \mathcal{R}_\text{He}] \nonumber \\
  & &&+ \delta[E_i - \mathcal{R}_\text{He} + \mathcal{R}] \sum_j  q^e_\text{He,b}(E_j) \mathbf{N}_\text{low}^\gamma[E_j] \,,
  \label{eqn:ionized_elec}
\end{alignat}
where $\delta[E_i - \mathcal{R}_\text{He} + \mathcal{R}]$ is one when the bin boundaries span the energy $\mathcal{R}_\text{He} - \mathcal{R}$ and is zero otherwise, and 
\begin{alignat}{1}
  q(E_i) \equiv \begin{cases}
  \frac{n_\text{HI} \sigma_\text{HI}(E_i)}{n_\text{HI} \sigma_\text{HI}(E_i) + n_\text{HeI} \sigma_\text{HeI} (E_i)}, & E_i > \mathcal{R}, \\
  0, & \text{otherwise},
  \end{cases}
  \label{eqn:p_def}
\end{alignat}
with the $\sigma$'s denoting the photoionization cross section of the appropriate species. $\mathbf{N}_\text{ion}^e$ is added to the low-energy electron spectrum, $\mathbf{N}^e_\text{low}$, which is then treated in the next section. The values of the $q$-coefficients depend on the method, and are shown in Table~\ref{tab:q_values}.

\setlength{\tabcolsep}{10pt}
\renewcommand{\arraystretch}{1.3}

\begin{table}
\centering
\begin{tabular}{r c c c c c}

\toprule
Method & $q_\text{H}^\gamma$ & $q_\text{H}^e$ & $q_\text{He}^\gamma$ & $q^e_\text{He,a}$ & $q^e_\text{He,b}$\\
\hline
  1 & 1 & 1 & 0 & 0 & 0 \\
  2 & 1 & $q$ & 0 & $1-q$ & $1-q$ \\
  3 & $q$ & $q$ & $1-q$ & $1-q$ & 0 \\
\botrule
\end{tabular}
\caption{List of $q$-coefficients for use in Eqs.~(\ref{eqn:H_ion_dep_phot})--(\ref{eqn:ionized_elec}). The variable $q$ is defined in Eq.~(\ref{eqn:p_def}).}
\label{tab:q_values}
\end{table}

\subsubsection{Electrons}
\label{sec:low_energy_elec}

To compute how low-energy electrons deposit their energy into the different channels, we use the results obtained by the MEDEA code \cite{Evoli:2012zz,Valdes:2009cq}, following a similar treatment to Ref. \cite{Galli:2013dna}. Although \texttt{DarkHistory} also includes a calculation of electron energy deposition, which we discussed in Sec.~\ref{sec:electron_cooling}, the MEDEA results are more accurate in the sub-\SI{3}{\kilo\eV} electron energy range, including a more detailed accounting of all possible atomic processes (such as $2s\to1s$ deexcitations) and with more up-to-date cross sections. However, at mildly nonrelativistic to mildly relativistic regimes, our calculation of ICS is more accurate, as argued in Sec.~\ref{sec:ICS}. Furthermore, the MEDEA results assume that hydrogen and helium are at similar ionization levels, which is not always a good assumption. In future versions of \texttt{DarkHistory}, an improved treatment of electrons may be a useful addition to the code.

The MEDEA code uses a Monte Carlo method to track high-energy electrons as they are injected into the IGM, and determines the fraction of the initial electron energy deposited into ionization, Lyman-$\alpha$ excitation, heating of the gas and sub-\SI{10.2}{\eV} photons. We use a table of these energy deposition fractions $p_c(E_i,x_{e,j})$~\cite{Galli:2013dna}, where $c \in $ \{`H$_\text{ion}$', `He$_\text{ion}$', `exc', `heat', `cont'\} as before, $x_{e,j}$ ranges between 0 and 1, and $E_i$ ranges between \SI{14}{\eV} and \SI{3}{\kilo\eV}, and perform an interpolation over these values. The energy deposition from electrons is then simply given by
\begin{alignat}{1}
    \left(\frac{dE^e}{dVdt}\right)^\text{dep}_\text{c} = \frac{1}{G(z)} \sum_i p_c(E_i,x_e)  E_i \, \mathbf{N}^e_\text{low}[E_i] \,,
    \label{eqn:f_elec}
\end{alignat}
keeping in mind that $\mathbf{N}^e_\text{ion}$ has already been added to $\mathbf{N}^e_\text{low}$. Between energies of \SI{10.2}{\eV} and \SI{13.6}{eV}, where collisional excitations of hydrogen are possible but not ionization, we use the result at \SI{14}{eV}, but setting the component into hydrogen ionization to zero and normalizing to unit probability. Below \SI{10.2}{eV}, electrons can only deposit energy through Coulomb heating. 

\subsubsection{High-Energy Deposition}
\label{sec:high_eng_dep}

Finally, the high-energy deposition component of the total energy deposited is given by:
\begin{alignat}{1}
  \left(\frac{dE^\text{high}}{dV \, dt}\right)^\text{dep}_c = \frac{1}{G(z)} E^\text{high}_c \,,
\end{alignat}
where $c \in $ \{ `ion', `exc', `heat' \}. Here, we add the high-energy excitation and ionization component to Lyman-$\alpha$ excitation and hydrogen ionization for simplicity, even though the high-energy deposition is computed for all atomic species. A more accurate computation of this together with a more consistent treatment of helium ionization will be a potential improvement in a future version of \texttt{DarkHistory}. 

\bigskip

With the rate of energy deposition through both low-energy photons and low-energy electrons computed, the total energy deposition rate is then straightforwardly given by
\begin{alignat}{1}
  \left(\frac{dE}{dV \, dt}\right)^\text{dep}_c = \sum_\alpha \left(\frac{dE^\alpha}{dV \, dt}\right)^\text{dep}_c \,,
\end{alignat}
where $\alpha \in \{\gamma, e, \text{high}\}$.

\subsection{TLA Integration and Reionization}
\label{sec:TLA_and_reionization}

\texttt{DarkHistory} offers several options for which set of assumptions should be used when integrating the ionization and thermal histories. In the simplest case, the user may integrate Eq.~(\ref{eqn:TLA_DarkHistory}) at each redshift step based on the $f_c(z,\mathbf{x})$ calculated above, with the reionization terms switched off. As we have discussed, including this backreaction is already a significantly better treatment compared to calculations which assume a standard recombination history, i.e.\ using $f_c(z,\mathbf{x}_\text{std}(z))$ (although backreaction can also be switched off within \texttt{DarkHistory}). 

The next significant improvement that is implemented within \texttt{DarkHistory} is the tracking of the neutral helium ionization fraction. Well before reionization, neglecting helium is a good approximation, since the number density of helium nuclei is only $\mathcal{F}_\text{He} \simeq 0.08$ of hydrogen, and we should expect only at most an 8\% correction to $x_e$ if we include helium.

However, tracking helium allows us to accomplish a self-consistent modeling of exotic energy injection together with the reionization of hydrogen and neutral helium. \texttt{DarkHistory} allows users to input a model of reionization, for the first time extending the validity of these energy injection calculations into a regime where hydrogen is fully ionized and helium is singly ionized. 

\subsubsection{Helium}
\label{sec:helium}

The \texttt{DarkHistory} evolution equation governing helium without any energy injection is identical to the \texttt{RECFAST} model, and is given by~\cite{Wong:2007ym}
\begin{alignat}{2}
  \dot{x}_\text{HeII}^{(0)} &=&& \,\, \mathcal{C}^s_\text{HeI} \big(x_\text{HeII} x_e n_\text{H} \alpha^s_\text{HeI} - \beta^s_\text{HeI} (\mathcal{F}_\text{He} - x_\text{HeII}) e^{-E^{s,\text{He}}_{21}/T_\text{CMB}}\big) \nonumber \\
   & &&+ \mathcal{C}^t_\text{HeI} \big( x_\text{HeII} x_e n_\text{H} \alpha^t_\text{HeI} - 3 \beta^t_\text{HeI} (\mathcal{F}_\text{He} - x_\text{HeII}) e^{-E^{t,\text{He}}_{21}/T_\text{CMB}} \big) \,.
   \label{eqn:helium_TLA}
\end{alignat}
The singlet and triplet ground states of helium must be treated separately, and terms relevant to the singlet or triplet state are represented with a superscript $s$ or $t$ respectively. Here, $\alpha_\text{HeI}$ and $\beta_\text{HeI}$ are the recombination and photoionization for HeI, $E_{21}^\text{He}$ represents the energy difference between the corresponding $n=1$ and $n=2$ states, and finally $\mathcal{C}_\text{HeI}$ is the analog to the Peebles-C coefficient found in Eq.~(\ref{eqn:TLA}), representing the probability of a helium atom in the $n=2$ state decaying to either the singlet or triplet ground state before photoionization can occur. The reader should refer to Refs.~\cite{Wong:2007ym,Kholupenko:2008gb,Kholupenko:2007qs} for details on the numerical values of the coefficients, as well as how to compute $\mathcal{C}_\text{HeI}$. 

We emphasize that although we have implemented all of the modifications to the standard TLA in Eq.~(\ref{eqn:TLA}), our code should not be used for high-precision cosmology, given that it has not been tested extensively, e.g.\ with different cosmological parameters from the central values used in \texttt{DarkHistory}. We find that our code agrees to within 3\% of the \texttt{RECFAST} $x_e$ values for the cosmological parameters used here, which is sufficient for computing the effects of exotic energy injection at this stage.

In the presence of exotic sources of energy injection, low-energy photons and electrons can also change the helium ionization level. Once again, we express the energy injection source term as
\begin{alignat}{1}
  \dot{x}^\text{inj}_\text{HeII} = \frac{f_\text{He ion}(z,\mathbf{x})}{\mathcal{R}_\text{He} n_\text{H}} \left(\frac{dE}{dV \, dt}\right)^\text{inj} \,,
\end{alignat}
where $\mathcal{R}_\text{He} = \SI{24.6}{\eV}$ is the ionization potential of neutral helium. As we discussed in Sec.~\ref{sec:low_energy_phot}, there are three different methods available to evaluate $f_{\text{He}_\text{ion}}$ which bracket the uncertainties involved in helium ionization.

To summarize, the user may opt to track the change in helium ionization levels. This means that in addition to Eq.~(\ref{eqn:TLA_DarkHistory}), we also include
\begin{alignat}{1}
  \dot{x}_\text{HeII} = \dot{x}_\text{HeII}^{(0)} + \dot{x}_\text{HeII}^\text{inj} + \dot{x}_\text{HeII}^\text{re} \,,
  \label{eqn:TLA_helium_darkhistory}
\end{alignat}
where $\dot{x}_\text{HeII}^\text{re}$ is the contribution from processes that are active during reionization.

\subsubsection{Reionization}

The evolution equations shown in Eqs.~(\ref{eqn:TLA_DarkHistory}) and~(\ref{eqn:TLA_helium_darkhistory}) can be integrated with all reionization terms switched off if the user is primarily interested in temperatures or ionization levels well before reionization starts at $z \sim 20$. In this regime, turning off helium is also a reasonable approximation. 

With reionization however, the helium ionization level should be solved as well for complete consistency. We solve the TLA differential equations shown in Eqs.~(\ref{eqn:TLA_DarkHistory}) and~(\ref{eqn:TLA_helium_darkhistory}) in two separate redshift regimes. Prior to some user-defined reionization redshift $1 + z_\text{re}$ ($z_\text{re} \leq 50$), we set $\dot{T}_m^\text{re}$, $\dot{x}_\text{HII}^\text{re}$ and $\dot{x}_\text{HeII}^\text{re}$ to zero. Once reionization begins, we set $\dot{x}_\text{HII}^{(0)}$ and $\dot{x}_\text{HeII}^{(0)}$ to zero for $z < z_\text{re}$ instead, switching over to the specified reionization model with its own photoionization and recombination rates.\footnote{We do not set $\dot{T}_m^{(0)} = 0$, since both adiabatic cooling and Compton scattering off the CMB remain active during reionization.} We also begin tracking doubly-ionized helium $x_\text{HeIII}$, which is always assumed to be zero before reionization. 

The $\dot{T}_m^\text{re}$, $\dot{x}_\text{HII}^\text{re}$ and $\dot{x}_\text{HeII}^\text{re}$ terms depend on the details of how reionization proceeds, which is still relatively uncertain. However, choosing a model for the formation of stars and active galactic nuclei (AGNs) and the associated photoionization and photoheating rates, these terms can be evaluated. \texttt{DarkHistory} by default includes the Puchwein+ model of Ref.~\cite{Puchwein:2018arm}. We also demonstrate how to implement the older Madau and Haardt model~\cite{Haardt:2011xv} in Example 8. Both models provide a photoionization rate $\Gamma_{\gamma X}^\text{ion}(z)$ and a photoheating rate $\mathcal{H}_{\gamma X}^\text{ion}(z)$ as a function of redshift and species $X$. 

Along with these energy injection rates, we must also include other relevant processes that alter the ionization fraction of each species. Since these processes generally convert kinetic energy to atomic binding energy, cooling or heating of the gas due to these processes must also be included in $\dot{T}_m^\text{re}$. The processes we include are:
\begin{enumerate}

    \item collisional ionization, occuring at a rate $\Gamma_{eX}^\text{ion}$ for each species $X$, and an associated cooling rate $-\mathcal{H}_{eX}^\text{ion}$;  

    \item case-A recombination, described by a rate coefficient $\alpha_{A,X}$ for each species $X$, and an associated cooling rate $-\mathcal{H}_X^\text{rec}$;

    \item collisional excitation cooling, with a rate $-\mathcal{H}_{eX}^\text{exc}$; and

    \item bremsstrahlung cooling, with a rate $-\mathcal{H}^\text{br}$. 

\end{enumerate}

The cooling rates here have been defined with a negative sign so that all quantities denoted by $\mathcal{H}$ contribute positively to any temperature change. Expressions for all of these rates can be found in Ref.~\cite{Bolton:2006pc}. They are explicitly dependent on the ionization fraction of all three of the relevant species, namely $x_\text{HI}$, $x_\text{HeI}$ and $x_\text{HeII}$.  The full expressions for the evolution of each of these fractions are as follows:
\begin{alignat}{2}
\label{eqn:tla_reion_xe}
    \dot{x}_\text{HII} &=&&\, x_\text{HI} \left(\Gamma_{\gamma \text{HI}}^\text{ion} + n_e \Gamma_{e \text{HI}}^\text{ion} \right) - x_\text{HII} n_e \alpha_{A, \text{HI}} \,, \nonumber \\
    \dot{x}_\text{HeII} &=&&\, x_\text{HeI} \left(\Gamma_{\gamma \text{HeI}}^\text{ion} + n_e \Gamma_{e \text{HeI}}^\text{ion}\right) + x_\text{HeIII} n_e \alpha_{A,\text{HeIII}} \nonumber \\
    & &&- x_\text{HeII} \left(\Gamma_{\gamma \text{HeII}}^\text{ion} + n_e \Gamma_{e \text{HeII}}^\text{ion} + n_e \alpha_{A,\text{HeII}}\right) \,, \nonumber \\
    \dot{x}_\text{HeIII} &=&&\, x_\text{HeII} \left(\Gamma_{\gamma \text{HeII}}^\text{ion} + n_e \Gamma_{e \text{HeII}}^\text{ion} - x_\text{HeIII} n_e \alpha_{A,\text{HeIII}} \right) \,,
\end{alignat}
with the temperature evolution given by
\begin{alignat}{2}
\label{eqn:tla_reion_T}
    \dot{T}_m^\text{re} &=&& \, \frac{2}{3(1 + \mathcal{F}_\text{He} + x_e) n_\text{H}} \sum_X \left(\mathcal{H}_{eX}^\text{ion} + \mathcal{H}_X^\text{rec} + \mathcal{H}_{eX}^\text{exc} + \mathcal{H}^\text{br} \right) \, .
\end{alignat}

Instead of specifying a full reionization model, the user may also choose the simpler alternative of fixing the value of $x_\text{HII}$ and $x_\text{HeII}$ as a function of redshift once reionization begins, and integrate the temperature evolution alone instead. We note that this approach is not self-consistent, since fixing the ionization levels forces us to neglect any additional contribution to ionization from exotic energy injection sources. However, if the contribution to ionization is known to be small, this can serve as a useful approximation.

\subsubsection{Numerical Integration}

To ensure that ionization fractions always remain appropriately bounded during integration, we introduce the variable
\begin{alignat}{1}
  \zeta_i \equiv \text{arctanh} \left[ \frac{2}{\chi_i} \left(\frac{n_i}{n_\text{H}} - \frac{\chi_i}{2}\right) \right] \,,
  \label{eqn:yi_variable}
\end{alignat}
where $\chi_i = 1$ for HI and $\chi_i = \mathcal{F}_\text{He}$ for HeI and HeII. This transformed equation is then integrated using the standard \texttt{odeint} integrator provided by SciPy.

At early times, the equations we are integrating are very stiff, and solving them directly with numerical integration can often run into difficulties. We therefore assume that when $x_\text{HII} > 0.99$ or $x_\text{HeII} > 0.99 \mathcal{F}_\text{He}$, either variable follows their Saha equilibrium values. 

In Sec.~\ref{sec:example_reionization}, we will show several thermal and ionization histories that showcase \texttt{DarkHistory}'s capabilities in tracking the helium ionization level, exotic energy injection and reionization all at the same time.

\section{Modules}
\label{sec:modules}
In this section we summarize the main modules in \texttt{DarkHistory}. We will pay particular attention to the modules shown in the flow chart in Fig.~\ref{fig:flowchart}, 
and as far as possible provide links between the code and the text.
Keep in mind that this is not a complete list and that it is subject to change in future versions of the code. There is more thorough documentation in \texttt{DarkHistory} itself that will be periodically updated at \href{https://darkhistory.readthedocs.io/en/development/}{https://darkhistory.readthedocs.io}, and will contain a more complete explanation of the code. In the interest of space, we only provide the full path of each module in the code when it is mentioned for the first time.

\subsection{Data}
\label{sec:data}

First, the user must download the data files found at \href{https://doi.org/10.7910/DVN/DUOUWA}{https://doi.org/10.7910/DVN/DUOUWA}. These files contain the photon propagation transfer function $\overline{\mathsf{P}}_\gamma$ and deposition transfer functions $\overline{\mathsf{D}}_\gamma$, $\overline{\mathsf{D}}_e$ and $\overline{\mathbf{D}}_c^\text{high}$, which have all been precomputed as discussed above. They also contain transfer functions for ICS calculations discussed in Appendix~\ref{app:ICS}, structure formation annihilation boost factors computed in Ref.~\cite{Liu:2016cnk}, the baseline thermal and ionization histories, data from \textsc{pppc4dmid}~\cite{Cirelli:2010xx} and $f_c(z)$ computed without backreaction for DM annihilation and decay, where photons and $e^+e^-$ are injected at a fixed set of energies. 

\subsection{\texttt{config}}
\label{sec:module_config}

The \texttt{config} module contains the code required to access the downloaded data, and to store them in memory for use. Users should ensure that the variable \texttt{data\_path} points to the directory containing the data files.

\subsection{\texttt{main}}
\label{sec:module_main}

The \texttt{main} module contains the function that implements the loop shown in Fig.~\ref{fig:flowchart}, \texttt{evolve()}. The usage of this function will be discussed in great detail in Sec.~\ref{sec:examples}. 

\subsection{\texttt{darkhistory.physics}}
\label{sec:module_physics}

This module contains physical constants and useful functions found in cosmology, particle physics and atomic physics. We use units of \SI{}{cm} for length, \SI{}{s} for time and \SI{}{\eV} for energy, mass and temperature. Some examples of functions that are included in this module include the Hubble parameter as a function of redshift, \texttt{physics.hubble()}, and the Peebles-C factor $\mathcal{C}$ found in Eq.~(\ref{eqn:TLA}), \texttt{physics.peebles\_C()}. Cosmological constants provided in this module are taken from central values of the Planck 2018 TT,TE,EE+lowE results~\cite{Aghanim:2018eyx} and the Particle Data Group review of particle physics~\cite{Tanabashi:2018oca}.

\subsection{\texttt{darkhistory.electrons}}
\label{sec:module_electrons}

The \texttt{electrons} module contains all of the functions necessary to perform the electron cooling calculation. The \texttt{positronium} submodule contains functions that return the spectrum of photons obtained during positronium annihilation, which we denoted as $\overline{\mathbf{N}}_\gamma^\text{pos}$ in Eq.~(\ref{eqn:positronium_photons}); Example 7 demonstrates how to use this module. The \texttt{ics} submodule contains all of the machinery necessary to compute the ICS scattered photon and electron spectra; for more details on how to use this submodule, refer to Example 4 in the code. 

\texttt{elec\_cooling} contains the code necessary to compute the transfer functions $\overline{\mathbf{R}}_c$, $\overline{\mathsf{T}}_\text{ICS}$ and $\overline{\mathsf{T}}_e$, as defined in Eqs.~(\ref{eqn:elec_cooling_dep_tf}),~(\ref{eqn:elec_cooling_ics}) and~(\ref{eqn:elec_cooling_lowengelec}) respectively; Example 6 shows how this module is used.

\subsection{\texttt{darkhistory.history}}
\label{sec:module_history}

This module contains our implementation of the TLA and reionization.
The submodule \texttt{tla} corresponds to Sec.~\ref{sec:histories} 
where the function \texttt{get\_history} implements the TLA, including all of the terms discussed in Eqs.~(\ref{eqn:TLA_DarkHistory}) and Eqs.~(\ref{eqn:TLA_helium_darkhistory})--(\ref{eqn:tla_reion_T}). 
The submodule \texttt{reionization} contains the Puchwein+ reionization model, and contains all of the coefficients found in Eqs.~(\ref{eqn:tla_reion_xe}) and~(\ref{eqn:tla_reion_T}). 

\subsection{\texttt{darkhistory.low\_energy}}
\label{sec:module_low_energy}

This module calculates $f_c(z)$. 
The \texttt{lowE\_photons} and \texttt{lowE\_electrons} submodules correspond 
to Sec.~\ref{sec:low_energy_phot} and Sec.~\ref{sec:low_energy_elec}, respectively, implementing Eqs.~(\ref{eqn:f_cont_phot})--(\ref{eqn:ionized_elec}) and Eq.~(\ref{eqn:f_elec}) respectively. 
The \texttt{lowE\_deposition} submodule then combines the energy deposited by photons, electrons (including photoionized electrons) and high-energy deposition to make $f_c(z, \mathbf{x})$.

\subsection{\texttt{darkhistory.spec}}
\label{sec:module_spec}

This module contains functions for handling and generating spectra and transfer functions. All one dimensional spectra in the code can be handled using the class \texttt{Spectrum}, which stores not just the data of the spectrum, but also the abscissa, and other relevant information like redshift or the injection energy of the particle that produced the spectrum. This class includes many convenience functions, such as the ability to rebin the spectrum into a new binning while conserving total number and energy, or the ability to quickly obtain the total number of particles within some energy range. Example 1 in our code gives a quick introduction to this class.

The user may also want to store closely related spectra in one object. This may be desirable for spectra of the same particle type over different redshifts, or if they correspond to spectra from the same injected particle but at different injection energies. The class \texttt{Spectra} has been written to do exactly this. Example 2 provides a good overview of what this class can do. 

\subsection{\texttt{darkhistory.spec.pppc}}
\label{sec:module_pppc}

Within the \texttt{spec} module, a dedicated submodule \texttt{pppc} has been written to calculate the electron and photon spectra from the injection of any arbitrary Standard Model particle, based on the \textsc{pppc4dmid} results. The function \texttt{pppc.get\_pppc\_spec()} is the main function to use for this end. See Example 4 for more information on how to use this function.

\section{Using the Code}
\label{sec:examples}

We will now apply \dhis to perform a variety of calculations in order to highlight the key functionalities of the code. Each of the subsections corresponds to an example Jupyter notebook that has been provided as part of the code; the user should refer to these examples for a deeper look at the full capability of the code, as well as to the online documentation. In this chapter, we will simply highlight capabilities and interesting physics results. 

Within the code and in this section, the word ``redshift'' and variables that represent redshift (usually called \texttt{rs} in the code) refer to the quantity $1+z$, since this is the physically relevant quantity in many cosmological calculations.

\subsection{A Simple Model: \texorpdfstring{$\chi \chi \to b\bar{b}$}{DM DM -> b bbar}}

As a first example, we will demonstrate how to compute the ionization and thermal history of a simple annihilation model. Consider a \SI{50}{\giga\eV} Majorana fermion DM particle that undergoes $s$-wave annihilation to a pair of $b \overline{b}$ quarks, with an annihilation cross section $\langle \sigma v \rangle = \SI{2e-26}{\centi\meter\cubed\per\second}$, close to the required thermal freezeout cross section for the correct relic abundance. Similar models have been considered as a possible dark matter explanation for the galactic center excess~\cite{Calore:2014nla} and the AMS-02 antiproton excess~\cite{Cui:2016ppb,Cuoco:2016eej}. We perform the calculation in a relatively simplified setting: with no reionization, no backreaction included, but with a boost to the annihilation rate from structure formation.  For more details, see Example 9 in the code.

The function that we use to compute histories is \texttt{main.evolve()}. There are many keyword parameters that can be used with this function, and the user should refer to the example notebooks and the online documentation for more information. To find the thermal history for this model, \texttt{evolve()} can be called in the following fashion:

\begin{lstlisting}{language=python}
    import main
    import darkhistory.physics as phys

    bbbar_noBR = main.evolve(
        DM_process='swave', mDM=50e9, sigmav=2e-26, primary='b', 
        start_rs=3000., coarsen_factor=32, backreaction=False, 
        struct_boost=phys.struct_boost_func()
    )
\end{lstlisting}
\begin{figure}
\centering
\includegraphics[scale=.43]{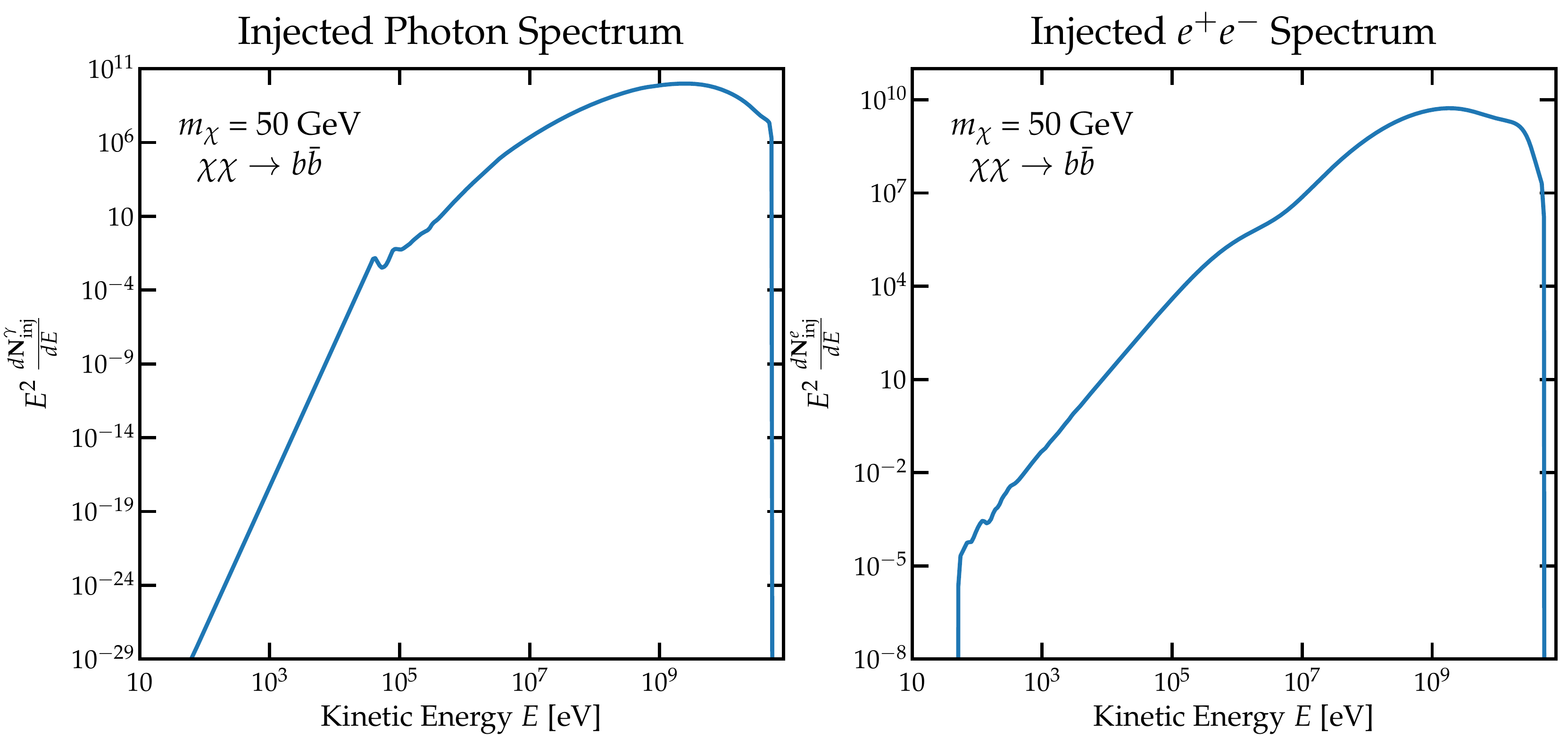}
\caption{Photon (left) and $e^+e^-$ (right) spectra produced by a single annihilation event, $\chi \chi \to b \overline{b}$, with $m_\chi = \SI{50}{\giga\eV}$. These spectra are based on the raw data provided by \textsc{pppc4dmid}.}
\label{fig:bbbar_spectra}
\end{figure}

The keyword parameters are as follows:

\begin{enumerate}
 \item \lstinline|DM_process='swave'| -- specifies the DM process of interest. \texttt{DarkHistory} can handle $s$-wave annihilating and decaying DM models (\lstinline|DM_process='decay'|) with this keyword; 

 \item \lstinline|mDM=50e9| -- specifies the DM mass, in \SI{}{\eV}; 

 \item \lstinline|sigmav=2e-26| -- specifies the velocity averaged annihilation cross section, in \SI{}{\centi\meter\cubed\per\second}; 

 \item \lstinline|primary='b'| -- specifies the annihilation channel. The options include all of those offered by \textsc{pppc4dmid}~\cite{Cirelli:2010xx}, and the spectra are extracted from the raw data provided by the cookbook. The $e^+e^-$ and photon spectra from the showering of a single $b \overline{b}$ pair are shown in Fig.~\ref{fig:bbbar_spectra}. These are proportional to the injection spectra $\mathbf{N}_\text{inj}^\alpha$ defined in Sec.~\ref{sec:input}, and can be generated using the function \texttt{pppc.get\_pppc\_spec()}; 

 \item \lstinline|start_rs=3000| -- the redshift at which to start the evaluation. $1+z = 3000$ is the highest redshift at which we have produced the photon cooling transfer functions, and represents the highest redshift that should be specified here. In this example, \texttt{start\_rs} fixes the initial conditions of the TLA in Eq.~(\ref{eqn:TLA}) at the baseline ionization and temperature values at this redshift; 

 \item \lstinline|coarsen_factor=32| -- the coarsening factor, defined in Sec.~\ref{sec:coarsening}. For a comparison between solutions with different coarsening factors, see Appendix~\ref{app:cross_checks}; 

 \item \lstinline|backreaction=False| -- this turns backreaction on and off; and

 \item \lstinline|struct_boost=phys.struct_boost_func()| -- specifies the structure formation prescription to use. Once dark matter halos start to collapse, the annihilation rate gets enhanced by the factor 
 \begin{alignat}{1}
   1 + \mathcal{B}(z) \equiv \frac{\langle \rho_\chi^2 \rangle}{\langle \rho_\chi \rangle^2} 
   \label{eqn:boost_factor}
 \end{alignat}
 compared to the smooth annihilation rate shown in Eq.~(\ref{eqn:energy_injection}). Here, the keyword \texttt{struct\_boost} specifies a function that takes redshift as the argument, and returns $1 + \mathcal{B}(z)$. The user can make use of the structure formation boosts that are saved by default in \texttt{DarkHistory} in the \texttt{physics} module, which include the boost factors computed in Ref.~\cite{Liu:2016cnk}, and is used as the default boost factor by \lstinline{struct_boost_func()}. 

\end{enumerate}

By default, the solver integrates the equations down to $1+z = 4$, and will not evolve the helium ionization levels. These choices can of course be changed with other keyword parameters. Note that the function is not limited to DM processes or \textsc{pppc4dmid} spectra; other keyword parameters allow the user to specify their own injection rates as a function of redshift (see the documentation for the keyword parameters \texttt{rate\_func\_N} and \texttt{rate\_func\_eng}), along with the spectra of photons and $e^+e^-$ injected (see the documentation for the keyword parameters \texttt{in\_spec\_elec} and \texttt{in\_spec\_phot}). 

The output of \texttt{evolve()}, stored in \texttt{bbbar\_noBR}, is a dictionary containing the redshift abscissa of the solutions, the ionization and temperature solutions, the propagating photon, low-energy photon and low-energy electron spectra, and the computed value of $f_c(z)$. To access the redshift, ionization and temperature, we can simply do: 
\begin{lstlisting}{language=python}
  # Redshift abscissa.
  rs_vec   = bbbar_noBR['rs']   
  # Matter temperature in eV.
  Tm_vec   = bbbar_noBR['Tm']
  # Ionization fraction. 
  # Stored as 1+z by {xHII, xHeII, xHeIII}.
  xHII_vec = bbbar_noBR['x'][:,0] 
\end{lstlisting}

In Fig.~\ref{fig:bbbar} we plot $T_m$ and $x_\text{HII}$ as a function of redshift for the $\chi\chi \to b\bar{b}$ model. For DM masses above $\gtrsim \SI{10}{\giga\eV}$, values of $\langle \sigma v \rangle$ required for thermal freezeout are unconstrained by the CMB anisotropy power spectrum energy injection constraints: the ionization fraction, which changes by approximately 25\% only at high redshifts, does not change enough to affect the power spectrum significantly. The sudden increase in ionization and temperature at $z \sim 30$ corresponds to an increase in the boost factor used (halos with an Einasto profile with halo substructure boost included~\cite{Liu:2016cnk}, found in \texttt{physics.struct\_boost\_func()}). 

We also show in Fig.~\ref{fig:bbbar} for completeness the effect of turning on backreaction, i.e.\ including the effect of the increased ionization level on the evolution of the ionization and thermal histories. This is conveniently done by setting \texttt{backreaction=True}. In this particular example, the effect of backreaction is small, but we will show more scenarios where backreaction has large effect on $T_m$, and explain why this can be significant in the next example. 

\begin{figure*}[t!]
\begin{tabular}{c}
\includegraphics[scale=.45]{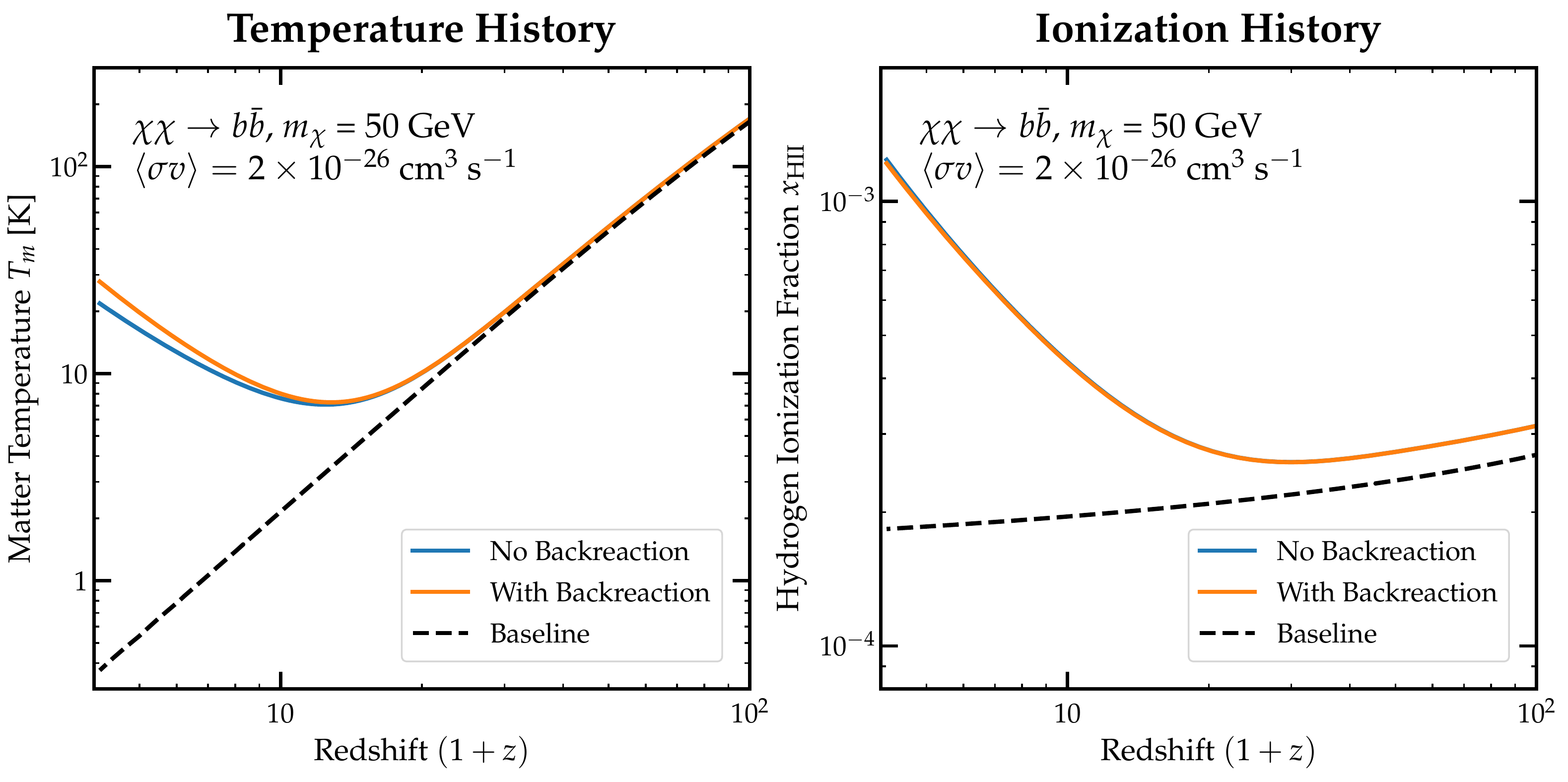}
\end{tabular}
\caption{Matter temperature $T_m$ (left) and hydrogen ionization fraction $x_\text{HII}$ (right) solved in the presence of dark matter annihilation into $b \bar{b}$ pairs using \texttt{DarkHistory}. Eq.~(\ref{eqn:TLA_DarkHistory}) is solved without dark matter energy injection to produce the baseline histories (black, dashed), with energy injection but without backreaction (blue), and with dark matter annihilation and backreaction (orange).  We assume a dark matter mass of \SI{50}{\giga\eV} and a velocity averaged annihilation cross section of \SI{2e-26}{\centi\meter\cubed\per\second}.
}
\label{fig:bbbar}
\end{figure*}
%

\subsection{Backreaction}
\label{sec:backreaction}

\begin{figure*}[t!]
 \centering
 \includegraphics[scale=0.45]{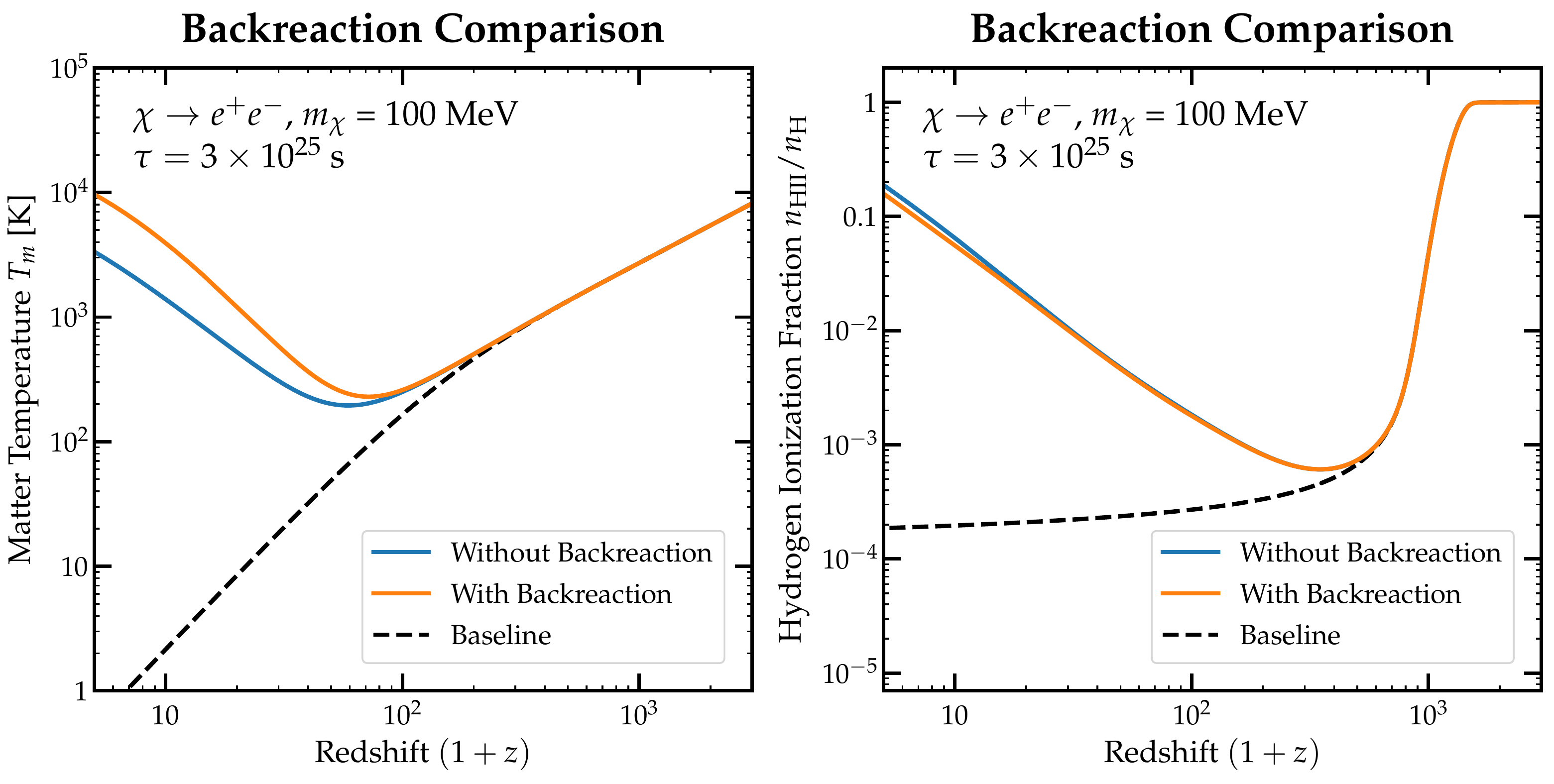}
 \caption{Temperature (left) and ionization (right) histories including the effects of dark matter decay to electrons and positrons.  We choose a lifetime of \SI{3e25}{\second}, which is consistent with the CMB constraints from Ref.~\cite{Slatyer:2016qyl}. We plot the baseline histories (black, dashed), the histories including dark matter energy injection but not backreaction (blue), and the histories including energy injection and backreaction (orange). These plots are a single vertical slice of the contour plots in Fig.~\ref{fig:backreact_mass_scan}. Additionally, these plots constitute a cross-check on \texttt{DarkHistory}, as they agree well with similar results obtained in Ref.~\cite{Liu:2016cnk}.}
 \label{fig:single_decay}
\end{figure*}

Let us explore the effects of backreaction a bit more using some of the code found in Example 10 of \texttt{DarkHistory}. As was described in Sec~\ref{sec:histories}, one of \texttt{DarkHistory}'s main improvements 
to ionization and temperature history calculations is its ability to include the effects of back-reaction. To see its importance, consider the example of \SI{100}{\mega\eV} dark matter decaying to a pair of $e^+e^-$, with a lifetime of $\tau = \SI{3e25}{\second}$, a value that is close to the minimum lifetime allowed by constraints from the CMB power spectrum~\cite{Slatyer:2016qyl}. The ionization and thermal histories can be evaluated in this way: 
\begin{lstlisting}{language=python}
  decay_BR = main.evolve(
    DM_process='decay', mDM=1e8, lifetime=3e25, primary='elec_delta', 
    start_rs=3000., coarsen_factor=16, backreaction=True
  )
\end{lstlisting}
The new keywords here are:
\begin{enumerate}

 \item \lstinline|DM_process='decay'| -- specifies the DM process of interest to be decays; 
 \item \lstinline|lifetime=3e25| -- specifies the decay lifetime in seconds; and
 \item \lstinline|primary='elec_delta'| -- the \texttt{primary} channel options \lstinline|'elec_delta'| and \lstinline|'phot_delta'| can be used to inject an $e^+e^-$ and $\gamma \gamma$ pair respectively, with no electroweak corrections applied. 

\end{enumerate}

To do the calculation without backreaction, we can simply set \lstinline|backreaction=False|. However, with \lstinline|primary='elec_delta'| or \lstinline|'phot_delta'|, \texttt{DarkHistory} can instead rely on tabulated results of $f_c(z)$ for these two channels, using the same method based on results from Ref.~\cite{Slatyer:2015kla}, to calculate the ionization and thermal histories without evolving the input spectrum, leading to a significant speed-up. This can be done using the function \texttt{tla.get\_history()}:

\begin{lstlisting}{language=python}
  import numpy as np
  from darkhistory.tla import get_history
  # get_history takes a redshift vector:
  rs_vec = np.flipud(np.arange(5, 3000, 0.1)) 

  result = get_history(
     rs_vec, baseline_f=True, mDM=1e8, lifetime=3e25, 
     DM_process='decay', inj_particle='elec_delta' 
  )
\end{lstlisting}
with the following parameters:

\begin{enumerate}

 \item \lstinline|rs_vec| -- the redshift vector, ordered from high to low, over which the temperature and ionization histories are to be evaluated; 
 \item \lstinline|baseline_f=True| -- this tells the code to use the baseline $f_c(z)$ computed by \texttt{DarkHistory} without backreaction. As we discussed in Sec.~\ref{sec:calculating_f}, these $f_c(z)$ agree with those computed in Ref.~\cite{Slatyer:2015kla} to within 10\%, and
 \item \lstinline|inj_particle='elec_delta'| --  used to specify one of two options: one of either \lstinline|'elec_delta'| or \lstinline|'phot_delta'|.

\end{enumerate}

The output \texttt{result} is an array of shape \lstinline|(len(rs_vec), 4)|, with the second dimension indexing $\{T_m, x_\text{HII}, x_\text{HeII}, x_\text{HeIII}\}$. The temperature (in \SI{}{\eV}) can be accessed through \lstinline|T_m = results[-1,0]|. 

Although only the $f_c(z)$ values for the injection for an $e^+e^-$ and $\gamma \gamma$ pair have been saved for use with \texttt{DarkHistory}, the $f_c(z)$ for any arbitrary channel can be computed from a weighted average of the electron and photon results~\cite{Slatyer:2015kla}. We stress once again, however, that this can only be done assuming no backreaction.

The histories are shown in Fig~\ref{fig:single_decay}, with and without backreaction turned on. First, even though the ionization level at $z \sim 10$ is three orders of magnitude larger than the baseline, such a scenario is actually still consistent with the CMB power spectrum constraints, owing to the fact that the ionization build-up occurs relatively late: the CMB constraints are sensitive to changes in $x_e$ near recombination, and become less sensitive at later times. 

Comparing the temperature histories with and without backreaction, we see that the main effect of this increase in $x_e$ on the energy deposition processes is to increase energy deposition into heating. Ionization and excitation rates depend on the neutral fraction, which is still close to 100\% even with energy deposition from DM. However, the energy rate into Coulomb heating is proportional to $x_e$, so taking into account the significantly elevated $x_e$ values leads to higher temperature levels. By about $z \sim 10$, $T_m$ with backreaction is larger than without backreaction by a factor of $\sim 4$, with the difference continuing to grow. Neglecting backreaction therefore leads to a severe underestimate of $T_m$, and including this effect consistently will certainly be important in understanding what measurements of $T_m$ at $z \simeq 20$ through the 21-cm signal or the Lyman-$\alpha$ power spectrum can tell us about exotic sources of energy injection.

We can perform the calculation over a range of DM masses by looping over values of \texttt{mDM}. For each value of $m_\chi$, we select the minimum lifetime $\tau$ which is consistent with the CMB power spectrum constraints, and compare the difference between the temperature history with backreaction ($T_{m,\text{BR}}$) and without ($T_{m,0}$) by computing the fractional change in temperature,
\begin{alignat}{1}
 \frac{\delta T_m}{T_{m,0}}(m_\chi, z) = \frac{T_{m,\text{BR}}(m_\chi, z) -T_{m,0}(m_\chi, z)}{T_{m,0}(m_\chi, z)} \,.
 \label{eqn:fractional_Tm_change}
\end{alignat}
In Fig~\ref{fig:backreact_mass_scan} we plot this variable over a range of redshifts and dark matter masses for this particular channel ($\chi \to e^+e^-$), but also for decay and annihilation into $e^+e^-$ and $\gamma \gamma$, taking the maximum $\langle \sigma v \rangle$ again allowed by the CMB power spectrum constraints. At a redshift of $z \sim 17$ near the end of the cosmic dark ages, $\delta T_m/T_{m,0} \sim 100\%$ (i.e.\ $T_m$ with backreaction is a factor of 2 larger than without) or more can easily be obtained. Even larger deviations are possible at lower redshifts, depending on the channel under consideration.

\begin{figure}[t!]
\centering
\begin{tabular}{cc}
 \includegraphics[scale=0.37]{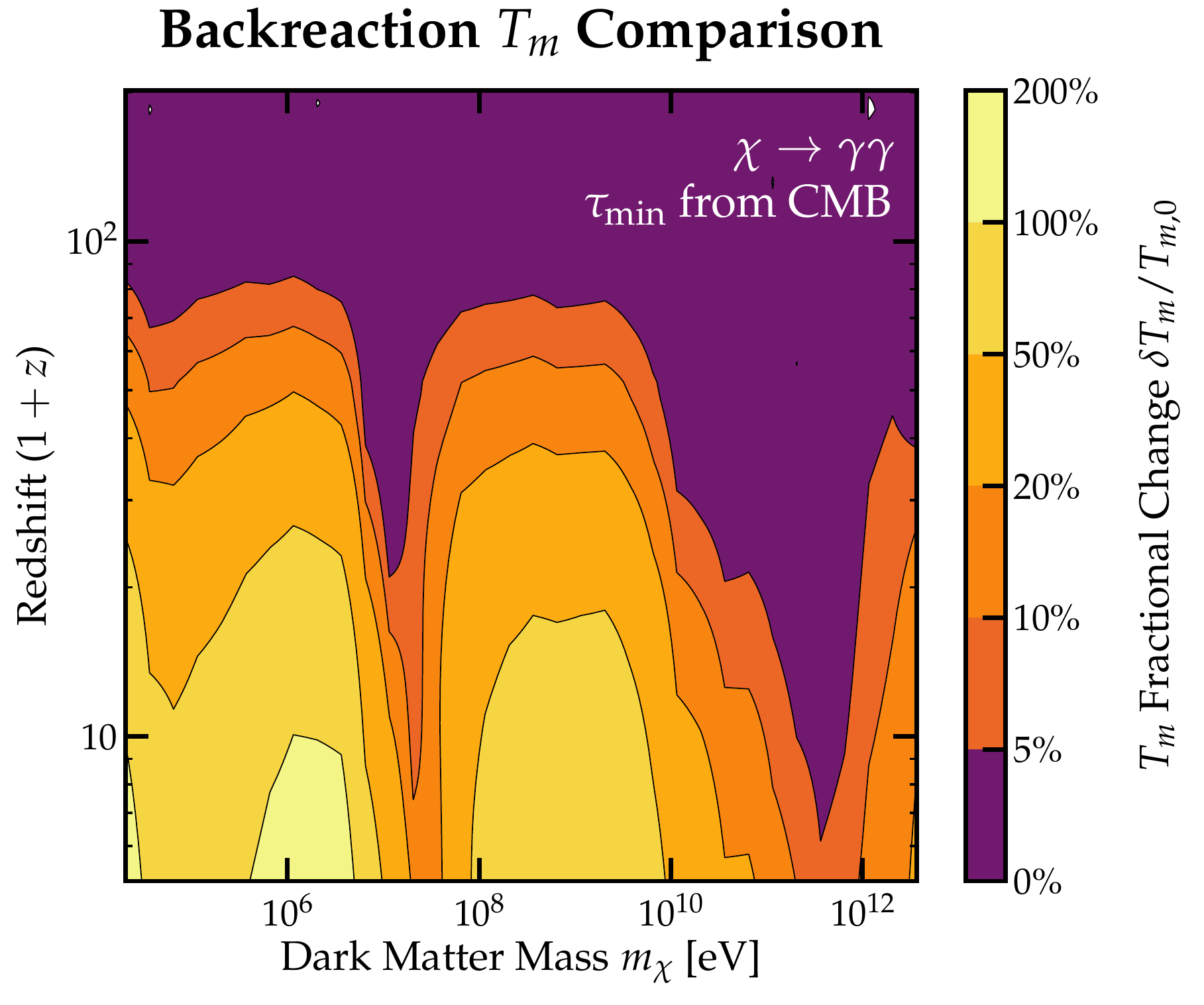} &
 \includegraphics[scale=0.37]{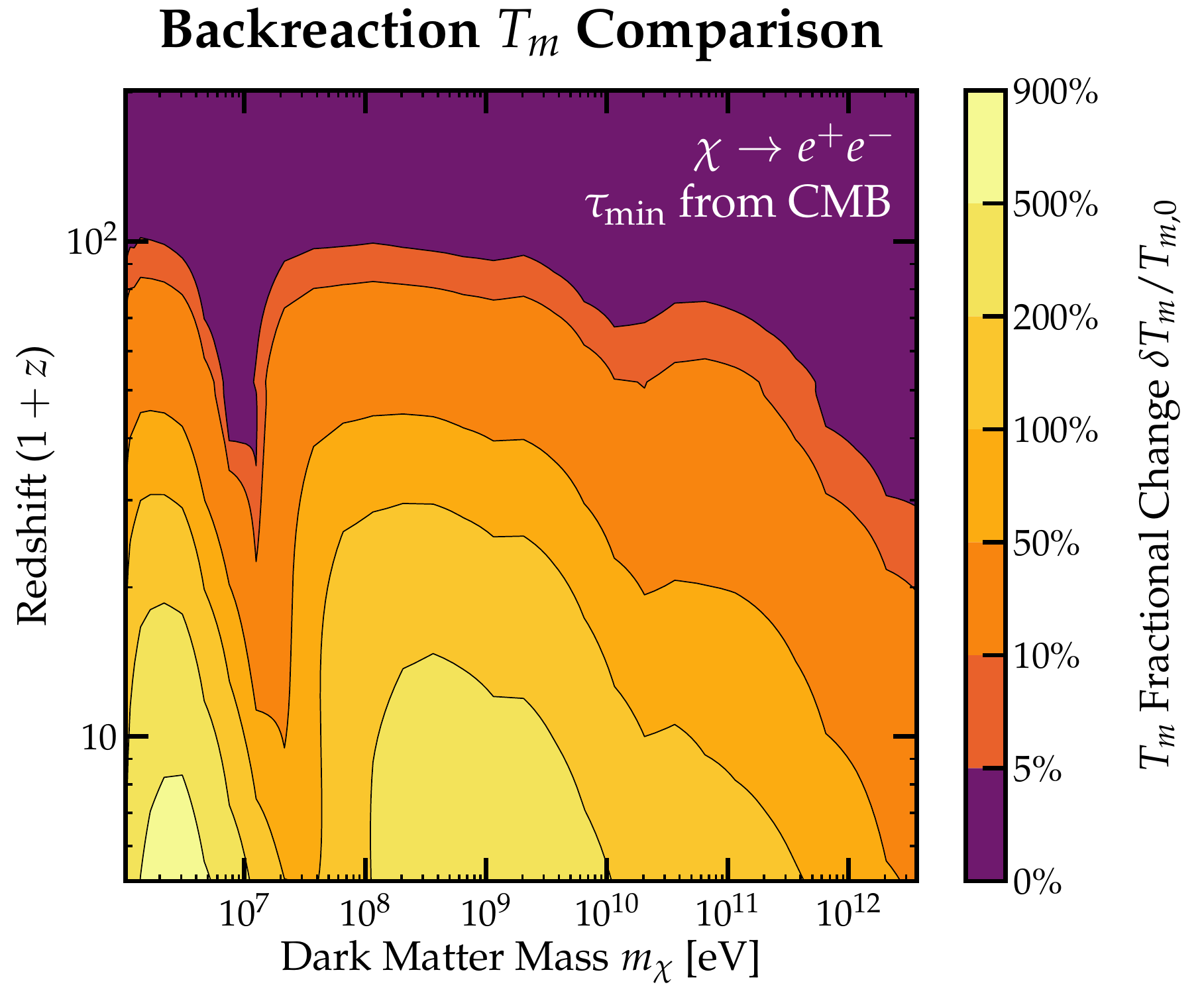} \\
 \includegraphics[scale=0.37]{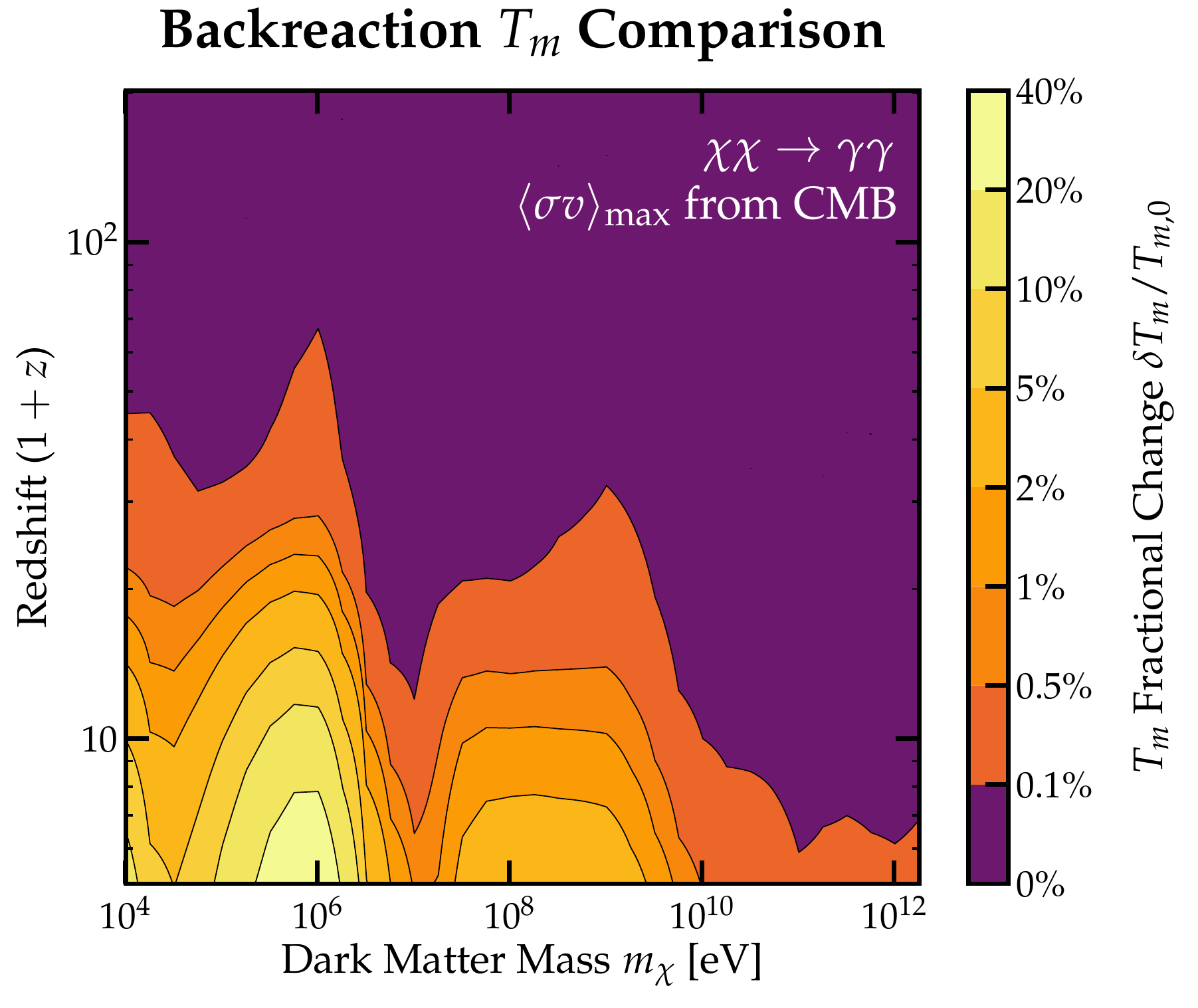} &
 \includegraphics[scale=0.37]{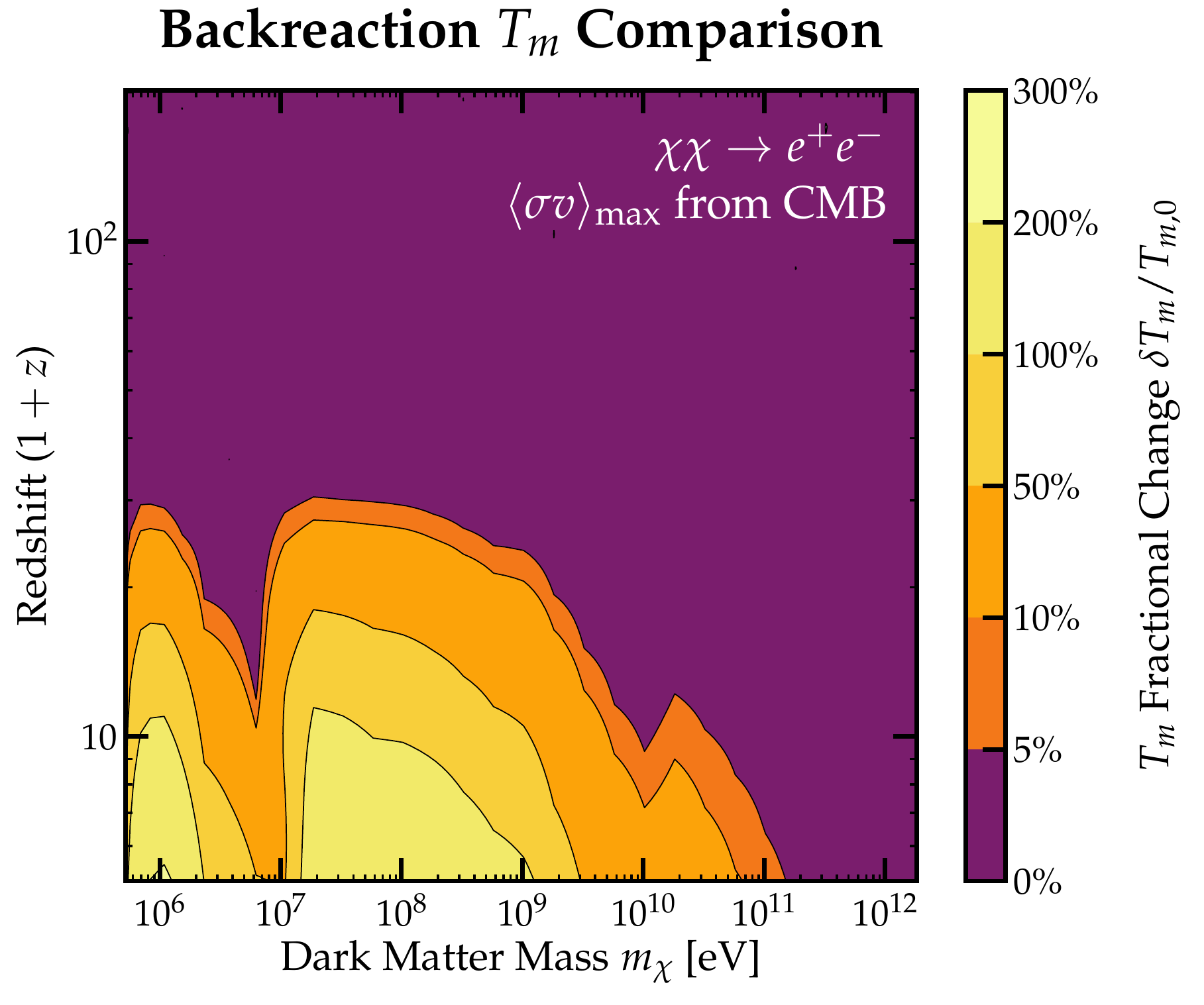} \\
\end{tabular}
\caption{Contour plots of the fractional change in temperature $\delta T_m/T_{m,0}$ caused by including the effects of backreaction, as a function of dark matter mass and redshift (See Eq.~(\ref{eqn:fractional_Tm_change})). For each dark matter mass, we choose the minimum $\tau$ or maximum $\langle \sigma v \rangle$ allowed by current CMB power spectrum constraints~\cite{Slatyer:2016qyl,Slatyer:2015kla}. 
}
\label{fig:backreact_mass_scan}
\end{figure}
%

\subsection{21-cm Sensitivity}
\label{sec:21_cm_sensitivity}

\begin{figure}[t!]
\centering
\begin{tabular}{cc}
 \quad \includegraphics[scale=.43]{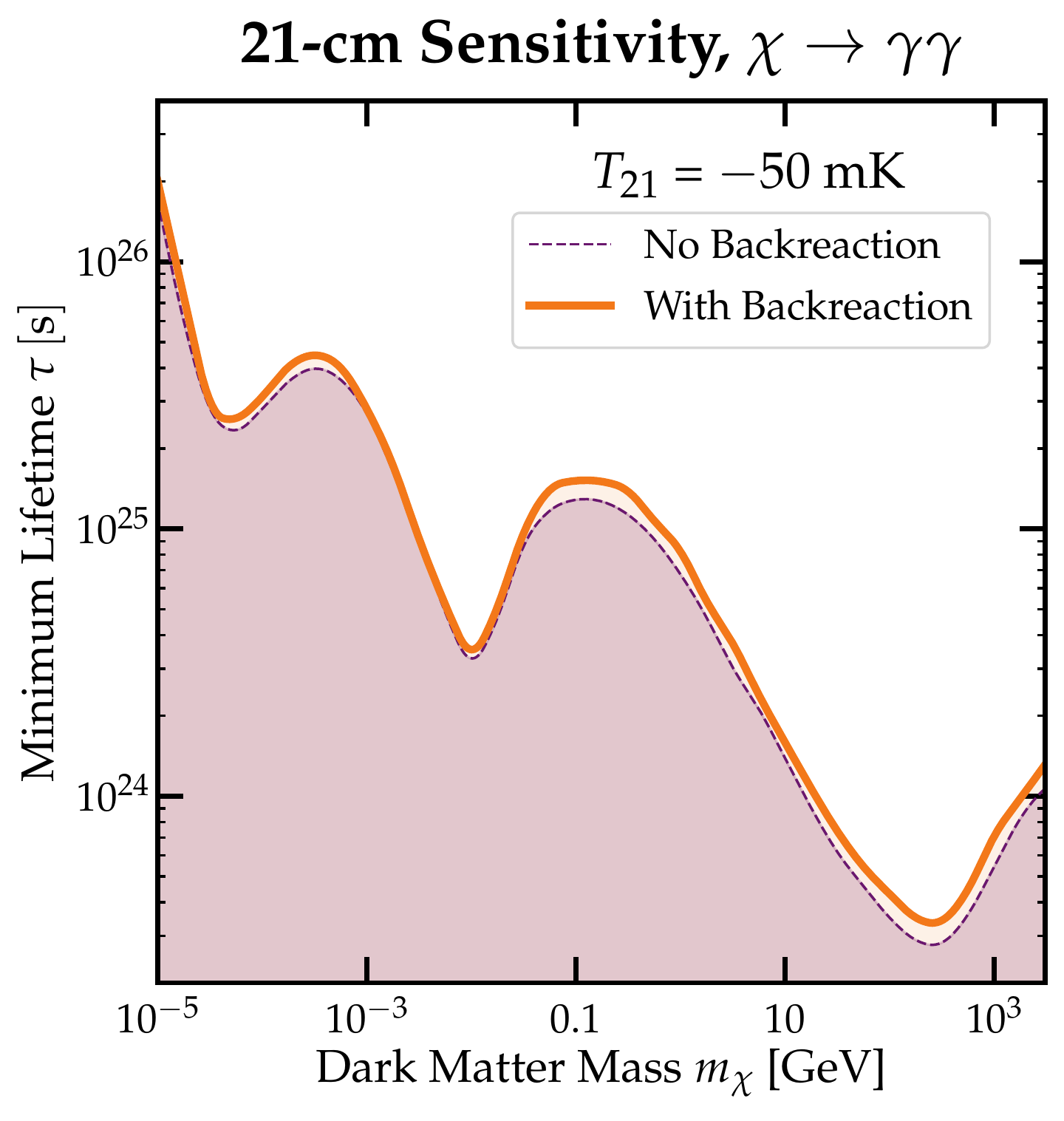}&  \
 \quad \includegraphics[scale=.43]{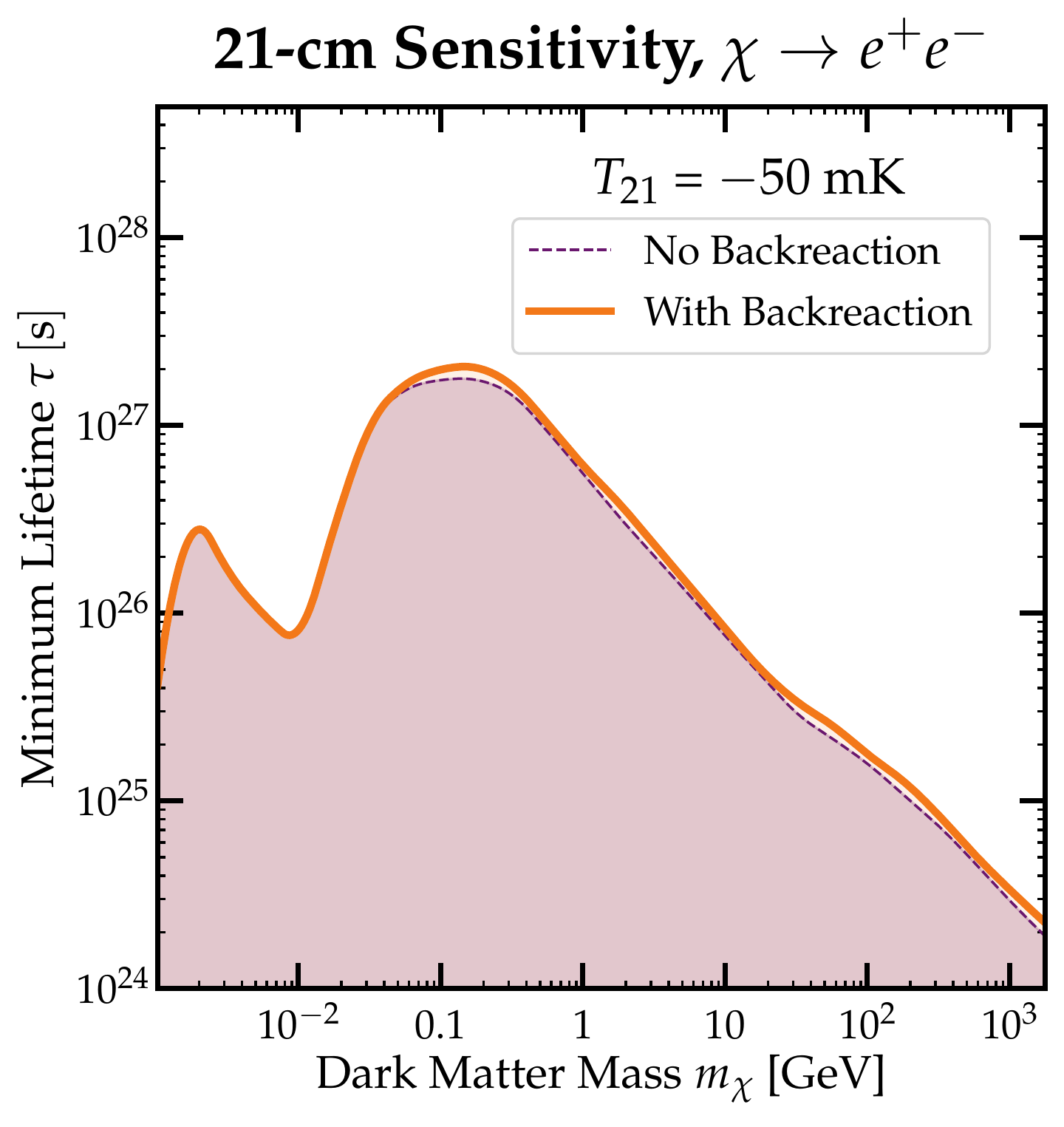} \\[6pt]
 \includegraphics[scale=.43]{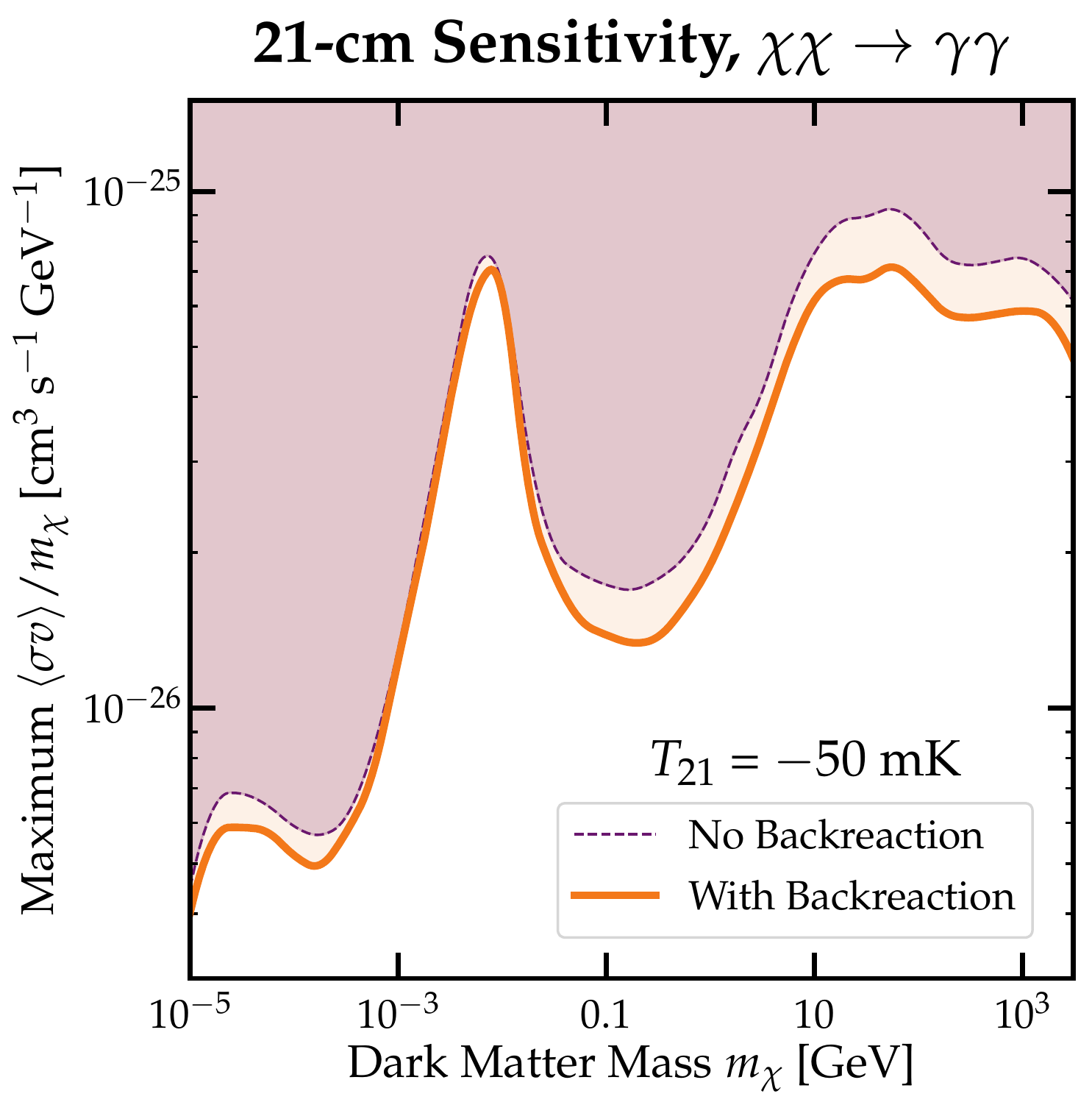} & \
 \includegraphics[scale=.43]{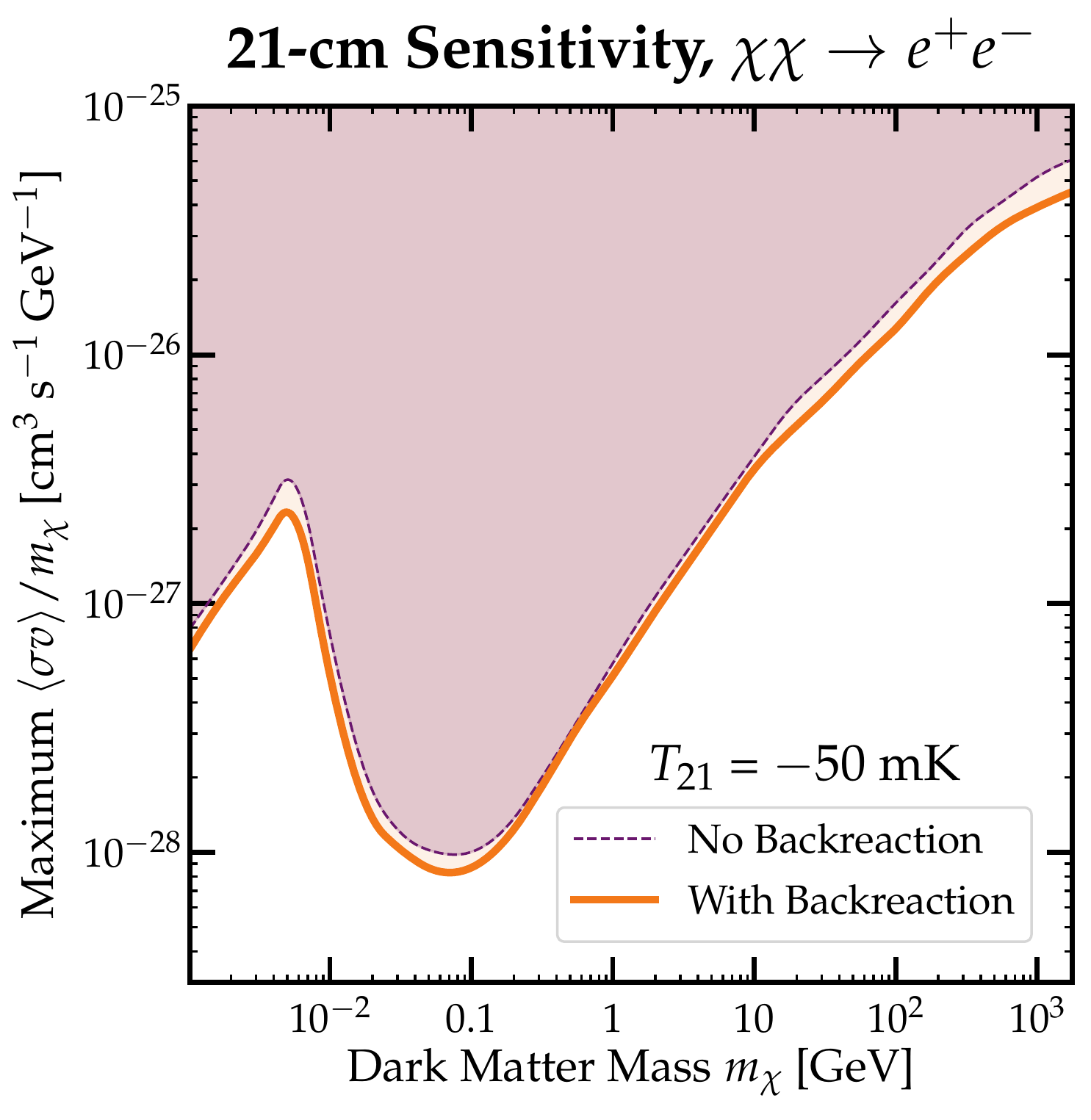} \\[6pt]
\end{tabular}
\caption{The minimum dark matter decay lifetime (top row) and maximum annihilation cross section (bottom row) bounds, derived from the global 21-cm signal. We assume a differential 21-cm brightness temperature of $T_\text{21} = -\SI{50}{\milli\kelvin}$, corresponding to a maximum $T_m$ of about \SI{20.3}{\kelvin} at $z \sim 17$. We consider decay and annihilation into $\gamma \gamma$ (left column) and $e^+e^-$ (right column) and compute the bounds with (orange, solid) and without (purple, dashed) backreaction.}
\label{fig:21cm}
\end{figure}

The global 21-cm signal is a measurement of the sky-averaged differential brightness temperature $T_{21}$ with respect to the background radiation. Measurements of this signal would open a window into the ionization and temperature histories of the universe at the cosmic dawn (see e.g.\ Ref.~\cite{Pritchard:2011xb} for a review of 21-cm cosmology). A first claim of such a measurement has already been made by the EDGES collaboration~\cite{Bowman:2018yin}. The brightness temperature of the 21-cm hydrogen absorption line relative to the background radiation temperature is given by~\cite{Pritchard:2011xb}:
\begin{alignat}{2}
	T_\text{21} &\approx&& \,\, x_\text{HI}(z) 
	 \left( \frac{0.15}{\Omega_m}\right)^{1/2} 
	 \left( \frac{\Omega_b h}{0.02}\right) \left( \frac{1+z}{10} \right)^{1/2} 
	 \left[ 1 - \frac{T_R(z)}{T_S(z)} \right]  \; \SI{23}{\milli \kelvin} \,,
	\label{eqn:T21}
\end{alignat}
where $\Omega_b$ is the baryon energy density today as a fraction of the critical density, $h$ is the Hubble parameter today in \SI{}{\kilo\meter \per \second \per \mega\parsec}, $T_R$ is the background radiation temperature (typically assumed to be the CMB temperature) and $T_S$ is the spin temperature of neutral hydrogen as a function of redshift, which determines the relative population of neutral hydrogen in the two hyperfine states. Due to the presence of an intense Lyman-$\alpha$ radiation field once stars begin to form, it is expected that $T_S \approx T_m$ at the cosmic dawn. This fact allows us to turn the 21-cm global signal into a limit on $T_m$ itself, assuming that $T_R = T_\text{CMB}$. 

We will focus on $1+z \approx 18$, roughly the central value of the absorption trough measured by EDGES~\cite{Bowman:2018yin}. At this redshift, almost all hydrogen is neutral, i.e.\ $x_\text{HI} \approx 1$, and we can invert Eq.~(\ref{eqn:T21}) to find $T_S$ as a function of $T_\text{21}$. Since $T_m < T_S$, this yields the bound
\begin{alignat}{1}
	T_m(z=17) < \left( 1 - \frac{T_\text{21}}{\SI{35}{\milli\kelvin}} \right)^{-1} \SI{49}{\kelvin} \,.
	\label{eqn:Tm_bound}
\end{alignat}
This temperature bound in turn puts a limit on the DM decay lifetime or cross-section 
because too much dark matter decay/annihilation would heat up $T_m$ past this point.

In contrast to the CMB power spectrum energy injection bounds, which is most sensitive to changes in $x_e$ around the time of recombination, the 21-cm global signal constraints are more sensitive to energy injection processes that are more active at late times, and are dependent primarily on $T_m$ instead. Since $T_m$ is significantly impacted by including the effects of backreaction, the calculation performed by \dhis becomes important for setting accurate constraints using the 21-cm global signal.

\begin{figure}[t]
    \centering
    \includegraphics[scale=0.45]{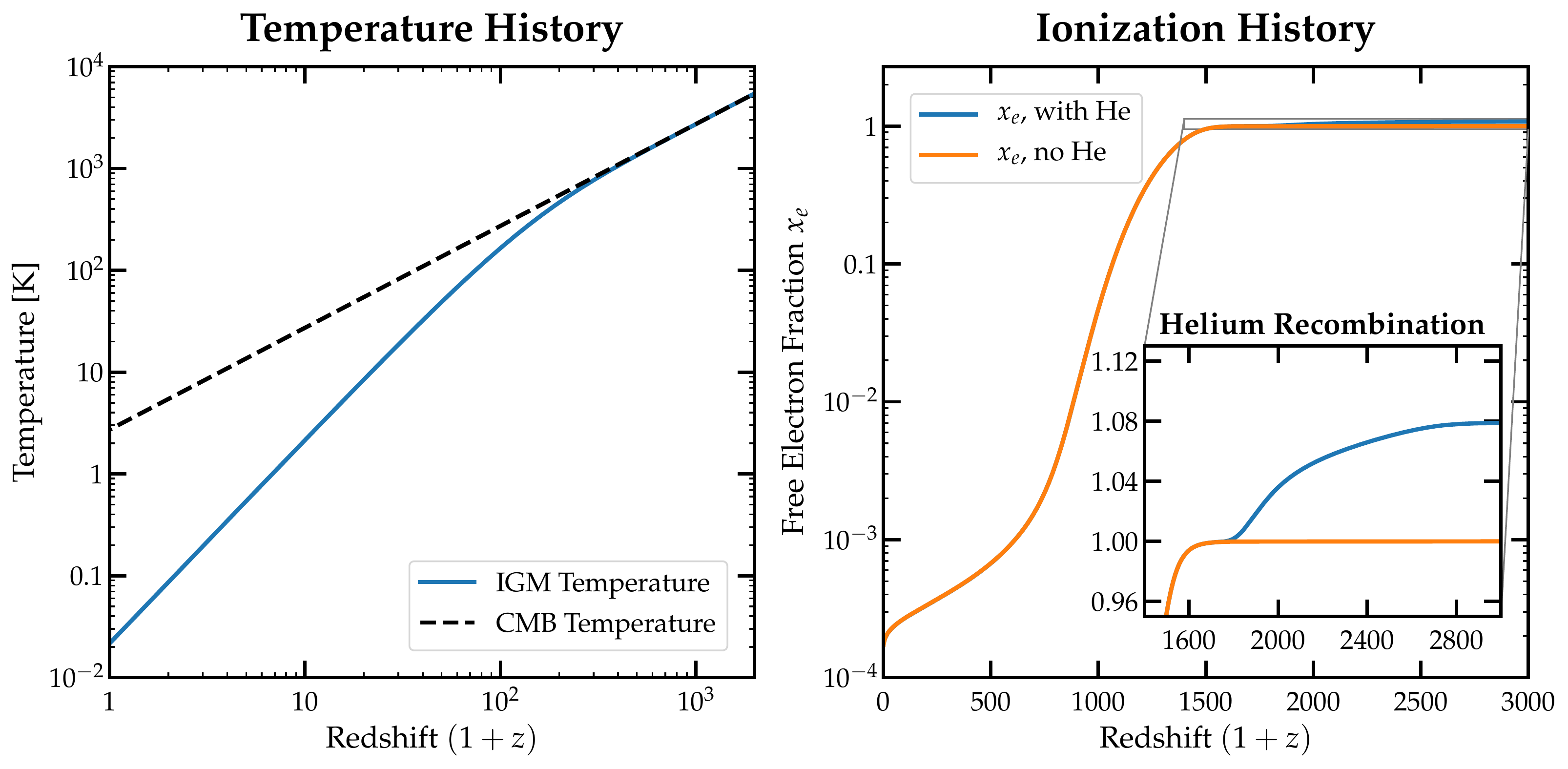}
    \caption{Temperature (left) and free electron fraction $x_e$ (right) as a function of redshift. $x_e$ is solved in \texttt{DarkHistory} with (blue) and without (orange) helium; both options lead to a similar temperature history (blue). With helium, helium recombination is correctly computed (inset). The CMB temperature is shown (black, dashed) for reference.}
 \label{fig:He_recomb}
\end{figure}

To illustrate this, we perform a simple sensitivity study by obtaining the constraints for a measured $T_\text{21}$ of \SI{-50}{\milli\kelvin}, and compare the constraints with and without backreaction taken into account. Although this value of $T_{21}$ is inconsistent with the EDGES experiment, it is impossible to interpret the EDGES result without proposing new physics that may be at play during the cosmic dark ages~\cite{Liu:2018uzy}, which is a more complicated task and less relevant to helping users understand the code.  The following analysis is worked out in more detail within the code in Example 11.

$T_{21} = \SI{-50}{\milli\kelvin}$ means that we require $T_m < \SI{20.3}{\kelvin}$ according to Eq.~(\ref{eqn:Tm_bound}). We once again scan over a grid of dark matter masses and lifetimes/cross-sections decaying/annihilating into $e^+e^-$ and $\gamma\gamma$, using \texttt{get\_history()} for the case with no backreaction and \texttt{evolve()} for the case with backreaction, as explained in the previous section, to find where in parameter space dark matter energy injection leads to a violation of Eq.~(\ref{eqn:Tm_bound}).

The resulting exclusion plots are shown in Fig.~\ref{fig:21cm}.  
We see that in each case the calculation with backreaction can be between 10\%--50\% stronger than without backreaction, which we would expect because backreaction leads to larger temperatures. We emphasize that this is the result for just one chosen value of $T_{21}$; for larger (less negative) $T_{21}$, we expect that the importance of backreaction will increase, since the energy injection is less constrained, allowing for larger values of $x_e$.

\subsection{Helium, Dark Matter and Reionization}
\label{sec:example_reionization}

\begin{figure*}[t]
   \centering
 \includegraphics[scale=0.45]{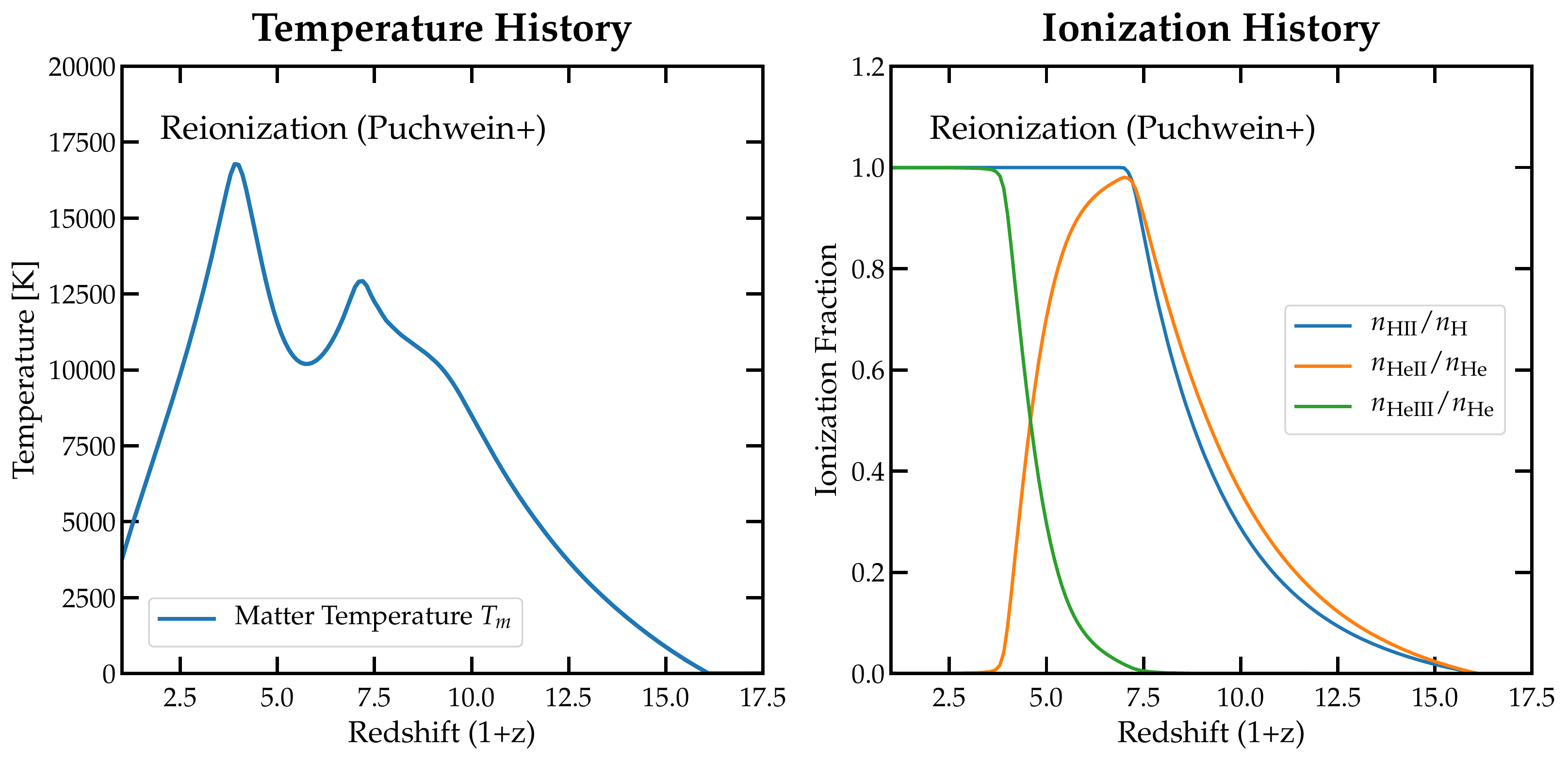}
 \caption{Temperature (left) and free electron fraction (right) as a function of redshift, solved in \texttt{DarkHistory} with the default Puchwein+ reionization model~\cite{Puchwein:2018arm}. The IGM temperature (blue) is shown on the left, while the ionization fractions $n_\text{HII}/n_\text{H}$ (blue), $n_\text{HeII}/n_\text{He}$ (orange) and $n_\text{HeIII}/n_\text{He}$ (green) are shown as well. These results agree very well with the same plots shown in Ref.~\cite{Puchwein:2018arm}.}
 \label{fig:Puchwein_reion}
\end{figure*}

\begin{figure*}[t]
 \centering
 \includegraphics[scale=0.45]{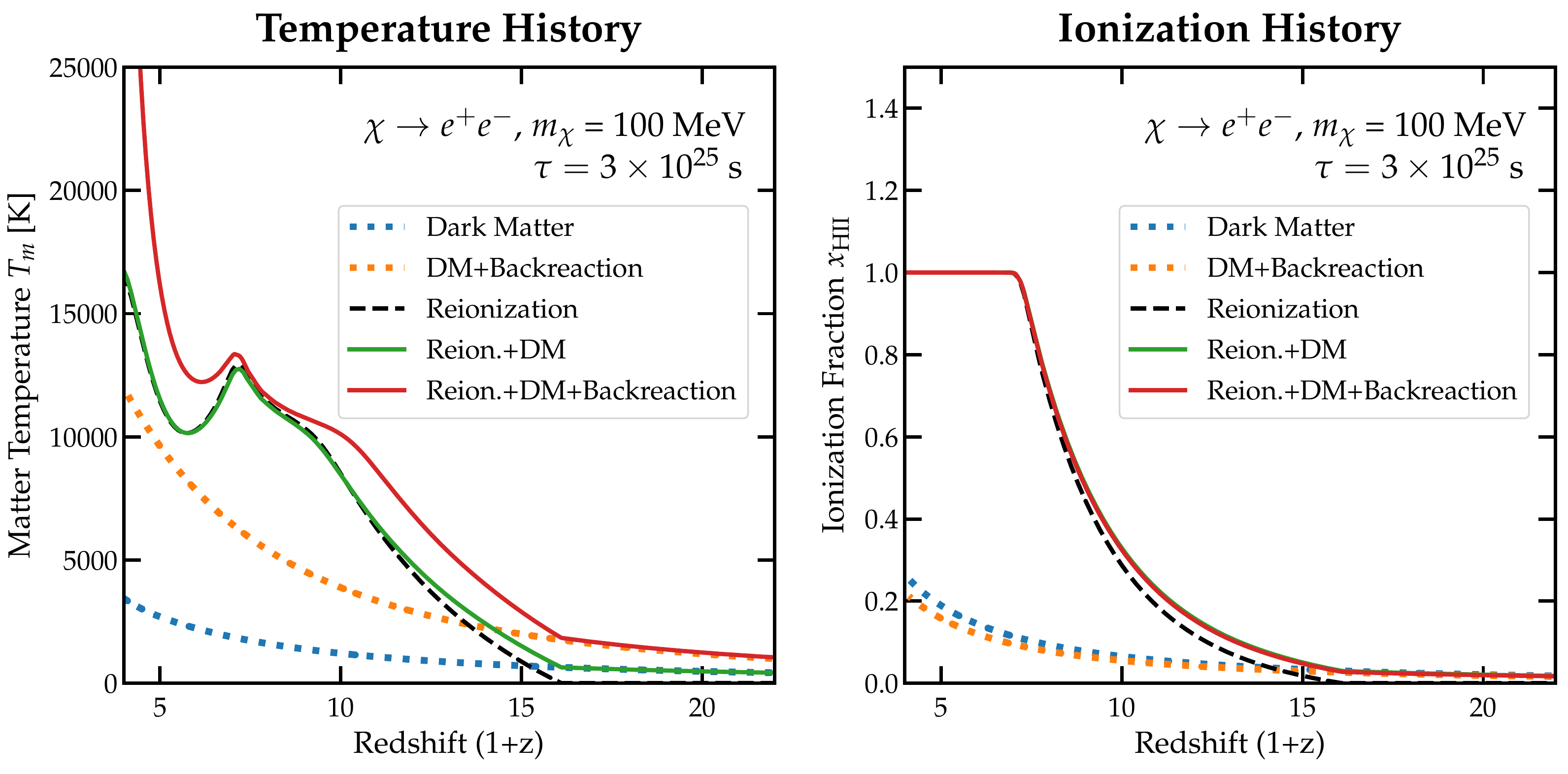}
 \caption{Temperature (left) and hydrogen ionization (right) history of the universe with DM decay and the default reionization model. The DM has a mass of $m_\chi = \SI{100}{\mega\eV}$ and decays to $e^+e^-$ with a lifetime of \SI{3e25}{s}. The temperature and ionization with DM decay alone is shown without (blue, dotted) and with (orange, dotted) backreaction included. The combined effect of DM decay and reionization without (green) and with (red) backreaction can be compared to the reference reionization model (black, dashed).}
 \label{fig:DM_reion}
\end{figure*}

Finally, we will take a closer look at the different options one can use within the code to evaluate temperature and ionization histories. Throughout this section, we will demonstrate these different options mostly using \texttt{get\_history()}, but similar options are also available in \texttt{evolve()}, which calls \texttt{get\_history()} with all of the relevant options provided. We refer the reader to the online documentation and to Example 8 in the code for more details.

Without any exotic energy injection or reionization, the function \texttt{get\_history()} accepts a redshift vector, and simply returns the baseline ionization and temperature histories, obtained by solving Eq.~(\ref{eqn:TLA_DarkHistory}):  
\begin{lstlisting}[language=python]
  import numpy as np
  from darkhistory.tla import get_history

  # Redshift vector in decreasing order. 
  rs_vec = np.flipud(np.arange(1., 3000., 0.1))
  soln_baseline = get_history(rs_vec)
\end{lstlisting}
Helium evolution within \texttt{get\_history()} is controlled by the flag \texttt{helium\_TLA}, i.e.
\begin{lstlisting}
  soln_He = get_history(rs_vec,helium_TLA=True)
\end{lstlisting}
Fig.~\ref{fig:He_recomb} shows the solution to Eq.~(\ref{eqn:TLA}) with just the ``$(0)$'' terms, i.e.\ without any energy injection or reionization, and compares that solution to one with Eq.~(\ref{eqn:TLA_helium_darkhistory}) added as well. This is simply the standard ionization history with helium recombination ($z \sim 1800$) and hydrogen recombination ($z \sim 1100$), eventually leading to the residual ionization fraction at redshifts well below hydrogen recombination of about $x_e \sim 2 \times 10^{-4}$. The inset of Fig.~\ref{fig:He_recomb} shows that \dhis is able to correctly reproduce helium recombination; the entire ionization history agrees with \texttt{RECFAST} results at the central cosmological parameters used by \dhis to within $\sim 3\%$. We recommend that helium ionization levels are tracked when used in combination with reionization.

The next important option is whether to include the effects of reionization. This option is controlled by the flag \texttt{reion\_switch}: 
\begin{lstlisting}
  soln_default_reion = get_history(
   rs_vec, helium_TLA=True, reion_switch=True
  )
\end{lstlisting}
With no other options set, setting \texttt{reion\_switch} to \lstinline|True| causes \dhis to use the standard reionization model, which is based on the photoionization and photoheating rates provided in~\cite{Puchwein:2018arm}. Fig.~\ref{fig:Puchwein_reion} shows the IGM temperature as well as the ionization levels of the different atomic species as a function of redshift. Both of these results agree well with the same result shown in Ref.~\cite{Puchwein:2018arm}. Reionization of hydrogen and neutral helium is complete by about $z \sim 6$; soon after, HeII starts to become doubly ionized, leading to a decrease in $n_\text{HeII}$ and a corresponding increase in $n_\text{HeIII}$. Dips in $T_m$ correspond to a decrease in photoheating rates once a species becomes completely ionized and the production of high-energy electrons from photoionization off these species ceases.

Aside from the default reionization model, the user may also supply their own reionization models in two different ways: by either providing their own photoionization and photoheating rates on each atomic species (e.g. based on a model that is different from the default, e.g.~\cite{Haardt:2011xv}), or by fixing the ionization history below a certain redshift, e.g.\ with a $\tanh$ model~\cite{Lewis:2008wr,Adam:2016hgk}. We leave a discussion of how to use these options to Example 8 in the code. 

With the ability to include both helium and reionization, we can now add a new source of energy injection and compute the effects on ionization and temperature levels. We remind the reader that this means we are solving Eq.~(\ref{eqn:TLA_DarkHistory}) together with Eq.~(\ref{eqn:TLA_helium_darkhistory}). This is accomplished in the code with both \texttt{reion\_switch} and \texttt{helium\_TLA} set to \lstinline|True|, and supplying the same keyword parameters used to inject energy from DM shown in Sec.~\ref{sec:backreaction}. We can add decaying DM with mass \SI{100}{\mega\eV} into an $e^+e^-$ pair with a lifetime of \SI{3e25}{\second}, like so (using \texttt{evolve()} in this example):
\begin{lstlisting}
  main.evolve(
    DM_process='decay', mDM=1e8, lifetime=3e25, primary='elec_delta',
    start_rs=3000., coarsen_factor=1, backreaction=True, 
    helium_TLA=True, reion_switch=True
  )
\end{lstlisting}
By turning on and off the flags \texttt{backreaction}, \texttt{helium\_TLA} and \texttt{reion\_switch}, we can produce histories including or excluding these various effects. 

The results from different combinations of these switches are summarized in Fig.~\ref{fig:DM_reion}. The dashed lines shows the histories with DM decay only, and illustrates the significant difference that can arise after taking into account backreaction, which we have already seen in Fig.~\ref{fig:single_decay}. Combining the DM energy injection with the reionization model gives the solid lines in Fig.~\ref{fig:DM_reion}. These curves should be compared to the default reionization model temperature and ionization histories, shown in the black, dashed lines. When computing the DM energy deposition without taking into account backreaction, we find that the amount of energy deposited into heating from DM is much smaller than heating from reionization processes once they begin in earnest, and so adding the DM decays on top of reionization produces only a small perturbation in the temperature history relative to $T_m$ for just the reionization model alone. In some cases, the addition of DM actually decreases $T_m$: this can happen due to reionization proceeding at a faster rate, leaving fewer atoms to photoionize and thus suppressing photoheating. 

It is clear, however, that neglecting backreaction leads to a severe underestimation of the energy deposition into heating. Performing the full calculation with DM, reionization and backreaction correctly accounted for produces the line in red, which shows that the addition of DM significantly increases $T_m$ compared to both the reionization model and the case where DM energy deposition is added without backreaction. Reionization greatly enhances the energy deposition rate into heating of the IGM by increasing the number of free charged particles available for Coulomb heating, and properly accounting for backreaction using \dhis is critical to predicting the IGM temperature growth due to energy injection once reionization begins. 

\section{Conclusion}
\label{sec:conclusion}

We have developed and made public a new code package for mapping out the effects of arbitrary exotic energy injections --- including dark matter annihilation and decay to arbitrary Standard Model final states --- on the temperature and ionization history of the early universe. \texttt{DarkHistory} is capable of self-consistently including the effects of conventional astrophysical sources of ionization and heating, and of including feedback effects that can significantly enhance the degree of heating. Additionally, the ICS module can be employed independently of the rest of the code, as an accurate and efficient numerical calculator of ICS across a very wide range of electron and photon energies. We have outlined here a number of worked examples, and provide more examples with the online code at \href{https://github.com/hongwanliu/DarkHistory}{https://github.com/hongwanliu/DarkHistory}.

\texttt{DarkHistory} has a modular framework and can in the future be improved in several different directions, while keeping the same underlying structure. In this first version we have focused on the homogeneous signal, and neglected the possible effect of new radiation backgrounds and/or gas inhomogeneities on the cascade of secondaries produced by injected high-energy particles. Such effects may become important in the late cosmic dark ages and the epoch of reionization. The spectrum of low-energy photons produced by energy injection, and the resulting distortion to the spectrum of the CMB, is a possible observable in its own right; the current version of \texttt{DarkHistory} provides only a partial calculation of this spectral distortion, due to our approximate treatment of low-energy electrons, but we intend to improve this aspect in future work. The effects of other new physics on the temperature/ionization evolution -- in particular, scattering between baryons and DM -- can be incorporated within the same framework. We also intend to explore the possibility of interfacing DarkHistory with existing public codes for computing the recombination history, perturbations to the CMB, and 21cm signals. 

The tools we have developed in this work can be used to understand the visible imprints of exotic energy injections that could appear in the CMB and the 21cm line of neutral hydrogen, and hence to place precise constraints on dark matter annihilation and decay. We hope they will help pave the way for a comprehensive description of the ways in which dark matter interactions, and other physics beyond the Standard Model, could reshape the early history of our cosmos.

\chapter{Conclusion}

In this thesis, I have discussed several contributions to our theoretical understanding of dark matter production and energy deposition across a large range of length scales. These ideas may give us some insight into the nature of dark matter itself, or at least ensure that we leave no stone unturned in our search. 

Much still remains to be done based on the work in this thesis. In particular, our work on \texttt{DarkHistory} can form the basis for accurately computing the effects of dark matter energy deposition on the 21-cm global signal and power spectrum with the help of existing 21-cm codes. Future improvements to \texttt{DarkHistory}, including a better treatment of electron cooling and integration with recombination codes like \textsc{hyrec}, will allow us to obtain the spectral distortion to the CMB due to dark matter processes, opening a new window into the dark sector.

We have only just begun to explore the myriad of ways in which dark matter may yet surprise us phenomenologically. The ideas contained in this thesis are only the beginning, and I hope there are many more exciting ideas about dark matter to come. 
\appendix

\chapter{Not-Forbidden Dark Matter}
\label{chap:app_NFDM}

\section{Coupled Boltzmann Equations and Prolonged Freezeout}

As mentioned above, an essential difference between NFDM and
conventional DM freezeout is the importance of tracking the evolution
of both the DM $\chi$ and the mediator particle (in our model, $A'$),
by solving the coupled Boltzmann equations (Eq.~(\ref{eq:boltz1}) and (\ref{eq:boltz2}) for relevant terms when $r \gtrsim 1$, Eq.~(\ref{eq:fullboltz1}) and (\ref{eq:fullboltz2}) for the complete equations)  for their respective densities.   The presence of two
equations implies that more than one scattering (or decay) process can
be important for determining the final abundance; hence both the
fastest and second fastest reactions are typically relevant.

This is in contrast to conventional DM freezeout based upon a single
Boltzmann equation, where the abundance depends upon the strongest
channel.  In the large $\epsilon$ limit of our model,
$A'$ decay is the fastest process, and enforces equilibrium of 
$A'$, $n_{A'} = n_{A',0}$. 
Hence smaller values of $\epsilon$ are necessary to realize the rich cosmology that comes from the interplay of the coupled Boltzmann equations of $\chi$ and $A'$. To simplify the subsequent discussion, we assume that these $\epsilon$-suppressed reactions are negligibly slow, i.e. we work in the secluded dark sector regime of the NFDM model.

It is useful to define the 
net rate of $3 \leftrightarrow 2$ or $2 \leftrightarrow 2$ interactions per $\chi$ or
$A'$ particle by considering the collision terms in the Boltzmann equations, 
written in the form $R_{\chi} \equiv d \log n_{\chi}/dt = -3H - R_{\chi}(3 \leftrightarrow 2) + R_{\chi}(2 \leftrightarrow 2)$ and $R_{A'} \equiv d \log n_{A'}/dt = -3H + R_{A'}(3 \leftrightarrow 2) - R_{A'}(2 \leftrightarrow 2)$, where
\begin{eqnarray}
\label{eq:Rnchi1}  
    R_{\chi} ( 3 \leftrightarrow 2 )  
         &\equiv&   2 \frac{n_{A'} }{ n_\chi} R_{A'} ( 3 \leftrightarrow 2 )   \\
      &=&
         \frac{1}{4} \langle \sigma v^2 \rangle_{\chi\chi\bar{\chi}  \to \chi A'  }
      \left(  n_\chi^2 - n_{\chi,0}^2 \frac{n_{A'}} {n_{A',0}}\right)\nonumber    \\
	&\equiv& R_\chi(\chi\chi\bar{\chi}  \to \chi A') - R_\chi(\chi A'\to\chi\chi\bar{\chi} ) \nonumber,
     \end{eqnarray}
\begin{eqnarray}
      \label{eq:Rnchi2}
    R_{\chi} ( 2  \leftrightarrow 2 )  &\equiv&
             \frac{n_{A'} }{ n_\chi} R_{A'} ( 2 \leftrightarrow 2 )  \\
      &=&
      \langle \sigma v \rangle_{A'A' \to \bar{\chi} \chi}
      \left(  \frac{n_{A'}^2}{n_\chi}  -  \frac{ n_{A', 0}^2  n_\chi}{n_{\chi,0}^2}\right)\nonumber\\
	&\equiv&  R_\chi(A'A' \to \bar{\chi} \chi) - R_\chi(\bar{\chi} \chi \to A'A') \nonumber.
\end{eqnarray}
Likewise, we define $2(n_{A'}/n_\chi)R_{A'}(\chi \chi \bar{\chi} \to \chi A') \equiv R_\chi(\chi \chi \bar{\chi} \to \chi A')$ and so on for the unidirectional rates. In this way, the signs for these definitions have been chosen so that all of the rates of individual sub-processes are now positive, although the overall rates $R_\chi$ and $R_{A'}$ can have any sign. When $m_{A'} > m_\chi$ and $T < m_\chi,m_{A'}$, the lower density of $A'$ relative to $\chi$
implies that $R_\chi (  3 \leftrightarrow 2) $ ($R_\chi (  2 \leftrightarrow 2) $) is generally smaller in magnitude than 
$R_{A'} (  3 \leftrightarrow 2) $ ($R_{A'} (  2 \leftrightarrow 2) $).
Thus the rates $R_{\chi}$ tend to fall below $H$ earlier than the corresponding rates $R_{A'}$.
This separation between freezeout of $\chi$ and $A'$ is the origin of the prolonged duration of the
overall freezeout process.

Suppose that only one channel, for example $2 \to 2$, occurs fast enough such that $R_\chi(A'A' \to \bar{\chi} \chi)\gg H$;
then this rate tends to be nearly equal to that of the reverse reaction, $R_\chi(\bar{\chi} \chi \to A'A')$, 
enforcing the condition $ n_{A'}^2  \simeq  n_{A', 0}^2\, {n_\chi^2}/{n_{\chi,0}^2}$
(though the cancellation is imperfect, so that the total rate $R_\chi(2\leftrightarrow 2)$ is also typically greater than $H$).
This by itself is not sufficient to force both the $\chi$ and $A'$ densities to track their
equilibrium values. For that, one generically needs
both $R_\chi(3\leftrightarrow 2)>H$ and $R_\chi(2\leftrightarrow 2)>H$ so that both independent combinations 
$n_\chi-n_{\chi,0}$ and $n_{A'}-n_{A',0}$
are driven to zero.\footnote{The typical behavior is that the strongest process is such that both the forward
and backward rates exceed $H$, as well as their difference.  For the second-strongest, only one of these
need be greater than $H$.}
This is always true at  sufficiently early times, allowing us to use equilibrium initial conditions for the
Boltzmann equations. The DM density $n_\chi$ starts to deviate from equilibrium when the rate of the weaker
annihilation channel becomes comparable to $H$; hence the second-strongest channel initiates the
freezeout process.

\begin{figure}
\centering
\subfigure[]{
\label{fig:rate1n}
\includegraphics[scale=0.54]{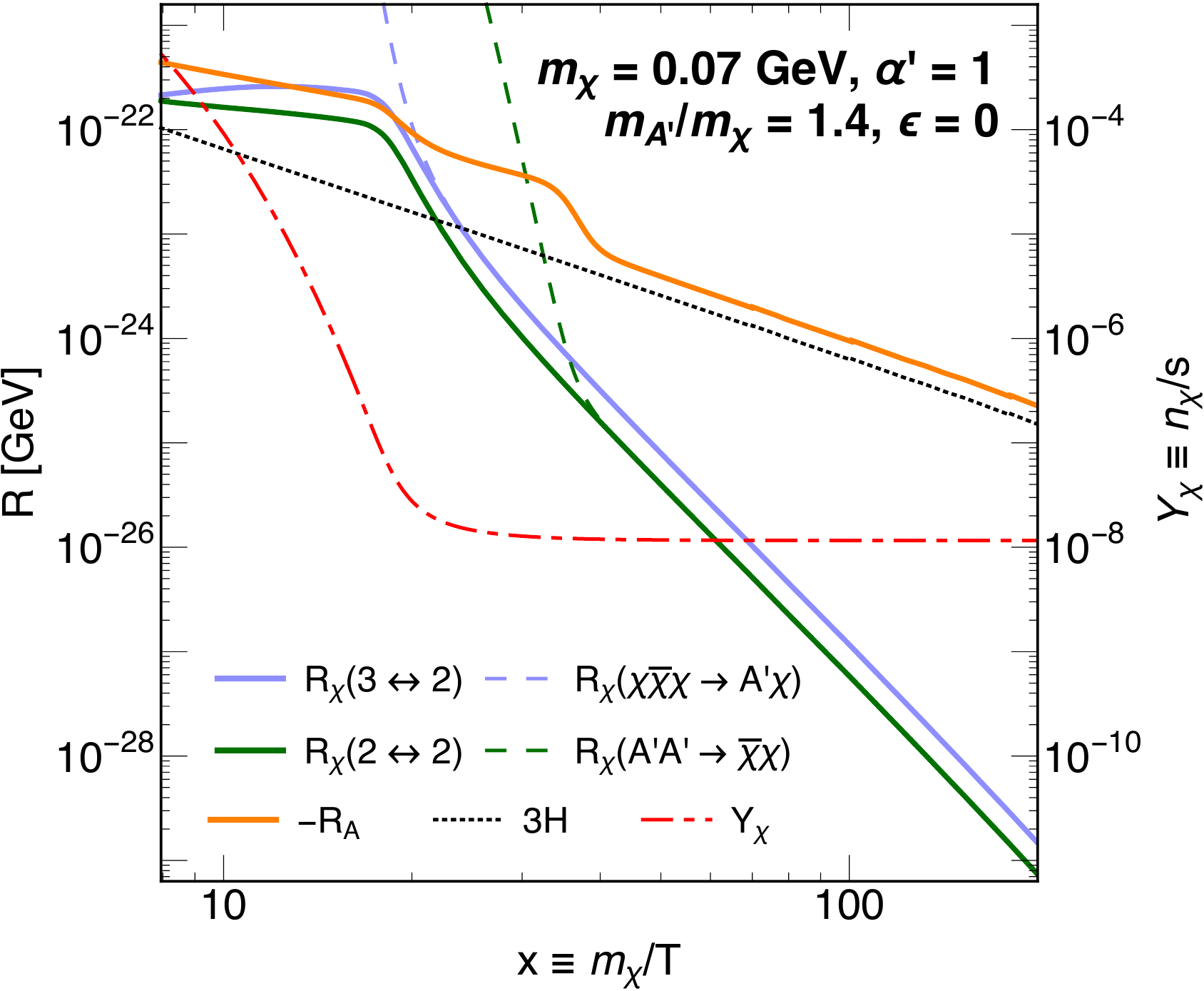}
} \\
\subfigure[]{
\label{fig:rate1A}
\includegraphics[scale=0.54]{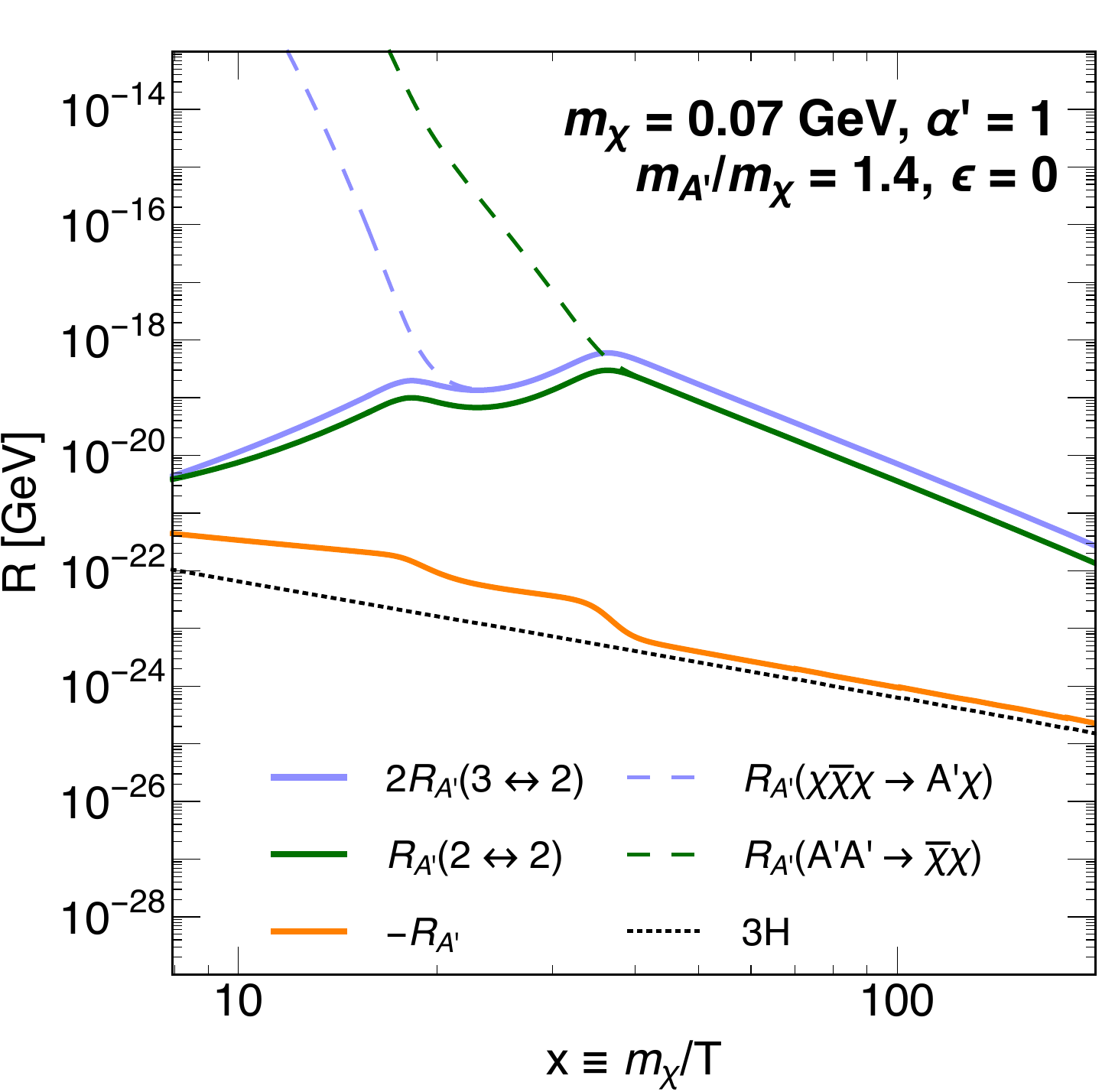}
}
\caption{Rates of different processes during freezeout for $m_{A'}/m_\chi = 1.4$: (a) evolution of $R_\chi(3
\leftrightarrow 2)$ (light blue), $R_\chi(2 \leftrightarrow 2)$ (green) and the $A'$ total rate $R_{A'}$
(orange) as a function of $x \equiv m_\chi / T$. Rates for processes in one direction, $R_\chi(\chi \chi
\bar{\chi} \to A' \chi)$ (light blue, dashed) and $R_\chi(A' A' \to \chi \bar{\chi})$ (green, dashed) are also
shown. The dark matter abundance $Y_\chi$ (red, dot-dashed) is plotted, with the appropriate (dimensionless) units
given on the right-hand axis; (b) evolution of $R_{A'}(3 \leftrightarrow 2)$ (light blue), $R_{A'}(2
\leftrightarrow 2)$ (green) and the $A'$ total rate $R_{A'}$ (orange) as a function of $x$. Rates for processes
in one direction, $R_{A'}(\chi \chi \bar{\chi} \to A' \chi)$ (light blue, dashed) and $R_{A'}(A' A' \to \chi
\bar{\chi})$ (green, dashed) are also shown. The evolution of the Hubble rate $H$ (black, dotted) is
shown in both plots for reference.}
\label{fig:rate1all}
\end{figure}

To illustrate this behavior,  we show some examples of the evolution of the rates in
Figs.~\ref{fig:rate1n}, \ref{fig:rate1A} and \ref{fig:rate2n}. Each example has  the same DM mass $m_\chi =
70~\mathrm{MeV}$, coupling $\alpha' =1 $, and kinetic mixing $\epsilon = 0$, but  different  values of $ r
= 1.4$, $ 1.7$, $ 1.9$. In these figures, the dot-dashed lines corresponding to  the evolution of DM number
density are shown to highlight the time of DM freezeout.  For $ r = 1.4$, the freezeout period is relatively
short; for $ r= 1.7$, freezeout is prolonged; and  for $ r = 1.9$, the freezeout is prolonged further and may
indeed be thought of as two separated  freezeouts. 

In Fig.~\ref{fig:rate1A}, we show the two rates $R_{A'} ( 3
\leftrightarrow  2 )$ and $R_{A'} ( 2 \leftrightarrow  2 ) $, which are
much larger than $H$; these cancel each other to order $H$. The
behavior is similar for other values of $r$. Since  $R_{A'} ( 3
\leftrightarrow  2 )   \simeq  R_A ( 2 \leftrightarrow  2 ) $, 
Eqs.~(\ref{eq:Rnchi1}) and~(\ref{eq:Rnchi2}) implies that $  R_\chi (
3 \leftrightarrow  2 ) \simeq 2 R_\chi ( 2 \leftrightarrow  2 ) $. This
relation is demonstrated in Fig.~\ref{fig:rate1n} and Fig.~\ref{fig:rate2n}.

Comparing these net rates however does not tell us which process
controls freezeout. Instead, we should look at the dashed lines,
which indicate  the unidirectional rates, $R_\chi ( A'A' \to
\bar{\chi} \chi )$  and  $R_\chi( \chi\chi\bar{\chi}  \to \chi A' )$.
Processes are out of equilibrium when these dashed lines overlap with
the solid lines. From the unidirectional rates, we can identify the
weaker annihilation channel and thus which process initiates the
freezeout. For $r = 1.4$, the weaker process is $3 \to 2$, and 
Fig.~\ref{fig:rate1n} confirms that the freezeout is indeed triggered
by $ 3 \to 2 $. For $ r = 1.7$ and $ r = 1.9$, the dashed line for 
$R_\chi ( A'A' \to \bar{\chi} \chi )$ in Fig.\ \ref{fig:rate2n} merges
with the solid line, $R_\chi ( 2 \leftrightarrow  2 ) $, when the rate
is  about $3 H$. It is the weaker channel $ 2 \to 2$ that initiates
freezeout. 

One difference between $ r < 1.5 $ and $ r > 1.5$ in
Fig.~\ref{fig:rate1n} versus Fig.~\ref{fig:rate2n} is that  $ r > 1.5$ normally
has a longer freezeout. The duration depends upon
whether the rate of the weaker annihilation channel is sensitive to
$n_{A'}$. For $r > 1.5$, the weaker process $ 2 \to 2$ has the rate $R_\chi (A' A' \to \chi \chi)
\sim  \langle \sigma v  \rangle_{A'A' \to \bar{\chi} \chi} n_{A'}^2 /
n_\chi $. Prior to the final freezeout, the larger $ 3 \to 2 $ rate
imposes the constraint that $n_{A'} \simeq n_{A',0} n_\chi^2 /
n_{\chi,0}^2 \sim r^{3/2} x^{3/2} m^{-3}   \exp( (2- r )x ) n_\chi^2$.
Since $n_{A'}$ increases with time, this means that the $ 2 \to 2 $
rate  $R_\chi ( A'A' \to \bar{\chi} \chi )$ can  be kept at the same
order as $H$ for a long period.  For this reason, the freezeout is
prolonged. For $ r < 1.5$, the duration  is relatively
short, because the rate of the weaker $ 3 \to 2$ process  goes as  $R_\chi(
\chi\chi\bar{\chi}  \to \chi A' ) \propto n_\chi^2$,  where $n_\chi$
is decreasing with time.

\begin{figure}
\centering
\subfigure[]{
\includegraphics[scale=0.6]{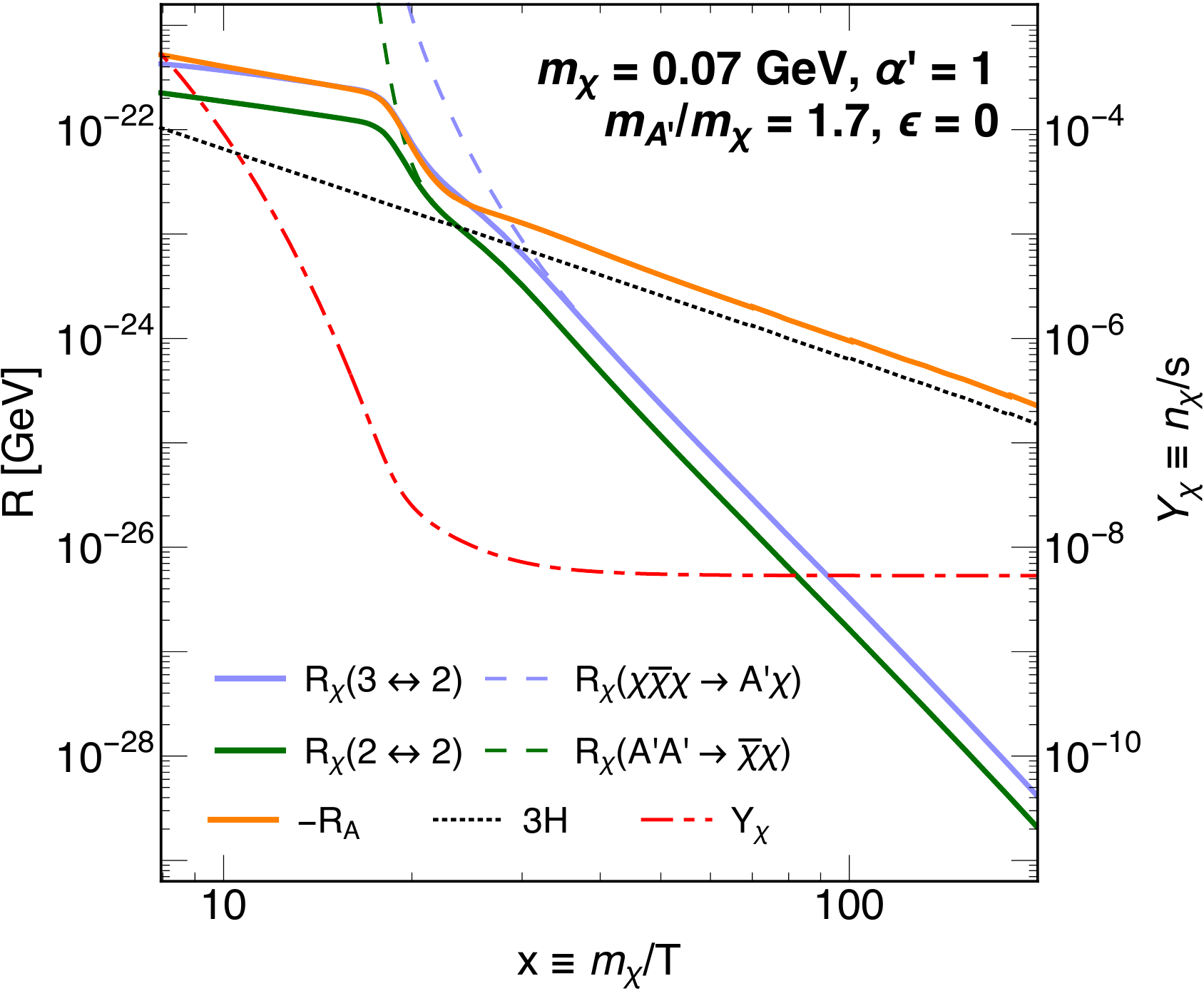}
}
\subfigure[]{
  \includegraphics[scale=0.6]{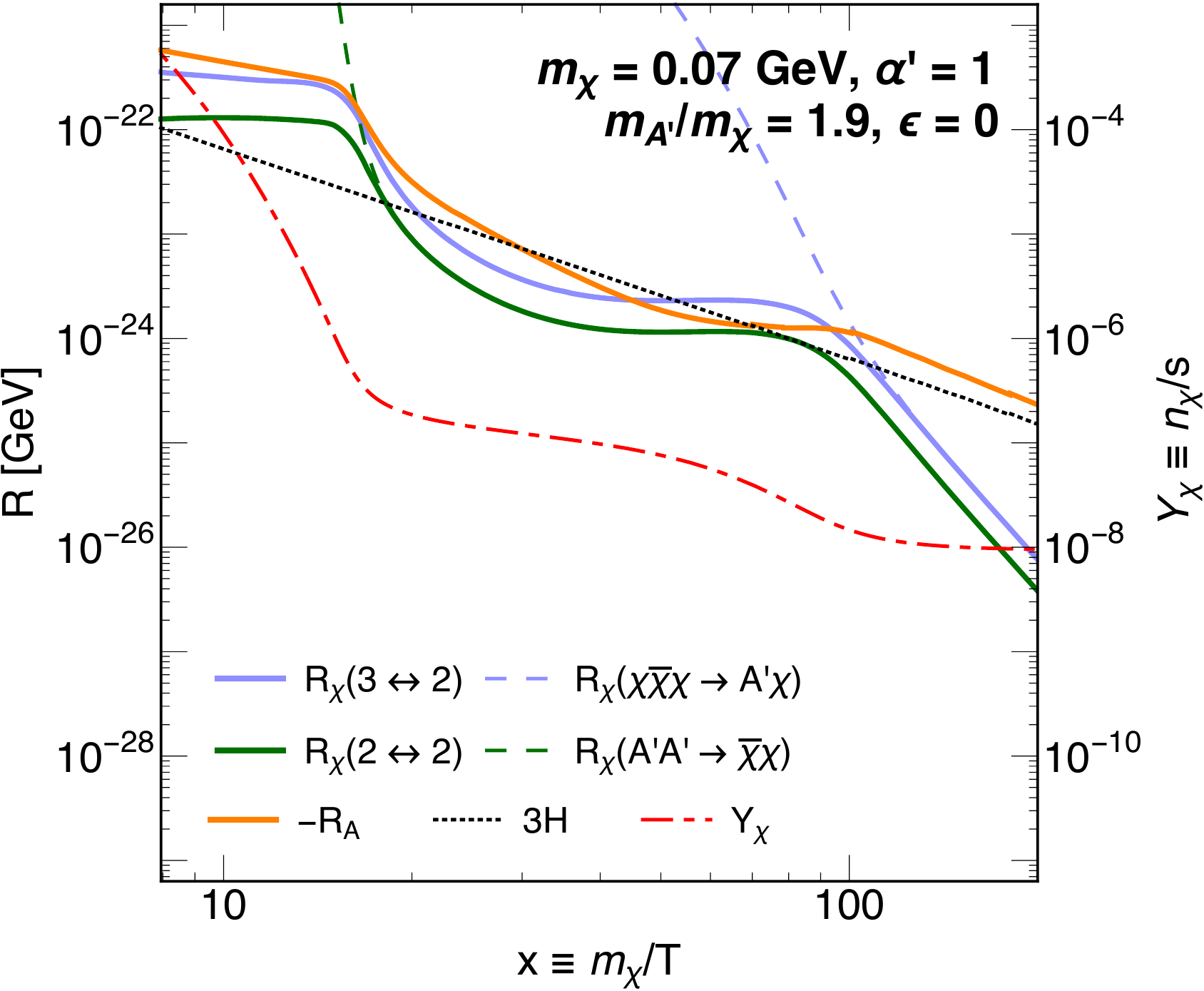}
}
\caption{Same as Fig. \ref{fig:rate1all}, but with (a) $m_{A'}/m_\chi = 1.7$, and (b) $m_{A'}/m_\chi = 1.9$.}
\label{fig:rate2n}
\end{figure}

Armed with our insight that the second-strongest channel matters
critically for freezeout, and having understood the reason that the
freezeout process is longer for $ r \gtrsim 1.5$, we can also explain the shape of the contours in Fig.~\ref{fig:23decaycontour} in Chapter~\ref{chap:nfdm}. There are several important regimes:

\begin{enumerate}
\item For $n_{A'} = n_{A',0}$, corresponding to large $\epsilon$, the two Boltzmann  equations are
reduced to one, and the shape of the contours can be understood using the usual parametrics of DM freezeout. This behavior occurs for the contours overlapping the dashed contours in Fig.~\ref{fig:23decaycontour}.

\item For $\epsilon=0$, we recover the secluded case discussed above, where the interplay of the $3\leftrightarrow 2$ and $2\leftrightarrow 2$ processes controls the freezeout. This behavior also occurs in the region where $r \sim 1.5$ and $ \alpha'  =10$,
because the $3 \to 2$ and $2 \to 2 $ rates for DM are significantly larger than $ R_\chi( A' \to e^+ e^-) \equiv \Gamma ( A' \to e^+ e^- ) n_{A'} / n_{\chi}$, so that  $\Gamma_{A' \to e^+ e^-}$ can be neglected.\footnote{We define the $A'$ decay rate $ R_{\chi} ( A' \to e^+ e^-)$ with
respect to the DM density; even though
this quantity does not appear in Boltzmann equation of DM, the
coupling to the $A'$ Boltzmann equation will cause $A' \to e^+ e^-$
to play an important role in determining the rate for DM processes in
some cases.}

\item For moderate $\epsilon$, the rates of the three processes should be
compared in order to determine which is weakest, and hence
irrelevant to the DM freezeout. The relevant rates to compare
are $R_\chi(\chi \bar{\chi} \chi \to A' \chi )$, $R_{\chi}( A' A' \to \chi \bar{\chi} )$ and 
$ R_{\chi} ( A' \to e^+ e^-)$,
evaluated at the Hubble crossing time of the
annihilation processes. Consider the case where $ 2 \to 2$ has a
lower rate than $3 \to 2$, such that it falls below $H$
first.  Whether the DM density freezes out or not at this time depends on
the relative sizes of $ R_{\chi} ( A' \to e^+ e^-)$    and $R_{\chi}( 
A' A' \to \chi \bar{\chi} )$. When $R_{\chi}( A' A' \to \chi \bar{\chi} )  <  R_{\chi} ( A' \to e^+ e^-) $, the
larger $R_{\chi} (A' \to e^+ e^-)$ rate in the coupled Boltzmann
equations provides enough constraints to keep $n_\chi$ and $n_{A'}$ 
near their equilibrium values.
An example of this more complex case is shown in
Fig. 2(a) of the main text, where the
freezeout starts when  $R_\chi( 3 \leftrightarrow 2 ) \sim H$.
Using the Boltzmann equation of $A'$, this rate is determined by the 
$A'$ decay, $R_\chi( 3 \leftrightarrow 2 ) \sim  2 R_\chi( A' \to e^+
e^-) $;  freezeout then terminates when $R_\chi( \chi
\bar{\chi} \chi \to A' \chi ) \sim H$. 

More broadly, this case is realized when $ \epsilon =
10^{-6}$, and either $r$ is close to $2$, or $\alpha'$ is large and $ r > 1.5 $. Fig.~\ref{fig:23decaycontour} shows that in this region the $\epsilon = 10^{-6}$ contours (solid lines) do not overlap with the dashed contours, 
for which the constraint $n_{A'} = n_{A',0}$ is imposed. In this regime the $ 3 \to 2 $ rate is the largest, and when 
$R_\chi( A' A' \to \chi \bar{\chi}) \sim H$,  $ R_{\chi} ( A' \to e^+ e^-)>R_\chi( A' A' \to \chi \bar{\chi} )$.
The freezeout is thus controlled by $ 3 \to 2$ processes and the decay of $A'$. In this case the $A'$ decay rate is not fast enough to keep the $A'$ abundance in equilibrium, and both $n_{A'}$ and $n_\chi$ are increased during freezeout relative to their values when the $A'$s remain in equilibrium. The $\chi$ annihilation rate thus needs to be increased to maintain the correct relic density, requiring lower $\chi$ masses; this is the reason that the contours in Fig.~\ref{fig:23decaycontour} bend toward lower masses as $\epsilon$ is decreased, for large $r$.

\end{enumerate}

\section{Dependence on \texorpdfstring{$T_d$}{Td}} 

\begin{figure}[t!]
\centering
\includegraphics[scale=0.73]{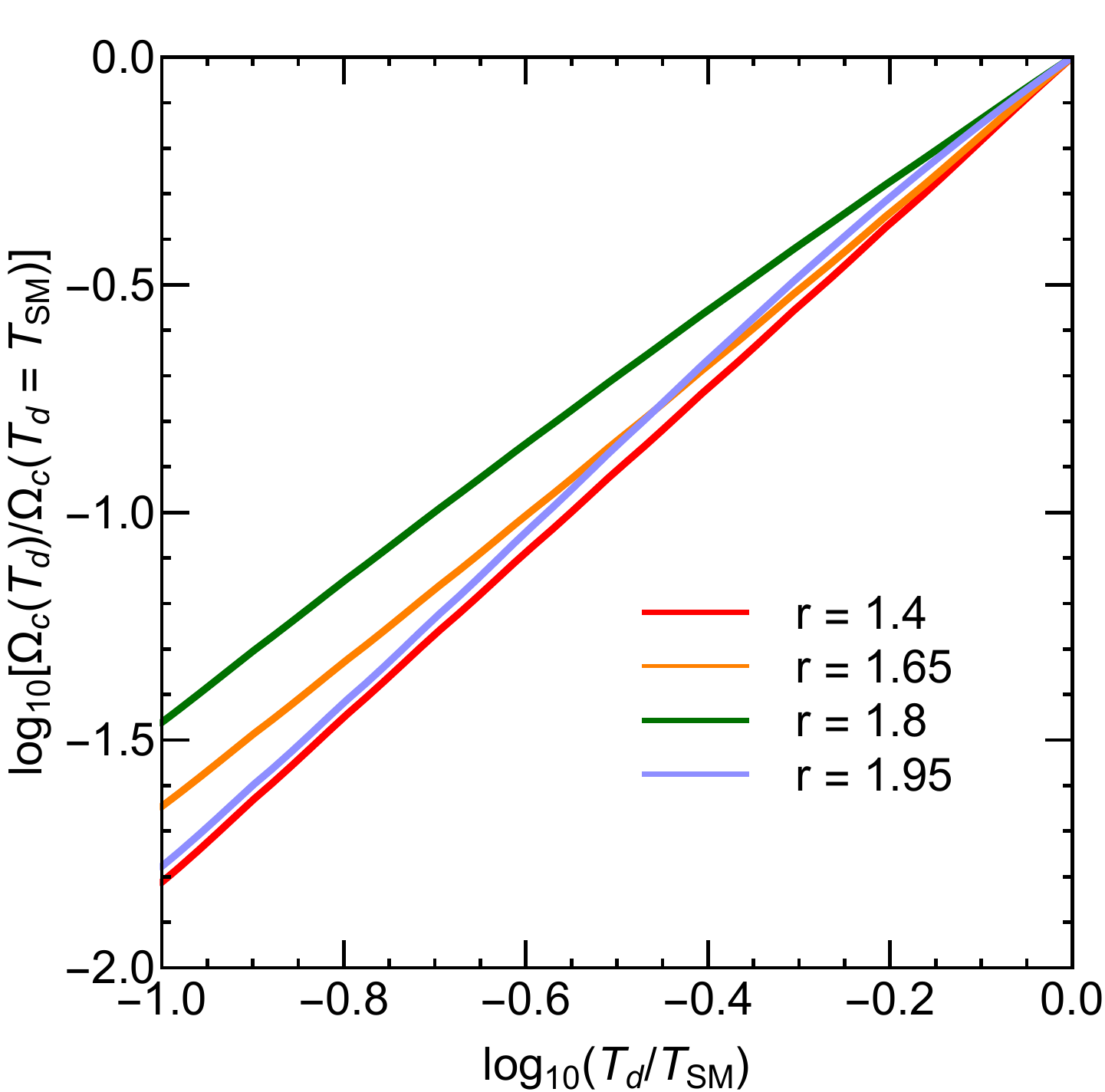}
\caption{Ratio of the relic abundance when $T_d < T_\text{SM}$ to the relic abundance with $T_d = T_\text{SM}$ as a function of $\gamma \equiv T_d/T_\text{SM}$ for $r \equiv m_{A'}/m_\chi = 1.4$ (red), 1.65 (orange), 1.8 (green) and 1.95 (light blue).}
\label{fig:omegaRatio_vs_Td}
\end{figure}

If the dark sector is secluded, its temperature $T_d$ may differ from
that of the visible sector, $T_{\text{SM}}$.  This difference
affects the evolution of the $\chi$ and $A'$ densities and ultimately
the DM relic abundance. In the Boltzmann equations
(Eqs.~(\ref{eq:boltz1}) and~(\ref{eq:boltz2}) in the main text), taking $T_d \neq T_\text{SM}$
changes the equilibrium densities, so that $n_{\chi,0} \sim \exp(-
m_\chi/T_d) = \exp(- x/\gamma)$, where we have defined $\gamma \equiv
T_d/T_\text{SM}$, and $x$ is still given by $x \equiv
m_\chi/T_\text{SM}$. Likewise, $n_{A',0} \sim \exp(-r x/\gamma)$.
Keeping in mind that $H$ is determined by $T_\text{SM}$, we can solve
the Boltzmann equations for $n_\chi(x)$ and $n_{A'}(x)$ with the 
$\gamma$-dependence coming from the equilibrium
densities. 

Fig.~\ref{fig:omegaRatio_vs_Td} shows the behavior of the ratio of
relic abundances $\Omega_c(T_d)/\Omega_c(T_d = T_\text{SM})$ as a
function of $T_d$ for $0.1\le\gamma\le 1$. Having $T_d < T_\text{SM}$
leads to an earlier freezeout, since the exponential decrease in the
equilibrium densities occurs more rapidly. For values of $r$ where the
backward and forward $3 \to 2$ processes fall out of equilibrium at
freezeout, we expect that $n_{\chi}^2 \sim H/ \langle \sigma v^2
\rangle_{\chi \chi \bar{\chi} \to \chi A'} \sim 1/x_f^2$, and
therefore that the relic abundance scales as $\Omega_c \sim x_f^2$.
For $1< r \lesssim 1.5$ where the $3 \to 2$ process determines the DM
abundance, the exponential dependence of $n_{\chi,0}$ with $x/\gamma$
results in $\Omega_c \sim \gamma^2$. On the other hand, in the case of
$r \lesssim 2$, the second freezeout occurs at $n_\chi \propto
n_{\chi,0}^4/n_{A',0}^2$, and a similar argument leads again to
$\Omega_c \sim \gamma^2$. At intermediate values of $r$, both the $3
\to 2$ and $2 \to 2$ processes freeze out at similar times. For a $2
\to 2$ freezeout, $n_\chi \sim H/\langle \sigma v \rangle_{\bar{\chi}
\chi \to A' A'}$, and as a result $\Omega_c \sim \gamma$.
Qualitatively, we expect the $\gamma$ dependence to lie between these
two regimes for intermediate values of $r$.    

\section{Constraints for Different \texorpdfstring{$r$}{r}}

\begin{figure}
\centering
\subfigure[]{
  \label{fig:constraints_r_14}
\includegraphics[scale=0.6]{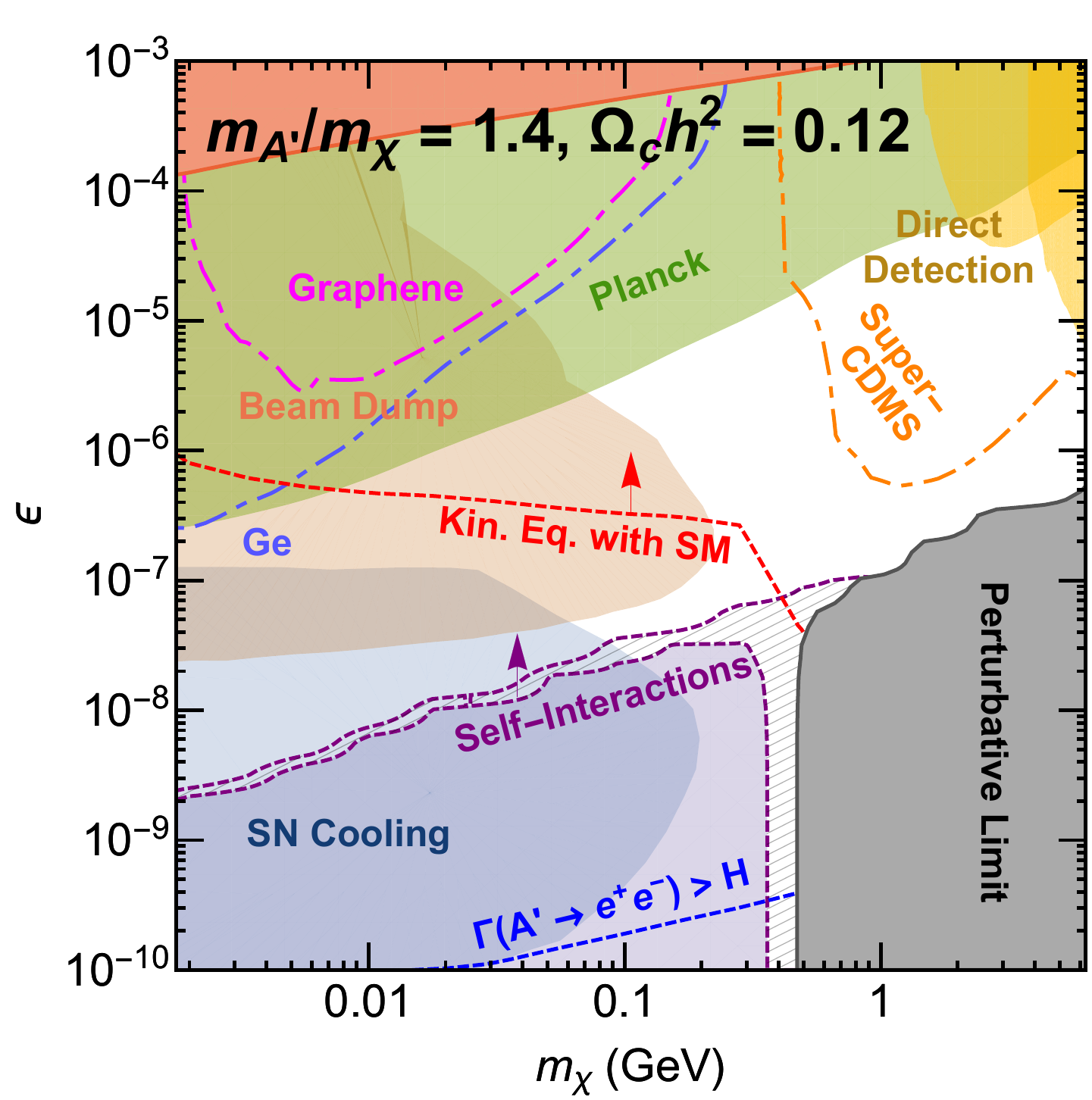}
} 
\subfigure[]{
  \label{fig:constraints_r_16}
  \includegraphics[scale=0.6]{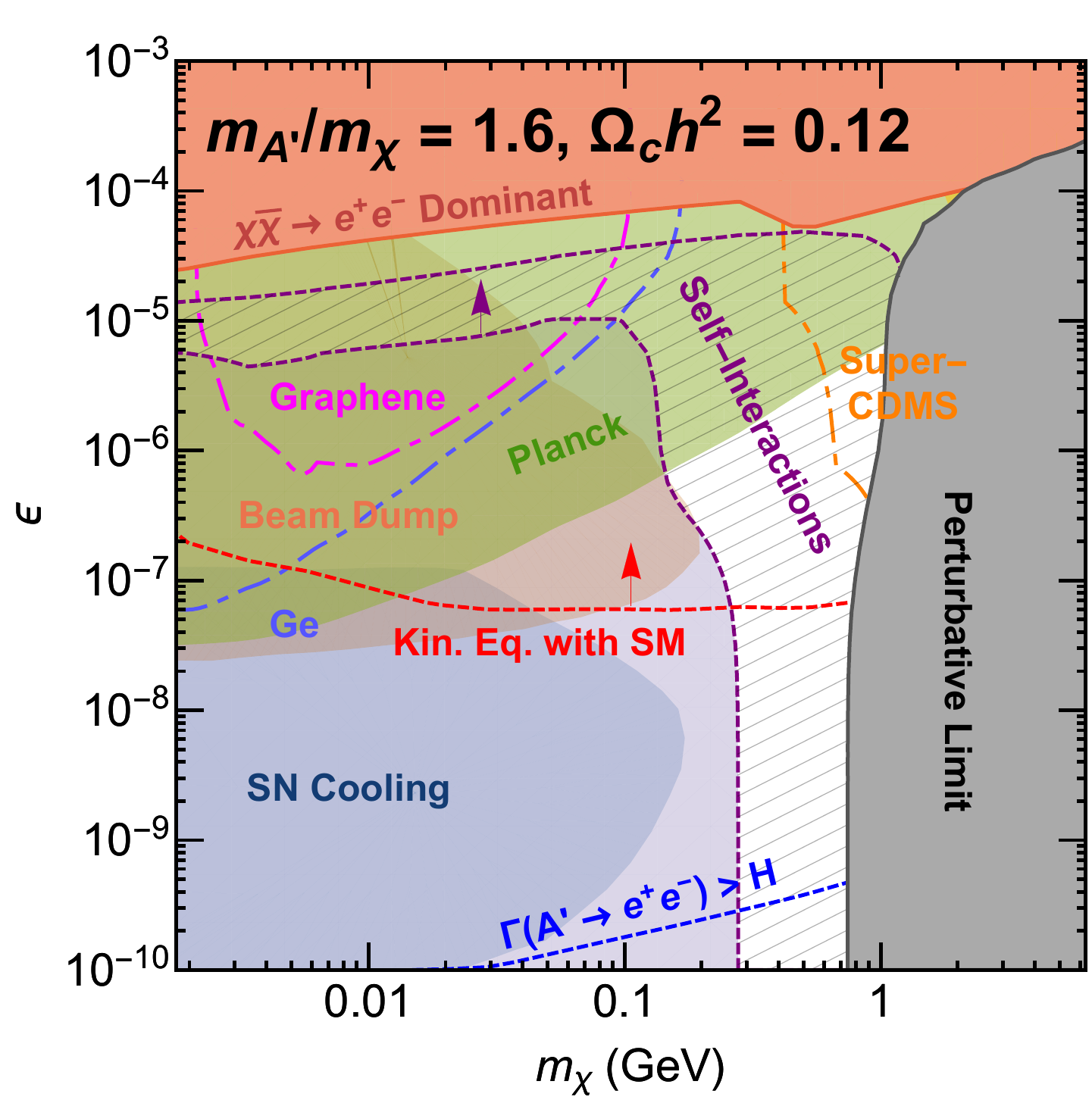}
}
\caption{Same as Fig.~\ref{fig:constraints} in Chapter~\ref{chap:nfdm}, but with (above) $m_{A'}/m_\chi = 1.4$, and (below) $m_{A'}/m_\chi = 1.6$.}
\label{fig:constraintsSupp}
\end{figure}

Fig.~\ref{fig:constraintsSupp} shows the constraints in the $m_\chi-\epsilon$ plane for
two representative values of $r$, with $\alpha'$ fixed to give the
correct present-day relic density. These have the same general
features as in Fig. 4 in the main text, but also exhibit several
distinct characteristics that we explain here. 

At $r = 1.4$, the transition from the secluded limit ($\epsilon \to
0$) to the kinetic equilibrium limit occurs in the range
$\epsilon \sim 10^{-9} - 10^{-6}$, which leads to a rapid decrease in $\alpha'$ between these two phases at a fixed value of $m_\chi$. This explains the rapid
change in the behavior of the region with suitable self-interaction for $m_\chi \lesssim 100\text{ MeV}$.

At $r = 1.6$, the most distinctive feature is the change in behavior
of the region where annihilations to $e^+e^-$ dominates at $m_\chi
\sim 100\text{ MeV}$. At masses smaller than this point, the $2 \to 2$
process is freezes out last, while at larger masses, it is the $3 \to
2$ process which does so. This difference accounts for the change in
the slope of the boundary. There is no such transition for the other
two cases, since at $r = 1.8$, the $3 \to 2$ process always freezes
out last, while for $r = 1.4$ it is the $2 \to 2$ process instead. 

For all values of $r$, a significant part of the $m_\chi - \epsilon$
parameter space is still consistent with the present-day relic density
while evading experimental constraints, showing that the NFDM scenario
is robust against taking different values of $r \gtrsim 1.5$.  
\vspace{2cm}

\section{Complete Boltzmann Equations}
\label{app:boltz}

The complete Boltzmann equations, including all relevant $2 \to 2$ and $3 \to 2$ processes for the full range of $r$ considered is:
\begin{alignat}{2}
    \frac{d n_{\chi}}{dt} + 3H n_{\chi} &=&& -\frac{1}{4} \langle \sigma v^2 \rangle_{\chi \chi \overline{\chi} \atop \to \chi A'} \left(n_\chi^3 - n_{\chi,0}^2 n_\chi \frac{n_{A'}}{n_{A',0}}\right) + \langle \sigma v \rangle_{A' A' \atop \to \overline{\chi} \chi} \left(n_{A'}^2 - n_{A',0}^2 \frac{n_\chi^2}{n_{\chi,0}^2}\right) \nonumber \\
    & &&- \frac{1}{2} \langle \sigma v^2 \rangle_{\chi \overline{\chi} A' \atop \to A' A'} \left(n^2 n_{A'} - n_{\chi,0}^2 \frac{n_{A'}^2}{n_{A',0}}\right) \nonumber \\
    & &&+ \frac{1}{3} \langle \sigma v^2 \rangle_{A'A'A' \atop \to \chi \overline{\chi}} \left(n_{A'}^3 - n_{A',0}^3 \frac{n_\chi^2}{n_{\chi,0}^2}\right) - \frac{1}{2} \langle \sigma v \rangle_{\chi \overline{\chi} \to e^+ e^-} \left(n_\chi^2 - n_{\chi,0}^2 \right),
 \label{eq:fullboltz1}
\end{alignat}
and
\begin{alignat}{2}
    \frac{d n_{A'}}{dt} + 3 H n_{A'} &=&& \,\, \frac{1}{8} \langle \sigma v^2 \rangle_{\chi \chi \overline{\chi} \atop \to \chi A'} \left(n_\chi^3 - n_{\chi,0}^2 n_\chi \frac{n_{A'}}{n_{A',0}} \right) - \langle \sigma v \rangle_{A' A' \atop \to \overline{\chi} \chi} \left(n_{A'}^2 - n_{A',0}^2 \frac{n_\chi^2}{n_{\chi,0}^2} \right) \nonumber \\
    & && - \frac{1}{4} \left( \langle \sigma v^2 \rangle_{\chi \overline{\chi} A' \atop \to \chi \overline{\chi}} \nonumber + \langle \sigma v^2 \rangle_{\chi \chi A' \atop \to \chi \chi} \right) \left(n_\chi^2 n_{A'} - n_\chi^2 n_{A',0} \right) \nonumber \\
    & &&+ \frac{1}{4} \langle \sigma v^2 \rangle_{\chi \overline{\chi} A' \atop \to A' A'} \left(n_\chi^2 n_{A'} - n_{\chi,0}^2 \frac{n_{A'}^2}{n_{A',0}} \right) \nonumber \\
    & && - \frac{1}{2} \langle \sigma v^2 \rangle_{\chi A' A' \atop \to \chi A'} \left(n_\chi n_{A'}^2 - n_\chi n_{A'} n_{A',0} \right) \nonumber \\
    & &&- \frac{1}{2} \langle \sigma v^2 \rangle_{A' A' A' \atop \to \chi \overline{\chi}} \left(n_{A'}^3 - n_{A',0}^3 \frac{n_\chi^2}{n_{\chi,0}^2} \right) - \Gamma_{A' \to f \overline{f}} \left(n_{A'} - n_{A',0} \right).
\label{eq:fullboltz2}
\end{alignat}

The symmetry factors preceding each term properly account for the number of identical particles in the initial state, the net number of particles created or destroyed in each annihilation process, as well as conjugate processes. These equations are used in all numerical calculations shown in the paper. 

\section{Cross Sections and Decay Rates}

\label{app:xsects}
The decay rate for $A'\to e^+e^-$ is
\begin{eqnarray}
   \Gamma (A' \to e^+ e^-)  &=&  \frac{ \epsilon^2 \alpha_{em}  }{3} m_{A'} 
   \left(  1 + 2 \frac{  m_e^2 }{ m_{A'}^2 }  \right)
      \sqrt{ 1 - 4 \frac{ m_e^2}{ m_{A'}^2} } \; .\nonumber\\
\end{eqnarray}

For scattering cross sections, the thermally-averaged $2 \to 2$ cross section for the process $1 + 2 \to 3 + 4$ is given by

\begin{alignat}{1}
    \langle \sigma v \rangle_{12 \to 34} = \frac{1}{S_f} \frac{1}{n_1 n_2} \int \prod_{i=1}^4 \frac{g_i d^3 \vec{p}_i}{(2\pi)^3 2E_i} (2\pi)^4 \delta^4(p_1 + p_2 - p_3 - p_4) f_1 f_2 \overline{|\mathcal{M}|^2},
\end{alignat}
where $g_i$ is the number of degrees of freedom and $f_i$ is the phase space distribution of species $i$. The averaged squared matrix element $\overline{|\mathcal{M}|^2}$ is averaged over both the initial and final state degrees of freedom. $S_f = \prod_i n_i!$ is a symmetry factor, where $n_i$ is the number of identical particles of species $i$ in the final state. Initial state symmetry factors are included explicitly in the Boltzmann equation, Eqs.~(\ref{eq:boltz1}) and (\ref{eq:boltz2}). This convention may differ from other sources in the literature. 

Similarly, the thermally-averaged $3 \to 2$ cross section for the process $1 + 2 + 3 \to 4 + 5$ is
\begin{alignat}{1}
    \langle \sigma v^2 \rangle_{123 \to 45} = \frac{1}{S_f} \frac{1}{n_1 n_2 n_3} \int \prod_{i=1}^5 \frac{g_i d^3 p_i}{(2\pi)^3 2E_i} (2\pi)^4 \delta^4(p_1 + p_2 + p_3 - p_4 - p_5) f_1 f_2 f_3 \overline{|\mathcal{M}|^2}.
\end{alignat} 

For simplicity and unless otherwise stated, we give cross sections at the kinematic threshold of the respective processes. In this limit, thermally averaged cross sections are
\begin{alignat}{1}
    \langle \sigma v \rangle_{12 \to 34} = \frac{g_3 g_4}{32 \pi S_f m_1 m_2} \lambda^{1/2} (m_1 + m_2, m_3, m_4) \overline{|\mathcal{M}|^2},
\end{alignat}
and
\begin{alignat}{1}
    \langle \sigma v^2 \rangle_{123 \to 45} = \frac{g_4 g_5}{64 \pi S_f m_1 m_2 m_3} \lambda^{1/2} (m_1 + m_2 + m_3, m_4, m_5) \overline{|\mathcal{M}|^2},
\end{alignat}
where $\lambda(x, y, z) \equiv (1 - (z+y)^2/x^2) (1 - (z-y)^2/x^2)$. This expression agrees with the result for the specific process of $3 \chi \to 2 \chi$ computed in \cite{Kuflik:2017iqs}.\footnote{Note that different conventions are used between this paper and \cite{Kuflik:2017iqs}.}

\renewcommand{\arraystretch}{2} 

\setlength{\tabcolsep}{15pt}

\begin{landscape}
\begin{table*}
\centering
\begin{tabular}{c c c} 
\toprule

Process & $\overline{|\mathcal{M}|^2}$ & Phase Space \\
\hline
$A' A' A' \to \chi \overline{\chi}$ & $\frac{g'^6(153r^6 - 47r^4 - 60r^2 + 24)}{9 m_\chi^2 r^8}$ & $\frac{\sqrt{9r^2 - 4}}{48 \pi m_\chi^3 r^4}$  \\
$\chi A' A' \to \chi A'$ & $\frac{2g'^6(195r^8 + 1156r^7 + 4670r^6 + 9444r^5 + 12214r^4 + 11192r^3 + 6732r^2 + 2272r + 320)}{9 m_\chi^2 (r+1)^2 (r+2)^4 (2r+1)(r^2 - 2r - 2)^2}$ & $\frac{3 \sqrt{3} \sqrt{3r^2 + 8r + 4}}{32 \pi m_\chi^3 r (2r+1)^2}$ \\
$\chi \chi A' \to \chi \chi$ & $\frac{2 g'^6 r(r+4)}{3 m_\chi^2(r+1)^2 (r+2)^2}$ & $\frac{\sqrt{r(r+4)}}{32 \pi m_\chi^3 r (r+2) }$  \\
$\chi \overline{\chi} A' \to \chi \overline{\chi}$ & $\frac{g'^6 (r+4) (9r^6 + 24r^5 + 4r^4 - 40r^3 + 168r^2 - 224r + 128)}{6 m_\chi^2 r^3 (r-2)^2 (r+1)^2 (r+2)^2}$ & $\frac{\sqrt{r(r+4)}}{16 \pi m_\chi^3 r (r+2)}$  \\
$\chi \overline{\chi} A' \to A' A'$ & $\frac{16g'^6(21r^6 - 4r^5 - 17r^4 + 24r^3 + 216r^2 + 288r + 112)}{27 m_\chi^2 (r-2)^4 (r+1)^4 (r+2)^2}$ & $\frac{9 \sqrt{-3r^2 + 4r + 4}}{128 \pi m_\chi^3 r(r+2)}$ \\
$\chi \overline{\chi} \chi \to A' \chi$ & $\frac{g'^6 (r-4)(r+4)(-32r^8 + 167r^6 - 534r^4 + 668r^2 - 512)}{36 m_\chi^2 (r^2 - 4)^4 (r^2 + 2)^2}$ & $\frac{\sqrt{r^4 - 20r^2 + 64}}{96 \pi m_\chi^3}$  \\
$\chi \overline{\chi} \to A' A' $ & $\frac{16g'^4(1-r^2)}{9(r^2 - 2)^2}$ & $\frac{9 \sqrt{1 - r^2}}{64 \pi m_\chi^2}$  \\
$A' A' \to \chi \overline{\chi}$ & $\frac{32g'^4(r^4 - 1)}{9r^4}$ & $\frac{\sqrt{r^2 - 1}}{8 \pi m_\chi^2 r^3}$ \\
$\chi \overline{\chi} \to e^+e^-$ & $\frac{4 e^2 \epsilon^2 g'^2 \left(2 + m_e^2/m_\chi^2\right)}{(r^2 - 4)^2}$ & $\frac{\sqrt{1 - m_e^2/m_\chi^2}}{8 \pi m_\chi^2}$ \\
\botrule
\end{tabular}
\caption{List of initial- and final-state averaged squared matrix element $\overline{|\mathcal{M}|^2}$ of each process, as well as the phase space factor $P$ such that $\langle \sigma v \rangle$ or $\langle \sigma v^2 \rangle = P \overline{|\mathcal{M}|^2}$. All values are evaluated at the kinematic threshold. For the last two processes, we use the expression for $\chi \overline{\chi} \to A' A'$ for $r < 1$ and $A' A' \to \chi \overline{\chi}$ for $r > 1$. }
\label{tab:xsec}
\end{table*}
\end{landscape}

In Table \ref{tab:xsec}, we list all of the number changing processes that are included in the Boltzmann equations Eqs.~(\ref{eq:fullboltz1}) and~(\ref{eq:fullboltz2}), the initial- and final-state averaged squared matrix element $\overline{|\mathcal{M}|^2}$ of each process as well as the phase space factor $P$ such that $\langle \sigma v \rangle$ or $\langle \sigma v^2 \rangle = P \overline{|\mathcal{M}|^2}$. We define $r \equiv m_{A'}/m_\chi$ throughout.

Two other processes that are important to our analysis are $\chi e^\pm \to \chi e^\pm$ which maintains kinetic equilibrium between the dark sector and the SM, and dark matter-dark matter scattering. 

\begin{itemize}

    \item{$\chi e^\pm \to \chi e^\pm$}: this cross section is important in determining if the DM is in kinetic equilibrium with the SM. In the limit where $T < \mu_{e\chi}$, where $\mu_{e\chi}$ is the electron-DM reduced mass, we have

    \begin{alignat}{1}
        \langle \sigma v \rangle = \frac{2(g' \epsilon e)^2 \mu_{e\chi}^2}{ \pi m_{A'}^4} \left(\frac{2T}{\pi \mu_{e\chi}}\right)^{1/2}.
    \end{alignat}
    
    At high temperatures, it approaches the limit
    
    \begin{alignat}{1}
        \langle \sigma v \rangle \to \frac{(g' \epsilon e)^2}{4 \pi m_{A'}^2}.
    \end{alignat}
    
    To get accurate results, however, we must use the exact thermal average over the cross section for $\chi e^\pm \to \chi e^\pm$, which is given by: 
    
    \begin{multline}
        \sigma = \frac{(g' \epsilon e)^2}{8\pi}  \left[ \frac{1}{s} + \frac{2}{r^2 m_\chi^2} + \frac{8m_e^2 + r^4 m_\chi^2}{r^2 [h(m_\chi,s) + r^2 m_\chi^2 s]} \right. \\
        \left. - \frac{2(r^2 m_\chi^2 + s)}{h(m_\chi,s)} \log \left(1 + \frac{h(m_\chi,s)}{r^2 m_\chi^2 s}\right)\right],
    \end{multline}
    
    where $h(m_\chi, s) = [s - (m_\chi + m_e)^2][s - (m_\chi - m_e)^2]$. The thermal average is then given by
    
    \begin{alignat}{1}
        \langle \sigma v \rangle = \int_{M^2}^\infty \frac{ds}{\sqrt{s}} \frac{h(m_\chi ,s) K_1(\sqrt{s}/T) \sigma}{8 T m_\chi^2 m_e^2 K_2(m_\chi/T) K_2(m_e/T)} \, ,
    \end{alignat}
    
    where $M = m_e + m_\chi$. 
    
    \item{$\chi \chi \to \chi \chi$}: the self-interaction cross section, averaged over particle-particle and particle-antiparticle scattering, is~\cite{{D'Agnolo:2015koa}}
    \begin{alignat}{1}
        \frac{\sigma_{\text{SI}}}{m_\chi} = \frac{3g'^4}{16\pi m_\chi^3} \frac{16 - 16r^2 + 5r^4}{r^4(r^2 - 4)^2} \; .
    \end{alignat}
    
\end{itemize}


\chapter{Dark Sector Bound States}
\label{chap:app_bound_states}

\section{Details of Kinetic Mixing}
\label{app:kineticMixing}

The kinetic mixing term in both the pseudo-Dirac U(1) model and the triple Higgs SU(3)$_D$ model is of the form
\begin{alignat}{1}
	\mathcal{L} \supset -\frac{1}{4} V_{\mu\nu}V^{\mu\nu} -\frac{\epsilon}{2} V_{\mu\nu} B^{\mu\nu} - \frac{1}{4} B_{\mu\nu} B^{\mu\nu} \, ,
\end{alignat}
where $V_{\mu\nu}$ is the dark sector gauge field strength. To diagonalize this, we define the new field $B'_\mu = B_\mu + \epsilon V_\mu$, and thus $B'_{\mu\nu} = B_{\mu\nu} + \epsilon V_{\mu\nu}^{\text{ab.}}$, where ab. indicates the abelian part of the field strength tensor. All terms of $\mathcal{O}(\epsilon^2)$ are neglected here. Then
\begin{alignat}{1}
	\mathcal{L} \supset -\frac{1}{4} V_{\mu\nu} V^{\mu\nu} - \frac{1}{4} B'_{\mu\nu} B'^{\mu\nu} - \frac{\epsilon g_D}{2} f^{abc} V_\mu^b V_\nu^c B'^{\mu\nu}  \, ,
\end{alignat}
The last term is an additional interaction that is unimportant in the diagonalization. In terms of the new field, we have the SM fields
\begin{alignat}{1}
	Z_\mu 
	&= c_W W^3_\mu - s_W B_\mu 
	= c_W W^3_\mu - s_W(B'_\mu - \epsilon V_\mu) = Z'_\mu + \epsilon s_W V_\mu, \nonumber \\
	A_\mu 
	&= s_W W^3_\mu + c_W B_\mu 
	= s_W W^3_\mu + c_W(B'_\mu - \epsilon V_\mu) = A'_\mu - \epsilon c_W V_\mu,
\end{alignat}
where we have defined $Z_\mu' = c_W W_\mu^3 - s_W B'_\mu$ and $A_\mu' = s_W W_\mu^3 + c_W B'_\mu$. 
In terms of these new fields, the mass terms for $Z$ and $V$ are:
\begin{alignat}{1}
	\mathcal{L} 
	&\supset 
	\frac{1}{2} m_Z^2 Z_\mu Z^\mu + \frac{1}{2} g^2 v^2 V_\mu V^\mu = \frac{1}{2} m_Z^2 Z'_\mu Z'^\mu + \epsilon s_W m_Z^2 Z_\mu' V^\mu  +  \frac{1}{2} g^2 v^2 V_\mu V^\mu.
\end{alignat}
Diagonalizing this and defining $r \equiv m_Z/m_V$ (we neglect the $\mathcal{O}(\epsilon^2)$ shift in the masses), we get the following mass eigenstates to first order in $\epsilon$ (marked by tildes):
\begin{alignat}{1}
	\tilde{Z}_\mu &= Z'_\mu - r^2 \frac{\epsilon s_W}{1 - r^2} V_\mu \, , \nonumber \\
	\tilde{V}_\mu &= V_\mu + r^2 \frac{\epsilon s_W}{1 - r^2} Z'_\mu \, ,
\end{alignat}
with $A'$ being the massless mode. 
With this, we see that the original SM fields become
\begin{alignat}{1}
	Z_\mu &= \tilde{Z}_\mu + \frac{\epsilon s_W}{1 - r^2} \tilde{V}_\mu, \\
	A_\mu &= A'_\mu - \epsilon c_W \tilde{V}_\mu,
\end{alignat}
and
\begin{alignat}{1}
    V_\mu &= \tilde{V}_\mu - r^2 \frac{\epsilon s_W}{1 - r^2} \tilde{Z}_\mu.
\end{alignat}
This is the result shown in Eq.~(\ref{eqn:currents}).

\section{Details of the Triple Higgs Model}
\label{app:triplehiggs}

\begin{table}
    \begin{center}
    \begin{tabular}{cc}
        \begin{tikzpicture}
            \begin{feynman}
                \vertex (chi2) {\(\chi_2\)};
                \vertex[right=1.4cm of chi2] (int);
                \vertex[above right=1cm and 1cm of int] (gluon) {\(V^{1,2}\)};
                \vertex[below right=1cm and 1cm of int] (chi1) {\(\chi_1\)};
                \diagram*{
                    (chi2) -- [fermion] (int);
                    (int) -- [boson] (gluon);
                    (int) -- [fermion] (chi1);
                };
            \end{feynman}
        \end{tikzpicture} &
        \begin{tikzpicture}
            \begin{feynman}
                \vertex (chi3) {\(\chi_3\)};
                \vertex[right=1.4cm of chi3] (int);
                \vertex[above right=1cm and 1cm of int] (gluon) {\(V^{4,5}\)};
                \vertex[below right=1cm and 1cm of int] (chi1) {\(\chi_1\)};
                \diagram*{
                    (chi3) -- [fermion] (int);
                    (int) -- [boson] (gluon);
                    (int) -- [fermion] (chi1);
                };
            \end{feynman}
        \end{tikzpicture} \\
        \begin{tikzpicture}
          \begin{feynman}
            \vertex (chi3) {\(\chi_3\)};
                \vertex[right=1.4cm of chi3] (int);
                \vertex[above right=1cm and 1cm of int] (gluon) {\(V^{6,7}\)};
                \vertex[below right=1cm and 1cm of int] (chi2) {\(\chi_2\)};
                \diagram*{
                    (chi3) -- [fermion] (int);
                    (int) -- [boson] (gluon);
                    (int) -- [fermion] (chi2);
                };
          \end{feynman} 
        \end{tikzpicture} & 
        \begin{tikzpicture}
          \begin{feynman}
            \vertex (chi123) {\(\chi_{1,2}\)};
                \vertex[right=1.8cm of chi123] (int);
                \vertex[above right=1cm and 1cm of int] (gluon) {\(V^{3}\)};
                \vertex[below right=1.cm and 0.8cm of int] (chiout) {\(\chi_{1,2}\)};
                \diagram*{
                    (chi123) -- [fermion] (int);
                    (int) -- [boson] (gluon);
                    (int) -- [fermion] (chiout);
                };
          \end{feynman} 
        \end{tikzpicture} \\
        \begin{tikzpicture}
          \begin{feynman}
            \vertex (chi123) {\(\chi_{1,2,3}\)};
                \vertex[right=1.8cm of chi123] (int);
                \vertex[above right=1cm and 1cm of int] (gluon) {\(V^{8}\)};
                \vertex[below right=1.cm and 0.8cm of int] (chiout) {\(\chi_{1,2,3}\)};
                \diagram*{
                    (chi123) -- [fermion] (int);
                    (int) -- [boson] (gluon);
                    (int) -- [fermion] (chiout);
                };
          \end{feynman} 
        \end{tikzpicture}
    \end{tabular}
    \end{center}
    
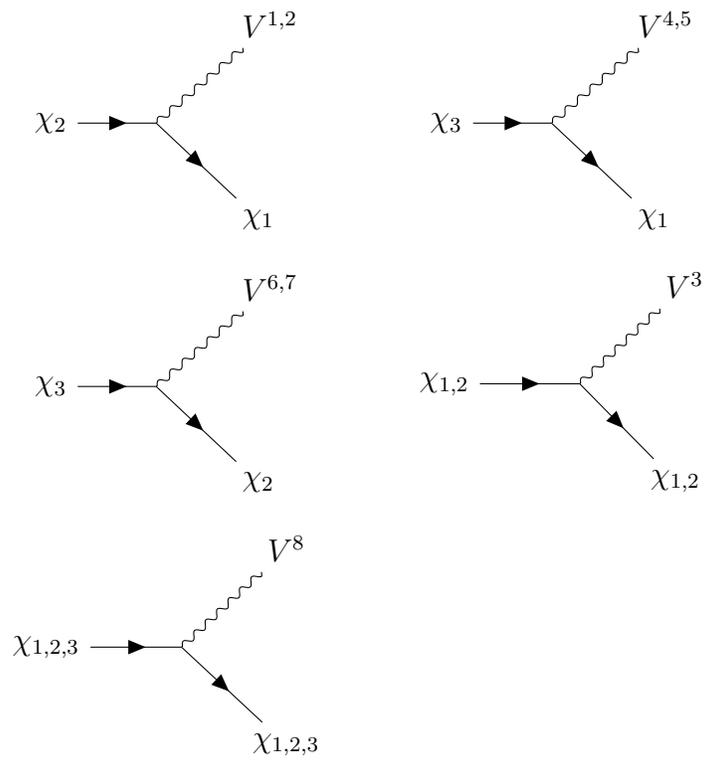
\captionof{figure}{Gauge-fermion-fermion vertices for the triple Higgs model.}
    \label{fig:feynmanTripleHiggs}
\end{table}

Figure \ref{fig:feynmanTripleHiggs} shows all the gauge-fermion-fermion interaction vertices in the triple Higgs model. These vertices are important in the discussion of the stability of the non-DM fermions and thermal freeze-out.

In the triple Higgs model of Sec.~\ref{sec:TripleHiggs}, $H_1$ and $H_2$ acquire a VEV $v_1$ and $v_2$ in the 1- and 2-directions respectively, at a scale above the bare Dirac mass $m_\chi$. This breaks the gauge symmetry down to a residual U(1). Subsequently, this remaining symmetry group is broken at a lower scale than $m_\chi$ by $H_8$, which acquires a VEV given by
\begin{alignat}{1}
	\langle H_8 \rangle &= \begin{pmatrix}
	v_8 \cos \theta \\ 0 \\ v_8 \sin \theta
\end{pmatrix},
\end{alignat}
with $v_8 \ll m_\chi$. This symmetry breaking pattern can be achieved with the following Higgs potential, which obeys the $\mathbb{Z}_2$ symmetry mentioned above: 
\begin{multline}
	V(H_1, H_2, H_8) = \sum_{i=1,2} \lambda_i (H_i^a H_i^a - v_i^2)^2 + \lambda_8 (H_8^\dagger H_8 - v_8^2)^2 \\
	+ \lambda_{12} (H_1^a H_2^a)^2 + \frac{1}{\Lambda_H^2} \sum_{i=1,2} (H_8^\dagger \tau^a H_8 H_i^a)^2.
\end{multline}
Note that terms like $H_1^a H_2^a$ and $H_1^a H_2^a H_8^\dagger H_8$ are forbidden by the $\mathbb{Z}_2$ symmetry that holds separately for $H_1$ and $H_2$. The last two terms forbid a VEV in the second component of $H_8$, which would make $H_8^\dagger \tau^a H_8$ non-zero for $a = 1$. This symmetry breaking pattern produces the mass hierarchy for the dark gluons and the mass splitting between the Dirac fields in $\chi$, which were discussed in the main text.

Note that the symmetry breaking pattern has been chosen for simplicity, and not because fine-tuning is required. If the VEVs of $H_1$ and $H_2$ are not orthogonal, but now lie in some plane in color space that we can take to be the 1- and 2-components without loss of generality, the mass hierarchy of the dark gluons remains unchanged as long as the VEVs are not close to parallel. We must now consider both $V_{\mu\nu}^1 B^{\mu\nu}$ and $V_{\mu\nu}^2 B^{\mu\nu}$ in the mixing with the SM, but as we point out in the footnote after Eq.~\eqref{eqn:tripleHiggsKineticMixing}, this does not affect the phenomenology significantly, since $V^1$ and $V^2$ couple in the same manner to the dark fermions.

Allowing a non-zero second component $\lambda$ in $\langle H_8 \rangle$ also does not alter the mass hierarchy of the gluons as long as $\lambda \sim v_8 \ll v_1, \, v_2$. $\lambda$ introduces mixing between $\chi_1$ and $\chi_2$, so that the mass splitting is altered and flavor eigenstates are no longer mass eigenstates; however, $V^1$ and $V^2$ couple in a similar manner to both $\chi_1$ and $\chi_2$, and so the phenomenology is once again not altered significantly, provided the mass hierarchy has a lowest energy state that is made up of $\chi_1$ and/or $\chi_2$. For the rest of the discussion, we will therefore only consider the special case introduced in the main body.

After symmetry breaking, the kinetic mixing terms with the SM become
\begin{alignat}{2}
	\mathcal{L}_{\text{mix}} &=&& -\frac{\epsilon}{2} V_{\mu\nu}^1 B^{\mu\nu} - \frac{\epsilon_8}{2} \left[ \cos^2 \theta \, V_{\mu\nu}^3 + \sin 2 \theta \, V_{\mu\nu}^4  \right] B^{\mu\nu} \nonumber \\
    & && - \frac{\epsilon_8}{2 \sqrt{3}} \left( \cos^2 \theta - 2 \sin^2 \theta \right) V_{\mu\nu}^8 B^{\mu\nu},
\end{alignat}
where we have defined $\epsilon_8/2 \equiv v_8^2/\Lambda_8^2$, taking $\epsilon_8 \ll \epsilon \ll 1$. The first term represents the kinetic mixing between dark sector and the SM discussed in the main text, while the remaining mixing terms are highly suppressed but non-zero for generic values of $\theta$: their existence guarantees that the gluons $V^3$, $V^4$ and $V^8$ decay to SM particles over cosmological timescales. 

With the $m_8 < 2m_\chi < m_{1,\cdots,7}$ mass hierarchy, the only stable dark sector particle is $\chi_1$, since we can assign a conserved dark baryon number to the fermions and $\chi_1$ is the lightest dark fermion. The heavy gluons $V^i$ for $i = 1, \cdots, 7$ can decay into a pair of dark fermions since their masses exceed $2m_\chi$. Decays into a pair of dark fermions for $V^i$ with $i = 2, 5, 6$ and 7 occur promptly for an $\mathcal{O}(1)$ coupling $g_D$. 

For $V^8$, which mixes directly into the SM and can decay into a pair of SM fermions, its lifetime is approximately
\begin{alignat}{1}
	\tau_{8} \sim \left(\frac{10^{-11}}{\epsilon_8}\right)^2 \left(\frac{100 \text{ MeV}}{m_8}\right) \times 10 \text{ s}.
\end{alignat}
Since $V^8$ couples the dark sector to the SM at tree level, $\epsilon_8$ must be sufficiently small to evade stringent direct detection constraints. Following the approximate limits on the couplings derived in Sec.~\ref{sec:DDL}, we require $\epsilon_8^2 g_D^2 e^2 \lesssim 10^{-5} m_8^4 m_\chi/\text{TeV}^5$, and for $m_8 \gtrsim \text{GeV}$, we require $\epsilon_8 \lesssim 10^{-8}$. Any choice of $10^{-11} \lesssim \epsilon_8 \lesssim 10^{-8}$ therefore ensures that $V^8$ decays to SM particles well before Big Bang nucleosynthesis (BBN).


For the fermions, both the $\chi_2$ and $\chi_3$ fermion can decay to $ \chi_1 f \overline{f} $ where $f$ is an SM fermion through an off-shell $V^1$ and $V^4$ respectively. The decay lifetime for the heavier dark fermions are approximately
\begin{alignat}{1}
	\tau_{\chi_2} &\sim \left(\frac{0.1}{\epsilon}\right)^2 \left(\frac{m_1}{\text{TeV}}\right) \left(\frac{\text{TeV}}{m_\chi}\right)^5 \left(\frac{10^{-6}}{\Delta m_\chi/m_\chi}\right)^5 \times 1 \text{ s}, \nonumber \\
	\tau_{\chi_3} &\sim \left(\frac{0.1}{\epsilon_8}\right)^2 \left(\frac{m_4}{\text{TeV}}\right) \left(\frac{\text{TeV}}{m_\chi}\right)^5 \left(\frac{10^{-6}}{\Delta m_\chi/m_\chi}\right)^5 \times 1 \text{ s}.
\end{alignat}
where $\Delta m_\chi \equiv v_8^2/2\Lambda_m$ is the mass splitting between $\chi_2$, $\chi_3$ and $\chi_1$. In order for $\epsilon_8$ to be small enough to evade direct detection constraints, $\chi_3$ can only decay significantly after BBN, so some care must be taken to ensure that such a long lifetime is not in contradiction with cosmic microwave background (CMB) \cite{Slatyer:2016qyl} or BBN constraints \cite{Slatyer:2012yq,Kawasaki:2017bqm,Cyburt:2009pg}. Given the suitable range of $\epsilon_8$ found earlier, $10^{-11} \lesssim \epsilon_8 \lesssim 10^{-8}$, it is easy to find $\Delta m_\chi \gtrsim 10^{-6}$ so that $\chi_3$ decays before recombination ($\sim 10^{13}$ s), avoiding the CMB limits. On the other hand, the BBN limits are only applicable when the fraction of the energy density released in the decay of $\chi_3$ is greater than $10^{-6} \Omega_\chi h^2$. Keeping in mind that after freeze-out, $\Omega_{\chi_3} / \Omega_{\chi_1} \sim 1/16$ owing to the larger coupling of $\chi_3$ to $V^8$, we can avoid the BBN constraints with $\Delta m_\chi / m_\chi \lesssim 16 \times 10^{-6} \sim 10^{-5}$. This limit should relax further in some cases given a more careful calculation, since a larger mass splitting rapidly decreases $\tau_{\chi_3}$, which significantly weakens the BBN constraints.

For sufficiently small values of $\epsilon_8$, the $V^8$ decay width into SM particles may be smaller than the Hubble rate during the freezeout of $\chi_1$, potentially taking it out of thermal equilibrium with the SM particles. However, we have checked that $\chi_1$ stays in kinetic equilibrium with the SM through the $\chi_1 e^\pm \to \chi_2 e^\pm$ scattering process, while $V^8$ stays in kinetic equilibrium with $\chi_1$ even long after the freezeout of $\chi_1$ through $\chi_1 V^8 \to \chi_1 V^8$. Number changing processes such as $\chi_1 V^8 \to \chi_1 V^8 V^8$ also do not freeze out until significantly later. These facts guarantee that the $V^8$ bath remains at the same temperature as the SM with zero chemical potential, and thus follows the standard equilibrium distribution with zero chemical potential through the freezeout of $\chi_1$.

\section{Decay Widths}
\label{app:widths}
Table~\ref{tab:BDecayWidths} shows the perturbative partial decay widths of the dark sector bound state $\cB$ in both models into SM final states, through mixing with the SM mediator $V$. 
The bound state wavefunction is given in Eq. (\ref{eqn:psi02}). 
In addition, Table~\ref{tab:VDecayWidths} shows the perturbative partial decay widths of the SM mediator $V$ into all possible final states.

\renewcommand{\arraystretch}{3}

\setlength{\tabcolsep}{15pt}

\begin{landscape}
\begin{table*}
\centering
\begin{tabular}{c c}
\toprule

Decay Process & $\frac{\Gamma(\mathcal{B} \to X)}{\alpha_D |\psi(0)|^2}$ \\
\hline
$\mathcal{B} \to f \overline{f}$ & $\frac{16 \pi N_m N_c r^4 \alpha \epsilon^2 m_\chi^2 \left[ \left(c_W^2 Q \left(m_Z^2 - 4m_\chi^2 \right) + 4 g_V m_\chi^2 \right)^2 + 16 g_A^2 m_\chi^4 + r^2 \Gamma_V^2 m_Z^2(g_V^2 + g_A^2)/(1-r^2)^2 \right]}{3 c_W^2 \left(m_Z^2 - 4m_\chi^2\right)^2 \left[ (m_Z^2 - 4 r^2 m_\chi^2)^2 + r^2 \Gamma_V^2 m_Z^2 \right]}$ \\
$\mathcal{B} \to W^+W^-$ & $\frac{4 \pi N_m r^4 c_W^2 \alpha \epsilon^2 \left(m_\chi^2 - m_W^2 \right)^{3/2} \left(4m_\chi^4 + 20 m_\chi^2 m_W^2 + 3m_W^4 \right) \left[ m_Z^4 + r^2 \Gamma_V^2 m_Z^2/(1 - r^2)^2\right] }{3 m_\chi m_W^4 \left(m_Z^2 - 4m_\chi^2 \right)^2 \left[(m_Z^2 - 4 r^2 m_\chi^2)^2 + r^2 \Gamma_V^2 m_Z^2 \right]}$ \\
$\mathcal{B} \to Z h^0$ & $\frac{\pi N_m r^4 \alpha \epsilon^2 \left[m_Z^4 + 2 m_Z^2 \left(20 m_\chi^2 - m_H^2 \right) + \left(m_H^2 - 4m_\chi^2\right)^2 \right] \sqrt{m_Z^4 - 2m_Z^2 \left(m_H^2 + 4m_\chi^2 \right) + \left(m_H^2 - 4m_\chi^2\right)^2} \left[16 m_\chi^4 + r^2 \Gamma_V^2 m_Z^2/(1 - r^2)^2 \right] }{192 c_W^2 m_\chi^4 \left(m_Z^2 - 4m_\chi^2 \right)^2 \left[ (m_Z^2 - 4 r^2 m_\chi^2)^2 + r^2 \Gamma_V^2 m_Z^2 \right]}$ \\
$\mathcal{B}_{\text{pD}} \to V h_D$ & $\frac{\pi \alpha_D \left[m_V^4 + 2 m_V^2 \left(20 m_\chi^2 - m_{h_D}^2 \right) + \left(m_{h_D}^2 - 4 m_\chi^2 \right)^2 \right] \sqrt{m_V^4 - 2 m_V^2 \left(m_{h_D}^2 + 4m_\chi^2 \right) + \left(m_{h_D}^2 - 4 m_\chi^2 \right)^2 }}{3 m_\chi^4 \left(m_V^2 - 4 m_\chi^2 \right)^2}$ \\
\botrule
\end{tabular}
\caption{Table of perturbative partial widths for the bound state $\mathcal{B}$ in both the dark sector models. $N_m = 4$ for the pseudo-Dirac model, and $N_m = 1$ for the triple Higgs model: this factor accounts for differences in the type of fermion in each theory, as well as the value of the coupling between the DM and the SM mediator. $N_c = 3$ for quarks and 1 otherwise, $g_V = g_{V,Z} \equiv \{0.25, -0.0189, 0.0959, -0.1730 \}$ and $g_A = g_{A,Z} \equiv \{0.25, -0.25, 0.25, -0.25\}$ are the vector and axial couplings to the $Z$-boson for $\{\nu_e, e, u, d\}$ and for the other 2 generations respectively. $Q$ is the electric charge of each species, and $\alpha$ electromagnetic fine structure constant; $\Gamma_V$ is the width of the SM mediator in each model ($V$ in the pseudo-Dirac model, $V^1$ in the triple Higgs), $m_H$ is the mass of the SM Higgs, and $m_{h_D}$ is the mass of the dark sector Higgs in the pseudo-Dirac model. The last expression is only applicable to the pseudo-Dirac model.}
\label{tab:BDecayWidths}
\end{table*}
\end{landscape}

\renewcommand{\arraystretch}{3}

\setlength{\tabcolsep}{15pt}

\begin{landscape}
\begin{table*}
\centering
\begin{tabular}{c c}
\toprule

Decay Process & Partial Width \\
\hline
$V \to \chi \overline{\chi}$ & $\frac{g_D^2}{12 \pi} \sqrt{m_V^2 - 4 m_\chi^2}$ \\
$V \to f \overline{f}$ & $\frac{N_c e^2 \epsilon^2}{12 \pi c_W^2(1 - r^2)^2}  \sqrt{m_V^2 - 4 m_f^2} \left[(g_V^2 + g_A^2) + \frac{2m_f^2}{m_V^2} (g_V^2 - 2g_A^2)\right]$ \\
$V \to W^+ W^-$ & $\frac{\epsilon^2 e^2 c_W^2 r^4 m_V}{192 \pi (1 - r^2)^2} \frac{(1 - 4x^2)^{3/2}}{x^4} (1 + 20x^2 + 12 x^4)$ \\
$V \to Zh^0$ & $\frac{\epsilon^2 e^2 m_V}{192 \pi c_W^2 r^2 (r^2 - 1)^2} \sqrt{(y^2 - 1)^2 r^4 - 2r^2(y^2 + 1) + 1} \left[r^4(y^2 - 1)^2 - 2r^2(y^2 - 5) + 1 \right]$ \\
\botrule
\end{tabular}
\caption{Table of perturbative partial widths for the SM mediator $V$ for both the dark sector models. $N_c = 3$ for quarks and 1 otherwise. $g_V = g_{V,Z} - c_W^2(1 - r^2)Q$ and $g_A = g_{A,Z}$, where $ g_{V,Z} \equiv \{0.25, -0.0189, 0.0959, -0.1730 \}$ and $g_{A,Z} \equiv \{0.25, -0.25, 0.25, -0.25\}$ are the vector and axial couplings to the $Z$-boson for $\{\nu_e, e, u, d\}$ and for the other 2 generations respectively, and $Q$ is the electric charge of the fermion. $r \equiv m_Z/m_V$, $x \equiv m_W/m_V$ and $y \equiv m_H/m_V$.}
\label{tab:VDecayWidths}
\end{table*}
\end{landscape}

\section{Bound State Formation via Initial/Final State Radiation}

In addition to the resonant formation process that has been our main focus in the body of this work, bound states can also form in conjunction with radiation of other particles in the initial or final state. This process is very important in the context of electron accelerators where the center-of-mass energy of the colliding particles is fixed and does not overlap the bound state resonance (as discussed e.g. in Refs.~\cite{Brodsky:2009gx,An:2015pva}), and so the resonant signal is absent. 

This process could also be critical if the decays of spin-0 bound states were much more observable than those of spin-1 bound states, and the mediator with the SM were a vector (or if the spin-1 states were more observable and the mediator were a scalar); emission of additional particles would then allow the production of the rarer but more observable bound state. However, in the examples we have studied, the latter situation does not hold; indeed, the spin-0 $s$-wave bound states can generally decay into light mediators and are thus likely to be more difficult to detect than their spin-1 counterparts.

Since initial and final state radiation inevitably involves extra powers of the coupling relative to the resonant case, we expect processes of this type to be suppressed relative to the resonant production. However, one might wonder whether threshold enhancements to the production and interaction cross section for unbound but slow-moving DM particles, in the presence of a light mediator, could modify this conclusion and lead to a large contribution from the threshold region. 

Note that this is a very different parameter regime to that considered for muonium production in Ref.~\cite{Brodsky:2009gx} and for light darkonium production in Ref.~\cite{An:2015pva}, where the beam energy is presumed to be large relative to the resonance energy, and the extra particle(s) emitted as initial/final state radiation carry away much of the beam energy; it is more similar to the situation in indirect detection, where slow-moving DM particles may emit a light particle and radiatively capture into a bound state (see e.g.~\cite{Pospelov:2008jd, MarchRussell:2008tu}). The rate for such radiative bound-state formation scales as $1/v$ close to threshold, for a massless mediator. However, we will show that in the case where the particles are produced near threshold and then form a bound state, this $1/v$ scaling is canceled out by the small phase space for the particle production near threshold.

Similar contributions to bound-state formation from soft gluon emission have been studied in the context of quarkonium formation using non-relativistic effective field theory techniques~\cite{Bodwin:1992qr, Bodwin:1994jh}. In that case, $p$-wave color-singlet quarkonia can be formed either directly or through an intermediate $s$-wave color-octet pair of heavy quarks; relatedly, the $s$-wave quarkonium state $|Q\bar{Q}\rangle$ can be described as having a small $\mathcal{O}(v^2)$ admixture of a Fock state $|Q\bar{Q} g\rangle$ containing an additional soft gluon. This approach suggests that the admixture term can be neglected to leading order when dealing with $s$-wave bound states, and should not experience large enhancements near threshold.

To see explicitly how this works in our case, note that we can write the matrix element for production of the bound state (plus a light mediator with momentum $\vec{l}$), via an intermediate state of two near-threshold (i.e. highly non-relativistic) but unbound DM particles, as:
\begin{alignat}{2} 
	i\mathcal{M}(i\rightarrow f) \nonumber &=&& \sqrt{\frac{2 M}{(2 m_\chi)(2 m_\chi)} } \int \frac{d^3 p_1}{(2\pi)^3}   \frac{2\pi \delta(E_\text{in} - E_{p_1} - E_{p_2})}{(2 E_{p_1})(2 E_{p_2})} \nonumber \\
    & &&\times \int \frac{d^3 a}{(2\pi)^3} \tilde{\psi}_{p_1,p_2}^*(\vec{a})i \bar{\mathcal{M}}(i \rightarrow \vec{a}_1, \vec{a}_2) \int \frac{d^3 b}{(2\pi)^3} \tilde{\psi}_{p_1,p_2}(\vec{b}) \nonumber \\
    & &&\times \int \frac{d^3 q}{(2\pi)^3} \tilde{\psi}^*_B(\vec{q}) \cdot i\bar{\mathcal{M}}(\vec{b}_1, \vec{b}_2 \rightarrow \vec{q}_1 \vec{q}_2 \vec{l})\,.
    \label{eq:fsrmaster} 
\end{alignat}
 Here $\bar{\mathcal{M}}(i \rightarrow \vec{a}_1, \vec{a}_2)$ is the hard matrix element describing production of two free DM particles with momenta $\vec{a}_1, \vec{a}_2$ from the initial state $i$, and likewise $\bar{\mathcal{M}}(\vec{b}_1, \vec{b}_2 \rightarrow \vec{q}_1 \vec{q}_2 \vec{l})$ is the hard matrix element describing the radiation of a light mediator with momentum $\vec{l}$ from the DM-DM state with particle momenta $\vec{b}_1, \vec{b}_2$, to produce final-state DM particles with momenta $\vec{q}_1, \vec{q}_2$. The wavefunctions convert the plane-wave states to the full intermediate and final states accounting for potential effects. $\vec{p}_1$ and $\vec{p}_2$ act as labels on the intermediate state with momentum-space wavefunction $\tilde{\psi}_{p_1, p_2}$, describing the momenta of the constituent particles at large separation. $\tilde{\psi}_B$ denotes the momentum-space wavefunction of the bound state (which in principle is labeled by the quantum numbers $n, l, m$; we suppress these indices). $m_\chi$ is the DM mass and $M \approx 2 m_\chi$ is the bound-state mass.

In the non-relativistic limit where the potential is neglected, the leading-order matrix element for light vector boson radiation from one of a pair of heavy fermions (with gauge coupling $g_\mathcal{B}$ and fermion masses $m_1$, $m_2$) is given by:
\begin{alignat}{2} 
 	i\mathcal{M}(\vec{b}_1, \vec{b}_2 \rightarrow \vec{q}_1 \vec{q}_2 \vec{l}) &=&& i g_\mathcal{B} \vec{\epsilon^*}(l) \left[ (\vec{b}_1 + \vec{q}_1) 2 m_2 (2\pi)^3 \delta^{(3)}(\vec{b}_2 - \vec{q}_2) \right. \nonumber \\
	& && \left. - (\vec{b}_2 + \vec{q}_2) 2 m_1 (2\pi)^3 \delta^{(3)}(\vec{b}_1 - \vec{q}_1)  \right]. 
\end{alignat}

Inserting this expression into Eq.~\eqref{eq:fsrmaster}, setting the masses of the two heavy fermions equal, $m_1 = m_2 = m_\chi$, working in relative momentum coordinates, and choosing the center-of-mass frame, we obtain:
\begin{alignat}{2} 
	i\mathcal{M}(i\rightarrow f) &=&& 2 i g_\mathcal{B} \sqrt{2 M} \vec{\epsilon^*}(l) \int \frac{d^3 p_1}{(2\pi)^3} \frac{2\pi \delta(E_\text{in} - E_{p_1} - E_{p_2})}{(2 E_{p_1})(2 E_{p_2})} \nonumber \\
 	& && \times \left[\int \frac{d^3 a}{(2\pi)^3} i \bar{\mathcal{M}}(i \rightarrow \vec{a}_1, \vec{a}_2) \tilde{\psi}_{p_1,p_2}^*(\vec{a})  \right]  \nonumber \\
 	& && \times \int \frac{d^3 q}{(2\pi)^3}        \tilde{\psi}^*_B(\vec{q}) \vec{q} \left[  \tilde{\psi}_{p_1,p_2} \left(\vec{q} + \vec{l}/2 \right) +  \tilde{\psi}_{p_1,p_2} \left(\vec{q} - \vec{l}/2 \right)  \right] .
\end{alignat}
 The integral over $d^3 q$ on the last line also appears in the matrix element for radiative bound state formation, and has been previously computed in the non-relativistic limit for massless vector mediators \cite{0953-4075-29-10-021,Asadi:2016ybp}. In the near-threshold regime, $l \lesssim \alpha^2 m_\chi$ (as the binding energy must provide the necessary energy to radiate the mediator), and the $l$-dependence of the integral can be neglected; in this case, the integral simply scales as $1/\sqrt{p}$, where $\vec{p} = (\vec{p}_1 - \vec{p}_2)/2$. (This factor, when squared, is responsible for the $1/v$ scaling of the radiative bound state formation cross section.)
 
 If we further suppose that the hard matrix element for production of the intermediate state from the initial state is independent of the final-state relative momentum $\vec{a}$, i.e. we can write $i \bar{\mathcal{M}}(i \rightarrow \vec{a}_1, \vec{a}_2) = i \bar{\mathcal{M}}(i \rightarrow \text{DM}, \text{DM})$  then the integral over $d^3 a$ simplifies to give $i \bar{\mathcal{M}}(i \rightarrow \text{DM}, \text{DM}) \psi^*_{p_1,p_2}(0)$, where $\psi$ denotes the position-space wavefunction. The wavefunction at the origin in a Coulomb-like potential scales as $\sqrt{\alpha_\mathcal{B} m_\chi/p}$ (e.g.~\cite{0953-4075-29-10-021}), which yields the usual Sommerfeld enhancement when squared.
 
 Putting these pieces together and performing the phase-space integral over $\int d^3 p_1$, writing $E_{p_1} = E_{p_2} = \sqrt{m_\chi^2 + |\vec{p}|^2}$ since we are working in the COM frame, we find that (keeping only scaling relationships, dropping order-1 factors):
 \begin{alignat}{2}  
 	i\mathcal{M}(i\rightarrow f) &\sim&& g_\mathcal{B} \sqrt{m_\chi} i \bar{\mathcal{M}}(i \rightarrow \text{DM}, \text{DM}) \nonumber \\
    & &&\times \frac{1}{E_\text{in}^2} \int \frac{d^3 p}{(2\pi)^3} 2\pi \delta\left(E_\text{in} - 2 \sqrt{m_\chi^2 + |\vec{p}|^2}\right) \sqrt{\frac{\alpha_\mathcal{B} m_\chi}{p}} \sqrt{\frac{1}{p}} \nonumber \\
 	&\sim&& \bar{\mathcal{M}}(i \rightarrow \text{DM}, \text{DM}) g_\mathcal{B} m_\chi \frac{\sqrt{\alpha_\mathcal{B}}}{E_\text{in}} \nonumber \\
 	&\sim&&  \bar{\mathcal{M}}(i \rightarrow \text{DM}, \text{DM}) \alpha_\mathcal{B} \,. 
\end{alignat}
 Note that as mentioned previously, the phase-space integral over the intermediate-state momentum $d^3 p$ has canceled out the $1/p$ scaling from the wavefunctions.
 
 Thus we see that the bound-state production cross section through this channel should scale as $|\bar{\mathcal{M}}(i \rightarrow \text{DM}, \text{DM})|^2 \alpha_\mathcal{B}^2$, multiplied by a 2-body phase space factor. Since the momenta in the final state are small, of order $l \sim \alpha_\mathcal{B}^2 m_\chi$, the overall scaling of the cross section with the couplings is $\alpha_\mathcal{B}^4 \times |\bar{\mathcal{M}}(i \rightarrow \text{DM}, \text{DM})|^2$.
 
 By comparison, we see that the resonant production cross section scales as $|\bar{\mathcal{M}}(i \rightarrow \text{DM}, \text{DM})|^2 \alpha_\mathcal{B}^3$, where the $\alpha_\mathcal{B}$ dependence arises from the $\mathcal{B}$ wavefunction. Thus the rate to produce an extra light mediator by emission from a near-threshold intermediate state, in conjunction with the bound state formation, is suppressed by one power of $\alpha_\mathcal{B}$ overall. This is the same suppression one would naively expect for emission of a hard photon from the initial or final state, with no small phase-space factors or threshold enhancements. We self-consistently neglect all such diagrams in the body of this work.
 
 Here we have neglected the mediator mass $m_Y$ in estimating the scalings; in particular, the intermediate-state position-space wavefunction may be steeply peaked near the origin for special values of $m_Y$, corresponding to the presence of near-zero-energy bound states (e.g.~\cite{Hisano:2004ds}). However, it seems likely that any apparent enhancement from this behavior can be reinterpreted as resonant capture into a near-zero-energy bound state, which is already accounted for in our formalism. We leave a more detailed study of the resonant regime to future work.

\chapter{Dark Matter and Reionization}
\label{chap:app_dm_reionization}

\section{Additional Constraints}
\label{app:additionalConstraints}

Figure \ref{fig:xeConstraintsTIGMPlot_sWave} shows the free electron fraction just prior to reionization $x_e(z=6)$ for the benchmark scenario of both $\chi \chi \to e^+e^-$ and $\chi \chi \to \gamma \gamma$ $s$-wave annihilations, as well as the excluded cross-sections due to constraints from the CMB power spectrum as measured by Planck and from the $T_{\text{IGM}}(z=4.8)$ constraints. The $T_{\text{IGM}}$ bounds alone can almost rule out a 10\% contribution from $\chi \chi \to e^+e^-$ above a mass of approximately \SI{1}{\giga\eV}, but are weaker for $\chi \chi \to \gamma \gamma$, since less energy goes into heating for this process. However, if the structure formation boost factor has been underestimated in our paper, these bounds will become stronger. This effectively sets a limit on how large the boost can be.

\begin{figure*}[t!]
    \centering
	\subfigure{
		\includegraphics[scale=0.63]{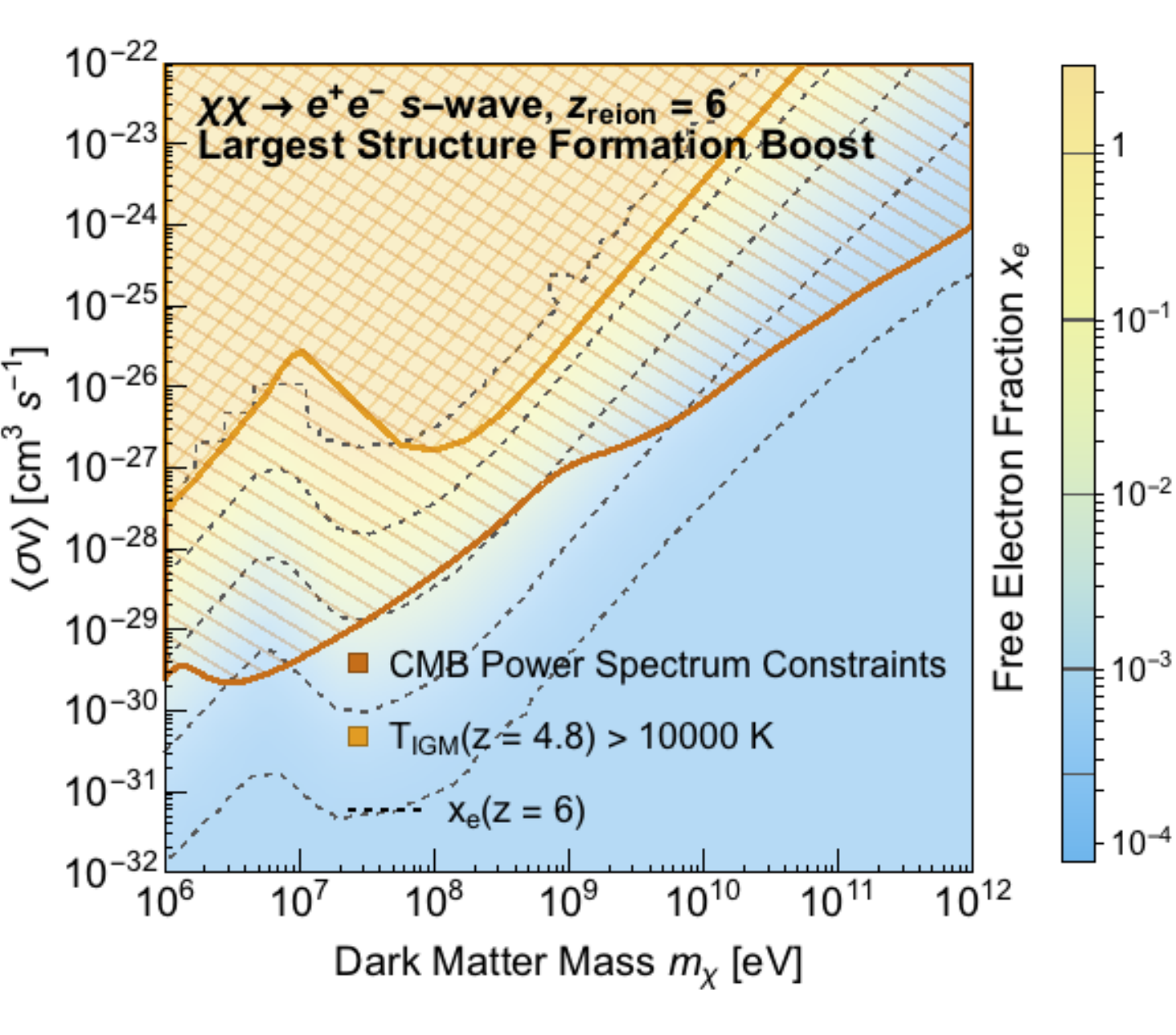}
	}
	\subfigure{
		\includegraphics[scale=0.63]{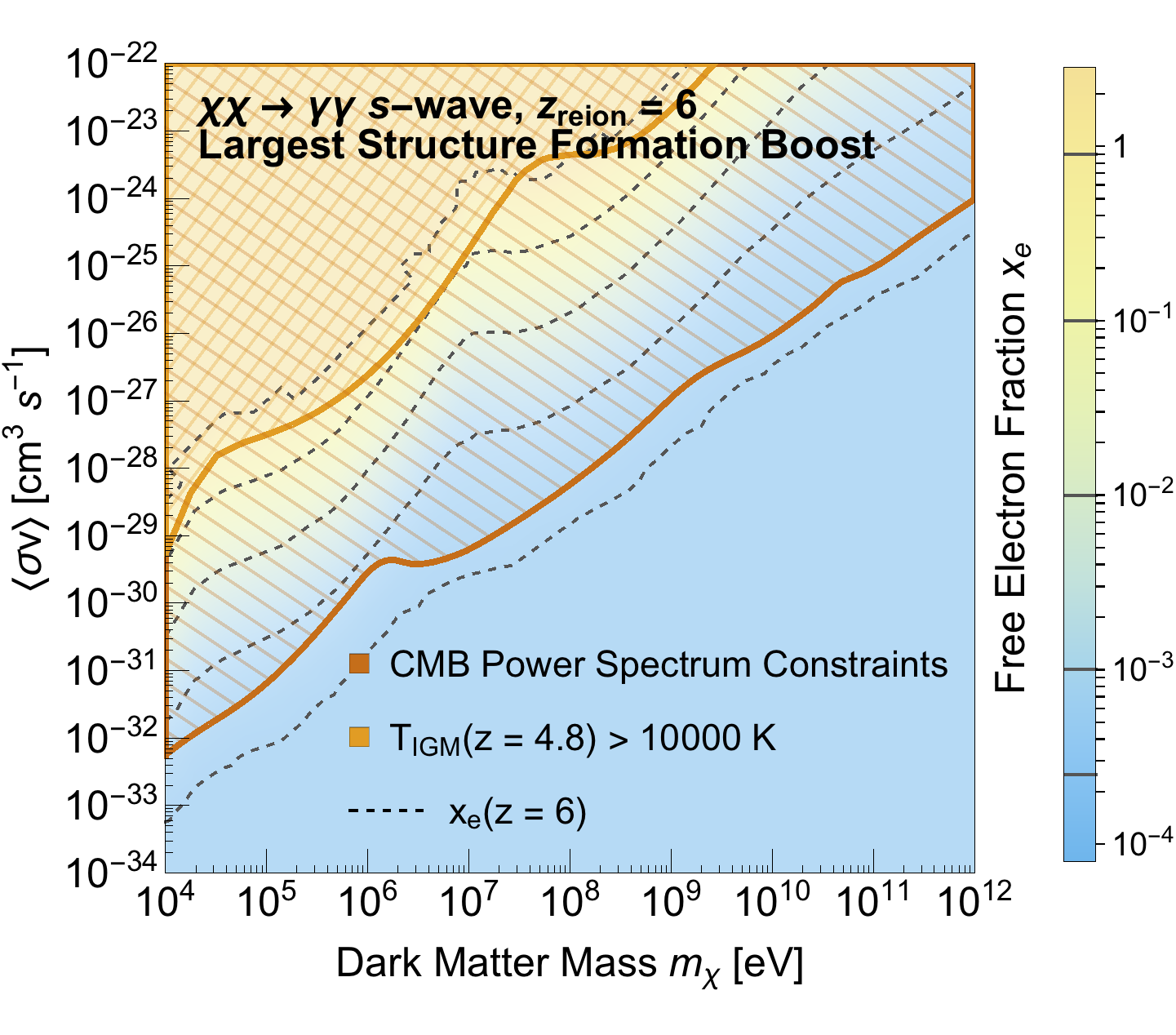}
	}
	\caption{DM contribution to reionization for $\chi \chi \to e^+e^-$ (left) and $\chi \chi \to \gamma \gamma$ (right) $s$-wave annihilation, benchmark scenario. The hatched regions correspond to parameter space ruled out by the CMB power spectrum constraints as measured by Planck (red) and $T_{\text{IGM}}(z = 4.8) < \SI{10000}{\kelvin}$ (orange) respectively. The color density plot shows the DM contribution to $x_e$ just prior to reionization at $z = 6$, with contours (black, dashed) shown for a contribution to $x_e(z = 6) = $ 0.025\%, 0.1\%, 1\%, 10\% and 90\% respectively.}
	\label{fig:xeConstraintsTIGMPlot_sWave}
\end{figure*}

Throughout this paper, we have obtained the limits on the contribution to reionization from DM in the case of $s$- and $p$-wave annihilation by considering the processes $\chi \chi \to e^+e^-$ and $\chi \chi \to \gamma \gamma$ with each annihilation product having fixed, identical total energy $E = m_\chi$. This allowed us to set limits on $\langle \sigma v \rangle$ or $(\sigma v)_{\text{ref}}$ as a function of $m_\chi$. However, the constraints that we set here extend beyond these two annihilation scenarios. The energy injection rate from annihilations is set only by the quantity $\langle \sigma v \rangle/m_\chi$, and is independent of the annihilation products produced; only the energy deposition rate is dependent on the species and energies of the annihilation products. 

Thus, if we were to recast the $\langle \sigma v \rangle - m_\chi$ parameter space in Figures \ref{fig:xeConstraintsPlot_sWave} and \ref{fig:xeConstraintsPlot_pWave} as a $\langle \sigma v \rangle/m_\chi - m_\chi$ parameter space, the latter parameter actually corresponds to the injection energy of the annihilation products, which is not necessarily equal to the DM mass. 

 Figures \ref{fig:xeConstraintsPlotSigmavOverMChi_sWave} and \ref{fig:xeConstraintsPlotSigmavOverMChi_pWave} present the same set of constraints and results for $x_e(z = 6)$ as a function of $\langle \sigma v \rangle/m_\chi$ or $(\sigma v)_{\text{ref}}/m_\chi$ and the injection energy of the $s$- or $p$-wave annihilation products, which in general can be very different from $m_\chi$. Table \ref{tab:Constraints} gives the $s$-wave CMB power spectrum constraints and the $p$-wave $T_{\text{IGM}}(z = 4.80) > \SI{10000}{\kelvin}$ constraints in table form for the convenience of the reader. For any arbitrary annihilation process, the total contribution to $x_e$ prior to reionization is strictly less than the highest contribution to $x_e$ possible among the different particles with different energies produced from the annihilation. This implies that for a given injection rate, the only dependence on the spectrum of the annihilation products enters through $f_c(z)$, and as a result, the CMB power spectrum constraints are relatively insensitive to the details of the injection spectrum from DM annihilations \cite{Elor2015a}.  

\begin{figure*}[t!]
    \centering
	\subfigure{
		\includegraphics[scale=0.63]{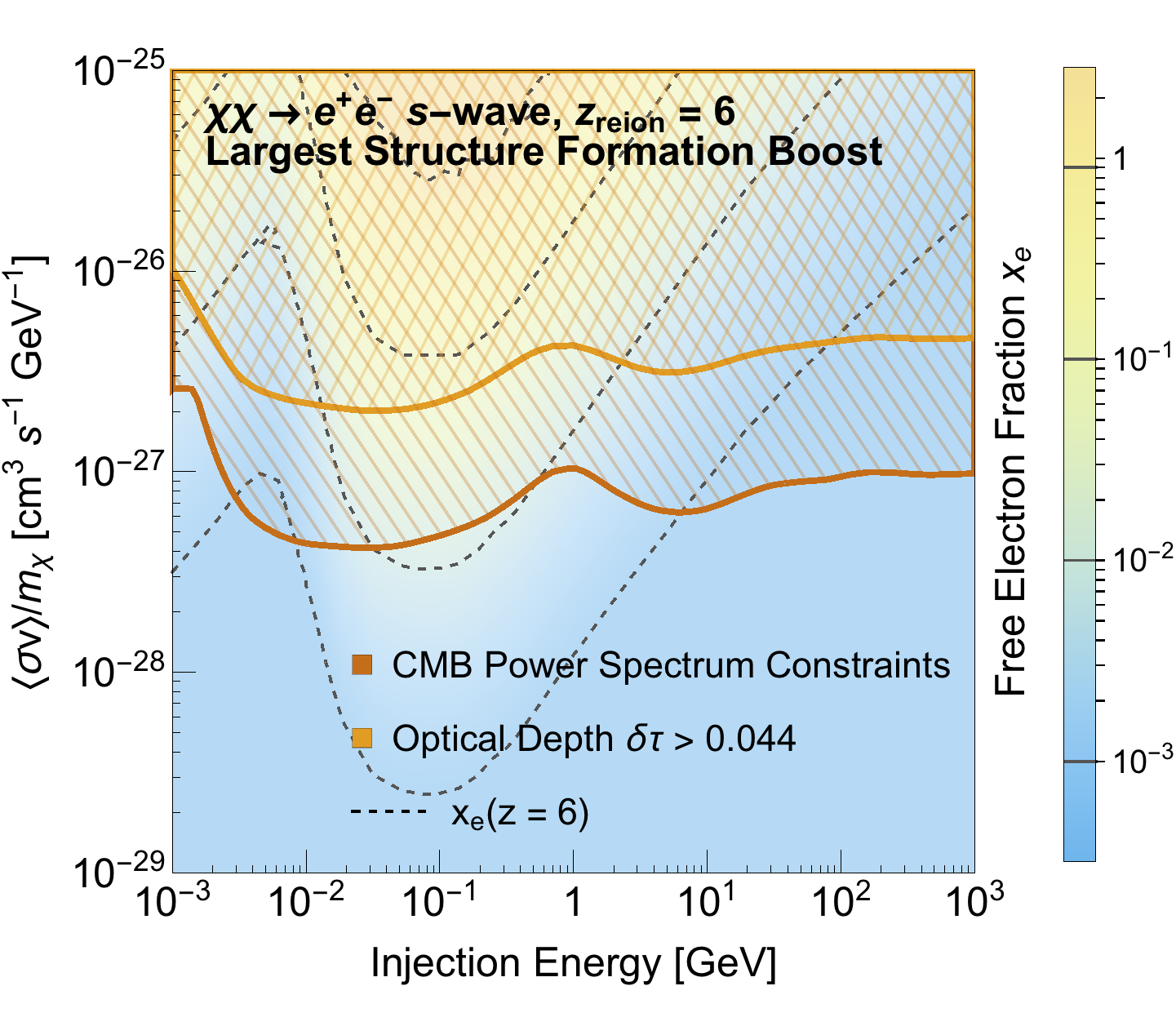}
	}
	\subfigure{
		\includegraphics[scale=0.63]{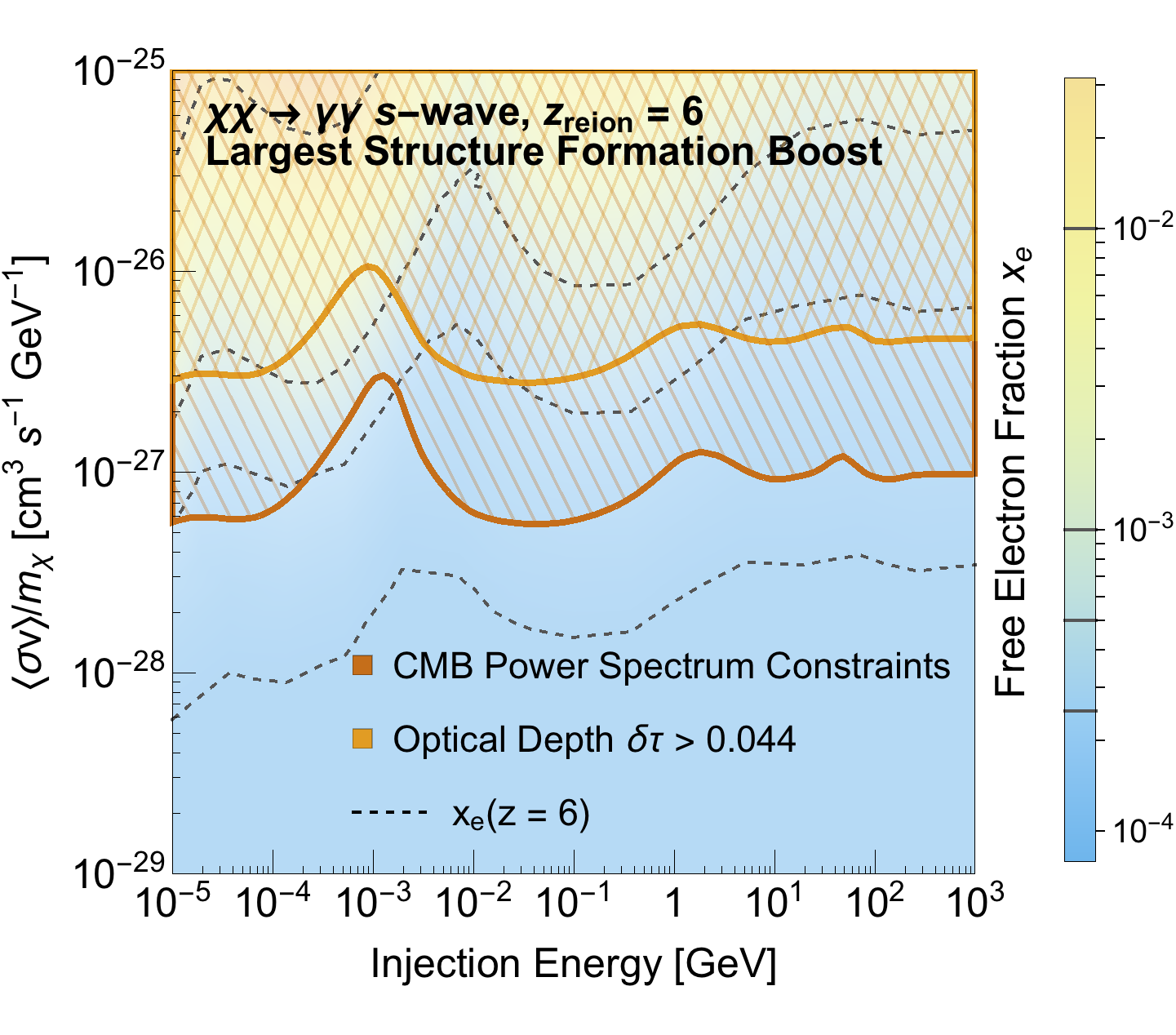}
	}
	\caption{DM contribution to reionization for $\chi \chi \to e^+e^-$ (left) and $\chi \chi \to \gamma \gamma$ (right) $s$-wave annihilation, plotted as a function of $\langle \sigma v \rangle / m_\chi$ and injection energy, benchmark scenario. The hatched regions correspond to parameter space ruled out by the CMB power spectrum constraints as measured by Planck (red) and optical depth constraints (orange) respectively. The color plot indicates the DM contribution to $x_e$, with contours drawn for a contribution of 0.025\%, 0.1\%, 1\%, 10\% and 90\% respectively. }
	\label{fig:xeConstraintsPlotSigmavOverMChi_sWave}
\end{figure*}

\begin{figure*}[t!]
    \centering
	\subfigure{
		\includegraphics[scale=0.63]{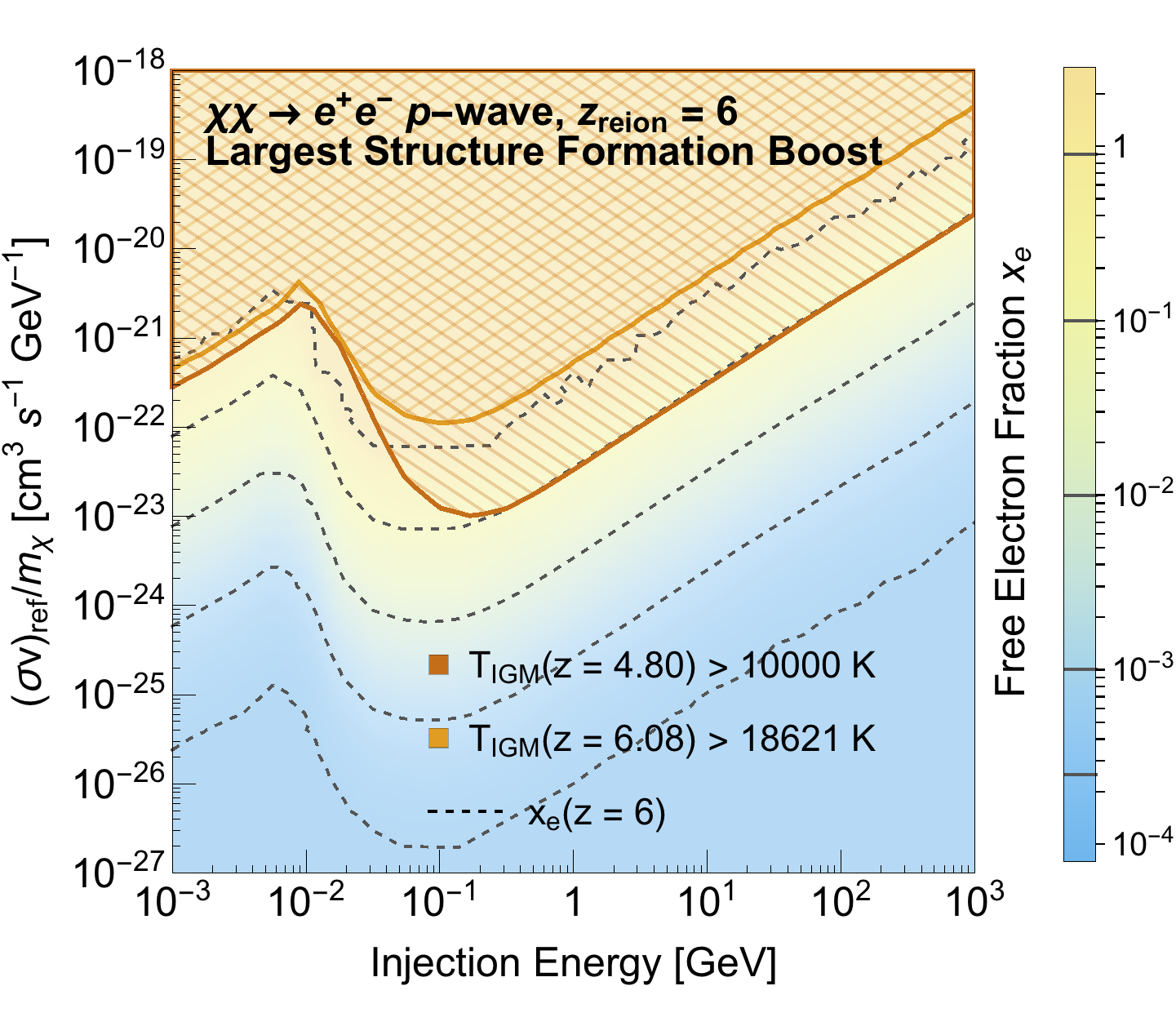}
	}
	\subfigure{
		\includegraphics[scale=0.63]{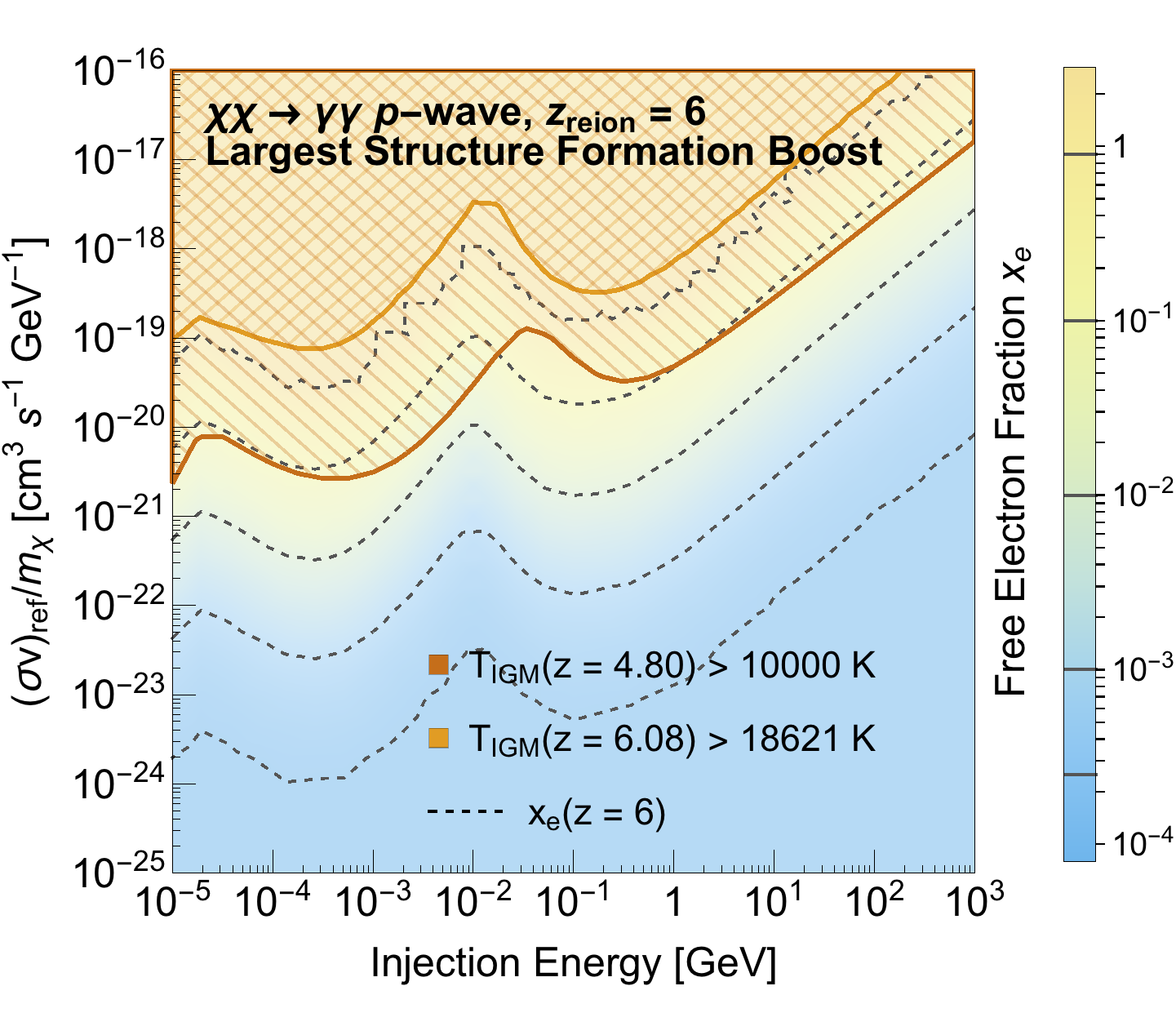}
	}
	\caption{DM contribution to reionization for $\chi \chi \to e^+e^-$ (left) and $\chi \chi \to \gamma \gamma$ (right) $p$-wave annihilation, plotted as a function of $\langle \sigma v \rangle / m_\chi$ and injection energy, benchmark scenario. The hatched regions correspond to parameter space ruled out by the CMB power spectrum constraints as measured by Planck (red) and optical depth constraints (orange) respectively. The color plot indicates the DM contribution to $x_e$, with contours drawn for a contribution of 0.025\%, 0.1\%, 1\%, 10\% and 90\% respectively. }
	\label{fig:xeConstraintsPlotSigmavOverMChi_pWave}
\end{figure*}

\renewcommand{\arraystretch}{1.1} 
\begin{table}
\begin{tabularx}{\textwidth}{c >{\centering\arraybackslash}X >{\centering\arraybackslash}X >{\centering\arraybackslash}X >{\centering\arraybackslash}X} 
	\toprule
	\multirow{3}{*}{$\log_{10}[m_\chi (\SI{}{\giga\eV})]$} & \multicolumn{2}{c}{$s$-wave} & \multicolumn{2}{c}{$p$-wave} \\ \cline{2-5}
	& \multicolumn{2}{c}{$\log_{10} \left[ \langle \sigma v \rangle/m_\chi \right]$} & \multicolumn{2}{c}{$\log_{10} \left[ (\sigma v)_{\text{ref}}/m_\chi \right]$ } \\ \cline{2-3} \cline{4-5}
	& $\chi \chi \to e^+e^-$ & $\chi \chi \to \gamma \gamma$ & $\chi \chi \to e^+e^-$ & $\chi \chi \to \gamma \gamma$ \\ \hline
	-5.00 & & -27.2502 & & -20.6327 \\
	-4.75 & & -27.2243 & & -20.1114 \\
	-4.50 & & -27.2311 & & -20.1027 \\ 
	-4.25 & & -27.2326 & & -20.2672 \\ 
	-4.00 & & -27.1866 & & -20.4146 \\
	-3.75 & & -27.0830 & & -20.5190 \\
	-3.50 & & -26.9280 & & -20.5746 \\
	-3.25 & & -26.7415 & & -20.5746 \\
	-3.00 & -26.5871 & -26.5424 & -21.5524 & -20.5075 \\
	-2.75 & -26.7722 & -26.6038 & -21.3538 & -20.3684 \\
	-2.50 & -27.1549 & -26.9224 & -21.1154 & -20.1486 \\
	-2.25 & -27.3000 & -27.1003 & -20.8725 & -19.8619 \\
	-2.00 & -27.3572 & -27.2023 & -20.5468 & -19.5262 \\
	-1.75 & -27.3727 & -27.2421 & -21.0758 & -19.1676 \\
	-1.50 & -27.3787 & -27.2574 & -21.8876 & -18.8817 \\
	-1.25 & -27.3611 & -27.2570 & -22.5907 & -18.9666 \\
	-1.00 & -27.3186 & -27.2409 & -22.9054 & -19.2229 \\
	-0.75 & -27.2587 & -27.2056 & -23.0043 & -19.4243 \\
	-0.50 & -27.1635 & -27.1489 & -22.9120 & -19.4912 \\
	-0.25 & -27.0370 & -27.0626 & -22.7140 & -19.4418 \\
	 0.00 & -26.9831 & -26.9568 & -22.4788 & -19.3185 \\
	 0.25 & -27.0701 & -26.9007 & -22.2346 & -19.1527 \\
	 0.50 & -27.1613 & -26.9332 & -21.9916 & -18.9624 \\
	 0.75 & -27.2024 & -27.0015 & -21.7520 & -18.7597 \\
	 1.00 & -27.1837 & -27.0369 & -21.5127 & -18.5503 \\
	 1.25 & -27.1212 & -27.0208 & -21.2700 & -18.3361 \\
	 1.50 & -27.0662 & -26.9702 & -21.0248 & -18.1182 \\
	 1.75 & -27.0467 & -26.9416 & -20.7816 & -17.8968 \\
	 2.00 & -27.0246 & -27.0247 & -20.5460 & -17.6747 \\
	 2.25 & -27.0014 & -27.0301 & -20.3158 & -17.4536 \\
	 2.50 & -27.0101 & -27.0116 & -20.0852 & -17.2340 \\
	 2.75 & -27.0139 & -27.0102 & -19.8505 & -17.0141 \\
	 3.00 & -27.0090 & -27.0089 & -19.6115 & -16.7924 \\
	 \botrule
\end{tabularx}
\caption{Tabulated $s$-wave CMB power spectrum constraints and $p$-wave $T_{\text{IGM}}(z = 4.80) > \SI{10000}{\kelvin}$ constraints. Cross sections divided by mass are in units of \SI{}{\centi\meter\cubed\per\second\per\giga\eV}.}
\label{tab:Constraints}
\end{table}

\section{\texorpdfstring{$p$}{p}-wave \texorpdfstring{$J$}{J}-Factor}
\label{app:JFactor}

The photon flux per unit energy due to DM annihilations from DM in the galaxy is given by \cite{Essig2013}
\begin{alignat}{1}
	\frac{d\Phi}{dE} = \frac{1}{2} \frac{r_\odot}{4\pi} \frac{\rho_\odot^2}{m_\chi} \frac{\langle \sigma v \rangle_\odot}{m_\chi} \frac{dN_\gamma}{dE} J,
\end{alignat}
where $dN_\gamma/dE$ is the annihilation photon yield, and $r_\odot$ and $\rho_\odot$ are the distance from the Sun to the galactic center and the local DM density respectively. $J$ is a dimensionless factor that encapsulates the averaging of the DM density along the line-of-sight of the entire field of observation, and is given by
\begin{alignat}{1}
	J = \int d\Omega \frac{ds}{r_\odot} \left(\frac{\rho(s)}{\rho_\odot} \right)^2.
\end{alignat}
For $s$-wave annihilations, $J$ contains all of the dependence of the photon flux on the DM distribution in the galaxy. In $p$-wave annihilations, however, the rate of DM annihilations also depends on the velocity dispersion of DM, and thus both the density and the velocity of DM along each line-of-sight must be averaged. We should therefore replace $J$ with
\begin{alignat}{1}
	J_p = \int d\Omega \frac{ds}{r_\odot} \left(\frac{\rho(s)}{\rho_\odot} \right)^2 \frac{v^2(s)}{v_\odot^2},
\end{alignat}
and now $\langle \sigma v \rangle_\odot$ is explicitly the local annihilation cross-section due to the velocity dependence of $\langle \sigma v \rangle$. 

Previous studies have implicitly assumed that $J$ and $J_p$ are equal. To assess the significance of this assumption, we consider a pure NFW DM profile given by equation (\ref{rho_smooth}) with $\alpha = 1$, with a corresponding velocity dispersion profile given by the following relation \cite{Zavala2014}:
\begin{alignat}{1}
	\frac{\rho(r)}{\sigma_{\text{1D}}^3(r)} \propto r^{-1.9}.
\end{alignat}
where $\sigma_{\text{1D}}$ is the 1D velocity dispersion that we use as a proxy for $v$. The constant of proportionality of this equation is determined by setting $\rho(r_\odot) = \SI{0.3}{\giga\eV\per\centi\meter\cubed}$ and assuming a Maxwellian distribution of the dark matter particles in the halo with a peak value set equal to the rotation velocity of the Sun given by $v = \SI{220}{\kilo\meter\per\second}$. With these assumptions, we find a difference between $J_p$ and $J$ of about 5 - 10\%, after averaging over the solid angle within some typical galactic diffuse gamma-ray background survey regions. This result has also been confirmed using DM particle dispersion velocities as a function of radius \cite{Necib2016} derived from the Illustris $N$-body simulation \cite{Vogelsberger2014}, which models both DM and baryons. 

We have therefore assumed throughout our analysis that $J_p = J$, and anticipate an error of about 10\% in translating the $\langle \sigma v \rangle$ constraints assuming and $s$-wave distribution directly into constraints for $(\sigma v)_{\text{ref}}$ in $p$-wave annihilations. Since the $p$-wave constraints that we have used rule out regions of parameter space with a contribution to reionization exceeding 10\% by more than 2 orders of magnitude, we do not expect this assumption to change our conclusions in any significant way.

\chapter{
Implications of a 21-cm Signal for Dark Matter Annihilation and Decay
}
\label{app:21cm_annihilation_decay}

\section{Astrophysical Systematics}
\label{app:systematics}


\setcounter{figure}{0}

\subsection{Uncertainties from Annihilation in Dark Matter Halos}

\begin{figure}
    \centering
    \subfigure{
        \label{fig:cooling_struct_form_sys_erfc}
        \includegraphics[scale=0.28]{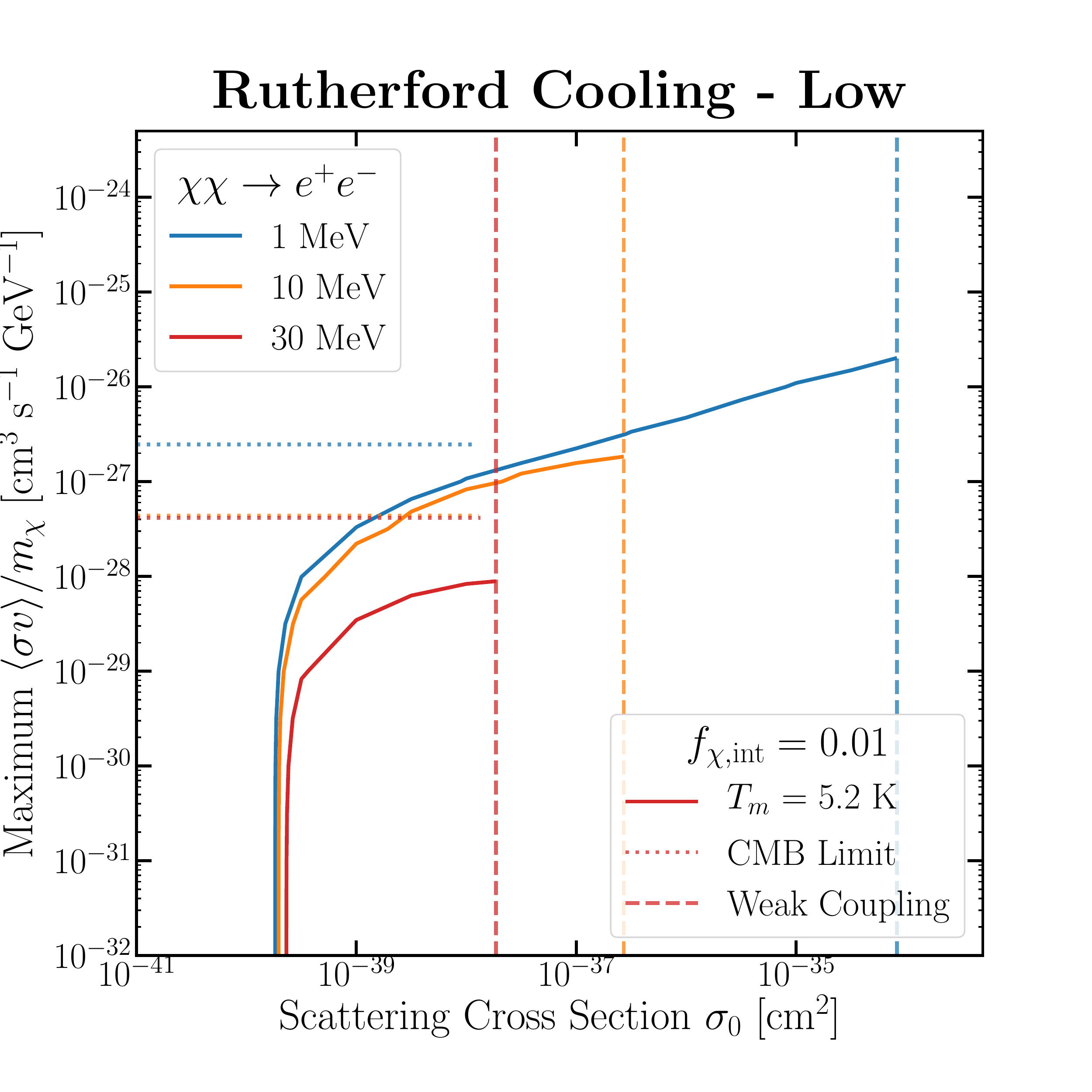}
    }
    \subfigure{
        \label{fig:cooling_struct_form_sys_NFW}
        \includegraphics[scale=0.28]{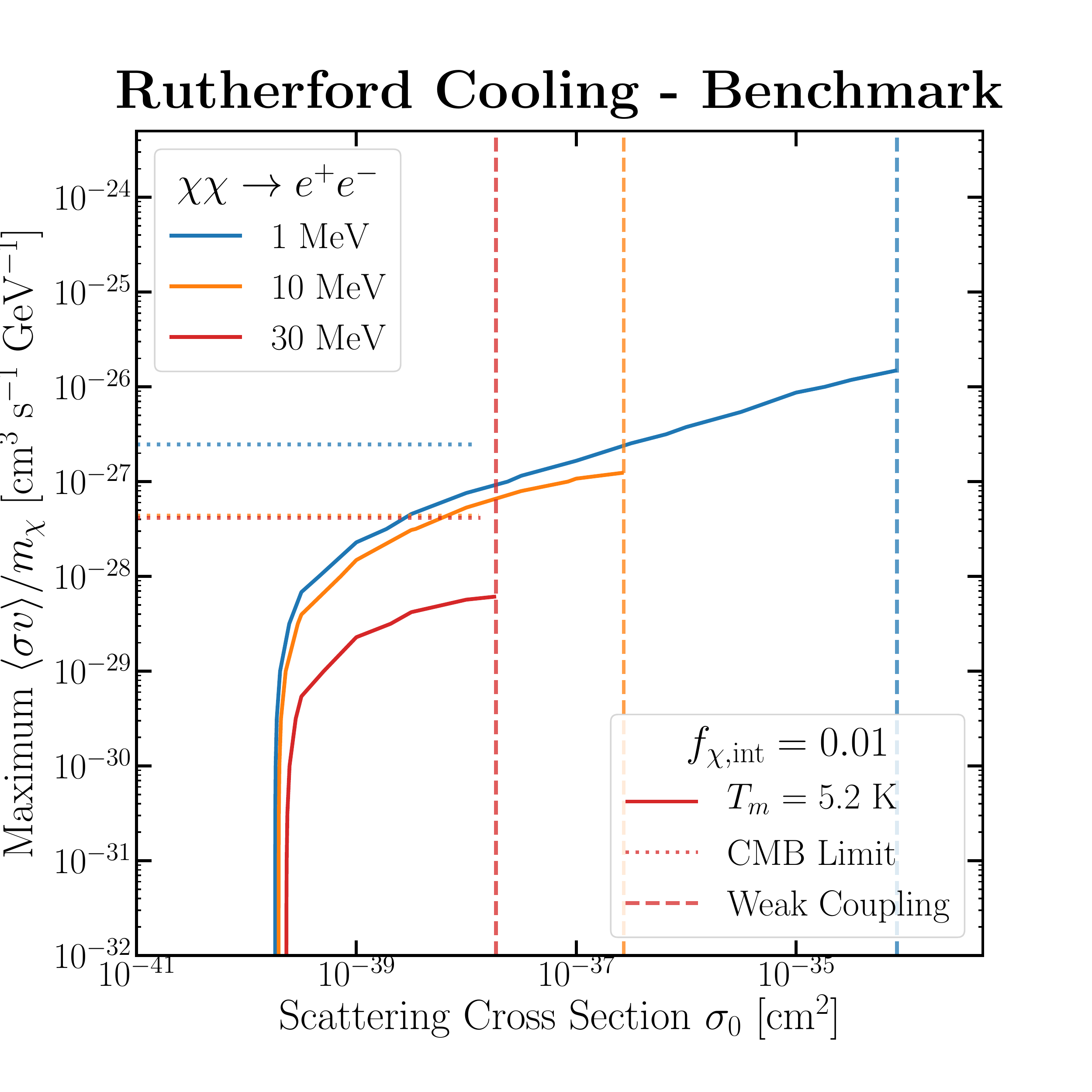}
    } \\
    \subfigure{
        \label{fig:cooling_struct_form_sys_einasto}
        \includegraphics[scale=0.28]{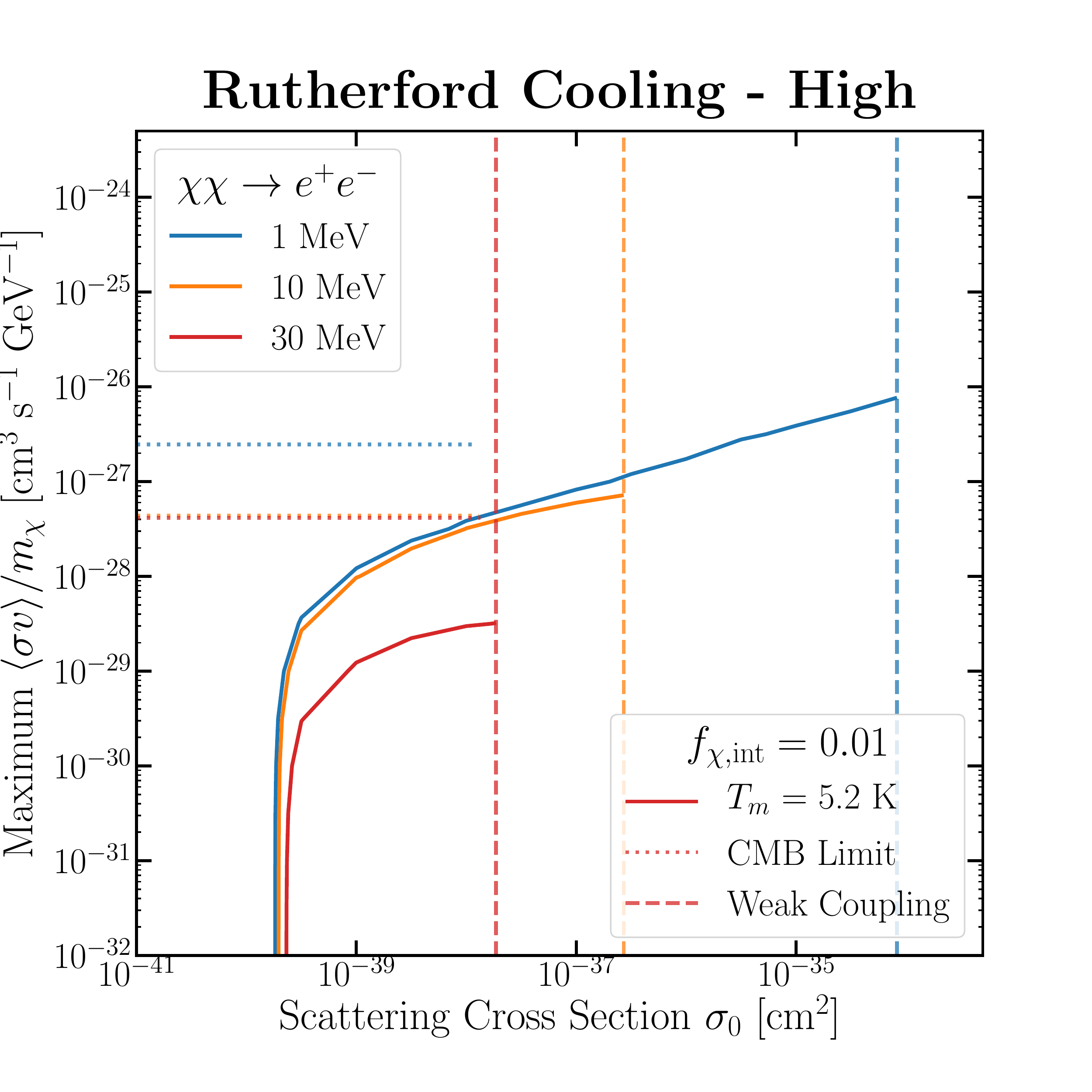}
    }
    \subfigure{
        \label{fig:cooling_struct_form_sys_no_boost
        }
        \includegraphics[scale=0.28]{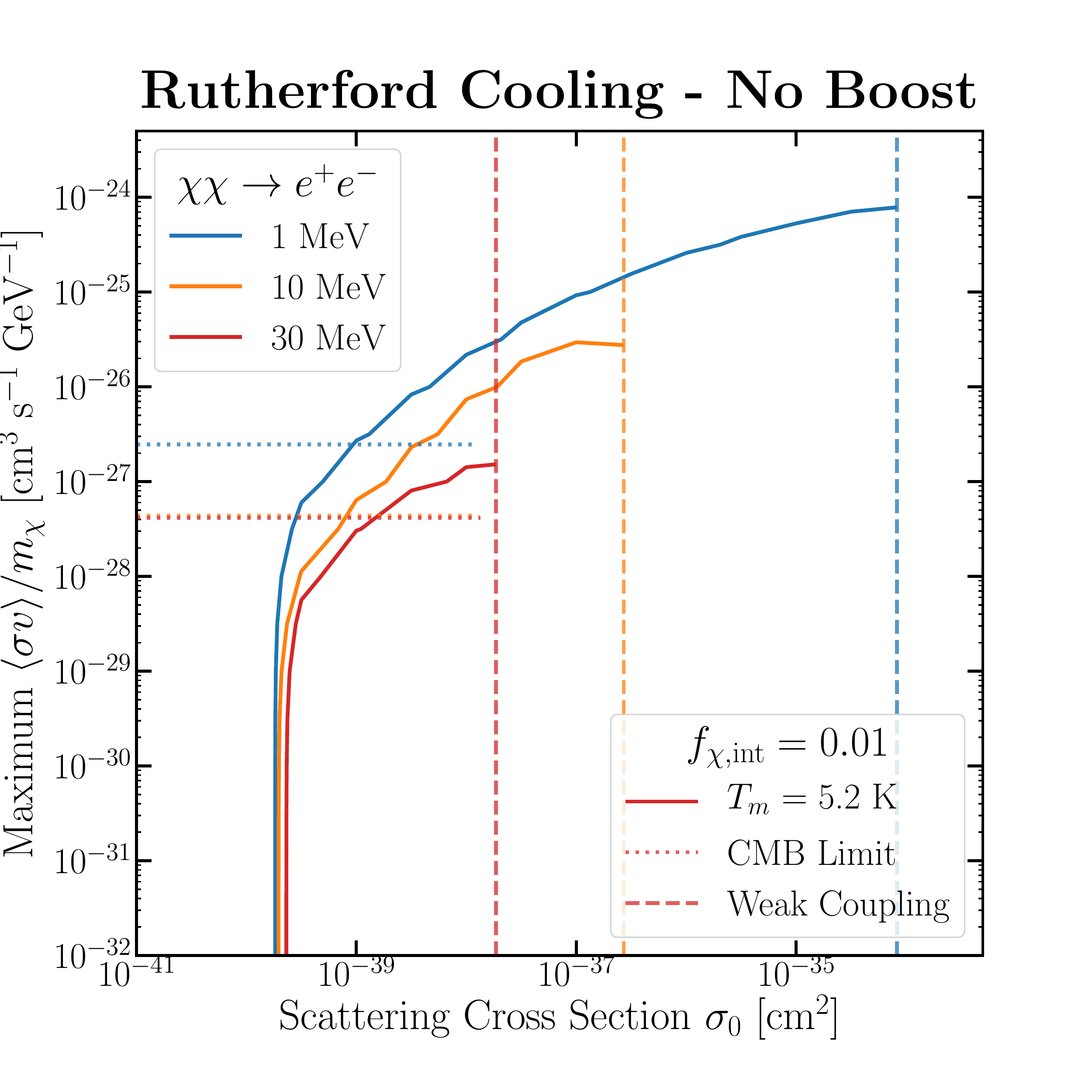}
    }
  \caption{Comparison of DM annihilation constraints when the gas is cooled by Rutherford scattering, where the HIGH (upper left), BENCHMARK (upper right) and LOW (lower left) and NO BOOST (lower right) models for the DM structure formation history are employed (see text for details).}
  \label{fig:cooling_swave_struct_form_sys}
\end{figure}

\begin{figure}
    \centering
    \subfigure{
        \label{fig:cooling_V_pec_0}
        \includegraphics[scale=0.34]{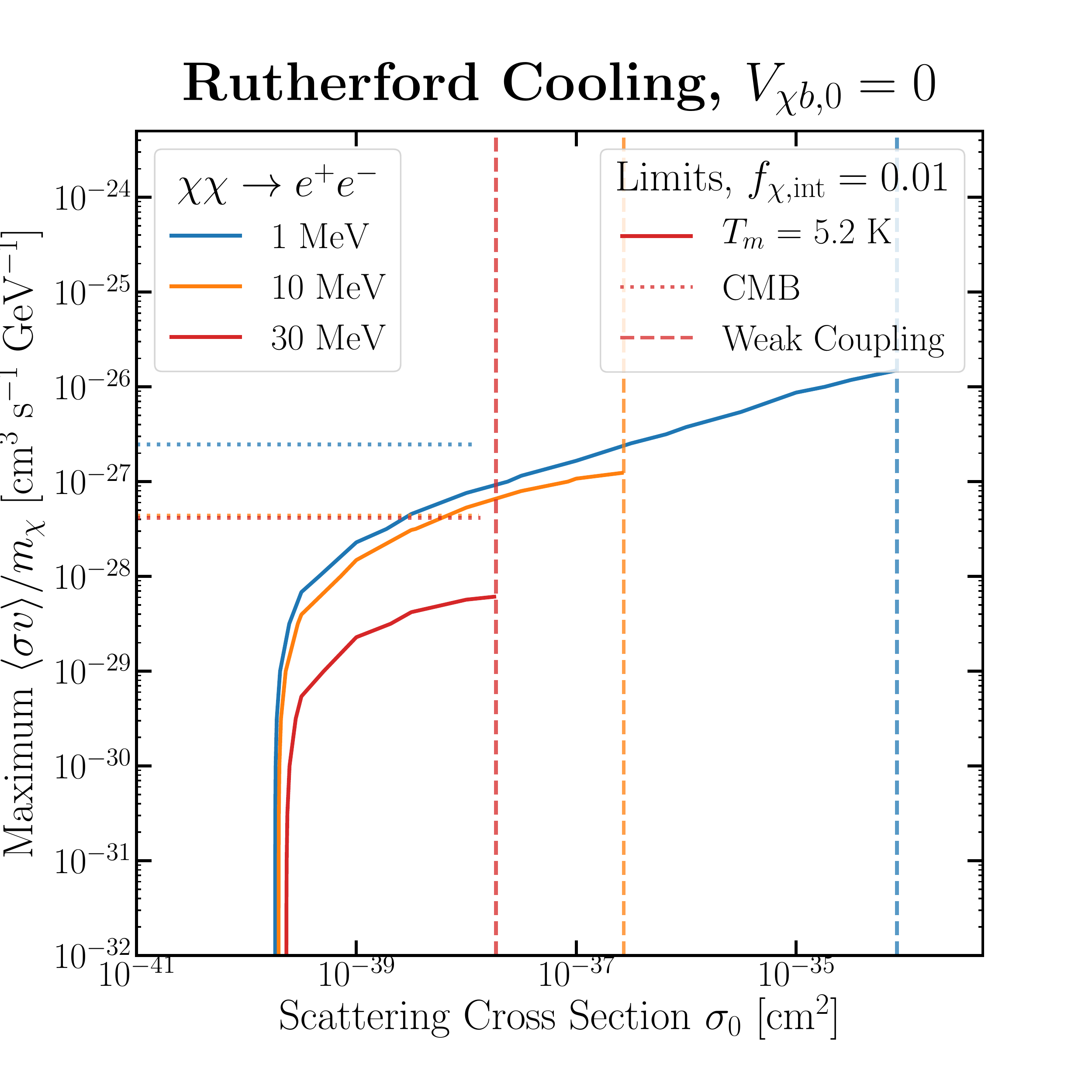}
    }
    \subfigure{
        \label{fig:cooling_V_pec_rms}
        \includegraphics[scale=0.34]{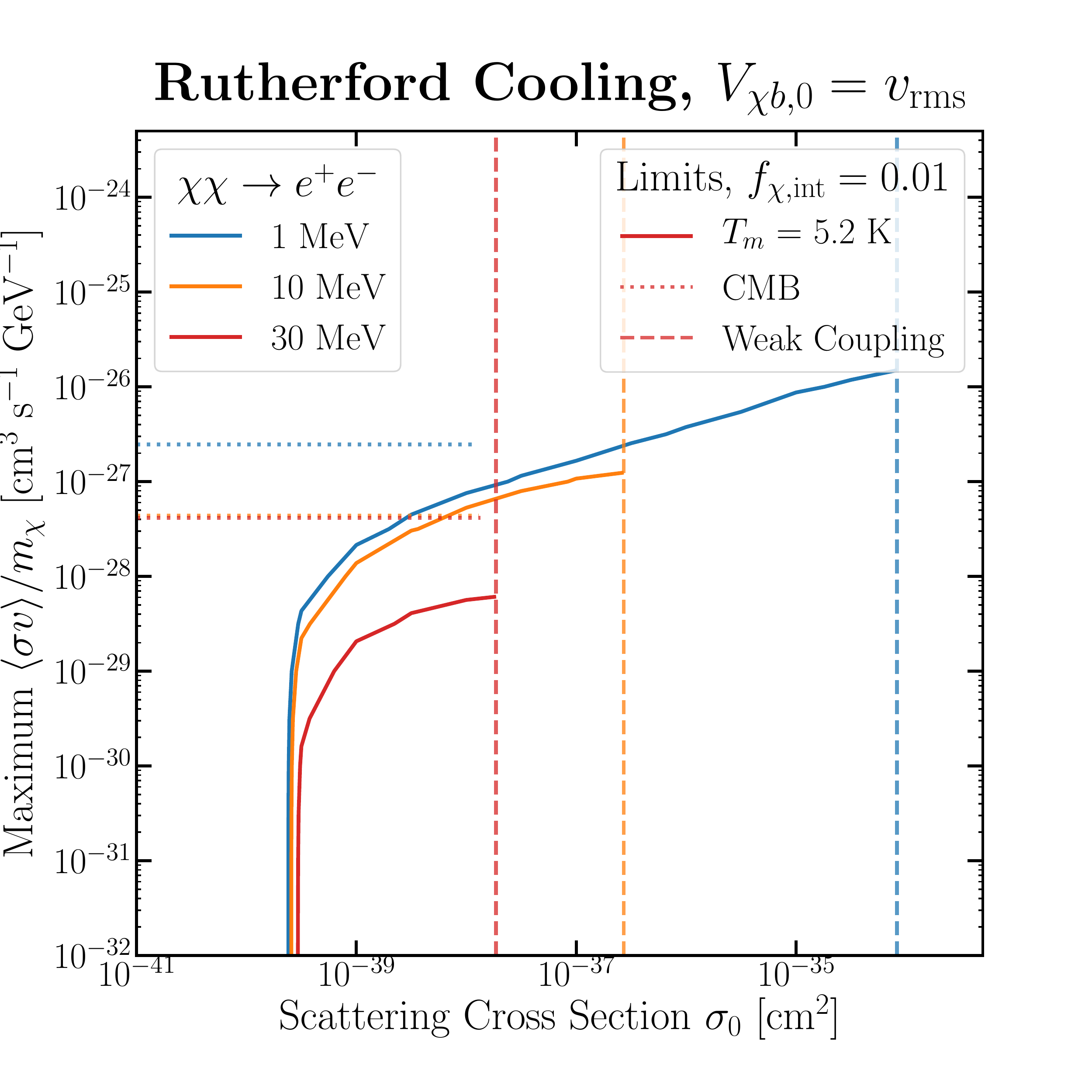}
    }
  \caption{Comparison of DM annihilation constraints when the gas is cooled by Rutherford scattering, where the bulk DM-baryon relative velocity at recombination is taken to be $V_{\chi b,0} = 0$ (left) or $V_{\chi b,0} = v_\text{rms}$ (right).}
  \label{fig:cooling_swave_vchib_init_sys}
\end{figure}

Once structure formation begins, the DM annihilation rate is no longer set purely by the well-measured cosmological average DM density, but instead becomes dominated by annihilation in over-dense regions, which can enhance $\langle \rho^2\rangle$ greatly over $\langle \rho \rangle^2$. A model for this ``boost factor'' $\langle\rho^2(z)\rangle/\langle\rho(z)\rangle^2$ is included in the $f_c(z)$ factors that determine the amount of heating and ionization from $s$-wave DM annihilation (as defined in Eq.~(\ref{eqn:f_z_structure_formation})), as discussed in \cite{Liu:2016cnk}.

On one hand, the size of this enhancement is quite uncertain, due mostly to large uncertainties in the abundance and concentration of low-mass halos that cannot be resolved by cosmological simulations. On the other hand, at $z\sim 17$ the enhancement factor is expected to still be relatively modest; furthermore the heating and ionization at that epoch are determined by the integral over DM annihilation at all previous times, not only at $z\sim 17$, which also dilutes the effect of late-time overdensities (e.g. \cite{Slatyer:2012yq}). This last effect is stronger for secondary particles that take longer to cool and deposit their energy; in particular, for most energies of injection, photons take longer to deposit their energy than electrons, and thus the systematic uncertainty in $f_c(z)$ due to structure formation is expected to be smaller (as the typical photon contributing to late-time ionization/heating was injected at an earlier epoch where structure formation was less important).

To quantitatively estimate the uncertainty in our quoted annihilation limits due to uncertainties in the contribution from DM overdensities, we repeat the analysis using three different models from the literature for the boost factor. The first two are limiting cases from \cite{Liu:2016cnk}: they correspond to (1) DM halos with Einasto density profiles \cite{Einasto}, and including an estimate of substructure within main halos, and (2) DM halos with Navarro-Frenk-White profiles \cite{Navarro:1995iw}, without substructure included. The second model, with the lowest boost factor of those considered in \cite{Liu:2016cnk}, is the benchmark we use for the plots in the main text. Finally, the third model is (3) a simple analytic form proposed as a conservative model for the boost factor by \cite{DAmico:2018sxd}. We label these models as HIGH, BENCHMARK and LOW respectively.

We show the limits on the annihilation cross section for DM annihilating to $e^+ e^-$ in the presence of Rutherford cooling, for these three models, in Fig.~\ref{fig:cooling_swave_struct_form_sys}. Since the boost factor is approximately degenerate with the annihilation cross section (changing it can also lead to a slight modification of the redshift dependence of the annihilation rate), we expect the changes in the limits on the cross section to be similar for the other scenarios (early decoupling and additional radiation). 

We find that, as expected, the constraints are more stringent for the HIGH model, by roughly a factor of 2. However, the BENCHMARK and LOW models agree closely, with the constraints differing on the $15 - 20$\% level. We have performed the same check for DM annihilating to photons, and the difference between the models is even smaller. Thus we expect our benchmark constraints to be similar to others set using a conservative structure formation model. 

If we completely ignore structure formation and consider only annihilations in the smooth DM density, then the constraints weaken considerably, by about a factor of 50, as shown in the fourth panel of Fig.~\ref{fig:cooling_swave_struct_form_sys}. This is the maximally conservative case, and is probably unrealistic; we leave a detailed study of the minimum possible boost factor to further work.

We note these uncertainties do not apply to DM decay, which probes the average DM density rather than the average DM density-squared.

\subsection{Uncertainties in the Initial Value of \texorpdfstring{$V_{\chi b}$}{Vchib}}

In the scenario where Rutherford scattering cools the gas, the scattering rate depends on the relative velocity between the DM and baryons, and hence on the initial value of the bulk relative velocity $V_{\chi b,0}$ at recombination. As argued in Sec.~\ref{sec:rutherford_cooling}, we expect the cooling effect to be strongest for $V_{\chi b,0} = 0$, thus leaving the maximum amount of room for heating from annihilation/decay products.

We test this hypothesis in Fig.~\ref{fig:cooling_swave_vchib_init_sys}, comparing constraints on DM annihilating to $e^+e^-$ for $V_{\chi b,0} = 0$ and $V_{\chi b,0} = v_\text{rms} = 29$ km s$^{-1}$, in the case where a 1\% fraction of the DM participates in the scattering. We find that once the scattering cross section $\sigma_0$ is well above the value required to cool the gas to 5.2 K, the constraints on annihilation are unaffected by this change in the initial conditions, because the large baryon-DM scattering cross section induces a drag force that drives $V_{\chi b}$ to zero (for the interacting DM component). However, the minimum $\sigma_0$ needed to cool the baryons to that temperature does increase modestly (by a few tens of percent) when $V_{\chi b,0}=v_\text{rms}$.

Accordingly we conclude that in the regime where the constraints are not very rapidly varying as a function of $\sigma_0$, away from the minimum $\sigma_0$ needed to achieve the required cooling, the systematic error due to neglecting the distribution of initial relative velocities $V_{\chi b}$ is small.

\section{Supplemental Plots}
\label{app:supplemental_plots}


\setcounter{figure}{0}

In Figs.~\ref{fig:source_decay_zoom} and~\ref{fig:source_swave_zoom}, we show zoomed-in versions of Figs.~\ref{fig:source_decay} and~\ref{fig:source_swave}, to highlight the region where the additional radiation source is comparable or smaller to the CMB, in terms of number density at a wavelength of 21 cm. 

\begin{figure}
    \centering
    \subfigure{
        \label{fig:source_elec_decay_zoom}
        \includegraphics[scale=0.34]{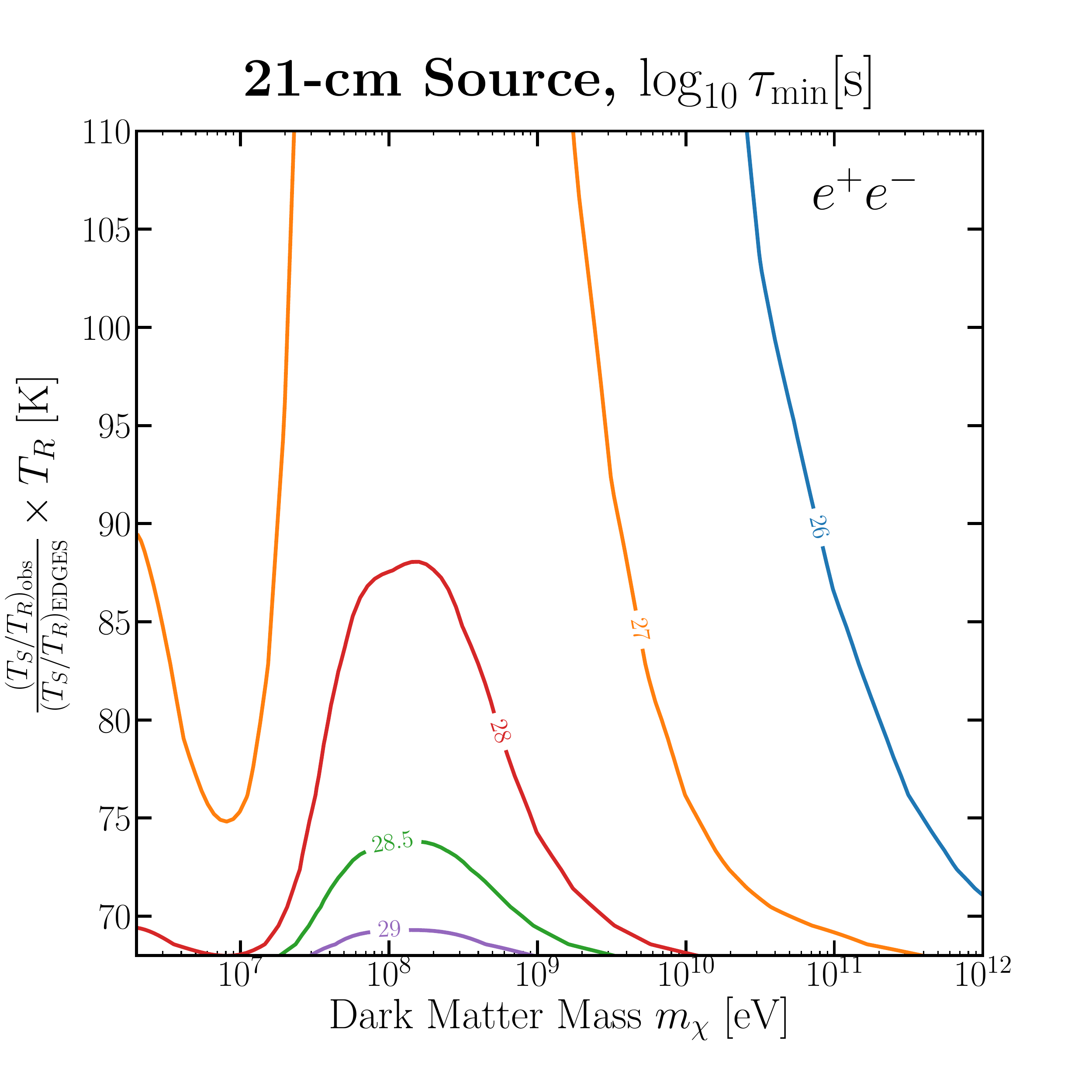}
    }
    \subfigure{
        \label{fig:source_phot_decay_zoom}
        \includegraphics[scale=0.34]{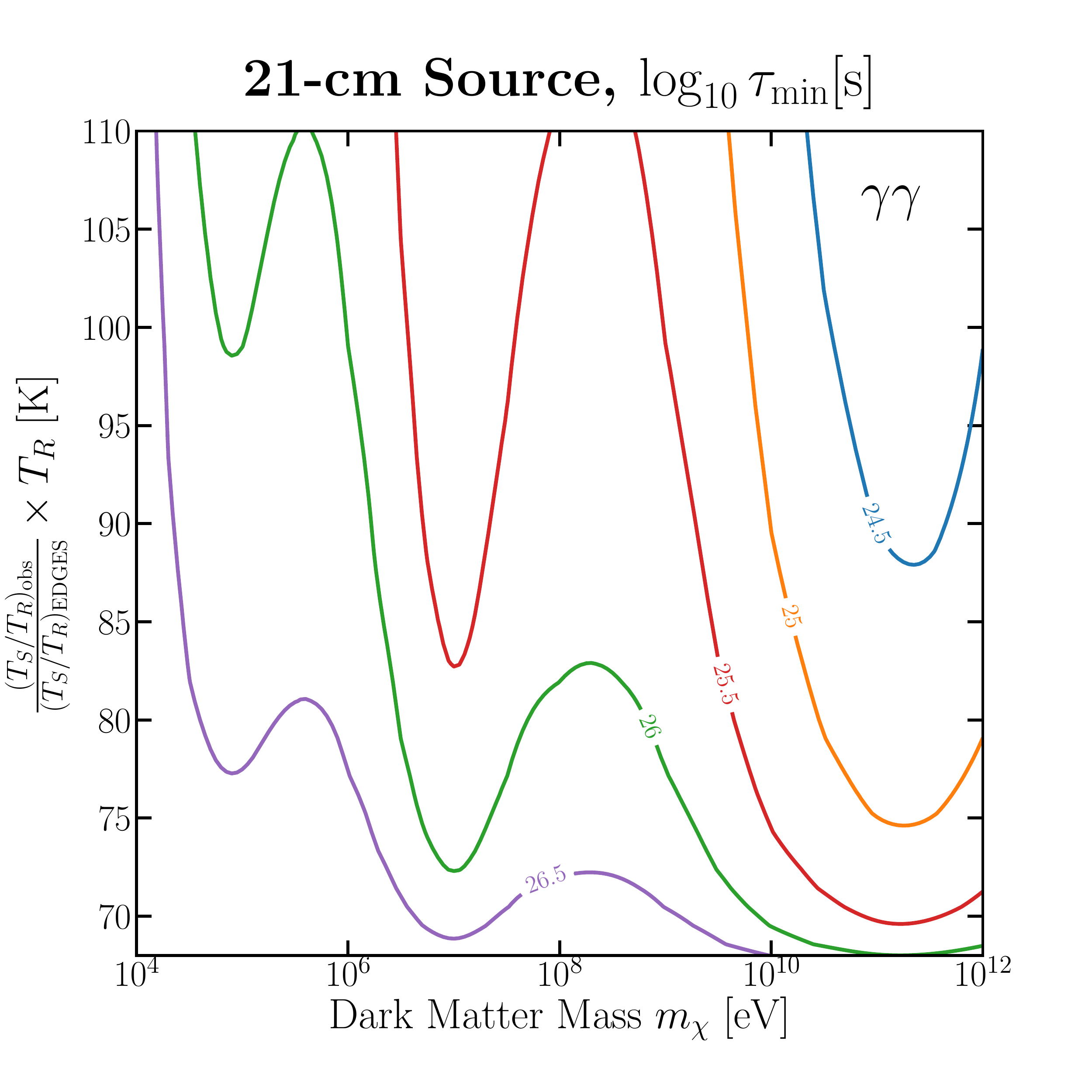}
    }
  \caption{Decay lifetime constraints with an additional 21-cm source with $\chi \to e^+e^-$ (left) and $\chi \to \gamma \gamma$ (right), as a function of $m_\chi$ and $(T_S/T_R)_\text{obs}/(T_S/T_R)_\text{EDGES} \times T_R$. This is a zoomed-in version of Fig.~\ref{fig:source_decay}. Contour lines of constant minimum $\log_{10}\tau$ (in seconds) are shown. 
  }
  \label{fig:source_decay_zoom}
\end{figure}

\begin{figure}
    \centering
    \subfigure{
        \label{fig:source_elec_swave_zoom}
        \includegraphics[scale=0.34]{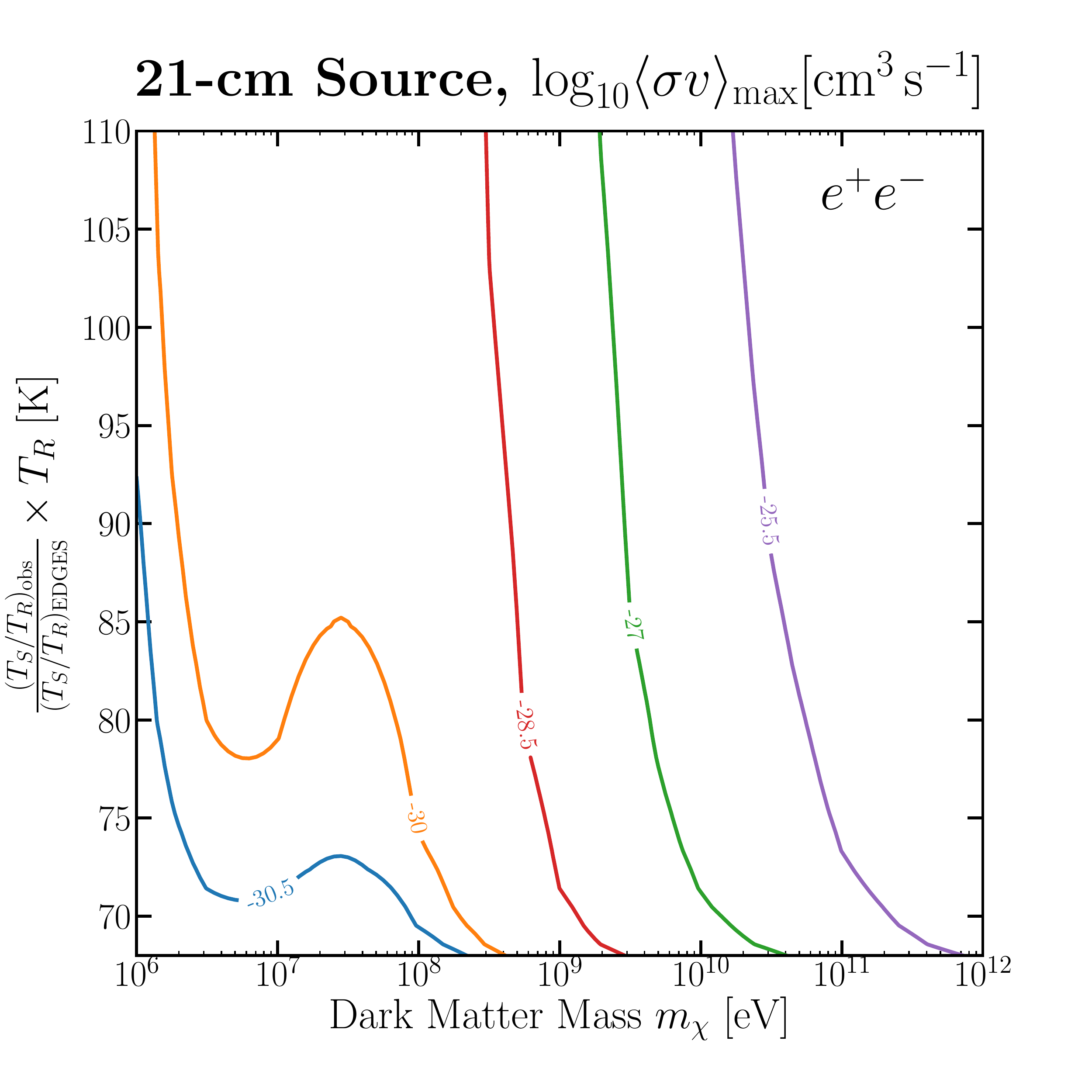}
    }
    \subfigure{
        \label{fig:source_phot_swave_zoom}
        \includegraphics[scale=0.34]{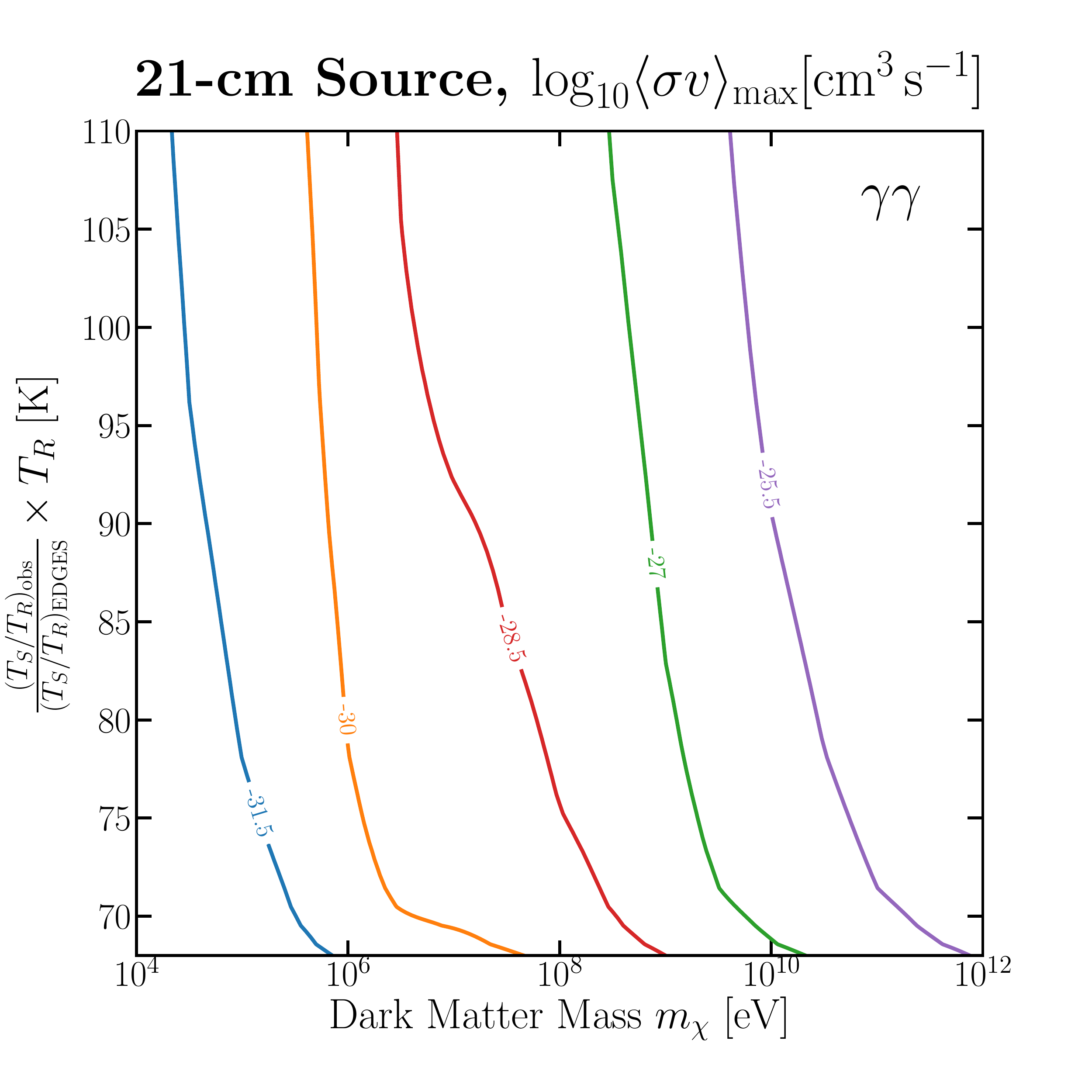}
    }
  \caption{Annihilation cross section constraints with an additional 21-cm source with $\chi \chi \to e^+e^-$ (left) and $\chi \chi \to \gamma \gamma$ (right), as a function of $m_\chi$ and $(T_S/T_R)_\text{obs}/(T_S/T_R)_\text{EDGES} \times T_R$. Contour lines of constant maximum $\log_{10}\langle \sigma v \rangle$ (in $\text{ cm}^3 \text{ s}^{-1}$) are shown. This is a zoomed-in version of Fig.~\ref{fig:source_swave} The green contour corresponds to the canonical relic abundance cross section of $\SI{3e-26}{cm^3 \, s^{-1}}$.
  }
  \label{fig:source_swave_zoom}
\end{figure}

\begin{figure}
    \centering
    \subfigure{
        \label{fig:recomb_elec_decay_low}
        \includegraphics[scale=0.34]{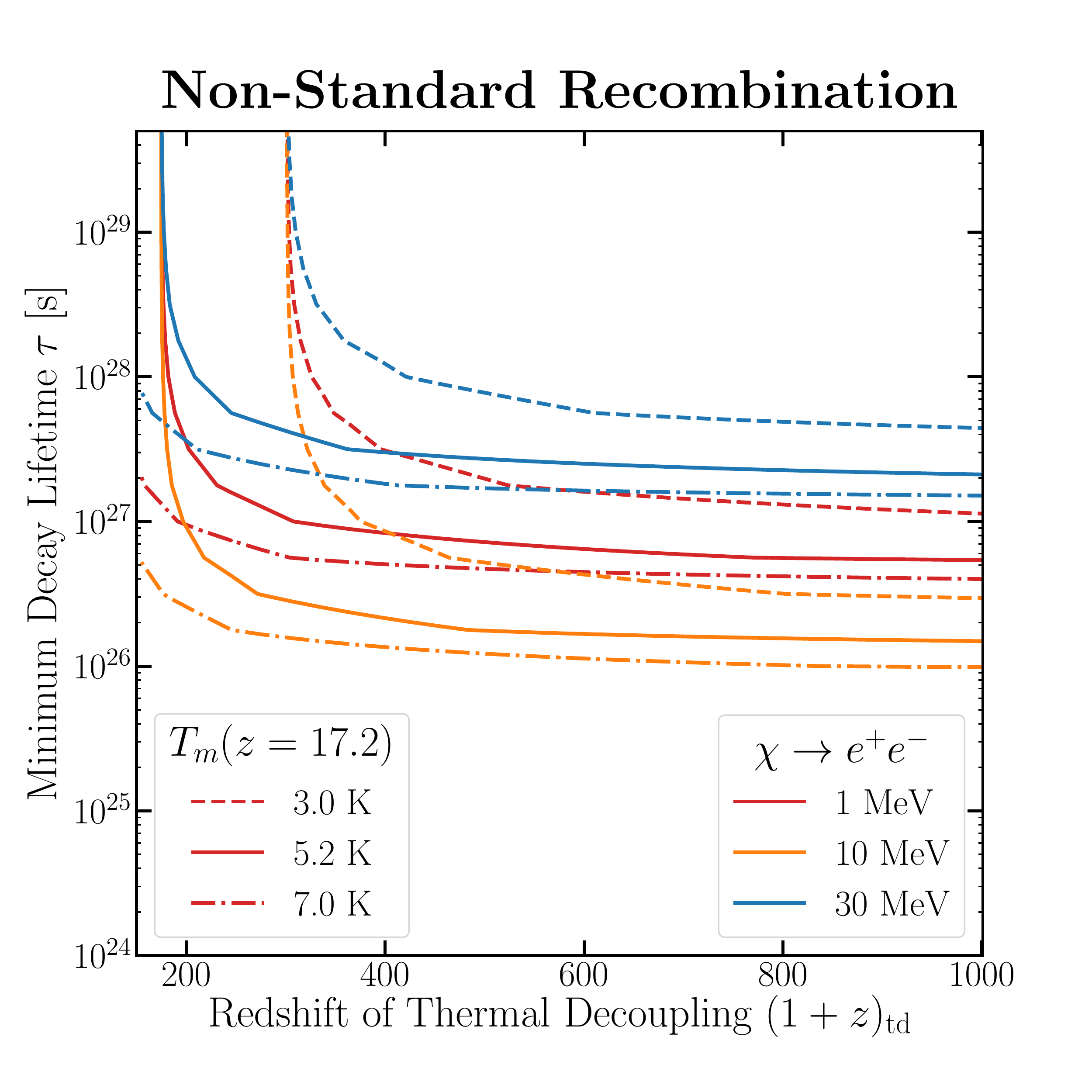}
    }
    \subfigure{
        \label{fig:recomb_phot_decay_low}
        \includegraphics[scale=0.34]{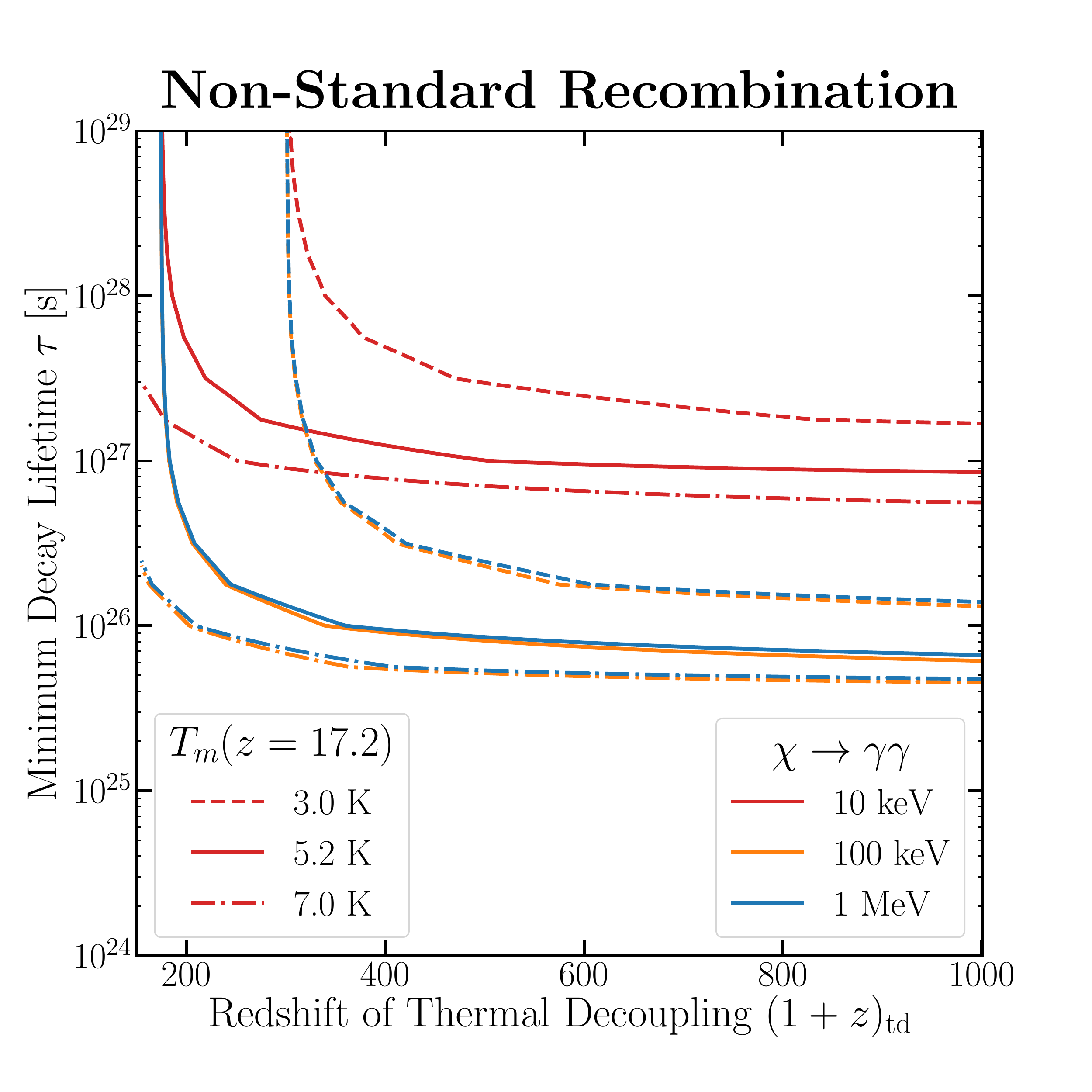}
    }
  \caption{As Fig.~\ref{fig:recomb_decay}, but extended to lower masses.}
  \label{fig:recomb_decay_low}
\end{figure}

In Figs.~\ref{fig:recomb_decay_low} and~\ref{fig:recomb_swave_low}, we show constraint plots for DM masses below 100 MeV in the presence of non-standard recombination, for $s$-wave annihilation and decay respectively. These analyses are otherwise performed as discussed in the main text.

\begin{figure}
    \centering
    \subfigure{
        \label{fig:recomb_elec_swave_low}
        \includegraphics[scale=0.34]{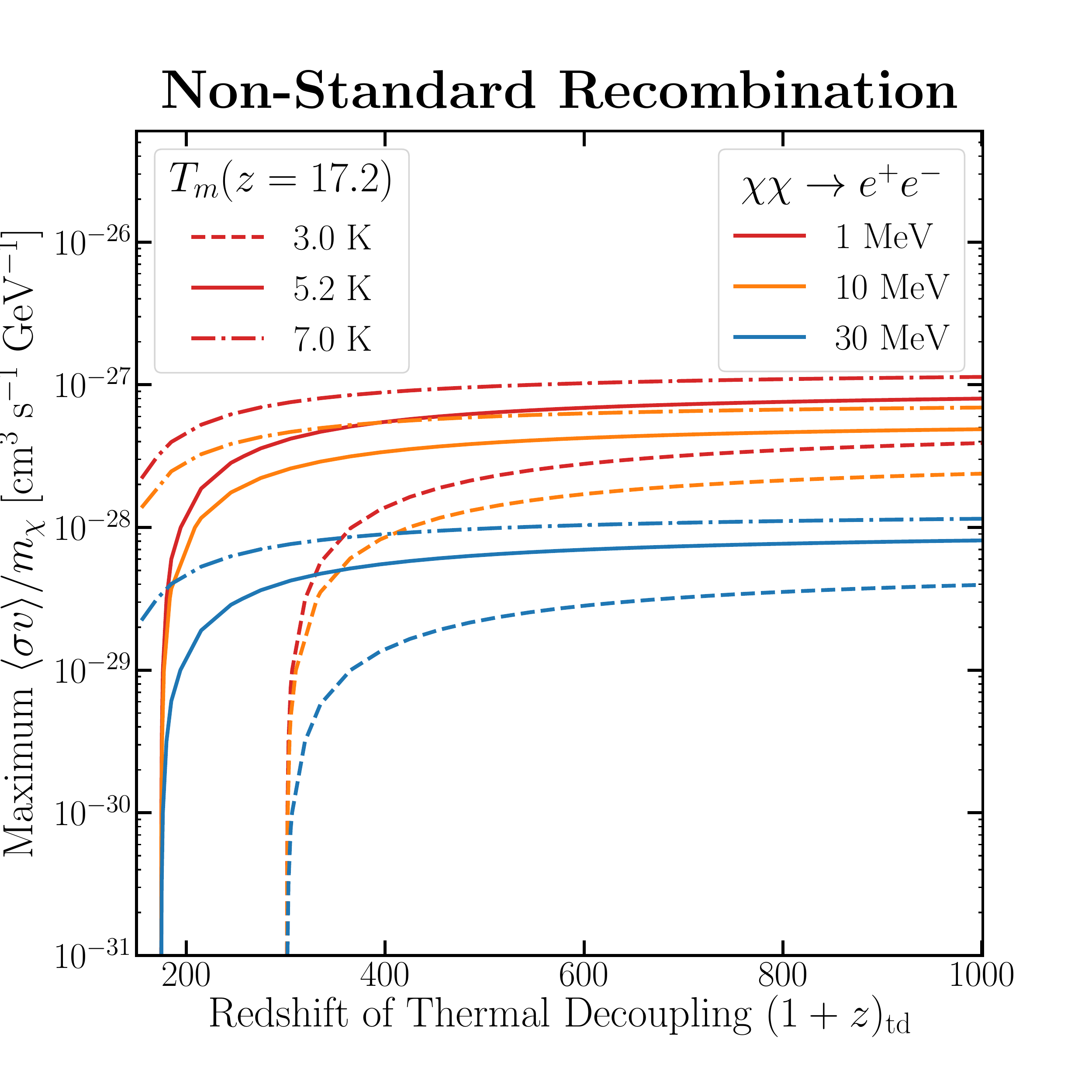}
    }
    \subfigure{
        \label{fig:recomb_phot_swave_low}
        \includegraphics[scale=0.34]{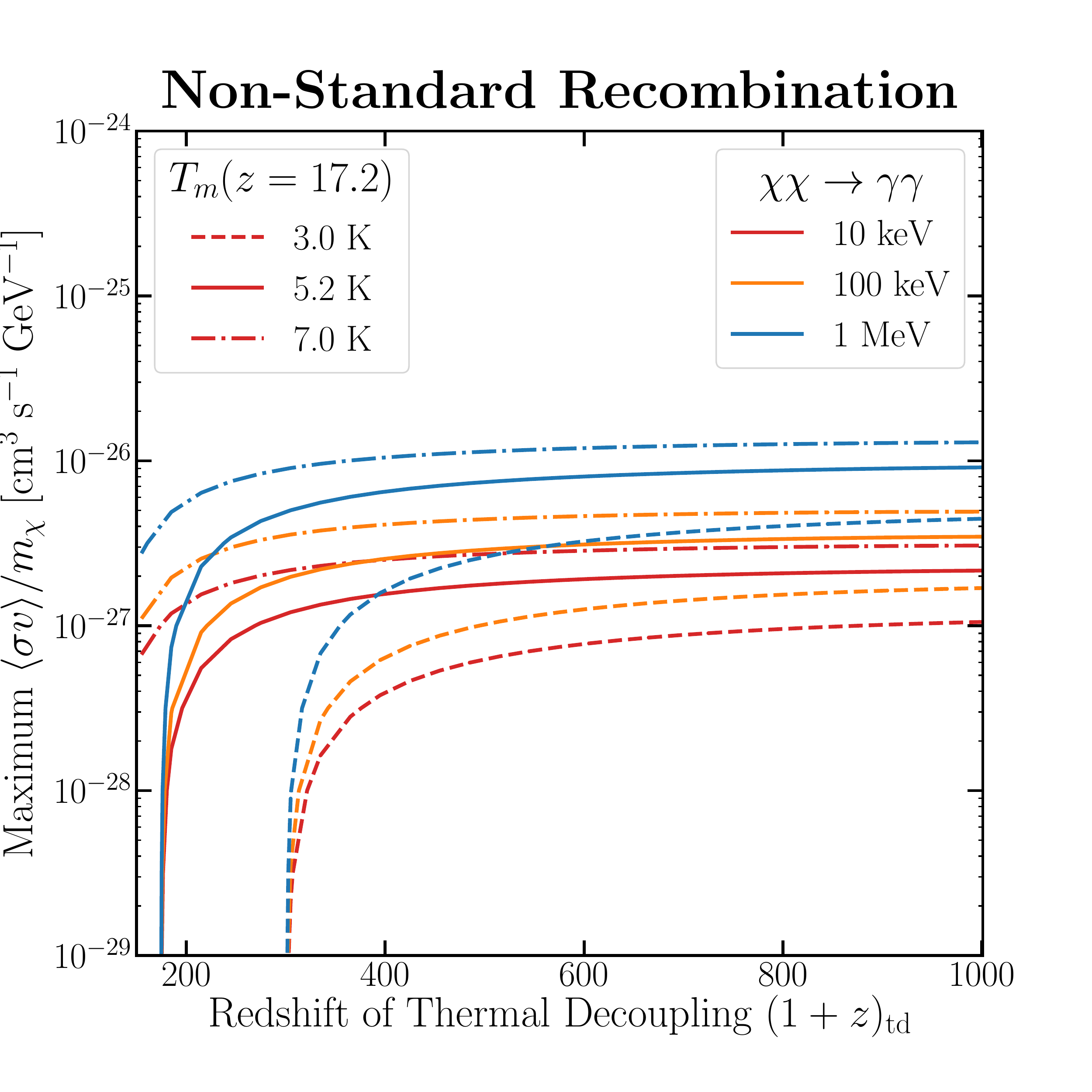}
    }
  \caption{As Fig.~\ref{fig:recomb_swave}, but extended to lower masses.}
  \label{fig:recomb_swave_low}
\end{figure}

\begin{figure}
    \centering
    \subfigure{
        \label{fig:cooling_elec_decay_f_1}
        \includegraphics[scale=0.34]{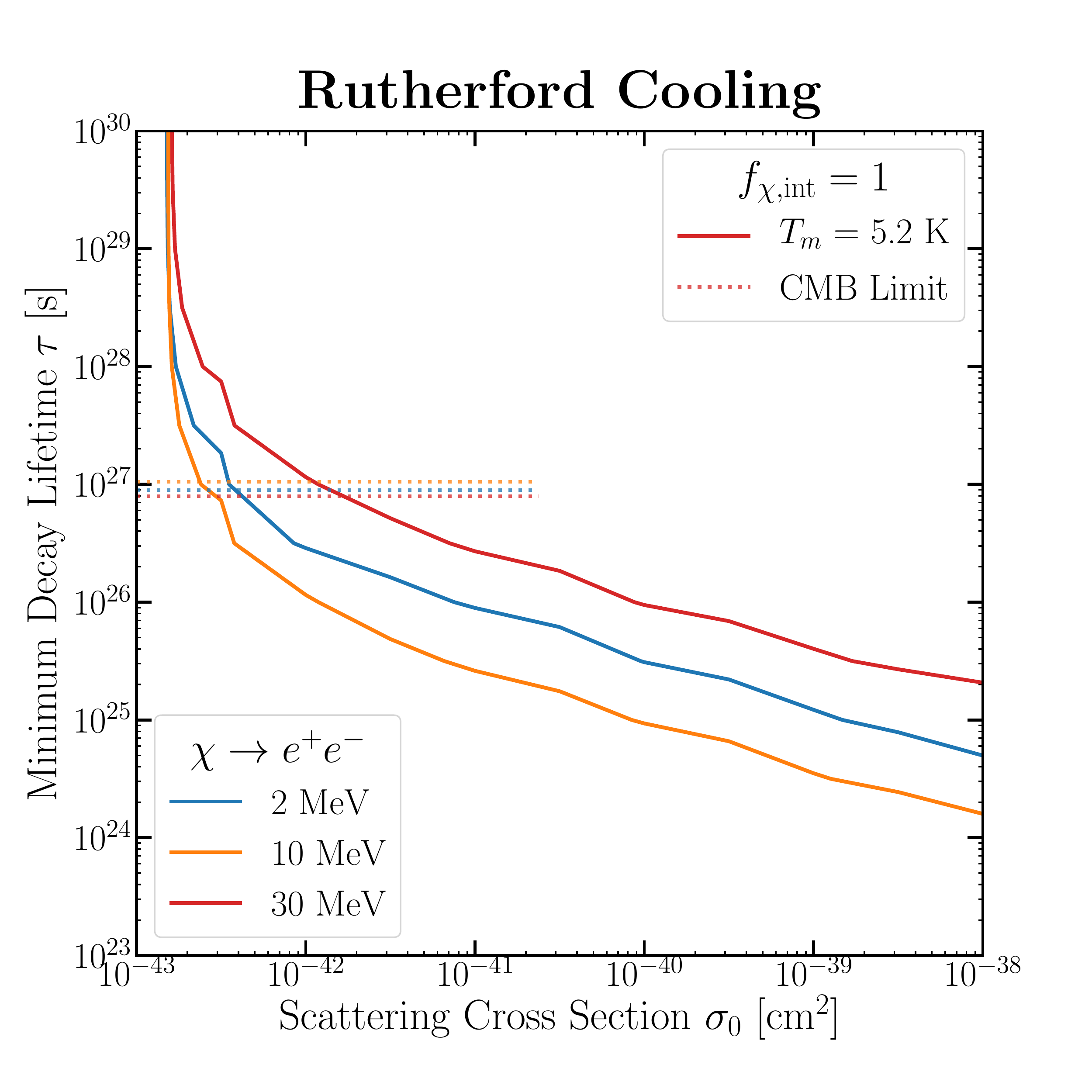}
    }
    \subfigure{
        \label{fig:cooling_phot_decay_f_1}
        \includegraphics[scale=0.34]{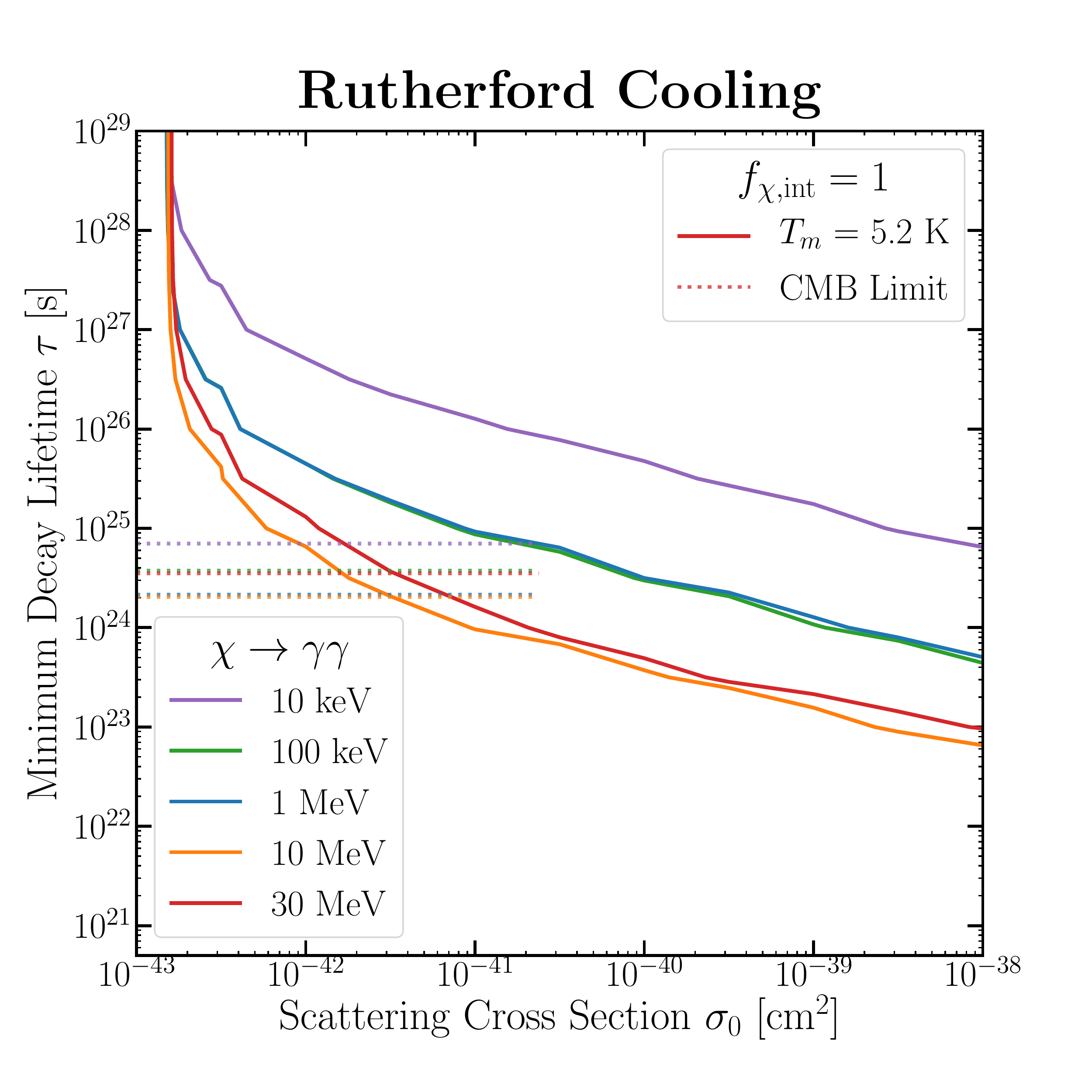}
    }
  \caption{Rutherford cooling constraints on the minimum decay lifetime for $\chi \to e^+e^-$ (left) and $\chi \to \gamma \gamma$ (right) from the matter temperature $T_m(z = 17.2) = $5.2 K (solid), $f_{\chi,\text{int}} = 1$. Limits from the Planck measurement of the CMB power spectrum are also shown up to $\sigma_0 = \sigma_{0,\text{td}}(z = 300)$ (dotted). 
  }
  \label{fig:cooling_decay_f_1}
\end{figure}

\begin{figure}
    \centering
    \subfigure{
        \label{fig:cooling_elec_swave_f_1}
        \includegraphics[scale=0.34]{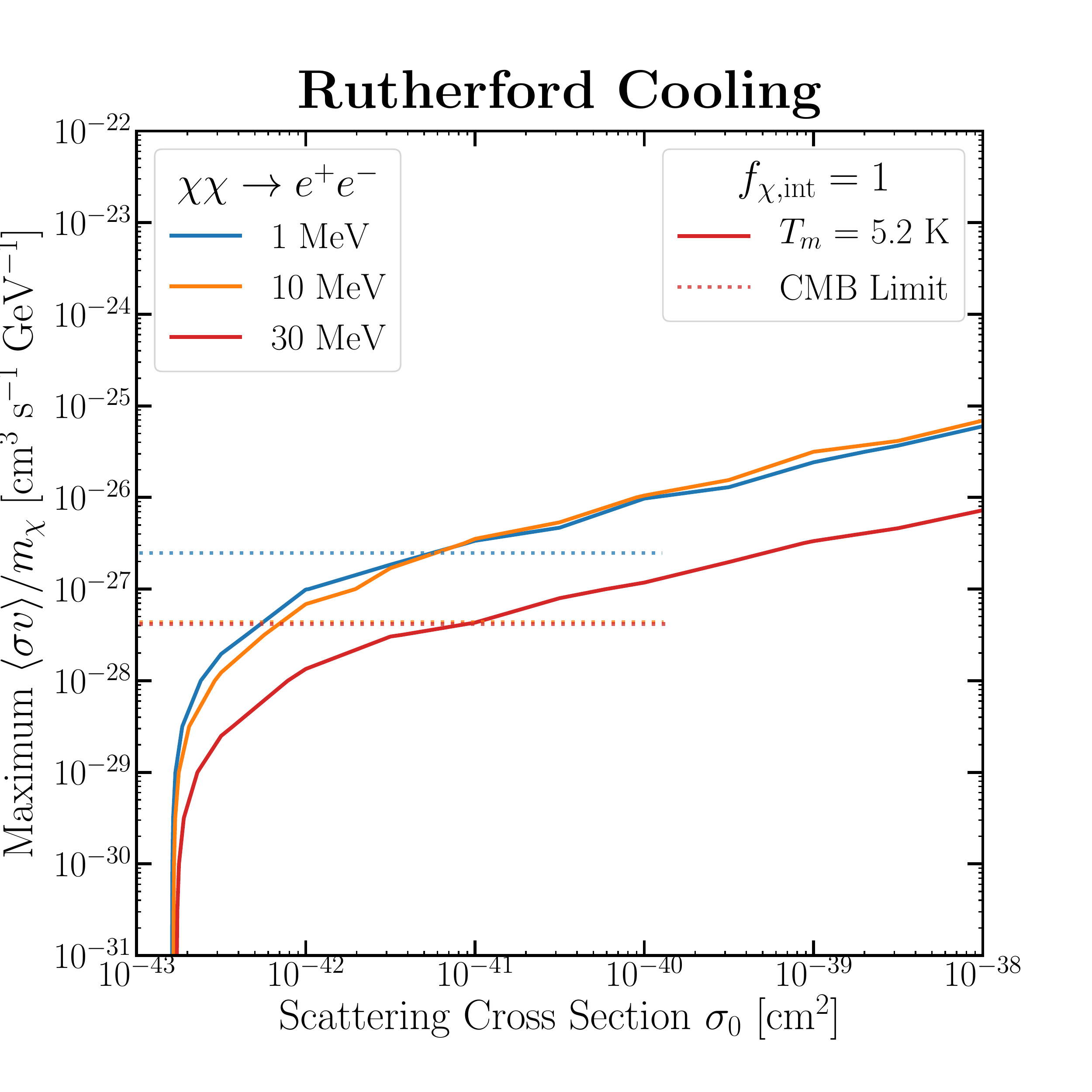}
    }
    \subfigure{
        \label{fig:cooling_phot_swave_f_1}
        \includegraphics[scale=0.34]{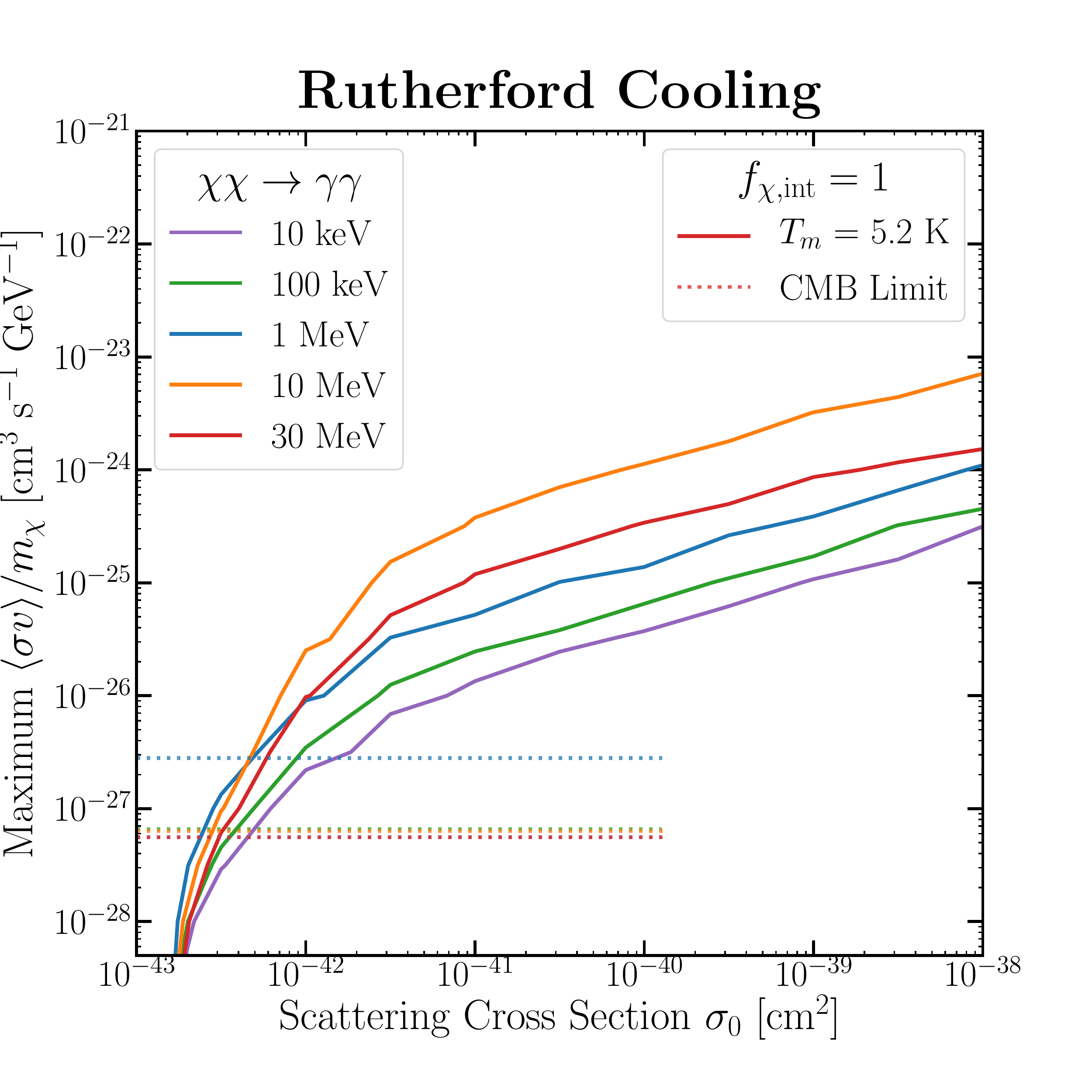}
    }
  \caption{Rutherford cooling $s$-wave annihilation constraints for $\chi \chi \to e^+e^-$ (left) and $\chi \chi \to \gamma \gamma$ (right) from the matter temperature $T_m(z = 17.2) = $ 5.2 K (solid), $f_{\chi,\text{int}} = 1$. Limits from the Planck measurement of the CMB power spectrum are also shown up to $\sigma_0 = \sigma_{0,\text{td}}(z = 600)$ (dotted). 
  }
  \label{fig:cooling_swave_f_1}
\end{figure}

Finally, Figs.~\ref{fig:cooling_decay_f_1} and~\ref{fig:cooling_swave_f_1} show the limits on the minimum decay lifetime and maximum annihilation cross section for Rutherford cooling with $f_{\chi,\text{int}} = 1$. Values of $\sigma_0$ exceeding $\sim 10^{-42} \text{ cm}^2$ affect the CMB power spectrum significantly and are ruled out by Planck \cite{Slatyer:2018aqg}; for models that are consistent with this limit, the value of $T_m$ at $z \sim 20$ is a more powerful constraint on additional energy injection on models than the high-redshift CMB limits on annihilation and decay. 

\chapter{\texttt{DarkHistory}}
\label{chap:app_DarkHistory}

\section{Inverse Compton Scattering}
\label{app:ICS}

In this appendix, we discuss in detail the methods used to compute the spectra of photons that are produced by the cooling of electrons through ICS. We restore $\hbar$, $c$ and $k_B$ in this appendix, since the exact numerical value of these spectra is important. 

\subsection{Scattered Spectra}

We begin with some preliminaries that will be important in understanding our subsequent discussion of ICS. The goal is to determine the secondary photon spectrum produced on average by multiple scatterings of a single electron.

Consider an electron with energy $E_e$ and corresponding Lorentz factor $\gamma$ incident on some distribution of photons $n(\epsilon)$ with initial energy $\epsilon$ in the comoving frame. Since we are only interested in ICS off the CMB, we will only consider an isotropic photon bath in the co-moving frame, distributed as a blackbody with some temperature $T$. The electron has some probability per unit time of scattering the photons into some outgoing energy $\epsilon_1$, with some probability distribution $dN_\gamma/(d\epsilon \, d\epsilon_1 \, dt)$, which we call the ``differential scattered photon spectrum''. This quantity is proportional to the number density per unit energy of the photon bath $n(\epsilon)$, so that integrating over $\epsilon$ also integrates over the distribution of these photons. This can be interpreted as a normalized scattered photon spectrum for ICS by many electrons with the same energy. Integrating the differential scattered photon spectrum with respect to $\epsilon$ gives us the ``scattered photon spectrum'', 
\begin{alignat}{1}
    \frac{dN_\gamma}{d \epsilon_1 \, dt} (E_e, T, \epsilon_1) = \int_{\epsilon_\text{min}}^{\epsilon_{\text{max}}} d\epsilon \frac{dN_\gamma}{d\epsilon \, d\epsilon_1 \, dt} (E_e, T, \epsilon, \epsilon_1) \,,
    \label{eqn:ics_scat_phot_spec}
\end{alignat}
with $\epsilon_\text{min}$ and $\epsilon_\text{max}$ determined by the kinematics of ICS.

We further define the ``scattered photon energy loss spectrum'',
\begin{alignat}{1}
    \frac{dN_\gamma}{d\Delta \, dt} (E_e, T, \Delta) &= \int d\epsilon \frac{dN_\gamma}{d\epsilon \, d\epsilon_1 \, dt} (E_e, T, \epsilon, \epsilon_1 = \epsilon + \Delta),
  \label{eqn:scattered_phot_engloss_spec}
\end{alignat}
where $\Delta$ is the change in energy of a photon scattering by a single electron. This is simply the distribution of scattered photons as a function of the energy gained or lost by the photon during the scattering. 

Now, consider some arbitrary injection spectrum of electrons $d \tilde{N}_e/dE_1$. The tilde serves to remind the reader that this is a distribution of electrons, and not a normalized quantity. From the definition of Eq.~(\ref{eqn:scattered_phot_engloss_spec}), we define the ``scattered electron spectrum'' as
\begin{alignat}{1}
  \frac{d \tilde{N}_e}{dE_1\, dt} = \int_0^\infty dE \, \frac{d \tilde{N}_e}{dE} \frac{dN_\gamma}{d \Delta \, dt} (E, T, \Delta = E - E_1) \,,
  \label{eqn:scattered_elec_spec_exact}
\end{alignat}
where $E_1$ is the energy of the scattered electron. However, this result allows some electrons to gain energy after scattering, significantly complicating our calculations. Intuitively, we expect electrons that upscatter from $E \to E_1$ to partially cancel with downscatters from $E_1 - E$, justifying an approximate treatment where we simply cancel out photons that downscatter (and upscatters an electron) with photons that upscatter (and downscatters an electron). We leave a full justification of this to the end of this section, but for now, we will accordingly define the ``scattered electron net energy loss spectrum'',
\begin{alignat}{1}
    \frac{dN_e}{d \Delta \, dt} (\beta, T, \Delta) = \frac{dN_\gamma}{d\Delta\, dt}(\beta, T, \Delta) - \frac{dN_\gamma}{d\Delta \, dt}(\beta, T, -\Delta),
    \label{eqn:electron_eng_loss_spec}
\end{alignat}
with $\Delta \geq 0$ in the expression above. For relativistic electrons, the average energy lost due to an upscattering a photon is much larger than the average energy gained due to downscattering a photon, and it is therefore a good approximation to consider only scattering events where electrons lose their energy~\cite{Blumenthal:1970gc}. The upscattered photons also have outgoing energy $\epsilon_1 \gg \epsilon$, and so a reasonable approximation to make in the relativistic limit is
\begin{alignat}{1}
    \left. \frac{dN_e}{d\Delta \, dt} \right|_{\beta \to 1} \approx \left. \frac{dN_\gamma}{d\epsilon_1 \, dt}\right|_{\beta \to 1} \, .
    \label{eqn:engloss_approx}
\end{alignat}

We now turn our attention to justifying the approximation laid out in Eq.~(\ref{eqn:electron_eng_loss_spec}). First, we split the exact integral in Eq.~(\ref{eqn:scattered_elec_spec_exact}) into an integral from 0 to $E_1$, and from $E_1$ to $\infty$. The first integral can be rewritten as (dropping the $T$ dependence for clarity)
 \begin{multline}
  \int_0^{E_1} dE \, \frac{d \tilde{N}_e}{dE} \frac{dN_\gamma}{d \Delta \, dt} (E, \Delta = E - E_1) \\
  = - \int_{E_1}^{2E_1} dx \frac{d \tilde{N}}{dx} \frac{dN_\gamma}{d\Delta \, dt} (E = 2E_1 - x, \Delta = E_1 - x) \,,
 \end{multline}
where we have simply made the substitution $x = 2E_1 - E$. In this part of the integral, we are dealing with upscattered electrons and downscattered photons, and so we know that $dN_\gamma / (d \Delta \, dt)$ only has support when $E - E_1 \sim T_\text{CMB} \ll E, E_1$, since ICS is only included for electrons with $E > \SI{3}{\kilo\eV}$~\cite{Slatyer:2015kla}. This implies that the integral only has support near $x = E_1$, and we can therefore make the following approximation:  
 \begin{multline}
  \int_0^{E_1} dE \, \frac{d \tilde{N}_e}{dE} \frac{dN_\gamma}{d \Delta \, dt} (E, \Delta = E - E_1) \\
  \approx - \int_{E_1}^{\infty} \, dx \frac{d \tilde{N}_e}{dx} \frac{dN_\gamma}{d\Delta \, dt} (E = x, \Delta = E_1 - x) \\
  = - \int_{E_1}^\infty dE \, \frac{d\tilde{N}_e}{dE} \frac{dN_\gamma}{d \Delta \, dt} (E, \Delta = E_1 - E) \,,
 \end{multline}
where in the last step we have trivially relabeled $x \to E$. We have therefore shown that
\begin{alignat}{2}
  \frac{d \tilde{N}_e'}{dE_1 \, dt} &\approx&& -\int_{E_1}^\infty dE \, \frac{d \tilde{N}_e}{dE} \frac{dN_\gamma}{d \Delta \, dt} (E, \Delta = E_1 - E) \nonumber \\
  & && \,\, + \int_{E_1}^\infty dE \, \frac{d\tilde{N}_e}{dE} \frac{dN_\gamma}{d \Delta \, dt} (E, \Delta = E - E_1) \,,
\end{alignat}
and that is a good approximation due to the relatively low temperature of the CMB.

With these definitions in mind, we are now ready to understand how to compute these scattered spectra when the electron is in two limits. For $\gamma > 20$, the spectra are computed in the relativistic limit, while below that, scattering with the CMB at all relevant redshifts lie well within the Thomson regime. Together, they cover all relevant kinematic regimes that we consider in our code.

\subsubsection{Relativistic Electrons}

The differential upscattered photon spectrum produced by ICS between an electron and the CMB blackbody spectrum in the relativistic regime ($\gamma \gg 1$) is given by~\cite{Blumenthal:1970gc}
\begin{alignat}{2}
    \frac{dN_\gamma}{d\epsilon \, d\epsilon_1 \, dt} &=&& \frac{2\pi r_0^2 c}{\gamma^2} \frac{n(\epsilon, T)}{\epsilon} \left[ 2q \log q + (1 + 2q)(1 - q) + \frac{1}{2} \frac{(\Gamma(\epsilon) q)^2}{1 + \Gamma(\epsilon) q}(1 - q) \right],
    \label{eqn:ICS_phot_spec_rel}
\end{alignat}
 where $r_0$ is the classical electron radius, $m_e$ is the electron mass, $\epsilon$ is the incident photon energy in the comoving frame, and $\epsilon_1$ is the scattered photon energy in the same frame, and we have defined
\begin{alignat}{1}
    \Gamma(\epsilon) = \frac{4 \epsilon \gamma}{m_e c^2}\, , \quad q = \frac{\epsilon_1}{\gamma m_e c^2 - \epsilon_1} \frac{1}{\Gamma(\epsilon)} \, .
\end{alignat}
We stress that Eq.~(\ref{eqn:ICS_phot_spec_rel}) is strictly only correct when photons are upscattered by the incoming electron, which corresponds to the kinematic regime $\epsilon \leq \epsilon_1 \leq 4 \epsilon \gamma^2/(1 + 4 \epsilon \gamma/m)$. In the opposite regime where $\epsilon/(4\gamma^2) \leq \epsilon_1 < \epsilon$ and photons get downscattered, we have~\cite{Jones:1968zza}
 \begin{alignat}{1}
   \frac{dN_\gamma}{d\epsilon \, d\epsilon_1 \, dt} = \frac{\pi r_0^2 c}{2 \gamma^4 \epsilon} \left(\frac{4\gamma^2 \epsilon_1}{\epsilon} - 1 \right) n(\epsilon, T) \,.
   \label{eqn:Jones_correction}
 \end{alignat}
For ICS off CMB photons, the $n(\epsilon)$ is the number density of photons per unit energy; for a blackbody, this is
\begin{alignat}{1}
    n_\text{BB}(\epsilon, T) = \frac{1}{\pi^2 \hbar^3 c^3} \frac{\epsilon^2}{\exp(\epsilon/k_B T) - 1} \, ,
\end{alignat}
where $T$ is the temperature of the CMB. 

The complete upscattered photon spectrum for ICS off the CMB is therefore obtained by performing the integral in Eq.~(\ref{eqn:ICS_phot_spec_rel}) over $\epsilon$, with the kinematic limits given by $1/4\gamma^2 \leq q \leq 1$~\cite{Blumenthal:1970gc}. Since the CMB photons at $z \lesssim 3000$ have energies less than \SI{1}{\eV}, the amount of energy transferred by an electron is always completely dominated by Eq.~(\ref{eqn:ICS_phot_spec_rel}). Furthermore, one can check that at $q = 1/4\gamma^2$, $\epsilon \gg T$. We can therefore make the approximation that Eq.~(\ref{eqn:ICS_phot_spec_rel}) gives the full ICS spectrum while neglecting Eq.~(\ref{eqn:Jones_correction}), and take the integral limits to be $0 \leq q \leq 1$ instead. This assumption is made in the ICS transfer functions provided as part of the downloaded data, but options are available in the \texttt{ics} module to turn these various assumptions off.

The quantity $\Gamma(\epsilon)$ separates the two kinematic regimes of Compton scattering: $\Gamma \gg 1$ for the Klein-Nishina regime, where Compton scattering in the electron rest frame is highly inelastic, and $\Gamma \ll 1$ for the Thomson regime, where it is almost elastic instead.\footnote{Although the scattering process is almost elastic in the initial electron rest frame, it is certainly not elastic in the co-moving frame. In the co-moving frame, the electron loses a small fraction of its energy per collision, but each collision can upscatter a CMB photon by a significant factor.} Eq.~(\ref{eqn:ICS_phot_spec_rel}) applies to both regimes, with the only assumption being $\gamma \gg 1$.

To avoid computing the scattered photon spectrum repeatedly in the code, we use the following relation between spectra at different temperatures:
\begin{alignat}{1}
    \frac{dN_\gamma}{d\epsilon_1 \, dt} (E_e, yT, \epsilon_1) = y^4 \frac{dN_\gamma}{d\epsilon_1 \, dt}(yE_e, T, y\epsilon_1) \,,
    \label{eqn:rel_temp_relation}
\end{alignat}
for any real positive number $y$, even if $yE_e$ is unphysical.\footnote{This trick can only be performed by integrating over $0 \leq q \leq 1$, and is the key reason for making such an approximation.} In \texttt{DarkHistory}, we evaluate the scattered photon spectrum at $1+z = 400$, and use this relation to compute the subsequent spectra at lower redshifts by a straightforward interpolation.

\subsubsection{Thomson Regime}

In the Thomson regime, the rate at which photons are scattered is given by~\cite{Blumenthal:1970gc}
\begin{alignat}{1}
    \frac{dN_\gamma}{dt} = \sigma_T c N_\text{rad},
    \label{eqn:thomson_scattering_rate}
\end{alignat}
where $N_\text{rad}$ is the total number density of incident photons, with $\sigma_T = 8 \pi r_0^2/3$ being the Thomson cross section. Note that the scattering rate is independent on the incident photon energy. The energy loss rate of the electron is~\cite{Blumenthal:1970gc}
\begin{alignat}{1}
    \frac{dE_e}{dt} = \frac{4}{3} \sigma_T c \gamma^2 \beta^2 U_\text{rad},
    \label{eqn:thomson_energy_loss_rate}
\end{alignat}
where $\beta$ is the velocity of the electron, with $U_\text{rad}$ being the total energy density of the incident photons.

While Eqs.~(\ref{eqn:thomson_scattering_rate}) and~(\ref{eqn:thomson_energy_loss_rate}) are well-known, the actual spectrum of scattered photons in the Thomson regime is much less so. The complete expression for the differential scattered photon spectrum with no further assumptions is, as far as the authors know, first given in Ref.~\cite{Fargion:1996xr}, and we reproduce their final result here for completeness. For $(1-\beta)\epsilon_1/(1+\beta) < \epsilon < \epsilon_1$, we have
\begin{alignat}{2}
    \left. \frac{dN_\gamma}{d \epsilon \,d\epsilon_1\, dt}  (\beta, T, \epsilon, \epsilon_1) \right|_{\epsilon < \epsilon_1} &=&& \frac{\pi r_0^2 c n(\epsilon, T)}{4 \beta^6 \gamma^2 \epsilon} \Bigg\{ \frac{1}{\gamma^4} \frac{\epsilon}{\epsilon_1} - \frac{1}{\gamma^4} \frac{\epsilon_1^2}{\epsilon^2} \nonumber \\
    & &&+ (1 + \beta) \left[\beta(\beta^2 + 3) + \frac{1}{\gamma^2}(9 - 4\beta^2) \right] \nonumber \\
    & &&+ (1-\beta) \left[\beta(\beta^2 + 3) - \frac{1}{\gamma^2}(9 - 4\beta^2) \right] \frac{\epsilon_1}{\epsilon} \nonumber \\
    & &&- \frac{2}{\gamma^2}(3 - \beta^2) \left(1 + \frac{\epsilon_1}{\epsilon}\right) \log \left(\frac{1+\beta}{1-\beta} \frac{\epsilon}{\epsilon_1} \right) \Bigg\},
    \label{eqn:thomson_scattered_diff_phot_spec}
\end{alignat}
and for $\epsilon_1 < \epsilon < (1+\beta) \epsilon_1 / (1-\beta)$, 
\begin{alignat}{1}
    \left. \frac{dN_\gamma}{d\epsilon \, d\epsilon_1 \, dt} (\beta, T, \epsilon, \epsilon_1) \right|_{\epsilon \geq \epsilon_1} = -\left. \frac{dN_\gamma}{d\epsilon \, d\epsilon_1 \, dt} (-\beta, T, \epsilon, \epsilon_1) \right|_{\epsilon < \epsilon_1} \!\!\!\!\!.
    \label{eqn:thomson_spectrum_upp_low_relation}
\end{alignat}
All other values of $\epsilon$ outside of the ranges specified are kinematically forbidden, and so to find the spectrum, we need to integrate over $\epsilon$ with $n(\epsilon) = n_\text{BB}(\epsilon)$ in the finite range specified above, i.e.\
\begin{alignat}{2}
    \frac{dN_\gamma}{d \epsilon_1 \, dt}(\beta, T, \epsilon_1) &=&& \int_{\frac{1-\beta}{1+\beta}\epsilon_1} ^{\epsilon_1} d\epsilon \left. \frac{dN_\gamma}{d\epsilon \, d\epsilon_1 \, dt} (\beta, T, \epsilon, \epsilon_1) \right|_{\epsilon < \epsilon_1} \nonumber \\
    & &&- \int_{\epsilon_1}^{\frac{1+\beta}{1-\beta} \epsilon_1} d\epsilon \left. \frac{dN_\gamma}{d\epsilon \, d\epsilon_1 \, dt} (-\beta, T, \epsilon, \epsilon_1) \right|_{\epsilon < \epsilon_1} \!\!\!\!\!.
    \label{eqn:thomson_scattered_phot_spec}
\end{alignat}
The relationship between spectra at different temperatures is given by
\begin{alignat}{1}
    \frac{dN_\gamma}{d\epsilon_1 \, dt}(\beta, yT, \epsilon_1) = y^2 \frac{dN_\gamma}{d\epsilon_1 \, dt}(\beta, T, \epsilon_1/y) \,.
    \label{eqn:thomson_temp_relation}
\end{alignat}

The scattered photon energy loss spectrum $dN_\gamma/(d\Delta \, dt)$ is similarly given by
 \begin{alignat}{1}
    \frac{dN_\gamma}{d\Delta dt}(\beta, T, \Delta) = \begin{cases} 
        \left. \int_{\frac{1-\beta}{2\beta}\Delta}^\infty d \epsilon \, \frac{dN_\gamma}{d \epsilon \, d\epsilon_1 \, dt}(\beta, T, \epsilon, \epsilon + \Delta) \right|_{\epsilon < \epsilon_1}, & \Delta > 0, \\
        \left. \int_{-\frac{1+\beta}{2\beta}\Delta}^\infty d \epsilon \, \frac{dN_\gamma}{d \epsilon \, d\epsilon_1 \, dt}(\beta, T, \epsilon, \epsilon + \Delta) \right|_{\epsilon \geq \epsilon_1}, & \Delta \leq 0.
    \end{cases}
 \end{alignat}
The relation shown in Eq.~(\ref{eqn:thomson_temp_relation}) between scattered photon spectra of different temperatures also holds for the energy loss spectrum, with $\epsilon_1 \to \Delta$. 

\subsection{Numerical Methods}

Computationally, to evaluate all of the scattered spectra, we need to perform numerical quadrature over a large range of electron and scattered photon energies; using a standard grid of $5000 \times 5000$ energy values, the grid would take the standard \texttt{numpy} integrator over a day to populate. While a substantial speed-up may be obtained by using packages like \texttt{Cython}~\cite{behnel2010cython}, numerical quadrature for ICS in the Thomson regime is also subject to significant numerical errors when the electron is nonrelativistic due to the existence of catastrophic cancellations. A semi-analytic approach provides both a faster method and a way to avoid such errors in a robust manner. 

\subsubsection{Thomson and Relativistic Regime: Large \texorpdfstring{$\beta$}{beta}}

For $\beta \gtrsim 0.1$, we can obtain the scattered photon spectrum in Eq.~(\ref{eqn:thomson_scattered_diff_phot_spec}) in the Thomson regime or Eq.~(\ref{eqn:ICS_phot_spec_rel}) in the relativistic regime, as well as the scattered electron energy loss spectrum in the Thomson regime in Eq.~(\ref{eqn:electron_eng_loss_spec}), by direct integration.

The problem of integrating these expressions reduces to obtaining an expression for indefinite integrals over the Bose-Einstein distribution of the form
\begin{alignat}{1}
    P_{f} (y) \equiv \int \frac{f(y)\, dy}{e^y - 1} \, .
    \label{eqn:planck_integral}
\end{alignat}
Throughout this appendix, we ignore the constant of integration for such indefinite integrals, since we will ultimately be taking differences of such expressions to find definite integrals. For $f(y) \equiv y^n$ with integer $n \geq 0$, the indefinite integral is known explicitly:
\begin{alignat}{1}
    P_{y^n} (x) = -n! \sum_{s=0}^n \frac{x^s}{s!} \text{Li}_{n-s+1}(e^{-x}) \quad (n = 0, 1, 2, \cdots),
\end{alignat}
where $\text{Li}_{m}(z)$ is the polylogarithm function of order $m$ with argument $z$ (see Appendix~\ref{app:ICS_integrals_series} for the definition). Note however that NumPy does not have a numerical function for the polylogarithm of order $m > 2$, and so the semi-analytic method that we describe below is still necessary for $P_{y^n}$, $n \geq 2$ due to this limitation. 

For other functions $f(y)$, closed-form solutions do not exist. However, an expression for the indefinite integral as an infinite series can be obtained~\cite{Zdziarski:2013gza}. Importantly, more than one series expression exists for all of the integrals $P_{f}(x)$ of interest in both the relativistic and nonrelativistic regimes, so that it is always possible to find a series expression that converges quickly for any integration limit. We tabulate the series expressions already found in Ref.~\cite{Zdziarski:2013gza} for completeness, together with the many new series expressions derived in this paper required for the nonrelativistic limit in Appendix~\ref{app:ICS_integrals_series}. 

\subsubsection{Thomson Regime: Small \texorpdfstring{$\beta$}{beta}}

In the Thomson regime for $\beta \lesssim 0.1$, catastrophic cancellations between terms in the integral make even the method described above insufficient. After integrating Eq.~(\ref{eqn:thomson_scattered_phot_spec}) over $\epsilon$ to get the scattered photon spectrum, for example, the final result must be $\mathcal{O}(\beta^0)$, even though the prefactor in Eq.~(\ref{eqn:thomson_scattered_diff_phot_spec}) is $\mathcal{O}(\beta^{-6})$. The integrals of all of the terms in the curly braces of Eq.~(\ref{eqn:thomson_scattered_diff_phot_spec}) and their analog from Eq.~(\ref{eqn:thomson_spectrum_upp_low_relation}) must therefore cancel among themselves to 1 part in $\beta^{-6}$; such a computation is impossible to perform for $\beta \lesssim 0.003$ due to floating point inaccuracy, even with double precision. 

We avoid this problem by expanding the scattered photon spectrum in Eq.~(\ref{eqn:thomson_scattered_phot_spec}) and the mean electron energy loss spectrum in Eq.~(\ref{eqn:electron_eng_loss_spec}). Eq.~(\ref{eqn:thomson_scattered_phot_spec}) can be expanded straightforwardly in $\beta$, but Eq.~(\ref{eqn:electron_eng_loss_spec}) must be expanded in both $\beta$ and $\xi \equiv \Delta/T$, since catastrophic cancellations occur when either variable is small. In \texttt{DarkHistory}, we expand these expressions up to $\mathcal{O}(\beta^6)$ and $\mathcal{O}(\xi^6)$, but the precision of this calculation is systematically improvable by adding more terms to the code as desired. The exact expressions for the expansions, details of their derivations and several consistency checks for these expressions can be found in Appendix~\ref{app:ICS_integrals_series}. 

\subsection{Results}

Figs.~\ref{fig:ics_thomson_scattered_phot_spec} and~\ref{fig:ics_rel_scattered_phot_spec} show the scattered photon spectrum in the Thomson and relativistic regimes respectively as a function of electron energy, at a CMB temperature of \SI{0.25}{\eV}, corresponding to a redshift of $z \approx 1065$ that is near recombination. By default, \texttt{DarkHistory} transitions between these two limits at $\gamma = 20$. Fig.~\ref{fig:ics_thomson_engloss_spec} shows the mean electron energy loss spectrum in the Thomson regime. Above $\gamma = 20$, \texttt{DarkHistory} uses the approximation shown in Eq.~(\ref{eqn:engloss_approx}). Finally, the computed secondary photon spectrum after completely cooling of all electrons and positrons through ICS is shown in Fig.~\ref{fig:ics_sec_phot_spec}. 

\begin{figure}[t]
    \centering
    \includegraphics[scale=0.55]{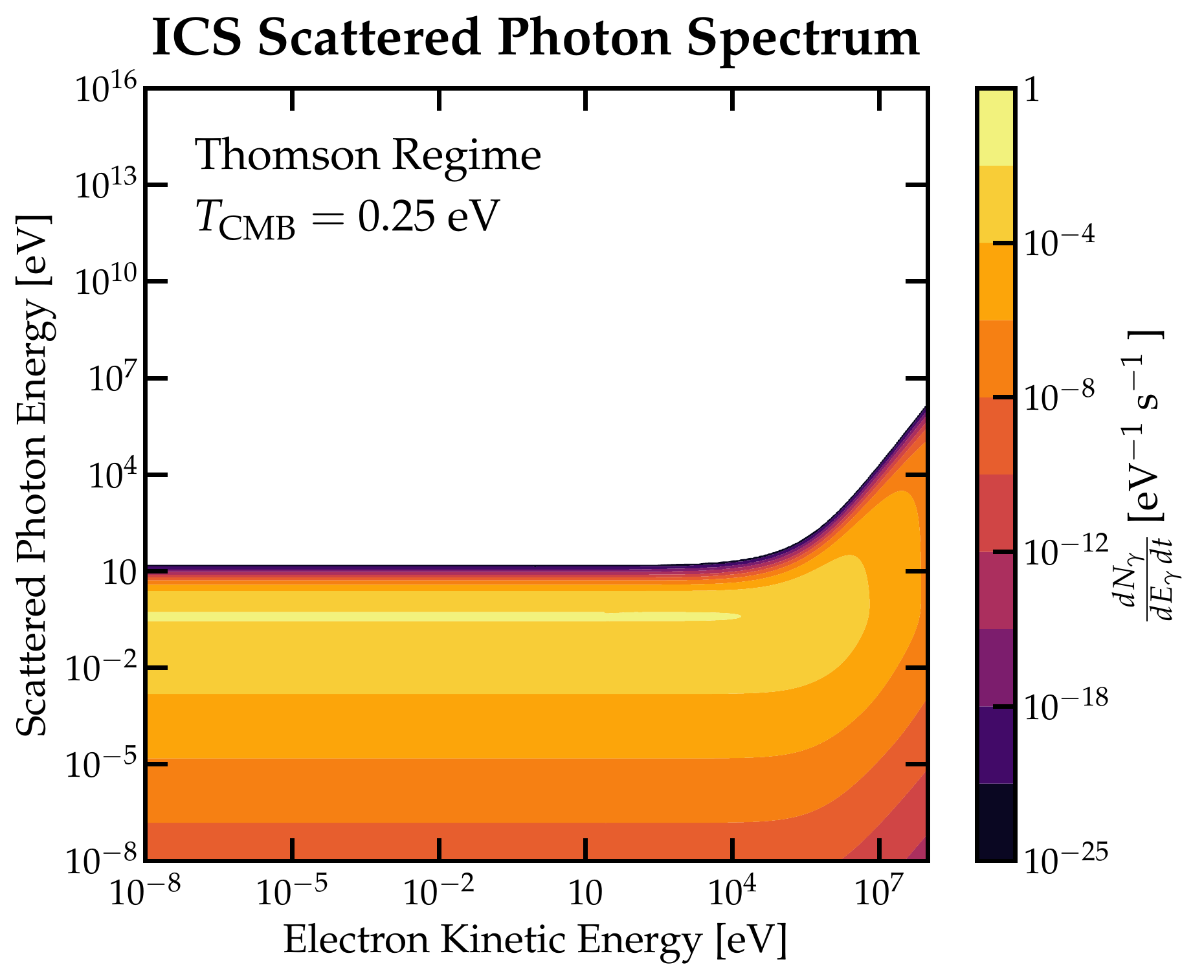}
    \caption{The ICS scattered photon spectrum in the Thomson regime, with $T_\text{CMB}$ = \SI{0.25}{\eV}. }
    \label{fig:ics_thomson_scattered_phot_spec}
\end{figure}

\begin{figure}[t]
    \centering
    \includegraphics[scale=0.55]{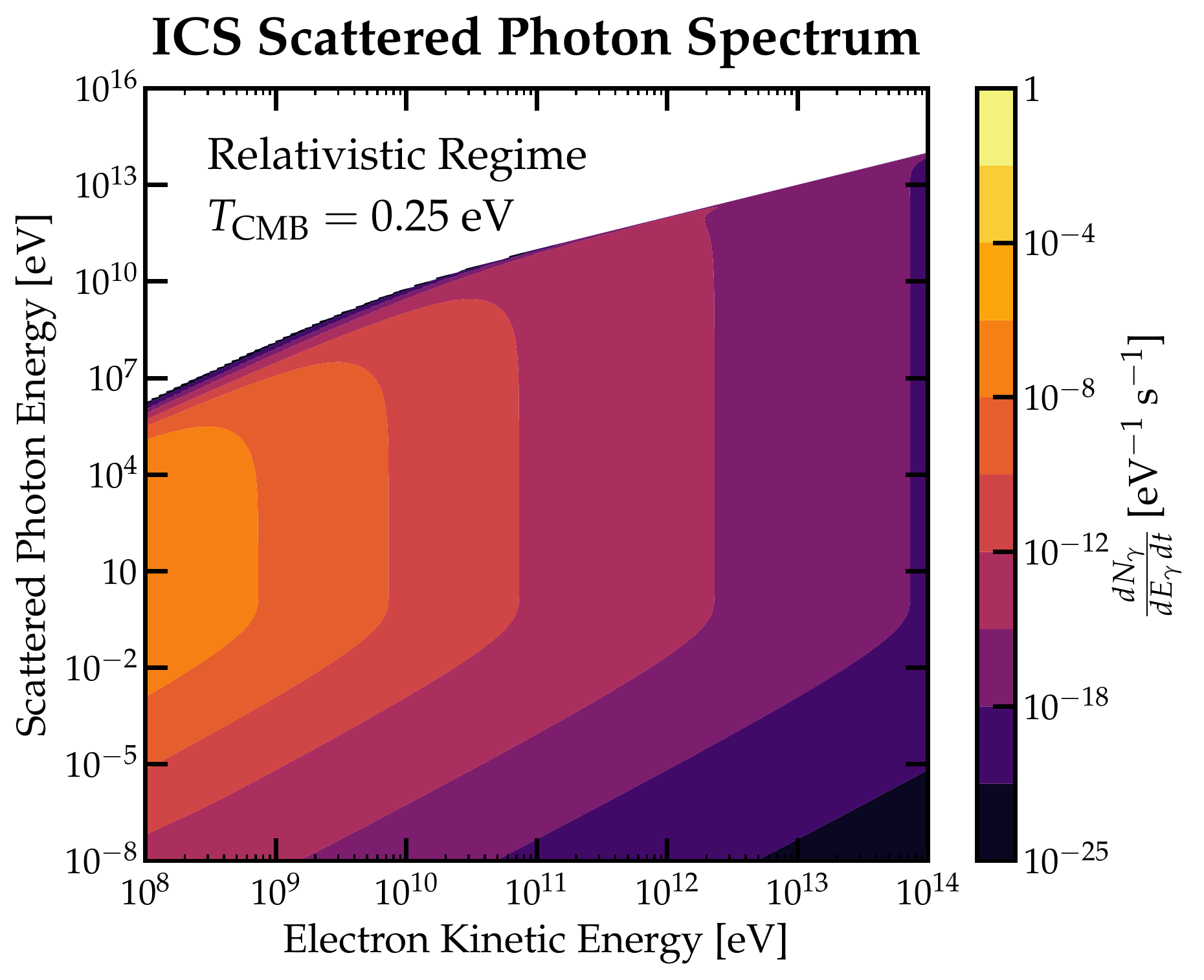}
    \caption{The ICS scattered photon spectrum in the relativistic regime, with $T_\text{CMB}$ = \SI{0.25}{\eV}. }
    \label{fig:ics_rel_scattered_phot_spec}
\end{figure}

All results shown here are computed using a $500 \times 500$ grid of electron and photon energies/energy loss, and each can be completed under ten seconds on a typical personal computer.

\begin{figure}[t]
    \centering
    \includegraphics[scale=0.55]{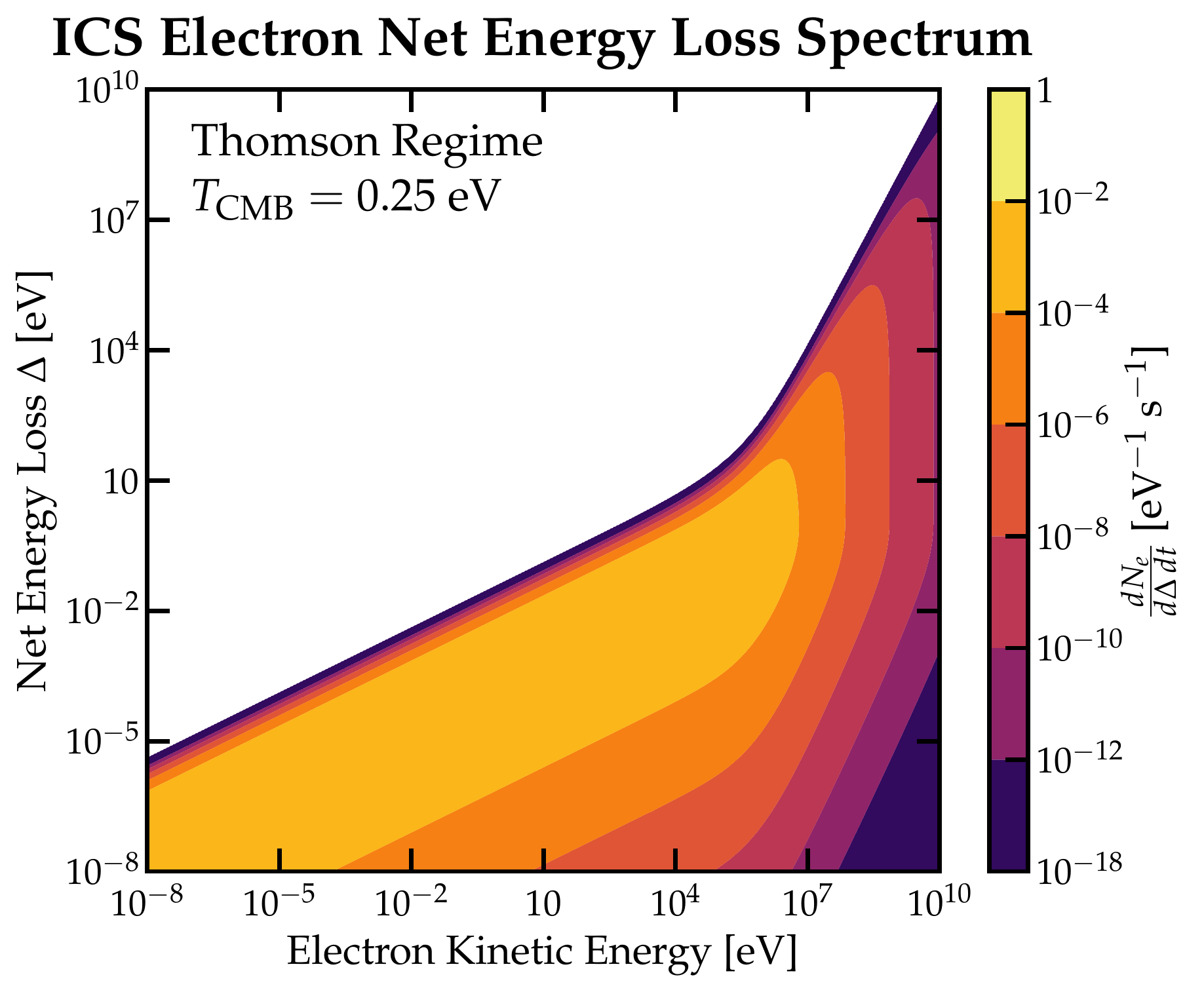}
    \caption{The ICS mean electron energy loss spectrum in the Thomson regime, with $T_\text{CMB}$ = \SI{0.25}{\eV}. }
    \label{fig:ics_thomson_engloss_spec}
\end{figure}

\begin{figure}[t]
   \centering
   \includegraphics[scale=0.55]{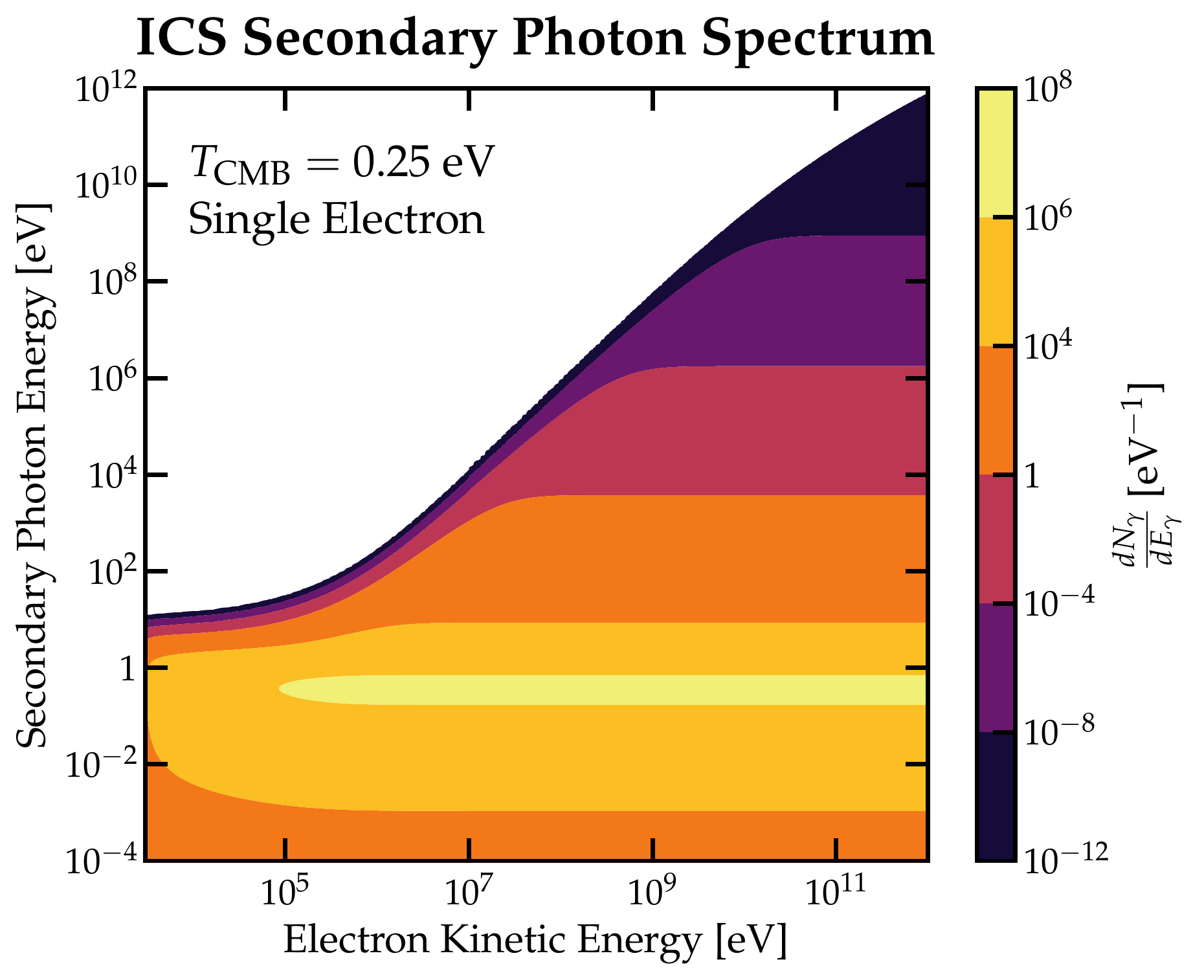}
   \caption{The ICS secondary photon spectrum after complete cooling of a single electron, with $T_\text{CMB}$ = \SI{0.25}{\eV}.}
   \label{fig:ics_sec_phot_spec}
\end{figure}

\subsection{Integrals and Series Expansions}
\label{app:ICS_integrals_series}

We are now ready to detail the integrals and series expansions used in the numerical methods described above.

\subsubsection{Bose Einstein Integrals}

Each $P_f(x)$ that is of interest has a series that converges quickly for small values of $x$, and another that converges quickly for large values of $x$. \texttt{DarkHistory} by default chooses $x = 2$ as the value to switch between the two expressions. 

Suppose we approximate the indefinite integral in Eq.~(\ref{eqn:planck_integral}) $P_f(x)$ by the first $N$ terms of its series expression, which we denote $S_N(x)$. Let $S_N^s(x)$ and $S_N^l(x)$ be the series expressions we obtain for $x < 2$ and $x \geq 2$ respectively. In all of the cases we are interested in, $S^l_{N \to \infty}(x \to \infty) = 0$ (with the constant of integration taken to be zero) due to the exponential function in the denominator of the original integral, and so 
\begin{alignat}{1}
	S^l_{N\to \infty}(b > 2) = -\int_b^\infty \frac{f(y) \, dy}{e^y - 1} \,.
\end{alignat}
Then defining $\Delta S_N^{s,l} (a, b) = S_N^{s,l}(b) - S_N^{s,l}(a)$, the definite integral is evaluated as
\begin{alignat}{1}
	\int_a^b \frac{f(y) \, dy}{e^y - 1} = \begin{cases}
		\Delta S_{N\to\infty}^s(a, b), & a < 2, b < 2; \\
		\Delta S_{N\to\infty}^s(a, 2) + \Delta S_{N\to\infty}^l(2, b), & a < 2, b \geq 2; \\  
		\Delta S_{N\to\infty}^l(a, b), & a \geq 2, b \geq 2.
	\end{cases}
\end{alignat}
Terms are added sequentially until the next contribution to the full integral falls below a given relative tolerance; the default value for this tolerance used by \texttt{DarkHistory} is $10^{-10}$. 

Before listing the series expressions, we must first introduce some notation that will be relevant. The numbers and analytic functions defined below are all well-known, but are often defined with different normalizations or given different names. We explicitly define all relevant functions used here for clarity.

$B_n$ are the Bernoulli numbers, defined through the following exponential generating function:
\begin{alignat}{1}
    \frac{x}{e^x - 1} \equiv \sum_{n=0}^\infty \frac{B_n x^n}{n!},
    \label{eqn:bernoulli_numbers}
\end{alignat}
with $B_{0,1,2,\cdots} = 1, -1/2, 1/6, \cdots$. Note that $B_{2j+1} = 0$ for all integers $j> 0$. 

Next, we define the generalized exponential integrals
\begin{alignat}{1}
	E_n(x) \equiv \int_1^\infty \frac{e^{-xt}}{t^n} \, dt
\end{alignat}
and the closely related incomplete gamma function
\begin{alignat}{1}
    \Gamma(n, x) \equiv \int_x^\infty t^{n-1} e^{-t}\, dt.
\end{alignat}
The polylogarithm of order $m$, denoted $\text{Li}_m(z)$, is defined as
\begin{alignat}{1}
    \text{Li}_m(z) = \sum_{k=1}^\infty \frac{z^k}{k^m} \,.
\end{alignat}
Finally, we define $_2F_1(a,b;c;z)$, the Gaussian hypergeometric function, as
\begin{alignat}{1}
    _2F_1(a, b; c; z) &\equiv 1 + \frac{ab}{1! c} z + \frac{a(a+1) b(b+1)}{2!c(c+1)} z^2 + \cdots =\sum_{n=0}^\infty \frac{(a)_n (b)_n}{(c)_n} \frac{z^n}{n!},
\end{alignat}
where $(x)_n \equiv x(x+1)\cdots(x+n-1)$ is the Pochhammer symbol. This function only appears in the form $R(n, x) \equiv \Re[_2 F_1(1, n+1; n+2; x)]$, where $\Re$ denotes the real part; to avoid the slow evaluation of the \texttt{hyp2f1} function in NumPy, we use instead the following relation:
\begin{alignat}{2}
    R(n, x) &\equiv&& \,\, \Re[_2 F_1(1, n+1; n+2; x)] \nonumber \\ 
    &=&& - (n+1) x^{-(n+1)} \log \left(\left| 1 - x \right| \right) - \sum_{j=1}^n \frac{n+1}{j} x^{j-n-1} \,.
\end{alignat} 

The list of all of the series expressions that we use, including those already derived in~\cite{Zdziarski:2013gza}, are shown in Tables~\ref{tab:series_expressions_low} and~\ref{tab:series_expressions_high} for $x < 2$ and $x\geq 2$ respectively.

\afterpage{
    \renewcommand{\arraystretch}{2.2}

    \setlength{\tabcolsep}{15pt}

    \begin{landscape}
    \begin{table*}[t!]
    \centering
    \begin{tabular}{
            >{$\displaystyle}c<{$}   
            >{$\displaystyle}c<{$}}

    \toprule
    f(y) & P_{f} ,\, x < 2 \\
    \hline
    y^n, \, n \geq 1 
        & \sum_{k=0}^\infty \frac{B_k x^{k+n}}{k!(k+n)} \\
    y \log y
        & x \log x - x + \sum_{k=1}^\infty \frac{B_k x^{k+1}}{k!(k+1)} \left[\log x - \frac{1}{k+1} \right] \\
    y \log(y+a),\, a > -x
        & \sum_{k=0}^\infty \frac{B_k x^{k+1}}{k!(k+1)} \left[\log(x + a) + \frac{R(k, -x/a)}{k+1} - \frac{1}{k+1} \right] \\
    1
        & \log(1 - e^{-x}) \\
    \log y 
        & \frac{1}{2}\log^2 x + \sum_{k=1}^\infty \frac{B_k x^k}{k! k} \left[ \log x - \frac{1}{k} \right] \\
    \log(y+a), \, a > 0
        & \log x \log a - \text{Li}_2(-x/a) + \sum_{k=1}^\infty \frac{B_k x^k}{k! k} \left[\log(x + a) - \frac{x}{a(k+1)} R(k, -x/a) \right] \\
    \log(y+a), \, -x < a < 0
        & \log(-x/a) \log(x+a) - \text{Li}_2(1+x/a) + \sum_{k=1}^\infty \frac{B_k x^k}{k! k} \left[\log(x+a) - \frac{x}{a(k+1)} R(k, -x/a) \right] \\
    \frac{1}{y+a}, \, a > -x
        & \frac{1}{a} \log \left(\frac{x}{x+a}\right) + \sum_{k=1}^\infty \frac{B_k x^k}{k! k} \left[\frac{1}{a} - \frac{k x }{(n+1)a^2} R(k, -x/a) \right]\\
    y^{-n},\, n \geq 1 
        & \sum_{k=0}^{n-1} \frac{B_k}{k!} \frac{x^{k-n}}{k-n} + \frac{B_n}{n!} \log x + \sum_{k=1}^\infty \frac{B_{k+n}}{(k+n)!} \frac{x^k}{k} \vspace{0.3cm} \\
        
    \botrule
    \end{tabular}
    \caption{Series expressions for the relevant indefinite integrals of the form shown in Eq.~(\ref{eqn:planck_integral}). Here, $y$ is the integration variable, and $x$ denotes the integration limit of interest. These expressions are used for $x < 2$.}
    \label{tab:series_expressions_low}
    \end{table*}
    \end{landscape}
}
\afterpage{
    \renewcommand{\arraystretch}{2.2}

    \setlength{\tabcolsep}{15pt}

    \begin{landscape}
    \begin{table*}[t]
    \centering
    \begin{tabular}{
            >{$\displaystyle}c<{$}    
            >{$\displaystyle}c<{$}}

    \toprule
    f(y) & P_{f},\, x \geq 2 \\
    \hline
    (y+a)^n, \, \forall n \in \mathbb{Z},\, a > -x
        & \sum_{k=1}^\infty \frac{e^{ka} \Gamma \big(n+1, k(x+a) \big)}{k^{n+1}} = \sum_{k=1}^\infty \frac{e^{ka} E_{-n}\big( k(x+a) \big)}{(x+a)^{-k-1}} \\
    y \log(y+a), \, a > -x
        &  \sum_{k=1}^\infty \frac{e^{ka}}{k^2} \left[(1+kx) e^{-k(x+a)} \log(x+a) + (1+kx)E_1\big(k(x+a)\big) + E_2\big(k(x+a)\big) \right] \\
    \log(y+a), \, a > -x
        &  \sum_{k=1}^\infty \frac{e^{ka}}{k} \left[e^{-k(x+a)} \log(x + a) + E_1 \big( k(x+a) \big) \right] \vspace{0.3cm} \\
        
    \botrule
    \end{tabular}
    \caption{Series expressions for the relevant indefinite integrals of the form shown in Eq.~(\ref{eqn:planck_integral}). Here, $y$ is the integration variable, and $x$ denotes the integration limit of interest. These expressions are used for $x \geq 2$.}
    \label{tab:series_expressions_high}
    \end{table*}
    \end{landscape}
}

\subsubsection{Nonrelativistic Thomson Limit: Small Parameter Expansion}

The expression for the scattered photon spectrum in the Thomson limit, shown in Eq.~(\ref{eqn:thomson_scattered_phot_spec}), can be expanded in the small $\beta$ limit, to obtain
\begin{alignat}{1}
    \frac{dN_\gamma}{d\epsilon_1 dt_1} = \frac{3 \sigma_T k_B^2 T^2}{32 \pi^2 \hbar^3 c^2} \sum_{n=0}^\infty \sum_{j=1}^{2n}\frac{A_n \beta^{2n} x_1^3 P_{n,j}(x_1) e^{-jx_1}}{(1 - e^{-x_1})^{2n+1}} \, ,
    \label{eqn:thomson_scattered_phot_spec_expansion}
\end{alignat}
where $x_1 = \epsilon_1/T$, $A_n$ is a constant, and $P_{j,n}(x_1)$ is some rational or polynomial function in $x_1$. These quantities are as follows. For $n = 0$ i.e.\ $\mathcal{O}(\beta^0)$, 
\begin{alignat}{1}
    A_0 = \frac{32}{3}, \quad P_{0, 1}(x) = \frac{1}{x} \,.
\end{alignat}
For $n = 1$, i.e.\ $\mathcal{O}(\beta^2)$,
\begin{alignat}{1}
    A_1 = \frac{32}{9}, \quad P_{1,1}(x) = x - 4, \quad P_{1,2}(x) = x + 4 \, .
\end{alignat}
For $n = 2$, i.e.\ $\mathcal{O}(\beta^4)$, 
\begin{alignat}{1}
    A_2 &= \frac{16}{225}, \nonumber \\
    P_{2,1}(x) &= 7x^3 - 84x^2 + 260x - 200, \nonumber \\
    P_{2,2}(x) &= 77x^3 - 252x^2 - 260x + 600, \nonumber \\
    P_{2,3}(x) &= 77x^3 + 252x^2 - 260x - 600, \nonumber \\
    P_{2,4}(x) &= 7x^3 + 84x^2 + 260x + 200,
\end{alignat}
and finally for $n = 3$, i.e.\ $\mathcal{O}(\beta^6)$, 
\begin{alignat}{2}
    A_3 &=&& \,\, \frac{16}{4725}, \nonumber \\
    P_{3,1}(x) &=&& \,\, 11x^5 - 264x^4 + 2142x^3 - 7224x^2 + 9870x - 4200, \nonumber \\
    P_{3,2}(x) &=&& \,\, 3(209x^5 - 2200x^4 + 6426x^3 -2408x^2 - 9870x + 7000), \nonumber \\
    P_{3,3}(x) &=&& \,\, 2(1661x^5 - 5280x^4 - 10710x^3 +28896x^2 + 9870x - 21000), \nonumber \\
    P_{3,4}(x) &=&& \,\, 2(1661x^5 + 5280x^4 - 10710x^3 -28896x^2 + 9870x + 21000), \nonumber \\
    P_{3,5}(x) &=&& \,\, 3(209x^5 + 2200x^4 + 6426x^3 +2408x^2 - 9870x - 7000), \nonumber \\
    P_{3,6}(x) &=&& \,\, 11x^5 + 264x^4 + 2142x^3 +7224x^2 + 9870x + 4200.
\end{alignat}
Furthermore, when $x_1$ is small, it becomes numerically advantageous to expand Eq.~(\ref{eqn:thomson_scattered_phot_spec_expansion}) in $x_1$ as well, leaving a simple polynomial in $x_1$ and $\beta$, i.e.\
\begin{alignat}{1}
    \frac{dN_1}{d\epsilon_1 dt_1} = \frac{3 \sigma_T k_B^2 T^2}{32 \pi^2 \hbar^3 c^2} \sum_{n=0}^\infty \sum_{j=1}^\infty C_{n,j} \beta^{2n} x_1^j \, .
    \label{eqn:thomson_scattered_phot_spec_expansion2} 
\end{alignat}
The values of $C_{n,j}$ are shown in Table~\ref{tab:C_n_j}.

\afterpage{
\renewcommand{\arraystretch}{2}

\setlength{\tabcolsep}{7pt}

\begin{landscape}
\begin{table}
\centering
\begin{tabular}{c c c c c c c c}

\toprule
$C_{n,j}$ & $x_1$ & $x_1^2$ & $x_1^3$ & $x_1^5$ & $x_1^7$ & $x_1^9$ & $x_1^{11}$ \\
\hline
    $\beta^0$ & 32/3 & -16/3 & 8/9 & -2/135 & 1/2835 & -1/113400 & 1/4490640 \\
    $\beta^2$ & -64/9 & 0 & 32/27 & -4/45 & 8/1701 & -1/4860 & 1/124740 \\
    $\beta^4$ & -256/225 & 0 & 32/27 & -296/1125 & 1208/42525 & -64/30375 & 389/3118500 \\
    $\beta^6$ & -832/1575 & 0 & 32/27 & -1828/3375 & 31352/297675 & -10669/850500 & 10267/9355500 \vspace{0.2cm}\\
\botrule
\end{tabular}
\caption{List of coefficients $C_{n,j}$ for use in Eq.~(\ref{eqn:thomson_scattered_phot_spec_expansion2}).}
\label{tab:C_n_j}
\end{table}
\end{landscape}
}

Three checks can be performed to verify that this is indeed the correct expansion in $\beta$. First, taking $\beta \to 0$, the scattered photon spectrum simply becomes $dN_\gamma/(d\epsilon_1 \, dt_1) = n_{\text{BB}}(\epsilon_1, T) \sigma_T c$, which is exactly the expected result for Thomson scattering in the rest frame of the electron: all photons simply scatter elastically at a rate governed by the Thomson scattering cross section, thus remaining in a blackbody distribution. Second, a more non-trivial check is to integrate Eq.~(\ref{eqn:thomson_scattered_phot_spec_expansion}) with respect to $\epsilon_1$, giving the total Thomson scattering rate given in Eq.~(\ref{eqn:thomson_scattering_rate}). Since the scattering rate is independent of $\beta$, the $\mathcal{O}(\beta^0)$ term in the series should integrate to exactly $\sigma_T c N_\text{rad}$ where $N_\text{rad}$ is the number density of the blackbody photons, while the other higher order terms should integrate to exactly zero. This is indeed the case for the series expansion shown here. Lastly, one can check that Eq.~(\ref{eqn:thomson_scattered_phot_spec_expansion}) agrees with the energy loss expression Eq.~(\ref{eqn:thomson_energy_loss_rate}), by noting that
\begin{alignat}{1}
  \int \frac{dN_\gamma}{d\epsilon_1 \, dt_1} \epsilon_1 \, d\epsilon_1 = \sigma_T c u_\text{BB}(T) + \frac{4}{3} \sigma_T c \beta^2 \gamma^2 u_\text{BB}(T) \,,
\end{alignat}
where $u_\text{BB}(T)$ is the blackbody energy density with temperature $T$, i.e.\ the produced secondary photon spectrum must have the same energy as the upscattered CMB photons plus the energy lost from the scattering electron. This check has also been performed for the series expansions shown here.

For the scattered electron energy loss spectrum shown in Eq.~(\ref{eqn:electron_eng_loss_spec}), the small $\beta$ and $\xi$ expansion can be written as
\begin{alignat}{1}
    \frac{dN_e}{d\Delta \, dt} = \frac{3 \sigma_T k_B^2 T^2}{32 \pi^2 \hbar^3 c^2}  \sum_{n=0}^\infty \left[ \sum_{j=1}^{2n} \frac{A^{j+1} Q_{n,j}(e^{-A}) }{(1-e^{-A})^j \beta^{-2n}} + R_n(A) \right],
    \label{eqn:electron_engloss_spec_expansion}
\end{alignat}
where $Q_{n,j}(x)$ is a polynomial, $A \equiv \Delta/(2\beta T) = \xi/(2 \beta)$, and $R_n(A)$ is a sum of integrals of the form
\begin{alignat}{1}
    P_k(A) = A^{k+1} \int_A^\infty \frac{x^{-k} \,dx}{e^x - 1}\, .
\end{alignat}
These integrals can be evaluated using the same methods detailed in Appendix~\ref{app:ICS_integrals_series}. The list of polynomials $Q_{n,j}$ and of $R_n(A)$ is given below. All expressions not listed should be taken to be zero. For $n = 0$, 
\begin{alignat}{1}
    R_0(A) = \frac{176}{15}P_0 - \frac{64}{3} P_3 + \frac{128}{5} P_5 \,.  
\end{alignat}
For $n = 1$, 
\begin{alignat}{1}
    Q_{1,1}(x) &= - \frac{32}{3} x \,, \qquad Q_{1,2} = \frac{8}{3} x \,, \nonumber \\
    R_1(A) &= -\frac{1168}{105} P_0 + \frac{128}{3}P_3 - \frac{2176}{15} P_5 + \frac{1280}{7} P_7 \,. 
\end{alignat}
For $n = 2$, 
\begin{alignat}{1}
    Q_{2,1}(x) &= -\frac{512}{15}x \,, \qquad Q_{2,2}(x) = \frac{8}{5}x \,, \nonumber \\
    Q_{2,3}(x) &= -\frac{8}{15} x(1+x)\,, \nonumber \\
    Q_{2,4}(x) &= \frac{2}{15}(x + 4x^2 + x^3) \,, \nonumber \\
    R_2(A) &= -\frac{64}{3}P_3 + \frac{640}{3}P_5 - 768 P_7 + \frac{14336}{15} P_{9}\,. 
\end{alignat}
And finally for $n = 3$, 
\begin{alignat}{1}
    Q_{3,1}(x) &= -\frac{416}{3}x \,, \qquad Q_{3,2}(x) = \frac{1184}{105}x \,, \nonumber \\
    Q_{3,3}(x) &= -\frac{256}{315}(x + x^2)\,, \nonumber \\
    Q_{3,4}(x) &= -\frac{2}{63}(x + 4x^2 + x^3) \,, \nonumber \\
    Q_{3,5}(x) &= - \frac{4}{315}(x + 11x^2 + 11x^3 + x^4) \,, \nonumber \\
    Q_{3,6}(x) &= \frac{1}{315}(x + 26x^2 + 66x^3 + 26x^4 + x^5) \,, \nonumber \\
    R_3(A) &= -\frac{512}{3465}P_0 - \frac{1408}{15}P_5 + \frac{6912}{7}P_7 - \frac{161792}{45}P_9 + \frac{49152}{11} P_{11} \,.
\end{alignat}
These are all the terms necessary to work at order $\mathcal{O}(\beta^6)$ and $\mathcal{O}(\xi^6)$. As before, if $A$ becomes small, we should expand Eq.~(\ref{eqn:electron_engloss_spec_expansion}) as
\begin{alignat}{1}
    \frac{dN_e}{d\Delta \, dt} = \frac{3 \sigma_T k_B^2 T^2}{32 \pi^2 \hbar^3 c^2} \sum_{n=0}^\infty \left[\sum_{j=0}^\infty D_{n,j} \beta^{2n} A^j + R_n(A)\right],
    \label{eqn:electron_engloss_spec_expansion2}
\end{alignat}
with the values of $D_{n,j}$ shown in Table~\ref{tab:D_n_j}.
\afterpage{
    \begin{landscape}
    \renewcommand{\arraystretch}{2}

    \setlength{\tabcolsep}{7pt}

    \begin{table*}[t!]
    \centering
    \begin{tabular}{c c c c c c c c}

    \toprule
    \hline
    $D_{n,j}$ & $A$ & $A^2$ & $A^3$ & $A^5$ & $A^7$ & $A^9$ & $A^{11}$ \\
    \hline
        $\beta^2$ & -8 & 16/3 & -10/9 & 7/270 & -1/1260 & 11/453600 & -13/17962560\\
        $\beta^4$ & -164/5 & 256/15 & -134/45 & 161/2700 & -19/9450 & 359/4536000 & -289/89812800 \\
        $\beta^6$ & -40676/315 & 208/3 & -1312/105 & 4651/18900 & -416/59535 & 989/4536000 & -173/22453200 \vspace{0.2cm}\\
    \botrule
    \end{tabular}
    \caption{List of coefficients $D_{n,j}$ for use in Eq.~(\ref{eqn:electron_engloss_spec_expansion2}).}
    \label{tab:D_n_j}
    \end{table*}
    \end{landscape}
}
These expressions are complicated, but can be checked in a similar fashion as the scattered photon spectrum by integrating over $\Delta \, d\Delta$ to obtain the mean energy loss rate of electrons scattering of a blackbody spectrum, given exactly in Eq.~(\ref{eqn:thomson_energy_loss_rate}). Using the fact that
\begin{alignat}{1}
    \int_0^\infty d\Delta \,\Delta \, P_k(A) = \frac{4 \pi^4 \beta^2 T^2}{15(n+2)}\,, 
\end{alignat}
one can verify that integrating the $\mathcal{O}(\beta^6)$ expansion gives
\begin{alignat}{1}
    \frac{dE_e}{dt} = \frac{4}{3} \sigma_T c U_\text{rad} \beta^2 (1 + \beta^2 + \beta^4 + \beta^6) \,,
\end{alignat}
which is precisely the Taylor expansion of Eq.~(\ref{eqn:thomson_energy_loss_rate}) in powers of $\beta$. 

\section{Positronium Annihilation Spectra}
\label{app:positronium_annihilation_spec}

The spin-triplet $^3S_1$ state of positronium annihilates to three photons, producing a photon spectrum per annihilation given by \cite{Ore:1949te}
\begin{alignat}{1}
    \left. \frac{dN_\gamma}{dE_\gamma} \right|_{^3S_1} = \frac{6}{(\pi^2 - 9) m_e} \bigg\{ \frac{2-x}{x} + \frac{x(1-x)}{(2-x)^2} + 2 \log(1 - x) \left[\frac{1-x}{x^2} - \frac{(1 - x)^2}{(2 - x)^3}\right] \bigg\} \,,
\end{alignat}
where $x \equiv E_\gamma/m_e$. The kinematically allowed range is $0 \leq x \leq 1$. Assuming that the formation of positronium by low energy positrons populates all of the degenerate ground states equally, the averaged photon spectrum per annihilation is 
\begin{alignat}{1}
    \left. \frac{dN_\gamma}{dE_\gamma} \right|_\text{Ps} = \frac{1}{4} \delta(E_\gamma - m_e) + \frac{3}{4}  \left. \frac{dN_\gamma}{dE_\gamma} \right|_{^3S_1}.
\end{alignat}

\section{Cross Checks}
\label{app:cross_checks}

\subsection{Helium Deposition}
\label{app:helium_deposition}

\begin{figure}[t]
  \centering
  \includegraphics[scale=0.63]{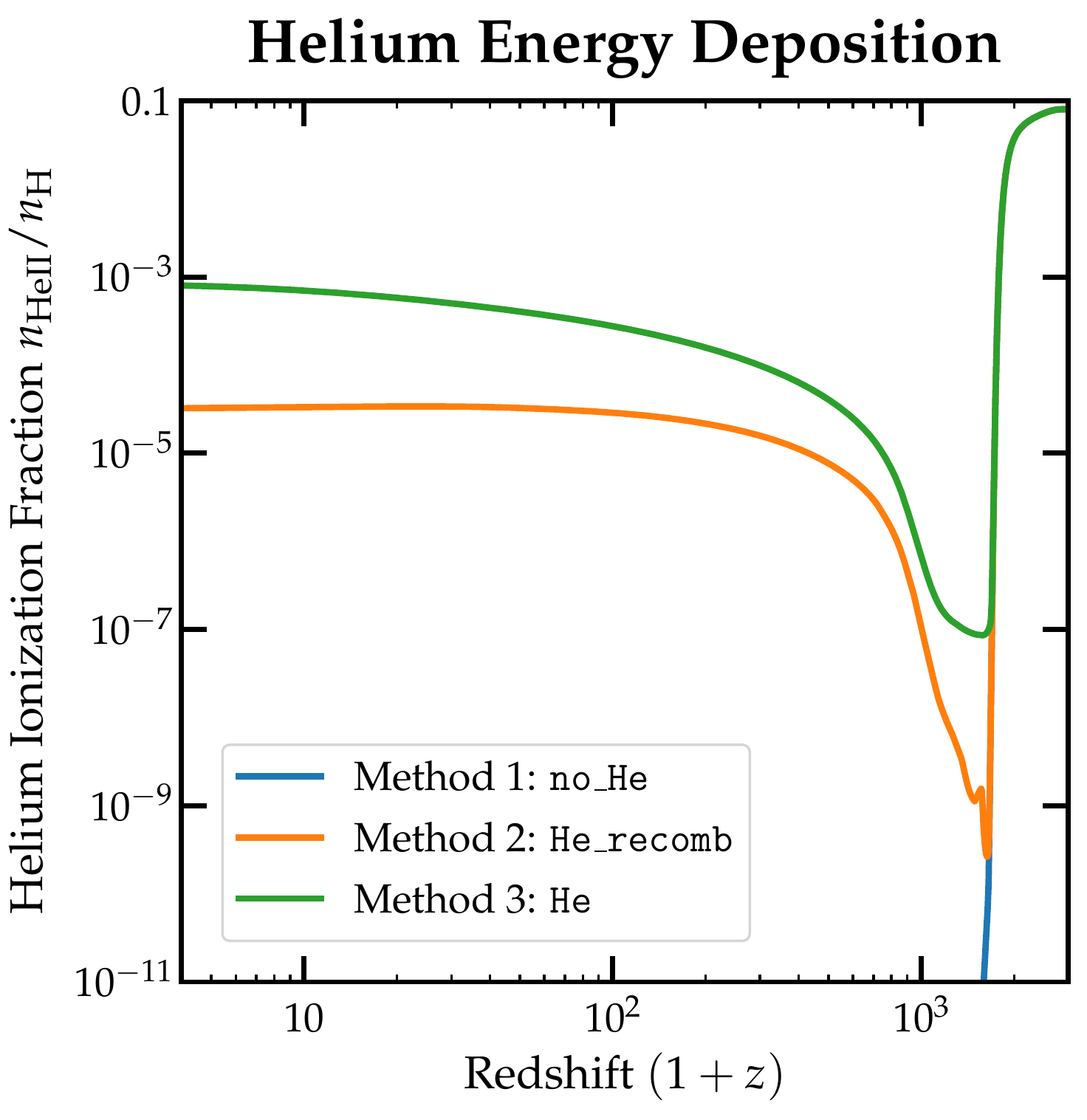}
  \caption{Helium ionization fraction with different helium energy deposition methods: (1) no tracking of the helium evolution (i.e.\ $x_\text{HeII}$ takes its baseline value) (blue) (2) all photoionized HeI atoms recombine, producing a photon that photoionizes hydrogen (orange), and (3) photoionized HeI atoms remain photoionized (green). The energy injection corresponds to \SI{100}{\mega \eV} DM decaying through $\chi \to \gamma \gamma$ with a lifetime of \SI{3e24}{s}.}
  \label{fig:He_f_method_xHeII}
\end{figure}

\begin{figure*}[t!]
  \centering
  \begin{tabular}{cc}
      \includegraphics[scale=0.44]{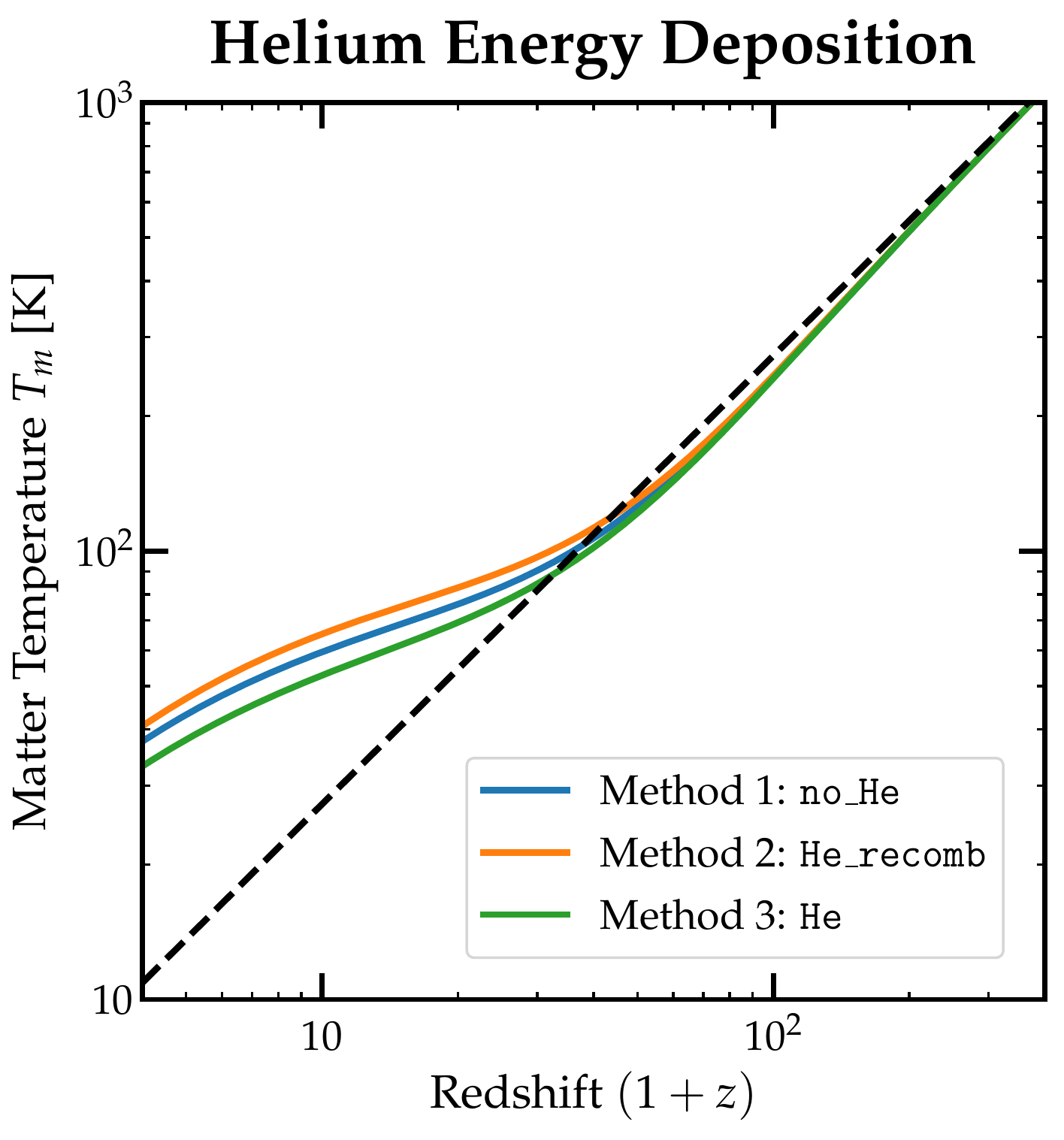} &
      \includegraphics[scale=0.44]{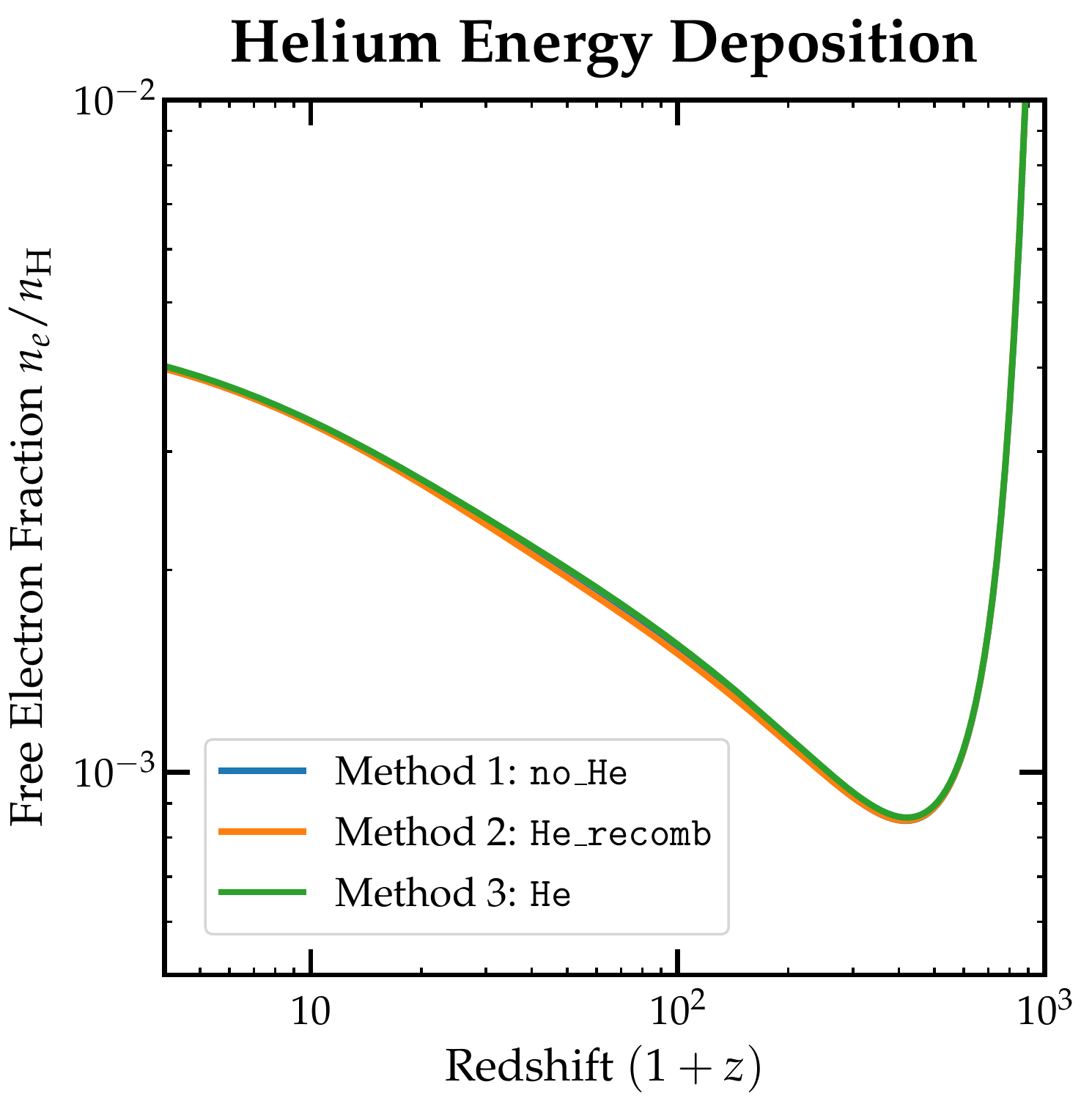} 
  \end{tabular}
  \caption{Matter temperature (left) and free electron fraction (right) evolution with different helium energy deposition methods: (1) no tracking of the helium evolution (i.e.\ $x_\text{HeII}$ takes its baseline value) (blue) (2) all photoionized HeI atoms recombine, producing a photon that photoionizes hydrogen (orange), and (3) photoionized HeI atoms remain photoionized (green). The CMB temperature is shown for reference (black, dashed). The energy injection corresponds to \SI{100}{\mega \eV} DM decaying through $\chi \to \gamma \gamma$ with a lifetime of \SI{3e24}{s}.}
\label{fig:He_f_method_histories}
\end{figure*}

In this section, we compare the various helium energy deposition methods discussed in Sec.~\ref{sec:photon_cooling}. We pick a dark matter candidate which decays to two photons with a lifetime of \SI{3e24}{\second} as an example, but the results are similar across different dark matter masses and energy injection rates. 

 Switching between methods can be done with the parameter \lstinline|compute_fs_method| passed to \texttt{evolve()}, with the following strings for each method:  (1) \lstinline|'no_He'|, (2) \lstinline|'He_recomb'| and (3) \lstinline|'He'|, e.g.
\begin{lstlisting}
  helium_method_alt = main.evolve(
      DM_process='decay', mDM=1e8, lifetime=3e24, 
      primary='phot_delta', start_rs=3000., backreaction=True, 
      helium_TLA=True, compute_fs_method='He_recomb'
  )
\end{lstlisting}

Fig.~\ref{fig:He_f_method_xHeII} shows the helium ionization fraction $x_\text{HeII}$ as a function of redshift for each of the different methods. In method (1), $x_\text{HeII}$ is simply the baseline helium ionization fraction, which is almost entirely neutral once helium recombination is complete. No energy is assigned to helium iondization at all. Method (2) has no contribution to helium ionization from photons, since every ionized helium atom is assumed to recombine, producing a photon that photoionizes hydrogen instead (i.e.\ setting $q^\gamma_\text{He} = 0$ in Eq.~(\ref{eqn:He_ion_dep_phot})). The helium ionization level therefore deviates from the baseline only from energy injection in the $\text{He}_\text{ion}$ channel from low-energy electrons. On the other hand, method (3) assumes that all helium atoms that get photoionized stay ionized, maximizing the amount of energy into $\text{He}_\text{ion}$ from photons (i.e.\ setting $q^\gamma_\text{He} = 1-q$ in Eq.~(\ref{eqn:He_ion_dep_phot})). This explains the higher $x_\text{HeII}$ obtained. 

Despite these differences in $x_\text{HeII}$, the evolution of $x_e$ remains almost identical, due to the fact that the total number of ionization events between both hydrogen and helium remains the same regardless of method used. This in turn ensures only a small difference in $T_m$ between the methods. The ionization and temperature histories for all three methods for the particular channel we have chosen are shown in Fig.~\ref{fig:He_f_method_histories}. Users may bracket the uncertainty in the treatment of helium with methods (2) and (3). 

\subsection{Coarsening}

\begin{figure}
  \centering
  \includegraphics[scale=0.63]{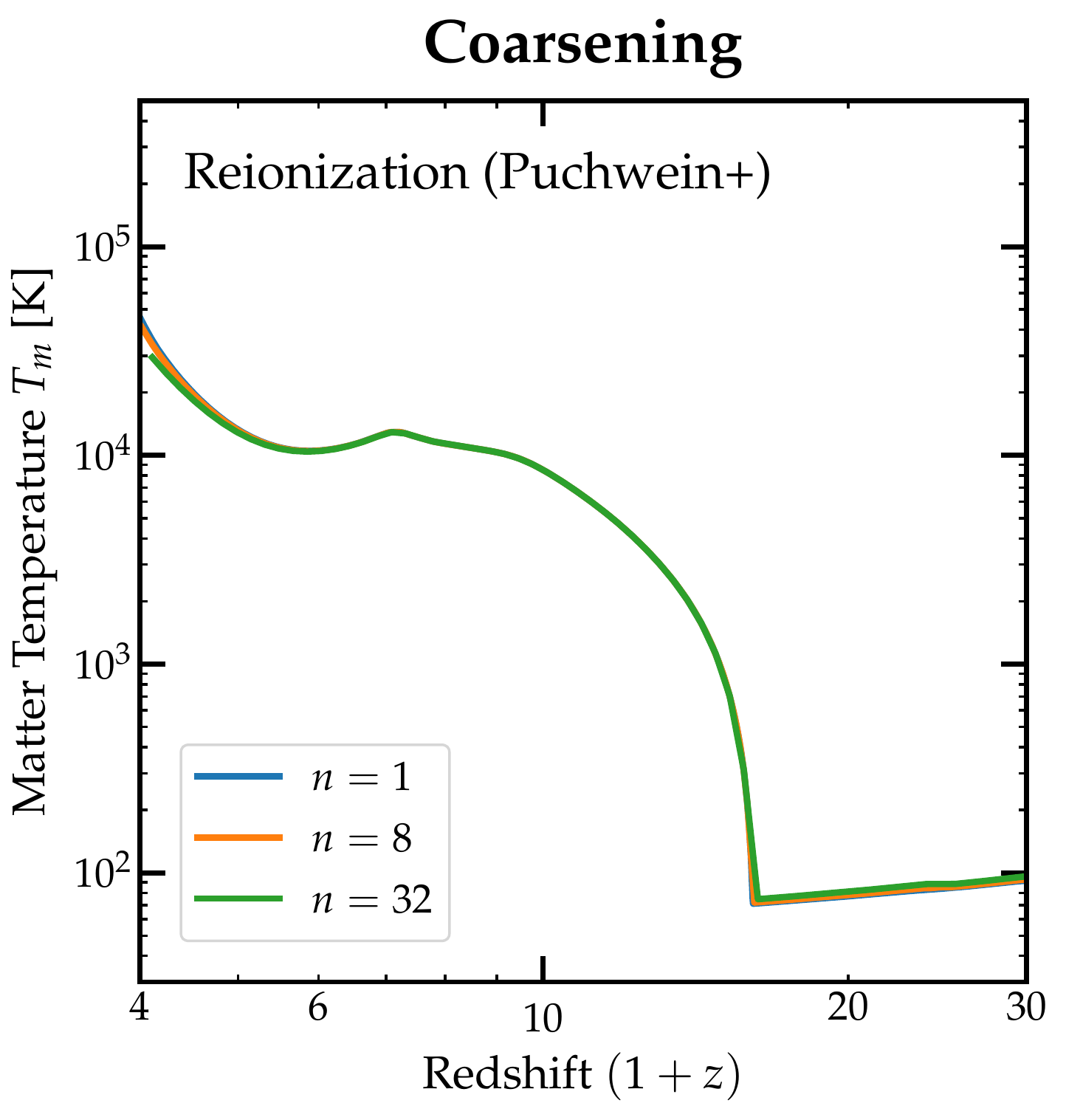}
  \caption{Matter temperature evolution with the default reionization model with no coarsening (blue), a coarsening factor of 8 (orange) and 32 (green).}
  \label{fig:coarsening}
\end{figure}

In the absence of reionization, a coarsening factor of up to 32 has been found to yield a small relative difference of between 5--10\% in the values of $f_c(z)$ across the full range of redshifts used in \texttt{DarkHistory}. With reionization, however, $T_m$ evolves more rapidly and attains larger values, and too much coarsening can lead to large absolute differences and somewhat larger relative differences in $T_m$, since we are averaging over the $T_m$ evolution over many redshift points. Fig.~\ref{fig:coarsening} shows the resultant temperature evolution as a function of redshift for the same model used in the previous section but with the default reionization model turned on, with coarsening factors of 1, 8 and 32. Once reionization starts, the difference in $T_m$ is $\sim 15\%$ for $n = 32$ compared to the uncoarsened result at $z \sim 4$, corresponding to an absolute error of $\sim \SI{5000}{\kelvin}$. Prior to reionization, the relative errors are slightly smaller at $\lesssim 10\%$. 

We therefore recommend using a coarsening factor of up to 32 if no reionization models are used, depending on the level of precision desired, and to use coarsening with care once reionization is included. We also emphasize that when using coarsening, it is best to check for convergence by comparing the result with less coarsening. 

\subsection{\texorpdfstring{$f_c(z)$ Contours}{f\_c(z) Contours}}

\begin{figure*}[t!]
  \centering
  \includegraphics[scale=0.18]{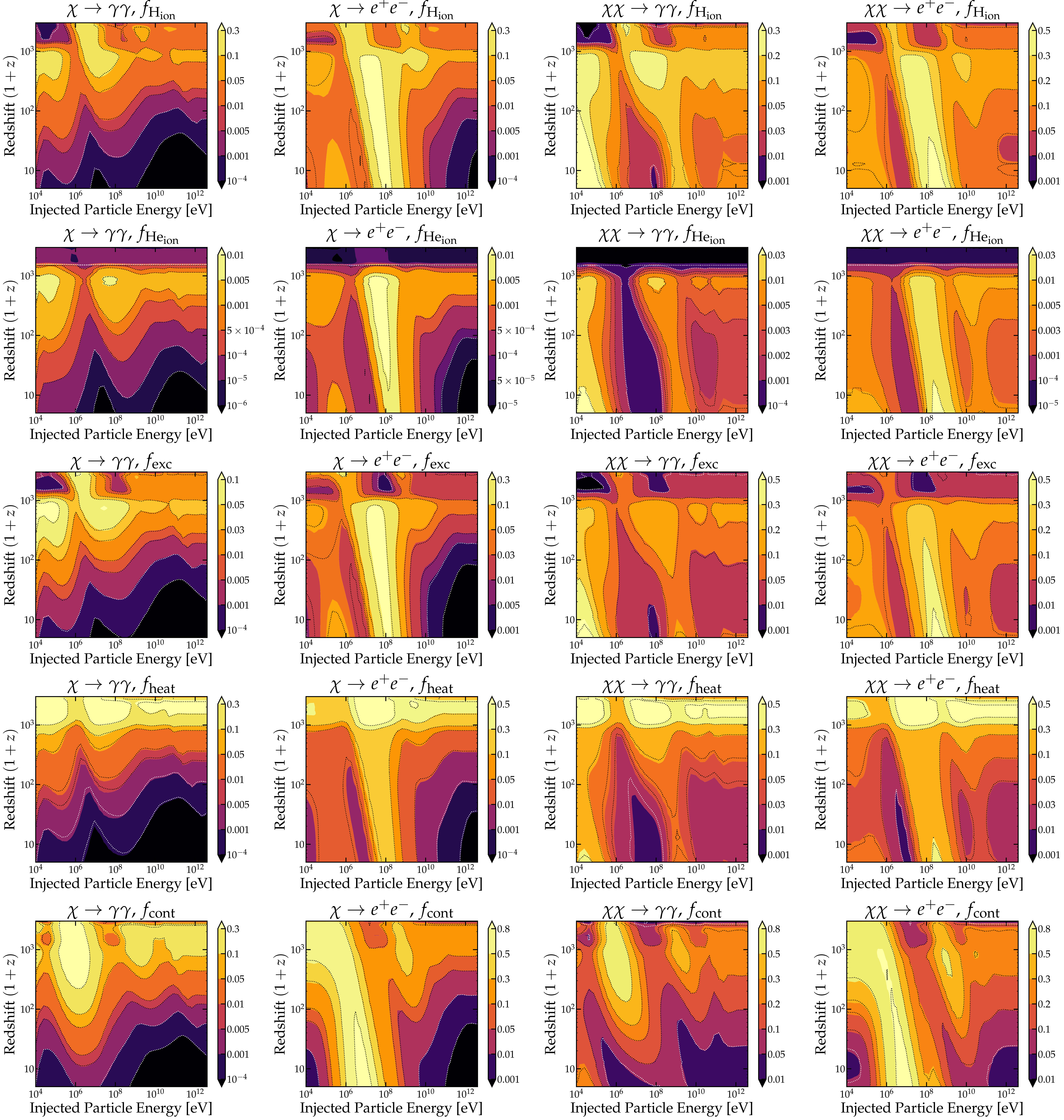}
  \caption{Computed $f_c(z)$ values without backreaction with \texttt{DarkHistory} for (from left to right) $\chi \to \gamma \gamma$ decays, $\chi \to e^+e^-$ decays, $\chi \chi \to \gamma \gamma$ annihilations and $\chi \chi \to e^+e^-$ annihilations (with no boost factor). The results from Refs.~\cite{Slatyer:2015kla,Liu:2016cnk} are shown for comparison (dashed lines). These contour plots agree with the previous results to within 10\% if all calculation methods are standardized between \texttt{DarkHistory} and Refs.~\cite{Slatyer:2015kla,Liu:2016cnk}, and represent an improved calculation of $f_c(z)$ neglecting backreaction.}
  \label{fig:f_contours_all}
\end{figure*}

Fig.~\ref{fig:f_contours_all} show the computed $f_c(z)$ contours within \texttt{DarkHistory} for all channels of interest without any backreaction.  The new $f_c(z)$ calculation by \texttt{DarkHistory} makes several small physics and numerical improvements over the previous calculation of these results~\cite{Slatyer:2015kla,Liu:2016cnk}, but still agree to within less than 10\% when methodologies (cosmological parameters, methods of interpolation etc.) are standardized between the code used in Ref.~\cite{Slatyer:2015kla} and \texttt{DarkHistory}. The new calculation also corrects a bug in earlier work in the treatment of prompt energy deposition from nonrelativistic and mildly relativistic injected electrons.  This accounts for the bulk of the visible differences in Fig.~\ref{fig:f_contours_all} between the current contours and those of Refs.~\cite{Slatyer:2015kla,Liu:2016cnk}, which are most pronounced for DM annihilation/decay to electrons and low injected particle energies.

\section{Table of Definitions}
\label{app:table}

Table~\ref{tab:variables} shows a list of variables and their definitions for reference. 

\clearpage

\begin{longtable}{C{0.11 \textwidth}  C{0.23 \textwidth} L{0.45 \textwidth}}
  \toprule
  \hline
  \textbf{Category} & \textbf{Symbol} & \textbf{Definition}\\
  \hline
  \hline
  General 

  &$y$ & log-redshift, $y \equiv \log(1+z)$. \\
  
  & $\Delta y$, $\Delta t$ & log-redshift step size and associated time step size. \\
  &$\mathbf{x}$ & Ionization levels: $\mathbf{x} \equiv (x_\text{HII}, x_\text{HeII}, x_\text{HeIII}) \
\equiv (n_\text{HII}/n_\text{H}, n_\text{HeII}/n_\text{H}, n_\text{HeIII}/n_\text{H})$ i.e.\ the fractional abundance of ionized hydrogen atoms, singly-ionized helium atoms and doubly-ionized helium atoms with respect to the number of hydrogen atoms (both neutral and ionized). \\

& $\zeta_i$ & $\text{arctanh} \left[ (2/\chi_i) \left(n_i/n_\text{H} - \chi_i/2\right) \right]$ where $i \in \{\text{HII}, \text{HeII}, \text{HeIII}\}$, convenient reparametrization of $\mathbf{x}$ introduced for numerical purposes.\\

& $T_m$ & Temperature of the IGM. \\

& $T^{(0)}_m$, $x^{(0)}_\text{HII}(z)$ & Baseline temperature and ionization histories, obtained from Eq.~(\ref{eqn:TLA}). \\

& $m_\chi$, $\tau$, $\langle \sigma v \rangle$ & Dark matter mass, lifetime, and velocity-averaged annihilation cross section. \\

& $\left(\frac{dE}{dV \,dt}\right)_\text{inj}$ & Energy injection rate per volume for exotic forms of energy injection, given for dark matter annihilation/decay in Eq.~(\ref{eqn:energy_injection}). \\
\hline

Spectra 

& $G(z)$ & Conversion factor between the rate of injected events per volume to the number of injected events per baryon within a log-redshift step, as defined in Eq.~(\ref{eqn:per_baryon_to_dVdt}). \\

&$\overline{\mathbf{N}}_\text{inj}^\alpha[E_{\alpha,i}']$				
& Spectrum containing number of particles of type $\alpha \in \{\gamma, e\}$ injected into energy bin $E'_{\alpha, i}$ per annihilation event.\\ 

&$\mathbf{N}_\text{inj}^\alpha[E_{\alpha,i}', y']$						
& Spectrum containing the number of particles per baryon in a log-redshift step of type $\alpha$ injected into energy bin $E'_{\alpha, i}$ at log-redshift $y'$, as defined in Eq.~(\ref{eqn:injected_discretized_spec}).\\

&$\overline{\mathbf{N}}^\gamma_\text{pos}[E_{\gamma,i}']$		
& Spectrum of photons produced from a single positronium annihilation event.\\

&$\mathbf{N}^\gamma_\text{new}[E'_{\gamma,i}, y']$					
& Sum of the spectra of primary injected photons, and secondary photons produced by the cooling of electrons, as defined in Eq.~(\ref{eqn:new_inj_photons}).
\\
&$\mathbf{N}_\text{prop}^\gamma [E_{\gamma,i}', y']$				
& Spectrum of propagating photons with energies greater than $13.6$ eV that do not photoionize or get otherwise deposited into low-energy photons.
\\
&$\mathbf{N}^\gamma [E_{\gamma,i}', y']$						
& $\mathbf{N}_\text{prop}^\gamma [E_{\gamma,i}', y'] + \mathbf{N}^\gamma_\text{new}[E_{\gamma,i}', y']$, as defined in Eq.~(\ref{eqn:N_prop_plus_new}). 
\\
&$\mathbf{N}^\alpha_\text{low} [E_{\alpha,i}, y]$
& Low-energy photons ($\alpha = \gamma$) or electrons ($\alpha = e$) at log-redshift $y$.
\\
\hline
Photon Cooling

& $\overline{\mathsf{P}}^\gamma[E'_{\gamma,i}, E_{\gamma,j}, y', \Delta y, \mathbf{x}]$ & Transfer function for propagating photons, which multiplies $\mathbf{N}^\gamma[E_{\gamma,i}', y']$ and produces $\mathbf{N}^\gamma_\text{prop} [E_{\gamma,j}, y' - \Delta y]$, as defined in Eq.~(\ref{eqn:discretized_prop_tf}).  \\

&$\overline{\mathsf{D}}^e[E'_{\gamma,i}, E_{e,j}, y', \Delta y, \mathbf{x}]$ & Low-energy electron deposition transfer function, which multiplies $\mathbf{N}^\gamma[E_{\gamma,i}', y']$ and produces $\mathbf{N}^e_\text{low} [E_{e,j}, y'-\Delta y]$, as defined in Eq.~(\ref{eqn:lowengelec_tf}).\\

&$\overline{\mathsf{D}}^\gamma[E'_{\gamma,i}, E_{\gamma,j}, y', \Delta y, \mathbf{x}]$						
& Low-energy photon deposition transfer function, which multiplies $\mathbf{N}^\gamma[E_{\gamma,i}', y']$ and produces $\mathbf{N}^\gamma_\text{low} [E_{\gamma,j}, y'-\Delta y]$, as defined in Eq.~(\ref{eqn:lowengphot_tf}).\\

&$\overline{\mathsf{D}}_c^\text{high}[E'_{\gamma,i}, y', \Delta y, \mathbf{x}]$		
& High-energy deposition transfer matrix, which multiplies $\mathbf{N}^\gamma[E_{\gamma,i}', y']$ and returns the total energy that greater than \SI{3}{\kilo\eV} electrons produce during the cooling process deposit into channel $c \in \{$`ion', `exc', `heat'$\}$, as defined in Eq.~(\ref{eqn:highengdep_tf}) in the next log-redshift step at $y' - \Delta y$.\\

&$\left( \overline{\mathsf{P}}^\gamma_{1/2} \right)^n$		
& Coarsened propagating photon transfer function with a coarsening factor of $n$, as defined in Eq.~(\ref{eqn:prop_tf_coarsening}), which multiplies $\mathbf{N}^\gamma[E_{\gamma,i}', y']$ and produces $\mathbf{N}^\gamma_\text{prop} [E_{\gamma,j}, y' - n \Delta y]$. 
\\
\hline
Electron Cooling

& $\overline{\mathsf{N}}[E'_{e,i}, E_{e,j}]$ 
& Spectrum of secondary electrons produced due to the cooling of a single injected electron with initial energy $E'_{e,i}$. \\

&$\overline{\mathbf{R}}_c[E'_{e,i}]$ 		
& High-energy deposition vector containing the total energy deposited into channel $c \in $\{`ion', `exc', `heat'\} by a single injected electron with kinetic energy $E'_{e,i}$, as defined in Eq.~(\ref{eqn:elec_cooling_dep_tf}). \\

&$\overline{\mathbf{R}}_\text{CMB}[E_{e,i}']$				
& Total initial energy of CMB photons that are upscattered via ICS due to the cooling of a single electron of energy $E_{e,i}'$. \\

&$\overline{\mathsf{T}}_\text{ICS,0}[E_{e,i}', E_{\gamma,j}]$
& Spectrum of photons produced with energy $E_{\gamma,j}$ due to the cooling of a single electron of energy $E_{e,i}'$, as defined in Eq.~(\ref{eqn:ics_photons}). \\

&$\overline{\mathsf{T}}_\text{ICS}[E_{e,i}', E_{\gamma,j}]$ & The same as $\overline{T}_\text{ICS,0}$, but with the pre-scattering spectrum of upscattered CMB photons subtracted out, as defined in Eq.~(\ref{eqn:elec_cooling_ics}).\\

&$\overline{\mathsf{T}}_e [E'_{e,i}, E_{e,j}]$			
& Low-energy electron spectrum produced due to the cooling of a single electron of energy $E_{e,i}'$, as defined in Eq.~(\ref{eqn:elec_cooling_lowengelec}).\\
\hline
Low-Energy Deposition 
& $f_c(z, \mathbf{x})$ 
& Ratio of deposited to injected energy, as a function of redshift $z$ and the ionization level $\mathbf{x}$, into channels $c \in \{$`H ion', `He ion', `exc', `heat', `cont$\}$, as defined in Eq.~(\ref{eqn:fz}).\\

& $\left(\frac{dE^\alpha}{dV \,dt}\right)_c$				
& Energy deposited per volume and time by low-energy photons ($\alpha = \gamma$) or electrons ($\alpha = e$) into channel $c$.\\

&$E_c^\text{high}[y]$								
& Total amount of high-energy deposition into channels $c \in $\{`ion', `exc', `heat'\} at log-redshift $y$. \\

\botrule
\caption{A list of the important definitions used in \texttt{DarkHistory}.  In this table, all spectra are discretized spectra as described in Sec.~\ref{sec:discretization}.  Spectra without overlines are normalized so that their entries contain number (per baryon) of particles produced in a redshift step.  A primed energy denotes the energy of an injected particle, and by energy we mean kinetic energy.  In this table, when we refer to electrons we will always mean electrons plus positrons.}
\label{tab:variables}
\end{longtable}

\begin{singlespace}
\nocite{apsrev41Control}
\bibliography{thesis}
\bibliographystyle{apsrev4-1}
\end{singlespace}

\end{document}